%% file: proceedings.tex
\documentclass[titlepage=true, parskip=half, DIV12]{scrbook}
\usepackage[noblocks, affil-it]{authblk}

\usepackage{arev}
\usepackage{helvet}
\usepackage{multicol}
\usepackage{amsmath, mathrsfs, amssymb, wasysym, gensymb, multirow}
\usepackage{graphicx, booktabs, tabularx}
\usepackage[numbers,sort&compress]{natbib}
\usepackage{chapterbib}
\usepackage{slashed}
\usepackage[utf8]{inputenc}
\usepackage[pdfa=true]{hyperref}
\usepackage{microtype}
\usepackage{subfigure}
\hypersetup{
bookmarks=false,
pdfborder={0 0 0},
pdftitle={Proceedings of the 2nd Workshop on Flavor Symmetries and Consequences in Accelerators and Cosmology (FLASY12)},
pdfauthor={},
pdfsubject={High energy physics},
pdfcreator={PDFLaTeX with hyperref},
pdfkeywords={}}
\makeatletter
\renewcommand\@author{\ifx\AB@affillist\AB@empty\AB@author\else
      \ifnum\value{affil}>\value{Maxaffil}\def\rlap##1{##1}
    \begin{flushleft}\AB@authlist\end{flushleft}\ \\[\affilsep]\setlength{\columnsep}{1cm}\begin{multicols}{2}\begin{flushleft}\AB@affillist\end{flushleft}\end{multicols}
    \else  \AB@authors\fi\fi}
\makeatother

\newcommand{\MHMoreRep}[2]{\ensuremath{\underline{\mbox{\textbf{#1}}} _{\mbox{\textbf{#2}}}}}
\newcommand{\MHRep}[1]{\ensuremath{\underline{\mbox{\textbf{#1}}}}}
\newcommand{\MHbraket}[1]{\ensuremath{\left<#1\right>}}

\def\SMvev#1{\left\langle #1\right\rangle}

\def\SNbeq{\begin{equation}}
\def\SNeeq{\end{equation}}

\def\SNbea {\begin{eqnarray}}
\def\SNeea {\end{eqnarray}}

\newcommand{\SNnn}{\nonumber}

\def\SNbsbar{{\bar B}_s^0}

\def\SNbs{{ B}_s^0}

\newcommand{\TWbea}{\begin{eqnarray}}
\newcommand{\TWeea}{\end{eqnarray}}
\newcommand{\TWrf}[1]{(\ref{#1})}

\begin{document}

\frontmatter
\title{\begin{flushleft}Proceedings of the 2nd Workshop on Flavor Symmetries and Consequences in Accelerators and Cosmology (FLASY12)\end{flushleft}}
\subtitle{\begin{flushleft}30 June 2012 -- 4 July 2012, Dortmund, Germany \end{flushleft}
\begin{flushright} {\footnotesize DO-TH 12/33} \end{flushright}
}
\date{}

\author[a]{\textbf{Editors:} \\ I. de Medeiros Varzielas}
\author[a]{C. Hambrock}
\author[a]{G. Hiller}
\author[a]{M. Jung}
\author[a]{P. Leser}
\author[a]{H. P\"as}
\author[a]{S. Schacht}
 \author[b,c]{\\\textbf{Authors:}\\ M. Aoki}
 \author[c]{J. Barry}
 \author[d]{G. Bhattacharyya}
 \author[e]{G. Blankenburg}
 \author[f]{A. J. Buras}
 \author[g,h]{L. Calibbi}
 \author[i]{L. Covi}
 \author[j]{D.~Das} 
 \author[k]{F. F. Deppisch}
 \author[j]{S. Descotes-Genon}
 \author[l]{G.-J. Ding}
 \author[c]{M. Duerr}
 \author[m]{T. Feldmann}
 \author[n, o]{M.~Freytsis} 
 \author[f]{J. Girrbach}
 \author[p]{F. Gonz\'alez Canales}
 \author[m]{F. Hartmann}
 \author[c]{J. Heeck}
 \author[q]{J. C. Helo}
 \author[r]{M. Hirsch}
 \author[s]{C.~M.~Ho}
 \author[c]{M. Holthausen}
 \author[a]{M. Jung} 
 \author[t]{A. Kadosh}
 \author[u,v]{J. F. Kamenik}
 \author[m]{W. Kilian}
 \author[w]{S. F. King}
 \author[x]{P.~Ko}
 \author[q]{S. Kovalenko}
 \author[y]{M. B. Krauss}
 \author[z]{M. Kreps}
 \author[b]{J. Kubo}
 \author[a]{P. Leser} 
 \author[n]{Z. Ligeti}
 \author[aa]{P. O. Ludl}
 \author[ab]{E.~Ma}
 \author[ac]{J. Matias}
 \author[w]{A. Merle}
 \author[ad]{A. Meroni}
 \author[ae]{A. Mondrag\'on}
 \author[ae]{M. Mondrag\'on}
 \author[r]{S. Morisi}
 \author[m]{S. Nandi}
 \author[x]{Y.~Omura} 
 \author[a]{H. P\"as}
 \author[r,af]{E. Peinado} 
 \author[ag]{F. Sala}
 \author[ae]{U. Salda\~na Salazar}
 \author[a]{S. Schacht}
 \author[c]{D. Schmidt}
 \author[m]{K.~Schnitter}
 \author[ah]{H. Ser\^{o}dio}
 \author[ah]{C. Sim\~oes}
 \author[ad]{M. Spinrath}
 \author[b]{H. Takano}
 \author[ai]{M. Tanimoto}
 \author[r]{M.~T{\'o}rtola}
 \author[n]{S. Turczyk} 
 \author[j]{A. Vicente}
 \author[ac]{J. Virto}
 \author[m]{Y.-M. Wang}
 \author[s]{T. Weiler}
 \author[ai]{K. Yamamoto}
 \author[aj]{M.~J.~S.~Yang}
 \author[x]{C. Yu}
\author[w,ak]{R. Zwicky}

\affil[a]{Fakult\"at f\"ur Physik, Technische Universit\"at Dortmund, 44221 Dortmund, Germany}
\affil[b]{Institute for Theoretical Physics, Kanazawa University, Kanazawa 920-1192, Japan}
\affil[c]{Max-Planck-Institut f\"ur Kernphysik, Saupfercheckweg 1, 69117 Heidelberg, Germany}
\affil[d]{Saha Institute of Nuclear Physics, 1/AF Bidhan Nagar, Kolkata 700064, India}
\affil[e]{Dipartimento di Fisica, Universit\`a di Roma Tre,  Via della Vasca Navale 84, I-00146 Roma, Italy}

\affil[f]{Institute for Advanced Study, Technical University Munich (TUM)}
\affil[g]{Service de Physique Th\'eorique, Universit\'e Libre de Bruxelles,
Bld du Triomphe, CP225, 1050 Brussels, Belgium}
\affil[h]{Max-Planck-Institut f\"ur Physik (Werner-Heisenberg-Institut), F\"ohringer Ring 6, D-80805 M\"unchen, Germany}
\affil[i]{Institut f\"ur theoretische Physik - Georg-August Universit\"at G\"ottingen}
\affil[j]{Laboratoire de Physique Th\'eorique, CNRS -- UMR 8627, Universit\'e Paris-Sud 11, F-91405 Orsay Cedex, France}

\affil[k]{Department of Physics and Astronomy, University College London, London WC1E 6BT, United Kingdom}
\affil[l]{Department of Modern Physics, University of Science and Technology of China, Hefei, Anhui 230026, China}
\affil[m]{Theoretische Physik 1 -- Universität Siegen, Germany}
\affil[n]{Ernest Orlando Lawrence Berkeley National Laboratory, University of California, Berkeley, CA 94720, USA}
\affil[o]{Berkeley Center for Theoretical Physics, Department of Physics, University of California, Berkeley, CA 94720, USA}

\affil[p]{Facultad de Ciencias de la Electr\'onica, Benem\'erita Universidad Aut\'onoma de Puebla, Apdo. Postal 157, 72570, Puebla, Pue., M\'exico.}
\affil[q]{Universidad T\'ecnica Federico Santa Mar\'\i a,  Centro-Cient\'\i fico-Tecnol\'{o}gico de Valpara\'\i so, Casilla 110-V, Valpara\'\i so,  Chile}
\affil[r]{AHEP Group, Instituto de F\'isica Corpuscular-C.S.I.C./Universidad de Valencia.  Edificio de Institutos de Paterna, Apartado 22085, E-46071 Valencia, Spain}
\affil[s]{Department of Physics \& Astronomy, Vanderbilt University, Nashville TN 37235}
\affil[t]{Centre for Theoretical Physics, University of Groningen}

\affil[u]{Institut Jo\v zef Stefan, Jamova 39, P. O. Box 3000, 1001 Ljubljana, Slovenia} 
\affil[v]{Department of Physics, University of Ljubljana, Jadranska 19, 1000 Ljubljana, Slovenia}
\affil[w]{School of Physics and Astronomy, University of Southampton, Southampton, SO17 1BJ, U.K.}
\affil[x]{School of Physics, KIAS, Seoul 130-722, Korea}
\affil[y]{Institut f{\"u}r Theoretische Physik und Astrophysik, Universit{\"a}t W{\"u}rzburg, Am Hubland, 97074 W{\"u}rzburg, Germany}

\affil[z]{University of Warwick, Coventry, United Kingdom}
\affil[aa]{University of Vienna, Faculty of Physics, Boltzmanngasse 5, A--1090 Vienna, Austria}
\affil[ab]{Department of Physics and Astronomy, University of California, Riverside, California 92506, USA}
\affil[ac]{Universitat Aut\`onoma de Barcelona, 08193 Bellaterra, Barcelona, Spain}
\affil[ad]{SISSA/ISAS and INFN, Via Bonomea 265, I-34136 Trieste, Italy}

\affil[ae]{Instituto de F\'{\i}sica, Universidad Nacional Aut\'onoma de M\'exico, Apdo. Postal 20-364, 01000, M\'exico D.F., M\'exico.}
\affil[af]{INFN, Laboratori Nazionali di Frascati, Via Enrico Fermi 40, I-00044 Frascati, Italy}
\affil[ag]{Scuola Normale Superiore and INFN, Pisa, Italy}
\affil[ah]{Departamento de F\'{\i}sica and CFTP, Instituto Superior T\'{e}cnico, Universidade T\'ecnica de Lisboa, Av. Rovisco Pais, 1049-001 Lisboa, Portugal}
\affil[ai]{Niigata University, Japan}
\affil[aj]{Department of Physics, University of Tokyo, Tokyo 113-0033, Japan}
\affil[ak]{School of Physics and Astronomy, University of Edinburgh, Edinburgh EH9 3JZ, Scotland}

\maketitle
\tableofcontents

\include{preface}
\include{participants}

\include{timetable}

\mainmatter
\include{Papers/jamesbarry}

\include{Papers/bhattacharyya}

\include{Papers/blankenburg}

\include{Papers/buras}

\include{Papers/calibbi}
\include{Papers/covi}
\include{Papers/debottamdas}
\include{Papers/deppisch}
\include{Papers/dinggj}
\include{Papers/feldmann}
\include{Papers/girrbach}
\include{Papers/hartmann}
\include{Papers/julianheeck}
\include{Papers/Helo}

\include{Papers/holthausen}
\include{Papers/martinjung}

\include{Papers/AvihayKadosh}
\include{Papers/jernejkamenik}
\include{Papers/king}
\include{Papers/krauss}

\include{Papers/michalkreps}
\include{Papers/kubo}
\include{Papers/ludl}
\include{Papers/Ma}

\include{Papers/Merle}
\include{Papers/meroni}

\include{Papers/mondragon}
\include{Papers/morisi}
\include{Papers/soumitra}
\include{Papers/omura}
\include{Papers/peinado}

\include{Papers/FilippoSala}
\include{Papers/schacht}
\include{Papers/DanielSchmidt}
\include{Papers/Hserodio}
\include{Papers/Simoes}

\include{Papers/spinrath}
\include{Papers/tanimoto}
\include{Papers/tortola}
\include{Papers/turczyk}
\include{Papers/vicente}
\include{Papers/JavierVirto}

\include{Papers/wang}
\include{Papers/tomweiler}
\include{Papers/yamamoto}
\include{Papers/yang}

\end{document}

%% file: preface.tex
\chapter{Preface}
FLASY12 is the second international workshop on flavor symmetries in a series first held in 2011 in Valencia.
With the exciting flavor and CP data from the LHC in heavy flavors
and the advent of a large $\theta_{13}$ in 2011 and many new theory ideas emerging ``it was just about time'' to host a workshop on quark and lepton flavor physics with our research groups in Dortmund.
The major incentive for the event was to bring together international experts on the flavor and beyond the Standard Model frontiers to discuss the status of the fields and further new theory and phenomenology driven avenues. These proceedings represent a snapshop of this enterprise as of summer/fall 2012.
We look forward to future FLASY workshops!

We are most happy to thank the members of our international advisory board,
Guido Altarelli, Andrzej Buras, Tom Kephart, Manfred Lindner, Ernest Ma, Stefano Morisi, Yossi Nir and Jose Valle for their valuable inputs, and J\"urgen Kroseberg for his impromptu talk
updating us on the LHC Higgs searches. We are indebted to our groups' secretary, Susanne Laurent, for efficiently handling paperwork and removing all kinds of administrative obstacles.
We would like to further thank the strategic Helmholtz alliance ``Physics at the Terascale'' for financial support, the local FLASY team of
Arnd Behring, Daniel Pidt, Christoph Rahmede, Dario Schalla, Peter Schuh, Henning Sedello and Danny van Dyk,
and, of course, our local co-organizers and co-editors, 
Ivo de Medeiros Varzielas (scientific secretary, almost everything), Christian Hambrock (graphics), Martin Jung (Indico),
Philipp Leser (scientific secretary, everything) and Stefan Schacht (Extraschicht tour). They made FLASY12, and these 
proceedings, happen.

Gudrun Hiller and Heinrich P\"as

{\centering
\begin{figure}
\includegraphics[width=\textwidth]{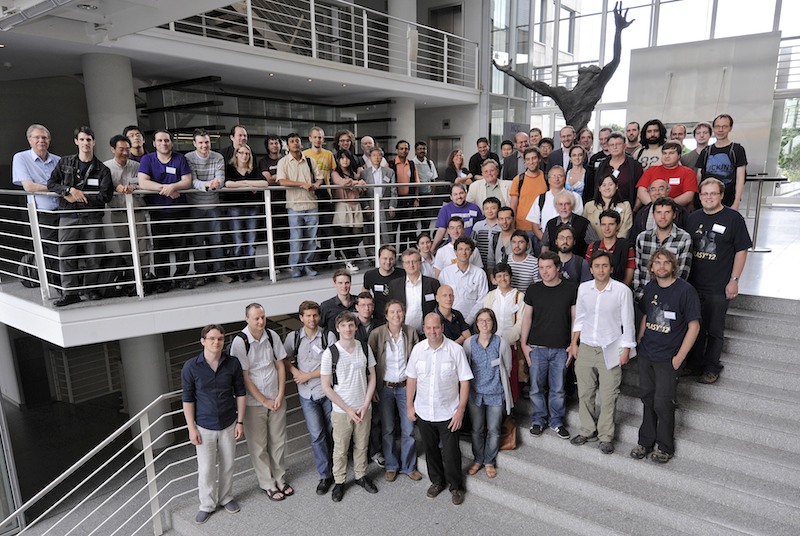} 
\end{figure}
}

%% file: participants.tex
\chapter{List of Participants}
{\footnotesize

\begin{tabularx}{1.1\textwidth}{XX|XX}
Barry, James & MPI Heidelberg 				                   &       Meroni, Aurora & SISSA, Trieste\\
Behring, Arnd & TU Dortmund 				                     &     Mondrag\'on, Myriam & Mexico U \\
Beneke, Martin & TU Munich 				                      &    Morisi, Stefano & IFIC, Valencia\\
Bhattacharyya, Gautam & SINP, Kolkata 				           &           Mueller, Michael & Bonn U\\    
Blankenburg, Gianluca & Roma Tre U 				              &            Nandi, Soumitra & Siegen U\\
Boucenna, Sofiane & IFIC, Valencia 				              &            Omura, Yuji& KIAS, Seoul\\
Branco, Gustavo & CFTP, Lisbon 				                  &        P\"as, Heinrich& TU Dortmund\\
Buras, Andrzej & TU Munich 				                      &    Peinado, Eduardo& IFIC, Valencia\\
Calibbi, Lorenzo & MPP Munich 				                   &       Perez, Gilad& Weizmann Inst., Rehovot\\
Chen, Mu-Chun & California U, Irvine 				            &     Pidt, Daniel& TU Dortmund\\         
Covi, Laura & Goettingen U 				                      &    Piepke, Andreas& Alabama U\\
Das, Debottam & LPT, Orsay 				                      &    Porod, Werner& Wuerzburg U\\
de Medeiros Varzielas, Ivo & TU Dortmund 				        &     Rahmede, Christoph& TU Dortmund\\             
Deppisch, Frank & U College, London 				             &            Ratz, Michael& TU Munich\\ 
Ding, Gui-Jun & Hefei U 				                         & Redi, Michele& CERN, Geneve\\
Faller, Sven & Siegen U 				                         & Sala, Filippo& SNS \& INFN, Pisa\\
Feldmann, Thorsten & Siegen U 				                   &     Schacht, Stefan& TU Dortmund\\  
Girrbach, Jennifer & TU Munich 				                  &     Schalla, Dario& TU Dortmund\\   
Hambrock, Christian& TU Dortmund				                 &    Schmidt, Daniel& MPI Heidelberg\\     
Hartmann, Florian& Siegen U				                      &    Schuh, Peter& TU Dortmund\\
Heeck, Julian& MPI Heidelberg				                    &      Sedello, Henning& TU Dortmund\\
Helo Herrera, Juan Carlos& Santa Maria U, Valparaiso	&			         Ser\^odio, Hugo& CFTP, Lisbon\\                 
Hiller, Gudrun& TU Dortmund				                      &    Shadmi, Yael& Technion, Haifa\\
Hirsch, Martin& IFIC, Valencia				                   &    Sim\~oes, Catarina& CFTP, Lisbon\\   
Hollenberg, Sebastian& TU Dortmund				               &   Spinrath, Martin& SISSA, Trieste\\        
Holthausen, Martin& MPI Heidelberg				               &   Tanimoto, Morimitsu& Niigata U\\        
Jung, Martin& TU Dortmund				                        &  T\'ortola, Mariam& IFIC, Valencia\\
Kadosh, Avihay& Groningen U				                      &  Turczyk, Sascha& LBNL, Berkeley\\  
Kamenik, Jernej& Jozef Stefan Inst., Ljubljana				   &     Valle, Jose& IFIC, Valencia\\                  
King, Stephen& Southampton U				                     &     van Dyk, Danny& TU Dortmund\\
Krauß, Martin& Wuerzburg U				                       & Vicente, Avelino& LPT, Orsay\\  
Kreps, Michal& Warwick U, Coventry				               &    Virto, Javier& UAB Barcelona\\       
Kroseberg, Jürgen & Bonn U				                       &   Wang, Yuming& Siegen U\\
Kubo, Jisuke & Kanazawa U 				                      &   Weiler, Thomas& Vanderbilt U, Nashville\\ 
Leser, Philipp& TU Dortmund 				                     &    Wingerter, Akin& LPSC, Grenoble\\ 
Lindner, Manfred& MPI Heidelberg 				                &   Yamamoto, Kei& Niigata U\\       
Ludl, Patrick& Vienna U				                          &Yang, Masaki& Tokyo U\\
Luhn, Christoph& Durham U				                        &  \\
Ma, Ernest& UC Riverside				                         & \\
Merle, Alexander& Southampton U				                  &    \\    
\end{tabularx}
}

%% file: timetable.tex
\chapter{Timetable}
\section{Saturday}
{
	\centering
        \begin{tabularx}{\textwidth}{>{\hsize=.5\hsize}X>{\hsize=1.7\hsize}X>{\hsize=.8\hsize}X}\toprule
            Time & Talk & Speaker\\
            \midrule
             14.05--14.25 & $A_4$, $\theta_{13}$, and $\Delta_{CP}$ & Ma, Ernest\\
             14.25--14.45 & Vacuum Alignment from Group Theory & Holthausen, Martin\\
             14.45--15.05 & Discrete flavor symmetries and geometrical $CP$ violation & Leser, Philipp\\
             15.05--15.25 & (Non-)Abelian discrete anomalies & Ratz, Michael\\
             15.25--15.45 & Models of Flavor Symmetries and the Stability of their Predictions & Chen, Mu-Chun\\ \midrule
             16.15--16.35 & Neutrino Masses and Mixing: Evidences and Implications & Valle, José\\
             16.35--16.55 & Embedding Models Generating Neutrino Masses via Higher-Dimensional Effective Operators in an $SU(5)$ GUT & Krauss, Martin\\
             16.55--17.15 & Relating Neutrino Mixing Angles to Neutrino Masses & Tanimoto, Morimitsu\\
             17.15--17.35 & A Minimal Model of Neutrino Flavour & Wingerter, Akin\\
             17.35--17.55 & Search for Neutrinoless Double-Beta Decay in 136Xe with EXO-200 & Piepke, Andreas\\
            \bottomrule
        \end{tabularx}
}\newpage
\section{Sunday}
{
	\centering
        \begin{tabularx}{\textwidth}{>{\hsize=.5\hsize}X>{\hsize=1.7\hsize}X>{\hsize=.8\hsize}X}\toprule
            Time & Talk & Speaker\\
            \midrule
             09.30--09.53 & Enhanced Higgs Mediated Lepton Flavour Violating Processes in the Supersymmetric Inverse Seesaw Model & Das, Debottam\\
             09.53--10.15 & New physics beyond flavour dogmas & Branco, Gustavo\\
             10.15--10.38 & Radiative Penguins in RS Models and $g-2$ & Beneke, Martin\\
             10.38--11.00 & Solving Flavor Problems in Composite Higgs Models & Redi, Michele\\ \midrule
             11.30--11.53 & Gravitino Dark Matter with colored NLSPs & Covi, Laura\\
             11.53--12.15 & Dark Matter and Flavor Symmetry & Morisi, Stefano\\
             12.15--12.38 & Phenomenological aspects of discrete Dark Matter & Boucenna, Sofiane\\
             12.38--13.00 & Multi-Component Dark Matter System with Non-Standard Annihilation Processes of Dark Matter & Kubo, Jisuke\\ \midrule
             14.30--14.53 & Family Symmetries in the Light of Large $\theta_{13}$ & Luhn, Christoph\\
             14.53--15.15 & TFH Mixing Patterns, Large $\theta_{13}$ and $\Delta(96)$ Family Symmetry & Ding, Gui-Jun\\
             15.15--15.38 & Spontaneous $CP$ Violation and Nonzero $\theta_{13}$ & Ser\^{o}dio, Hugo\\
             15.38--16.00 & Flavour Symmetry Models after Daya Bay and RENO & King, Steve\\ \midrule
             16.30--16.53 & Pathways to small neutrino masses: The one-loop story & Hirsch, Martin\\
             16.53--17.15 & Enhancing $\ell_i\to 3\ell_j$ with the $Z^0$ Penguin & Vicente, Avelino\\
             17.15--17.38 & Neutrino Masses and LFV from $U(3)^5\to U(2)^5$ in SUSY & Blanckenburg,\newline Gianluca\\
             17.38--18.00 & RS-$A_4$, $\theta_{13}$ and LFV & Kadosh, Avihay\\
            \bottomrule
        \end{tabularx}
}\newpage
\section{Monday}
{
	\centering
        \begin{tabularx}{\textwidth}{>{\hsize=.5\hsize}X>{\hsize=1.7\hsize}X>{\hsize=.8\hsize}X}\toprule
            Time & Talk & Speaker\\
            \midrule
             09.30--09.53 & keV Sterile Neutrinos as Dark Matter & Lindner, Manfred\\
             09.53--10.15 & Direct Detection of Leptophilic Dark Matter in a Model with Radiative Neutrino Masses & Schmidt, Daniel\\
             10.15--10.38 & Sterile Neutrinos for Warm Dark Matter and the Reactor Anomaly in Flavor Symmetry Models & Barry, James\\
             10.38--11.00 & 2012 Status of Neutrino Oscillations: $\theta_{13}$ and Beyond & T{\'o}rtola, Mariam\\ \midrule
             11.30--11.53 & Local Flavour Symmetries & Heeck, Julian\\
             11.53--12.15 & The quark NNI textures rising from $SU(5) \times Z_4$ symmetry & Sim\~oes, Catarina\\
             12.15--12.38 & Higgs mediators, $\theta_{13}$ and light familons & de Medeiros Varzielas, Ivo\\
             12.38--13.00 & Explicit and spontaneous breaking of $SU(3)$ into its finite subgroups & Merle, Alexander\\ \midrule
             14.30--14.53 & On the Messenger Sector of SUSY Flavour Models & Calibbi, Lorenzo\\
             14.53--15.15 & Flavour in SUSY Models with Extended Gauge Symmetries & Porod, Werner\\
             15.15--15.38 & Squark Flavour Implications from $B \to K^*\ell^+\ell^-$ & Schacht, Stefan\\
             15.38--16.00 & Flavoured Gauge Mediation & Shadmi, Yael\\ \midrule
             16.30--16.53 & Renormalization Group Effects on the Chromoelectric Dipole Moment via $CP$ violating Four Fermion Operators & Yang, Masaki\\
             16.53--17.15 & Novel Higgs signatures from $S_3$ flavor symmetry & Bhattacharyya,\newline Gautam\\
             17.15--17.38 & A SUSY $SU(5) \times T'$ Unified Model of Flavour with large $\theta_{13}$ & Meroni, Aurora\\
             17.38--18.00 & The ``Singlet Flavor'' and Particle Time Travel & Weiler, Thomas\\
            \bottomrule
        \end{tabularx}
}\newpage
\section{Tuesday}
{
	\centering
        \begin{tabularx}{\textwidth}{>{\hsize=.5\hsize}X>{\hsize=1.7\hsize}X>{\hsize=.8\hsize}X}\toprule
            Time & Talk & Speaker\\
            \midrule
             09.30--09.53 &Correlations in Minimal $U(2)^3$ models and an $SO(10)$ SUSY GUT model facing new data & Girrbach, Jennifer\\
             09.53--10.15 & Flavour Symmetry in Pati-Salam GUTs & Hartmann, Florian\\
             10.15--10.38 & Neutrinoless Double Beta Decay Mechanisms at LHC & Helo Herrera,\newline Juan Carlos\\
             10.38--11.00 & Probing the Right-Handed Neutrino Sector of Left-Right Symmetric Models at the LHC & Deppisch, Frank\\ \midrule
             11.30--11.55 & LHCb results now and tomorrow & Kreps, Michal\\
             11.55--12.18 & Quark Flavour Physics Facing Recent LHCb Data & Buras, Andrzej\\
             12.18--12.40 & (Buried) Non-degenerate squarks, from flavor precision to colliders & Perez, Gilad\\
              \midrule
             14.30--14.53 & The $S_3$ Flavour Symmetry: Quarks, Leptons and Higgs & Mondragón, Myriam\\
             14.53--15.15 & Reactor mixing angle and flavor symmetries & Peinado, Eduardo\\
             15.15--15.38 & The finite subgroups of $SU(3)$ & Ludl, Patrick\\
             15.38--16.00 & Two Approaches for Flavour Models with Large $\theta_{13}$ & Spinrath, Martin\\ \midrule
             16.30--16.48 & Semileptonic $B$ Meson Decays in QCD & Wang, Yu-Ming\\
             16.48--17.06 & Flavor Constraints from $b\to s$ Transitions & Virto, Javier\\
             17.06--17.24 & Squark Flavor Mixing and $CP$ Violation of Neutral $B$ Mesons at LHCb & Yamamoto, Kei\\
             17.24--17.42 & Constraining $CP$ Violation in Neutral Meson Mixing with Theory Input & Turczyk, Sascha\\
             17.42--18.00 & Extraction of the $CP$ phase $2\beta_s$ and the life time difference ($\Delta\Gamma_s$) from penguin free tree level $B_s$ decays & Nandi, Soumitra\\
            \bottomrule
        \end{tabularx}
}\newpage
\section{Wednesday}
{
	\centering
        \begin{tabularx}{\textwidth}{>{\hsize=.5\hsize}X>{\hsize=1.7\hsize}X>{\hsize=.8\hsize}X}\toprule
            Time & Talk & Speaker\\
            \midrule
             09.30--09.53 &Top Forward-Backward Asymmetry in Chiral $U(1)'$ Models & Omura, Yuji\\
             09.53--10.16 & Determing Weak Phases from $B\to J/\Psi P$ Decays & Jung, Martin\\
             10.16--10.39 & Flavour Physics from an approximate $U(2)^3$ Symmetry & Sala, Filippo\\
              \midrule
             11.10--11.33 & $\Delta A_{CP}$ and ``Old Physics'' in $D$ Decays & Feldmann, Thorsten\\
             11.33--11.55 & Implications of the $\Delta A_{CP}$ Measurement for New Physics & Kamenik, Jernej\\
             11.55--12.55 & Higgs results from the LHC & Kroseberg, Jürgen\\             
            \bottomrule
        \end{tabularx}
}\newpage

%% file: Papers/jamesbarry.tex

%
%
%
%
%
%

\chapter[Sterile neutrinos for warm dark matter and the reactor anomaly in flavor symmetry models (Barry)]{Sterile neutrinos for warm dark matter and the reactor anomaly in flavor symmetry models}
\vspace{-2em}
\paragraph{J. Barry}
\paragraph{Abstract}

Although existing theoretical models generally prefer extremely heavy right-handed neutrinos to generate light neutrino masses via the seesaw mechanism, there are several observed phenomena that point to the existence of both keV- and eV-scale sterile neutrinos. We present two $A_4$ flavor symmetry models that can accommodate light sterile neutrinos and remain compatible with neutrino data. Higher order seesaw terms and higher-dimensional operators are studied, and the phenomenological implications of light sterile neutrinos on neutrinoless double beta decay are discussed.

\section{Introduction: motivations for sterile neutrinos}

The existence of neutrino masses requires physics beyond the standard model, as does the presence of Dark Matter (DM). In the well-studied seesaw model \cite{Minkowski:1977sc,Yanagida:1979as,GellMann:1980vs,Mohapatra:1979ia} the light active neutrino masses come from integrating out heavy GUT-scale right-handed (sterile) neutrinos. On the other hand, keV-scale sterile neutrinos with small mixing to the active ones are good candidates for Warm Dark Matter (WDM) \cite{Boyarsky:2009ix,Kusenko:2009up,deVega:2011si}, which addresses some of the unsolved problems of the Cold Dark Matter model, i.e., the abundance of Dwarf satellite galaxies or cuspy DM halos. Furthermore, there are several experimental hints for the existence of eV-scale sterile neutrinos, such as the apparent neutrino flavor transitions at LSND and MiniBooNE and the ``reactor anomaly'' \cite{Mention:2011rk,Huber:2011wv}. Recent results from precision cosmology and Big Bang Nucleosynthesis \cite{Cyburt:2004yc,Izotov:2010ca,Hamann:2010bk,Giusarma:2011ex} also point to an additional relativistic degree of freedom, for which light steriles are a good candidate. Another phenomenological impact of light sterile neutrinos is in the amplitude for neutrinoless double beta decay ($0\nu\beta\beta$), which can show markedly different behaviour (see Section~\ref{JB_sect:0nubb}).

Reconciling these observed phenomena within a theoretical model is the aim of the present work. The required mass hierarchy in the sterile sector is achieved with the Froggatt-Nielsen mechanism~\cite{Froggatt:1978nt}\footnote{Other approaches include the ``split seesaw''~\cite{Kusenko:2010ik}, softly broken flavor symmetries~\cite{Shaposhnikov:2006nn,Lindner:2010wr} or extended seesaw models~\cite{Chun:1995js,Zhang:2011vh}.}, which suppresses both the Dirac and Majorana mass terms while leaving the leading order seesaw formula invariant~\cite{Merle:2011yv}; the neutrino flavor structure is provided by an $A_4$ flavor symmetry. Higher order seesaw terms are non-negligible in a model with eV-scale sterile neutrinos, and in our case lead to deviations from tribimaximal mixing (TBM). In addition, higher-dimensional operators can give values of $\theta_{13}$ compatible with recent fits~\cite{Tortola:2012te}, but their effect on the active-sterile mixing, defined as a ratio of mass scales, is negligible.

\section{Double beta decay with sterile neutrinos} \label{JB_sect:0nubb}

\begin{figure}[tb]
\centering 
\includegraphics[width=0.65\textwidth]{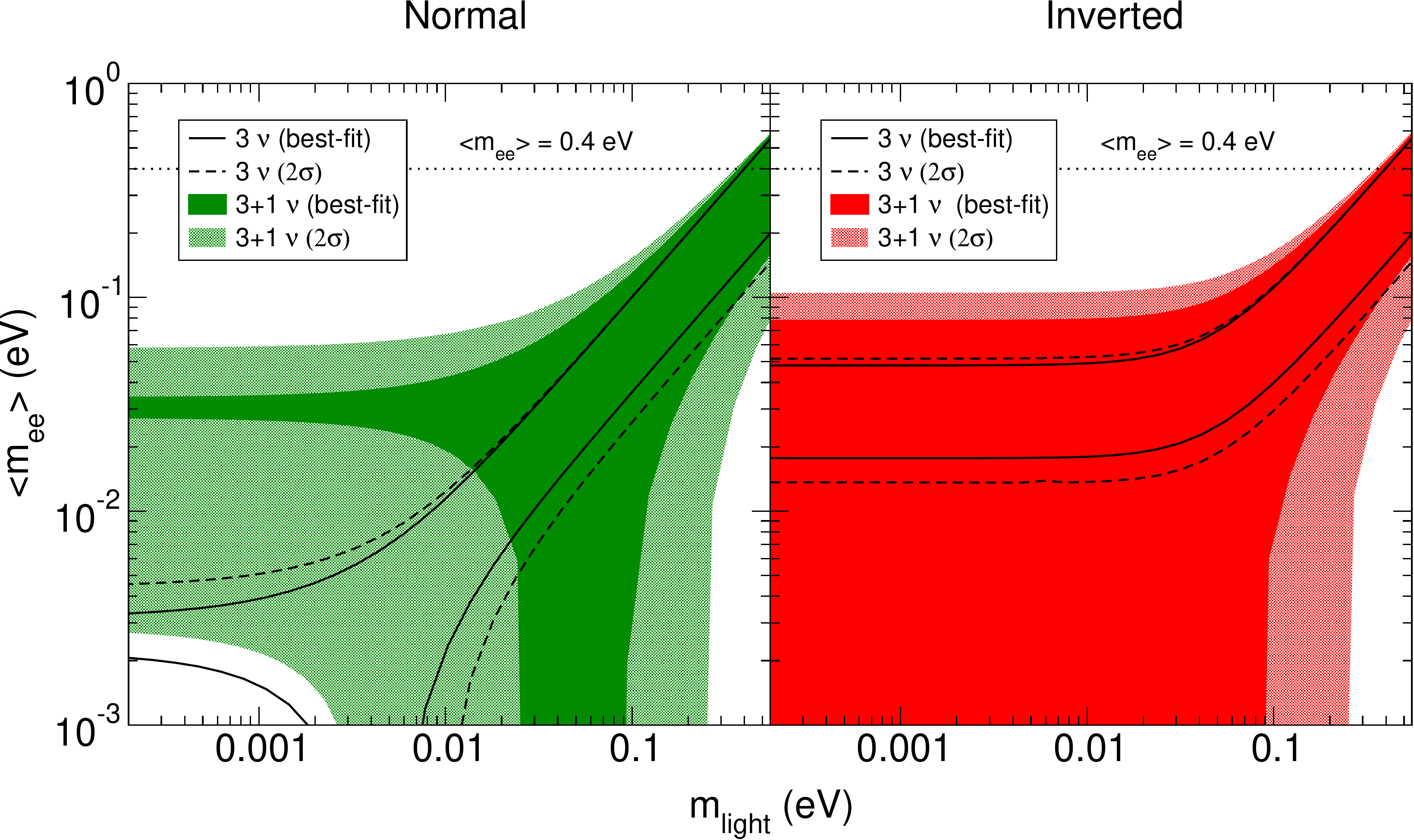}
\includegraphics[width=0.65\textwidth]{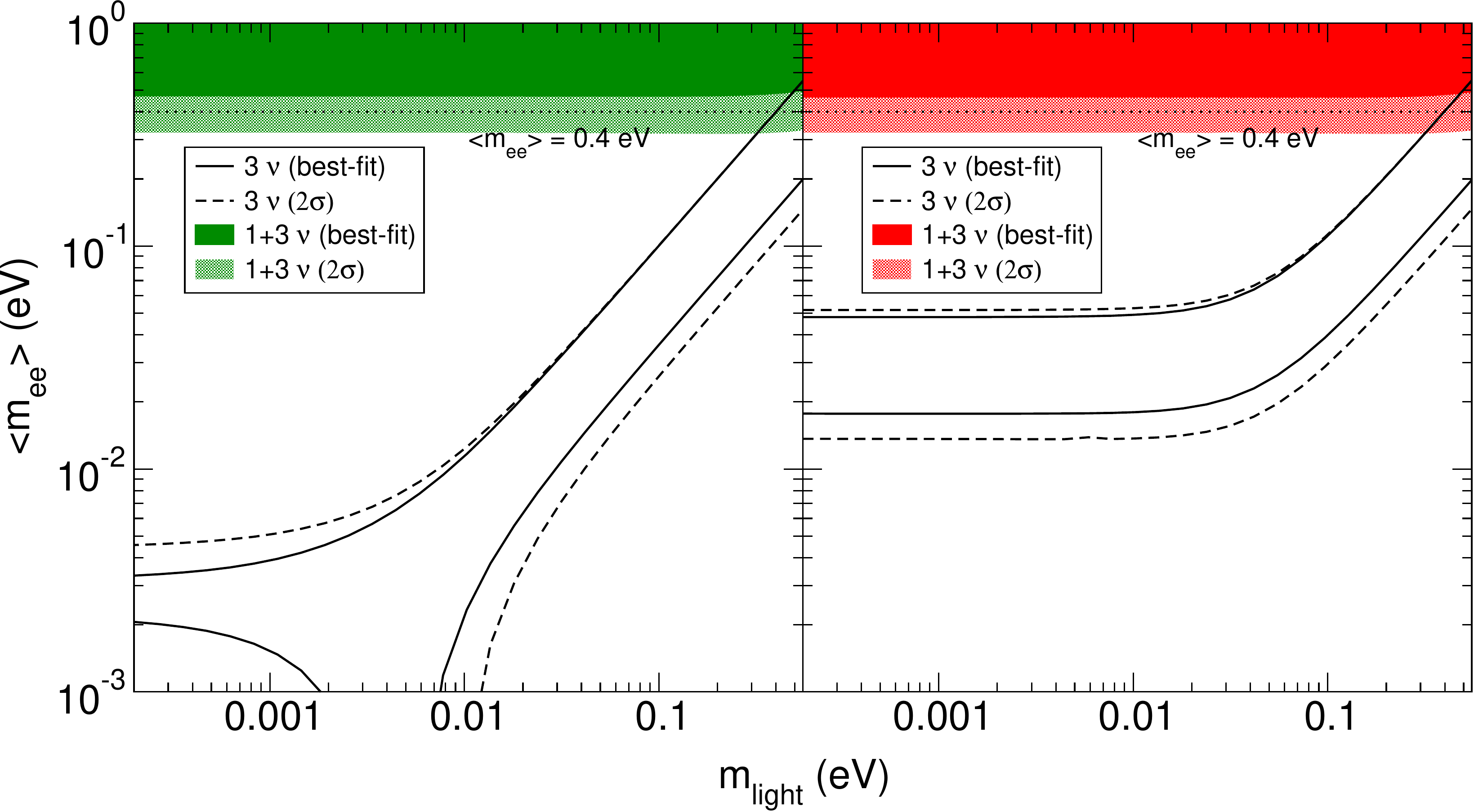}
\caption{The allowed ranges in the $\langle m_{ee}\rangle-m_{\rm light}$ parameter space, both in the standard three-neutrino picture (unshaded regions) and with one sterile neutrino (shaded regions), for the 3+1 (top) and 1+3 (bottom) cases. The current experimental upper bound for $\langle m_{ee}\rangle$ is indicated by the horizontal dotted line.}
\label{JB_fig:mee_4nu}
\end{figure}

The presence of light sterile neutrinos can have a significant effect on the amplitude for $0\nu\beta\beta$. In the case of one sterile neutrino, there are three additional mixing angles, $\theta_{i4}$, ($i$ = 1,2,3), as well as four additional phases (two Dirac and two Majorana)\footnote{See Refs.~\cite{Maltoni:2007zf,Barry:2011wb} for a detailed discussion of mixing parameterizations with sterile neutrinos.}. The effective neutrino mass in $0\nu\beta\beta$ is given by
\begin{equation}\label{JB_eq:meff4}
\langle m_{ee} \rangle_{4\nu} =
\left|c_{12}^2c_{13}^2c_{14}^2m_1+s_{12}^2c_{13}^2c_{14}^2m_2e^{i\alpha}
+s_{13}^2c_{14}^2m_3e^{i\beta}+s_{14}^2m_4e^{i\gamma}\right|,
\end{equation}
with $c_{ij} = \cos\theta_{ij}$, $s_{ij} = \sin\theta_{ij}$ and $\alpha$, $\beta$ and $\gamma$ the Majorana phases.
Figure~\ref{JB_fig:mee_4nu} displays the allowed range of the effective mass as a function of the lightest mass $m_{\rm light}$, using data from
Refs.~\cite{Schwetz:2011qt,Kopp:2011qd}. In the upper panel, with the sterile neutrino heavier than the active ones (the 3+1 case), the additional term proportional to $m_4$ in Eq.~\eqref{JB_eq:meff4}
means that the whole effective mass is larger, and can even vanish in the inverted hierarchy case, which is completely different to the standard picture. The lower panel of Fig.~\ref{JB_fig:mee_4nu} shows the effective mass in the 1+3 scenario when the sterile neutrino is lighter than the active ones; in this case there are three quasi-degenerate neutrinos with mass $\sqrt{\Delta m_{41}^2} \simeq 1.3$ eV, which sets the scale of the effective mass.
This situation is relatively disfavored by cosmological bounds on the sum of neutrino masses \cite{Hamann:2010bk}, since $\sum m_i \gtrsim 3\sqrt{\Delta m_{41}^2} \simeq 4$ eV. Parts of the allowed parameter space are also excluded from the latest limit on the effective mass, $\langle m_{ee} \rangle \gtrsim 0.4$~eV, as shown in Fig.~\ref{JB_fig:mee_4nu}. If taken at face value, this means that $\sqrt{1 -\sin^2 2 \theta_{12} \, \sin^2 \alpha/2} \lesssim  0.4 $, thus already putting strong constraints on the solar neutrino mixing angle and the Majorana phase $\alpha$. 

Things are different if the seesaw mechanism is at play. Neutrinos with mass below $|q| \simeq 100\,\,{\rm MeV}$ contribute to $0\nu\beta\beta$ via the effective mass, in analogy to Eq.~\eqref{JB_eq:meff4}, where one sums over all the light neutrino mass eigenstates. The direct contribution of right-handed neutrinos with masses much larger than $|q|$ is strongly suppressed by the inverse of their mass. Therefore, if all the right-handed neutrinos are light, i.e.~$M^2_i \ll q^2$, one obtains
\begin{eqnarray}\label{JB_eq:mee_6by6}
\langle m_{ee} \rangle = \left|\sum^3_{i=1} U^2_{ei} m_i +
\sum^3_{i=1}U^2_{e,3+i} M_{i}\right| = \left[M_\nu^{6 \times 6}\right]_{ee} = 0\; ,
\end{eqnarray}
showing that the effective mass cancels exactly, since the the ($1,1$) entry of the full $6\times6$ neutrino mass matrix is zero. However, if at least one of the right-handed neutrinos is very heavy one should decouple this heavy neutrino in computing the amplitude for $0\nu\beta\beta$, so that the cancellation condition is not valid anymore.

\section{Sterile neutrinos in flavor symmetry models}

We present two different sterile neutrino models based on the discrete group $A_4$, a popular choice for model builders (see the classification tables in Refs.~\cite{Barry:2010zk,Ding:2011gt}). In the effective approach an additional sterile neutrino singlet is added to an existing model, whereas in the seesaw approach we build a new model and discuss the effect of NLO seesaw terms. In both cases higher-dimensional operators are required to generate nonzero $\theta_{13}$.

\subsection{Effective $A_4$ model}

\begin{table}[ht]
\centering \caption{Particle assignments of the $A_4$ model, modified from Ref.~\cite{Altarelli:2005yp} to include a sterile
neutrino $\nu_s$.} 
\label{JB_table:afmodel} \vspace{8pt}
\begin{tabular}{c|cccccccc|c}
  \hline \hline  Field & $L$ & $e^c$ & $\mu^c$ & $\tau^c$ & $h_{u,d}$ & $\varphi$ & $\varphi'$ & $\xi$ & $\nu_s$ \\
\hline  $SU(2)_L$ & $2$ & $1$ & $1$ & $1$ & $2$ & $1$ & $1$ & $1$ & $1$ \\
 $A_4$ & $\underline{3}$ & $\underline{1}$ & $\underline{1}''$ & $\underline{1}'$ & $\underline{1}$ & $\underline{3}$ & $\underline{3}$ & $\underline{1}$ & $\underline{1}$ \\
 $Z_3$ & $\omega$ & $\omega^2$ & $\omega^2$ & $\omega^2$ & 1 & 1 & $\omega$ & $\omega$ & 1 \\
 $U(1)_{FN}$ & - & $F_e$ & $F_\mu$ & $F_\tau$ & - & - & - & - & $F_\nu$ \\[1mm] \hline \hline
\end{tabular}
\end{table}

The Altarelli-Feruglio model of Ref.~\cite{Altarelli:2005yp} can be modified by adding a sterile neutrino $\nu_s$; the 
relevant particle assignments are shown in Table~\ref{JB_table:afmodel}. The right-handed charged leptons and the sterile neutrino are charged under an additional $U(1)_{\rm FN}$ symmetry, which will be used to generate the correct mass hierarchy via the Froggatt-Nielsen mechanism; the $Z_3$ symmetry separates the neutrino and charged lepton sectors.

With the usual VEV alignments $\langle \varphi \rangle = (v,0,0)$ and $\langle \varphi' \rangle = (v',v',v')$, the $4\times 4$ neutrino mass matrix is 
\begin{equation}
 M^{4\times4}_\nu = \begin{pmatrix} a+\frac{2d}{3} & -\frac{d}{3} & -\frac{d}{3} & e \\ \cdot & \frac{2d}{3} & a-\frac{d}{3} & e \\ \cdot & \cdot & \frac{2d}{3} & e \\ \cdot & \cdot & \cdot & m_s \end{pmatrix},
\label{JB_eq:m4by4}
\end{equation}
where $a = 2x_a\frac{u v_u^2}{\Lambda^2}$, $d = 2x_d\frac{v' v_u^2}{\Lambda^2}$ and $e = \sqrt{2}x_e \frac{u v'v_u}{\Lambda^2}$ have dimensions of mass. The first three elements of the fourth row of $M^{4\times4}_\nu$ are identical because of the VEV alignment $\langle \varphi' \rangle = (v',v',v')$, which was necessary to generate TBM in the 3 neutrino case.

The mixing matrix is (to second order in the small ratio $e/m_S$)
\begin{equation}
   U \simeq \begin{pmatrix} \frac{2}{\sqrt{6}} & \frac{1}{\sqrt{3}} & 0 & 0 \\ -\frac{1}{\sqrt{6}} & \frac{1}{\sqrt{3}} & -\frac{1}{\sqrt{2}} & 0 \\ -\frac{1}{\sqrt{6}} & \frac{1}{\sqrt{3}} & \frac{1}{\sqrt{2}} & 0 \\ 0 & 0 & 0 & 1 \end{pmatrix} + \begin{pmatrix} 0 & 0 & 0 & \frac{e}{m_s} \\ 0 & 0 & 0 & \frac{e}{m_s} \\ 0 & 0 & 0 & \frac{e}{m_s} \\ 0 & -\frac{\sqrt{3}e}{m_s} & 0 & 0 \end{pmatrix} + \begin{pmatrix} 0 & -\frac{\sqrt{3} e^2}{2m_s^2} & 0 & 0 \\ 0 & -\frac{\sqrt{3} e^2}{2m_s^2} & 0 & 0 \\ 0 & -\frac{\sqrt{3} e^2}{2 m_s^2} & 0 & 0 \\ 0 & 0 & 0 & -\frac{3e^2}{2 m_s^2}\end{pmatrix} ,
\label{JB_eq:v4}
\end{equation}	
with the eigenvalues
\begin{equation}
 m_1 = a+d\, ,~~  m_2 = a - \frac{3e^2}{m_s}\, , 
~~  m_3 = -a+d\, ,  ~~m_4 = m_s + \frac{3e^2}{m_s}\, . 
\end{equation}

The reactor mixing angle retains its TBM value, $\sin \theta_{13}=0$, whereas
$\sin^2\!\theta_{12}$ and $\sin^2\!\theta_{23}$ receive small corrections:
\begin{equation}
 \sin^2\!\theta_{12} \simeq \frac{1}{3}\left[1 - 2\sin^2\!\theta_{14}\right]\, , \quad 
 \sin^2\!\theta_{23} \simeq \frac{1}{2}\left[1 + \sin^2\!\theta_{14}\right]\, .
\label{JB_eq:sinsq12_23}
\end{equation}
Non-zero $\theta_{13}$ can be generated by adding higher order terms, as discussed in Ref.~\cite{Altarelli:2005yx}; the active-sterile mixing is hardly affected.

\subsection{Seesaw model}

In this case we introduce three right-handed neutrinos with different FN charges, three $A_4$ triplets to generate the columns of the Dirac mass matrix, as well as three singlets for the right-handed sector. The particle assignments of the model are shown in Table~\ref{JB_table:afssmodel_a}, and the $A_4 \times Z_3 \times U(1)_{\rm FN}$ invariant Lagrangian is
\begin{align}
\nonumber -{\cal L}_{\rm Y} &=  \frac{y_e}{\Lambda}\lambda^3 \left(\varphi L h_d\right) e^c
+ \frac{y_\mu}{\Lambda} \lambda\left(\varphi L
h_d\right)' \mu^c + \frac{y_\mu}{\Lambda} \left(\varphi L h_d\right)''
\tau^c  \\
&+  \frac{y_1}{\Lambda}\lambda^{F_1}(\varphi L h_u) \nu^c_1 +
\frac{y_2}{\Lambda}\lambda^{F_2} (\varphi' L h_u)'' \nu^c_2 +
\frac{y_3}{\Lambda}\lambda^{F_3} (\varphi''L h_u) \nu^c_3
\label{JB_eq:seesaw_lag} \\
\nonumber
 &+ \frac{1}{2} \left[w_1 \lambda^{2F_1}\xi \nu^c_1 \nu^c_1 + w_2\lambda^{2F_2} \xi' \nu^c_2 \nu^c_2 + w_3 \lambda^{2F_3}\xi'' \nu^c_3 \nu^c_3
 \right] + {\rm h.c.}
\end{align}
\begin{table}[tp]
\centering \caption{Particle assignments of the $A_4$ type I seesaw
 model, with three right-handed sterile neutrinos.}
\label{JB_table:afssmodel_a} \vspace{8pt}
\begin{tabular}{c|ccccc|ccccccc|ccc}
\hline \hline  Field & $L$ & $e^c$ & $\mu^c$ & $\tau^c$ & $h_{u,d}$ & $\varphi$ & $\varphi'$ & $\varphi''$ & $\xi$ & $\xi'$ & $\xi''$ & $\Theta$ & $\nu^c_{1}$ & $\nu^c_{2}$ & $\nu^c_{3}$ \\
\hline $SU(2)_L$ & $2$ & $1$ & $1$ & $1$ & $2$ & $1$ & $1$ & $1$ & $1$ & $1$ & $1$ & $1$ & $1$ & $1$ & $1$ \\
$A_4$ & $\underline{3}$ & $\underline{1}$ & $\underline{1}''$ & $\underline{1}'$ & $\underline{1}$ & $\underline{3}$ & $\underline{3}$ & $\underline{3}$ & $\underline{1}$ & $\underline{1}'$ & $\underline{1}$ & $\underline{1}$ & $\underline{1}$ & $\underline{1}'$ & $\underline{1}$  \\
$Z_3$ & $\omega$ & $\omega^2$ & $\omega^2$ & $\omega^2$ & $1$ & $1$ & $\omega$ & $\omega^2$ & $\omega^2$ & $\omega$ & $1$ & $1$ &  $\omega^2$ & $\omega$ & $1$  \\
$U(1)_{\rm FN}$ & - & $F_e$ & $F_\mu$ & $F_\tau$ & - & - & - & - & - & - & - & $-1$ & $F_1$ & $F_2$ & $F_3$ \\[1mm] \hline \hline
\end{tabular}
\end{table}

In order to have a keV sterile neutrino as WDM, we choose $F_1=9$, so that $M_1 \simeq 1$~keV and $\theta_1^2\simeq10^{-8}$, which effectively decouples $\nu^c_1$ from the seesaw mechanism. The remaining $5\times 5$ mass matrix can be diagonalized using the formalism in Refs.~\cite{Schechter:1981cv,Grimus:2000vj,Hettmansperger:2011bt}, taking care to include higher order corrections, proportional to $M_D M_R^{-1}$. Note that the active-sterile mixing is defined as
\begin{equation}
 \theta_i^2 \equiv \sum_{\alpha = e,\mu,\tau} \left|\frac{[M_D
V^*_R]_{\alpha i}}{M_{i}}\right|^2\; ,
\label{JB_eq:as_mixing2}
\end{equation}
in seesaw models, which is just the ratio of two scales, $M_D$ and $M_R$. In the normal ordering case, for example, the VEV alignments $\langle \varphi' \rangle = (v',v',v')$ and $\langle \varphi''
\rangle = (0,v'',-v'')$ give TBM at leading order.
\begin{table}[tbp]
\centering \caption{Summary of the different scenarios discussed in the $A_4$ seesaw model. In each case the WDM sterile neutrino has a mass $M_{1} = {\cal O}({\rm keV})$, and the corresponding active neutrino is approximately massless. 
$\Delta m_{\rm S}^2$ and $\Delta m_{\rm A}^2$ are the solar and atmospheric mass squared differences, respectively.} \label{JB_table:seesaw_summary}
\vspace{8pt}\begin{footnotesize}
\begin{tabular*}{0.98\textwidth}{@{\extracolsep{\fill}}c|cccccc|c}
\hline \hline \multirow{2}{*}{} & \multirow{2}{*}{$F_1$,
$F_2$, $F_3$} & \multirow{2}{*}{Mass spectrum} &
\multirow{2}{*}{$|U_{\alpha 4}|$} & \multirow{2}{*}{$|U_{\alpha
5}|$} & \multicolumn{2}{c|}{$\langle m_{ee} \rangle$} & \multirow{2}{*}{Phenomenology}
\\ & & & & & NO & IO & \\ \hline I & $9$, $10$, $10$ & $M_{2,3} =
{\cal O}({\rm eV})$ & ${\cal O}(0.1)$ & ${\cal O}(0.1)$ & 0 & 0 &
$3+2$ mixing
\\[1mm] \hline \multirow{2}{*}{IIA} &  \multirow{2}{*}{$9$, $10$, $0$} & $M_{2}
= {\cal O}({\rm eV}) $ &  \multirow{2}{*}{${\cal O}(0.1)$} &
\multirow{2}{*}{${\cal O}(10^{-11})$} & \multirow{2}{*}{0} &
\multirow{2}{*}{$\dfrac{2\sqrt{\Delta m_{\rm A}^2}}{3}$ 
}  & \multirow{5}{*}{$3+1$
 mixing}\\[1mm] && $M_{3} = {\cal O}(10^{11}\,{\rm
GeV})$ &&&& \\[1mm] \cline{1-7} \multirow{2}{*}{IIB} &
\multirow{2}{*}{$9$, $0$, $10$} & $M_{2} = {\cal
O}(10^{11}\,{\rm GeV})$ & \multirow{2}{*}{${\cal O}(10^{-11})$} &
\multirow{2}{*}{${\cal O}(0.1)$} & \multirow{2}{*}{$\dfrac{\sqrt{\Delta m_{\rm S}^2}}{3}$
} &
\multirow{2}{*}{$\dfrac{\sqrt{\Delta m_{\rm A}^2}}{3}$
} &  \\[1mm] && $M_{3} = {\cal O}({\rm eV})$ &&&&  \\[1mm] \hline III & $9$, $5$, $5$ & $M_{2,3} = {\cal O}(10\,{\rm GeV})$ & ${\cal O}(10^{-6})$ & ${\cal O}(10^{-6})$ & $\dfrac{\sqrt{\Delta m_{\rm S}^2}}{3}$
& $\sqrt{\Delta m_{\rm A}^2}$
& Leptogenesis\\[3mm]
\hline \hline
\end{tabular*}
\end{footnotesize}
\end{table}

\begin{figure}[tbp]
\centering \vspace{-0.cm}
\includegraphics[width=0.5\textwidth]{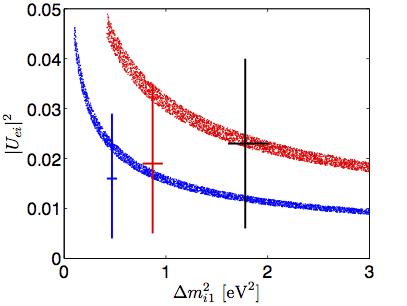}
\vspace{-0.cm}
\caption{The allowed ranges of $|U_{e4}|^2-\Delta m^2_{41}$ (blue) and $|U_{e5}|^2-\Delta m^2_{51}$ (red) in the inverted ordering, requiring that the oscillation parameters lie in their currently allowed 2$\sigma$ ranges. The blue and red vertical and horizontal error bars indicate the allowed $2\sigma$ range for the $3+2$ mass and mixing parameters from Ref.~\cite{Kopp:2011qd}, their intersection is the best-fit point. The black errors bars are for the $3+1$ case from
Ref.~\cite{Kopp:2011qd}. \label{JB_fig:mass_mix}}
\end{figure}

Different phenomenological scenarios are possible, depending on the FN charges $F_2$ and $F_3$.  Table~\ref{JB_table:seesaw_summary} summarizes the different cases studied. In scenarios I and II, higher order seesaw terms give deviations from TBM, but NLO operators are required to give nonzero $\theta_{13}$. Figure~\ref{JB_fig:mass_mix} compares the model predictions to data from the global fit in Ref.~\cite{Kopp:2011qd}, for the inverted ordering. Here it is possible obtain mass and mixing parameters compatible with the data, whereas in the normal ordering case the active-sterile mixing is too small.

\section{Conclusion}

Light sterile neutrinos exhibit distinct phenomenological signatures, both in terrestrial experiments and in cosmology. From the theoretical point of view, one requires a mechanism to suppress the sterile neutrino mass, while keeping the light neutrinos at the sub-eV scale. Two different $A_4$ models for sterile neutrinos have been discussed, and in both cases it is possible to accommodate light sterile neutrinos by assigning the appropriate FN charges. Higher order seesaw terms result in deviations from TBM. Nonzero $\theta_{13}$ can be generated from NLO operators, suppressed by additional powers of the cutoff scale.

\section*{Acknowledgments}

We thank W. Rodejohann and H. Zhang for fruitful collaboration. This work was supported by the ERC under the Starting Grant MANITOP and by the Deutsche Forschungsgemeinschaft in the Transregio 27 ``Neutrinos and beyond -- weakly interacting particles in physics, astrophysics and cosmology''.


\bibliography{jamesbarry}
\bibliographystyle{apsrev4-1}

%% file: Papers/bhattacharyya.tex

\chapter[Exotic Higgs phenomenology from $\mathsf{S}_3$ flavor symmetry (\textit{Bhattacharyya}, Leser, Päs)]{Exotic Higgs phenomenology from $\mathsf{S}_3$ flavor symmetry}
\vspace{-2em}
\paragraph{\textit{G.~Bhattacharyya}, P.~Leser, H.~Päs}
\paragraph{Abstract}
We consider an $\mathsf{S}_3$ flavor symmetry model, and by imposing
this global discrete symmetry in the scalar potential we observe some
interesting decay signatures of a light scalar and a pseudo-scalar
which might be buried in the existing collider data.

\section{Introduction}
Discrete flavor symmetries are often used to explain the masses and
mixing of quarks and
leptons\,\cite{Altarelli:2010gt,Ma:2007ia,Ma:2004zd,Ishimori:2010au}.
These scenarios predict nonstandard decay signatures involving scalars
and gauge bosons, and flavor changing neutral currents (FCNC).  The
flavor group $\mathsf{S}_3$ was introduced early in
Ref.\,\cite{Pakvasa:1977in} and has since been used in many different
scenarios\,\cite{Dong:2011vb,Grimus:2005mu,Harrison:2003aw,Ma:1990qh,
  Ma:1999hh,Mohapatra:1999zr,Mohapatra:2006pu,Morisi:2005fy,Teshima:2005bk,
  Teshima:2012cg,Cao:2011df}. Our analysis is based
on the realization in Ref.\,\cite{Chen:2004rr}. The group structure of
$\mathsf{S}_3$ favors maximal atmospheric mixing angle which still
gives a good fit after the recent measurements of non-zero
$\theta_{13}$\,\cite{Meloni:2012ci,Dev:2012ns,Zhou:2011nu}.
$\mathsf{S}_3$ has three irreducible representations: $\mathbf{1},
\mathbf{1'}$, and $\mathbf{2}$. The invariants $\mathbf{1}$ can be
constructed using the multiplication rules
$\mathbf{2}\otimes\mathbf{2}=\mathbf{1}\oplus\mathbf{1'}\oplus\mathbf{2}$
and $\mathbf{1'}\otimes\mathbf{1'}=\mathbf{1}$. We take the particle
assignments\,\cite{Chen:2004rr}, which we have followed also in
Ref.\,\cite{Bhattacharyya:2010hp,Bhattacharyya:2012ze}:
\begin{equation}
\begin{aligned}
    (L_\mu,L_\tau)&\in \mathbf{2}\, , & L_e, e^c,\mu^c
    &\in \mathbf{1}\, ,
    & \tau^c&\in \mathbf{1'} \, ,\\
    (Q_2,Q_3)&\in\mathbf{2}\, , & Q_1, u^c,c^c,d^c,s^c
    &\in\mathbf{1}\, ,
    & b^c,t^c&\in\mathbf{1'} \, , \\
    (\phi_1,\phi_2)&\in \mathbf{2}\, , & \phi_3&\in\mathbf{1} \,.
\end{aligned}
\end{equation}
The fields $Q_{1/2/3}$ and $L_{e/\mu/\tau}$ are the quark and lepton
$\mathsf{SU}(2)$ doublets of the three generations. This assignment
was motivated in Ref.\,\cite{Chen:2004rr} to have a reasonably
successful reproduction of quark and lepton masses and mixing.  All
the three scalar $\mathsf{SU}(2)$ doublets $\phi_{\{1,2,3\}}$ take
part in electroweak symmetry breaking.  The general structure of the
model allows for tree-level FCNC due to the absence of natural flavor
conservation\,\cite{Glashow:1976nt}, although those are too suppressed
by the Yukawa couplings to cause any problem even for scalar masses of
the electroweak
scale\,\cite{Bhattacharyya:2010hp,Bhattacharyya:2012ze,Kubo:2003iw}. However, in models
where the flavor symmetry does not apply on Yukawa couplings, the
scalar masses are pushed beyond the TeV
scale\,\cite{Yamanaka:1981pa}. In our
analysis\,\cite{Bhattacharyya:2010hp,Bhattacharyya:2012ze} we observe noteworthy decay
properties of a scalar and a pseudo-scalar: ($i$) Two of the three
scalars $h_{b,c}$ have standard model (SM)-like gauge and Yukawa
couplings, and they can dominantly decay into the third absolutely
non-SM-like scalar $h_a$; ($ii$) The scalar (pseudo-scalar) $h_a$
($\chi_a$) has no $(h_a/\chi_a)VV$-type vertices, where $V\equiv
W^\pm,Z$; ($iii$) $h_a/\chi_a$ has \emph{only} flavor off-diagonal
Yukawa couplings with one fermion from the third generation.  We have
included all scalar degrees of freedom: three CP-even neutral scalars,
two CP-odd neutral scalars and two sets of charged scalars. The
special features of our analysis are: ($i$) determination of the mass
spectrum of the neutral scalars/pseudoscalars and the charged scalars
following an improved potential minimization method, ($ii$)
calculation of their gauge and Yukawa couplings, and ($iii$)
identification of a novel decay channel of a scalar (pseudoscalar)
which can be experimentally tested.

\section{Mass spectrum}
The explicit form of the general $\mathsf{S}_3$ invariant scalar potential, which we do not give here for brevity, is given in Refs.~ 
\cite{Bhattacharyya:2010hp,Bhattacharyya:2012ze,Kubo:2004ps}. It has eight dimensionless couplings $\lambda_i$ and two mass-squared dimensional parameters $m^2$ and $m_3^2$.

The replacement $\phi_i \to \left(h_i^+, v_i + h_i +
\mathrm{i}\chi_i\right)^\intercal$ is done, assuming $v_1 = v_2 = v$
and $v_3$, which allow for maximal atmospheric neutrino mixing, where
$2v^2+v_3^2=v_\text{SM}^2$ has to hold with
$v_\text{SM}=246$\,GeV. After diagonalizing the mass matrices the
physical CP-even, CP-odd and charged scalars are denoted by
$h_{a,b,c},\chi_{a,b}$ and $h_{a,b}^+$, respectively. 

Note that by imposing $v_1=v_2$ on the potential $m^2$ and $m_3^2$ are related to the couplings $\lambda_i$ and the VEVs $v$ and $v_3$.
To make sure that this point is actually a minimum of the potential,
the determinant of the Hessian has to be positive, which is equivalent
to imposing the condition of positive squared scalar masses.  As a
first step towards potential minimization, we first try to provide an
analytical feel. We identify some simple-looking relations of the
coefficients that keep the potential always bounded from below. To do
this we factorize the scalar potential into a simplified polynomial in
$\phi_1,\phi_2$ and $\phi_3$. Three distinct types of terms emerge
with power four: $\phi_i^4$, $\phi_i^2\phi_j^2$ and
$\phi_i^2\phi_j\phi_k$, where $i,j,k=1\ldots 3$. Of the nine terms,
only six have independent coefficients, which we call $c_{\{1\ldots
  6\}}$:
\begin{equation}
    \label{eqn:exppotential}
    c_1\phi_1^4 + c_1\phi_2^4 + c_2\phi_3^4 + c_3\phi_1^2\phi_2^2 
    + c_4\phi_1^2\phi_3^2 + c_4\phi_2^2\phi_3^2 + 
    c_5\phi_1^2\phi_2\phi_3 + c_5\phi_1\phi_2^2\phi_3 
    + c_6\phi_1\phi_2\phi_3^2\, .
\end{equation}
It follows that
\begin{equation}
\begin{aligned}
    c_1 &= \lambda_1/2 + \lambda_2/2, &c_2 &= \lambda_4/2, &c_3 &= \lambda_1 
    - \lambda_2 + \lambda_3, &c_4 &= \lambda_5 + \lambda_6, &c_5 
    &= 2\lambda_8, &c_6 &= 2\lambda_7\, .
\end{aligned}
\end{equation}
By inspection, the following conditions emerge:
\begin{equation}
\begin{aligned}
    \label{eqn:condbyinspection}
    c_1,c_2&>0, & 2c_3,2c_4 &\geq  -c_1, & 2c_3,2c_4 &\geq  -c_2, 
    & 2c_4&\geq -c_1,\, ,\\
    -1/2c_1&\leq c_5,c_6\leq c_1, 
    & -1/2c_2&\leq c_5,c_6\leq c_2\, .
\end{aligned}
\end{equation}
Then we get an acceptable mass spectrum for all types of scalars, and
the potential turns out to be globally stable. However, this method
overlooks a large part of the otherwise valid parameter, and an
uncomfortable feature is that none of the masses exceeds 300 GeV when
${\left|\lambda_{\{1\ldots 8\}}\right|\leq \pi}$.

Now we propose a better method for ensuring global stability. We
transform Eq.\,(\ref{eqn:exppotential}) into spherical coordinates
$(\rho,\theta,\phi)$, which splits the potential into a radial and an
angular part. Global stability then means the positivity of the
angular part:
\begin{multline}
    \label{eqn:angularpart}
    \sin^4\theta  \bigl\{(2 c_1 -c_3 ) \cos (4 \phi )+6 c_1 +c_3 \bigr\}
    +8 c_2  \cos^4\theta +\sin^2(2 \theta ) \bigl(2 c_4  \sin^2\phi 
    +c_6  \sin (2 \phi )\bigr)\\
    +8 c_4  \cos ^2\phi  \sin^2\theta  \cos ^2\theta 
    +4 c_5  \sin (2 \phi ) \sin ^3\theta  \cos \theta  \bigl(\sin \phi 
    +\cos \phi\bigr)>0 
\end{multline}
\begin{figure}[t!]
    \centering
    \includegraphics[height=2in]{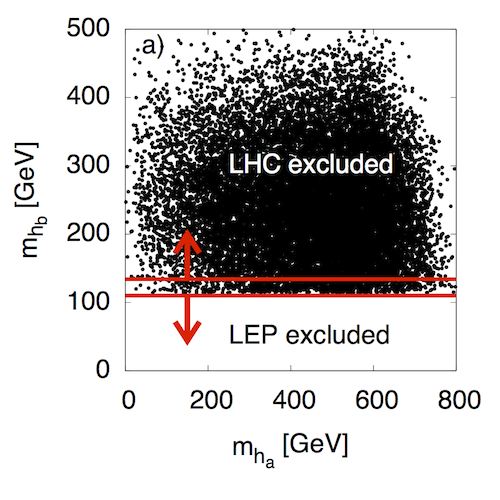}
    \includegraphics[height=2in]{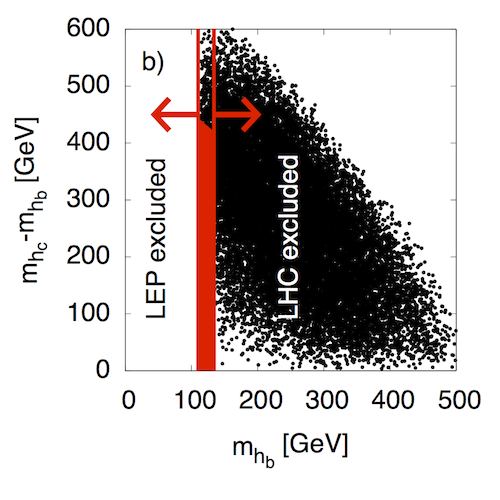}  
    \includegraphics[height=2in]{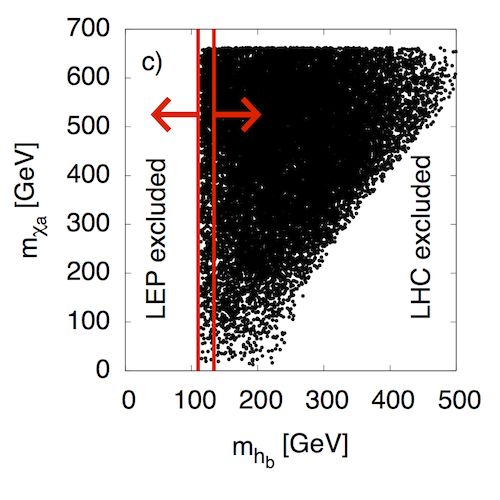}    
    \caption{\label{fig:scatterplots}\textit{Scatter plots of masses
        of $h_a,h_b,h_c$ and $\chi_a$, fixing $v_3/v=0.6$. The lines
        give the window between $114 - 130$ GeV. The highlighted strip
        in the middle plot is disfavored by LHC which disfavors a
        second SM-like Higgs within 550 GeV.}}
\end{figure}
As it is a transcendental inequality, no simple-looking analytic
solutions emerge by solving Eq.\,(\ref{eqn:angularpart}).  We
therefore check the positivity of this function numerically at each
point of the parameter space.  This method allows us to explore the
so-far inaccessible territory of the stable parameter space that could
not be reached by Eq.\,(\ref{eqn:condbyinspection}). Interestingly,
the heavy scalar and pseudoscalar masses can be pushed well above 300
GeV even for $\left|\lambda_{\{1\ldots 8\}}\right|\leq\pi$.

We express the physical pseudo-scalar ($\chi$) and scalar ($h$) states
denoted by {\em roman alphabets} as subscripts in terms of their weak eigenstates 
distinguished by {\em hindu numerals}: 
\begin{equation}
\begin{aligned}
    \label{eqn:pseudoscalarmixing}
    \chi_{1(2)} &= \left(v/v_\text{SM}\right) G^0 \mp \left(1/\sqrt{2}\right) 
    \chi_a - v_3/\left(\sqrt{2} v_\text{SM}\right)\chi_b, 
    & \chi_3 &= \left(v/v_\text{SM}\right) G^0 
    + \sqrt{2}\left(v_3/v_\text{SM}\right)\chi_b\, ; \\
 h_{1(2)}&= U_{1(2)b}~h_b + U_{1(2)c}~h_c \mp \left(1/\sqrt{2}\right) ~h_a, & 
 h_3&= U_{3b}~h_b + U_{3c}~h_c \, , 
\end{aligned}
\end{equation}
where $U_{ib}$ and $U_{ic}$ are complicated functions of the
$\lambda_{\{1\ldots 8\}}$, $v$ and $v_3$. The corresponding mixing
relations for $h_{a,b}^+$ are obtained by substituting $\chi\to h^+$
and $G^0\to G^+$ in Eq.\,(\ref{eqn:pseudoscalarmixing}). The masses
for the CP-even scalars are\,\cite{Bhattacharyya:2010hp,Bhattacharyya:2012ze}
\begin{equation}
\begin{aligned}
    m_{h_a}^2 &= 4\lambda_2 v^2 - 2\lambda_3 v^2 - v_3
    \left(2\lambda_7 v_3 + 5\lambda_8 v\right)\, ,\\
    m_{h_{b(c)}}^2 &= \frac{1}{2v_3}\left[4 \lambda_1 v^2 v_3+2
    \lambda_3 v^2 v_3+2 \lambda_4 v_3^3-2 \lambda_8 v^3+3
    \lambda_8 v v_3^2 \mp \Delta m^3\right]\, ,
    \label{eqn:msqscalars}
\end{aligned}
\end{equation}
where $\Delta m^3$ is a complicated expression of the $\lambda_i$ and the VEVs given in Refs.~\cite{Bhattacharyya:2010hp, Bhattacharyya:2012ze}.

The pseudo-scalar squared masses are
\begin{equation}
\begin{aligned}
    m_{\chi_a}^2 &= -9 \lambda_8 v v_3, & m_{\chi_b}^2 &= -v_\text{SM}^2 
    \left(2\lambda_7 + \lambda_8 v/v_3\right)\, ,
\end{aligned}
\end{equation}
while the charged scalars' squared masses are
\begin{equation}
\begin{aligned}
    m_{h_a^+}^2 &= -2\lambda_3 v^2 - v_3^2 \left(\lambda_6 + \lambda_7 \right) 
    + 5 \lambda_8 v v_3, 
    & m_{h_b^+}^2 &= -v_\text{SM}^2\left(\lambda_6+\lambda_7 
    + \lambda_8 v/v_3\right)\, .
\end{aligned}
\end{equation}

The allowed ranges for the masses obtained by varying
$\lambda_{\{1\ldots 8\}}\in[-\pi,\pi]$ and keeping the ratio $v_3/v$
fixed to $0.6$ (chosen in Ref.\,\cite{Bhattacharyya:2010hp,Bhattacharyya:2012ze} for
compatibility with the quark masses) are shown in Fig.~1. In view of
the recent LHC results\,\cite{ATL2012, CMS2012}
that claims discovery of a Higgs-like boson at around $125$\,GeV with
a large excluded region above and below, the mass spectrum in this
model fits well with the following scenario:
\begin{enumerate}
    \item $h_b$ is identified with the 125 GeV Higgs boson
      \cite{ATL2012, CMS2012}. The Yukawa and
      gauge couplings of $h_b$ and $h_c$ are like those of the SM Higgs. $h_c$ is somewhat
      heavier than $h_b$.

    \item $h_a$ and $\chi_a$ have nonstandard 
    interactions that hide them from standard searches.

    \item All other masses, including the charged 
    scalars, can be above $550$\,GeV, although from the experimental
    point of view the charged scalars need not be that heavy.
\end{enumerate}

\section{Couplings}
\label{sec:three}
\begin{table}[h]
    \begin{center}
        \begin{tabular}{c||c|c|c|c|c|c}
            & $\mathbf{h_a^\pm W^\mp}$ & $\mathbf{h_b^\pm W^\mp}$ 
            & $\mathbf{\chi_aZ}$ & $\mathbf{\chi_bZ}$ & $\mathbf{W^\pm W^\mp}$  
            & $\mathbf{ZZ}$\\
            \hline\hline
            $\mathbf{h_a}$ & $\checkmark$ & -- & $\checkmark$ &--&--&--\\
            $\mathbf{h_b}$ &-- &$\checkmark$ & --&$\checkmark$ & $\checkmark$ 
            & $\checkmark$\\
            $\mathbf{h_c}$ &-- &$\checkmark$ &-- &$\checkmark$ & $\checkmark$ 
            & $\checkmark$
        \end{tabular}\quad\quad
        \begin{tabular}{c||c|c}
            & $\mathbf{h_a^\pm W^\mp}$ & $\mathbf{h_b^\pm W^\mp}$\\
            \hline\hline
            $\mathbf{\chi_a}$ & $\checkmark$ &--\\
            $\mathbf{\chi_b}$ &  -- & $\checkmark$\\
            &
        \end{tabular}
    \end{center}
    \caption{\label{tab:gaugecouplings}\textit{3-point vertices with
        at least one $h$ (or $\chi$) and $W/Z$ boson. A checkmark
        means that the vertex exists.}}
\end{table}

The couplings involving $h_a$ do not depend on the scalar potential
parameters, while those of of $h_b$ and $h_c$ do and that too in a
complicated way, which we refer by putting checkmark signs in Tables
\ref{tab:gaugecouplings} and \ref{tab:othercouplings} without
displaying their expressions explicitly. The $h_a\chi_aZ$ coupling is
$\frac{\mathrm{i}}{2} G q_\mu$, where $G=\sqrt{g^2+g'^2}$ and $q_\mu$
is the momentum transfer.  Since neither $h_a$ nor $\chi_a$ couples to
pairs of gauge bosons via the three-point vertex, their masses are not
constrained from direct searches at LEP2 or by electroweak precision
tests. In fact, the conventional LHC Higgs search strategy would not
apply on them either.
\begin{table}[h]
    \begin{center}
        \begin{tabular}{c||c|c|c|c}
            & $\mathbf{h_a^\mp  \gamma}$ & $\mathbf{h_a^\mp Z }$ &  
            $\mathbf{h_b^\mp  \gamma}$ & $\mathbf{h_b^\mp Z }$\\
            \hline\hline
            $\mathbf{h_a^\pm}$ &
            $\checkmark$ & $\checkmark$ & -- &--\\
            $\mathbf{h_b^\pm}$ &
            --&-- &$\checkmark$ & $\checkmark$\\
            &&&&
        \end{tabular}\quad\quad
        \begin{tabular}{c||c|c|c|c|c|c|c|c|c}
            & $\mathbf{h_{a}h_{a}}$& $\mathbf{h_ah_b}$ 
            & $\mathbf{h_ah_c}$&$\mathbf{h_{a}^\pm 
            h_{a}^\mp}$&$\mathbf{h_{b}^\pm h_{b}^\mp}$
            &$\mathbf{h_{a}^\pm h_{b}^\mp}$
            &$\mathbf{\chi_{a}\chi_{a}}$
            &$\mathbf{\chi_{b}\chi_{b}}$
            &$\mathbf{\chi_{a}\chi_{b}}$\\
            \hline\hline
            $\mathbf{h_a}$ &--& $\checkmark$ & $\checkmark$ & -- 
            &--&$\checkmark$ &-- & --&$\checkmark$\\
            $\mathbf{h_b}$ & $\checkmark$ &-- &-- &$\checkmark$ & $\checkmark$ 
            &--& $\checkmark$ & $\checkmark$&--\\
            $\mathbf{h_c}$ & $\checkmark$ &-- &-- &$\checkmark$ & $\checkmark$ 
            &-- &$\checkmark$ & $\checkmark$&--
        \end{tabular}
    \end{center}
    \caption{\label{tab:othercouplings}\textit{Other 3-point
        vertices. A checkmark indicates that the vertex exists.}}
\end{table}
Now we come to Yukawa interaction, whose explicit form is given in Refs.~\cite{Bhattacharyya:2010hp, Bhattacharyya:2012ze}.

The scalars are rotated to their physical basis
$\{h_a,h_b,h_c\}$, and we obtain the Yukawa matrices
$Y_{\{a,b,c\}}$.  The individual mixing matrices for up- and down-type
quarks contain large mixing angles as a consequence of $S_3$ symmetry
and the particle assignments\,\cite{Chen:2004rr}. Specifically, the
doublet representation of $S_3$ generates maximal mixing when we set
$v_1=v_2$. Now, the CKM matrix involves a relative alignment of those
two matrices which yields small mixing for quarks. Similarly, the PMNS
matrix is given by the relative orientation of the mixing matrices for
the charged leptons and neutrinos. Since we assume that the neutrino
mass matrix is diagonal being generated by a type-II seesaw mechanism,
the large mixing angles in the lepton sector survive. There are two
generic textures of Yukawa couplings in our
model\,\cite{Bhattacharyya:2010hp,Bhattacharyya:2012ze}:
\begin{equation}
\begin{aligned}
    Y_{a} &= \begin{pmatrix}
                0 & 0 & Y_{13} \\
                0 & 0 & Y_{23} \\
                Y_{31} & Y_{32} & 0
             \end{pmatrix},
    & 
    Y_{b,c} &= \begin{pmatrix}
                Y_{11} & Y_{12} & 0\\
                Y_{21} & Y_{22} & 0\\
                0 & 0 & Y_{33}
               \end{pmatrix}\, .
\end{aligned}
\end{equation}
Here $Y_{a}$ symbolically describes the Yukawa couplings for
$h_a,\chi_a$ and $h_a^+$, while $Y_{b,c}$ describe the couplings for
$h_b, h_c, \chi_b$ and $h_b^+$, and the pattern holds both for leptons
and quarks. The off-diagonal couplings in $Y_{b,c}$ are numerically
small and can be controlled by one free parameter which keeps
dangerous FCNC processes under control. The largest off-diagonal
coupling in $Y_{a}$ is $(h_a/\chi_a) c t \sim 0.8$; it leads to viable
production channel of $h_a$ via $t$ decays. The next largest couplings
are $(h_a/\chi_a) s b \approx 0.02$ and $(h_a/\chi_a) \mu \tau \approx
0.008$. The $\chi_a \mu \tau$ coupling leads to an interesting decay
channel that can lead to observable signatures at the LHC.  

\section{How to search for $h_a$ at the LHC?}
\begin{figure}[t!]
    \centering
    a) \includegraphics[height=1in]{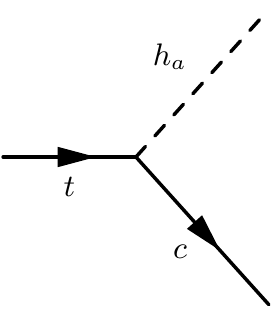}\quad\quad\quad
    b) \includegraphics[height=1in]{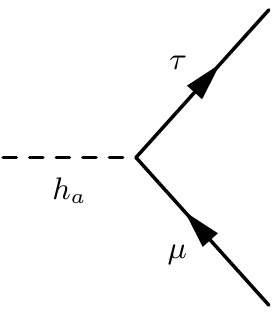}\quad\quad\quad 
    c) \includegraphics[height=1in]{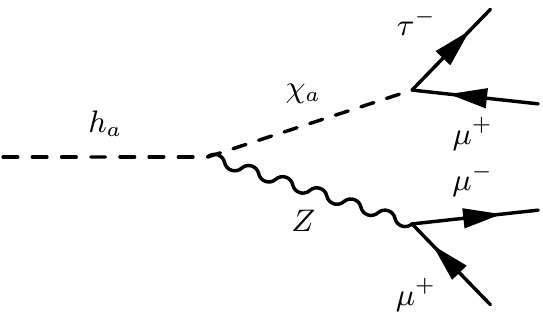}
    \caption{\label{fig:feynmangraphs}\textit{Feynman graphs for dominant 
    sources of $h_a$ production and decays which might be relevant at the LHC.
    }}
\end{figure}

If kinematically allowed, $h_a$ can be produced e.g. through $t\to h_a
c$ [Fig.~\ref{fig:feynmangraphs}(a)]. After that, if
$m_{h_a}<m_{\chi_a}$, $h_a$ decays dominantly into $b$ and $s$ quarks,
or into $\tau$ and $\mu$ [see Fig.~\ref{fig:feynmangraphs}(b)]. The
branching ratio (BR) for $t\to h_ac$ is $0.17 (0.06)$ for $m_{h_a}=
130 (150)$\,GeV. Then $h_a\to\mu\tau$ occurs with a BR of $10\%$ and
$h_a\to b s$ with $90\%$.
\begin{figure}[t!]
    \centering
    \includegraphics[height=2in]{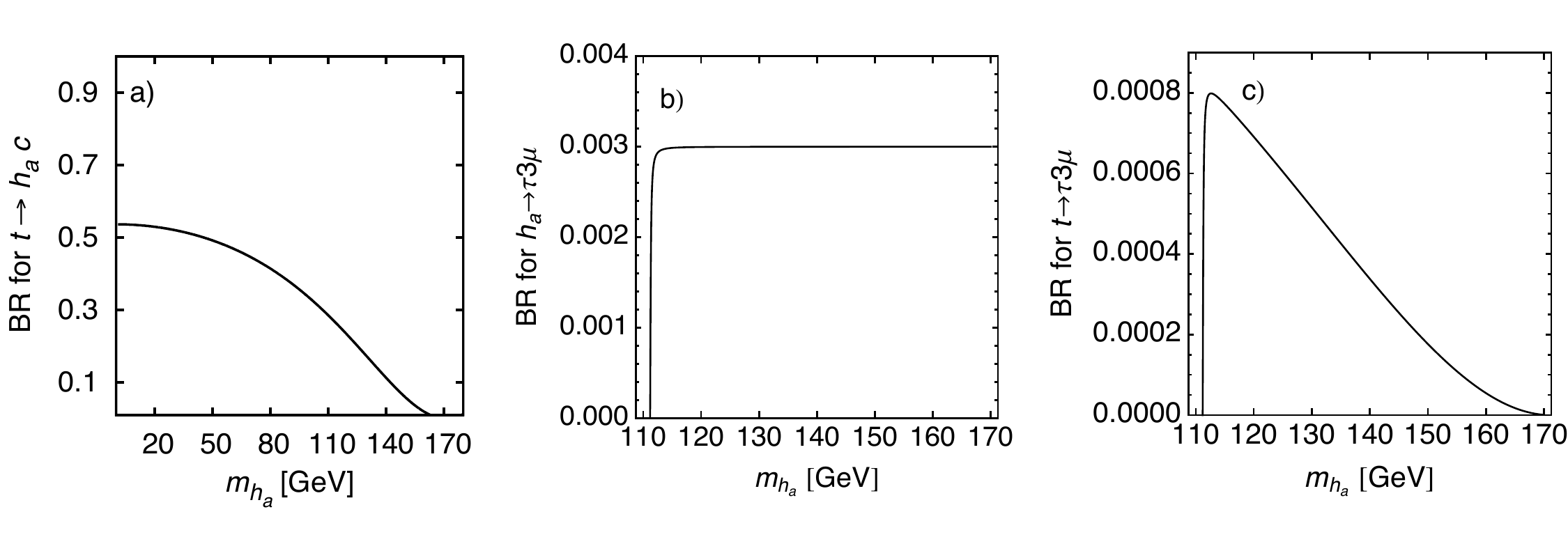}
    \caption{\label{fig:hadecay}\textit{Different branching ratios involving 
    the production and decay of $h_a$. In all cases, $m_{\chi_a}=20$\,GeV is 
    assumed.
    }} 
\end{figure}

We stress on a spectacular channel that opens up when $h_a\to \chi_a
Z$ is kinematically accessible [Fig.~\ref{fig:feynmangraphs}(c)]. The
corresponding BR is almost $100\%$ since gauge couplings dominate over
the light fermion Yukawa couplings. Then $\chi_a\to\tau\mu$ proceeds
with a BR of $10\%$, and $Z\to \mu\mu$ occurs with a BR of $3\%$. If
two $h_a$ are produced from $t\bar{t}$ pairs, this could lead to a
characteristic signal with up to six muons with the taus used as
tags. The relevant BRs are plotted in
Figs.~\ref{fig:hadecay}(a)--(c). Throughout we have assumed that
$m_{\chi_a}=20$\,GeV.

\section{Conclusions and outlook}
We have analyzed the complete scalar/pseudoscalar sector of an
$\mathsf{S}_3$ flavor model. We simultaneously handle three CP-even,
two CP-odd and two sets of charged scalar particles. We followed a
novel technique of potential minimization which allowed to us to
explore the parameter space better.  The scalar $h_b$ mimicks the
standard Higgs-like object weighing around $125$\,GeV, while $h_a$ and
$\chi_a$ evade conventional collider searches at LEP/Tevatron/LHC and
hence can be rather light. The other scalars/pseudoscalars can stay
beyond the current LHC reach (e.g., $550$\,GeV). We stressed on a
promising channel for $h_a$ search at LHC involving up to six muons in
the final state to be searched with the tau tags.  We urge our
experimental colleagues to look for this channel.

%% file: Papers/blankenburg.tex
\chapter[Neutrino Masses and LFV from $U(3)^5 \to U(2)^5$ in SUSY (Blankenburg)]{Neutrino Masses and LFV from $U(3)^5 \to U(2)^5$ in SUSY}
\label{chap:blankenburg}
\vspace{-2em}
\paragraph{G. Blankenburg}
\paragraph{Abstract}
We analyze neutrino masses and Lepton Flavor Violation (LFV) in charged leptons 
with a minimal ansatz about the breaking of the  $U(3)^5$ flavor symmetry, consistent with the 
$U(2)^3$ breaking pattern of quark Yukawa couplings, in the context of supersymmetry. 
Neutrino masses are expected to be almost degenerate, close to present bounds from cosmology 
and $0\nu\beta\beta$ experiments.
We also predict $s_{13} \approx  s_{23} |V_{td}|/|V_{ts}| \approx 0.16$, in perfect agreement
with the recent results. For slepton masses below 1 TeV
we expect $\mathcal B(\mu \to e \gamma) > 10^{-13}$ and $\mathcal B(\tau \to \mu \gamma) > 10^{-9}$,
within the reach of future experimental searches.

\section{Introduction}

A TeV extension of the SM aimed to address, at least in part, both the stability of the electroweak sector 
and the flavor problem is supersymmetry with heavy squark masses for the first two 
families, in short {\em split-family} SUSY \cite{Papucci:2011wy}.  

A hierarchical squark spectrum is not enough to suppress flavor violation to a level consistent with experiments.
This is why split-family SUSY with a minimally broken $U(2)^3=U(2)_q \times U(2)_d \times U(2)_u$ 
flavor symmetry, acting on the first two generations of quarks (and squarks),  has been considered in Ref.~\cite{Barbieri:2011ci}.
This set-up has the following advantages:  i) it provides some insights about the structures of the Yukawa couplings; 
ii) it ensures a sufficient protection of flavor-changing neutral currents; iii) it leads to an improved 
CKM fit with tiny and correlated non-standard contributions to $\Delta F=2$ observables. 

The purpose of this article is to extend the idea of a minimally broken flavor symmetry acting on the first two generations to the lepton sector \cite{Blankenburg:2012nx}. The extension is straightforward in the case of  charged leptons, enlarging the flavor symmetry from $U(2)^3$ to $U(2)^5=U(2)^3\times U(2)_l \times U(2)_e$. However, the situation is more involved in the neutrino sector, whose mass matrix has a rather different flavor structure: no large hierarchies in the eigenvalues, and large mixing angles. 
A simple ansatz to circumvent this problem is to a assume a two-step breaking in the neutrino sector: first, a leading breaking of the maximal flavor symmetry, $U(3)_l \times U(3)_e$, that includes the total lepton number (LN), giving rise to a fully degenerate neutrino spectrum. 
This would be followed by a sub-leading LN-conserving breaking with a hierarchical structure similar to the one occurring in the charged-lepton sector. 

\section{Flavor symmetries  and symmetry  breaking}

The $U(2)^2 = U(2)_l \times U(2)_e$ flavor symmetry, under which the 
lepton superfields transform as 
\begin{eqnarray}
  L_L  \equiv ( L_{L1}  , L_{L2} )^{\phantom{T}}  &\sim& (\bar 2,1)~, \qquad L_{L3} \sim (1,1)~, \\
  e^{c}\equiv ( e_{1}^c , e_{2}^c )^T    &\sim& (1,2)~, \qquad e^c_{3} \sim (1,1)~,
\end{eqnarray}
offers a natural framework to justify the hierarchal structure of the charged-lepton Yukawa coupling, in close analogy to the $U(2)^3$ symmetry introduced in Ref.~\cite{Barbieri:2011ci} for the quark sector: 
\begin{equation}
 \label{GB_eq:Yquark}
Y_{u,d} = y_{t,b} \left(\begin{array}{ccc}
 \Delta Y_{u,d} &\vline& V \\\hline
 0 &\vline& 1
\end{array}\right)~.
\end{equation}
However in the neutrino sector the large $\theta_{23}$ angle impose a strong connection between the second and third family (large 2-3 mixing or strong degenerancy between second and third eigenvalues), ie between the doublet and the singlet of $U(2)$. This seems to be incompatible with the breaking of $U(2)_l \times U(2)_e$, but it can be easily obtained embedding 
$U(2)_l$ in $U(3)_l$.

In fact starting from a $U(3)_l \times U(3)_e$ symmetry  a degenerate configuration for $m_\nu$ is achieved assuming that  $U(3)_l$ and the total 
lepton number, 
$
U(1)_{\rm LN}=U(1)_{l+e} 
$
,are broken by a spurion $m_\nu^{(0)}$ transforming
as a ${\mathbf 6}$ of $U(3)_l$ and leaving invariant a subgroup of $U(3)_l$ that we denote $O(3)_l$.
By a proper basis choice in the $U(3)_l$ flavor space we can set 
\begin{equation}
m_\nu^{(0)} \propto \left(\begin{array}{ccc}
 I &\vline& 0 \\ \hline 
0 & \vline& 1
 \end{array}\right)~.
\end{equation}

We shall also require that $U(3)_l \times U(3)_e$ is broken by $U(1)_{\rm LN}$ invariant spurions
to the subgroup $U(2)_l \times U(2)_e$ relevant to the charged-lepton Yukawa coupling.
However, it is essential
for our construction that this (sizable) breaking  does not spoil the Majorana sector, at least in first approximation. 
This can be achieved in a supersymmetric context introducing a new spurion $Y^{(0)} \sim ({\mathbf 3}, \bar {\mathbf 3})$
that breaks $U(3)_l\times U(3)_e$ to $U(2)_l\times U(2)_e$ 
By means of $Y^{(0)}$ we can have a non-vanishing Yukawa coupling for the third generation 
in the superpotential
\begin{equation}
{\mathbf L_L} Y^{(0)} {\mathbf e^c} \to  y^{(0)}_\tau  L_3  e^c_3~.
\label{GB_eq:Y0}
\end{equation}
and, in first approximation, the Majorana mass matrix is unchanged. 
Note that supersymmetry is a key ingredient for the the latter statement to hold.  
Indeed, if the mass operator was not holomorphic, a Majorana term of the type
${\mathbf L_L} Y\,Y^\dag\, m_\nu^{(0)}\, {\mathbf L_L^T}$ could also be included and this would spoil the degenerate configuration.

Summarizing, introducing the two spurions $m_\nu^{(0)}$ and  $Y^{(0)}$ we recover phenomenologically viable first approximations to 
both the neutrino and the charged-lepton mass matrices and we are left with an exact $O(2)_l \times U(2)_e$ symmetry
that leaves invariant both $m_\nu$ and  $Y_e$. Moreover, thanks to supersymmetry, the two sector considered separately are
invariant under larger symmetries: $O(3)_l$ for the neutrinos and $U(2)_l\times U(2)_e$ for the charged leptons.

We can then proceed introducing the small $O(2)_l \times U(2)_e$ breaking terms responsible for the subleading terms in $Y_e$
in Eq.~(\ref{GB_eq:Yquark}). 
This spurion $V$ should be regarded as the $O(2)_l$ component of an appropriate ${\mathbf 8}$ of  $U(3)_l$ with the following structure
\begin{equation}
X= \left(
\begin{array}{ccc}
\Delta_L & \vline&V \\ \hline 
V^\dagger & \vline&x 
\end{array}\right)~.
\end{equation}
This allows to write the additional Yukawa interaction ${\mathbf L_L} X\,Y^{(0)}\, {\mathbf e^c}$ that, 
combined with the leading term in (\ref{GB_eq:Y0}) and with a proper redefinition of $y_\tau$ and $V$ 
implies
\begin{equation}
Y^{(1)}_e = y_\tau\,\left(
\begin{array}{ccc}
0 & \vline&V \\ \hline 
0 & \vline&1
\end{array}\right)~.
\end{equation}
All the components of $X$ do appear in the Majorana sector, via the terms
${\mathbf L_L} X\, m_\nu^{(0)}\, {\mathbf L_L^T}$
and ${\mathbf L_L} m_\nu^{(0)} \,X^T {\mathbf L_L^T}$. These imply the following structure 
\begin{equation}
m_\nu= m_{\nu_1}^{(0)} \left[ I + a\,  \left( 
\begin{array}{ccc}
 \Delta_L & \vline&V \\ \hline  
V^T &  \vline&x 
\end{array}\right) \right]~,
\end{equation}
where $a$ is a  $O(1)$ complex coupling. Assuming that all the entries of $X$ are at most of $O(\epsilon)$
does not spoil the degenerate configuration of $m_\nu$ in first approximation. 
In addition, pursuing the analogy with the squark sector and protecting against too strong FCNCs, we are 
forced to assume $(\Delta_L)_{12}$ at most of $O(\epsilon^2)$ in the basis where $V_1=0$:
\begin{equation}
V= \left( \begin{array}{c} 0 \\  O(\epsilon) \end{array} \right) , \quad 
\Delta_L=\left(\begin{array}{cc}
0 & O(\epsilon^2) \\
O(\epsilon^2) & O(\epsilon)\\
\end{array}\right), \quad x=O(\epsilon).
\end{equation}
In the same basis, redefining the unknown parameters, we arrive to the following parametric expression
\begin{equation}
m_\nu= \bar m_{\nu_1}\left[ I + e^{i\phi_\nu} \left(
\begin{array}{ccc}
-\sigma \epsilon & \ \gamma\epsilon^2 &  0 \\
\ \gamma\epsilon^2 & -\delta \epsilon &  \  r\epsilon \\
0 & \ r\epsilon & 0
\end{array}\right)\right]~, 
\end{equation}
where $\phi_\nu$, $\sigma$, $\delta$, $\gamma$, and $r$ are 
real parameters expected to be of $O(1)$.

The final step for the construction of a realistic charged-lepton Yukawa coupling 
is the introduction of the $U(2)_l \times U(2)_e$ bi-doublet $\Delta Y_e$. The most economical way 
to achieve this goal in the context of $U(3)_l \times U(3)_e$ is to introduce a bi-triplet 
with the following from,
\begin{equation}
\Delta \hat Y_e = \left(
\begin{array}{ccc}
\Delta Y_e & \vline&0 \\ \hline 
0 & \vline&0
\end{array}\right)~,
\end{equation}
which provides the desired correction to $Y_e$ and has no relevant 
impact on $m_\nu$.


\section{Predictions for neutrino masses and mixings}

Given the above neutrino mass matrix we are ready to show few simple analytic results, valid to leading order in $\epsilon$, for masses and mixings. These results are also tested with a systematic numerical scan of the four O(1) 
free parameters \cite{Blankenburg:2012nx}.

We obtain almost degenerate eigenvalues
\begin{eqnarray}
 m_{\nu_1}^2 &=& \bar m^2_{\nu_1}\left(1-2\sigma\,\epsilon\right)~, \\
 m_{\nu_2}^2 &=& \bar m^2_{\nu_1}\left[1-\delta\,\epsilon-(\delta^2+4r^2)^{1/2}\epsilon\right]~, \\
 m_{\nu_3}^2 &=& \bar m^2_{\nu_1}\left[1-\delta\,\epsilon+(\delta^2+4r^2)^{1/2}\epsilon\right]~,
\end{eqnarray}
with $\epsilon$ that controls the overall scale of neutrino masses, whose natural scale is 
$
O[ (\Delta m^2_{\rm atm})^{1/2} \epsilon^{-1/2} ] = O(0.3~{\rm eV}),
$
just below the existing bounds \cite{Komatsu:2010fb}.

For the mixing angles we obtain in a first approximation
\begin{eqnarray}
 t_{23} &=& \frac{s_{23}}{c_{23}}  = \frac{\delta \pm [\delta^2+4r^2]^{1/2}}{2r}~,\\
s_{13} e^{i \delta_P }  &=& s_e s_{23} e^{\alpha_e+\pi}~,\\
\tan2\theta_{12} &=& \frac{4\gamma\,\epsilon}{2\sigma-\delta-[\delta^2+4r^2]^{1/2}}c_{23} =O(1) \times \frac{\epsilon}{\zeta^2}~.
\end{eqnarray}
that means that $t_{23}$ and $t_{23}$ are generically $O(1)$ while we obtain $s_{13}= 0.16 \pm 0.02$ assuming in analogy to the quark sector $s_e = s_d =|V_{td}|/|V_{ts}|$ ($s_e$ and $\alpha_e$ are the mixing parameters in the 1-2 sector of the charged leptons).

\section{The slepton sector and LFV}

Having identified the minimal set of spurions necessary to build the lepton Yukawa coupling and the neutrino mass matrix, 
we can now turn to study the consequences of this symmetry-breaking pattern in the slepton sector. 

Let's start from the $LL$ soft slepton mass matrix,
which transforms as  ${\bf 8}\oplus{\bf 1}$ under $U(3)_l$ and is invariant under $U(3)_e$.
Expanding to the first non-trivial  order in the spurions, it assumes the following form
\begin{eqnarray}
 \tilde m^2_{LL}&=&\left(\begin{array}{ccc}
 (m^2_{L})_{\rm hh} &\vline& c_3 V^* \\ \hline
c_3 V^T &\vline& (m^2_{L})_{\rm 33}
\end{array}\right)\tilde m^2_{L}~,  \\
 (m^2_{L})_{\rm hh} &=& I+c_3\Delta^*_L + c_4 \Delta Y_e^*\Delta Y_e^T~,    \\
 (m^2_{L})_{33}&=& 1+c_2 |y_\tau|^2 +c_3 x~,
\end{eqnarray}
with all constants being real and $O(1)$. 

In the sfermion sector, the main difference between the $U(3)^5$ set-up we are considering, and that based on a $U(2)^5$ symmetry, lies on the fact that in the latter case one can naturally have sfermions of the first two families considerably heavier than those of the third family, 
while in the $U(3)^5$ set-up this can be obtained at the price of some fine-tuning of the $O(1)$ symmetry-breaking terms ($1+c_2 |y_\tau|^2 \ll 1 $ for $\tilde m^2_{LL}$). However, it is worth to stress that in the slepton sector the requirement of a sizable mass 
splitting among the families is less motivated: the sleptons play a minor role in the hierarchy problem and there are no stringent direct experimental bounds on any of the slepton families. 

The $RR$ soft slepton mass matrix transforms as  ${\bf 8}\oplus{\bf 1}$ under $U(3)_e$ and is invariant under $U(3)_l$.
Proceeding similarly to the $LL$ case we find 
\begin{eqnarray}
\tilde m^2_{RR}&=&\left(\begin{array}{ccc}
 (m^2_{RR})_{\rm hh} &\vline& \Delta Y_e^T V^* y_\tau^* \\ \hline
 y_\tau V^T\Delta Y_e^* &\vline& ( m^2_{RR})_{\rm 33} 
\end{array}\right) 
 \tilde m^2_{e_R^c}~,   \\
(m^2_{RR})_{\rm hh} & = & I + c_4' \Delta Y_e^T\Delta Y_e^* + c_5 \Delta Y_e^T\Delta_L^*\Delta Y_e^*~,   \\
 (m^2_{RR})_{\rm 33} &=& 1+y_\tau^*y_\tau(c_2'+c_3' x)~.  
\end{eqnarray}
Here all off-diagonal terms are heavily suppressed by the first and second generation 
Yukawa couplings and, to a good approximation, can be neglected.

Finally, let's consider the trilinear soft-breaking term $A_e$,
responsible for the  $LR$ entries in the slepton mass matrices. 
The symmetry breaking structure of $A_e$ is identical to that of the Yukawas, 
albeit with different complex $O(1)$ factors $a_i$:
\begin{equation}
 A_e=\left(\begin{array}{ccc}
 a_1 \Delta Y_e &\vline& a_2 V \\ \hline
 0 &\vline& a_3
\end{array}\right)y_\tau A_0~.
\end{equation}

\begin{figure}[t]
\centering
\includegraphics[width=0.5\textwidth]{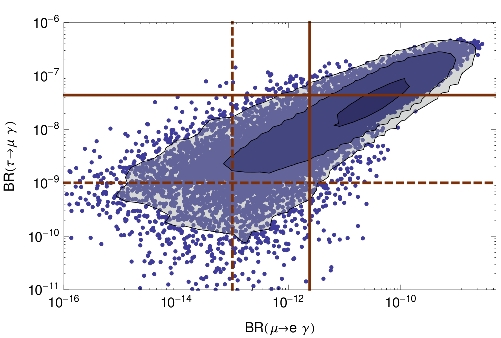}
\label{GB_fig:lfv}
\end{figure}

We perform a analytical and numerical analysis of the main LFV processes \cite{Blankenburg:2012nx}. Here we report the main results where we take the $(3,3)$ and $(6,6)$ elements in the range $(200~{\rm GeV})^2-(1000~{\rm GeV})^2$, while we assume values between $5^2$ and $100^2$ times heavier for the other mass eigenvalues. The $A_0$ parameter is assumed to be proportional to the heavy sfermion mass with a proportionality constant in the range $[-3,3]$. The chargino soft mass is fixed to $M_2=500$~GeV, 
and we use gaugino unification arguments to set $M_1=0.5M_2$. We also fix $\tan\beta=10$, and $\mu=600$~GeV.
In Figure~\ref{GB_fig:lfv} we show the correlation between $\mathcal{B}(\tau\to\mu\gamma)$ and $\mathcal{B}(\mu\to e\gamma)$. We show the current bounds for each branching ratio with solid brown lines \cite{Adam:2011ch,Aubert:2009ag}, while the expected sensitivity of the relevant experiment (MEG for $\mu\to e\gamma$ , Belle II and SuperB for $\tau\to \ell\gamma$) is shown using dashed brown lines.
Figure~\ref{GB_fig:lfv} shows that, although a small part of the parameter space is ruled out already, there exist a significant number of points that can be probed by $\mu\to e\gamma$, $\tau\to\mu\gamma$ and possibly also  $\mu\to e$ conversion experiments in the near future.

\section{Conclusion}
We have proposed an ansatz for the neutrino mass matrix and the charged lepton Yukawa coupling
based on a minimal breaking of the  $U(3)^5$ flavor symmetry, consistent with the $U(2)^3$ 
breaking pattern of the quark Yukawa couplings discussed in Ref~\cite{Barbieri:2011ci}. 
The key hypothesis that allows us to relate the non-hiearchical neutrino sector to the Yukawa sector
is the assumption of a two-step breaking structure in the neutrino case:  a leading breaking of the maximal flavor symmetry, 
$U(3)_l \times U(3)_e$, giving rise to a fully degenerate neutrino spectrum, 
followed by a sub-leading hierarchical breaking similar to the one occurring in the Yukawa sector. 

This framework is able to reproduce all the neutrino oscillation parameters without particular tuning of the free parameters and it can naturally be implemented in supersymmetric extensions of the SM and, more explicitly, within  the well-motivated 
set-up with heavy masses for the first two generations of squarks. LFV processes are expeted to be measured in the near future.

\bibliography{blankenburg.bib}
\bibliographystyle{apsrev4-1}

%% file: Papers/buras.tex

%
%
%
%
%
%

\chapter[On the Roles of $V_{ub}$ and Correlations between Flavour Observables in Indirect Searches for New Physics (Buras)]{On the Roles of $V_{ub}$ and Correlations between
Flavour Observables in Indirect Searches for New Physics}
\vspace{-2em}
\paragraph{A. J. Buras}
\paragraph{Abstract}

We emphasize the important roles of $|V_{ub}|$ and of correlations between 
flavour observables  
in indirect searches for New Physics (NP) by 
means of FCNC processes. 
We illustrate with few examples how different scenarios of NP can be 
distinguished through the value of  $|V_{ub}|$ favoured phenomenologically by 
them and through correlations between different flavour observables.
Precise lattice calculations of the relevant non-perturbative parameters are 
 essential in this context.
\section{Introduction}
In this short presentation I would like to emphasize 
the important roles of $|V_{ub}|$ and of correlations between 
flavour observables  
in indirect searches for New Physics (NP) by 
means of FCNC processes. I will also reemphasize
the important role of lattice calculations in this context. Several of 
the points made below appeared already in the recent long 
review \cite{Buras:2012ts} but I think it is useful to exhibit them here 
in isolation.

In indirect searches for NP through
particle-antiparticle mixings in $K$, $B_{s,d}$ and $D$
meson systems and rare decays of $K$, $B_{s,d}$ and $D$ mesons it is crucial
to know the background to NP: the predictions for various flavour observables within 
the Standard Model (SM). If these predictions suffer from large uncertainties 
also the room left for NP is rather uncertain and if a given NP model 
contains many free parameters, the characteristic flavour violating features 
of this model cannot be transparently seen. They are simply often washed out 
by hadronic and parametric uncertainties even in the presence of accurate 
data. Most prominent examples of this type are the mass differences $\Delta M_{d,s}$ and the parameter $\varepsilon_K$.

While the important role of lattice calculations in the search for NP 
is well known in the literature, it appears to me that the important role of 
  $|V_{ub}|$ in this context is underestimated in most papers. Most people 
would agree that the ultimate precise value for $|V_{ub}|$ extracted one day 
from tree-level decays will be found from present perspective 
at any place in the range
\begin{equation}
\label{Vubrange}
2.8\times 10^{-3}\le |V_{ub}|\le 4.6\times 10^{-3}.  
\end{equation}
%

The determinations from exclusive semi-leptonic B-decays, supported by lattice, cluster around the value of $3.1\times 10^{-3}$, while the inclusive 
semi-leptonic B-decays imply values more like  $4.3\times 10^{-3}$. 
Even slightly higher value was favoured by $B^+\to\tau^+\nu_\tau$ until 
ICHEP 2012, but the recent Belle data do not seem to require it 
anymore \cite{BelleICHEP}. The new world 
average provided by the UTfit collaboration \cite{Tarantino:2012mq} is
$\mathcal{B}(B^+ \to \tau^+ \nu)_{\rm exp} = (0.99 \pm 0.25) \times 10^{-4}~$,
which is consistent with the SM.

This situation is really a problem for testing NP scenarios. After a lengthy 
calculation of tree, one-loop and sometimes more-loop diagrams, including 
often QCD corrections at NLO and NNLO level and derivation of very 
elegant expressions for various observables in a given NP model, 
one is faced with the choice of input parameters, one of them being $|V_{ub}|$. 
This parameter is special, like $\theta_{13}$ in neutrino physics. It is the 
smallest element in the  CKM matrix and if it was vanishing, there would 
be no CP violation in the SM. Therefore its value is crucial for knowing 
the size of CP violation in this model.

There are two extreme strategies one could adopt in this situation:
\begin{itemize}
\item
Follow the advices of professionals like UTfitters, CKMfitters or PDG on CKM 
parameters and in view 
of several new parameters in a given model combine this information with 
sophisticated Markov-chain Monte Carlos, in particular improved versions 
of the  classical Metropolis algorithm \cite{Barbieri:2011ci,Altmannshofer:2011gn,Altmannshofer:2012az,Botella:2012ju} in order to find the allowed 
ranges for different observables. 
 While this strategy is clearly 
legitimate and many people would claim that there is no other way out, 
I do not want to follow this route here. The main reason is that in
the outcome of such analyses, in which some average between 
exclusive and inclusive determinations of $|V_{ub}|$, with a sizable 
uncertainty is used, 
many of the features seen in the elegant expressions of a given NP 
model are often washed out.
\item
Simplified approach by studying how a given NP model would face the 
future more precise data with more precise input, in particular a precise 
value of $|V_{ub}|$. In this manner some of the characteristic features 
of the particular NP model are not washed out and one discovers patterns of flavour 
violations, in particular correlations between different observables, 
that could distinguish between different NP  models.
It is the second approach, already used in several papers in my group at TUM 
\cite{Buras:2012ts}, that I will follow here.
\end{itemize}

\section{Setting the Scene}
Our goal is to find out how the pattern of flavour violation in a given model, required to cure the problems of the SM, depends on $|V_{ub}|$. To this end 
we set all non-perturbative parameters relevant for $\Delta F=2$ processes at 
their present central values \cite{Laiho:2009eu}
and the remaining three parameters of the CKM matrix at 
\begin{align}
|V_{us}|=0.2252, \qquad |V_{us}|=0.0406, \qquad \gamma=68^\circ.
\end{align}
The first two are the central values from tree-level decays. The value of 
$\gamma$ is fully consistent with its known determinations, in particular by
using the ratio $\Delta M_d/\Delta M_s$ and also tree-level decays.


We next consider two scenarios for  $|V_{ub}|$ 
\begin{itemize}
\item
{\bf Exclusive (small)  $|V_{ub}|$  Scenario 1:}
Here the SM value of $|\varepsilon_K|$ is visibly 
smaller than its experimental determination. On the other hand $S_{\psi K_S}$ agrees well with experiment.
\item
{\bf Inclusive (large)  $|V_{ub}|$ Scenario 2:}
Now the SM value of $|\varepsilon_K|$ is consistent with its 
experimental value.  On the other hand 
$S_{\psi K_S}$ is significantly higher than its  experimental value.
\end{itemize}

In Table~\ref{tab:SMpred} we illustrate the SM predictions for these 
observables and $\Delta M_{s,d}$ for different values of  $|V_{ub}|$. 
The properties stated above are clearly seen. Moreover,
with the present lattice input 
$\Delta M_s$  and $\Delta M_d$, although
slightly above the data,  are both in a good agreement with the latter
independently of $|V_{ub}|$.
Yet, one should emphasize that these results depend significantly on the lattice input and in the case
of $\Delta M_d$ on the value of $\gamma$, the (-)phase of $V_{ub}$. 
Therefore to get a better insight
both lattice input and the tree level determination of $\gamma$
have to improve. Fortunately this is expected to happen in coming years.


\begin{table}[!tb]
\centering
\begin{tabular}{|c||c|c|c|c|c|}
\hline
 $|V_{ub}| \times 10^3$  & $3.1$ & $3.4$ & $4.0$ & $4.3$  & Experiment\\
\hline
\hline
  \parbox[0pt][1.6em][c]{0cm}{} $|\varepsilon_K|\times 10^3$ & $1.72(26)$ & $1.87(26)$  & $2.15(32)$ & $2.28(32)$ &$ 2.228(11)$\\
 \parbox[0pt][1.6em][c]{0cm}{}$\mathcal{B}(B^+\to \tau^+\nu_\tau)\times 10^4$&  $0.62(14)$& $0.74(14)$ &  $1.02(20)$ & $1.19(20)$ & $0.99(25)$\\
 \parbox[0pt][1.6em][c]{0cm}{}$(\sin2\beta)_\text{true}$ & $0.623(25)$ & $0.676(25)$ & $0.770(23)$  & 
$0.812(23)$ & $0.679(20)$\\
 \parbox[0pt][1.6em][c]{0cm}{}$\Delta M_s\, [\text{ps}^{-1}]$ & $19.0(21)$ & $19.0(21)$ & $19.0(21)$ & 
$19.1(21)$ &$17.77(12)$ \\
 \parbox[0pt][1.6em][c]{0cm}{} $\Delta M_d\, [\text{ps}^{-1}]$ & $0.56(6)$ & $0.56(6)$ & $0.56(6)$ & $0.56(6)$   &  $0.507(4)$\\
 \hline
\end{tabular}
\caption{SM prediction for various observables as functions of 
$|V_{ub}|$ and $\gamma =
68^\circ$. 
}\label{tab:SMpred}
\end{table}

Table~\ref{tab:SMpred} illustrates the main point of this note clearly. 
There are tensions between various observables within the SM which gives 
some signals for the presence of NP. However,
dependently which scenario for  $|V_{ub}|$ is considered,  
this NP has to remove different 
discrepancies of the SM with the data. In particular it has
to provide
{\it constructive} NP contributions to $|\varepsilon_K|$ (Scenario 1)
or {\it destructive} NP contributions to  $S_{\psi K_S}$ (Scenario 2)
without spoiling
the agreement with the data
for $S_{\psi K_S}$ (Scenario 1) and for $|\varepsilon_K|$ (Scenario 2).

While  models with many new parameters can face successfully both scenarios
removing the deviations from the data for certain range of their parameters,
in  simpler models, with a definite structure of flavour violation and/or 
small numbers of free parameters,
only one scenario for $|V_{ub}|$ can be admitted 
as only in that scenario  a given
model has a chance to fit $\varepsilon_K$ and $S_{\psi K_S}$ simultaneously. 
Let us then summarize how
 five simple extensions of the SM select the 
scenario for $|V_{ub}|$ in order to remove the tension 
between  $\varepsilon_K$ and $S_{\psi K_S}$. Constrained Minimal 
Flavour Violation (CMFV) \cite{Buras:2012ts}
and maximally gauged flavour models (MGFM)  \cite{Buras:2011wi}, both favour
 Scenario 1. The absence of new phases in these scenarios requires the
 exclusive $|V_{ub}|$ in order  to reach agreement 
of  $S_{\psi K_S}$ with the data. On the other hand 
 the 2HDM with MFV and flavour blind phases,
${\rm 2HDM_{\overline{MFV}}}$ \cite{Buras:2010mh,Buras:2012ts}, 
selects  Scenario 2 for
 $|V_{ub}|$ as the contributions in this model to  $\varepsilon_K$ are tiny 
and there are new phases in the $B_d-\bar B_d$ mixing which allow to 
obtain good agreement with the data for  $S_{\psi K_S}$ in spite of a large 
value of $|V_{ub}|$. Similar solution is offered by a particular model 
with extended gauge group $SU(3)_c\times SU(3)_L\times U(1)_X$ ($\overline{331}$) in which NP contributions are governed by tree-level 
heavy neutral gauge boson ($Z'$) exchanges \cite{Buras:2012xx}. 
On the other hand 
models with a global $U(2)^3$ flavour symmetry, representing
 simple non-MFV extensions of the SM can face successfuly both scenarios 
for  $|V_{ub}|$ with interesting consequences for the $S_{\psi\phi}$ asymmetry
as we will see below  \cite{Buras:2012sd}.

Having made strong statements on the important role of $V_{ub}$ in the 
indirect searches for NP, it should be stressed that its precise 
determination one day
can only give us some direction towards the class of successful NP models. 
Even more important are the correlations between various flavour 
observables. This is the next topic I want to discuss.
\section{Correlations Between Flavour Observables}
It is my strong believe that searching for correlations 
between the measured observables is a very  powerful tool in the 
indirect searches for NP. Extensive studies of 
correlations between various observables in concrete models performed in 
my group in the last ten years illustrate 
very clearly the power of this strategy. Quite often only a qualitative 
behaviour of these correlations is sufficient to eliminate the model 
as a solution to observed anomalies or to select models as candidates 
for a new SM. A detailed review of such explicit studies can be found in 
\cite{Buras:2012ts,Buras:2010wr,Blanke:2009pq,Blanke:2012he}. They 
include in particular correlations in CMFV models, LHT, RS and SUSY 
flavour models \cite{Altmannshofer:2009ne}. 
See also \cite{Barbieri:2011ci,Altmannshofer:2011gn,Altmannshofer:2012az,Botella:2012ju}.
With improved data and theory all these results will be increasingly useful.

In view of space limitations I just would like to list my 
favourite correlations that I hope will be tested precisely in the coming 
years.  To this end let me just quote the LHCb data for some of the observables
discussed below \cite{Aaij:2012ac,LHCb:2012py}:
\begin{equation}\label{LHCb1}
S_{\psi\phi}=0.002\pm 0.087, \quad S^{\rm SM}_{\psi\phi}=0.035\pm 0.002,
\end{equation}
\begin{equation}\label{LHCb2}
\mathcal{B}(B_{s}\to\mu^+\mu^-) \le 4.2\times 10^{-9}, \quad
\mathcal{B}(B_{s}\to\mu^+\mu^-)^{\rm SM}=(3.23\pm0.27)\times 10^{-9},
\end{equation}
\begin{equation}\label{LHCb3}
\mathcal{B}(B_{d}\to\mu^+\mu^-) \le 8.2\times 10^{-10}, \quad
\mathcal{B}(B_{d}\to\mu^+\mu^-)^{\rm SM}=(1.07\pm0.10)\times 10^{-10}.
\end{equation}
Here we have also shown the SM predictions  for these
observables with details on 
$\mathcal{B}(B_{q}\to\mu^+\mu^-)$ 
given in \cite{Buras:2012ts,Buras:2012ru}.
In quoting these results
we did not include the correction from $\Delta\Gamma_s$ 
\cite{deBruyn:2012wj,deBruyn:2012wk,Fleischer:2012fy}  but it has to be taken into account when the data improve.

\begin{figure}[!tb]
 \centering
\includegraphics[width = 0.45\textwidth]{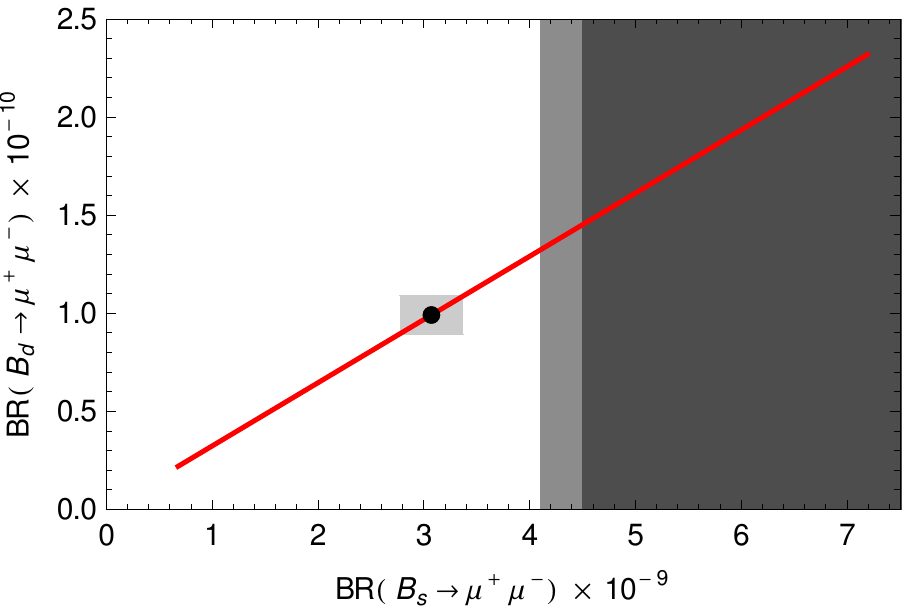}
\includegraphics[width = 0.45\textwidth]{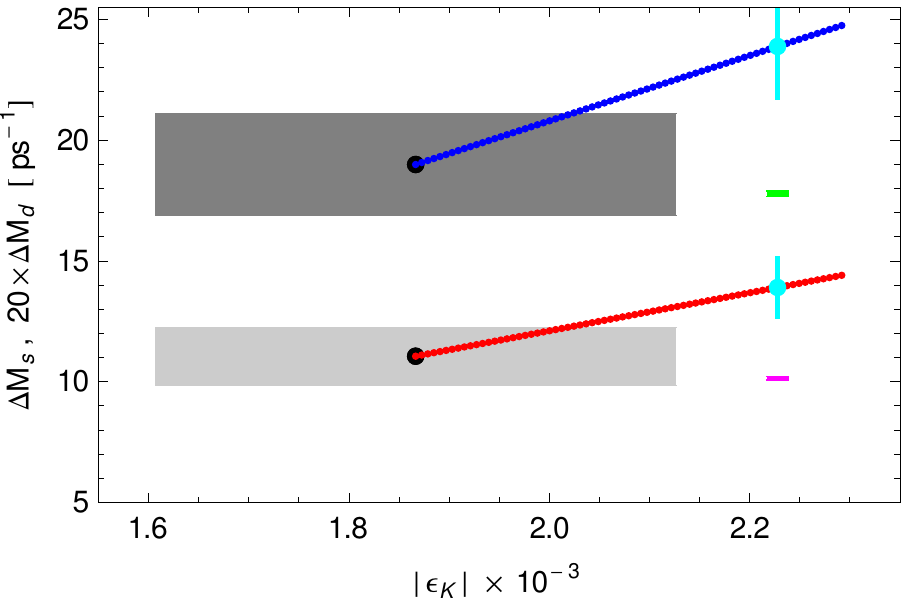}
\caption{ Correlations within models 
with CMFV. See text and \cite{Buras:2012ts} for explanations.}\label{fig:BsBdmumu}
\end{figure}

\subsection{Correlations in CMFV Models}
In this class of models the absence of new sources of flavour violation 
beyond the CKM matrix implies stringent relations among various observables
 \cite{Buras:2003jf}. The most important at present is this one
\begin{equation}\label{bmumu}
\frac{\mathcal{B}(B_d\to\mu^+\mu^-)}{\mathcal{B}(B_s\to\mu^+\mu^-)}=
\frac{\tau({B_d})}{\tau({B_s})}\frac{m_{B_d}}{m_{B_s}}
\frac{F^2_{B_d}}{F^2_{B_s}}
\left|\frac{V_{td}}{V_{ts}}\right|^2  
=\frac{\hat B_{d}}{\hat B_{s}}
\frac{\tau( B_{s})}{\tau( B_{d})}
\frac{\Delta M_{s}}{\Delta M_{d}},
\end{equation}
where $\hat B_{d,s}$ are non-perturbative parameters.
The first relation is valid in all models with Minimal 
Flavour Violation (MFV), while the second one only in CMFV models. We show 
this strict CMFV correlation on the left in Fig.~\ref{fig:BsBdmumu} taken 
from \cite{Buras:2012ts}.

Within CMFV models there is also a unique correlation between
$|\varepsilon_K|$, $\Delta M_s$ and $\Delta M_d$. In fact it can 
be shown that only enhancements over the SM values of 
$|\varepsilon_K|$, $\Delta M_s$ and $\Delta M_d$ are possible in CMFV models
\cite{Blanke:2006yh} and
the enhancement of
one of these observables implies uniquely the enhancements of other
two. A look at Table~\ref{tab:SMpred} shows that this 
correlation is a problem for CMFV models.
The solution to the $|\varepsilon_K|-S_{\psi K_S}$ tension in these models 
can only be provided by enhancement of $|\varepsilon_K|$ which in turn 
enhances $\Delta M_{s,d}$, that are already larger than the data. Thus 
this solution generates a new tension:
$\Delta M_{s,d}-|\varepsilon_K|$ tension, which is shown on the right in
Fig.~\ref{fig:BsBdmumu}. The same difficulty is found in MGFM 
\cite{Buras:2011wi}.

\boldmath
\subsection{Triple Correlation: $S_{\psi K_S}-S_{\psi\phi}-|V_{ub}|$ in $U(2)^3$ Models}
\unboldmath
In models with new sources of CP violation the mixing induced asymmetries 
$S_{\psi K_S}$ and $S_{\psi\phi}$ are modified by new phases $\varphi_{B_d}$ and 
$\varphi_{B_s}$, respectively:
\begin{equation}
S_{\psi K_S} = \sin(2\beta+2\varphi_{B_d})\,,
\qquad
S_{\psi\phi} =  \sin(2|\beta_s|-2\varphi_{B_s})\,.
\label{eq:3.42}
\end{equation}

 \begin{figure}[!tb]
  \centering
 \includegraphics[width = 0.6\textwidth]{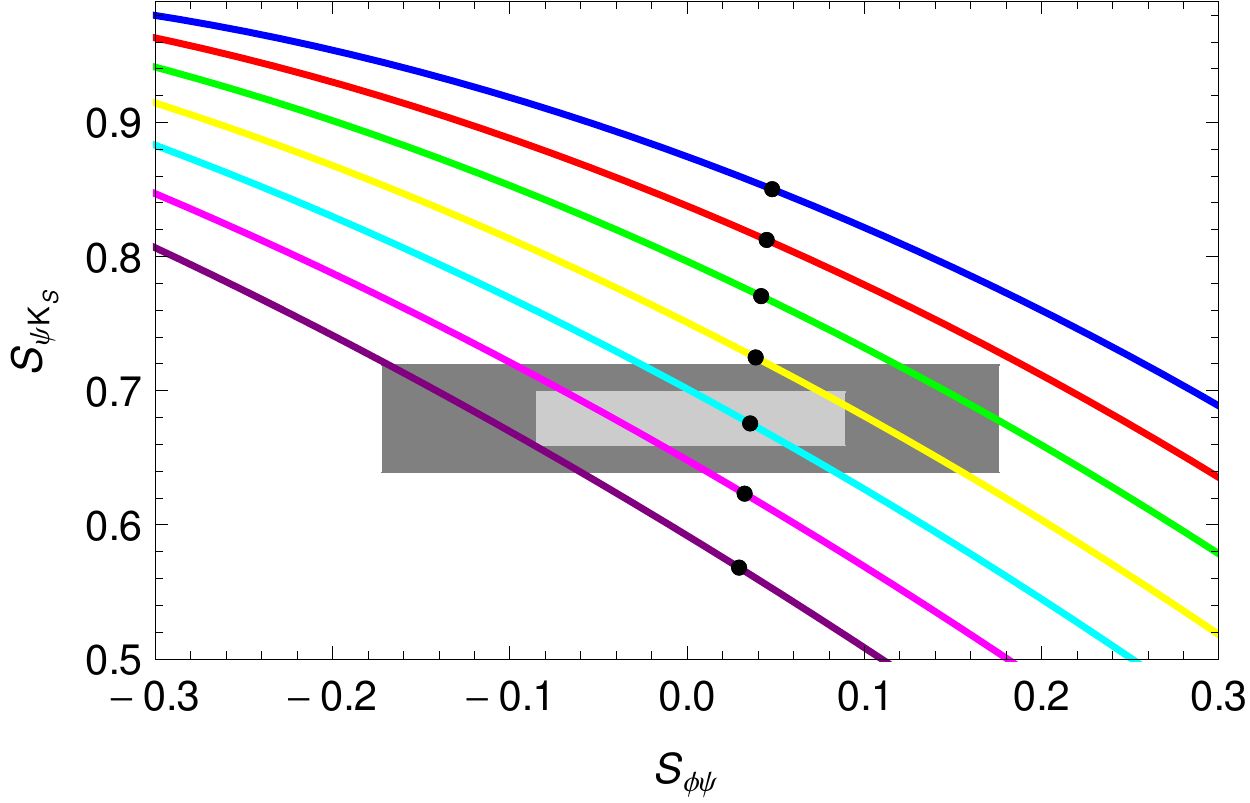}
 \caption{$S_{\psi K_S}$ versus $S_{\psi \phi}$ in  models with 
 $U(2)^3$ symmetry for different values of $|V_{ub}|$. From top to bottom: $|V_{ub}| =$ $0.0046$ (blue), $0.0043$ (red), $0.0040$
(green),
 $0.0037$ (yellow), $0.0034$ (cyan), $0.0031$ (magenta), $0.0028$ (purple). Light/dark gray: experimental $1\sigma/2\sigma$
 region \cite{Buras:2012sd}.}\label{fig:SvsSb}
\end{figure}

In models with $U(2)^3$ symmetry \cite{Barbieri:2011ci,Barbieri:2011fc,Barbieri:2012uh,Barbieri:2012bh,Crivellin:2011fb,Crivellin:2011sj,Crivellin:2008mq} these 
new phases are equal to each other: $\varphi_{B_d}=\varphi_{B_s}$.
As pointed out in \cite{Buras:2012sd} this equality of new phases implies 
not only  the correlation between these two asymmetries but also 
 a  triple 
$S_{\psi K_S}-S_{\psi\phi}-|V_{ub}|$ correlation which will provide a crucial test of this NP scenario.
This is shown in Fig.~\ref{fig:SvsSb} for fixed $\gamma
= 68^\circ$. Varying $\gamma$ between $63^\circ$ and $73^\circ$ does not 
change the result significantly.
We note that negative $S_{\psi\phi}$ is only possible for small $|V_{ub}|$, in the ballpark of the exclusive value, while 
for inclusive $|V_{ub}|$, $S_{\psi\phi}$ is always larger than the SM prediction. The latter case is the only possibility in the ${\rm 2HDM_{\overline{MFV}}}$ 
model for which the correlation shown in  Fig.~\ref{fig:SvsSb} also applies, 
but only for inclusive values of  $|V_{ub}|$. Therefore, in the latter 
model a satisfactory description of the data for $S_{\psi K_S}$ requires 
$S_{\psi\phi}\ge 0.15$, that is $2\sigma$ above the present central LHCb value.

The plot in  Fig.~\ref{fig:SvsSb} indicates that if 
the $U(2)^3$ flavour symmetry in the minimal version turns out to be true, one can determine $ |V_{ub}|$ by means of precise measurements
of $S_{\psi K_S}$ and $S_{\psi \phi}$ with small hadronic uncertainties.
For more details see  \cite{Buras:2012sd,Girrbach:2012gz}

\section{Summary}
In this short note I have emphasized the important roles of  $|V_{ub}|$, 
lattice calculations and in particular of correlations between various 
observables. Our simplified analysis shows that once  $|V_{ub}|$ will be 
precisely determined and non-perturbative parameters calculated precisely 
by lattice simulations, the different patterns of flavour violation in 
various NP models will be clearly seen.

\section*{Acknowledgments}
I thank the organizers of FLASY12 for the opportunity to give this talk
 and my collaborators for an enjoyable and fruitful time we spent together.
I thank Jennifer Girrbach for  comments on
this manuscript. I acknowledge financial support by  ERC Advanced Grant project ``FLAVOUR''(267104). This report carries the number ERC-28.


\bibliography{buras}
\bibliographystyle{apsrev4-1}


%% file: Papers/calibbi.tex

\chapter[On the messenger sector of (SUSY) flavour models (Calibbi)]{On the messenger sector of (SUSY) flavour models}
\vspace{-2em}
\paragraph{L. Calibbi}
\paragraph{Abstract}
We discuss the phenomenological consequences of the messenger fields that constitute the UV completion of generic flavour models, with 
particular emphasis on their contribution to flavour changing operators and their impact on the soft SUSY breaking terms. 

\section{Introduction}

Models based on new ``horizontal'' symmetries of flavour represent a common approach to account for the observed hierarchies 
of fermion masses and mixing \cite{Froggatt:1978nt,Leurer:1992wg,Leurer:1993gy}. 
In this kind of models, the Standard Model (SM) fermions transform under the flavour symmetry, 
which is spontaneously broken by the vevs of scalar fields called flavons.
Small Yukawa couplings are forbidden at the renormalisable level and only arise from higher-dimensional operators involving suitable powers of the flavons as determined by the symmetry. The flavour hierarchies are then explained by small order parameters given by ratios of the flavon vevs and the UV cutoff scale. This scale itself remains undetermined and can in principle be as large as the Planck scale. In case it is smaller, one can interpret this cutoff as the typical mass scale of new degrees of freedom, the so-called ``flavour messengers''. The dynamics of this sector may have important impact on low-energy physics. 
If its characteristic scale is relatively small, the unavoidable contributions to flavour changing 
and CP violating operators can give sizeable deviations from the SM predictions and strongly constrain the messenger scale and/or the structure
of the Yukawa matrices \cite{Calibbi:2012at}.
On the other hand, if the messenger scale is very high, such direct effects are irrelevant, but the messenger sector 
can still have important consequences both for the Yukawa couplings and the flavour structure of the sfermion masses 
in supersymmetric models \cite{Calibbi:2012yj}.

\section{The messenger sector}

Instead of working in a specific model, we consider a generic flavour symmetry group $G_F$ spontaneously broken by the vevs of 
the flavon fields $\phi_I$. The SM Yukawa couplings arise from higher-dimensional $G_F$-invariant operators involving 
the flavons \cite{Froggatt:1978nt,Leurer:1992wg,Leurer:1993gy}:
\begin{align}
\mathcal{L}_{yuk} & = y_{ij}^U ~\overline{q}_{L i} u_{Rj}\, \tilde{h} + y_{ij}^D~ \overline{q}_{L i} d_{Rj} \, h +{\rm h.c.} & 
 y_{ij}^{U,\,D} & \sim \prod_{I} \left( \frac{\langle\phi_I\rangle}{M} \right)^{n^{U,\,D}_{I,ij}}, 
\label{LC-eq:yuk}
\end{align}
where the suppression scale $M \gtrsim \langle\phi_I\rangle $ is the typical scale of the flavour sector dynamics. The coefficients of the effective operators are assumed to be $\mathcal{O}(1)$, so that the hierarchy in the Yukawa matrices arise exclusively from the small order parameters $\epsilon_I \equiv {\langle\phi_I\rangle}/{M}$. 
The transformation properties of the SM fields and the flavons under $G_F$ are chosen such that the parameters $\epsilon_I$ together with the exponents $n^{U,D}_{I,ij}$ reproduce the observed flavour hierarchies.
\begin{figure}[t]
\centering
\includegraphics[scale=0.65]{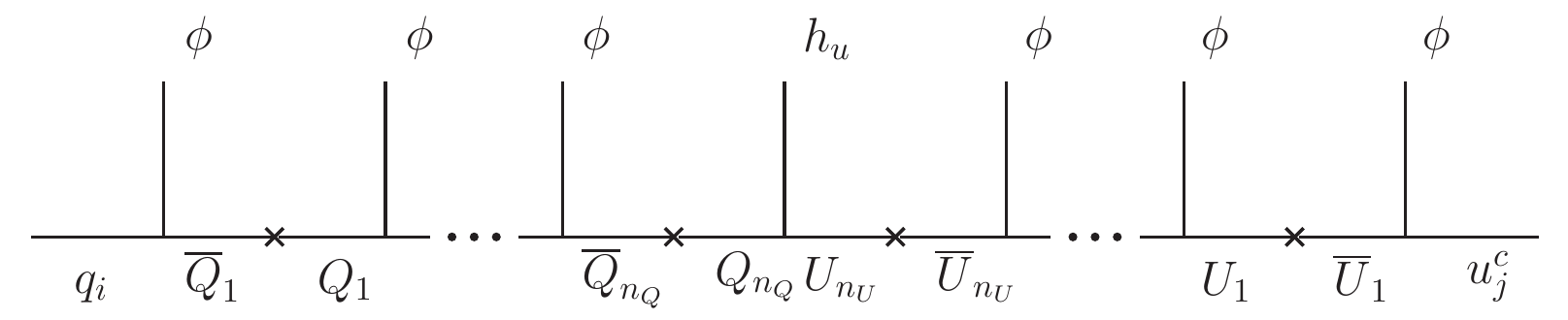}
\includegraphics[scale=0.65]{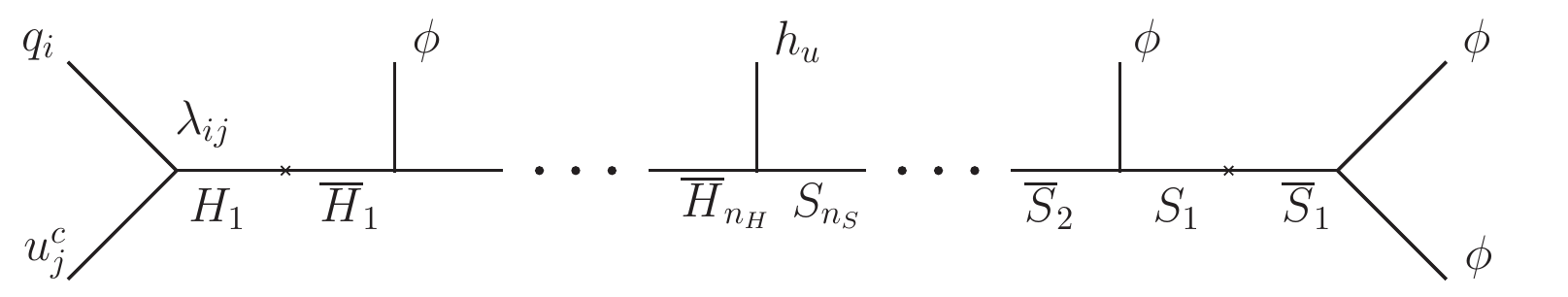}
\caption{Schematic supergraphs for Fermion (up) and Higgs (down) UV completions.\label{LC-chains}}
\end{figure}

In order to UV-complete models of this kind, one has to introduce new fields at the scale $M$. 
These messenger fields are in vectorlike representations of the SM gauge group and charged under $G_F$. In order
to generate the effective Yukawas of Eq.~(\ref{LC-eq:yuk}), the messengers must couple to SM fermions and flavons. 
Depending on their nature, they mix either with the SM fermions or with the SM Higgs. 
The first possibility corresponds to introducing vectorlike fermions with the quantum numbers of the SM fermions 
($Q_\alpha + \overline{Q}_\alpha,~U_\alpha + \overline{U}_\alpha,~D_\alpha   +\overline{D}_\alpha~\dots$), 
the second case to scalar fields with the quantum numbers of the SM Higgs field, possibly together with heavy SM singlets 
($H_\alpha + \overline{H}_\alpha,~S_\alpha +\overline{S}_\alpha$). 
The two possibilities are illustrated by the supersymmetric graphs of Fig.~\ref{LC-chains}.
In the fundamental theory, small fermion masses arise from a small mixing of light and heavy fermions for the first possibility, while they arise from small vevs of the new scalars in the second case. We refer to the two cases as ``Fermion UV completion'' (FUVC) and ``Higgs UV completion'' (HUVC), respectively.

More details on the construction of viable sets of messengers are provided in \cite{Calibbi:2012yj}. 
Here, we just want to highlight an interesting feature of HUVC that allows to enforce texture zeros in the Yukawa matrices in a very simple way. 
From the second graph of Fig.~\ref{LC-chains} it is clear that a specific Yukawa entry can only arise if the corresponding coupling to a heavy Higgs is present.
Although a certain Yukawa entry would be allowed by the flavour symmetry, if the Higgs field with the correct transformation properties under $G_F$ is missing, 
such entry vanishes in the fundamental theory and remains zero in the low-energy effective theory.\footnote{This elegant possibility to produce texture zeros has been first outlined in~\cite{Ramond:1993kv}.}

\section{Phenomenology of low-energy messengers}

\begin{figure}[t]
\centering
\includegraphics[scale=0.55]{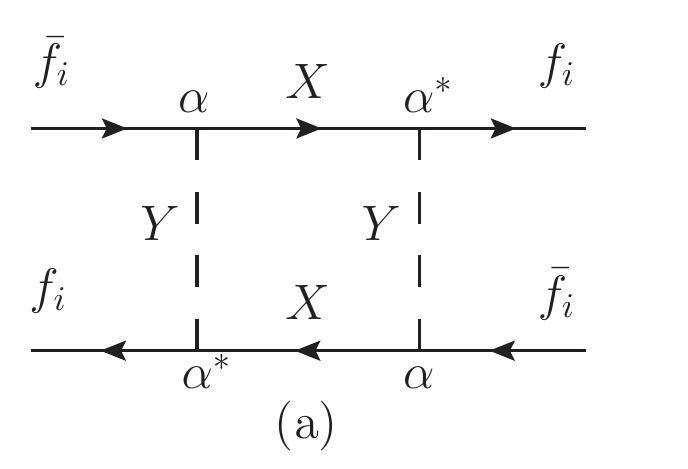}
\includegraphics[scale=0.5]{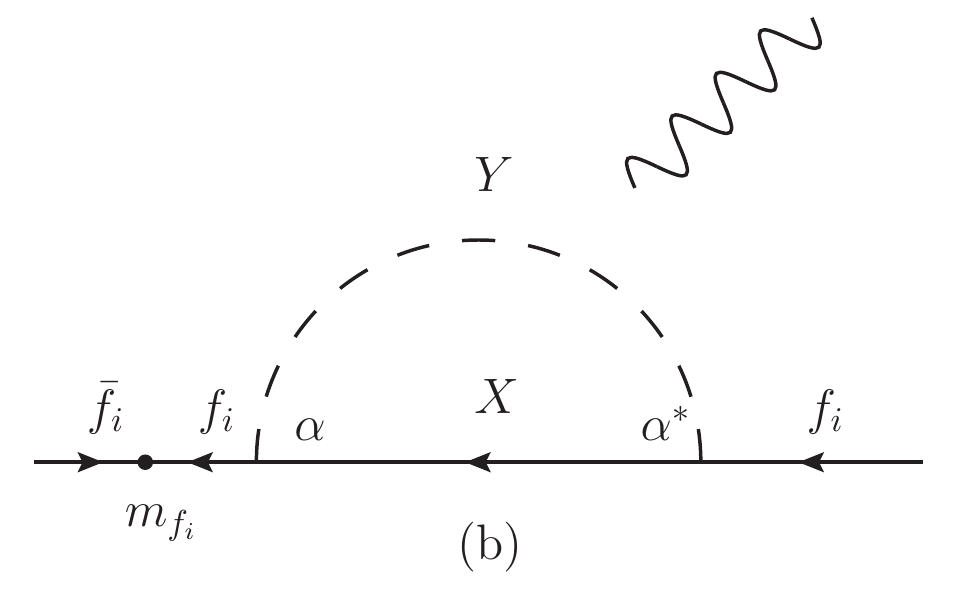}
\includegraphics[scale=0.55]{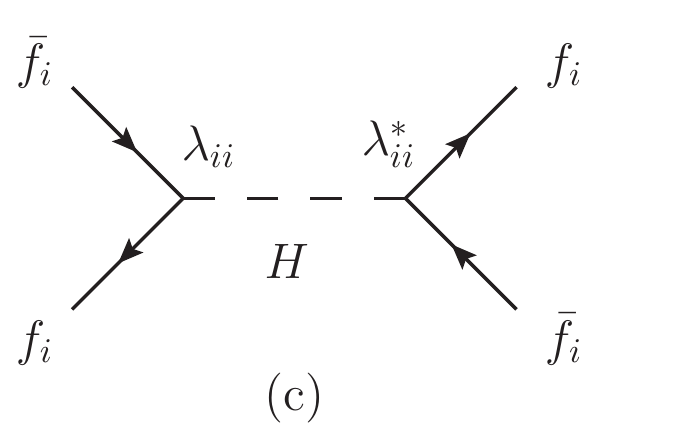}
\caption{Schematic diagrams responsible for the arising of flavour-violating operators\label{LC-fig:diag}.}
\end{figure}

We now discuss the effective flavour-violating operators that arise from messenger exchange independently of the details of the particular flavour model \cite{Calibbi:2012at}. 
From Fig.~\ref{LC-chains} we see that, for both kinds of UVC, couplings of the form 
$\alpha \, \overline{f}_{i} X Y$ must be present in the messenger Lagrangian. Here $\alpha \sim \mathcal{O}(1)$, $f_{i}$ is a (mainly) light fermion 
and $X$ and $Y$ are a fermion and a scalar of which at least one is a heavy messenger. 
As an example, we see that from box diagrams as in Fig.~\ref{LC-fig:diag}a one obtains effective {\it flavour-conserving} operators of the kind
$|\alpha|^4 (\overline{f}_{i} \gamma^\mu f_{i})^2 /({16 \pi^2 M^2})$,
where $M$ is the heaviest mass in the loop and $\mathcal{O}(1)$ factors were neglected. 
Similarly, the same coupling enters a penguin diagram with a mass insertion in the external fermion line (see Fig.~\ref{LC-fig:diag}b) that generates a dipole operator of the form $m_i \, \overline{f}_{Li} \sigma^{\mu \nu} f_{Ri} F_{\mu \nu}$, where $m_i$ is the light fermion mass.
Finally there are also certain tree-level operators which unavoidably arise in HUVC as shown in Fig.~\ref{LC-fig:diag}c.

The unavoidable flavour-conserving operators generated by the diagrams in Fig.~\ref{LC-fig:diag} give rise to flavour-violating operators, after rotating the
light fermions to the mass basis.
In abelian models there is no reason for cancellations among different contributions to a FCNC operator generated by such rotation, 
as the messengers have different $\mathcal{O}(1)$ couplings to light fermions by construction.
In non-abelian models these couplings can be universal (controlled by the symmetry). Each flavour transition is then additionally suppressed 
by a factor that depends on the flavon vevs responsible for universality breaking (see \cite{Calibbi:2012at} for further details). 
In summary we can obtain minimal predictions for the coefficients of certain flavour-violating effective operators, which do not depend on the details of the flavour model but only on the rotation angles that connect flavour and fermion mass basis. 
They can be compared with the experimental bounds on flavour-violating operators \cite{Calibbi:2012at} to constrain the messenger scale.

As an illustration, we show here 
the results in the hadronic 1-2 sector. In Table~\ref{LC-KDtab} the lower bounds on the messenger scale (in TeV) are shown for different combinations
of the left-handed and right-handed quark rotations (with $\epsilon$ being of the order of the Cabibbo angle).
Since the left-handed rotation must be $\mathcal{O}(\epsilon)$ either in the up or in the down sector or both, to account for the Cabibbo angle, 
the messenger scale must be larger than the smallest entry in Table~\ref{LC-KDtab}. Up to unknown $\mathcal{O}(1)$ coefficients, 
one can therefore obtain an overall minimal bound on the messenger scale, $M\gtrsim 20$ TeV. 
Since in non-abelian models there are additional suppression factors, the minimal effects alone do not exclude the possibility that the messenger fields of such models could be as light as a TeV and therefore in the reach of the LHC. 

Let us finally mention that the minimal bound discussed above does not prevent effects in $B_q - \overline{B}_q$ mixing and LFV decays
in the reach of currently running and future experiments, with the possibility of peculiar correlations such as 
${\rm BR}(\mu \to eee )$/${\rm BR}(\mu \to e \gamma) \sim \mathcal{O}(10)$ in the HUVC case, where $\mu \to eee$ arises at tree-level from Higgs messengers exchange \cite{Calibbi:2012at}.
\begin{table}[t]
\centering
\footnotesize
\begin{tabular}{|c c || c | c |c |c | }
$\theta^{DL}_{12}$ & $\theta^{DR}_{12}$ & HUVC & HUVC$^*$ & FUVC &FUVC$^*$ \\
\hline
\hline
$\epsilon$ & 0 & $ 19  $& $310 $ & $ 19 $ & $ 310$ \\
$\epsilon$ & $\epsilon$ & $ 3,400  $& $ 54,000 $ & $ 19 $ & $ 310$ \\
$\epsilon$ & $1$ & $4,900 $& $ 80,000 $ & $42 $ & $680 $ \\
0 & $1$ & $ 42  $& $680 $ & $42 $ & $680 $ \\
\hline
\hline
\end{tabular}
\vspace{0.3cm}
\begin{tabular}{|c c || c | c |c |c | }
$\theta^{UL}_{12}$ & $\theta^{UR}_{12}$ & HUVC & HUVC$^*$ & FUVC &FUVC$^*$ \\
\hline
\hline
$\epsilon$ & 0 & $ 27 $& $ 51 $ & $ 27 $ & $ 51 $ \\
$\epsilon$ & $\epsilon$ & $ 1,100 $& $ 2,200 $ & $27$ & $51 $ \\
$\epsilon$ & $1$ & $1,700 $& $ 3,200$ & $58 $ & $110  $ \\
0 & $1$ & $ 58    $& $110 $ & $58 $ & $ 110  $ \\
\hline
\hline
\end{tabular}
\caption{\label{LC-KDtab}Constraints from $K - \overline{K}$ (up) and  $D - \overline{D}$ (down) mixing  on the messenger scale in TeV for Higgs and fermion UV completions with real and complex (*) rotations angles.}.
\end{table}

\section{Messenger-induced radiative effects in SUSY flavour models}

Given the large number of messengers that one typically has to introduce, flavour theories require very heavy messengers ($M \gtrsim 10^{10}$ GeV ) to remain perturbative up to $M_{\rm Planck}$~\cite{Calibbi:2012yj}. This implies that all direct low-energy effects discussed above vanish. However, in SUSY even such high scales can have an impact on TeV scale physics through SUSY particles. Sfermion masses are determined by the underlying mechanism of SUSY breaking, and are usually generated at very high scales as well. This means that the messenger sector can interfere with the SUSY breaking sector. As the messenger sector strongly violates flavour universality by construction, it can easily induce flavour violation in the sfermion masses with drastic consequences for low-energy observables 
(for a review see \cite{Altmannshofer:2009ne}).

The most interesting consequence of the messenger sector on the sfermion masses is the radiative breaking of flavour universality.
Even in presence of a mechanism of SUSY breaking that generates universal sfermion masses at a scale $M_S$ (as is the case of Gauge Mediation), 
if $M_S$ is above the messenger scale, universality is spoiled by messenger loop corrections \cite{Calibbi:2012yj}, as in general happens in presence 
of flavour-dependent couplings of sfermions with new fields beyond the MSSM \cite{Hall:1985dx}.
The starting point is a universal sfermion mass matrix at $M_S$. When this matrix is evolved down to the scale $M$ where the messengers decouple, all entries receive RG corrections (for simplicity we restrict to 1st and 2nd generation):
\begin{align}
\tilde{m}_{ij}^2 (M)= \begin{pmatrix} \tilde{m}_0^2 + \Delta \tilde{m}^2_{11}& \Delta \tilde{m}^2_{12} \\ \Delta \tilde{m}^2_{21} &  \tilde{m}_0^2 + \Delta \tilde{m}^2_{22} \end{pmatrix}.
\end{align}
The final evolution to the soft SUSY breaking scale scale is determined by gauge (hence flavour universal) terms and by Yukawa couplings that can be neglected in the case of the first two generations. The 1-2 entry in the super-CKM basis is then approximately given by: 
\begin{align}
\label{LC-2contrib}
\tilde{m}_{12}^2 \approx \Delta \tilde{m}^2_{12} + \left( \Delta \tilde{m}^2_{22} - \Delta \tilde{m}^2_{11} \right) \theta_{12},
\end{align}
where $\theta_{12}$ denotes the (complex) rotation angle in the fermion sector under consideration. For simple flavour models like $U(1), U(1)^2$ or $SU(3)$ it is easy to see that the second term is always larger or equal than the first one, provided the rotation angle does not vanish \cite{Calibbi:2012yj}.
Therefore in the following we restrict our attention on the second term in Eq.~(\ref{LC-2contrib}). 

The sfermion mass splitting depends on the RG running. Since the sfermion-messenger couplings are $\mathcal{O}(1)$ 
the RG coefficients are in general large ($\sim 10$ is a conservative estimate \cite{Calibbi:2012yj}). 
In abelian models there is no extra suppression, because different generation sfermions couple with different $\mathcal{O}(1)$ couplings to the messengers. Instead in non-abelian models there can be additional suppression as above, since different generation sfermions can be embedded 
in the same representation under the flavour group, which implies universal couplings to the messengers \cite{Calibbi:2012yj}. 
To estimate the mass splitting we consider the case of abelian flavour symmetries, and keep in mind possible non-abelian suppressions.\footnote{The abelian case is also relevant in non-abelian models with (s)fermions transforming as singlets under $G_F$.} We can then estimate the off-diagonal sfermion mass at leading log by:
\begin{align}
\label{LC-m12est}
\tilde{m}^2_{12} \approx \left(  \Delta \tilde{m}^2_{22} - \Delta \tilde{m}^2_{11} \right) \theta_{12} \approx   \theta_{12}  \frac{\tilde{m}_0^2}{16 \pi^2} 10 \log\frac{M_S}{M} .
\end{align}
We notice that the radiatively generated $\tilde{m}^2_{12}$ is roughly of the same order as one would expect for a tree-level sfermion mass matrix 
only constrained by the flavour symmetry (see e.g.~\cite{Lalak:2010bk}). 
The corresponding mass insertion $\delta_{12} \equiv \tilde{m}^2_{12}/\sqrt{\tilde{m}^2_{11}\tilde{m}^2_{22}}$ then reads:
\begin{align}
\label{LC-predMI}
\delta_{12}^{\rm ab.} & \approx  \frac{ \theta_{12}}{16 \pi^2} 10~ \mathcal{R}\log\frac{M_S}{M}\,,    
\end{align}
where $\mathcal{R}$ is the suppression due to the gaugino-driven evolution of the diagonal entries.\footnote{$\mathcal{R}$ is typically $\mathcal{O}(1)$ in the case of sleptons, while for squarks it ranges from $\mathcal{O}(1)$ down to $\mathcal{O}(0.1)$.} 

In \cite{Calibbi:2012yj}, the above estimate is compared to the various bounds on the mass insertions obtained from FCNC and LFV processes.
Since the effect in Eq.~(\ref{LC-predMI}) depends only on the rotation angle and the ratio of SUSY and messenger scale, for a given ratio one gets 
an upper bound on the real and imaginary part of the rotation angle, which can be used to constrain the Yukawa matrices.
Such bounds are unavoidable whenever $M_S > M$, which includes mSUGRA. 
For instance, one typically finds $\theta_{12}\lesssim 10^{-2}$ both in the leptonic and hadronic sectors.
As either $\theta^{UL}_{12}$ or $\theta^{DL}_{12}$ must be $\mathcal{O}(\epsilon) \approx 0.2$, this puts abelian models in troubles, even under the strong assumption of universal soft masses at $M_S$. The constraints are less severe in non-abelian models (that gives an additional suppression $\lesssim \epsilon^2$ in the 1-2 sector), so that non-abelian models are preferred from what concerns the radiative effects discussed above.

\section{Conclusion}
We have shown in a model-independent way that the messenger sector can have a strong impact on the low-energy phenomenology of flavour models, as well as on the
structure of Yukawa matrices. In particular, low-energy flavour models are strongly constrained by flavour and CP violating processes induced by messenger
exchanges, while high-energy messengers can still affect the flavour structure of the sfermion masses in SUSY models.

\section*{Acknowledgments}
I am grateful to S.~Lalak, S.~Pokorski and R.~Ziegler for collaborations on which this talk is based. 
I would also like to thank the organisers of FLASY12 for giving me the opportunity of presenting my work in
a nice and fruitful atmosphere. 

\bibliography{calibbi}
\bibliographystyle{apsrev4-1}


%% file: Papers/covi.tex

%
%
%
%
%
%

\chapter[Gravitino Dark Matter with colored NLSP (Covi)]{Gravitino Dark Matter with colored NLSP}
\vspace{-2em}
\paragraph{L. Covi}
\paragraph{Abstract}
We will review the case for gravitino Light Supersymmetric Particle (LSP) 
and Dark Matter and discuss in detail the cosmological constraints 
for a colored Next-to-Lightest or Next-to-Next-to Lightest Supersymmetric 
Particle (NLSP or NNLSP respectively).

\section{Introduction}

The particle identity of Dark Matter is one of the still open questions of both 
cosmology and particle physics. Indeed the evidence for Dark Matter is surely 
one of the stronger hints for physics beyond the Standard Model, since no candidate
for Dark Matter is to be found among the known Standard Model particles 
(neutrinos are too light and constitute Hot Dark Matter and their density is 
bounded to be at most 10-20\% of the total Dark Matter density).
It is therefore imperative to look for interesting Dark Matter candidates in model
of physics beyond the Standard Model.  Since supersymmetry is a leading
candidate for such an extension, providing e.g. a solution for the hierarchy problem
and the possibility of Grand Unification at a high scale, we will here concentrate 
on a supersymmetric candidate, the gravitino, superpartner of the graviton~(for
a introduction to the graviton multiplet and supergravity, see e.g. \cite{Wess:1992cp}). 
Such scenario  is very attractive since the solution of the Dark Matter problem 
is then contained in the gravitational sector of the theory, in some sense in 
the supersymmetrization, of gravity.
In general the question of flavour is considered to be independent of the question 
of Dark Matter, but in certain cases the "flavour" of the NLSP can make a huge 
difference in the phenomenology of a gravitino Dark Matter scenario.
Also gravitino Dark Matter can be in some cases compatible with thermal 
leptogenesis (for a review see e.g. \cite{Davidson:2008bu}), which is connected 
to lepton flavour. Thermal leptogenesis in the simplest realization needs reheat 
temperatures above $10^9 $ GeV~\cite{Buchmuller:2004tu, Davidson:2008bu}.
We will in the following discuss in particular colored N(N)LSPs, like the lightest stop
or the gluino and see if some parameter space compatible with thermal
leptogenesis and present LHC searches can still be found.

\section{Gravitinos as Dark Matter}

Gravitinos are natural candidates for Dark Matter within supersymmetric models.
They were actually proposed as thermal Dark Matter even before the neutralino
by Pagels \& Primack in 1982 \cite{Pagels:1981ke}.  In such models though, 
the gravitinos  have to be very light since their number density as relativistic relics is large, i.e.
\begin{equation}
\Omega_{3/2} h^2 \sim 0.1 \left( \frac{m_{3/2}}{0.1 \mbox{keV}} \right)
\left( \frac{g_*}{106.75} \right)^{-1}
\end{equation}
where $m_{3/2}$ is the gravitino mass and $g_*$ are the effective number of degrees 
of freedom thermalized at the time of gravitino decoupling. Such a small gravitino
mass corresponds to Warm/Hot Dark Matter and it excluded by structure formation.

On the other hand, if the gravitinos never reach thermal equilibrium, 
they can be heavier and nevertheless produced in the right number density
to be Dark Matter by scattering processes in the plasma involving in particular 
the gauge interactions. Since those processes are mediated by a non-renormalizable
dimension 5 operator, the resulting particle density is linear in the
thermal bath temperature and the largest gravitino population is produced
at the highest temperature reached by the thermal plasma, $T_{RH} $.
The gravitino energy density from thermal scattering has been the study
of detailed work in the recent years \cite{Bolz:2000fu,Pradler:2006qh,Rychkov:2007uq} 
and the result is given by 
\begin{equation}
\Omega_{3/2} h^2 \sim 0.3  \left( \frac{m_{3/2}}{1 \mbox{GeV}} \right)^{-1}
 \left( \frac{T_{RH}}{10^{10} \mbox{GeV}} \right)
\sum_i c_i  \left( \frac{M_i}{100 \mbox{GeV}} \right)^2
\end{equation}
where $c_i $ are coefficients of order 1 and $ M_i $ denote the three 
gaugino masses at EW temperature (RGE effects to $T_{RH} $ are included 
in the coefficients $c_i$).

Note that if the gravitino is not the LSP, such a population of gravitinos decays
into the other supersymmetric particles quite late in the cosmological history, 
during or after Big Bang Nucleosynthesis causing the infamous "gravitino problem"
\cite{Khlopov:1984pf, Ellis:1984er}. Therefore strong constraints on the gravitino 
density and therefore the reheat temperature can be set in that 
case (for recent results see \cite{Kawasaki:2008qe}).

\section{Stable gravitino and colored NLSP}

BBN bounds on colored long-lived relics are even stronger than those
for charged or neutral particles since the colored particles can become
bounded within the nuclei and change the rates of the nuclear processes
during Nucleosynthesis. Such constraints have been recently computed
by \cite{Kusakabe:2009jt} and are very steep in the particle lifetime. They approximately
result in constraining the NLSP lifetime to be shorter than 200 s.
Even increasing the colored state annihilation cross-section thanks to
the Sommerfeld effect does not relax those bounds substantially.
The only possibility is to consider a sufficiently heavy NLSP so
that the decay happens early enough. This gives approximately,
assuming the dominant decay channel of the stop is to top and gravitino,
\begin{equation}
m_{\tilde t} \leq 1 \mbox{TeV} \left( \frac{m_{3/2}}{10 \mbox{GeV}} \right)^{2/5}.
\end{equation}
A slightly weaker bound can be obtained for the gluino NLSP, which
annihilates a bit more efficiently.

\begin{figure}
\begin{center}
\includegraphics[width=12cm]{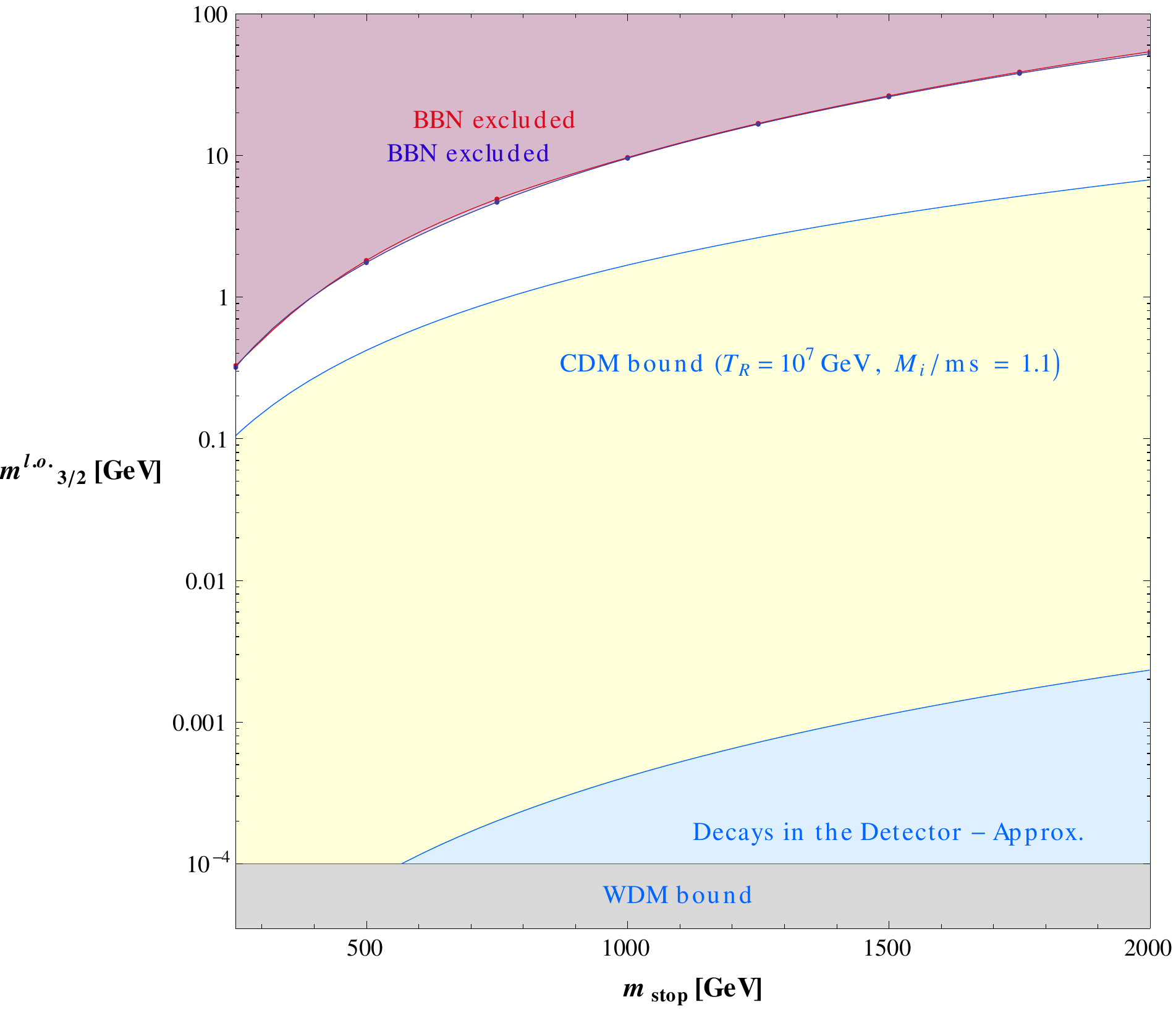}
\caption{The bounds on the scenario of gravitino stable LSP with 
a light stop NLSP in the plane of the stop versus gravitino mass: 
the upper region is excluded by the BBN constraints on the stop
lifetime as explained in the text, while the lower part of the plane
is excluded since the gravitino would be WDM. The yellow
region indicated by "CDM bound" gives a too large density
of gravitino Dark Matter for the reheat temperature indicated
and the ratio of the gaugino masses to the gravitino mass of $1.1$. 
In the blue corner on the right side, the stop lifetime is short
enough to allow for the stop to decay into top and gravitino in the
LHC detectors.
}
\end{center}
\end{figure}

The BBN bounds for the stop case are shown in the Figure~1.
 The two lines practically overlapping in the top part of the
 figure give the lifetime constraints from BBN depending on
 the stop relic density, computed either without or with the Sommerfeld 
 enhancement for the stop annihilation cross-section as given 
 in \cite{Berger:2008ti}.
The region indicated by "CDM bound" corresponds to a too large 
density of gravitino Dark Matter for the reheat temperature indicated
and a minimal ratio of gaugino to gravitino masses of 1.1.
In the white region just at the upper boundary of that line, the
gravitino has exactly the right density to be Dark Matter for a
reheat temperature of $10^{7}$ GeV. For larger reheat temperatures
the curve moves up and touches the BBN bounds for the range
of stop masses considered for a reheat temperature of 
about $10^8$ GeV, not quite compatible with "vanilla" thermal 
leptogenesis. 

The characteristic signature of this scenario at colliders is that of a
metastable stop: such a particle has been searched
by CMS with no evidence for any excess reaching a lower limit on 
the mass  at present of about  800 GeV \cite{Chatrchyan:2012sp}.

Another possibility, which relaxes the BBN constraints, is to have a 
small breaking of R-parity, allowing the stop NLSP to decay before
BBN and therefore relaxing the constraints \cite{Buchmuller:2007ui}.
In that case the dominant decay channel depends on the R-parity 
breaking model. For bilinear R-parity violation, the decay into
a b-quark and lepton is dominant and the lepton flavour gives
indication on the bilinear R-parity breaking pattern \cite{Covixxx}.

\section{Gluino NNLSP}

One way to avoid the BBN constraints for colored relics is to have
a neutral NLSP with a colored NNLSP. In this case, if the two
particles are sufficiently degenerate in mass, co-annihilation
between the neutral and colored particles strongly reduces the
NLSP density and relaxes substantially the BBN constraints.
It has been shown in \cite{Covi:2010au}, that the co-annihilation is particularly
effective for masses below 300 GeV and for degeneracy 
between the neutralino and the gluino of the order of 1-3\%
, so that then the neutralino density is so suppressed that 
even reheat temperatures above $10^9$ GeV are allowed.

In that particular case, the gluino decays promptly into
a neutralino and a gluon or a light quark-antiquark pair,
but the visible particles carry very small energy and have 
on average such very low $p_T$ that they escape from the
usual searches for Missing $E_T$ and jets.
Then the most sensitive channel is the one including
one single monojet, either from Initial State Radiation or
from the associate production of gluino and squark, with 
the squark providing a highly energetic jet in its decay into
gluino. This channel has been studied by the LHC collaborations
for the case of graviton production in extra-dimensional
scenarios and Dark Matter searches ~\cite{:2012fw, Chatrchyan:2012me}.
Recently such results have been also reinterpreted for the
case of a degenerate spectrum and they put severe
constraints on this case, excluding gluino masses
up to 450-500 GeV \cite{Dreiner:2012gx} and therefore also the preferred
region around 300 GeV.

\section{Conclusion}

The gravitino is a good DM candidate, which can reconcile a relatively high 
reheat temperature with supersymmetry, especially with colored NNLSP or NLSPs.
Big Bang Nucleosynthesis  constrains the lifetime and density of the NLSP, 
also in case of colored relics, and tends to point naturally to a heavy spectrum.

We discussed the case of stop or gluino NLSPs and the case of gluino NNLSP
with neutralino NLSP. In the first case it is difficult to find parameter space
in agreement with thermal leptogenesis up to masses of the NLSP of order
2 TeV, which could still be in the reach of the next phase of the LHC.
Still the option remains to add a small amount of R-parity to evade the
BBN constraints and allow for larger gravitino masses.
In the second case, higher reheat temperatures are in principle allowed for
degenerate masses of the NLSP/NNLSP around 300 GeVs, but such masses
are now excluded by the LHC searches.

In conclusion, gravitino Dark Matter is compatible also with a relatively heavy 
SUSY spectrum for low reheat temperature, but there is still the chance that
some "exotic" signal at the LHC, like a charged metastable particle or a 
displaced vertex, will show up soon and point us to this specific scenario.

\section*{Acknowledgments}
This project is supported by the German-Israeli Foundation for scientific research and
development(GIF). The author also acknowledges financial support by the 
EU FP7 ITN Invisibles (Marie Curie Actions, PITN-GA-2011-289442)

The author would like to thank the organizers for the very interesting meeting and
all her collaborators for the enjoyable work together. Special thanks go to 
Federico Dradi who produced the figures in time for the meeting.

\bibliography{covi}
\bibliographystyle{apsrev4-1}

%% file: Papers/debottamdas.tex

%
%
%
%
%
%

\chapter[Higgs Mediated Lepton Flavour Violation in the Supersymmetric Inverse Seesaw Model (Das)]{Higgs Mediated Lepton Flavour Violation in the Supersymmetric Inverse Seesaw Model}
\vspace{-2em}
\paragraph{D. Das}

\paragraph{Abstract}
We have investigated Higgs mediated lepton flavor violating observables in the inverse
seesaw framework of Minimal Supersymmetric Standard Model. We have shown that, 
lightness of the sterile (s)neutrinos can enhance the effective
coupling $H/A-l_i-l_j$. As a consequence, all Higgs mediated flavor violating
observables are enhanced by as much as two orders of magnitude.
\section{Introduction}
Neutrino oscillations have provided one of the most intriguing experimental 
evidence towards the beyond Standard Model (SM) physics. Minimal Supersymmetric Standard Model (MSSM), one of the most popular extension of the Standard Model
can also accommodate neutral flavor oscillation when it is extended to include
the right handed neutrino superfields. The additional supersymmetric (SUSY) 
states with masses in the
TeV scale can provide contributions to charged lepton flavor violations (cLFV),
such as $l_i\to l_j\gamma$ or three body decays $l_i\to 3l_j$. Thus, any
cLFV signal, if observed, would clearly convey the indirect evidence for 
new physics.

The introduction of the right handed neutrino superfields in the SUSY 
theories naturally invites seesaw mechanism to embed with it. In the 
SUSY-seesaw theories, neutrino Yukawa couplings can induce mixing term
in the SUSY soft-breaking slepton mass matrices through renormalisation 
group evolution (RGE) of the aforementioned parameters. This in turn generates 
observable effects in the charged lepton flavor violation through the mixings 
in the slepton
mass matrices. However, in this seesaw scheme, requirement of $O(1)$ neutrino 
Yukawa couplings leads the right handed 
neutrino mass scale or seesaw scale
very close to the gauge coupling unification scale   
which is impossible to probe experimentally. 

On the contrary, inverse seesaw scenarios \cite{Mohapatra:1986bd} offers an appealing
alternative, where one can retain $O(1)$ neutrino 
Yukawa couplings while the right handed neutrino mass scale can reside near
the TeV scale. 
This at one hand offers testability by directly producing 
the sterile neutrinos at the Large Hadron Collider, while on the other hand,
can enhance the charged lepton flavor violating processes through the
unsuppressed lepton number {\em conserving} dimension-6 operator 
$\left(Y_\nu^{\dagger}\frac{1}{\left|M\right|^{2}}Y_\nu\right)$ ($M$ represents
right handed neutrino mass scale). Indeed, in view
of this strong potential, several phenomenological studies have been carried
out in the recent past.
 
The singlet neutrinos with masses at the TeV scale may significantly 
contribute to cLFV observables, 
irrespective of the supersymmetric states .
Supersymmetric realisations of the inverse seesaw may enhance these cLFV rates 
even further \cite{Deppisch:2004fa,Deppisch:2005zm}.
This particular work is devoted to the Higgs mediated charged lepton
flavor violation processes in the supersymmetric inverse seesaw framework 
~\cite{Abada:2011hm}. We have shown that the effective coupling $H/A-l_i-l_j$ 
can be
enhanced significantly, thanks to the comparatively light right-handed 
neutrinos and sneutrinos (which provide negligible contribution 
in the framework of a type I SUSY-seesaw). We find that this new contribution, in particular leads
to a significant enhancement of the several cLFV observables. 

\section{Inverse Seesaw Mechanism in the MSSM}\label{sec:mod}
Here, the MSSM field contents is augmented  
by three pairs of singlet superfields, $\widehat{\nu}^c_i$ and $\widehat{X}_i$ ($i=1,2,3$)
with lepton numbers $-1$ and $+1$, respectively. 
Consequently, the superpotential for the supersymmetric inverse seesaw model 
can be defined by
\begin{eqnarray}
{\mathcal W}&=& \varepsilon_{ab} \left[
Y^{ij}_d \widehat{D}_i \widehat{Q}_j^b  \widehat{H}_d^a
              +Y^{ij}_{u}  \widehat{U}_i \widehat{Q}_j^a \widehat{H}_u^b 
              + Y^{ij}_e \widehat{E}_i \widehat{L}_j^b  \widehat{H}_d^a \right. \nonumber \\
              &+&\left. Y^{ij}_\nu 
\widehat{\nu}^c_i \widehat{L}^a_j \widehat{H}_u^b - \mu \widehat{H}_d^a \widehat{H}_u^b \right] 
+M_{R_i}\widehat{\nu}^c_i\widehat{X}_i+
\frac{1}{2}\mu_{X_i}\widehat{X}_i\widehat{X}_i  ~,
\label{eq:SuperPot}
\end{eqnarray}
The information of inverse seesaw are encoded in the last two terms in Eq:
\ref{eq:SuperPot}. Here, $M_{R_i}$ represents the right-handed 
neutrino
mass term that conserves lepton number while $\mu_{X_i}$ violates the same
by two units. The terms $\widehat {\nu}^c_i
\widehat X_i$ and $\widehat X_i \widehat X_i$ are assumed to be diagonal in generation space. \\
\noindent
The soft SUSY breaking Lagrangian can be written as
\begin{multline}
-{\mathcal L}_{\rm soft}=-{\mathcal L}^{\rm MSSM}_{\rm soft} 
         +  m^2_{\widetilde \nu^c} \widetilde\nu^{c\dagger}_i \widetilde\nu^c_i
         +m^2_X \widetilde X^{\dagger}_i \widetilde X_i\\
     + \left(A_{\nu}Y^{ij}_\nu \varepsilon_{ab}
                 \widetilde\nu^c_i \widetilde L^a_j H_u^b +
                B_{M_{R_i}} \widetilde\nu^c_i \widetilde X_i 
      +\frac{1}{2}B_{\mu_{X_i}}\widetilde X_i \widetilde X_i
      +{\rm h.c.}\right),
\label{eq:softSUSY}
\end{multline}
where ${\mathcal L}^{\rm MSSM}_{\rm soft}$ denotes the soft 
SUSY breaking terms of the MSSM. In the above, for the singlet scalar states 
we assume $m^2_{X_i}=m^2_{X}$ and $m^2_{\widetilde{\nu}^c_i}=m^2_{\widetilde{\nu}^c}$. The 
parameters $B_{M_{R_i}}$ and $B_{\mu_{X_i}}$ represent the bilinear couplings for
the sterile neutrino states. Note that while 
the former conserves lepton number, the latter generates the
lepton number violating $\Delta L=2$ term. 

\medskip
Now we illustrate the pattern of light neutrino masses in the 
inverse seesaw model considering only one-generation case.
In the 
$\{\nu,{\nu^c},X\}$ basis
the $(3 \times 3)$ neutrino mass matrix can be written as
\begin{eqnarray}
{\cal M}&=&\left(
\begin{array}{ccc}
0 & m_D & 0 \\
m_D & 0 & M_R \\
0 & M_R & \mu_X \\
\end{array}\right) \ ,
\label{nmssm-matrix}
\end{eqnarray}
with $m_D= Y_\nu v_u$, yielding the mass eigenvalues ($m_1 \ll m_{2,3}$):
\begin{eqnarray}
 m_1 = \frac{m_{D}^2 \mu_X}{m_{D}^2+M_{R}^2} \, , ~~~~ 
 m_{2,3} = \mp \sqrt{M_{R}^2+m_{D}^2} + 
\frac{M_{R}^2 \mu_X}{2 (m_{D}^2+M_{R}^2)} \, . 
\label{masses}
\end{eqnarray}
The advantage of the inverse seesaw is that, here the lightness of the
smallest eigenvalue $m_1$ can be attributed to the 
smallness of $\mu_X$ ($\mu_X\simeq m_1$). Technically, 
such small value of $\mu_X$ is natural in the sense of 't~Hooft 
since in the limit $\mu_X\to 0$, the total lepton number symmetry 
is restored
. Thus the lepton number conserving mass 
parameters ($m_D$ and $M_R$) are completely unconstrained in this model.

Finally, the effective right-handed sneutrino mass term (Dirac-like) 
can be expressed as
$M^2_{\widetilde \nu^c_i} = m^2_{\widetilde \nu^c} + M_{R_i}^2 + 
\sum_j { |Y^{ij}_\nu|^2 v_u^2}$. 
Assuming $M_R \sim {\mathcal{O}}$(TeV), the effective sneutrino mass term also 
assumes $O(1)~$TeV, in clear contrast
to what occurs in the standard (type I) SUSY seesaw where it takes masses
$O(M_R)$. Such a light sneutrino (i.e.  
$M^2_{\widetilde \nu^c} \sim M^2_\text{SUSY}$) leads to the enhancement of Higgs 
mediated contributions to 
lepton flavour violating observables.
\section{Lepton flavour violation: Higgs-mediated contributions}
\label{lfv:diag}
In the SUSY seesaw framework, 
the neutrino Yukawa couplings, which are non-diagonal to accommodate the neutrino oscillation data are the only sources for flavour violation.
The presence of right handed neutrino 
would drive the soft SUSY breaking slepton mass parameter $m^2_{{\tilde L}_{ij}}$
(for $i\ne j$) to acquire non vanishing contribution at the weak scale.
Considering cMSSM/mSUGRA like boundary condition at the GUT scale, in the leading logarithmic approximation this
radiative effect is proportional to $Y_\nu$ 
~\cite{Borzumati:1986qx,Hisano:1995cp} and can be expressed as 
\begin{eqnarray}
(\Delta m_{\widetilde{L}}^2)_{ij}&\simeq&
-\frac{1}{8\pi^2}(3m_0^2+A_0^2) 
(Y_\nu^\dagger L Y_\nu)_{ij} \,, ~~ L=\ln\frac{M_{GUT}}{M_{R}} \,
\nonumber \\ 
&=&\xi (Y^\dagger_\nu Y_\nu)_{ij},
\label{slepmixing}
\end{eqnarray}
where for simplicity, we assume 
degenerate right-handed neutrino spectrum, $M_{R_i}=M_{R}$. As can be guessed
from Eq:\ref{slepmixing}, the factor $\xi$ would be enhanced in the inverse
seesaw framework compared to the standard (type I) SUSY seesaw, thanks to smallness of the right handed neutrino mass term. 

On the other hand, Higgs-mediated flavor violating processes are induced 
by the non-holomorphic Yukawa interactions $\bar D_RQ_LH_u^*$ at the one-loop level. This was first pointed out in the context of quark families 
in~\cite{Hall:1993gn}. On a similar note, in the lepton sector, 
the Higgs-mediated flavour violating couplings
are also induced at the one loop level by the 
non-holomorphic Yukawa term $\bar E_RLH_u^*$~\cite{Babu:2002et}.
Consequently, its role has been studied in the context of 
several lepton flavor violating processes like $\tau \rightarrow 3\mu$~
\cite{Babu:2002et}, $B_s \rightarrow \mu\tau$, 
$B_s \rightarrow e\tau$~\cite{Dedes:2002rh}, 
$\tau \rightarrow \mu \eta$~\cite{Sher:2002ew}. 
A detailed analyses of the several $\mu-\tau$ lepton
flavour violating observables $\tau \rightarrow \mu X$ 
($X = \gamma,e^+e^-,\mu^+\mu^-,\rho,\pi,\eta,\eta^\prime$) 
can be found in~\cite{Brignole:2004ah}.

The effective Lagrangian that describes the coupling of the neutral Higgs fields
to the charged leptons can be expressed as
\begin{eqnarray}
-{\cal L}^\text{eff}=\bar E^i_R Y_{e}^{ii} \left[ 
\delta_{ij} H_d^0 + \left(\epsilon_1 \delta_{ij} + 
\epsilon_{2ij} (Y_\nu^\dagger Y_\nu)_{ij} \right) H_u^{0\ast }
\right] E^j_L + \text{h.c.}  \,. 
\label{Leff}
\end{eqnarray}
The first term represents the usual Yukawa interaction, 
while the coefficient $\epsilon_1$ encodes
the corrections to the charged lepton Yukawa couplings. In the basis 
for diagonal charged lepton Yukawa couplings, 
the last term in Eq.~(\ref{Leff}), i.e. 
$\epsilon_{2ij} (Y_\nu^\dagger Y_\nu)_{ij}$, is in general  
non-diagonal which introduces the flavor violating Higgs coupling 
$H/A-l_i-l_j$. 

In the standard seesaw mechanism, the co-efficient $\epsilon_{2ij}$ 
encodes the sole contribution to the cLFV processes 
where LFV 
 is introduced via a radiatively induced non-diagonal terms 
in the slepton masses $(\Delta m_{\widetilde{L}}^2)_{ij}$ (see Eq. (\ref{slepmixing})) (For details see ref:~\cite{Abada:2011hm}). 

\begin{figure}
\begin{center}
\includegraphics[width=3.5in]{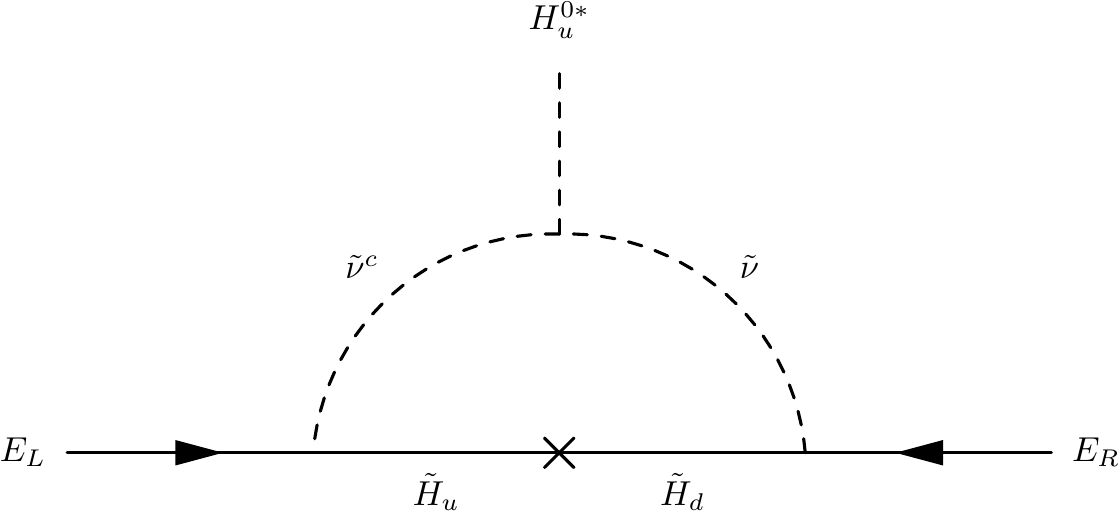}
\caption{Right-handed sneutrino contribution  to $\epsilon'_2$. 
}\label{2}
\end{center}
\end{figure}

Now, in the framework of the inverse SUSY seesaw,  
there is an additional diagram: the sneutrino-chargino mediated loop \footnote{Note that the 
large masses of $\widetilde \nu^c$ in the standard (type I) seesaw makes
this effect negligible, thus has not been taken into account in the literature.}, depicted in Fig.~\ref{2} which provides the leading contribution. 
This new contribution can be computed from
\begin{eqnarray}
\epsilon'_{2ij}= \frac{1}{16\pi^2} \mu A_\nu 
F_1(\mu^2,m^2_{\widetilde \nu_i},M^2_{\widetilde \nu^c_j}),
\label{newdiag}
\end{eqnarray}
\noindent
with
\begin{eqnarray}
F_1\left(x,y,z\right)&=& 
-\frac{xy\ln (x/y)+yz\ln (y/z)+zx\ln (z/x)}
{(x-y)(y-z)(z-x)} \,.
\end{eqnarray}
Here, we have parametrized the soft trilinear  term for the neutral leptons as $A_\nu Y_\nu$, and 
$A_\nu$ is a flavour independent real mass term. Consequently, the effective
Lagrangian is modified as
\begin{eqnarray}
-{\cal L}^\text{LFV} = \bar E^i_R Y_{e}^{ii}\epsilon^\text{tot}_{2ij} (Y_\nu^\dagger Y_\nu)_{ij} H_u^{0\ast }
 E^j_L + \text{h.c.}  \,, 
\label{Leff1}
\end{eqnarray}
with $\epsilon^\text{tot}_2=\epsilon_2+\epsilon_2'$.

Note that $\epsilon'_{2ij}$ does not require any LFV mass insertions, thus naturally dominate over 
$\epsilon_{2ij}$. This can easily be understood from a simple analysis
where we have assumed all dimensionful parameters as $m_{SUSY}$ and $M_R \sim 1$TeV. In this limit, 
the loop functions are given by 
$F_2\left(x,x,x,x\right) = \frac{1}{6x^2}$ and $F_1\left(x,x,x\right) = 
\frac{1}{2x}$. This leads to
\begin{eqnarray}
\epsilon_{2} \simeq -0.0007\,, ~~and ~~ 
\epsilon'_{2} \simeq 0.003\,\nonumber.
\end{eqnarray}
In the above, we have further
assumed that at $M_\text{GUT}$, one has $A_0=0$, taking for 
the gauge couplings $\alpha_2=0.03$ and
$\alpha'=0.008$. Thus, at the leading 
order in the inverse seesaw,  the lepton flavour violation
coefficient becomes 
$|\epsilon^\text{tot}_{2}|= |\epsilon_{2}+\epsilon'_{2}| \simeq 2 \times 10^{-3}$. 

On the contrary, 
in the standard seesaw model ($M_R\sim 10^{14}$ GeV), 
the coefficient $\xi$ would be small, thus one finds 
$|\epsilon^\text{tot}_{2}|= |\epsilon_{2}|\simeq 2 \times 10^{-4}$. 
This shows how in the inverse SUSY seesaw, $\epsilon^\text{tot}_{2}$ 
is enhanced by a factor of order $\sim 10$ compared to the standard seesaw.

The effective 
Lagrangian describing $\bar E^i_{R}E^j_{L}H_k$ (where $H_k = h,H,A$)
 can be derived from 
Eq.~(\ref{Leff}), and reads~\cite{Babu:2002et,Dedes:2002rh} as
\begin{eqnarray}
-{\cal L}^\text{eff}_{i\neq j} =
(2G_F^2)^{1/4} \,
\frac{m_{E_i} \kappa^E_{ij}}{\cos^2\beta}
\left(\bar E^i_{R}\,E^j_{L}\right)
\left[\cos(\alpha-\beta) h + \sin(\alpha-\beta) H - i A\right]+\text{h.c.}
\,,\,\,\,&&
\label{Leffl}
\end{eqnarray}
where $\alpha $ is the CP-even Higgs mixing angle and $\tan\beta=v_u/v_d$, and 
\begin{eqnarray}
\kappa^E_{ij} &=& \frac{\epsilon^\text{tot}_{2ij} (Y^\dagger_\nu Y_\nu)_{ij}
}{
\left[1+\left(\epsilon_1+\epsilon^\text{tot}_{2ii}
(Y^\dagger_\nu Y_\nu)_{ii}\right)\tan\beta\right]^2 }\ \label{kappa}.
\end{eqnarray}

As clear from the above equation, large values of 
$\epsilon^\text{tot}_2$ lead to large values  
of $\kappa^E_{ij}$. Since the cLFV branching ratios are proportional 
to $({\kappa^E_{ij}})^2$,
a sizeable enhancement, as large as two orders of magnitude, is expected for all Higgs-mediated LFV observables.
\section{Results and Discussion}
As can be seen from Eq:\ref{Leffl}, Higgs mediation would be more pronounced
for large values of $\tan\beta$ and small values Higgs boson masses. 
Similarly, the corresponding amplitude  
strongly depends on the chirality of the lepton. The cLFV observables would
be maximized if the right-handed particle is the heaviest lepton $\tau$. In view of this we particularly focus on the following observables:
\begin{enumerate}
\item
{$\text{Br}(\tau\to3\mu)$}
\item
{$\text{Br}(B_s\to \ell_i \ell_j)$} 
\item
{$\tau\to \mu P$ ($P=\pi, \eta, \eta'$)}.
\end{enumerate}
Analytical results for these observables can be found in ref:~\cite{Abada:2011hm}(also
 see the references therein). Here, we numerically evaluate the 
LFV observables where the benchmark points are selected 
from Ref:~\cite{AbdusSalam:2011fc}. Moreover, 
we also consider scenarios of Non-Universal Higgs Masses (NUHM), 
as this allows to explore the impact of the lightness of the CP-odd Higgs boson.
In Table~\ref{tab:sfp10.1}, we list the chosen points: CMSSM-A and CMSSM-B respectively 
correspond to 
the 10.2.2 and 40.1.1 benchmark points in~\cite{AbdusSalam:2011fc}, while NUHM-C is an example of a non-universal scenario.
\begin{table}[htb!]
\begin{center}
\begin{tabular}{|c||c||c|c||c|c|c|c|c|}
\hline
Point & $\tan\beta$ & $m_{1/2}$ & $m_0$ & $m^2_{H_U}$& $m^2_{H_D}$ &$A_0$  &$\mu$&$m_A$ \\
\hline
\hline
 CMSSM-A &10& 550 & 225 & $(225)^2$ &$(225)^2$ &0 &690 & 782 \\
\hline
\hline
CMSSM-B &40& 500 & 330 & $(330)^2$&$(330)^2$ &-500&698 &604 \\
\hline
NUHM-C &15& 550 & 225 &$(652)^2$ &$-(570)^2$ &0&478&150  \\
\hline
\hline
\end{tabular}
\caption{Benchmark points used in the numerical analysis (dimensionful parameters in GeV).
\label{tab:sfp10.1} }
\end{center}
\end{table}
For these points, the 
low-energy SUSY parameters were obtained using SuSpect~\cite{Djouadi:2002ze}. 
The flavour-violating  
slepton mass term $(\Delta m_{\widetilde{L}}^2)_{ij}$ or $\xi$, are calculated
at the leading order using Eq.~(\ref{slepmixing}). (For NUHM,
 we also use the same value of $\xi$ as for CMSSM-A.)
In addition, the (physical) right-handed sneutrino masses are assumed
$M_{\widetilde \nu^c} \approx 3$ TeV and $\left(Y_\nu^{\dagger}Y_\nu \right) = 0.7$, particularly in agreement with the Non-Standard Neutrino Interactions bounds 
~\cite{Antusch:2006vwa}. Moreover, in our numerical analysis, we have 
fixed the trilinear soft breaking parameter $A_\nu = - 500$ GeV (at the SUSY scale).
\begin{table}[htb]
\begin{center}
\hskip -1.5 cm
\small
 \begin{tabular}{|c|c|c|c|c|c|}
  \hline
    LFV Process & Present Bound & Future Sensitivity & CMSSM-A & CMSSM-B & NUHM-C\\
  \hline
    $\tau \rightarrow \mu \mu \mu$ & $2.1\times10^{-8}$ (Belle) & $8.2 \times 10^{-10}$ (SuperB)  & $1.4 \times 10^{-15}$ & $3.9 \times 10^{-11}$ & $8.0 \times 10^{-12}$\\
      $\tau \rightarrow \mu \eta$ & $2.3\times 10^{-8}$ (Belle) & $\sim 10^{-10}$ (SuperB) &$8.0 \times 10^{-15}$ & $3.3 \times 10^{-10}$ & $4.6 \times 10^{-11}$  \\
    $B^{0}_{s} \rightarrow  \mu \tau$ & & & $7.7 \times 10^{-14}$ & $2.5 \times 10^{-8}$ & $7.8 \times 10^{-10}$ \\
    $B^{0}_{s} \rightarrow e \mu$ & $2.0\times 10^{-7}$ (CDF) & $6.5\times 10^{-8}$(LHCB) & $3.4 \times 10^{-16}$ & $8.9 \times 10^{-11}$ & $3.4 \times 10^{-12}$
\\
  \hline
 \end{tabular}

\end{center}
  \caption{Higgs-mediated contributions to the branching ratios of several lepton flavour violating 
processes, for the different benchmark points of Table~\ref{tab:sfp10.1}. We also present the current experimental 
bounds and future sensitivities for the LFV observables.}\label{das_4}
\end{table}
As can be seen (From Table~\ref{das_4}) $\tau \rightarrow \mu \eta$ 
is the most promising concerning the next generation of $B$ factories. 
The $B^{0}_{d,s} \rightarrow  \mu \tau$ decay is also interesting, 
but there is not much hope concerning the future sensitivities.
\section{Conclusion}
Lepton flavor violation, if observed in the charged lepton sector
would (i) manifest the presence of new physics and (ii) could provide a hint 
for the origin of neutrino masses and mixings. Assuming inverse seesaw framework
in the Minimal Supersymmetric Standard Model we have studied the impact of the
Higgs mediation to the cLFV observables. We have argued that TeV scale right-handed 
(s)neutrinos offer the possibility to enhance the Higgs-mediated contributions. 
Consequently, different LFV branching ratios can be enhanced by as much as 
two orders of magnitude when compared to the standard (type I) SUSY seesaw 

\section*{Acknowledgments}
I am grateful to Asmaa Abada and Cedric Weiland for collaboration. I
acknowledge the FLASY12 organisers for their warm hospitality.

\bibliography{debottamdas}
\bibliographystyle{apsrev4-1}


%% file: Papers/deppisch.tex


\chapter[Probing the Flavour Structure of Right-Handed Neutrinos in Left-Right Symmetry at the LHC (Deppisch)]{Probing the Flavour Structure of Right-Handed Neutrinos in Left-Right Symmetry at the LHC}
\vspace{-2em}
\paragraph{F. F. Deppisch}
\paragraph{Abstract}
Lepton flavour couplings can be probed at the LHC, complementing searches for lepton flavour violation in low energy experiments. This can be used to shed light on the flavour structure of new physics models and the presence of possible flavour symmetries. We highlight this possibility in the context of left-right symmetry through the production and decay of heavy right-handed neutrinos at the LHC and discuss the expected sensitivity on the right-handed neutrino mixing matrix, as well as on the right-handed gauge boson and heavy neutrino masses. By comparing the sensitivity of the LHC with that of searches for low energy lepton flavour violating processes, favourable areas of the parameter space are identified where the complementarity between lepton flavour violation at low and high energies can be explored.

\section{Introduction}
It is natural to expect that the violation of lepton flavour observed in neutrino oscillations~\cite{fukuda:1998mi, ahmad:2002jz, eguchi:2002dm} should also show up in charged lepton flavour violating (LFV) processes such as the decay $\mu^-\to e^-\gamma$, and possibly also at the high energies accessible at the Large Hadron Collider (LHC). Oscillation experiments also demonstrate that neutrinos have small but finite masses, and many mechanisms of generating light neutrino masses have been discussed, the most popular example being the seesaw mechanism. Here, heavy right-handed Majorana neutrinos produce the light Majorana masses of the observed neutrinos through their mixing with the left-handed neutrinos. The Majorana character of the light neutrinos can then be traced to the breaking of lepton number symmetry at a very high energy scale~\cite{minkowski:1977sc, gell-mann:1980vs, yanagida:1979, mohapatra:1979ia, schechter:1980gr, schechter:1981cv, lazarides:1980nt}.

Despite its theoretical attractiveness, the standard type-I seesaw mechanism has phenomenological shortcomings: The right-handed neutrinos have masses close to the unification scale and can therefore not be directly observed. In addition, the right-handed neutrinos are gauge singlets, and even if they are light enough to be produced at colliders, the heavy neutrinos only couple through their mixing with the left-handed neutrinos which is tightly constrained by the small\-ness of neutrino masses as well as electroweak precision data and searches for LFV~\cite{Forero:2011pc, Abada:2007ux, delAguila:2008pw}. This means that the standard seesaw mechanism is difficult to test at the LHC~\cite{delAguila:2007em}.

A well known alternative of the standard Seesaw scheme is the left-right symmetrical model (LRSM) which extends the electroweak Standard Model (SM) gauge symmetry to the group SU(2)$_L~\otimes$ SU(2)$_R~\otimes$ U(1)$_{B-L}$~\cite{Pati:1974yy, Mohapatra:1974gc, Senjanovic:1975rk, Duka:1999uc}. Right-handed neutrinos are a necessary ingredient and they appear as part of an SU(2)$_R$ doublet. Consequently, heavy neutrinos can be produced with gauge coupling strength, with promising discovery prospects.

This opens up the possibility to test the naturally expected presence of lepton flavour couplings in the right-handed charged currents of the left-right symmetrical model. In fact, the large mixing observed in oscillation experiments could suggest a similar pattern among right-handed neutrinos in this framework. The observation of LFV processes mediated by the heavy neutrinos would provide important information on the flavour structure of the model and could help distinguish between models such as emerging from different flavour symmetries. On the other hand, the non-observation of lepton flavour violating processes in low-energy experiments so far puts stringent constraints on the strength of flavour violating couplings and the spectrum of the mediating particles. It is therefore interesting to understand how searches for lepton flavour violation in high-energy processes at the LHC can complement low-energy searches, and how these can be combined to shed light on the flavour structure of new physics models.

\section{Left-Right Symmetry}

We will here highlight the sensitivity of LHC searches to LFV couplings in the minimal Left-Right symmetric model, and the following results are mostly based on the analysis~\cite{Das:2012ii}. In the LRSM, a generation of leptons is assigned to the multiplet $L_i = (\nu_i, l_i)$ with the quantum numbers $Q_{L_L} = (1/2, 0, -1)$ and $Q_{L_R} = (0, 1/2, -1)$ under SU(2)$_L~\otimes$ SU(2)$_R~\otimes$ U(1)$_{B-L}$. The Higgs sector of the model contains a bidoublet $\phi$ and two triplets $\Delta_{L,R}$. The vacuum expectation value (VEV) $v_R$ of $\Delta_R$ breaks SU(2)$_R~\otimes$ U(1)$_{B-L}$ to U(1)$_Y$ and generates the masses of the right-handed $W_R$ boson, the right-handed $Z_R$ boson and the heavy right-handed neutrinos. Since significant deviations from SM predictions and new heavy particles have not been observed, $v_R$ is required to be sufficiently large. The VEVs of the neutral component of the bidoublet break the SM symmetry and are therefore of the order of the electroweak scale.

The LRSM accommodates a general $6\times 6$ neutrino mass matrix in the basis $(\nu_L, \nu^c_L)^T$ of the form
\begin{equation}
	\mathcal{M} =
	\begin{pmatrix}
		M_L & M_D \\ 
		M_D^T & M_R
	\end{pmatrix},
\label{eq:matr}
\end{equation}
with Majorana and Dirac mass entries of the order $M_{L,R} \approx y_M v_{L,R}$ and $M_D=y_D v$, respectively. Here, $y_{M,D}$ are Yukawa couplings, $v$ denotes the electroweak mass scale and $v_L$ is the VEV of the left-handed triplet $\Delta_L$ satisfying $v^2 = v_L v_R$. The Dirac mass term $M_D$ leads to a mixing between left- and right-handed neutrinos which is constrained to be $M_D/M_R \lesssim 10^{-2}$. The following results are reported in the regime with a small Dirac mass term to accommodate the light neutrino masses $m_\nu = M_D^2/M_R$ and right-handed neutrino masses at the TeV scale. With $M_D \lesssim 10^{-4}$~GeV, an admixture between the light and heavy neutrinos is negligible. 

\begin{figure}[t!]
\includegraphics[clip,width=0.32\textwidth]{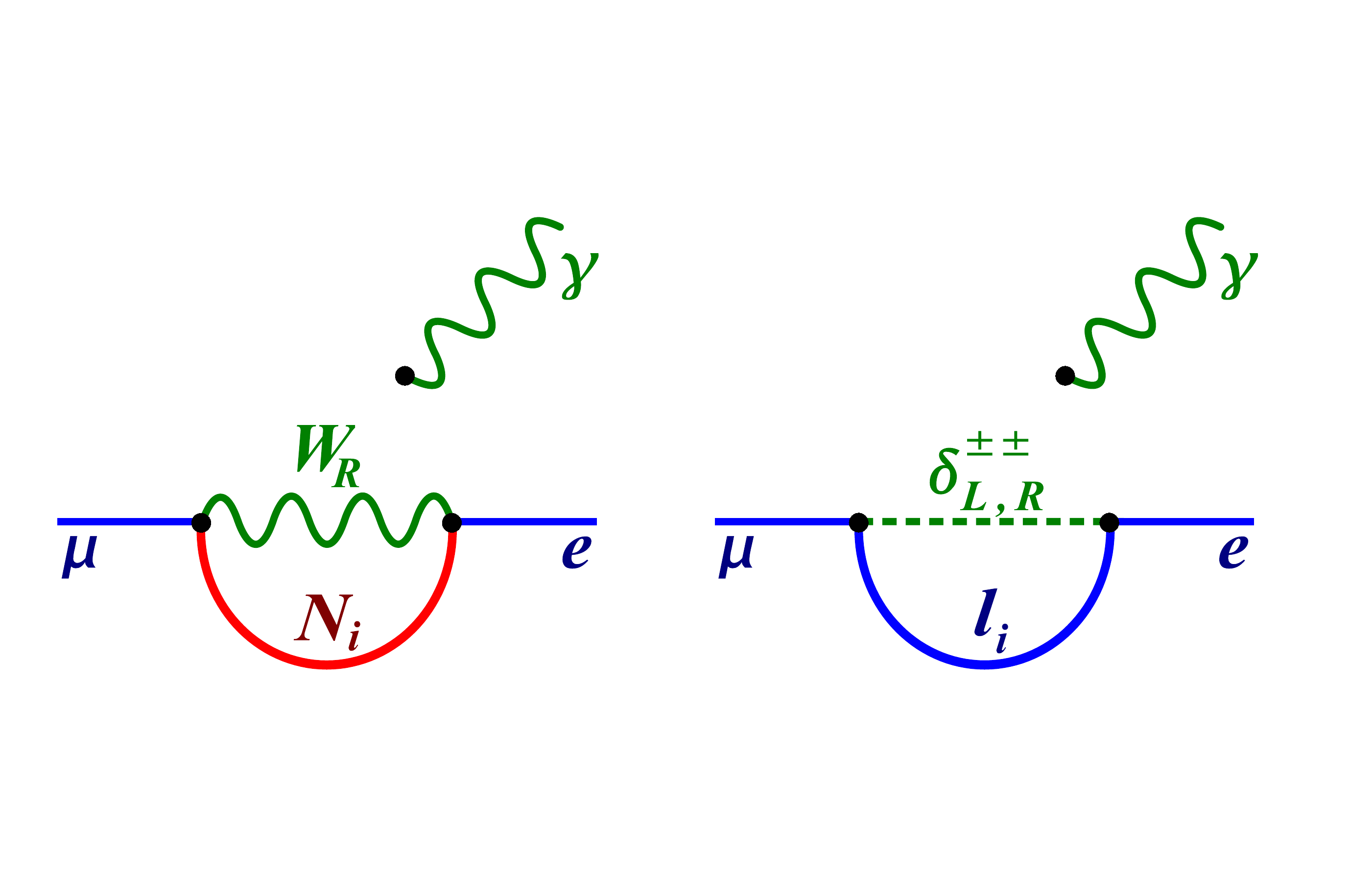}
\includegraphics[clip,width=0.32\textwidth]{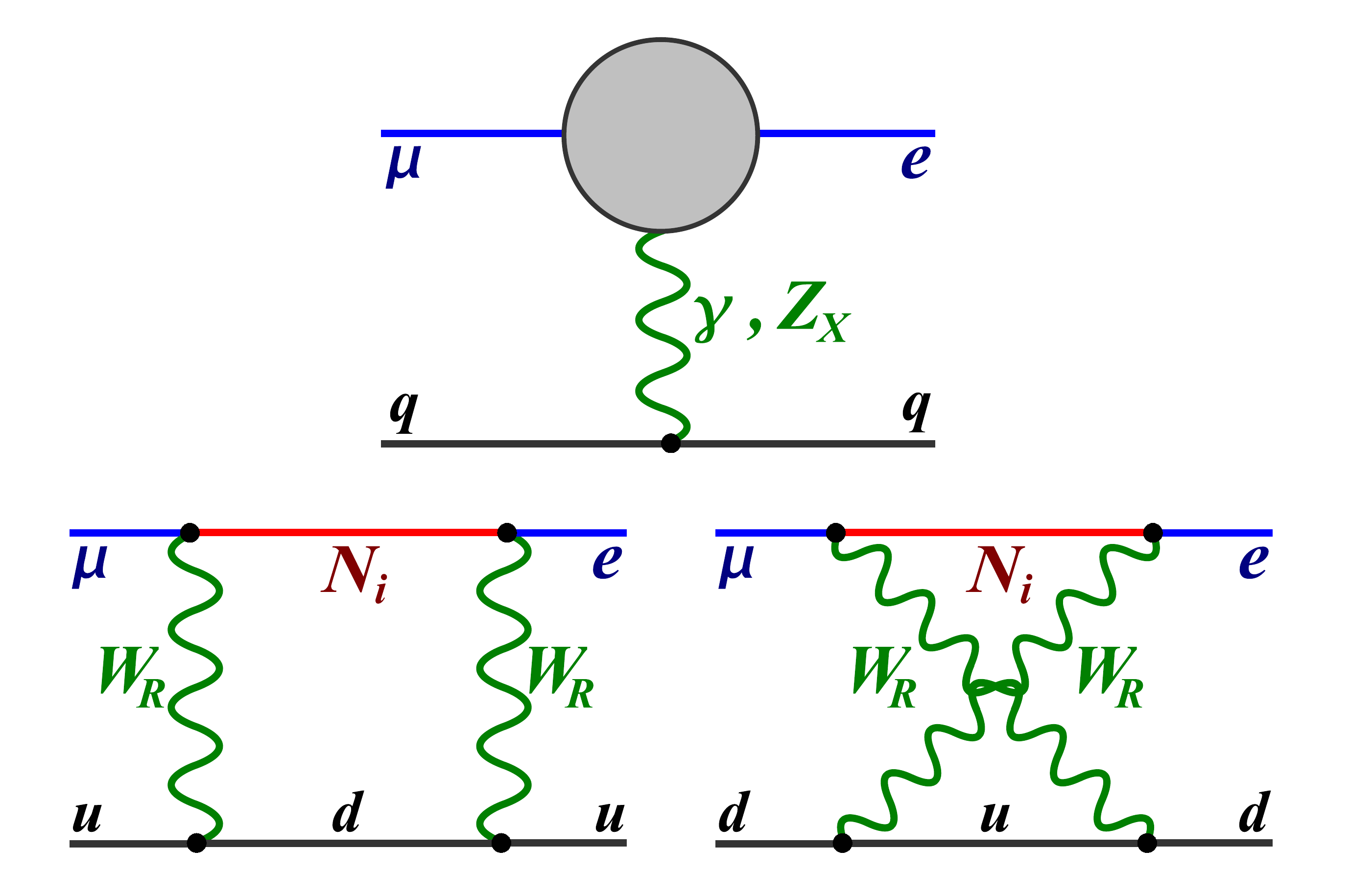}
\includegraphics[clip,width=0.32\textwidth]{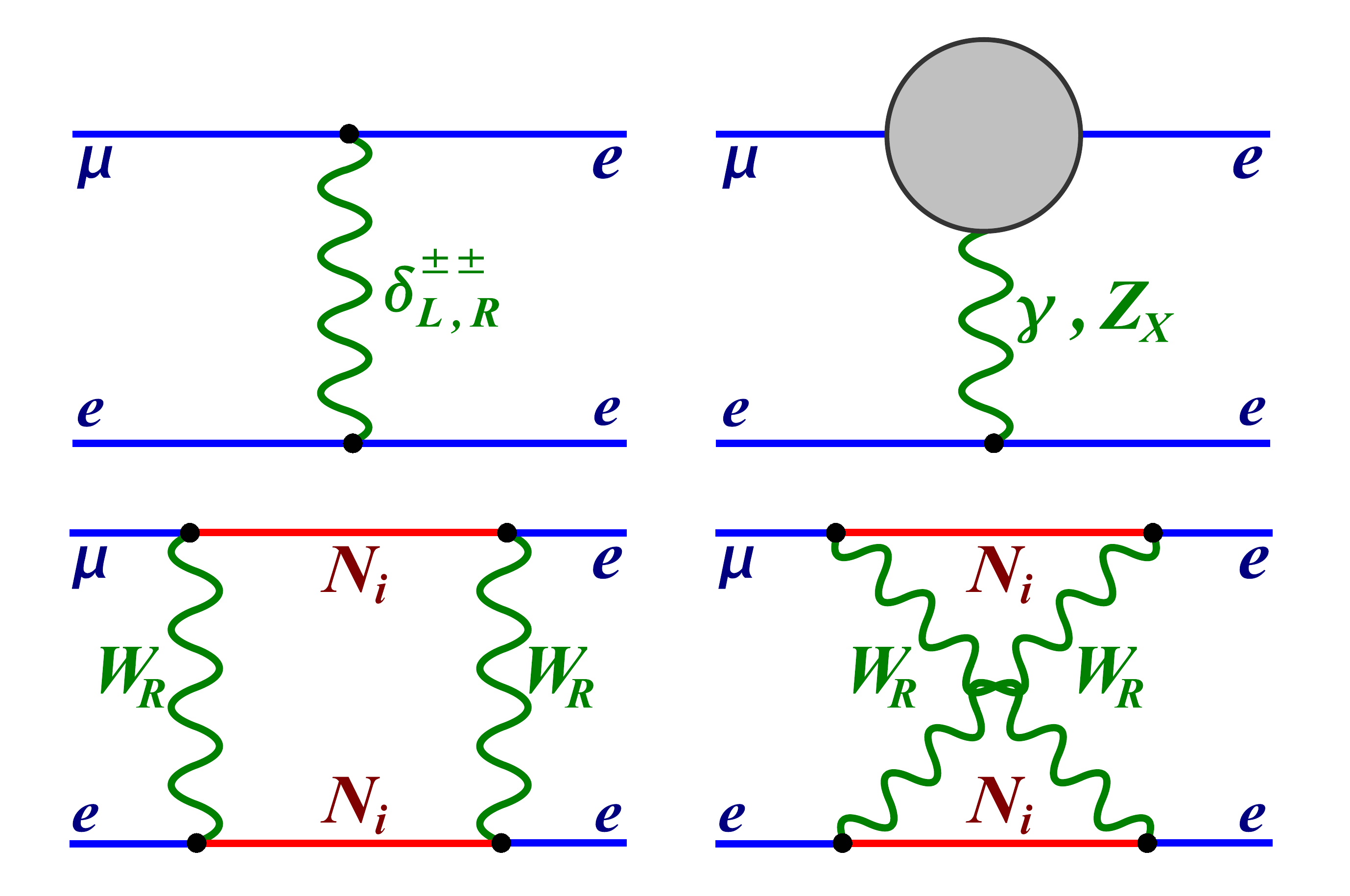}
\caption{Contributions to $\mu\to e\gamma$ (left, the photon line may be attached to any charged particle line), $\mu\to e$ conversion in nuclei (center) and $\mu\to eee$ (right) in left-right symmetry (from \cite{Das:2012ii}). The grey circle represents the effective $\mu-e-$gauge boson vertex of $\mu\to e\gamma$.}
\label{FD_fig:diagrams_LFV_mue} 
\end{figure}
Fig.~\ref{FD_fig:diagrams_LFV_mue} shows the contributions to the LFV processes $\mu\to e\gamma$, $\mu\to e$ conversion in nuclei and $\mu\to 3e$, which are mediated by heavy neutrinos and doubly charged bosons $\delta_{L,R}$. In general, the rates of these processes depend on many parameters, but under the assumption of similar mass scales of the heavy LRSM particles, $m_{N_i} \approx m_{W_R} \approx m_{\delta_{L,R}}$, simple approximations can be derived~\cite{Cirigliano:2004mv}. As all these masses are generated through the breaking of right-handed symmetry, such a spectrum is naturally expected, and in this case the branching ratio of $\mu\to e\gamma$ can be approximated as~\cite{Cirigliano:2004mv}
\begin{align}
\label{FD_eq:BrmuegammaSimplified}
	Br(\mu\to e\gamma) 
	&\approx 1.5 \times 10^{-7} |g_{e\mu}|^2 \left(\frac{1\text{ TeV}}{m_{W_R}}\right)^4,
	\text{ with}\quad
	g_{e\mu} = 
		\sum_{n=1}^3 V^\dagger_{en} V^{\phantom{\dagger}}_{n\mu}
		\left(\frac{m_{N_n}}{m_{W_R}}\right)^2.
\end{align}
Here, $V$ is the $3\times 3$ flavour mixing matrix in the right-handed charged current interaction between the heavy neutrinos with masses $m_{N_n}$ and the right-handed charged leptons. The other lepton flavour violating processes have the following properties in the chosen regime: (i) Both Br$(\mu\to e\gamma)$ and the $\mu-e$ conversion rate in nuclei $R_{\mu e}$ are proportional to the LFV factor $|g_{e\mu}|^2$, and their ratio is $R_{\mu e}/Br(\mu\to e\gamma) = \mathcal{O}(1)$. (ii) Unless there are cancellations among the flavour couplings, one has $Br(\mu\to eee) / R_{\mu e} = \mathcal{O}(300)$ ($m_{\delta_{L,R}} \approx$~1~TeV). These findings are in stark contrast to many other new physics models such as SUSY seesaw models, where the photon penguin contribution dominates.

\section{Dilepton Signals at the LHC}

\begin{figure}[t]
\includegraphics[clip,width=0.49\textwidth]{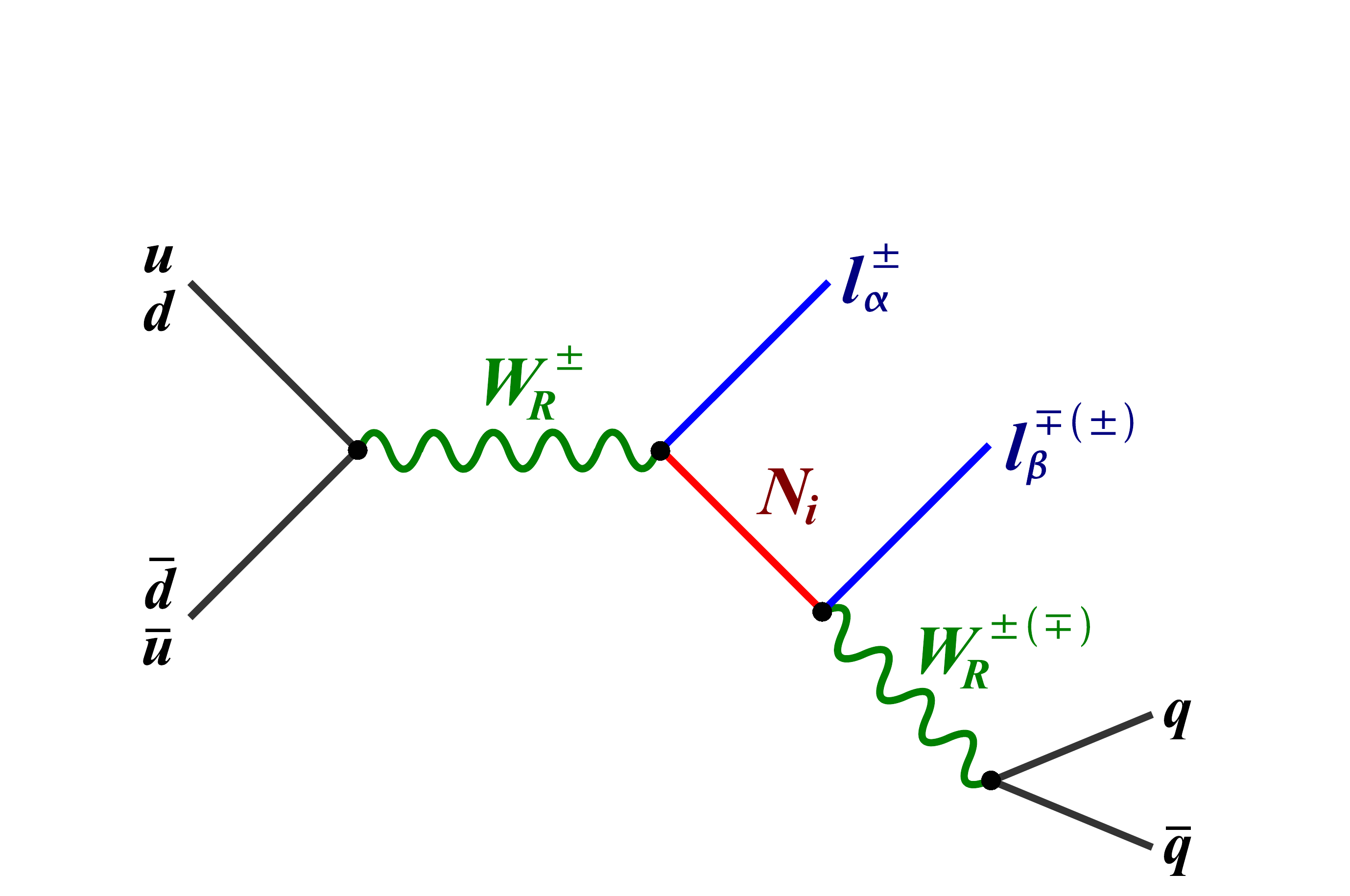}
\caption{Production and decay of a heavy right-handed $W$ boson and neutrino with dilepton signature at the LHC (from \cite{Das:2012ii}).}
\label{FD_fig:diagramsLHC}
\end{figure}
In \cite{Das:2012ii}, the discovery potential of flavour violating signals $pp\to W_R \to e^\pm \mu^{\pm,\mp} +2$~jets at the LHC via a heavy right-handed neutrino \cite{Keung:1983uu} was assessed (cf. Fig.~\ref{FD_fig:diagramsLHC}), with opposite sign (lepton number conserving) and same sign (lepton number violating) leptons in the final state. If the masses of the three heavy neutrinos are sufficiently different, only one neutrino in the intermediate state has to be taken into account. This is because either only one right-handed neutrino is light enough to be produced in this process or the neutrino mass resonances can be individually reconstructed.

\begin{figure}[t]
\includegraphics[clip,width=0.478\textwidth]{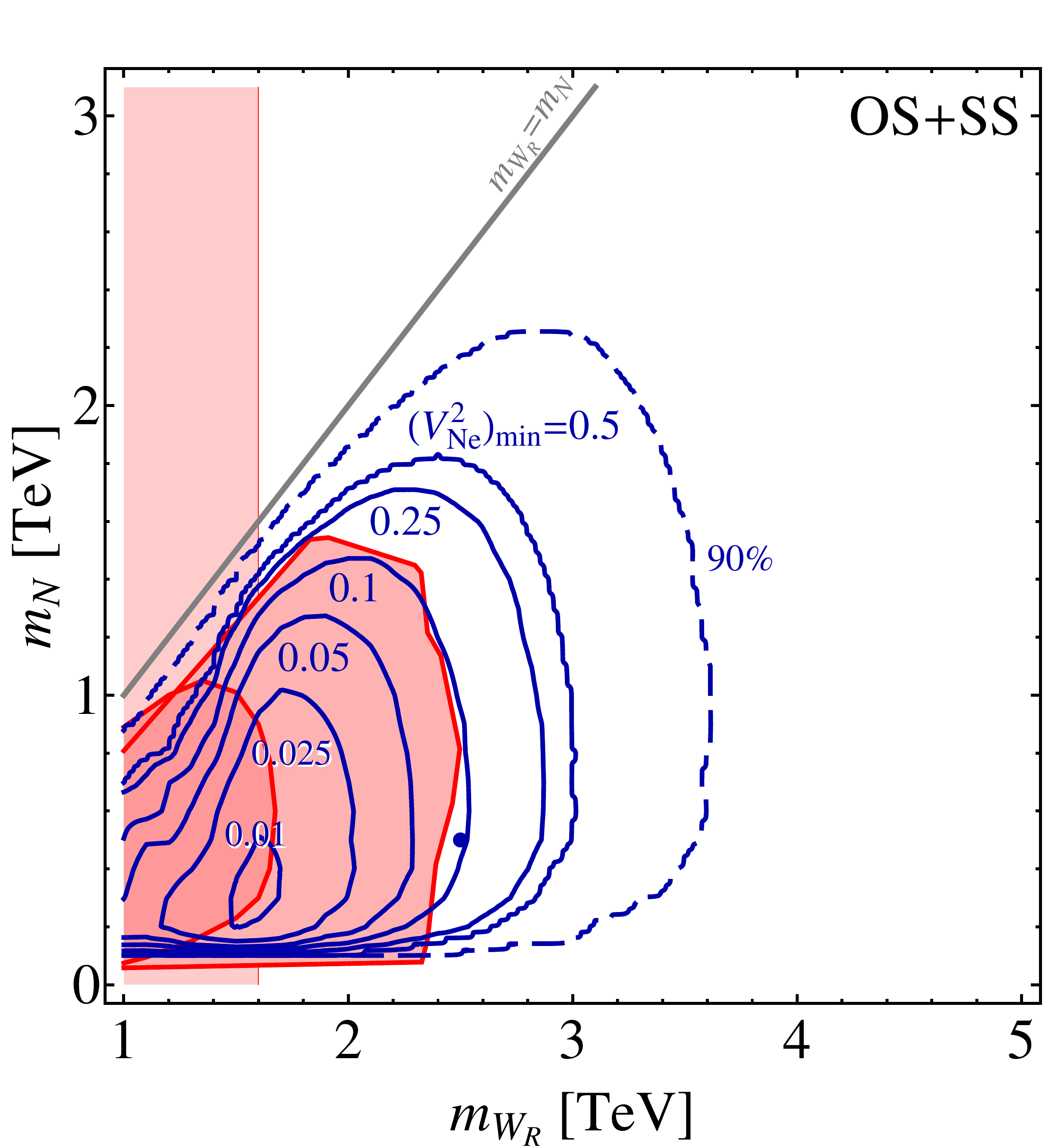}
\includegraphics[clip,width=0.502\textwidth]{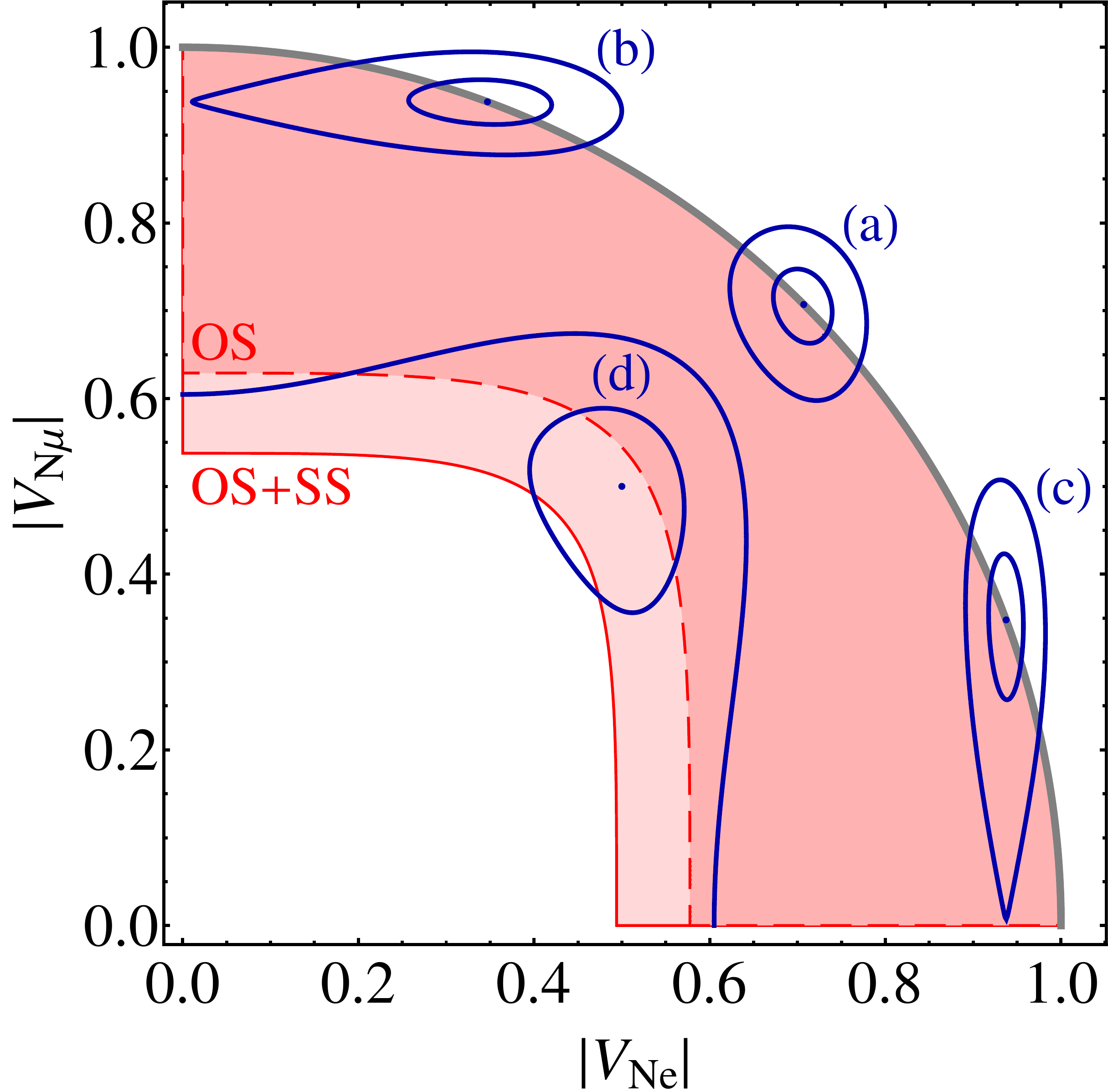}
\caption{Sensitivity to the coupling $(V_{Ne})^2 = 1 - (V_{N\mu})^2$ as function of $m_{W_R}$ and $m_{N_R}$ at the LHC with 14~TeV and $\mathcal{L}=30\text{ fb}^{-1}$ using both opposite and same sign LFV lepton events (left, from \cite{Das:2012ii}). The solid contours indicate a $5\sigma$ discovery. The shaded red areas are excluded by indirect (vertical bar) and direct LHC searches. Sensitivity to the potentially non-unitary couplings $|V_{Ne}|$ and $|V_{N\mu}|$ with $(m_{W_R},m_{N_R})=(2.5,0.5)$~TeV at the LHC with 14~TeV and $\mathcal{L}=30\text{ fb}^{-1}$ (right, from \cite{Das:2012ii}). The shaded areas would be excluded at 90\% CL using opposite sign (OS) or both sign (OS+SS) signatures whereas the blue contours give the $1\sigma$ and $5\sigma$ uncertainty contours in measuring the couplings in four hypothetical scenarios.}
\label{FD_fig:Ve_MWMN} 
\end{figure}
Fig.~\ref{FD_fig:Ve_MWMN}~(left) shows the smallest coupling $|V_{Ne}|$ of the heavy neutrino with an electron that results in a signal at $5\sigma$. Here, unitary flavour mixing in the $e-\mu$ sector is assumed, $|V_{Ne}|^2 + |V_{N\mu}|^2 = 1$. With the direct limits from $W_R$ and $N_R$ searches, flavour violating heavy neutrino-lepton couplings as small as $|V_{Ne(\mu)}| \approx 10^{-1}$ can be tested at the LHC with 14~TeV and $\mathcal{L} = 30\text{ fb}^{-1}$. This can be generalized to non-unitary mixing, and the sensitivity to the couplings $|V_{Ne}|$ and $|V_{N\mu}|$ can be assessed. This is shown in Fig.~\ref{FD_fig:Ve_MWMN}~(right) which gives both the excluded parameter space in the case of non-observation as well as the expected precision in measuring the couplings in four hypothetical scenarios. For LFV signals with a $\tau$ lepton, a 30\% reduction of the signal efficiency is expected~\cite{AguilarSaavedra:2012fu, AguilarSaavedra:2012gf}.

If two heavy neutrinos are light enough to be produced at the LHC, a potentially small squared mass difference $\Delta m_N^2$ leads to interference effects and as the heavy neutrinos become more and more degenerate, all LFV processes will suffer a GIM-like suppression if the flavour mixing is unitary. A crucial difference to the radiative rare decays is that the heavy neutrinos are produced on-shell at the LHC. Because of their small decay width, this results in a quick decoherence of the right-handed neutrino oscillations, and the mass difference suppression is $\propto\Delta m_N^2/(m_N \Gamma_N)$, rather than $\Delta m_N^2/m_N^2$~\cite{Deppisch:2003wt}.
\begin{figure}[t!]
\includegraphics[clip,width=0.467\textwidth]{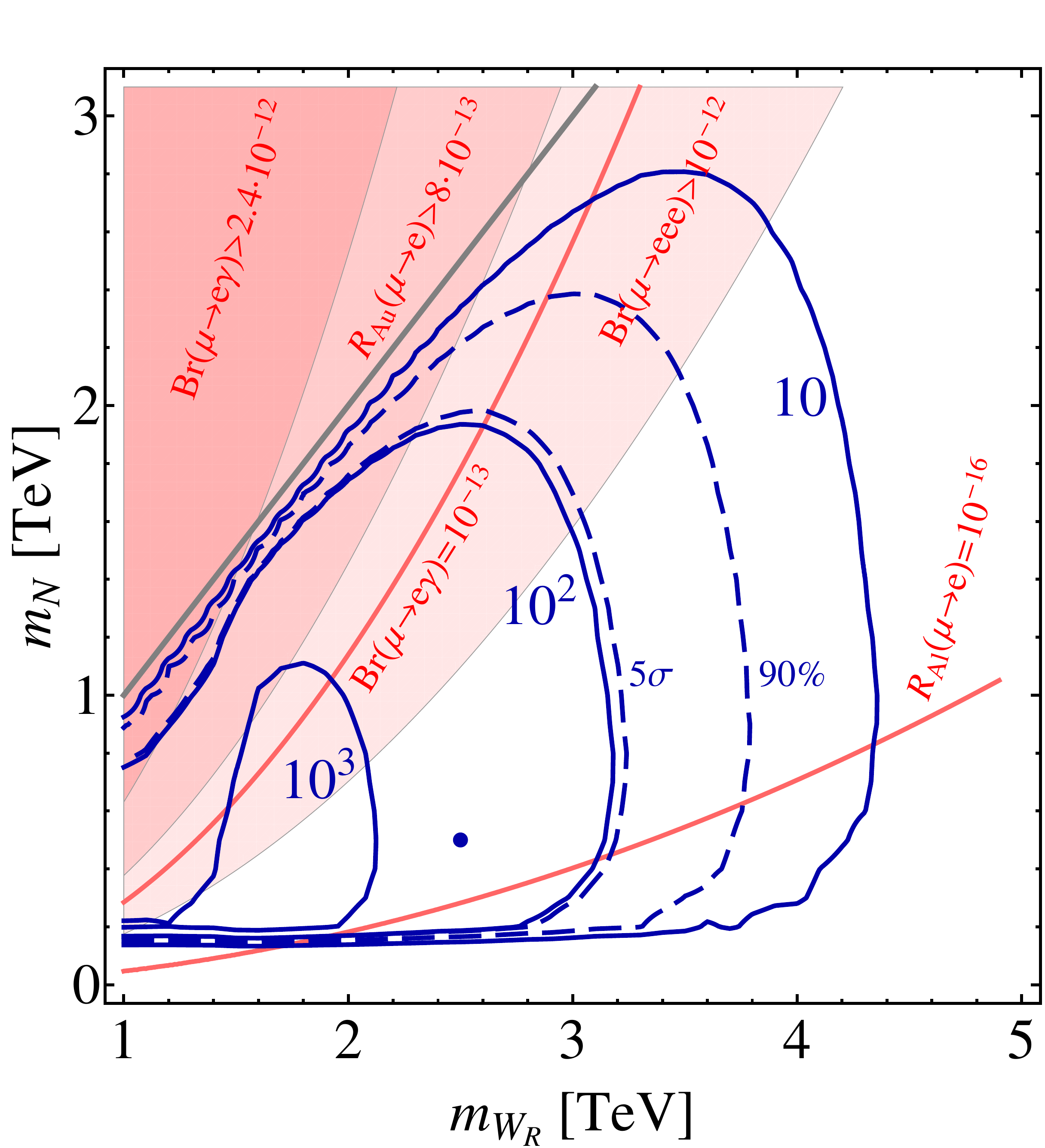}
\includegraphics[clip,width=0.513\textwidth]{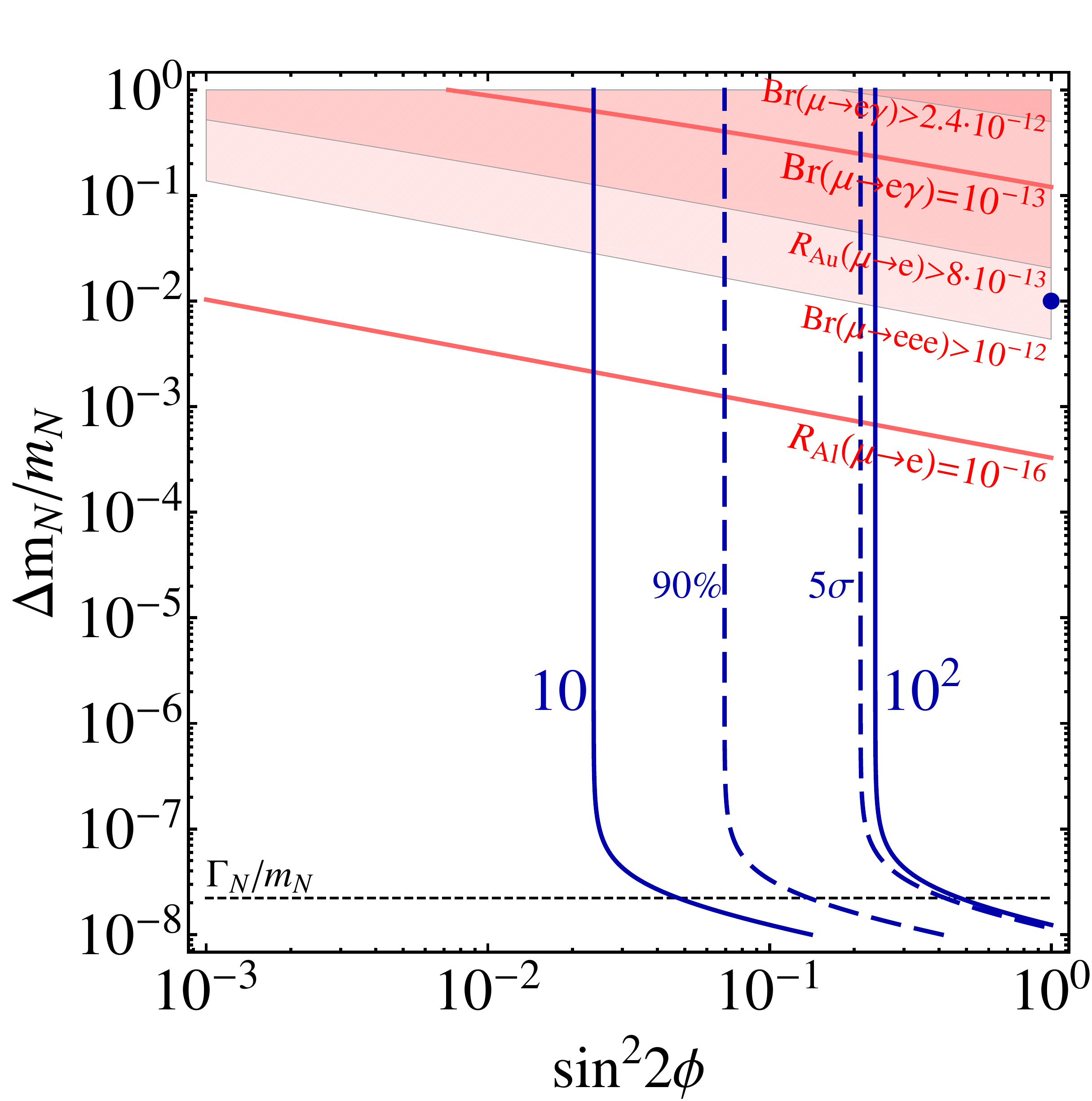}
\caption{Dependence of the rates of low energy LFV processes (red contours and shaded areas) and the LFV signature $e^\pm \mu^{\pm,\mp} + 2j$ at the LHC (blue solid contours) with 14~TeV and $\mathcal{L}=30\text{ fb}^{-1}$ (from \cite{Das:2012ii}). The calculation assumes a doubly charged boson spectrum of $m_{\delta_{L,R}} = m_{W_R}$.
(Left) Dependence on the $W_R$ boson mass and the heavy neutrino mass scale $m_N$ for maximal flavour mixing and 1\% neutrino mass splitting, $\phi=\pi/4$, $\Delta m_N/m_N = 0.01$. (Right) Dependence on the mixing angle parameter $\sin^2 2\phi$ and the heavy neutrino mass splitting $\Delta m_N/m_N$ for $(m_{W_R},m_N) = (2.5,0.5)$~TeV.}
\label{FD_fig:ComparisonLowEnergy} 
\end{figure}
Fig.~\ref{FD_fig:ComparisonLowEnergy} demonstrates this complementarity as it compares the sensitivity of LHC searches and $\mu-e$ LFV processes, either as a function of the heavy particle mass scales (left plot) or as a function of the heavy neutrino mass difference and flavour mixing angle (right plot). The current limits on the rare processes put strong constraints on the parameter space, with $\mu\to eee$ proving to be most stringent due to the tree-level doubly charged bosons contribution (cf. Fig.~\ref{FD_fig:diagrams_LFV_mue}~(right)). As a consequence of the decoherence in on-shell production, the LFV process rate at the LHC is independent of the neutrino mass splitting until it becomes comparable to or smaller than the heavy neutrino decay width. On the other hand, the low energy LFV processes exhibit the typical GIM-suppressed behaviour $\propto \sin^2(2\phi)(\Delta m_N^2)^2$ and can only test much larger mass differences $\Delta m_N / m_N \gtrsim 10^{-3} - 10^{-4}$.

\section{Conclusion}

Our discussion in the context of left-right symmetry highlights that under favourable conditions, lepton flavour violation can be probed directly at the LHC, complementary to low-energy searches. Because the particles mediating the flavour violation can be produced on-shell, the LHC has the potential to pin-point individual couplings and probe much smaller heavy neutrino mass splittings than low energy LFV processes. In the scenario considered here, with the resonant production of heavy right-handed neutrinos, the most optimistic scenario would be that all three neutrinos are accessible with mass differences large enough so that they can be individually reconstructed and their coupling strengths may be measured. This would provide detailed information on the flavour structure of the model, directly complementary to light neutrino oscillations, which is not accessible through the observation of rare LFV processes.

\section{Acknowledgments}
The author would like to thank J.~A. Aguilar-Saavedra, S.~P. Das, O. Kittel and J.~W.~F. Valle for a 
fruitful collaboration, and he is grateful to the organizer of FLASY12 for the opportunity to participate 
in the workshop. The author also acknowledges financial support by an IPPP associateship.

\bibliography{deppisch}
\bibliographystyle{apsrev4-1}


%% file: Papers/dinggj.tex

%
%
%
%
%
%

\chapter[TFH Mixing Patterns, Large $\theta_{13}$ and $\Delta(96)$ Flavor Symmetry (Ding)]{TFH Mixing Patterns, Large $\theta_{13}$ and $\Delta(96)$ Flavor Symmetry}
\vspace{-2em}
\paragraph{G.-J. Ding}
\paragraph{Abstract}

We perform a comprehensive analysis of the Toorop-Feruglio-Hagedorn (TFH) mixing pattern within the family symmetry $\Delta(96)$. The possible realizations of the TFH mixing in $\Delta(96)$ are analyzed in the minimalist framework. The dynamical model which naturally produces the TFH mixing pattern at leading order is constructed based on flavor symmetry $\Delta(96)\times Z_3\times Z_3$, and the next to leading order terms introduce corrections of order $\lambda^2_c$ to the three mixing angles. The allowed mixing patterns are studied under the condition that the Klein four subgroups and the cyclic $Z_N$ subgroups with $N\geq3$ are preserved in the neutrino and the charged lepton sector respectively. We suggest that the deformed tri-bimaximal mixing is a good leading order approximation to understanding a largish reactor angle.

\section{Introduction}

Recently the T2K \cite{Abe:2011sj} and MINOS \cite{minos} collaborations reported the evidence for a relatively large $\theta_{13}$ at the level of $2.5\sigma$ and $1.7\sigma$ respectively, this has been confirmed by the Daya-Bay \cite{An:2012eh} and RENO \cite{RENO} experiments at 5.2 $\sigma$ and 4.9 $\sigma$ confidence level respectively. The global fitting including all the current neutrino oscillation data further support that $\theta_{13}$ is somewhat large, Valle and Fogli's groups find the 3 $\sigma$ ranges of $\theta_{13}$ are 0.015 (0.016)$\leq\sin^2\theta_{13}\leq 0.036 (0.037)$ \cite{Tortola:2012te} and $0.0149(0.015)\leq\sin^2\theta_{13}\leq0.0344(0.0347)$ \cite{Fogli:2012ua} respectively. The the important question is whether and how can understanding this relative large $\theta_{13}$ from symmetry. By analyzing the symmetry breaking of the finite modular group $\Gamma_{N}$, Feruglio et al. suggested that the attractive mixing texture with $\sin^2\theta_{13}=(2-\sqrt{3})/6$, $\sin^2\theta_{12}=(8-2\sqrt{3})/13$, $\sin^2\theta_{23}=(5+2\sqrt{3})/13$, $\delta_{CP}=\pi$ can be generated if we choose $\Delta(96)$ as the flavor symmetry and further break it into the Klein four ($K_4$) and $Z_3$ subgroups in the neutrino and charged lepton sectors respectively \cite{Toorop:2011jn,deAdelhartToorop:2011re}. In a particular phase convention, the corresponding Pontecorvo-Maki-Nakagawa-Sakata (PMNS) matrix is given by
\begin{equation}
\label{eq2}U_{TFH}=\left(\begin{array}{ccc}
\frac{1}{6}(3+\sqrt{3})  &  \frac{1}{\sqrt{3}}  & \frac{1}{6}(-3+\sqrt{3})\\
\frac{1}{6}(-3+\sqrt{3}) & \frac{1}{\sqrt{3}}   & \frac{1}{6}(3+\sqrt{3}) \\
-\frac{1}{\sqrt{3}}      &  \frac{1}{\sqrt{3}}  & -\frac{1}{\sqrt{3}}
\end{array}\right),
\end{equation}
This mixing pattern will be denoted as THF henceforth, obviously it is an excellent approximation to the current neutrino mixing data, especially a large $\theta_{13}$. In this work, we shall investigate whether we can and how to consistently derive the TFH textures with the $\Delta(96)$ family symmetry. This proceeding is based on the work \cite{Ding:2012xx}

\section{Pathway to TFH mixing within $\Delta(96)$}

The $\Delta(96)$ is a non-abelian finite subgroup of SU(3) of order 96, it is isomorphic to $(Z_4\times Z_4)\rtimes S_3$, and it can be conveniently defined by four generators $a$, $b$, $c$ and $d$ obeying the relations:
\begin{eqnarray}
\nonumber&a^3 ~=~ b^2 ~=~ (ab)^2 ~=~ c^4 ~=~ d^4 ~=~1,\quad cd ~=~ dc\\
\nonumber&a c a^{-1}=c^{-1} d^{-1},\quad a d a^{-1} =c,\quad b c b^{-1}=d^{-1}, \quad b d b^{-1}=c^{-1} 
\end{eqnarray}
Note that the generator $d$ is not independent. The structure of $\Delta(96)$ group is rather complex, it has 10 irreducible representations: two singlets $\mathbf{1}$ and $\mathbf{1'}$, one doublet $\mathbf{2}$, six triplets $\mathbf{3_1}$, $\mathbf{3'_1}$, $\mathbf{\overline{3}_1}$, $\mathbf{\overline{3}'_1}$, $\mathbf{3_2}$ and $\mathbf{3'_2}$, and one sextet $\mathbf{6}$. The basic properties of $\Delta(96)$ such as the conjugate classes, Kronecker product and Clebsch-Gordan coefficients have been presented in detail in Ref. \cite{Ding:2012xx}.

Now we investigate how to produce the TFH mixing from $\Delta(96)$ flavor symmetry. To simplify the problem, we work in the so-called minimalist framework \cite{Zee:2005ut}, where the charged lepton masses are generated by the operator of the following form
\begin{equation}
\label{eq16}\mathcal{O}_{\ell}=E^c\ell h_d\phi_{\ell}\,,
\end{equation}
where $E^c$ is the right-handed charged lepton field, $\ell$ is the lepton doublet field, $h_d$ is the down-type Higgs doublet, and $\phi_{\ell}$ is the flavon field which breaks $\Delta(96)$ in the charged lepton sector at LO. Neutrino masses are generated by the Weinberg operator
\begin{equation}
\label{eq17}\mathcal{O}_{\nu}=\ell h_u \ell h_u\phi_{\nu}\,,
\end{equation}
where $h_u$ is the up-type Higgs doublet, and $\phi_{\nu}$ is the flavon field in the neutrino sector. We assign the fields $E^c$, $\ell$, $\phi_{\ell}$ and $\phi_{\nu}$ to various representations of $\Delta(96)$, then write down all the symmetry allowed forms of the operators $\mathcal{O}_{\ell}$ and $\mathcal{O}_{\nu}$. We find that there are numerous ways to produce TFH mixing within $\Delta(96)$. The possible assignments leading to TFH1 mixing are listed in Table \ref{tab:pathway}, it is remarkable that the lepton doublet $\ell$ can not be assigned to the triplet $\mathbf{3_2}$ or $\mathbf{3'_2}$. To generate TFH mixing, the vacuum expectation value (VEV) of $\phi_{\ell}$ should be aligned as follows:
\begin{equation}
\label{eq18}\langle\phi_{\ell}\rangle=\left\{\begin{array}{ll}
(0,0,v),&~~\quad\quad\phi_{\ell}\sim\mathbf{3_1}, \mathbf{3'_1}, \mathbf{\overline{3}_1}, \mathbf{\overline{3}'_1}, \mathbf{3_2}, \mathbf{3'_2}\\
(0,0,v_3,0,0,v_6),&~~\quad\quad\phi_{\ell}\sim\mathbf{6}\,.
\end{array}
\right.
\end{equation}
It breaks the flavor symmetry $\Delta(96)$ into the $Z_3$ subgroup generated by $a^2cd$. The vacuum configuration of $\phi_{\nu}$ is
\begin{equation}
\label{eq19}\langle\phi_{\nu}\rangle=\left\{
\begin{array}{ll}
(1,1,1)u,&~~\quad\quad\phi_{\nu}\sim\mathbf{3'_1}, \mathbf{\overline{3}'_1}\\
(u_1,u_2,(u_1+u_2)/2),&~~\quad\quad\phi_{\nu}\sim\mathbf{3'_2}
\end{array}
\right.
\end{equation}
$\Delta(96)$ is broken into the $K_4$ subgroup generated by the elements $a^2bd$ and $d^2$. This mismatch of the symmetry breaking induced by the VEV of $\phi_{\ell}$ and $\phi_{\nu}$ is exactly the origin of the TFH mixing. In the realizations listed in Table \ref{tab:pathway}, the three charged lepton masses are given in terms of three independent parameters. However, in order to match the observed masses $m_e$, $m_{\mu}$ and $m_{\tau}$, we need to tune the parameters such that some sort of cancellation between them happens. Further fine tuning is required if subleading corrections are included. To improve upon this situation, we should further break the remnant $Z_3$ symmetry in the charged lepton sector, a explicit model is presented in the following section.

\begin{table}[t!]
\begin{center}
\begin{tabular}{|c|c|c|c|}  \hline\hline

$\ell$       &      $E^c$      &     $\phi_{\ell}$    &      $\phi_{\nu}$ \\  \hline

\multirow{8}{*}{$\mathbf{3_1}$}  & $\tau^{c}\sim\mathbf{1}$, $(\mu^c, e^c)\sim\mathbf{2}$  &  $\mathbf{\overline{3}_1}$, $\mathbf{\overline{3}'_1}$  &  \multirow{8}{*}{$\mathbf{3'_1}$ , $\mathbf{3'_2}$}  \\ \cline{2-3}

   & $\tau^{c}\sim\mathbf{1'}$, $(\mu^c, e^c)\sim\mathbf{2}$   &    $\mathbf{\overline{3}_1}$, $\mathbf{\overline{3}'_1}$  &   \\  \cline{2-3}
   & $(\mu^c, e^c, \tau^c)\sim\mathbf{3_1}$   &  $\mathbf{3_1}$, $\mathbf{3'_1}$, $\mathbf{3'_2}$   &    \\  \cline{2-3}
   & $(\mu^c, e^c, \tau^c)\sim\mathbf{3'_1}$  &  $\mathbf{3_1}$, $\mathbf{3'_1}$, $\mathbf{3_2}$    &    \\   \cline{2-3}
   & $(e^c, \mu^c, \tau^c)\sim\mathbf{\overline{3}_1}$   &   $\mathbf{1}$, $\mathbf{6}$  &    \\   \cline{2-3}
   & $(e^c, \mu^c, \tau^c)\sim \mathbf{\overline{3}'_1}$  &  $\mathbf{1'}$, $\mathbf{6}$  &    \\   \cline{2-3}
   &  $(\mu^c, e^c, \tau^c)\sim\mathbf{3_2}$    &   $\mathbf{3'_1}$, $\mathbf{6}$  &   \\    \cline{2-3}
   &  $(\mu^c, e^c, \tau^c)\sim\mathbf{3'_2}$    &   $\mathbf{3_1}$, $\mathbf{6}$  &   \\    \hline

\multirow{8}{*}{$\mathbf{3'_1}$}  &  $\tau^{c}\sim\mathbf{1}$, $(\mu^c, e^c)\sim\mathbf{2}$  &  $\mathbf{\overline{3}_1}$, $\mathbf{\overline{3}'_1}$  & \multirow{8}{*}{$\mathbf{3'_1}$ , $\mathbf{3'_2}$}  \\  \cline{2-3}

 & $\tau^{c}\sim\mathbf{1'}$, $(\mu^c, e^c)\sim\mathbf{2}$   &    $\mathbf{\overline{3}_1}$, $\mathbf{\overline{3}'_1}$  &   \\  \cline{2-3}
 & $(\mu^c, e^c, \tau^c)\sim\mathbf{3_1}$   &  $\mathbf{3_1}$, $\mathbf{3'_1}$, $\mathbf{3_2}$   &    \\  \cline{2-3}
 & $(\mu^c, e^c, \tau^c)\sim\mathbf{3'_1}$  &  $\mathbf{3_1}$, $\mathbf{3'_1}$, $\mathbf{3'_2}$    &    \\   \cline{2-3}
 & $(e^c, \mu^c, \tau^c)\sim\mathbf{\overline{3}_1}$   &   $\mathbf{1'}$, $\mathbf{6}$  &    \\   \cline{2-3}
 & $(e^c, \mu^c, \tau^c)\sim \mathbf{\overline{3}'_1}$  &  $\mathbf{1}$, $\mathbf{6}$  &    \\   \cline{2-3}
 &  $(\mu^c, e^c, \tau^c)\sim\mathbf{3_2}$    &   $\mathbf{3_1}$, $\mathbf{6}$  &   \\    \cline{2-3}
 &  $(\mu^c, e^c, \tau^c)\sim\mathbf{3'_2}$    &   $\mathbf{3'_1}$, $\mathbf{6}$  &   \\    \hline  \hline

\end{tabular}
\caption{\label{tab:pathway}Possible assignments of the fields $E^c$, $\ell$, $\phi_{\ell}$ and $\phi_{\nu}$ for the TFH mixing, where $\ell=(\ell_1,\ell_2,\ell_3)$ is the lepton doublet field. The assignments by performing complex conjugation to all the involved fields are also admissible.  }
\end{center}
\end{table}

\section{Model for TFH Mixing}

\begin{table}[hptb]
\begin{center}
\begin{tabular}{|c|c|c|c|c|c|c||c|c|c|c|c|c|c||c|c|c|c|c|c|c|}\hline\hline
{\tt Fields} & $\ell$ & $e^c$  &  $\mu^c$ &  $\tau^c$  & $\nu^{c}$ &$h_{u,d}$ &  $\chi$  & $\phi$ & $\eta$ &$\xi$  &  $\rho$ & $\varphi$ & $\psi$
\\\hline

$\Delta(96)$  & $\mathbf{3_1}$ &  $\mathbf{1}$  &  $\mathbf{1'}$ & $\mathbf{1}$  & $\mathbf{\overline{3}_1}$ & $\mathbf{1}$ & $\mathbf{3_1}$ &  $\mathbf{\overline{3}_1}$  &  $\mathbf{2}$ & $\mathbf{3'_1}$  &  $\mathbf{2}$  &  $\mathbf{\overline{3}'_1}$  & $\mathbf{3'_2}$  \\

$Z_3$ & 0 & 2  & 2& 2 & 0  & 0  &  1 & 1 & 0 & 0  &  0  &  0  & 0  \\

$Z_3$ & 0 & 1  &  2  & 0 & 0 & 0 &  0 &  0  &  1  & 1 &  0 &   0  & 0   \\\hline\hline

\end{tabular}
\caption{\label{tab:field1} The transformation properties of the matter fields, the electroweak Higgs doublets, the flavon fields and the driving fields under the flavor symmetry $\Delta(96)\times Z_3\times Z_3$.}
\end{center}
\end{table}

We formulate our model in the framework of type I see-saw mechanism, and supersymmetry (SUSY) is introduced to simplify the discussion of the vacuum alignment. In our model, the full flavor symmetry is $\Delta(96)\times Z_3\times Z_3$. The fields in the model and their classifications under the flavor symmetry are summarized in Table \ref{tab:field1}. We use the now-standard supersymmetric driving field method to arrange the vacuum alignment, as is shown in Ref. \cite{Ding:2012xx}, the following vacuum configuration can be achieved naturally
\begin{eqnarray}
\nonumber&&\langle\chi\rangle=(0,0,v_{\chi}),~~~\langle\phi\rangle=(0,0,v_{\phi}),~~~\langle\eta\rangle=(v_{\eta},0),~~~\langle\xi\rangle=(v_{\xi},0,0) \\
\label{eq:vacuum}&&\langle\rho\rangle=(1,\omega)v_{\rho},~~~\langle\varphi\rangle=(1,1,1)v_{\varphi},~~~\langle\psi\rangle=(v_1,v_2,(v_1+v_2)/2)
\end{eqnarray}
This LO vacuum alignment is stable under small perturbations, as usual we assume all the VEVs scaled by the cutoff $\Lambda$ are of order $\lambda^2_c$, where $\lambda_c\simeq0.23$ is the well-known Cabibbo angle. In this model, the charged lepton masses are described by the following Yukawa superpotential at LO
\begin{eqnarray}
\nonumber&&w_{\ell}=\frac{y_{\tau}}{\Lambda}\tau^c(\ell\phi)h_d+\frac{y_{\mu_1}}{\Lambda^2}\mu^c(\ell(\chi\xi)_{\mathbf{\overline{3}'_1}})'h_d
+\frac{y_{\mu_2}}{\Lambda^2}\mu^{c}(\ell(\eta\phi)_{\mathbf{\overline{3}'_1}})'h_d+\frac{y_{e_1}}{\Lambda^3}e^c(\ell\phi)(\eta\eta)h_d \\
\nonumber&&\quad\quad+\frac{y_{e_2}}{\Lambda^3}e^c(\ell((\eta\eta)_{\mathbf{2}}\phi)_{\mathbf{\overline{3}_1}})h_d+\frac{y_{e_3}}{\Lambda^3}e^c(\ell(\chi(\eta\xi)_{\mathbf{3_1}})_{\mathbf{\overline{3}_1}})h_d
+\frac{y_{e_4}}{\Lambda^3}e^c(\ell(\chi(\eta\xi)_{\mathbf{3'_1}})_{\mathbf{\overline{3}_1}})h_d\\
\label{eq29}&&\quad\quad+\frac{y_{e_5}}{\Lambda^3}e^c(\ell(\chi(\xi\xi)_{\mathbf{3'_2}})_{\mathbf{\overline{3}_1}})h_d\,,
\end{eqnarray}
It is remarkable that the electron, muon and tau mass terms are suppressed by $1/\Lambda$, $1/\Lambda^2$ and $1/\Lambda^3$ respectively. At LO, only the tau mass is generated, the flavor symmetry $\Delta(96)$ is broken into $Z_3$ by the VEV of $\phi$, the remaining terms further break this $Z_3$ symmetry completely. With the vacuum alignment in Eq.(\ref{eq:vacuum}), $w_{\ell}$ leads to a diagonal charged lepton mass matrix:
\begin{equation}
\label{eq30}m_{\ell}=\left(\begin{array}{ccc}
\omega^2y_{e_2}\frac{v^2_{\eta}v_{\phi}}{\Lambda^3}+(y_{e_4}-y_{e_3})\frac{v_{\eta}v_{\xi}v_{\chi}}{\Lambda^3}+y_{e_5}\frac{v^2_{\xi}v_{\chi}}{\Lambda^3} &0  &0 \\
0 & y_{\mu_1}\frac{v_{\xi}v_{\chi}}{\Lambda^2}+y_{\mu_2}\frac{v_{\eta}v_{\phi}}{\Lambda^2}  &  0  \\
0 & 0  & y_{\tau}\frac{v_{\phi}}{\Lambda}
\end{array}\right)v_{d}\,,
\end{equation}
Obviously the mass hierarchies of the charged leptons are naturally recovered. The superpotential for the neutrino sector can be written as
\begin{eqnarray}
\label{eq31}&&w_{\nu}=y(\nu^c\ell)h_u+x_{\nu_1}((\nu^c\nu^c)_{\mathbf{3'_1}}\varphi)+x_{\nu_2}((\nu^c\nu^c)_{\mathbf{3'_2}}\psi)+\ldots
\end{eqnarray}
We can straightforwardly read the Dirac and Majorana neutrino mass matrices as following,
\begin{eqnarray}
\nonumber&m_D=yv_u\mathbb{1}&\\
\nonumber&m_M=\left(\begin{array}{ccc}
-4x_{\nu_1}v_{\varphi}+2x_{\nu_2}v_2  & 2x_{\nu_1}v_{\varphi}+x_{\nu_2}(v_1+v_2)   &   2x_{\nu_1}v_{\varphi}+2x_{\nu_2}v_1  \\
2x_{\nu_1}v_{\varphi}+x_{\nu_2}(v_1+v_2)  &  -4x_{\nu_1}v_{\varphi}+2x_{\nu_2}v_1   &  2x_{\nu_1}v_{\varphi}+2x_{\nu_2}v_2 \\
2x_{\nu_1}v_{\varphi}+2x_{\nu_2}v_1   &  2x_{\nu_1}v_{\varphi}+2x_{\nu_2}v_2    &  -4x_{\nu_1}v_{\varphi}+x_{\nu_2}(v_1+v_2)
\end{array}\right)&
\end{eqnarray}
The effective light neutrino mass matrix is given by the see-saw formula
\begin{equation}
\label{eq34}m_{\nu}=-m^T_Dm^{-1}_Mm_D=U_{TFH}\mathrm{diag}(m_1,m_2,m_3)U^{T}_{TFH}
\end{equation}
where $m_{1,2,3}$ are the light neutrino masses
\begin{eqnarray}
\nonumber m_1=\frac{y^2v^2_u}{6x_{\nu_1}v_{\varphi}+\sqrt{3}\,x_{\nu_2}(v_1-v_2)},~m_2=-\frac{y^2v^2_u}{3x_{\nu_2}(v_1+v_2)},~m_3=\frac{y^2v^2_u}{6x_{\nu_1}v_{\varphi}-\sqrt{3}\,x_{\nu_2}(v_1-v_2)}
\end{eqnarray}
Obviously the leptonic mixing matrix is exactly the desired TFH matrix. There are no correlations among the above three light neutrino masses, the neutrino mass spectrum can be both normal and inverted order hierarchy. After including the subleading order terms allowed by the symmetry, the above LO predictions for lepton masses and flavor mixing are corrected by both the shifted vacuum and the higher dimensional operators in the Yukawa superpotentials. Detailed and lengthy analysis in Ref. \cite{Ding:2012xx} showed that all the three leptonic mixing angle receive corrections of order $\lambda^2_c$, the agreement between the theoretical predictions and the experimental data remains.

\section{Beyond TFH mixing within $\Delta(96)$}

In discrete flavor symmetry model building, the flavor symmetry is spontaneously broken by the flavons into $G_{\ell}$ and $G_{\nu}$ subgroups in the charged lepton and neutrino sectors. The mismatch between $G_{\ell}$ and $G_{\nu}$, which results in the mismatch between the neutrino and charged lepton mass matrices, could leads to some interesting mass-independent textures. There is a direct group-theoretical connection between lepton mixing and the horizontal symmetry \cite{Lam:2011ag}. Given the whole flavor symmetry and the surviving subgroups $G_{\ell}$ and $G_{\nu}$, one can carry out a purely group-theoretical analysis to obtain all the possible mixings, without the presence of flavon fields nor the help of the Lagrangian. If neutrinos are Majorana particles, it could be shown that the remnant symmetry of the left-handed neutrinos forms a $K_4$ group for the TFH mixing. Consequently we choose $G_{\nu}$ to be the $K_4$ subgroups of $\Delta(96)$, and $G_{\ell}$ is taken to be the cyclic $Z_{N}$ subgroups of $\Delta(96)$ with $N\geq3$, since the resulting three charged lepton masses would be completely or partially degenerate if $G_{\ell}$ is some non-abelian subgroups. $\Delta(96)$ has seven $K_4$ subgroups, sixteen $Z_3$ subgroups, twelve $Z_4$ subgroups and six $Z_8$ subgroups. All these subgroups are listed in Ref. \cite{Ding:2012xx} in terms of the generators $a$, $b$, $c$ and $d$. By considering the large number of combinatorial choices of $G_{\nu}$ and $G_{\ell}$, all the possible lepton mixing matrices and the group structures generated by $G_{\nu}$ and $G_{\ell}$ are listed in Table 5 and Table 6 of Ref. \cite{Ding:2012xx}. These tables are too lengthy to be included in this proceeding, please refer to Ref. \cite{Ding:2012xx} for detail. It is clear that seven mixing patterns including the tri-bimaximal, bimaximal and TFH mixings can be reproduced within $\Delta(96)$. If we require that the elements of $G_{\ell}$ and $G_{\nu}$ generate the full group $\Delta(96)$, only the TFH and the bimaximal mixing patterns are admissible. These results are consistent with those obtained in Ref. \cite{deAdelhartToorop:2011re}.

It is worth noting that we are able to derive the tri-bimaximal mixing matrix, if $\Delta(96)$ is broken into $K_4$ and $Z_3$ in the neutrino and charged lepton sectors respectively, one can refer to Ref. \cite{Ding:2012xx} for concrete choices of $G_{\ell}$ and $G_{\nu}$. However, the group generated subgroup is $S_4$ instead of $\Delta(96)$. This result is consistent with the claim that the minimal flavor symmetry capable of yielding the tri-bimaximal mixing without fine tuning is $S_4$ from group theory point of view \cite{Lam:2008rs}. By exchanging the rows and columns of the tri-bimaximal mixing matrix, we find another interesting mixing pattern,
\begin{equation}
\label{eq76}||U_{PMNS}||=\left(\begin{array}{ccc}
\frac{1}{\sqrt{2}}  &  \frac{1}{\sqrt{3}}  &  \frac{1}{\sqrt{6}}  \\
0 & \frac{1}{\sqrt{3}}  &  \sqrt{\frac{2}{3}}  \\
\frac{1}{\sqrt{2}}  &  \frac{1}{\sqrt{3}}  &  \frac{1}{\sqrt{6}}
\end{array}\right)\,.
\end{equation}
This texture will be called deformed tri-bimaximal (DTB) mixing, and the resulting mixing angles are
\begin{equation}
\label{eq77}\sin^2\theta_{13}=\frac{1}{6}\,,\quad\quad \sin^2\theta_{12}=\frac{2}{5}\,,\quad\quad \sin^2\theta_{23}=\frac{4}{5}\,.
\end{equation}
In order to be compatible with experimental data, all the three mixing angles should undergo large corrections of order $0.1\sim0.2$, which is roughly the size of the Cabibbo angle. This mixing pattern is an interesting alternative for explaining largish $\theta_{13}$ and current atmospheric neutrino mixing angle. Obviously the appropriate framework to derive DTB mixing is the $S_4$ horizontal symmetry.

\section{Conclusion}

The TFH mixing pattern is a good approximation to the current neutrino flavor mixing data, especially it can help us to understand largish $\theta_{13}$. We show that the TFH texture can be naturally derived with the flavor symmetry $\Delta(96)$. Within the so-called minimalist framework, we study the possible ways to produce TFH mixing in $\Delta(96)$, the assignments of the matter fields under $\Delta(96)$ and the associated flavons are presented. However, in these assignments, we need to tune the involved parameters to account for the tiny masses of electron and muon. This defect can be overcame by further breaking the remnant $Z_3$ symmetry of the charged lepton sector, and a consistent model realization of this scenario is presented. Furthermore we investigate the possible mixing patterns if the family symmetry $\Delta(96)$ is broken into $K_4$ in the neutrino sector and the cyclic group $Z_N$ ($N\geq3$) in the charged lepton sector. We find that the TFH mixing can be accommodated if certain $Z_3$ subgroups are preserved in the charged lepton sector. We suggest the so-called DTB mixing is an another good LO texture to understand current leptonic flavor mixing. Finally we note that $\Delta(96)$ has doublet representation which can be utilized to describe the quark sector, moreover we could combine the grand unification theory with $\Delta(96)$ flavor symmetry \cite{King:2012in}.

\section*{Acknowledgments}
We would like to thank the organizers of FLASY 2012 where this talk was presented for their hospitality and a stimulating workshop. This work is supported by the National Natural Science Foundation of China under Grant No~10905053, Chinese Academy KJCX2-YW-N29 and the 973 project with Grant No.~2009CB825200.

\bibliographystyle{apsrev4-1}


%% file: Papers/feldmann.tex

%
%
%
%
%
%

\chapter[$\Delta A_{\rm CP}$ in $D$-Decays and ``Old Physics'' (Feldmann)]{$\Delta A_{\rm CP}$ in $D$-Decays and ``Old Physics''}
\vspace{-2em}
\paragraph{T. Feldmann}
\paragraph{Abstract}
We investigate to what extent the recently measured value for
a non-vanishing direct CP asymmetry in $D^0 \to K^+K^-$ and $D^0 \to \pi^+\pi^-$
decays can be accommodated in the Standard Model (SM).

\section{Introduction}

\begin{figure}[ht]
\centering \includegraphics[width=0.5\textwidth]{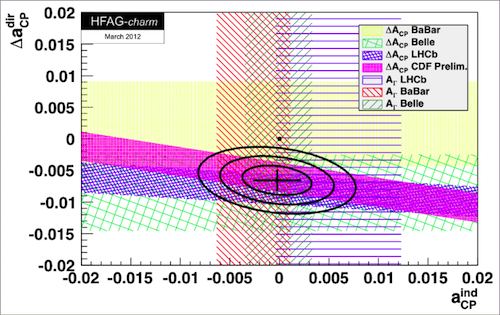}
\caption{\label{fig:tf_hfag} HFAG average for direct and indirect CP asymmetries in $D$-meson decays
\cite{Amhis:2012bh}.}
\end{figure}

CP-violating asymmetries in $D^0$ decays have been recently measured by 
different experiments, including LHCb \cite{Aaij:2011in,Aaij:2011ad}, CDF \cite{Aaltonen:2011se} and the B-factories
BaBar \cite{Aubert:2007en,Aubert:2007if} and Belle \cite{Staric:2007dt,Staric:2008rx}, with different sensitivities to direct and indirect
contributions. The present situation (as of march 2012) has been summarized by HFAG, see Fig.~\ref{fig:tf_hfag},
and corresponds to a direct CP asymmetry in the difference of two-body decays into charged kaons or pions
of
\begin{align}
 \Delta A_{\rm CP}^{\rm dir} &=   A_{\rm CP}^{\rm dir}(D^0\to K^+K^-) -  A_{\rm CP}^{\rm dir}(D^0\to \pi^+\pi^-) 
= (- 0.656 \pm 0.154 )\% 
\end{align}
The size of the experimental average (several permille to almost a percent),
is larger than the naive SM expectation (for details, see below), where in
the limit of exact U-spin symmetry between $s$ and $d$-quarks, one has
\begin{align}
 A_{\rm CP}^{\rm dir}(D^0\to K^+K^-) =  -  A_{\rm CP}^{\rm dir}(D^0\to \pi^+\pi^-) \,,
\label{eq:tf_naiverel}
\end{align}
and both decays contribute equally to $\Delta A_{\rm CP}^{\rm dir}$ with
an estimated size of 
\begin{align}
 \Delta A_{\rm CP}^{\rm dir} = 2 \, A^2 \, {\lambda^4} \, \eta \cdot r \, \sin \Delta \phi_{\rm strong}
\simeq 0.11\%  \cdot r \, \sin \Delta \phi_{\rm strong} \,,
\end{align}
where $r \, e^{i\phi}$ parameterizes the relevant hadronic amplitude ratio which is expected to
be smaller than $1$ in magnitude if the naive-factorization approximation of long-distance QCD effects
holds.

In this write-up, we will report about our
analysis in \cite{Feldmann:2012js} which focuses on a SM interpretation of $\Delta A_{\rm CP}$
and tries to answer the following questions:
What is the size of U-spin breaking? ---
What is the magnitude to expect for amplitude ratios and strong-phase differences
      beyond the factorization approximation? ---
Does a simple relation to non-factorizable effects in $B$-meson decays exist?
Related work can be found, for instance, in
\cite{Pirtskhalava:2011va,Brod:2011re,Brod:2012ud,Bhattacharya:2012ah,Bhattacharya:2012kq,Franco:2012ck,Li:2012cfa,Cheng:2012wr,Isidori:2011qw}. 
For possible NP signatures from $\Delta A_{\rm CP}$,
we refer the reader to the contribution of J.~Kamenik in chapter \ref{chap:kamenik}.

\section{Standard Model Analysis}

If we include first-order U-spin breaking effects,
the amplitudes for the various $D^0$ decays to charged pions or
kaons can be parameterized as follows,
 \begin{align}
{\cal A}[D^0 \to K^- \pi^+ ] & =
 2 \,V_{cs}^* V_{ud} \, B_{U=1} \left[  1 - {r_1' \, e^{i \, \phi_1'}} \right] \,,
\nonumber \\[0.2em]
{\cal A}[D^0 \to \pi^+ \pi^-] &= B_{U=1}\left[(\lambda_d + \lambda_s)\,\left( {r\, e^{i\,\phi}} + {r_1\,e^{i\,\phi_1}} \right) 
                + (\lambda_d -\lambda_s) \, \left(1 + {r_0\, e^{i\,\phi_0}} \right)\right] \,, \nonumber 
\\[0.2em] 
{\cal A}[D^0 \to K^+ K^-] &= B_{U=1}\left[(\lambda_d + \lambda_s)\, \left({r\, e^{i\,\phi}} - {r_1\,e^{i\,\phi_1}} \right) 
             - (\lambda_d -\lambda_s) \, \left(1 - {r_0\, e^{i\,\phi_0}}\right)\right]  \,,
\nonumber \\[0.2em]
 {\cal A}[D^0 \to K^+ \pi^-] &= 2 \,V_{cd}^* V_{us} \, B_{U=1} \left[  1 + {r_1' \, e^{i \, \phi_1'}} \right]  \,,
\label{eq:tf_uspin}
\end{align}
where $\lambda_q = V_{cq}^* V_{cq}$ in the standard notation for CKM factors.
The two complex amplitude ratios $r_0,r_1'$ (with their corresponding strong phases) describe the 
U-spin breaking in Cabbibo-favoured terms ($\sim 1,\lambda,\lambda^2$).
The two complex amplitude ratios $r,r_1$ parameterize  U-spin symmetric and U-spin breaking 
effects in the Cabbibo-suppressed terms {($\sim\lambda^5$)}. The supposed-to-be leading amplitude 
for $\Delta U=1$ transitions, $B_{U=1}$, has been factored out.

\begin{figure}[ht!]
 \centering
%
\fbox{\includegraphics[height=0.2\textwidth]{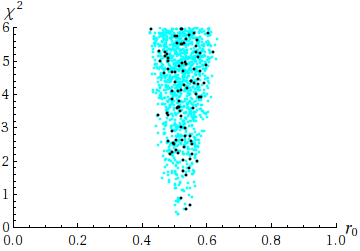}} \
\fbox{\includegraphics[height=0.2\textwidth]{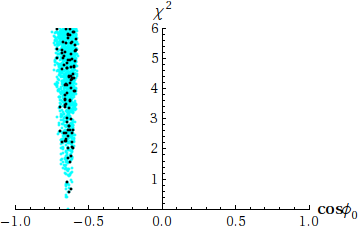}}\\
\fbox{\includegraphics[height=0.2\textwidth]{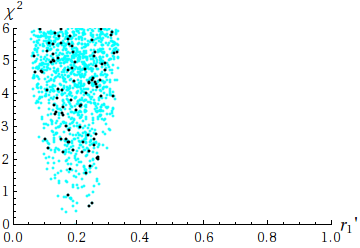}} \ 
\fbox{\includegraphics[height=0.2\textwidth]{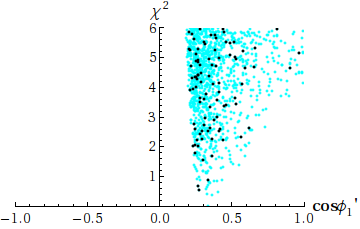}}
\caption{\label{fig:tf_uspin} $\chi^2$ distribution for the amplitude parameters
$r_0$, $\cos\phi_0$,$r_1'$, $\cos\phi_1'$ (top left to bottom right).}
\end{figure}
From this parameterization, the amount of U-spin breaking
can  be quantified from the experimental measurements of the 4 individual branching ratios
together with the experimentally fitted strong-phase difference in $D^0 \to K^\pm \pi^\mp$
decays. A poor man's $\chi^2$ analysis as explained in \cite{Feldmann:2012js}
then leads to, see also Fig.~\ref{fig:tf_uspin},
$$
r_0 \simeq 0.52\,, \quad \cos\phi_0 \simeq -0.64\,,\qquad r_1' \simeq 0.19\,,\quad \cos\phi_1' \gtrsim 0.18 \,.
$$

As concerns the size of sub-leading amplitude ratios,
the experimentally measured value of $ \Delta A_{\rm CP}^{\rm dir}$ alone
is not sufficient to determine the individual amplitude parameters.
However, we find that a certain average of amplitude ratios can be constrained,
\begin{align} 
\bar r \equiv \sqrt{r^2/2+r_1^2/2} & \quad {\gtrsim \, 2 (3)} \ \quad \mbox{\small (with $2\sigma$ ($1\sigma$))}\,,
\end{align}
which implies that at least one of the relevant amplitude ratios has to be larger than naively expected.
As an immediate consequence of the sizable amount of U-spin breaking, we also find that the 
relation (\ref{eq:tf_naiverel}) between individual
 CP asymmetries can be violated by ${\cal O}(1)$ effects. For instance,
assuming universal strong phases
between $U=0$ and $U=1$ amplitudes,  one obtains
$\Sigma A_{\rm CP}^{\rm dir}/\Delta A_{\rm CP}^{\rm dir}\simeq -50\%$.


The theoretical interpretation of the BRs and CP asymmetries in $D^0\to P^+P^-$ decays
in the SM shows that the factorization approximation in non-leptonic $D$-meson decays
is badly violated, inducing large amplitude ratios and strong phases.
This implies that none of the expansion parameters $\Lambda/m_c$, $\alpha_s(m_c)/\pi$, $1/N_C$, etc.\
is sufficiently small. It also allows for significant contributions from both, $\Delta U=0$ and $\Delta U=1$ operators
to $\Delta A_{\rm CP}$.
A particular scenario, where the hadronic enhancement is associated to long-distance penguin contractions,
has been discussed in \cite{Brod:2012ud}. The parametrization in that paper amounts to setting
\begin{align*}
 \frac{r_0}{r_1'} &= \frac{{\epsilon} \, |2 {s_1}|}{{\epsilon} \, |t_1|} \gg 1  \,,
\qquad
 r_1 = \frac{{\epsilon} \, |{p_1}|}{|t_0|}  \sim 1 \,, \qquad 
 r = \frac{2 { p_0}}{t_0} \gg 1 \,,
\end{align*}
and the power-counting assumes small U-spin breaking of order $\epsilon \ll 1$, 
while penguin amplitudes $s_1$ and $p_1$ are enhanced with respect to tree amplitudes $t_{0,1}$.

To validate/falsify this or alternative assumptions, including NP explanation, 
further experimental tests of other charm decay modes have to be performed.
In particular, since the net CP asymmetries in non-leptonic decays arise as
the consequence of a rather involved interference of several 
hadronic effects, one could expect that -- within the SM -- 
there will also be modes with small CP violation due to destructive 
interference, for instance for decays with vector mesons instead of pseudoscalars 
in the final state.
On the other hand, if the observed $\Delta A_{\rm CP}$ arises dominantly from
a NP source with definite flavour structure, one would expect 
correlated deviations of several independent CP asymmetries from the SM expectation,
or significant violation of certain SM sum rules 
\cite{Grossman:2012eb} between related decay modes.
In extreme cases, the presence of new sources of CP violation in the charm sector
can even lead to sizeable enhancement of electric dipole moments \cite{Mannel:2012hb},
or CP violation in tree decays.
Below, we will briefly discuss an example for a NP model with constrained flavour
sector. More on the NP interpretation of $\Delta A_{\rm CP}$ can be found in J.~Kamenik's contribution in chapter \ref{chap:kamenik}.


\section{Sequential 4th Generation -- A new physics example with constrained flavour coefficients}

\begin{figure}[ht!]
 \centering
 \includegraphics[height=0.3\textwidth]{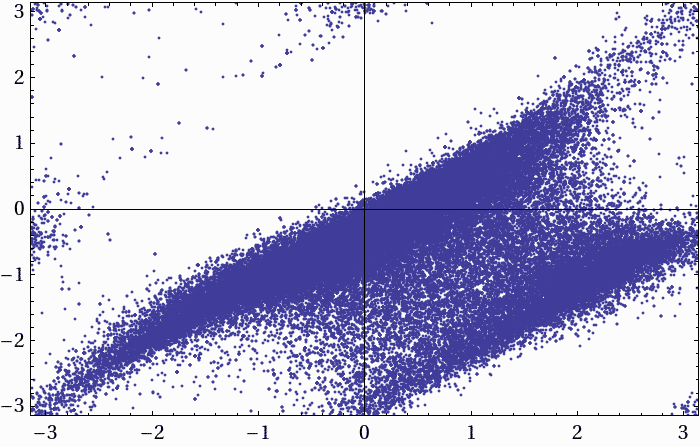} 
\qquad
 \includegraphics[height=0.3\textwidth]{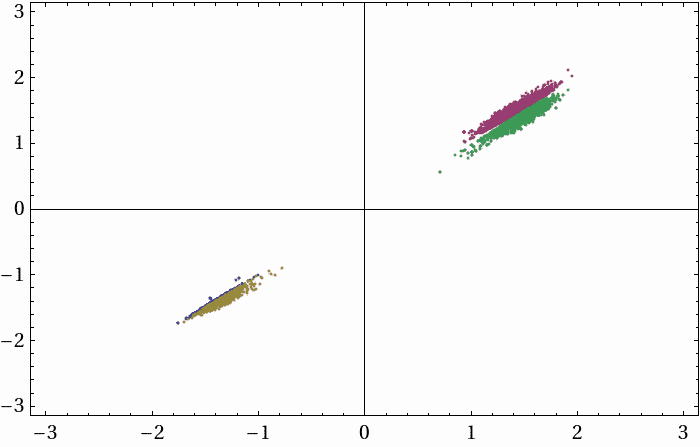} 
\caption{\label{fig:tf_4Gmix} Two examples for implications of phenomenological
flavour constraints on 4G mixing parameters. Left: Allowed values for the new CP phases 
$\delta_{14}$ and $\delta_{24}$ for small mixing angles $\theta_{i4}$. Right: The same for 
large mixing angles, which requires tuned values for CP phases, satisfying 
$\delta_{14} \simeq \delta_{24}$. Taken from \cite{Buras:2010pi}.}
\end{figure}

Extensions of the SM by a fourth generation of quarks (and also leptons) have
recently received a lot of attention. From the flavour-symmetry perspective,
they represent examples for NP with ``next-to-minimal flavour violation'' (nMFV)
\cite{Feldmann:2006jk}, where the additional flavour mixing through the 4th generation (4G)
is assigned to an additional complex ``spurion field'' which transforms as a fundamental
triplet under $SU(3)$ rotations of the left-handed quark doublets, and the mixing angles
fulfill consistency relations in forms of inequalities,\footnote{In the SM with 3 generations, one has $\theta_{12}\theta_{23} \sim \theta_{13}$, \
                            $\theta_{12}\theta_{13} \ll \theta_{23}$, \  $\theta_{13} \theta_{23} \ll \theta_{12}$.}
\begin{align} \theta_{i4} \theta_{j4} \lesssim \theta_{ij} \,, \qquad
 \theta_{ij} \, \theta_{j4} \lesssim \theta_{i4}\,.
\end{align}
In this sense, the 4G flavour sector represents a whole class of nMFV models,
where the experimentally explored flavour phenomenology already severely constrains
the NP flavour parameters. In particular, the 
constraints from $B$-meson and kaon observables imply that
for large new CP phases  
one has to require small 4G mixing angles,
and vice versa --- two examples are shown in Fig.~\ref{fig:tf_4Gmix}.
As a consequence, the presence of a fourth quark generation alone \emph{cannot}
lead to a large parametric enhancement of $\Delta A_{\rm CP}$ (which receives
additional contributions proportional to $\sin (\delta_{14}-\delta_{24})$
and $\sin\theta_{i4}$). Therefore, in principle, the presence of new 4G penguin 
contributions to $\Delta A_{\rm CP}$ can constructively or destructively interfere 
with the SM effects, leading to moderate enhancement or reduction.



\section{Relation to Non-Leptonic $B$-Decays?}

In contrast to $D$-meson decays, QCD factorization works 
reasonably well in non-leptonic $B$-meson thanks to the fact
that the $b$-quark mass is sufficiently large. Still, the
non-zero sensitivity to non-factorizable effects in penguin-dominated
decays, like in certain $B\to \pi K$ modes, can be taken as an
indicator for enhanced non-perturbative effects in non-leptonic $D$-decays.

For a quantitative comparison, we consider the academic 
(i.e.\ not-expected-to-be-realistic) example of 
non-leptonic $b \to s$ transitions with additional up-
and charm-quarks in the final state which we will relate
by ``W-spin'' symmetry ($c\leftrightarrow u$) and its
breaking in complete analogy to the U-spin analysis
of non-leptonic $D$-meson decays (\ref{eq:tf_uspin}). We thus parameterize
\begin{align}
{\cal A}[\bar B^0 \to D^+ K^-] &
= 2 \,V_{cb} V_{us}^* \, B_{W=1}
\left[  1 - r_1' \, e^{i \, \phi_1'} \right] \,,
\nonumber\\[0.25em]
{\cal A}[\bar B^0 \to K^- \pi^+] 
&= B_{W=1}\left[(\lambda_u + \lambda_c)\,\left(r\, e^{i\,\phi} + r_1\,e^{i\,\phi_1} \right) 
  + 
                    (\lambda_u -\lambda_c) \, \left(1 + r_0\, e^{i\,\phi_0}\right)\right]\,, 
\nonumber \\[0.1em] 
{\cal A}[\bar B^0 \to D_s^- D^+] 
&= B_{W=1}\left[(\lambda_u + \lambda_c)\, \left(r\, e^{i\,\phi} - r_1\,e^{i\,\phi_1} \right) 
             - (\lambda_u -\lambda_c) \, \left(1 - r_0\, e^{i\,\phi_0}\right)\right] \,,
\nonumber\\[0.1em]
 {\cal A}[\bar B^0 \to D_s^- \pi^+] & 
= 2 \,V_{ub} V_{cs}^*\, B_{W=1}
\left[  1 + r_1' \, e^{i \, \phi_1'} \right] \,.
\end{align}
Perhaps surprisingly, the fit to the available experimental data yields
qualitatively similar results as for the U-spin analysis of $D \to P^+P^-$
which may be due to the fact that $m_c \ll m_b$ is as good/bad an approximation
for $B$-decays as $m_s \ll m_c$ for $D$-decays. 
The essential features are compared in Table~\ref{tab:tf_comp}. 
In particular, the fit result for the amplitude ratios in the W-spin analysis
could be taken as a ``guesstimate'' for an upper bound on the amplitude ratios
in the U-spin analysis. 

\begin{table}[hb!]
 \centering
\caption{\label{tab:tf_comp} Comparison between non-leptonic
$B$- and $D$-decays.}
  \begin{tabular}{|c| c |}
\hline
   W-spin analysis of $B \to P^+P^-$ & U-spin analysis of $D \to P^+P^-$ 
\\
\hline \hline
  solutions with $r_1'<1$ & $r_1' \simeq 0.19$ 
\\
  $1 \lesssim \sqrt{\frac{r^2+r_0^2+r_1^2}{2}} \lesssim 6$ &  $2-3 \lesssim \sqrt{\frac{r^2+r_1^2}{2}}$ 
\\
  $A_{\rm CP}(\bar B^0 \to D_s^- D^+) < 12\%$ & $A_{\rm CP}(D^0\to \pi^+\pi^-) \neq-A_{\rm CP}(D^0\to K^+K^-)$ 
\\
\hline
  \end{tabular}
\end{table}


\section{(Inconclusive) Conclusions}

At the moment, the situation concerning the theoretical understanding
of $\Delta A_{\rm CP}$ is still rather unclear, and evidently 
our conclusions will be somewhat vague with some question marks
left open for future studies:
\begin{itemize}
 \item If the central value of $\Delta A_{\rm CP}$ is confirmed with higher experimental precision, can this be used to
    rule out the SM?  --- The answer is: ``Probably NO!'', because the factorization approximation for the hadronic
    dynamics is clearly insufficient, and therefore a solid SM prediction for both, the central value and the
    hadronic uncertainties is theoretically out of reach.

\item Can one understand the presently measured value of $\Delta A_{\rm CP}$ within the SM? ---
     The answer is: ``Maybe.'' As we have seen, an enhancement of non-factorizable effects compared to 
      non-leptonic $B$-decays like $B \to \pi K$ appears quite natural, and also the similarities 
      between the pattern of U-spin breaking in $D \to PP$ and W-spin breaking in $B \to PP$
      does not rule out amplitude ratios as large as being needed for the SM explanation of $\Delta A_{\rm CP}$.  

\item Can we still hope to see NP emerging in D-decays? --- The answer is: ``Let's see\ldots''.
     We have shown an example where the NP effects in the charm sector are already constrained 
     from flavour observables in $B$- and $K$-decays, such that the SM and the NP effects interfere with similar
     magnitudes which makes it notoriously difficult to draw definite conclusions.
     On the other hand, a global analysis of many independent D-decay modes can help to identify the short-distance
     sources responsible for the observed CP violation with less ambiguities. 
 \end{itemize}
In any case, a continuation of the charm-physics program with more experimental data on
various decay modes will shed more light on these issues in the future, see for instance
the recent discussions in \cite{Gersabeck:2012xk,Gersabeck:2012rp,Bediaga:2012py}.

\section*{Acknowledgments}

I would like to congratulate the organizers of FLASY 2012 in Dortmund
for a very stimulating and exciting workshop.
I would also like to thank Soumitra Nandi and Amarjit Soni for a 
very fruitful collaboration.



\bibliographystyle{apsrev4-1}

%% file: Papers/girrbach.tex
\chapter[Correlations in Minimal $U(2)^3$ models and an  $SO(10)$ SUSY GUT model facing new data (Girrbach)]{Correlations in Minimal $U(2)^3$ models and an  $SO(10)$ SUSY GUT model facing new data}
\label{chap:girrbach}
\vspace{-2em}
\paragraph{J. Girrbach}
\paragraph{Abstract}

Models with an approximate $U(2)^3$ flavour symmetry represent simple non-MFV extensions
of the SM. We compare correlations of $\Delta F = 2$ observables in CMFV and in a minimal version of $U(2)^3$ models, $MU(2)^3$.
Due to the
different treatment of the third generation $MU(2)^3$ models avoid the $\Delta M_{s,d}-|\varepsilon_K|$ correlation of CMFV which
precludes to
solve the $S_{\psi K_S}- |\varepsilon_K|$ tension present in the flavour data. While the
flavour structure in $K$ system is the same for CMFV and $MU(2)^3$ models, CP violation in $B_{d,s}$ system can deviate
in $MU(2)^3$ models from CMFV. We point out a triple correlation between $S_{\psi\phi}$, $S_{\psi K_S}$ and $|V_{ub}|$ that can
provide a distinction between different $MU(2)^3$ models.

GUTs open the possibility to transfer the neutrino mixing matrix $U_\text{PMNS}$ to the quark sector which leads to correlations
between leptonic and hadronic observables. This is accomplished in a
controlled way in an $SO(10)$ SUSY GUT model proposed by Chang, Masiero and Murayama (CMM model) whose flavour structure differ
significantly from the CMSSM.  We present a summary of a global analysis of several flavour processes containing
$B_s-\overline{B}_s$ mixing, $b\to s\gamma$ and $\tau\to\mu\gamma$. Furthermore we comment on the implications on the model due
to the latest data of $S_{\psi\phi}$, $\theta_{13}$ and the Higgs mass.

\section{Current situation of the flavour data}

With the start of the LHCb experiment a new era in precision measurements in flavour physics started.
The present 95\% C.L. upper bound  $\mathcal{B}(B_s\to \mu^+\mu^-)\leq 4.5\cdot 10^{-9}$ \cite{Aaij:2012ac} is already close to
the
SM prediction  $\mathcal{B}(B_s\to \mu^+\mu^-)^\text{SM} = (3.1\pm 0.2)\cdot 10^{-9}$
\cite{Buras:2012ts,LHCb:2012py}\footnote{In \cite{Buras:2012ru} the ``non-radiative'' branching ratio that corresponds to the
branching ratio fully inclusive of bremsstrahlung radiation was calculated to  $\mathcal{B}(B_s\to \mu^+\mu^-) =(3.23 \pm 0.27)
\cdot 10^{-9}$. When the corrections from $\Delta \Gamma_s$, pointed out in \cite{deBruyn:2012wj,deBruyn:2012wk} are taken into
account the experimental upper bound is reduced to $4.1\cdot 10^{-9}$.}. New data on mixing induced CP
violation in $B_s-\overline{B}_s$ mixing measured by $S_{\psi\phi} = 0.002\pm 0.0087$ \cite{Clarke:1429149} is consistent with the SM
prediction of $S_{\psi\phi}^\text{SM} =0.0035\pm 0.002 $  and excludes ranges from CDF and D\O\ with large $S_{\psi\phi}$.
Thus there is not much room left for new physics (NP).

However a slight tension in the flavour data concerns $|\varepsilon_K|$, $B^+\to \tau^+\nu$ and $S_{\psi K_S}$ which can be
related with the so-called
$|V_{ub}|$-problem. Both $|\varepsilon_K|\propto \sin2\beta |V_{cb}|^4$ and $S_{\psi K_S}$ can be used to determine $\sin2\beta$.
 In
Fig.~\ref{fig:DeltaMvsepsKv2} (left) one can see that the $\sin2\beta$ derived from the experimental value of $S_{\psi K_S}$
is
much smaller that the one derived from $|\varepsilon_K|$. This issue was discussed in \cite{Lunghi:2008aa,Buras:2008nn}.
 The ``true'' value of $\beta$ depends on the value of $|V_{ub}|$ and $\gamma$.  However there is a tension between the exclusive
and
inclusive determinations of  $|V_{ub}|$ \cite{Nakamura:2010zzi}:
\begin{align}
 &|V_{ub}^\text{incl.}| = (4.27\pm 0.38)\cdot 10^{-3}\,,\qquad |V_{ub}^\text{excl.}| = (3.38\pm 0.36)\cdot
10^{-3}\,.
\end{align}
 Now one can distinguish between these two benchmark scenarios: If one uses the
exclusive value of $|V_{ub}| $ to derive $\beta_\text{true}$ and then calculates $S_{\psi
K_S}^\text{SM}=\sin2\beta_\text{true}$ one
finds agreement with the data whereas $|\varepsilon_K|$ stays below the data. Using the inclusive  $|V_{ub}| $ as input for 
$\beta_\text{true}$, $S_{\psi K_S}$ is above the measurements while $|\varepsilon_K|$ is in agreement with the
data. However in such considerations one has to keep in mind the error on  $|\varepsilon_K|$ coming dominantly from the 
error of $|V_{cb}|$ and the error of the QCD factor $\eta_1$~\cite{Brod:2011ty}.

The branching ratio $\mathcal{B}(B^+\to \tau^+\nu)$ can also be used to  measure $|V_{ub}|$. The SM
prediction  $\mathcal{B}(B^+ \to \tau^+ \nu)_{\rm SM}= (0.80 \pm 0.12)\cdot
10^{-4}$ as calculated in \cite{Altmannshofer:2009ne} where  one eliminates
the uncertainties of $F_{B^+}$ and $|V_{ub}|$ by using $\Delta M_d$,  $\Delta M_d/\Delta M_s$ and
$S_{\psi K_S}$  is about a factor 2 below the experimental world avarage based
on results by BaBar \cite{Aubert:2009wt} and Belle \cite{Ikado:2006un}: $\mathcal{B}(B^+ \to \tau^+ \nu)_{\rm exp} = (1.67 \pm
0.30) \cdot 10^{-4}$ \cite{HFAG}. Consequently this favors a large $|V_{ub}|$ and leads to a $S_{\psi
K_S}-\mathcal{B}(B^+\to\tau^+\nu)$ tension discussed for example in \cite{Nierste:2011na}. Recently  new results have been
provided by
BaBar $\mathcal{B}(B^+ \to \tau^+ \nu)_{\rm exp} = (1.79 \pm 0.48) \cdot 10^{-4} $ \cite{Lees:2012ju}
and by Belle $
\mathcal{B}(B^+ \to \tau^+ \nu)_{\rm exp} = (0.72 \pm^{0.27}_{0.25} \pm^{0.46}_{0.51}) \cdot 10^{-4}$
\cite{Adachi:2012mm} where the latter   value went down and is consistent with the SM prediction.

It is now interesting to see if a certain new physics model can solve these problems and if yes, which $|V_{ub}|$ scenario is
chosen.
In the following we will confront constraint minimal flavour violation (CMFV) and models with a global $U(2)^3$ symmetry to this
tension. At the end we discuss the CMM model  as an alternative to MFV.

\boldmath
\section{Correlations of $\Delta F = 2$ observables: CMFV vs. $MU(2)^3$}
\unboldmath

\begin{figure}
\centering
\includegraphics[width = 0.45\textwidth]{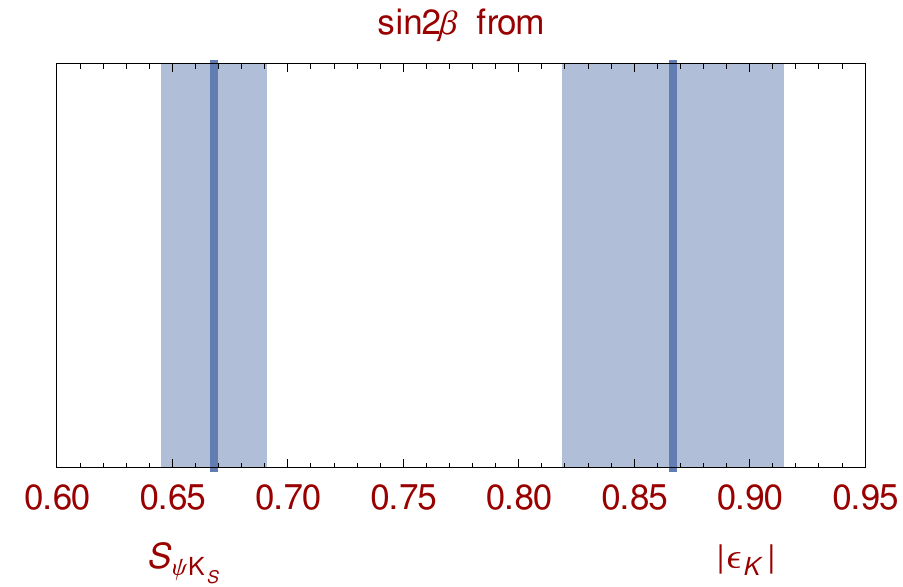}
 \includegraphics[width = 0.45\textwidth]{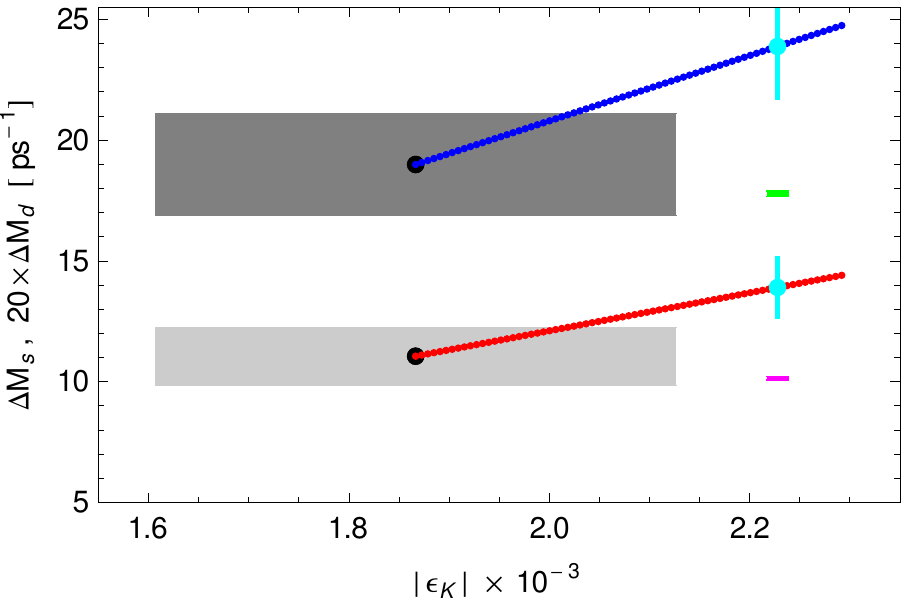}
\caption{Left: $\sin2\beta$ determined from $S_{\psi K_S}$  and $|\varepsilon_K|$. Right: $\Delta M_{s}$ (blue) and
$20\cdot\Delta M_{d}$ (red) as functions of $|\varepsilon_K|$ in models 
with CMFV for $|V_{ub}| = 0.0034$ chosen by these models. The short green and magenta  lines represent the data, while the large
gray regions corresponds to the SM 
predictions \cite{Buras:2012ts}.}\label{fig:DeltaMvsepsKv2}
\end{figure}

A very simple extension of the SM is CMFV, where the CKM matrix is the only source of flavour and CP violation and only SM
operators are
relevant below the electroweak scale.
Phenomenological consequences of CMFV concerning $\Delta F = 2$ observables are the following:
First, since there are no new CP violating phases the mixing induced CP asymmetries stay as in the SM: $S_{\psi K_S} =
\sin2\beta$, $ S_{\psi\phi} =
\sin2|\beta_s|$. Second, $\Delta M_{s,d}$ and $|\varepsilon_K|$ can only be enhanced  simultaneously relative to the
SM~\cite{Blanke:2006yh,Buras:2000xq}. Third, CMFV chooses exclusive $|V_{ub}|$ because $S_{\psi K_S}$ stays as in the SM
and $|\varepsilon_K|$ can be enhanced.
But if one wants to solve the $|\varepsilon_K|-S_{\psi K_S}$ tension one gets a problem with $\Delta M_{s,d}$. This
$\Delta M_{s,d}-|\varepsilon_K|$ tension is shown in
Fig.~\ref{fig:DeltaMvsepsKv2} (right).

 \begin{figure}[!tb]
  \centering
 \includegraphics[width = 0.45\textwidth]{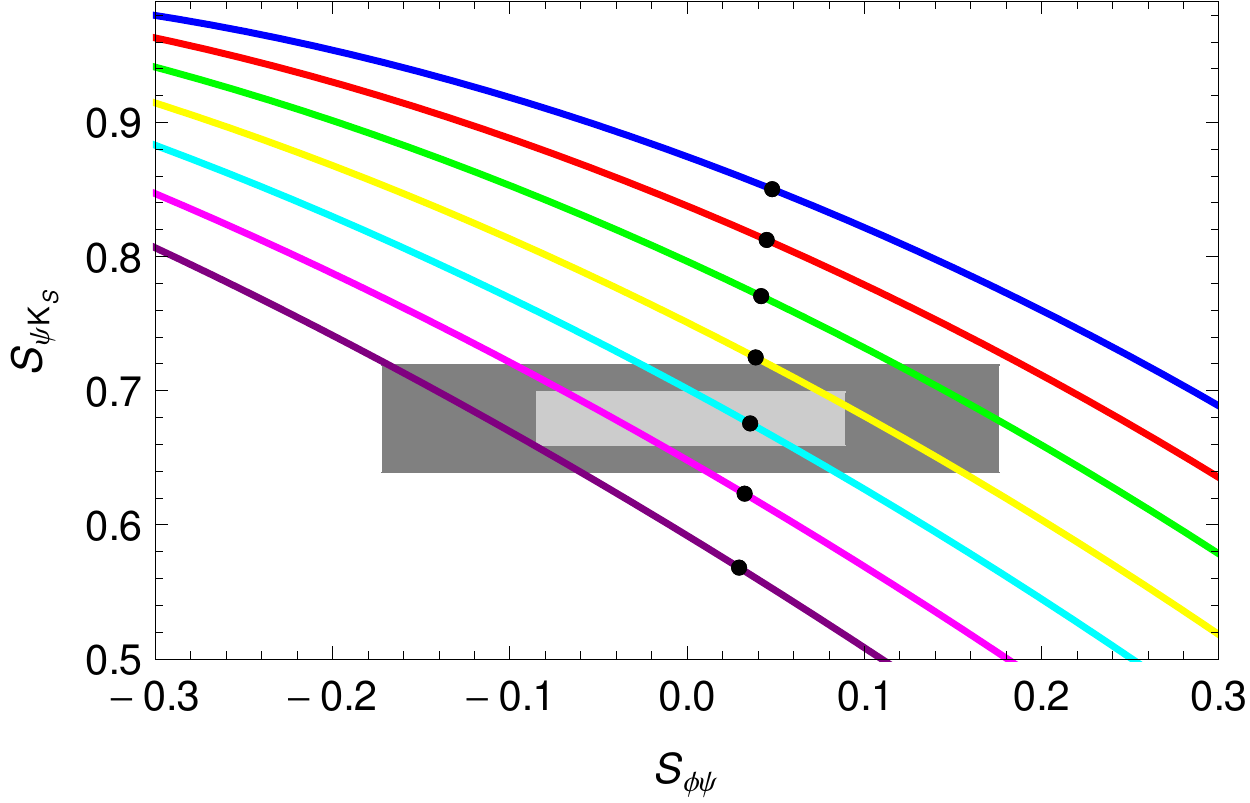}
 \includegraphics[width = 0.45\textwidth]{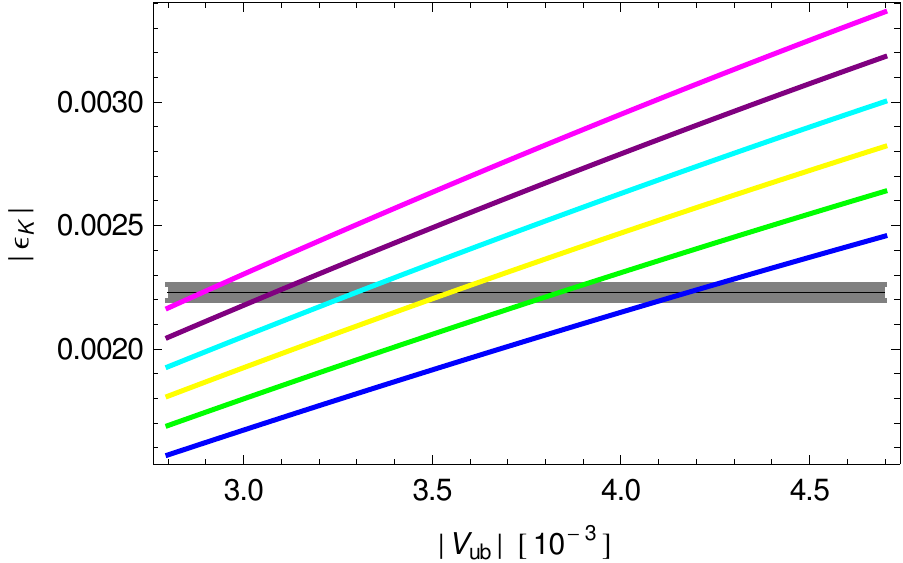}
 \caption{ Left: $S_{\psi K_S}$ vs $S_{\psi \phi}$ in $MU(2)^3$ for different values of $|V_{ub}|$. From top to bottom: $|V_{ub}|
=$ $0.0046$ (blue), $0.0043$ (red), $0.0040$
(green),
 $0.0037$ (yellow), $0.0034$ (cyan), $0.0031$ (magenta), $0.0028$ (purple). Light/dark gray: experimental $1\sigma/2\sigma$
 region. Right: $|\varepsilon_K|$ vs $|V_{ub}| $   for fixed $S_{\psi K_S}=0.679$ and different 
 values of $r_K$. From top to bottom: $r_K = 1.5$ (magenta), $1.4$ (purple), $1.3$ (cyan), $1.2$ (yellow),
 $1.1$
 (green), $1$ (blue, SM prediction)). Gray  region: experimental $3\sigma$ range of $|\varepsilon_K|$
\cite{Buras:2012sd}.}\label{fig:SvsS}
\end{figure}

Models with a global $U(2)^3$ flavour symmetry represent simple non-MFV extensions of the SM and can help avoiding this $\Delta
M_{s,d}-|\varepsilon_K|$ tension.
The $U(2)^3$ symmetry was first studied in 
\cite{Pomarol:1995xc,Barbieri:1995uv} and then in
 \cite{Barbieri:2011ci,Barbieri:2011fc,Barbieri:2012uh,Barbieri:2012bh,Crivellin:2011fb,Crivellin:2011sj,Crivellin:2008mq}
where  a detailed description of the model can be found (see also talk by F. Sala during this workshop).
A nice feature of $U(2)^3$ is that one can easily embed SUSY with heavy 1$^\text{st}$/2$^\text{nd}$ sfermion
generation and a light
3$^\text{rd}$
generation which is still consistent with current collider bounds on sparticle masses. In a minimal version of this model,
called $MU(2)^3$, the symmetry is broken minimally by three spurions and only SM
operators are relevant.
General consequences  of $MU(2)^3$ concerning $\Delta F = 2$ observables
are the following:
\begin{itemize}
 \item The flavour structure in the $K$-meson system is governed by MFV (no new phase $\varphi_K$).
\item Corrections in $B_{d,s}$ system are proportional to the SM CKM structure and  universal.
\item There exists one new universal phase that only appears in $B_{d,s}$ system: $\varphi_\text{new}$.
\end{itemize}
These properties lead to the following equations describing $\Delta F = 2$ observables where only three new parameters appear
\begin{align}
& S_{\psi K_S} = \sin(2\beta + 2 \varphi_\text{new})\,,\qquad
S_{\psi\phi} = \sin(2|\beta_s|-2\varphi_\text{new})\,,\\
&\Delta M_{s,d} = \Delta M_{s,d}^\text{SM} r_B\,,\qquad \qquad\quad \varepsilon_K =
r_K\varepsilon_K^\text{SM,tt} + \varepsilon_K^\text{SM,cc+ct}\,.
\end{align}
The parameters $r_{K,B}$ are real and positive definite and further $r_K\geq 1$. In contrast to CMFV $r_B$ and $r_K$ are in
principle unrelated. However in concrete realizations of the model, e.g. SUSY
 they both depend on SUSY masses.
In \cite{Buras:2012sd} we point out a  triple 
$S_{\psi K_S}-S_{\psi\phi}-|V_{ub}|$ correlation which will provide a crucial test of the
$MU(2)^3$ scenario once the three observables will be precisely known. This is shown in Fig.~\ref{fig:SvsS} (left) for fixed
$\gamma
= 68^\circ$\footnote{Varying $\gamma$ between $63^\circ$ and $73^\circ$ does not change the result significantly.}.
Negative $S_{\psi\phi}$ is  only possible for small $|V_{ub}|$ in the ballpark of the exclusive value. 
For inclusive $|V_{ub}|$, $S_{\psi\phi}$ is always larger than the SM prediction. $MU(2)^3$ models that are consistent with this
correlation should also describe the data for $|\varepsilon_K|$ and $\Delta M_{d,s}$. For example for $S_{\psi\phi}<0$
the particular $MU(2)^3$ model must provide a 25\% enhancement of $|\varepsilon_K|$ (see Fig.~\ref{fig:SvsS} right plot).
Moreover, if this $MU(2)^3$ flavour symmetry turns out to be true one can determine $ |V_{ub}|$ by means of precise measurements
of $S_{\psi K_S}$ and $S_{\psi \phi}$ with small hadronic uncertainties. The dependence of $|\varepsilon_K|$ (only central values)
on $|V_{ub}|$ for different values of $r_K$ is shown in the right plot of
Fig.~\ref{fig:SvsS}. 
Fixing $S_{\psi K_S}=0.679$ to its central experimental value we can use the triple correlation to get the connection between
$|\varepsilon_K|$ and $S_{\psi \phi}$ (see Fig.~4 in \cite{Buras:2012sd}). Thus we see that even in
$MU(2)^3$ models correlations between $B$- and $K$-physics are possible.

\section{$SO(10)$ SUSY GUT: CMM model }

In an $SO(10)$ SUSY GUT model proposed by Chang, Masiero and Murayama \cite{Chang:2002mq,Moroi:2000tk}  the
neutrino mixing matrix $U_\text{PMNS}$ is transfered to the right-handed down quark and charged lepton sector. In
\cite{Girrbach:2011an} we have performed a global analysis in the CMM
model including an extensive renormalization group (RG) analysis to connect Planck-scale and low-energy parameters. A short
summary  can be found in \cite{Buras:2012ts,Girrbach:2011wt,Nierste:2011na}. In view of the new 
knowledge about the Higgs mass and the latest measurements of the reactor neutrino mixing angle $\theta_{13}$ an updated analysis
of this model would be desirable.


The basic ingredient of the flavour structure is that not only the neutrinos are rotated with $U_\text{PMNS}$
but the whole $\mathbf{5}$-plets of SU(5) $\mathbf{5}_i = (d_{Ri}^c, \,\ell_{Li},\,-\nu_{\ell_i})^T$. Including SUSY
the 
atmospheric neutrino mixing angle $\theta_{23}\approx 45^\circ$ is responsible for large $\tilde b_R-\tilde s_R$- and $\tilde
\tau_L-\tilde\mu_L$-mixing which can then induce $b\to s$ and $\tau\to\mu$ transitions via SUSY loops. For a more detailed
derivation starting from an $SO(10)$ superpotential see \cite{Girrbach:2011an}. From the superpotential and the requirement of
perturbative couplings up to the Planck scale one can derive a range for $\tan\beta$: $2.7\lesssim\tan\beta\lesssim 10$.
Rotating from flavour to mass eigenstate basis the right-handed down squark mass matrix at $M_Z$ reads
\begin{align}\label{equ:msquark}
 m_{\tilde{D}}^2 = U_D m_{\tilde{d}}^2 U_D^\dagger\approx
m_{\tilde{d}_1}^2\left(\begin{array}{ccc}
1 & 0 & 0\\
0 & 1-\frac{1}{2}\Delta_{\tilde{d}} &
-\frac{1}{2}\Delta_{\tilde{d}}e^{i\xi}\\
0 & -\frac{1}{2}\Delta_{\tilde{d}}e^{-i\xi}&
1-\frac{1}{2}\Delta_{\tilde{d}}
                                \end{array}
\right)\,,
\end{align}
where the neutrino mixing enters through $ U_D =
U_\text{PMNS}^\ast\,\text{diag}(1,\, e^{i\xi},\,1)$ and $\Delta_{\tilde{d}}\in [0,\,1]$ defines the relative mass splitting
between the 1$^\text{st}$/2$^\text{nd}$ and
3$^\text{rd}$ down-squark
generation. It is generated by RG effects of the top Yukawa coupling and can reach $0.4$. Thus the CMM model shares the feature
of $U(2)^3$ models of heavy 1$^\text{st}$/2$^\text{nd}$ squark generations but a light 3$^\text{rd}$ generation.
The 23-entry $ \propto\Delta_{\tilde{d}}$ is responsible for $\tilde b_R-\tilde s_R$-mixing and a
new
CP violating phase $\xi$ enters that affects $B_s-\overline{B}_s$-mixing. The ``$\approx$'' sign in (\ref{equ:msquark}) gets a
``$=$'' if one uses
tribimaximal mixing in $U_\text{PMNS}$. However, the latest data show that $\theta_{13}$
is  non-zero \cite{Abe:2011fz,An:2012eh,Ahn:2012nd}. Including $\theta_{13}\neq 0$
the 12- and 13-entry in (\ref{equ:msquark}) are no longer zero, but still much smaller than the 23-entry. This gives small
corrections to $K-\overline{K}$- and $B_d-\overline{B}_d$-mixing.


%

Flavour processes where we expect large CMM contributions are $B_s-\overline{B}_s$ mixing, $b\to s\gamma$ and $\tau\to\mu\gamma$
since here the  angle $\theta_{23}\approx 45^\circ$ enters.
CMM effects in $\mathcal{B}(B_s\to\mu^+\mu^-)$ are  small and compatible with the LHCb bound because at the electroweak scale the
CMM model is a special
version of the MSSM  with small $\tan\beta$.
Due to the structure of (\ref{equ:msquark}) the contributions to $K-\overline{K}$ mixing, $B_d-\overline{B}_d$ mixing and $\mu\to
e\gamma$ are absent. However there are two sources of small corrections: a non-vanishing $\theta_{13}$  and corrections due to
dimension-5-Yukawa terms that are needed to fix  $\mathsf Y_d = \mathsf Y_\ell^\top$  for
the 1$^\text{st}/2^\text{nd}$  generation. The latter point was worked out in \cite{Trine:2009ns} where it was also
shown that the  $|\varepsilon_K|-S_{\psi K_S}$ tension can be removed with the help of
higher-dimensional Yukawa couplings.

Results from our global analysis are the following: $\tau\to \mu\gamma$ constrains the sfermion masses of the first two
generations to lie above 1~TeV while the third generation can be much lighter ($\tau\to\mu\gamma$ gives stronger bounds than
$b\to s\gamma$).   Concerning
$B_s-\overline{B}_s$ mixing the situation changed after the LHCb data for $S_{\psi\phi}$. Due to the free phase $\xi$ it is
possible to get large CP violation in the $B_s$ system in the CMM model while at the same time $\Delta M_s$ stays within its
experimental range.   In view of the data from CDF and D\O\ on $S_{\psi\phi}$ this property was very welcomed in 2010.
The new data on $S_{\psi\phi}$ implies new constraints on the model parameters, especially  on $\xi$ and on the ratio of gluino
and squark masses $m_{\tilde g}/M_{\tilde q}$ which must now be smaller than before. This was exemplarily shown in
\cite{Buras:2012ts}.   

\begin{figure}[t!]
\centering
 \includegraphics[width = 0.7\textwidth]{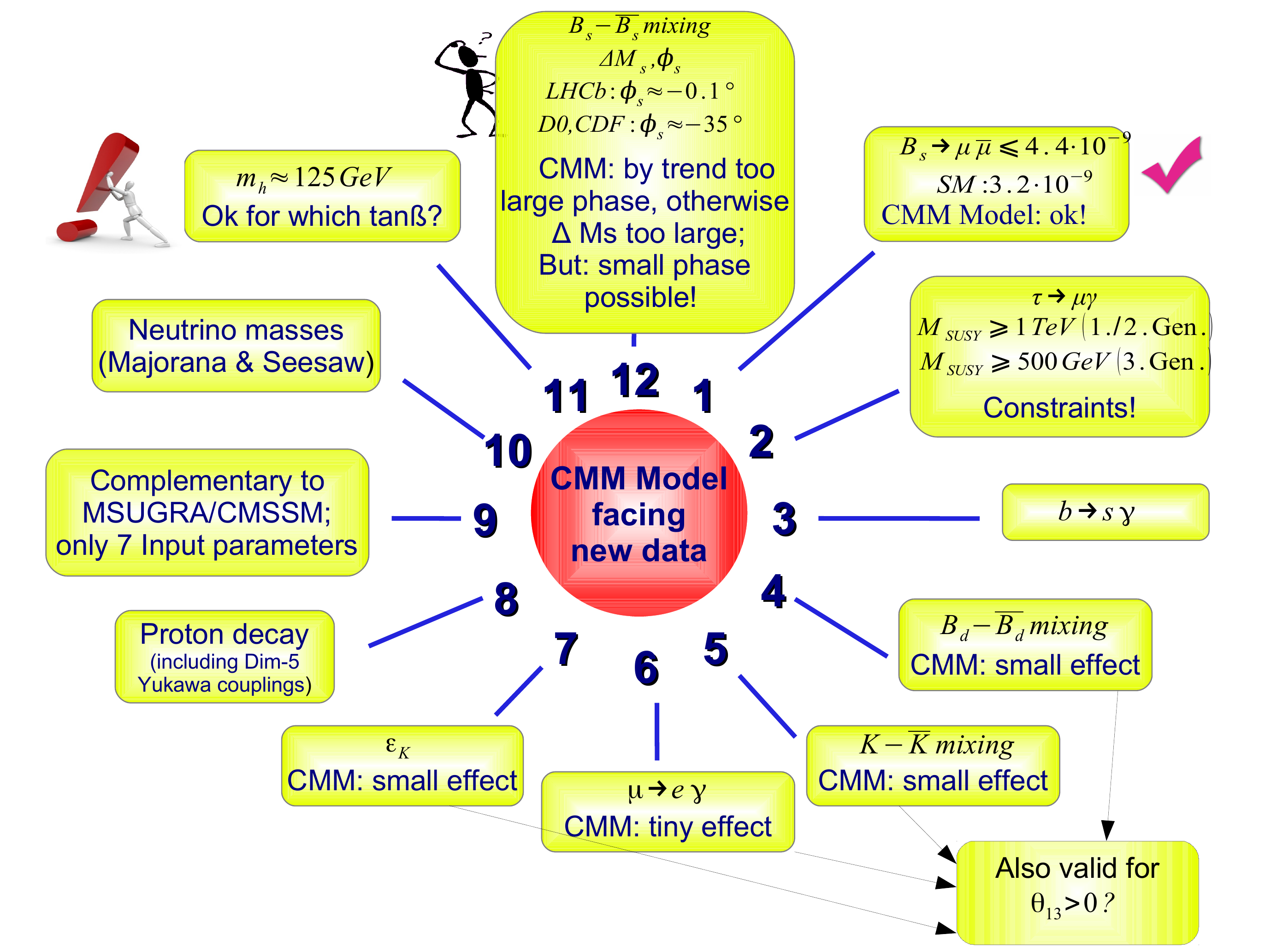}
\caption{Basic properties of the CMM model.}\label{fig:CMMUhr}
\end{figure}

Another observable that needs further investigation is the Higgs mass. In the CMM model the mass of the lightest neutral Higgs  is
very sensitive to $\tan\beta$\footnote{Decreasing $\tan\beta$ also decreases the Higgs mass because  a larger $y_t$
increases the mass splitting $\Delta_{\tilde d}$ in the RG running  which leads to smaller stop
masses.}. In \cite{Girrbach:2011an} we pointed
out
that $\tan\beta = 3$ is excluded due to the LEP bound. For  $\tan\beta = 6$ the Higgs mass can be up to 120~GeV in the parameter
range consistent with
flavour observables. Consequently one has to increase $\tan\beta$ further to accommodate a Higgs mass of 125~GeV.

\section{Summary}

In the first part we studied and compared correlations of $\Delta F = 2$ observables in CMFV and in a minimal version of
models with an approximate global  $U(2)^3$ flavour symmetry. These $MU(2)^3$ models are very simple non-MFV extensions
of the SM that avoid the  $\Delta M_{s,d}-\varepsilon_K$ tension present in CMFV.  We pointed out a triple correlation between
$S_{\psi\phi}$, $S_{\psi K_S}$ and $|V_{ub}|$ that constitutes an important test for $MU(2)^3$ models. In the last part an
$SO(10)$ SUSY GUT model, the CMM model was under consideration where a summary can be found in   Fig.~\ref{fig:CMMUhr}.

\section*{Acknowledgments}

I thank the organizers  for the opportunity to give this talk and my collaborators  A.~Buras,  S.~J\"ager, M.~Knopf,
W.~Martens, U.~Nierste, C.~Scherrer and
S.~Wiesenfeldt for an enjoyable collaboration.  I acknowledge financial support by  ERC Advanced Grant project
``FLAVOUR''(267104).


\bibliography{girrbach}
\bibliographystyle{apsrev4-1}


%% file: Papers/hartmann.tex

%
%
%
%
%
%

\chapter[$\textsf{SU}(3)$-Flavons and Pati-Salam-GUTs (\textit{Hartmann}, Kilian, Schnitter)]{$\textsf{SU}(3)$-Flavons and Pati-Salam-GUTs}
\vspace{-2em}
\paragraph{\textit{F. Hartmann}, W. Kilian, K. Schnitter}
\paragraph{Abstract}
We consider a multi step breaking of supersymmetric Pati-Salam GUT models. We investigate how this can be achieved with different GUT-Higgs contents and derive ranges for the associated energy scales.
Starting from these models we enlarge the model by an SU(3) flavour symmetry and demand that the SM Higgs boson should be a triplet under that symmetry. This enables us to provide the possibility of ``SM-matter-Higgs-unification''. Furthermore we show how SM-like Yukawa couplings can be derived from the vacuum expectation values of flavon fields that break the flavour symmetry.

\section{Introduction}

GUTs are often considered to unify all couplings at one scale. This leads to strongly constrained models. A possible GUT gauge group is the Pati-Salam (PS) symmetry, which unifies only to a semi simple group but can be further embedded into $SO(10)$. We consider such a PS model and allow for a intermediate left-right (LR) scale which corresponds to a right-handed neutrino mass scale. \\
In order to reach ``SM-matter-higgs-unification'' in flavour models one has to consider the implications of flavour triplet SM Higgs doublets. These lead to new invariant structures which can be accommodated by certain flavon representations.

\section{GUT}

We consider a supersymmetric Pati-Salam GUT model which is broken down to the SM in multiple steps \cite{Hartmann:2012}. All fields breaking the symmetry are allowed to have a different mass. The lowest scale is accociated with the usual ew breaking.
\vspace{-1ex}
 \begin{equation}
\label{FH_breaking1}
     SU(3)_C \otimes SU(2)_L \otimes U(1)_Y \quad\xrightarrow{\langle h \rangle}\quad SU(3)_C \otimes U(1)_\text{em}   
 \end{equation}
The SM should be valid up to the left-right unification scale which may be at around $10^{13}$~GeV. 
\vspace{-1ex}
 \begin{equation}
\label{FH_breaking2}
   SU(3)_C \otimes U(1)_{B-L} \otimes SU(2)_L \otimes SU(2)_R \otimes Z_2 \quad\xrightarrow{\langle H^R_{u/d} \rangle}\quad SU(3)_C \otimes SU(2)_L \otimes U(1)_Y 
  \end{equation}
PS unification should happen roughly at the usual unification scale of $10^{16}$~GeV which then may be valid up to the Planck scale where a complete gauge unification to e.g. $SO(10)$ or $E_6$ can be realized. Because physics near the Planck scale is not understood jet, we do not consider the breaking of the Planck scale symmetry.
\vspace{-1ex}
 \begin{equation}
\label{FH_breaking3}
    SU(4) \otimes SU(2)_L \otimes SU(2)_R \otimes Z_2 \quad\xrightarrow{\langle\Sigma \rangle}\quad SU(3)_C\otimes U(1)_{B-L} \otimes SU(2)_L \otimes SU(2)_R \otimes Z_2 
 \end{equation}

We start with an higgs field content consistent with complete $SO(10)$ representations, because we want to be consistent with complete unification. The full field content is given in table~\ref{FH_fullfieldcontent}. For these fields we have constructed the full renormalisable higgs sector superpotential and calculated the tree-level Higgs masses. For a large splitting of the LR and PS scales ($\langle\Sigma\rangle \gg \langle H^R\rangle$) a new intermediate mass scale $ M_F \sim m_F + \tfrac{\langle H^R\rangle^2}{\langle\Sigma\rangle}$ occurs. At this scale the colour triplets from the field $F$ are located. These light triplets can be useful for a unification with intermediate PS symmetry as was shown in \cite{Kilian:2006hh}. In addition we found that there is the possibility of massless (``at the order of the MSSM scale'') Higgs doublets. These stem from the breaking of the fields $H^L$ and are associated to the LR Goldstone bosons by the $Z_2$ symmetry. Thus they are similar to the MSSM higgs and make the field $h$ optional.
\begin{table}
\centering
  \begin{tabular}{lll}
  \textbf{PS breaking Higgs}  &\phantom{bobobobobo}&\\[1ex]
     $\Sigma = (\mathbf{15},\mathbf1,\mathbf1)$ to break $SU(4)$ & \picture(0,0)\put(0,-8){$\left.\rule{0pt}{3ex}\right\}$ $\ \mathbf{45}$}\endpicture &  \picture(0,0)\put(0,-36){$\left.\rule{0pt}{9ex}\right\}$ $\ \mathbf{78}$}\endpicture \\
     $E=(\mathbf6,\mathbf2,\mathbf2)$ and $T_{L}= (\mathbf1,\mathbf3,\mathbf1)$,$T_{R}= (\mathbf1,\mathbf1,\mathbf3)$ &&\\[1.5ex]
  \textbf{L-R breaking Higgs }&&\\[1ex]
    $H^R_u = (\mathbf4,\mathbf1,\mathbf2)$ and $H^R_d=(\mathbf{\overline{4}},\mathbf1,\mathbf2)$ to break $U(1)_{B-L}\times SU(2)_R$ & \picture(0,0)\put(0,-8){$\left.\rule{0pt}{3ex}\right\}$ $\ \mathbf{16} \oplus \mathbf{\overline{16}}$ }\endpicture &\\
    $H^L_u = (\mathbf4,\mathbf2,\mathbf1)$ and $H^L_d=(\mathbf{\overline{4}},\mathbf2,\mathbf1)$ because of $Z_2$  && \\[1.5ex]
  \textbf{optional MSSM Higgs} &&\\[1ex]
    $h=(\mathbf1,\mathbf2,\mathbf2)$: MSSM-higgs  &\picture(0,0)\put(0,-8){$\left.\rule{0pt}{3ex}\right\}$ $\ \mathbf{10}$}\endpicture &\picture(0,0)\put(0,-31){$\left.\rule{0pt}{7ex}\right\}$ $\ \mathbf{27}$}\endpicture\\
    $F=(\mathbf6,\mathbf1,\mathbf1)$: possibly light triplets && \\ [1.5ex]
  \textbf{matter} &&\\[1ex]
    $\Psi=(\mathbf4,\mathbf1,\mathbf2)$ and $\Psi^c=(\mathbf{\overline4},\mathbf2,\mathbf1)$   & $\mathbf{16}$ &
  \end{tabular}
 \caption{Full field content and unification to $SO(10)$ and $E_6$}
 \label{FH_fullfieldcontent}
\end{table}
The unification conditions are the following:
\begin{align}
\label{FH_LRcondition}
\alpha^{-1}_Y\left(M_\text{LR}\right) \;&=\; \alpha^{-1}_{R}\left(M_\text{LR}\right) \,+\, \tfrac{2}{3}\ \alpha^{-1}_{B-L}\left(M_\text{LR}\right) \\
\label{FH_GUTcondition}
\alpha^{-1}_4\left(M_\text{GUT}\right) \;&=\; \alpha^{-1}_{L} \left(M_\text{GUT}\right) \;=\; \alpha_R^{-1}\left(M_\text{GUT}\right) \\
\label{FH_PScondition}
 \alpha^{-1}_4 \left(M_\text{PS}\right)  \;&=\;   \alpha^{-1}_3\left(M_\text{PS}\right) \;=\; \alpha^{-1}_{B-L} \left(M_\text{PS}\right)
\end{align}
These three unification conditions build a system of equations which constrain the possible mass scales. They are not fixed completely, but depend on each other. 
Moreover we consider the intermediate scale $M_F$ to be a free parameter. This scale is in principle fixed by the potential, but it can be tuned over quite a large mass range by choosing the coefficients of the superpotential. \\
In addition to the unification conditions shown above we have the additional constraint that the mass scales have to be ordered $M_\text{SUSY}\,<\, M_\text{F} \,<\, M_\text{LR} \,<\, M_\text{PS} \,<\, M_\text{GUT} \,\leq\, M_\text{Planck}$.
With these inputs we calculate systematically different types of models from which the most important ones should be mentioned now.

\begin{figure}[ht]
\centering
\subfigure[Minimal Model]{
\includegraphics[width=.45\textwidth]{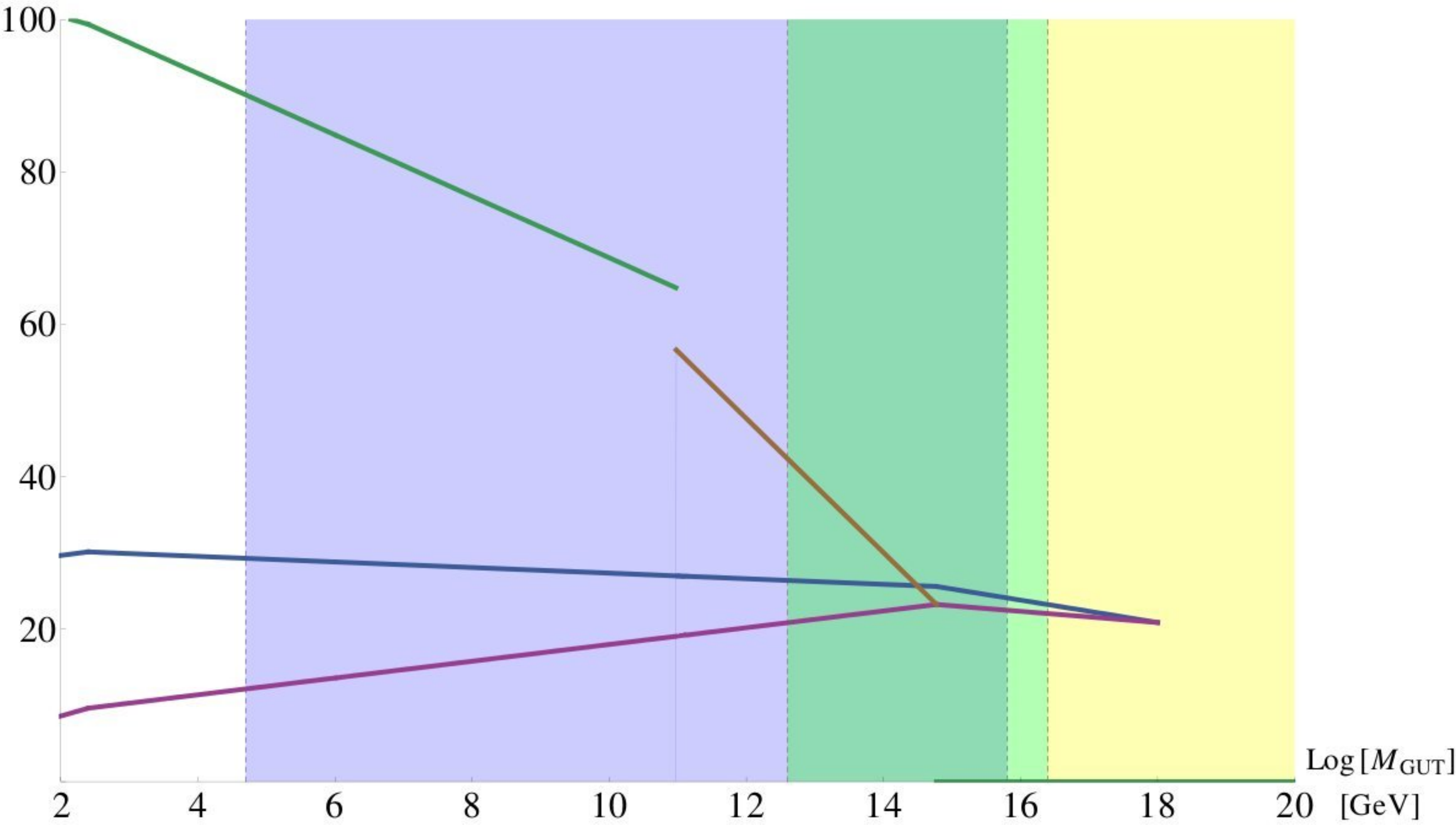}
\label{FH_fig:runningminmodel}}
\subfigure[Full Model]{
\includegraphics[width=.45\textwidth]{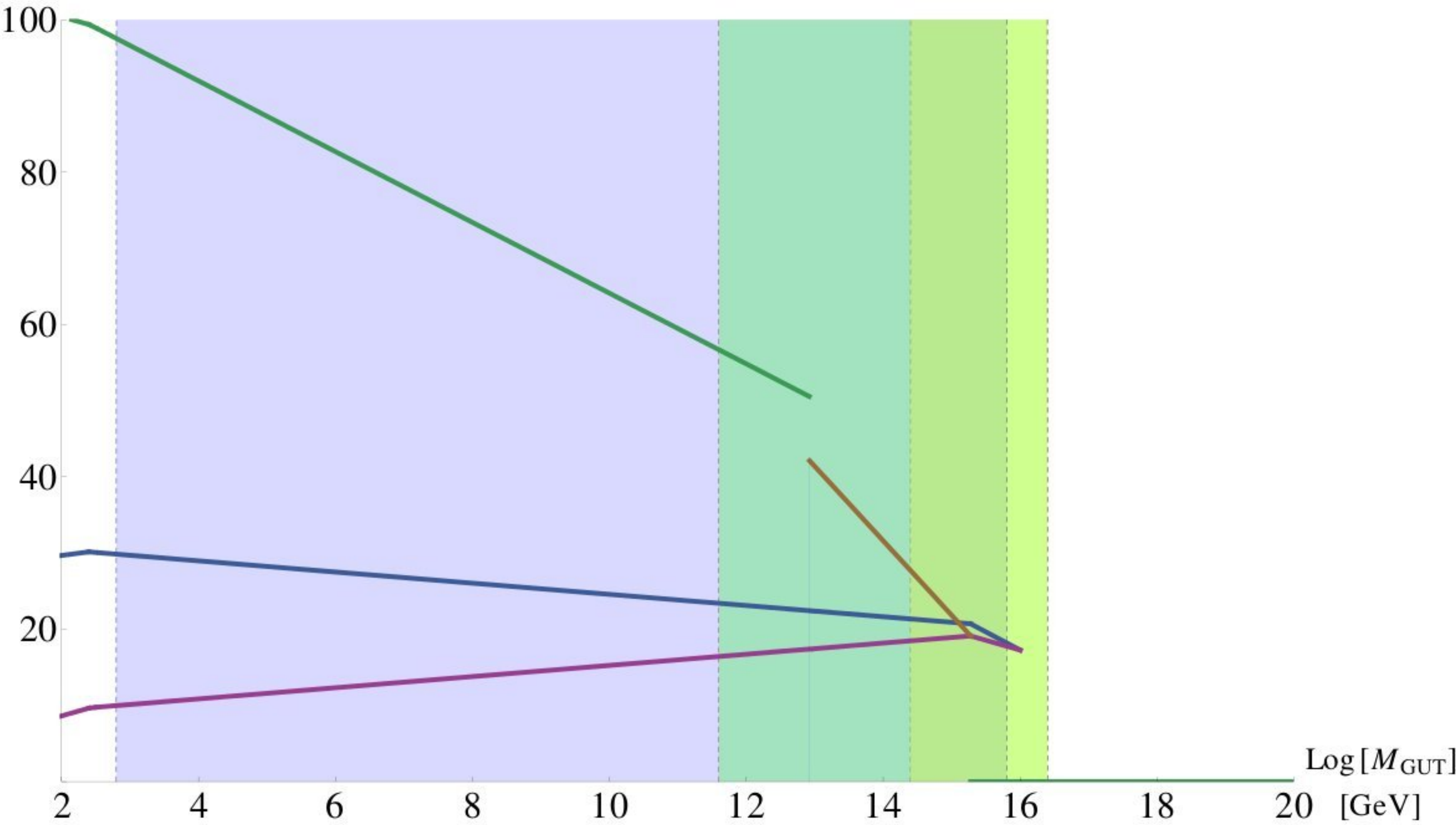}
\label{FH_fig:runningfullmodel}}
\subfigure[{$E_6$ inspired Model}]{
\includegraphics[width=.45\textwidth]{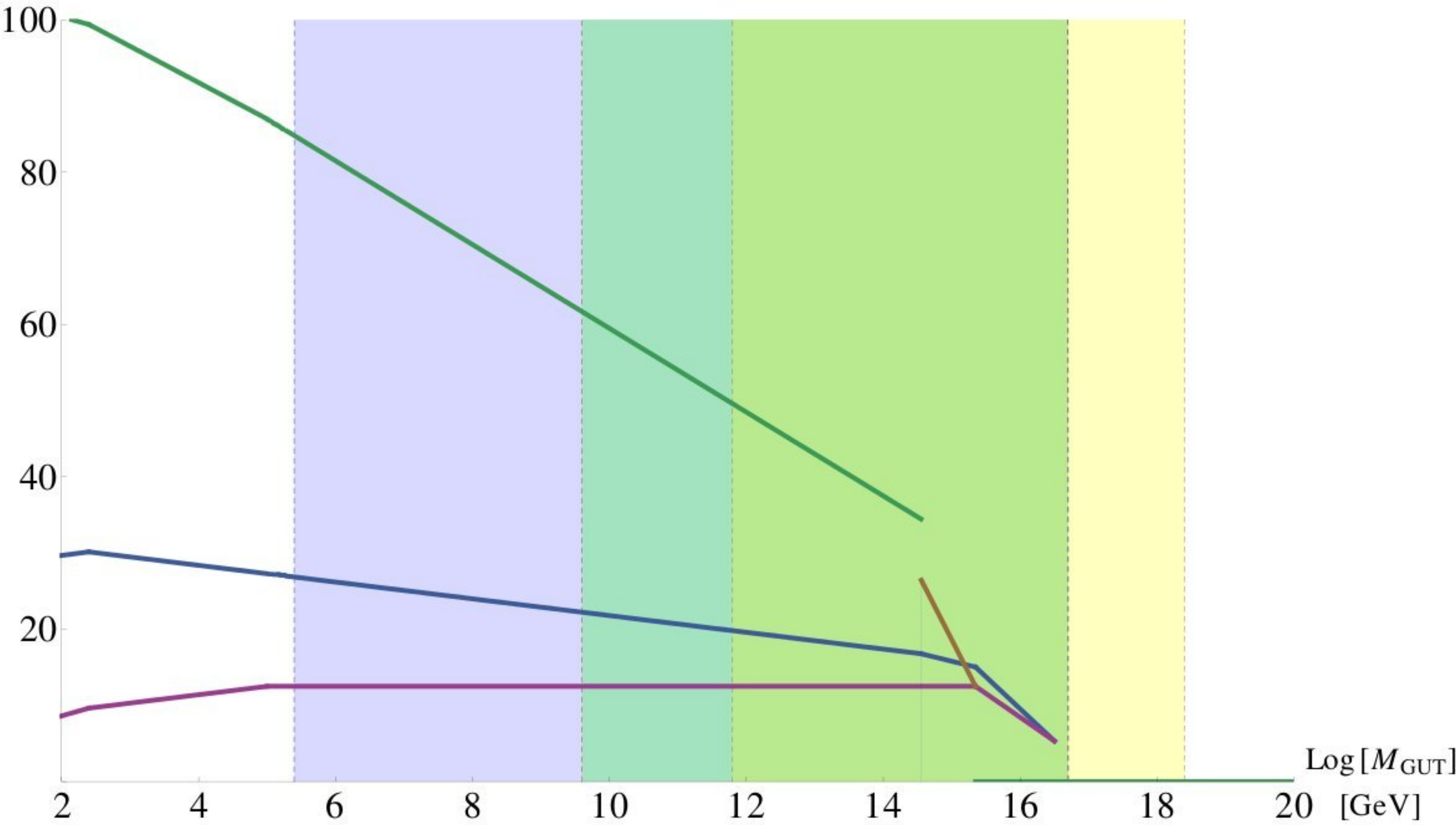}
\label{FH_fig:runninge6model}}
\caption{Running couplings in the different models including the possible variation of the scales (blue: $M_\text{LR}$, green: $M_\text{PS}$, yellow: $M_\text{GUT}$)}
\end{figure}

\paragraph{Minimal Model} $\;$ \\
In the minimal model only the fields $H_{u/d}^{L/R}$ and $\Sigma$ are included below the Planck scale. This allows for unification of the couplings as shown in fig.~\ref{FH_fig:runningminmodel}. The apparent jump in the $U(1)$ coupling is required by eqn~\ref{FH_LRcondition}.
 The possible variation of the scales is indicated by the colored areas. This leads to a (not independent) possible variation of the scales of
\begin{equation*}
  4.7 \leq \log\left(M_{LR}\right) \leq 15.8  \  ;\quad
  12.7 \leq \log\left(M_{PS}\right) \leq 16.4 \  ;\quad
  16.4 \leq \log\left(M_{GUT}\right) \leq 20.1
\end{equation*}
We find that the special case of a complete unification at a single mass scale ($M_\text{LR}=M_\text{PS} =M_\text{GUT}$) is not possible. Nevertheless, a unification of LR directly to a complete GUT ($M_\text{PS}=M_\text{GUT}$) is possible.
  
\paragraph{Full Model} $\;$ \\
In the full model we consider the full higgs field content shown in table~\ref{FH_fullfieldcontent}. These form the complete $SO(10)$ representation $\mathbf{10} \oplus \mathbf{16}\oplus\mathbf{\overline{16}}\oplus\mathbf{45}$.
The possible unification plot is shown in fig.~\ref{FH_fig:runningfullmodel}. The unification is possible for the following mass ranges
\begin{equation*}
  2.8 \leq \log\left(M_{LR}\right) \leq 15.8  \  ;\quad
  11.5 \leq \log\left(M_{PS}\right) \leq 16.4 \  ;\quad
  14.4 \leq \log\left(M_{GUT}\right) \leq 16.4
\end{equation*}
In such a model the unification to $SO(10)$ is at a rather low scale far below the Planck scale. Again it is possible for $M_\text{PS}$ and $M_\text{GUT}$ but not for $M_\text{LR}$ to coincide. 

\paragraph{$\mathbf{E_6}$ inspired Flavour Triplet Model} $\;$ \\
Here we consider a full model with three generations of MSSM higgs $h$. This enables a ``SM matter-higgs unification'' \`a la $E_6$ and can lead to interesting flavour models (see section~\ref{FH_flavourmodel}). The field content unifies to the complete $E_6$ representations $3\times\mathbf{27} \oplus \mathbf{78}$.
The possible unification plot is shown in fig.~\ref{FH_fig:runninge6model}. The unification is possible in the following mass ranges
\begin{equation*}
  5.4 \leq \log\left(M_{LR}\right) \leq 16.7  \  ;\quad
  9.6 \leq \log\left(M_{PS}\right) \leq 16.7 \  ;\quad
  11.8 \leq \log\left(M_{GUT}\right) \leq 18.4
\end{equation*}
In this model it is possible that all three mass scales coincide at $M\approx10^{16.7}$~GeV, leading to direct GUT unification. Below this scale it is again possible that the PS and GUT scale are equal.

\section{Flavour}
\label{FH_flavourmodel}

We now consider Flavour models which are motivated by the Pati-Salam models discussed above. These are only weakly dependent on the choice of an explicit model. The Flavour model reproduces the flavour structure of the SM in an indirect approach. We consider an $SU(3)_F$ flavour symmetry which is embedded in a supersymmetric Pati-Salam GUT framework. It is broken by flavons which either transform as triplets or decuplets under the flavour symmetry \cite{Hartmann:2012}.

It was shown that a Yukawa matrix with the structure shown in eqn~\ref{FH_Yukawastructure} can reproduce the quark-data for the CKM matrix quite well \cite{Roberts:2001zy}. In addition it has been shown that with such a Yukawa structure and sequential right-handed neutrino dominance (SRHND) also the neutrino data and thus the PMNS matrix can be reproduced \cite{King:2003jb}. Also values of $\theta_{13} \neq 0$ are possible. 

\begin{equation}
Y_{u/d} \approx  \left( \begin{matrix} 0 				& O\left(\epsilon^3\right) & O\left(\epsilon^3\right)	\\
					    O\left(\epsilon^3\right)	& O\left(\epsilon^2\right) & O\left(\epsilon^2\right)	\\
					    O\left(\epsilon^3\right)	& O\left(\epsilon^2\right) & O\left(1\right)		 \end{matrix}\right) m_t 
\quad \text{with } \epsilon_u \approx 0.05 \text{ and }\epsilon_d \approx 0.15
\label{FH_Yukawastructure}
\end{equation}

In our model we consider the left and right-handed matter representations $\Psi$ and $\Psi^c$ as well as the MSSM Higgs $h$ to transform as triplets under the flavour symmetry. This corresponds to a realization of the ``$E_6$ inspired Flavour Triplet model''. For the other types of GUT models mentioned above one Higgs doublet would be light and the two other very heavy ($\sim M_\text{Planck}$).
In such models a new trivial invariant $ \varepsilon^{ijk}\ \Psi^L_i\, \Psi^R_j\, h_k $ occurs. This is an antisymmetric combination of the fields and thus leads to off diagonal entries. These are of order one, because no flavon insertion is necessary. Therefore additional discrete symmetries are needed in order to forbid such and other unwanted contributions.

\subsection{Triplet Flavons}
In the case of triplet flavons we consider a model similar to \cite{King:2003rf}, but with three generations of higgs. Therefore we have to find a new set of quantum numbers of the additional symmetry $U(1)\times Z_2$ reproducing the structure mentioned above. The flavon vevs are aligned in the 3- and 2,3- direction ($\langle\phi_3\rangle = \langle\bar\phi_3\rangle = (0,0,1)\, M $ and $ \langle\phi_{23}\rangle = \langle\bar\phi_{23}\rangle = (0,\epsilon,{\scriptstyle \pm}\epsilon)\,M $). 

Such a model is possible but one encounters large corrections to the Yukawa structure from operators of next to leading order in flavon insertion. That is because an additional symmetry can only forbid field configurations and not single invariants. Thus if one forbids all unwanted terms one also forbids  most terms creating the wanted structure up to high orders. A different approach is to introduce some fine-tuning or different field configurations.

\subsection{Decuplet Flavons}
Flavons transforming as decuplet under the flavour symmetry are of special interest because the decuplet is the symmetric combination of three triplets
. Thus we can form at leading order in flavon insertion the type of invariant we are interested in. Therefore the leading order superpotential is particularly simple. Also the number of sub-leading terms is limited, because the structure of the invariants is more complex. The field content shown in table~\ref{FH_decupletfieldcontent} lead to the leading order superpotential of eqn~\eqref{FH_decupletleadordersuperpotential}. Together with the vev alignment shown in table~\ref{FH_decupletvevalignment} this potential generates the desired Yukawa structure of eqn~\eqref{FH_Yukawastructure}.

\begin{table}[ht]
 \begin{minipage}{.47\textwidth}
  \centering
  \begin{tabular}{c|c|c|r}
      Field				& $SU(3)_F$		& PS			& $U(1)$	\\
      \hline \hspace{5cm}&&&\\[-2.5ex]\hline &&&\\[-2ex]
      $\Psi$				&	$\mathbf{3}$	& $(4,2,1)$		& $1$		\\[0.1ex]
      $\Psi^c$			&	$\mathbf{3}$	& $(\bar{4},1,2)$	& $1$		\\[.1ex]
      $h$				&	$\mathbf{3}$	& $(1,2,2)$		& $1$		\\[1ex]
      $\bar{\phi}_1$			&	$\mathbf{\overline{10}}$	& $(1,1,1)$		& $-3$		\\[.1ex]	
      $\bar{\phi}_2$			&	$\mathbf{\overline{10}}$	& $(1,1,1)$		& $-3$		\\[.1ex]
      $\bar{\phi}_3$			&	$\mathbf{\overline{10}}$	& $(1,1,1)$		& $-3$		\\[.5ex]
      $H_d$				&	$\mathbf{\overline{3}}$	& $(4,1,2)$ 		& $-2$		
  \end{tabular}
  \caption{Field content of the decuplet model including all symmetries}
  \label{FH_decupletfieldcontent}
 \end{minipage}\hfill
  \begin{minipage}{.47\textwidth}
    \vspace{4.5ex}
    \centering
    \begin{eqnarray*}
	    \langle\bar{\phi}_1\rangle_{333} &\approx& M \\
	    \langle\bar{\phi}_2\rangle_{223} &\approx& \epsilon^2M \\
	    \langle\bar{\phi}_2\rangle_{233} &\approx& \epsilon^2M \\
	    \langle\bar{\phi}_3\rangle_{123} &\approx& \epsilon^3M \\
	    \langle\bar{\phi}_3\rangle_{133} &\approx& \epsilon^3 M 
    \end{eqnarray*}
  \caption{Vev alignment}
  \label{FH_decupletvevalignment}
  \end{minipage}
\end{table}
\vspace{-3ex}
\begin{equation}
	  W_\text{lead} = \sum_{i=1}^3 \frac{y_i}{M} \Psi \Psi^c h \bar{\phi}_i
  \label{FH_decupletleadordersuperpotential}
  \end{equation}
Together with additional anti-decuplets (which have to be introduced in order to guarantee D-term flatness) and some driving fields (additional gauge and flavour singlets), a renormalisable potential can be constructed leading to such a vev alignment. The potential for the right handed Majorana mass matrix is at leading order shown in eqn~\eqref{FH_decupletmajoranapotential}. This leads to an Majorana matrix with hierarchical eigenvalues of order $(1, \epsilon^2, \epsilon^4)$ and thus to sequential dominance.
\begin{equation}
  W_\text{Maj}= \frac{1}{M^3}{\Psi^c}^2 H^2_d \left(\tilde{C}_1\, \overline{\phi}_1\phi_1 + \tilde{C}_2\,\overline{\phi}_1\phi_2 + \tilde{C}_3\,\overline{\phi}_1\phi_3 + \tilde{C}_4\,\overline{\phi}_2\overline{\phi}_2 \right)
\label{FH_decupletmajoranapotential}
\end{equation}

\section{Conclusion}

We have shown that a multi-step breaking of a supersymmetric Pati-Salam model is possible.
 The scales at which the different sub-unifications are present depend on the higgs content and can vary over quite a large mass range. By doing the multi-step breaking we can naturally introduce an intermediate left-right scale. This scale can be located at the mass scale of right handed neutrinos. In addition, these models allow for a low scale of light colour triplets, due to a seesaw like mechanism. \\
In the second part we have shown that models with flavour triplet Higgs are possible in the framework of $SU(3)_F\otimes \text{PS}$ models. Here we provide two possible Ans\"atze how such a model can be realized, namely the breaking with triplet and decuplet flavons.

\bibliography{hartmann}
\bibliographystyle{apsrev4-1}


%% file: Papers/julianheeck.tex

%
%
%
%
%
%

\chapter[Local Flavor Symmetries (Heeck)]{Local Flavor Symmetries}
\vspace{-2em}
\paragraph{J. Heeck}
\paragraph{Abstract}

Augmenting the Standard Model by three right-handed neutrinos allows for an anomaly-free gauge group extension $G_\mathrm{max} = U(1)_{B-L}\times U(1)_{L_e-L_\mu} \times U(1)_{L_\mu-L_\tau}$. Simple $U(1)$ subgroups of $G_\mathrm{max}$ can be used to impose structure on the right-handed neutrino mass matrix, which then propagates to the active neutrino mass matrix via the seesaw mechanism. We show how this framework can be used to gauge the approximate lepton-number symmetries behind the normal, inverted, and quasidegenerate neutrino mass spectrum, and also how to generate texture-zeros and vanishing minors in the neutrino mass matrix, leading to testable relations among mixing parameters.

\section{Introduction}

A very nice explanation of the small neutrino masses (compared to the electroweak scale) comes from the (type-I) seesaw mechanism. For this, three right-handed neutrino partners are introduced, which can acquire a very large Majorana mass $\mathcal{M}_R$ because they are Standard Model singlets. Diagonalization of the neutral fermion mass matrix then leads to three light---mainly active---neutrinos, with masses suppressed by the heavy mass scale $\mathcal{M}_R$. Since this mechanism can not shed light on the peculiar lepton mixing angles, uncountable models have been brought forward imposing discrete non-abelian global symmetries, such as $A_4$, $S_4$, $\Delta (27)$, etc.~(see various articles in these proceedings). These models typically involve an untestable scalar sector at high energies, and in some cases suffer from other problems like vacuum alignment, so we propose a much simpler set of symmetries, based on continuous abelian local symmetries, i.e.~additional $U(1)'$. While such symmetries can not yield tri-bimaximal mixing, the very simple and economical scalar sector and the easily testable $Z'$ gauge boson make up for the lack of \emph{definite} mixing angle predictions.\footnote{$U(1)'$ flavor symmetries typically give vanishing, maximal, or undefined mixing angles.}

In order to only influence neutrino mixing, we assign the same universal $U(1)'$ charge to all quarks: $Y'(q_{L i}) = Y'(u_{R i}) = Y'(d_{R i})$ $\forall i = 1,2,3$. To allow at least diagonal Dirac mass matrices, we also set $Y'(\ell_{L i}) = Y'(e_{R i}) = Y'(N_{R i})$, but with different charges for the lepton generations in general, i.e.~$Y'(\ell_{L i}) \neq Y'(\ell_{L j})$.

Anomaly cancellation of the full gauge group $SU(3)_C \times SU(2)_L \times U(1)_Y \times U(1)'$ gives the sole constraint
\begin{align}
 9\,  Y' ( q_L )  +Y' (\ell_{L1}) +Y' (\ell_{L2})+Y' (\ell_{L3})= 0\,,
 \label{jh_anomalyconstraint}
\end{align}
which leads to $U(1)'$ groups generated by
\begin{align}
 &B - \sum_\alpha x_\alpha L_\alpha \text{ with } \sum_\alpha x_\alpha= 3\,, && 
\text{ or } &&
& \sum_\alpha y_\alpha L_\alpha \text{ with } \sum_\alpha y_\alpha= 0\,.
\end{align}
Special cases include the well-known $B-L$ symmetry---which contains no information about mixing---and the lepton number differences $L_\alpha - L_\beta$, which are anomaly-free in the SM alone and have been discussed extensively~\cite{Foot:1990mn,He:1991qd,Heeck:2011md}. Some of the $B - \sum_\alpha x_\alpha L_\alpha$ symmetries have been discussed already in the literature~\cite{Ma:1997nq,Lee:2010hf} (incomplete list).

Note that all solutions to Eq.~\eqref{jh_anomalyconstraint} can be viewed as subgroups of the abelian group $U(1)_{B-L} \times U(1)_{L_e - L_\mu} \times U(1)_{L_\mu - L_\tau}$, which itself is a subgroup of the non-abelian $U(1)_{B-L} \times SU(3)_\ell$~\cite{Araki:2012ip}. Consequently, all $U(1)'$ groups discussed here can be embedded into these larger groups, but we omit a discussion due to the necessary enlarged scalar sector.

\section{Neutrino Hierarchies}

\begin{figure}[t]
	\begin{center}
		\includegraphics[width=0.42\textwidth]{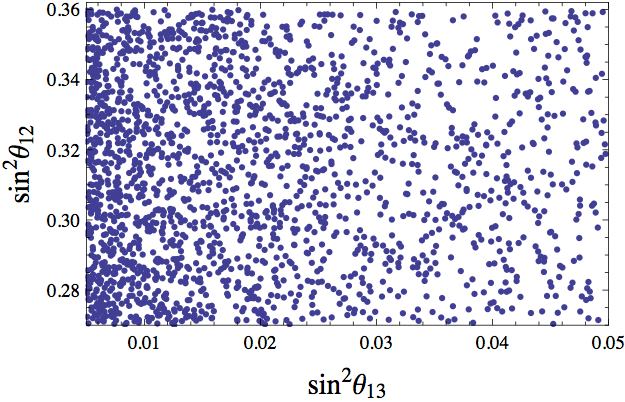}\hspace{1ex}
		\includegraphics[width=0.42\textwidth]{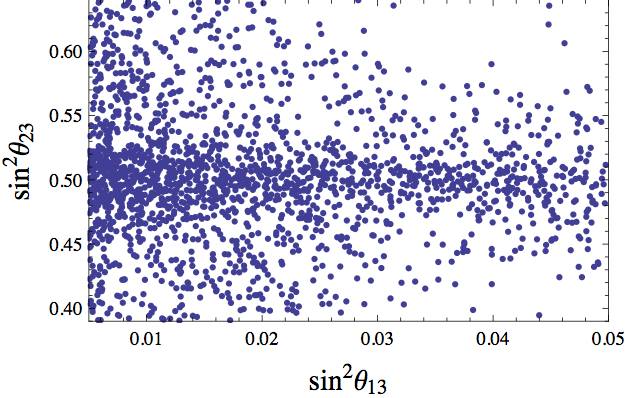}
	\end{center}
		\caption{Scatter plots for $L_\mu - L_\tau$, spontaneously broken by two scalars with vacuum expectation values $v_S/\mathcal{M}_R \sim 0.02$.}
	\label{jh_scatter}
\end{figure}

The $U(1)'$ groups generated by $B - \sum_\alpha x_\alpha L_\alpha$ or $\sum_\alpha y_\alpha L_\alpha$ lead to different neutrino phenomenology, depending on the coefficients $x_\alpha$ or $y_\alpha$. We will first show how the championed symmetries behind normal, inverted and quasidegenerate neutrino mass hierarchy can be promoted to local symmetries. As shown in Ref.~\cite{Choubey:2004hn} (see also Ref.~\cite{Branco:1988ex}), good zeroth-order Majorana mass matrices that conserve a lepton number are given by
\begin{align}
	\mathcal{M}_\nu^{L_e} \sim \begin{pmatrix} 0 & 0 & 0\\ 0 & \times & \times \\ 0 & \times & \times\end{pmatrix} , &&
	\mathcal{M}_\nu^{\bar{L}} \sim \begin{pmatrix} 0 & \times & \times \\ \times & 0 & 0\\ \times & 0 & 0\end{pmatrix} , &&	
	\mathcal{M}_\nu^{L_\mu-L_\tau} \sim \begin{pmatrix} \times & 0 & 0 \\ 0 & 0 &
\times \\ 0 & \times & 0\end{pmatrix} ,
\end{align}
where $\times$ denotes a non-zero entry and $\bar{L}\equiv L_e - L_\mu - L_\tau$. The $L_\mu-L_\tau$ symmetry---which is a good symmetry for quasidegenerate neutrinos---can be readily gauged and leads to numerous interesting effects, e.g.~the solution of the anomalous magnetic moment of the muon~\cite{Heeck:2011md}. As for the neutrino mixing angles, $L_\mu-L_\tau$ leads to maximal $\theta_{23}$ and vanishing $\theta_{13}$ and $\theta_{12}$, but small breaking terms suffice to generate valid angles~\cite{Choubey:2004hn}. To generate the observed $\theta_{13}$, the $L_\mu-L_\tau$ breaking scale should be $\sim 100$ times below the seesaw-scale $\mathcal{M}_R$ (see Fig.~\ref{jh_scatter}). A more detailed discussion can be found in the given references.

To promote the lepton number $L_e$---which leads to normal hierarchy solutions---to a local symmetry, we note that a broken $\bar{L}$ symmetry actually leads to an approximate $L_e$ symmetric $\mathcal{M}_\nu$ via seesaw~\cite{Heeck:2012cd}, so gauging the anomaly-free $B+3 \bar{L}$---and then breaking it spontaneously in the right-handed sector---will lead to an approximate $\mathcal{M}_\nu^{L_e}$. The $B+3 \bar{L}$ breaking scale can be $\sim 20$ times below the seesaw-scale $\mathcal{M}_R$ to generate the right mixing angles.

Having found local symmetries that lead to a quasi-degenerate spectrum ($L_\mu-L_\tau$) or normal hierarchy ($B+3 \bar{L}$), we turn to inverted hierarchy. As shown in Ref.~\cite{Heeck:2012cd}, it suffices to decouple one of the three right-handed neutrinos with an additional $\mathbb{Z}_2$ symmetry to flip the spectrum of $B+3 \bar{L}$ from normal to inverted. In this form, the reactor angle $\theta_{13}$ vanishes, even after spontaneously breaking $B+3 \bar{L}$, so the model is excluded by now. This can be easily amended, however, by introducing five instead of three right-handed neutrinos. The anomaly constraints can still be fulfilled, and after decoupling one of the five singlets by a $\mathbb{Z}_2$, we find an inverted hierarchy spectrum in the active neutrinos with $\theta_{13}\neq 0$.
The $\mathbb{Z}_2$ symmetry that we introduced to obtain inverted hierarchy of course makes the decoupled right-handed neutrino a (Majorana) dark matter candidate, which interacts with the Standard Model via the $Z'$ gauge boson and the Higgs portal. The arising phenomenology of this dark matter particle is similar to the recently studied $U(1)_{B-L}\times \mathbb{Z}_2$ model~\cite{Okada:2010wd} and can be found in Ref.~\cite{Heeck:2012cd}.

\section{Texture Zeros and Vanishing Minors}

As an extreme case of imposing structure on $\mathcal{M}_R$, we will show how to generate texture zeros. Texture zeros in our case refer to vanishing entries in the low-energy Majorana neutrino mass matrix $\mathcal{M}_\nu$, which lead to constraints on the three mixing angles $\theta_{ij}$, three masses $m_i$ and three phases $\delta$, $\alpha$ and $\beta$. More than two independent texture zeros are incompatible with observations, and out of the 15 possible two-zero textures, only seven are allowed by current data~\cite{Fritzsch:2011qv}. In a similar analysis, one can consider texture zeros in $\mathcal{M}_\nu^{-1}$, which are just vanishing minors in $\mathcal{M}_\nu$~\cite{Lavoura:2004tu,Lashin:2007dm}. Again, seven different two-zero textures are allowed, and four of them coincide with two-zero patterns in $\mathcal{M}_\nu$. So overall, there are ten different allowed two-zero textures in $\mathcal{M}_\nu$ and $\mathcal{M}_\nu^{-1}$.

Looking at the charge-matrices of $Y'(\overline{N}^c_i N_j)$ for $\sum_\alpha y_\alpha L_\alpha$ and $B - \sum_\alpha x_\alpha L_\alpha$,
\begin{align}
\begin{pmatrix}2 y_e & y_e + y_\mu & -y_\mu \\ y_e + y_\mu & 2 y_\mu & -y_e\\ -y_\mu & -y_e & -2 (y_e+y_\mu)\end{pmatrix} , &&
\begin{pmatrix}-2 x_e & -x_e -x_\mu & x_\mu-3\\ -x_e - x_\mu & -2 x_\mu & x_e -3\\ x_\mu -3 & x_e -3 & 2 x_e + 2 x_\mu -6\end{pmatrix} ,
\end{align}
shows that we can easily generate texture zeros in $\mathcal{M}_R$. If we restrict ourselves to completely family-non-universal charges, i.e.~$y_\alpha \neq y_\beta$ etc., the leptonic Dirac matrices $m_D$ and $m_\ell$ are strictly diagonal, so the texture zeros in $\mathcal{M}_R$ become texture zeros of $\mathcal{M}_\nu^{-1}$ via seesaw, i.e.~vanishing minors in $\mathcal{M}_\nu \simeq - m_D \mathcal{M}_R^{-1} m_D^T$.

All that is left is to determine the charges that lead to the allowed patterns and find the charges that the symmetry-breaking scalars should carry. As an example, consider $L_\mu-L_\tau$ again, but now with just one scalar $S$ with charge $Y'(S) = 1$. The symmetric right-handed neutrino mass matrix consists of a part symmetric under the $U(1)'$ and a part proportional to the vacuum expectation value of $S$:
\begin{align}
 \mathcal{M}_R = M_{L_\mu-L_\tau} \begin{pmatrix}\times & 0 & 0 \\ \cdot & 0 & \times \\ \cdot & \cdot & 0\end{pmatrix} + \langle S \rangle \begin{pmatrix}0 & \times & \times \\ \cdot & 0 & 0 \\ \cdot & \cdot & 0\end{pmatrix} \sim  \begin{pmatrix}\times & \times & \times \\ \cdot & 0 & \times \\ \cdot & \cdot & 0\end{pmatrix}.
\end{align}
We have therefore found a very economic realization of the allowed pattern $\boldsymbol{C}^R$~\cite{Araki:2012ip}. Four more allowed two-zero textures in $\mathcal{M}_R$ can be obtained with the $B - \sum_\alpha x_\alpha L_\alpha$ symmetries with just one symmetry-breaking scalar (Tab.~\ref{jh_symmetryzeros}). The remaining two allowed patterns can be obtained by introducing two instead of one scalar, or by using $\mathbb{Z}_N$ subgroups of the $B - \sum_\alpha x_\alpha L_\alpha$ symmetries~\cite{Araki:2012ip}. 

Since the $Z'$ gauge boson couples in all cases quite differently to electron, muon and tauon, this framework provides a new handle to distinguish these patterns outside of the neutrino sector.

\begin{table}[t]
\renewcommand{\baselinestretch}{1.2}\footnotesize
\centering
\begin{tabular}[t]{|l|c|c|l|c|}
\hline
			 Symmetry generator $Y'$ & $|Y' (S)|$ & $v_S = \sqrt{2}\, |\langle S\rangle|$ & Texture zeros in $\mathcal{M}_R$ & Texture zeros in $\mathcal{M}_\nu$\\
			 \hline\hline
			 $L_\mu-L_\tau$ &  $1$ & $\geq 160~\mathrm{GeV}$&  $(\mathcal{M}_R)_{33}$, $(\mathcal{M}_R)_{22}$ ($\boldsymbol{C}^R$) & --\\
			 $B - L_e + L_\mu - 3 L_\tau$ &  $2$ & $\geq 3.5~\mathrm{TeV}$& $(\mathcal{M}_R)_{33}$, $(\mathcal{M}_R)_{13}$ ($\boldsymbol{B}_4^R$)& $(\mathcal{M}_\nu)_{12}$, $(\mathcal{M}_\nu)_{22}$ ($\boldsymbol{B}_3^\nu$)\\	
			 $B - L_e - 3 L_\mu + L_\tau$ &  $2$ & $\geq 4.8~\mathrm{TeV}$& $(\mathcal{M}_R)_{22}$, $(\mathcal{M}_R)_{12}$ ($\boldsymbol{B}_3^R$) & $(\mathcal{M}_\nu)_{13}$, $(\mathcal{M}_\nu)_{33}$ ($\boldsymbol{B}_4^\nu$)\\
			 $B + L_e - L_\mu - 3 L_\tau$ &  $2$ & $\geq 3.5~\mathrm{TeV}$& $(\mathcal{M}_R)_{33}$, $(\mathcal{M}_R)_{23}$ ($\boldsymbol{D}_2^R$) & $(\mathcal{M}_\nu)_{12}$, $(\mathcal{M}_\nu)_{11}$ ($\boldsymbol{A}_1^\nu$)\\
			 $B + L_e - 3 L_\mu - L_\tau$ &  $2$ & $\geq 3.5~\mathrm{TeV}$& $(\mathcal{M}_R)_{22}$, $(\mathcal{M}_R)_{23}$ ($\boldsymbol{D}_1^R$) & $(\mathcal{M}_\nu)_{13}$, $(\mathcal{M}_\nu)_{11}$ ($\boldsymbol{A}_2^\nu$)\\		 
\hline
\end{tabular}
\renewcommand{\baselinestretch}{1.0}\normalsize 
\caption{
\label{jh_symmetryzeros}
Anomaly-free $U(1)$ gauge symmetries that lead to allowed two-zero textures in the right-handed Majorana mass matrix $\mathcal{M}_R$ with the addition of just one SM singlet scalar $S$. Some of the texture zeros propagate to $\mathcal{M}_\nu \simeq -m_D \mathcal{M}_R^{-1} m_D^T$ after seesaw. Classification of the two-zero textures according to Ref.~\cite{Araki:2012ip}.}
\end{table}

\section{Conclusion}

The type-I seesaw mechanism provides a fascinating explanation of small neutrino masses. The three additional right-handed neutrinos also significantly increase the number of anomaly-free symmetries, which can be used to explain the peculiar leptonic mixing parameters. We have shown how to use $U(1)'$ groups generated by $B - \sum_\alpha x_\alpha L_\alpha$ or $\sum_\alpha y_\alpha L_\alpha$ to promote the championed symmetries behind normal, inverted and quasidegenerate neutrino mass hierarchy to local symmetries, with the obvious implications for collider physics. We have further shown that all seven currently allowed two-zero textures in $\mathcal{M}_\nu^{-1}$ can be realized very economically by $U(1)'$ symmetries with at most two additional scalars.

\section*{Acknowledgments}
J.H.~thanks W.~Rodejohann, J.~Kubo and T.~Araki for collaboration on the work presented here.
The work of J.H.~was supported by the the ERC under the Starting Grant MANITOP and by the IMPRS-PTFS.

\bibliography{julianheeck.bib}
\bibliographystyle{apsrev4-1}


%% file: Papers/Helo.tex

%
%
%

%
%
%
%
%
%

\chapter[Neutrinoless double beta decay at LHC (\textit{Helo}, Hirsch, Kovalenko, Päs)]{Neutrinoless double beta decay at LHC}
\vspace{-3em}
\paragraph{\textit{J. C. Helo}, M. Hirsch, S. Kovalenko, H. Päs}
\paragraph{Abstract}
We analyze the possibility of discriminating different mechanisms of neutrinoless double beta decay in the LHC experiments from a general point of view.
We distinguish basic topologies of these mechanisms with one or two heavy intermediate particles on-mass-shell which can be 
accessible for observations in high-energy pp-collisions.

\section{Introduction}
Neutrinoless double beta decay is believed to be the most sensitive probe of lepton number
violation.   On the other hand observation of this rare decay will not be easily interpreted
as evidence for a specific model of new physics beyond the Standard Model (SM).
Several mechanisms including exchange of heavy neutrinos  \cite{Benes:2005hn}, \cite{Helo:2010cw}   right-handed
$W_{R}$ bosons \cite{Mohapatra:1980yp}, \cite{Maiezza:2010ic}, leptoquarks \cite{Davidson:1993qk}, SUSY partners \cite{Hirsch:1995cg}  etc. have been discussed in the literature (for recent reviews see e.g. \cite{Deppisch:2012nb}, \cite{ Rodejohann:2011mu}) 
besides the most popular mechanism with 
exchange of a light Majorana neutrino.
Although there exist in the literature several proposals of  how to discriminate different 
mechanisms \cite{Fogli:2009py}, \cite{Gehman:2007qg}, \cite{Deppisch:2006hb}, \cite{Ali:2007ec}, 
they typically lack sensitivity to at least some of the mechanisms and/or are difficult to observe experimentally.


If $0\nu\beta\beta$ decay is observed in the next generation experiments the question of which mechanism produce the signal will immediately arise.  What can the LHC say about this?
 To answer this question   we need first to  identify possible topologies of the diagrams at the level of  renormalizable interactions contributing both to $0\nu\beta\beta$-decay and to like-sign dilepton production at LHC. 



At the level of renormalizable interactions all tree-level diagrams for $0\nu \beta \beta$ decay fall just into two types of topologies, shown 
in fig. \ref{Diagrams}.
%
%
%
Same topologies that contribute to $0\nu \beta \beta$ may  contribute to like-sign dilepton events at LHC accompanied with 
jets 
$pp\rightarrow l^{\pm}l^{\pm}+ jets$ \cite{Allanach:2009iv}, \cite{Allanach:2009xx}.  Here we consider renormalizable interactions which correspond to  vertices in the diagrams of Topology I and II (Fig. \ref{Diagrams})
 with scalar $S$ and/or (axial-)vector $V_{\mu}$ and fermionic $N$  intermediate particles. A complete list of renormalizable  Lagrangian terms contributing  to $0\nu\beta\beta$  and LHC will be analyzed  in \cite{Soon}.
Here we consider the following interaction terms underlying the diagrams in Fig. \ref{Diagrams}:
 %
 %
 \begin{eqnarray}\label{Lagrangian-S}
 {\cal L}_{S} &=& g^{S}_q S \bar d (1 \pm \gamma_5 ) u  
+  g^{S}_l S \bar e (1 \pm \gamma_5 ) N  
+ \mbox{h.c.}\\
\label{Lagrangian-V}
 {\cal L}_{V} &=& \frac{g^{V}_q}{2 \sqrt{2}} V_{\mu} \bar d \gamma^\mu (1 \pm \gamma_5 ) u  
 +  \frac{g^{V}_l}{2 \sqrt{2}} V_{\mu}^- \bar e \gamma^\mu (1 \pm \gamma_5 ) N  
+ \mbox{h.c.}\\
\label{Lagrangian-S2}
{\cal L}_{S2} &=& g^{S}_q S  \bar d (1 \pm \gamma_5 ) u  
%
 + \
 g_{S} \ v S S S_2 +  g^{S_2}_l S_2\ \bar e  (1 \pm \gamma_5 ) e^{c} + h.c \\
 \
 \label{Lagrangian-V2}
 {\cal L}_{V2} &=& \frac{g^{V}_q}{2 \sqrt{2}} V_{\mu} \bar d \gamma^\mu (1 \pm \gamma_5 ) u  +
 g_{V} \ v V_{\mu}  V^\mu  S_2 +  \ g^{S_2}_l S_2 \ \bar e   (1 \pm \gamma_5 ) e^{c} +  h.c
 \end{eqnarray}
 Here $S=S^{-}, V_{\mu} = V_{\mu}^{-}$ and $S_{2} = S_{2}^{--}$ are single and doubly charged bosons while $N$ is a neutral fermion, $v$ is a parameter with dimension of mass.   The first two Lagrangians can generate same-sing dileptons if  $N^c=N$ is a Majorana fermion.
 Below we denote  masses of the $S, V$ and $N$ particles as $m_{S}$, $m_{V}$ and $m_{N}$.  Evidently the Largangians  (\ref{Lagrangian-S})-(\ref{Lagrangian-V}) lead to 
the diagrams of Topology I  while  (\ref{Lagrangian-S2})-(\ref{Lagrangian-V2}) to Topology II (see Fig. \ref{Diagrams}).

Following the notations of Ref.  \cite{Pas:2000vn} we calculated the contributions  of these diagrams to the half-life of neutrinoless double beta decay
$T^{0\nu \beta \beta}_{1/2} $ separately for the scalar and vector  cases of the above specified 
Lagrangians
Eqs. (\ref{Lagrangian-S}), (\ref{Lagrangian-V}), (\ref{Lagrangian-S2}), (\ref{Lagrangian-V2}).  The dependencies of the half-life on masses and couplings are :
%
%
%
\begin{eqnarray}
\label{TT-S}
\text{Topo I} : \ \ T^{0\nu \beta \beta}_{1/2(P)} \propto
 \frac{(M_{eff(P)}^{(I)})^{10}}{(g^{(I)}_{eff(P)})^8 }\ \  ; \ \ \text{Topo II} : \ \  T^{0\nu \beta \beta}_{1/2(P)} \propto 
 \frac{(M_{eff(P)}^{(II)})^{12}}{  (g^{(II)}_{eff(P)} v^{1/4})^8}
\end{eqnarray}
where  we have defined the effective masses and couplings
\begin{eqnarray}
\label{Efective-1}
M^{(I)}_{eff(P)} &=&  (m_{P}^4  m_N )^{1/5}, \ \ \ \ g_{eff(P)}^{(I)} = (g_q^{P}  g_l^{S_2})^{1/2}
 \\ \label{Efective-2} M^{(II)}_{eff(P)} &=& (m_{P}^2 m_{S_2} )^{1/3}, \ \ \ \  g_{eff(P)}^{(II)} = ((g^{P}_l)^2 g^{S_2}_l g_{P})^{1/4} 
 \end{eqnarray}
The current limit on $T^{0\nu \beta \beta}_{1/2}$ of ${}^{76} Ge$ \cite{KlapdorKleingrothaus:2000gr} excludes the region of these effective parameters shown in Figs. \ref{M5g2}, \ref{M3g4} as the black area. The red region of Figs. \ref{M5g2}, \ref{M3g4} represents the parametric region which could be excluded at the expected  sensitivity  $T^{0\nu \beta \beta }_{1/2} =1 \times 10^{27} yrs$ of  the  future $0\nu\beta\beta$ experiments
and the pink region in Fig. \ref{M5g2}  is inaccessible for these  experiments.
\subsection{Topolgy I}
\label{TopoI-S}
%
 We will assume that a particle $P \ (= S, V)$ is produced at LHC,  decays to $P \to N  e$ and than $N$ decays to $N\to e^- u d$ (See Fig. 9a), producing the events with same-sign lepton pairs and two jets, as a clear signature of lepton number violation. This situation is described by the Lagrangians (\ref{Lagrangian-S}), (\ref{Lagrangian-V}).



 The lepton-number violating character of the process  $p p \rightarrow e  e jj$ leads to a very clear signature at LHC.  The background for this process, mainly coming from $\bar t t$ 
 events, is very low   due to the presence of two isolated leptons of the same sign. In Refs.  \cite{Das:2012ii, Ferrari:2000sp} detailed simulations of this signal have been carried out. The kinematical region $m_N < m_P$ enables ideal decay kinematics with a large production cross section and isolated leptons except for $m_N \lesssim m_P$   where the two-body decay $P \rightarrow e N$  is kinematically forbidden,  and thus the LHC sensitivity is suppressed \cite{Das:2012ii}.
 %
%
 In this paper we  have assumed  a  sensitivity to the cross section up to $\sigma(p p \rightarrow e  e jj) = 10^{-2} fb $  for  $m_N < m_P$  when  the LHC  runs at $\sqrt{s} = 14 TeV $. This  is consistent with the different simulations of the signal  in Refs. \cite{Das:2012ii}, \cite{Ferrari:2000sp}, \cite{Dreiner:2000vf}. 
\begin{figure}[htbp]
\begin{minipage}[b]{.45\linewidth}
\includegraphics[width=\linewidth]{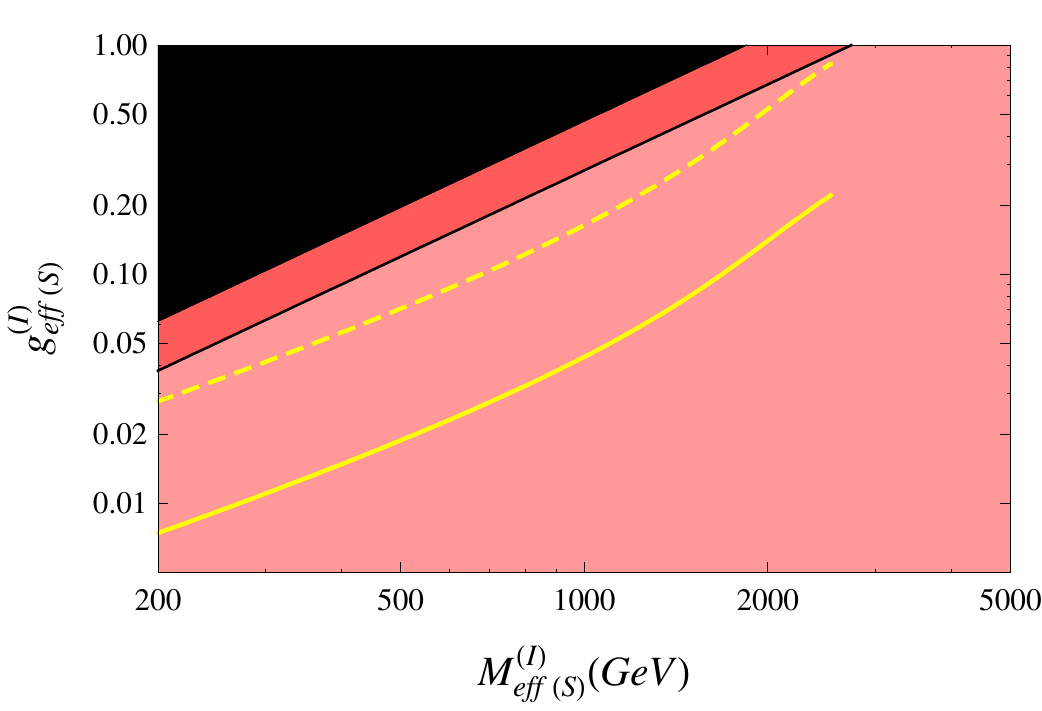}
\end{minipage}
\begin{minipage}[b]{.05\linewidth}
\hspace{1pt}
\end{minipage}
\begin{minipage}[b]{.45\linewidth}
\vspace{0pt}
\includegraphics[width=\linewidth]{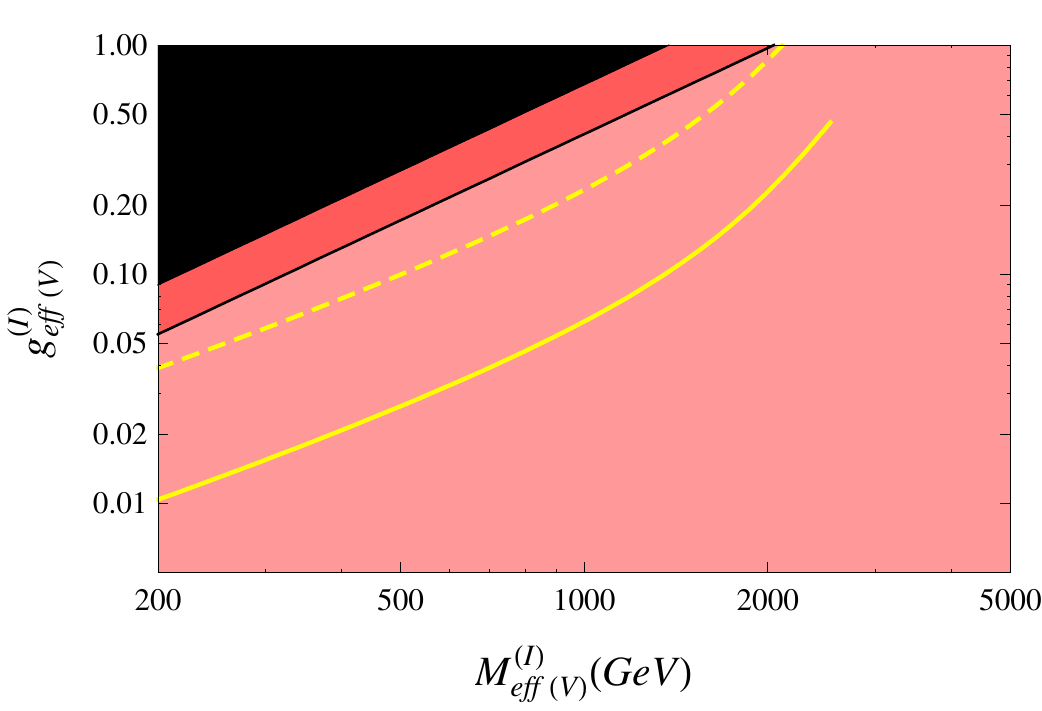}
\end{minipage}
\vspace{-0.2 cm}
\caption{ LHC and $0 \nu\beta \beta$ sensitivities on the LNV  Lagrangians (\ref{Lagrangian-S}), (\ref{Lagrangian-V}). See text for details.
 }
 \label{M5g2}
\end{figure}
 The cross section $\sigma(pp \rightarrow e e j j)$   can be written as
\begin{eqnarray} \label{Xsec}
 \sigma(pp \rightarrow e e j j) = F(m_P) \times (g^P_q)^2 \times Br(P \rightarrow e N)  \times Br(N \rightarrow e  j j) 
 \end{eqnarray} 
where the Branching ratios are, according to the Lagrangians (\ref{Lagrangian-S}), (\ref{Lagrangian-V}), equal to
\begin{eqnarray} \label{Br}
Br(P \rightarrow e N) = \frac{ a \ (g^S_l)^2}{ 3 (g^S_q)^2  + a \ (g^S_l)^2} \ \  \ \ ; \ \  \ \  Br(N \rightarrow e  j j) = 1/2.
\end{eqnarray}
Here $a < 1$ is a phase space factor $a= 1-(m_N/m_P)^2$. The function $F(m_P) =\sigma(pp \rightarrow e e j j) / (Br \ (g^P_q)^2) $ has been calculated using CALCHEP \cite{Pukhov:2004ca}. We can rewrite this cross section (\ref{Xsec}) in terms of the effective mass and couplings relevants for $0 \nu \beta \beta$ using Eqs. (\ref{Efective-1})
 \begin{eqnarray} \label{XsecEff}
 \sigma(pp \rightarrow e e j j) = F\left(  \sqrt[4] {  \frac{(M^{(I)}_{eff})^5}{m_N}} \right) \times  \frac{ \ (g_{eff}^{(I)})^4}{ 3  (g^P_q)^4 / a +   (g_{eff}^{(I)})^4 } \times \frac{1}{2}
 \end{eqnarray} 
Assuming that $g^P_q, g^P_l < 1$,  for  a fixed value of the effective coupling $g_{eff}^{(I)}$, the coupling $g^P_q$ is limited as  $(g_{eff}^{(I)})^2 < g^P_q < 1$.
%
%
Using this limits  we can put a lower limit on $ \sigma(pp \rightarrow e e j j) $ which depends  on the effective parameters (\ref{Efective-1}) and   the fermion mass $m_N$.  
 \begin{eqnarray} \label{XsecEffLim}
\sigma(pp \rightarrow e e j j) > F\left( \sqrt[4] {  \frac{(M^{(I)}_{eff})^5}{m_N}} \right) \times \frac{ \ (g_{eff}^{(I)})^4}{3/a +1  }  \times \frac{1}{2}
  \end{eqnarray} 
%
%
%
%
 %

If the LHC does not observe $pp \rightarrow e e j j$ it will put limits on its cross section which we have assumed to be  $\sigma(pp \rightarrow e e j j) < 10^{-2}fb$ according to our previous discussion. Then with this assumption  and the numerical values of the function $F(x)$ which have been  calculated using CALCHEP \cite{Pukhov:2004ca},  we can extract limits on the parametric region $(M^{(I)}_{eff(P)}, g^{(I)}_{eff(P)})$ from   Eq. (\ref{XsecEffLim}) for different values of $m_N$.  We have   drawn these limits as the solid yellow line  in Fig. \ref{M5g2} for $m_N = 200 GeV$,   and compered them with the limits expected  from the future  $0\nu\beta \beta$ decay experiments.  For larger masses $m_N > 200 GeV$ these limits on $(M^{(I)}_{eff(P)}, g^{(I)}_{eff(P)})$ will become even  more strength except for the threshold region  $m_N \lesssim m_P$ where the LHC sensitivity gets very low as we discussed above. As we can see from Fig. \ref{M5g2} the LHC is much more sensitive than $0\nu\beta\beta$ and has good perspective to rule out this mechanism, as a possible dominant contribution to $0\nu\beta\beta$ decay,  if after running at $\sqrt s = 14 TeV$ it does not find positive signals of $pp \rightarrow e e  j j$.

 It is important to emphasise that by construction of the Lagrangians (\ref{Lagrangian-S}), (\ref{Lagrangian-V}) we are not considering flavour mixing and the Branching ratio of the neutral fermion decay  is $Br(N \rightarrow e j j ) = 1/2$. However even for $Br(N \rightarrow e j j ) = 10^{-2}$ our LHC limits are  still more sensitive than $0\nu\beta\beta$ as can be seen from Fig. \ref{M5g2} where  dashed yellow line represents the LHC limits for $Br(N \rightarrow e j j ) = 10^{-2}$.
\subsection{Topology II}
In order to analyse the impact of the LHC experiments on Topology II contributions to $0\nu\beta\beta$ described by the Lagrangians (\ref{Lagrangian-S2}), (\ref{Lagrangian-V2}) we have studied the production of the like-sign dileptons $p p \rightarrow e e j j $ through the single production of $S_2^{++}$ via    a $SS$ or $VV$  fusion $p p \rightarrow S_2^{++} j j $ followed by the decay $S_2^{++} \rightarrow e e $ at LHC.
This process is experimentally interesting due to its clear like-sign dilepton signature 
from the decay of the doubly  charged scalar $S_2^{++}$.They can be easily  distinguished  from the SM background due its resonance contribution to the invariant mass distribution of the two leptons   which has a very small SM background for invariant masses larger than $100 GeV$ \cite{Ferrari:2000sp}. Then as an example we have estimated the LHC sensitivity to be  $\sigma(pp \rightarrow  e e j j ) \sim 1.6 \times 10^{-2} fb$ corresponding  to 5 events at luminosity of $300 fb^{-1}$.  
 
 The corresponding cross sections   have been calculated using CALCHEP. \cite{Pukhov:2004ca}. These cross sections are much smaller than the corresponding ones for the Topology I.
 Then the LHC is not as competitive with $0\nu\beta\beta$  experiments as in the case of Topology I, being actually less sensitive than $0\nu\beta\beta$ for a big part of the parametric space. However there are still some parametric regions where the LHC is competitive and even  more sensitive than $0\nu\beta\beta$. To give an example of this we  case will analyzed a scenario where 
  \begin{eqnarray} \label{RegionII}
   m_{S_2} < 2 m_P 
 \end{eqnarray}
\begin{figure}[htbp]
\begin{minipage}[b]{.45\linewidth}
\includegraphics[width=\linewidth]{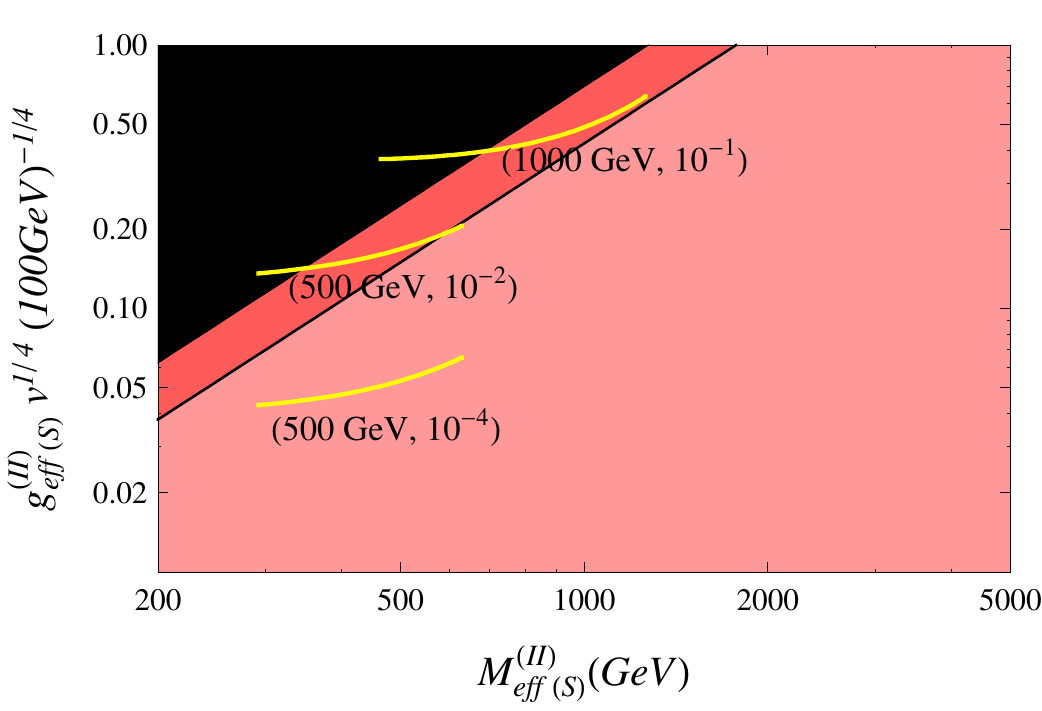}
\end{minipage}
\begin{minipage}[b]{.05\linewidth}
\hspace{1pt}
\end{minipage}
\begin{minipage}[b]{.45\linewidth}
\vspace{0pt}
\includegraphics[width=\linewidth]{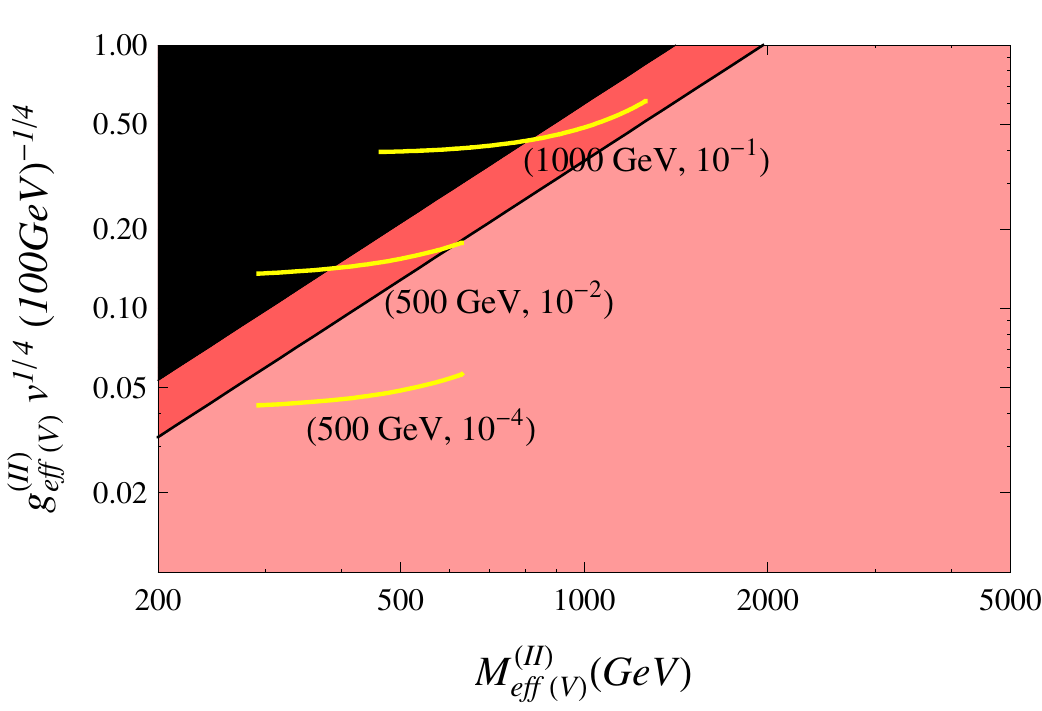}
\end{minipage}
\vspace{-0.2 cm}
\caption{ LHC and $0 \nu\beta \beta$ sensitivities on the LNV  Lagrangians (\ref{Lagrangian-S2}), (\ref{Lagrangian-V2}). See  text for details.
 }
 \label{M3g4}
\end{figure}
%
 %
%
 %
 %
 The cross section of the like-sign dileptons plus two jets process at LHC $\sigma(pp \rightarrow e e j j)$ can be written as 
  \begin{eqnarray} \label{XsecII}
 \sigma(pp \rightarrow e^+ e^+ j j) = H(m_P, m_{S_2}) \times   (g^P_q)^4 \times (v g_P)^2 \times Br(S_2^{++} \rightarrow e^+ e^+)
 \end{eqnarray} 
 
  In the scenario (\ref{RegionII}) the doubly charged scalar $S_2^{++} $ can decay into $S_2^{++} \rightarrow e^+ e^+$ or back to $S_2^{++} \rightarrow j j j j$. However this last channel will be very suppressed compared with the decay channel $S_2^{++} \rightarrow e^+ e^+$. Then, 
  we can approximately put  $Br(S_2^{++} \rightarrow e^+ e^+) =1 $.  The function $H(m_P, m_{S_2})$  has been calculated using CALCHEP \cite{Pukhov:2004ca} for $P = S, V$.  Using the numerical values of   $H(m_P, m_{S_2})$ and the assumed LHC sensitivity $\sigma(pp \rightarrow e e j j) < 1.6 \times 10^{-2} fb$ we can extract limits on the parameters $(m_{S_2},  \ (g^P_q)^2  v g^P)$ from Eq. (\ref{XsecII}) for fixed values of $m_{P}$. We can convert these limits on $(m_{S_2},  \ (g^P_q)^2  v g^P)$  into limits on the effective parameters $(M_{eff(P)}^{(II)},  g_{eff(P)}^{(II)} v^{1/4})$  of Eq. (\ref{Efective-2})  fixing  values of the coupling $g_l^{S_2}$. Then for  different  values of  $(m_P, g_l^{S_2})$ we have different LHC limits on the effective parameters $(M_{eff(P)}^{(II)},  g_{eff(P)}^{(II)} v^{1/4})$.  In Fig.  (\ref{M3g4})  we  have plotted these LHC  limits on the effective parameters (\ref{Efective-2})  as the yellow lines and compered them  with the $0\nu\beta\beta$ limits at the  future sensitivities. As we can see from Fig. (\ref{M3g4}) yellow lines cover only a segment of the  possibles values of $M_{eff(P)}^{(II)}$ in the plot. This is because we have used for this calculation $m_{S_2} > 100 GeV $ and the limit  $ m_{S_2} < 2 m_P $ defined by the scenario (\ref{RegionII}) . These limits on $m_{S_2}$ impose upper and lower limits on the effective mass $   (m_P^2 100 GeV)^{1/3} < M_{eff(P)}^{(II)} <  (2 m_P^3)^{1/3} $ which are reflected in the borders of the  yellow lines of Fig. (\ref{M3g4}) for $m_P = 500 GeV,  \ 1000 GeV$. As we can see from Fig. \ref{M3g4} the LHC is competitive to $0\nu\beta\beta$ in the scenario of Eq. (\ref{RegionII}) for $(m_P = 500 GeV , g_l^{S_2} = 10^{-2})$ and $(m_P = 1000 GeV , g_l^{S_2} = 10^{-1})$ and much more sensitive than $0\nu\beta\beta$ for smaller values of $g_l^{S_2}$. 
\begin{figure}[htbp]
\centering
\includegraphics[width=0.8\textwidth,bb=100 580 600 670] {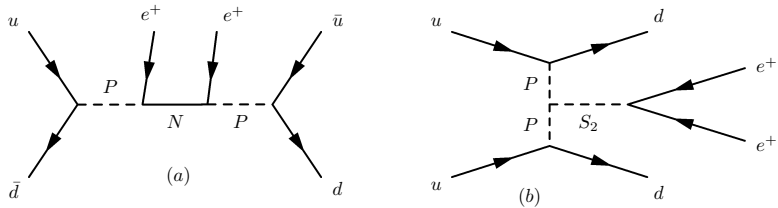}
\caption{ Two types of diagrams (Topology I, II) contributing both to like sign lepton plus two jets production at LHC $p p \to e e j j$ and 
to $0\nu \beta \beta$ decay. Here $P=S, V_{\mu}$. 
}  
\label{Diagrams}
\end{figure}
\section{Conclusion}
We have analyzed the possibility of discriminating different mechanisms of $0\nu\beta \beta$ decay at LHC. We have found that for the case of the mechanisms corresponding to the Topology I  the LHC is significantly more  sensitive than  $0\nu\beta \beta$ decay experiments and may be capable to probe or rule out these mechanisms. For the case of Topology II  the situation is more ambiguous and the LHC could be more sensitive  only in a particular part of the parametric space.
\section*{Acknowledgments}
This work was supported by FONDECYT (Chile) under projects 1100582, 1100287;
CONICYT(Chile) under projects 
Centro-Cientifico-Tecnologico de Valparaiso PBCT ACT-028 and 
Departamento de Relaciones Internacionales "Programa de Cooperacion Cientifica Internacional" CONICYT/DFG-648.  J.C.H. thanks the IFIC for hospitality during his stay.
M.H. acknowledges support from the Spanish MICINN grants FPA2008-00319/FPA, FPA2011-22975, MULTIDARK CSD2009-00064 and
2009CL0036 and by GV grant Prometeo/2009/091 and the
EU Network grant UNILHC PITN-GA-2009-237920.

\bibliography{Helo}
\bibliographystyle{apsrev4-1}


%% file: Papers/holthausen.tex


%
%
%
%
%
%


\chapter[Vacuum Alignment from Group Theory (Holthausen)]{Vacuum Alignment from Group Theory}
\vspace{-2em}
\paragraph{M. Holthausen}
\paragraph{Abstract}
Models based on non-abelian discrete symmetries that aim to explain mixing patterns such as tri-bi-maximal mixing(or perturbations thereof) usually require the symmetry group to be broken to different subgroups in the charged lepton and neutrino sectors by scalar fields with a particular configuration of vacuum expectation values. This configuration cannot be obtained from a straightforward minimization of the potential, but it requires an additional dynamical mechanism. Here we present a mechanism based on group theoretic considerations. 

\section{Introduction}
Until quite recently, the tri-bi-maximal(TBM) ansatz for lepton mixing that corresponds to the mixing angles 
$
\sin^2\theta_{12}=\frac13 , \;
\sin^2\theta_{23}=\frac12, \;
\sin^2\theta_{13}=0\;
$
was a quite good fit to the experimental data. This peculiar mixing pattern might be understood as the consequence of non-commuting remnant symmetries $G_e$ of the charged lepton and $G_\nu$ of the neutrino mass matrices\cite{Lam:2008fj}: 
\begin{align*}
\rho(g_e)^T M_e M_e^\dagger \rho(g_e)^*=M_e M_e^\dagger, \quad\rho(g_\nu)^T M_\nu \rho(g_\nu)=M_\nu\qquad \mathrm{and}\quad g_e \in G_e, g_\nu \in G_\nu.
\end{align*}
Indeed if one takes $G_\nu$ to be the Klein group $G_\nu=\langle S,U\vert S^2=U^2=E;SU=US\rangle\cong Z_2\times Z_2$ and $G_e$ to be $G_e=\langle T\vert T^3=E \rangle\cong Z_3$ with the 3-dimensional generators
\begin{align}
\rho(S)= \left(\begin{array}{ccc}
1&0&0\\
0&-1&0\\
0&0&-1
\end{array}\right), \quad \rho(U)=-\left(\begin{array}{ccc}
1&0&0\\
0&0&1\\
0&1&0
\end{array}\right),\quad
\rho(T)=\left(\begin{array}{ccc}
0&1&0\\
0&0&1\\
1&0&0
\end{array}\right)
\label{MH_eq:S;T}
\end{align}
the resulting mixing matrix is of TBM form. The symmetry group built up of $S$ and $T$ is $A_4$, the symmetry group formed out of $S,T$ and $U$ is $S_4$. In $A_4$ models that predict TBM at leading order(LO) the symmetry $U$ is accidental. Giving up on the accidental symmetry $U$, one has $G_\nu=\langle S\vert S^2=E\rangle\cong Z_2$ and the mixing matrix $U=U_{HPS}U_{13}$ is determined up to a 13-rotation. This is called trimaximal mixing(TMM) and can bring discrete family symmetry models of this type back in agreement with the experimental data(i.e. the large value of $\theta_{13}$).  
\section{Vacuum Alignment Problem}
To break the symmetry group $A_4$ to the subgroups S and T, scalar fields("flavons") $\chi$ and $\phi$  are introduced with VEVs along the directions $\langle\chi\rangle\sim (v',v',v')$, that conserves T, and  $\langle\phi\rangle\sim (v,0,0)$, that conserves S. The most general scalar potential formed out of these two scalar fields does not allow this VEV configuration, as may be seen from the fact that the number of algebraically independent minimization conditions is higher than the number of VEVs\cite{Altarelli:2005uq,He:2006dk} and therefore this VEV configuration requires special fine-tuned relations among the potential parameters.\footnote{This problem cannot be cured by introducing singlets, etc. as was shown in \cite{Holthausen:2012pb}}

\begin{figure}[tb]
\centering
\includegraphics[width=\textwidth]{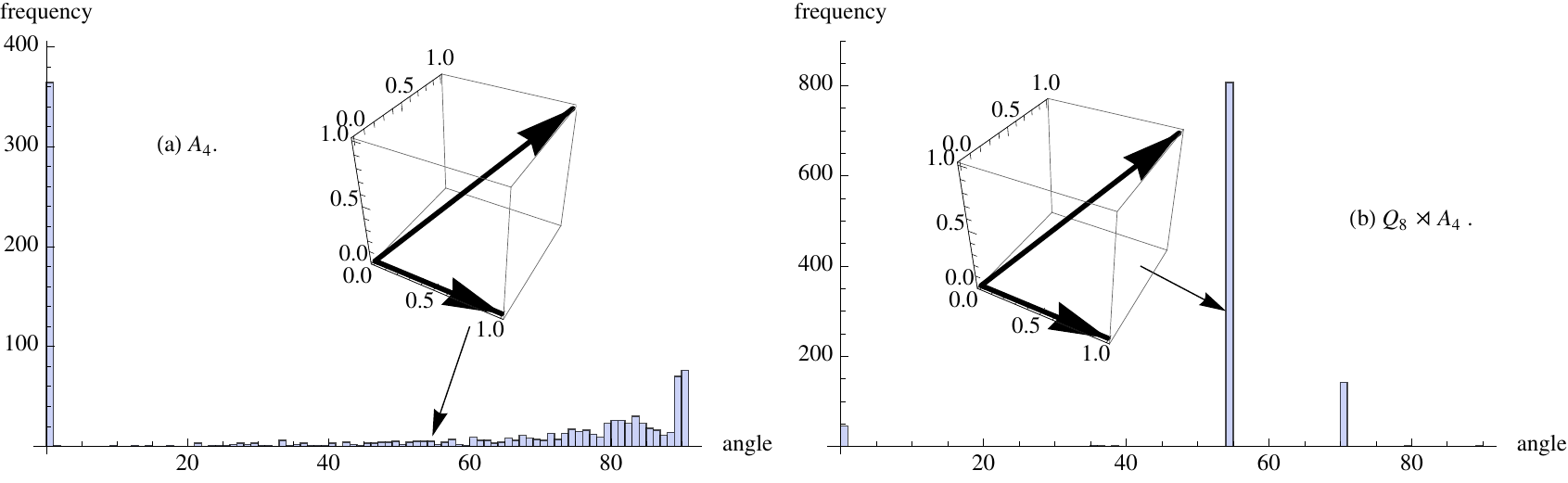} 
\caption[]{Distribution of the opening angle spanned by the two (effective) flavon fields that couple to neutrinos and charged leptons, respectively, for random values of potential parameters for the most general scalar potential of $A_4$(left) and $Q_8\rtimes A_4$(right). For  $Q_8\rtimes A_4$ the relevant effective flavon in the neutrino sector is $ (\phi_1 \phi_2)_{\MHMoreRep{3}{1}}$. The tri-bi-maximal vacuum configuration depicted in the inlay corresponds to an opening angle of $54.7^\circ$.   }
\label{fig:aligned-scan}
\end{figure}

Another way to see this is to take the full potential of the two scalar fields and scan over the global minima that one obtains. In Figure \ref{fig:aligned-scan}, on the left-hand side,  the distribution of opening angles between the two flavons is plotted for a random scan over order one potential parameters. One can see that there is no phase where the TBM vacuum configuration is realized, but rather two phases can be identified: one phase where both flavons conserve the same subgroup and point in the same direction(angle$=0^\circ$) and one phase where the symmetry is broken completely. The TBM vacuum is part of the later phase but it is not special. If there is no TBM phase in the potential, the whole discrete symmetry approach amounts to nothing more than replacing adjusting Yukawa couplings to adjusting potential parameters.
\section{A Solution via Group Extensions}
We have seen in the last section that the most general scalar potential does not allow the desired vacuum pattern. This can be traced back to the existence of couplings such as $(\chi \chi)_{\MHMoreRep{1}{2}} (\phi \phi)_{\MHMoreRep{1}{3}} $ that connect the $A_4$ transformations of $\chi$ and $\phi$\cite{Babu:2010bx,Holthausen:2012pb}. To solve the vacuum alignment problem we demand the following: 
\begin{itemize}
\item we want to extend successful flavour groups $H=A_4, S_4, T^\prime, \Delta(27), T_7$, therefore there should be an surjective homomorphism $\xi:G\rightarrow H$ from $G$ onto $H$. The homomorphism $\xi$ guarantees the existence of representations of $G$ that are inherited from $H$: $\rho^G=\rho^H\circ \xi$, to which we will assign leptons. Therefore the lepton structure is the same as in $H$.
\item we demand the existence of an irreducible representation $\phi$, whose product $\phi^n$ should contain $\MHRep{3}^G$ and the renormalizable potential formed out of $\phi$ and $\chi\sim \MHRep{3}^G$ should have an accidental symmetry $G\times A_4$ such that $\chi$ can be rotated by independent $A_4$ symmetry transformations.
\end{itemize}
Using the computer algebra program GAP, we have performed a scan over all groups of order smaller than 1000, and we have found a number of candidate groups. We do not repeat the entire catalogue of groups here, but rather refer the reader to \cite{Holthausen:2012pb} and present the smallest extensions of $A_4$, $S_4$ and $T^\prime$, of which only the first one was presented in detail in \cite{Holthausen:2012pb}.
\subsection{$Q_8\rtimes A_4$}
The group may be presented by the three generators $S,T,X$ fulfilling the relations
\begin{align}
 S^2=T^3=X^4=SXSX^{3}=(ST)^3=T^{2}XT^{2}X^{3}T^{2}X^{3}
= STX^{3}T^{2}STX^{3}T^{2}=E
\end{align}
One can see that the group element $X^2$ commutes with all other elements. This generates the center $Z(Q_8\rtimes A_4)=\{E,X^2\}$, representation can be classified according to $\rho(X^2)=\pm 1$.

The defining representation matrices for the representations are given in \cite{Holthausen:2012pb}. Of importance is the 3-dimensional representation with $\rho_{\MHMoreRep{3}{1}}(S),\rho_{\MHMoreRep{3}{1}}(T)$ given in Eq.\eqref{MH_eq:S;T} and $\rho_{\MHMoreRep{3}{1}}(X)={1}_3,$ which is exactly the inherited 3-dimensional $A_4$ representations. Obviously, this representation only knows about the $A_4$ subgroup generated by $S$ and $T$ and it is therefore not-faithful. The  other crucial ingredient we needed was a faithful representation of $G$ that did not contain any $A_4$ representation in its symmetric product. This representation can be readily identified to be $\MHMoreRep{4}{1}$
\begin{align*}
\rho_{\MHMoreRep{4}{1}}(S)=&
\sigma_3 \otimes\sigma_1,
&
\rho_{\MHMoreRep{4}{1}}(T)&=\mathrm{diag}(\rho(T),1),
&\rho_{\MHMoreRep{4}{1}}(X)=&-\mathrm{i} \sigma_2 \otimes\sigma_3.
\end{align*}
with $\rho(T)$ given in \eqref{MH_eq:S;T} and the crucial property
$
\MHMoreRep{4}{1}\times\MHMoreRep{4}{1}={\MHMoreRep{1}{1}}_S+{\MHMoreRep{3}{1}}_A+{\MHMoreRep{3}{2}}_S+{\MHMoreRep{3}{3}}_S+{\MHMoreRep{3}{4}}_S+{\MHMoreRep{3}{5}}_A.
$
\subsection{$Z_4.S_4$}
This group in the GAP notation [96,67] is the smallest extension of $S_4$ that allows for a solution of the vacuum alignment problem. It is generated by 2 generators  $A$ and $B$ that fulfil the relations:
\begin{align}
A^4=B^4=A B^{-1} A^{-1}BA^{-1}B^{-1}=E
\end{align}
and the faithful representation that solves the VEV alignment problem is given by
\begin{align}
A: \frac{1}{\sqrt{2}} \left(
\begin{array}{cccc}
 0 & 0 & z^{13} & z^{19} \\
 0 & 0 & z^{13} & z^7 \\
 z^{11} & z^{23} & 0 & 0 \\
 z^5 & z^5 & 0 & 0
\end{array}
\right) \qquad \mathrm{and} \qquad
B:
\frac{1}{\sqrt{2}}
\left(
\begin{array}{cccc}
 0 & 0 & z^5 & z^5 \\
 0 & 0 & z^{23} & z^{11} \\
 z^{19} & z & 0 & 0 \\
 z^7 & z & 0 & 0
\end{array}
\right)
\end{align}
with $z=e^{\mathrm{i}\pi/12}$.
\subsection{$Q_8\rtimes T^{\prime}$}
This group is generated by $S,T,R,X$ that fulfill the relations
\begin{small}
$$
S^2R= T^3=(ST)^3=R^2=X^4=SXSX^{3}=(ST)^3=T^{2}XT^{2}X^{3}T^{2}X^{3}
= STX^{3}T^{2}STX^{3}T^{2}=RXRX^3=E
$$
\end{small}
The generator $R$ therefore commutes with all group elements and the center is enlarged to $Z(Q_8\rtimes T^\prime)=\{E,R,R X^2,X^2\}\cong Z_2\times Z_2$. The relevant representations can be constructed from the homomorphism $g:Q_8\rtimes T^\prime\rightarrow Q_8\rtimes A_4$ defined by $g:\{ R,S,T,X \}\rightarrow \{E,S,T,X\}$. To solve the vacuum alignment problem, the leptons should be assigned to  $\ell\sim\rho_{\MHMoreRep{3}{1}}\circ g$ and the neutrino sector flavon $\phi\sim \rho_{\MHMoreRep{4}{1}}\circ g$. The additional representations may be used to describe the quark sector.

\section{TBM Model based on $Q_8\rtimes A_4$}
To give a concrete model that solves the vacuum alignment problem, we introduce lepton doublets $\ell \sim \MHMoreRep{3}{1}$ and lepton singlets $e^c+\mu^c+\tau^c\sim \MHMoreRep{1}{1}+\MHMoreRep{1}{2}+\MHMoreRep{1}{3}$ under $Q_8 \rtimes A_4$. Symmetry breaking is done via $\chi\sim \MHMoreRep{3}{1}$ and  $\phi_{1,2}\sim \MHMoreRep{4}{1}$\footnote{We also need a discrete subgroup of lepon number $L$: $i^L$ under which $\phi_2$ is odd.}. To lowest order, the charged lepton masses arise from the operators
\begin{equation}
\mathcal{L}_e^{(5)} = y_e (\ell \chi)_{\MHMoreRep{1}{1}} e^c \tilde{H}/\Lambda
+y_\mu (\ell \chi)_{\MHMoreRep{1}{3}} \mu^c  \tilde{H}/\Lambda
+y_\tau (\ell \chi)_{\MHMoreRep{1}{2}} \tau^c  \tilde{H}/\Lambda +\text{h.c.}\; ,
\end{equation}
with the Higgs field $\tilde{H}=\mathrm{i} \sigma_2 H^*$, and the neutrino masses are generated from the effective interactions
\begin{align}
\mathcal{L}_\nu^{(7)} &= x_{a} (\ell H\ell H)_{\MHMoreRep{1}{1}} (\phi_1 \phi_2)_{\MHMoreRep{1}{1}} /\Lambda^3 + x_d (\ell H\ell H)_{\MHMoreRep{3}{1}} \cdot(\phi_1 \phi_2)_{\MHMoreRep{3}{1}} /\Lambda^3 +\text{h.c.}\;. 
\end{align}
The resulting mass matrices lead to TBM at LO, as in the Altarelli-Feruglio model\cite{Altarelli:2005uq}. The vacuum configuration 
$
\langle\chi\rangle =(v',v',v')^T, \langle\phi_1\rangle =\frac{1}{\sqrt{2}}(a,a,b,-b)^T, \langle\phi_2\rangle=\frac{1}{\sqrt{2}}(c,c,d,-d)^T
$
can be obtained as a natural solution of the most general scalar potential involving the given scalars as has been shown in \cite{Holthausen:2012pb}. This can be seen in the right-hand side of Figure \ref{fig:aligned-scan} where the distribution of opening angles between the two flavon $\chi$ and the effective flavon $(\phi_1 \phi_2)_{\MHMoreRep{3}{1}}$ is plotted for a random scan over order one potential parameters. We see that this vacuum configuration is obtained for a finite portion of parameter space,i.e. there is a phase with the TBM vacuum. This is to be contrasted with the potential without the alignment mechanism on the left-hand side of Figure \ref{fig:aligned-scan}.
\section{TMM Model based on $Q_8\rtimes A_4$}
Another possibility to generate deviations from TBM is the introduction of an additional flavon $\tilde{\xi}\sim \MHMoreRep{1}{2}$ which transforms as $\mathrm{i}$ under the auxiliary $Z_4$ and breaks the accidental symmetry U in the neutrino sector by the VEV $\MHbraket{\tilde{\xi}}=\tilde{w}$. This scalar can couple to neutrinos via the effective operator 
\begin{align}
\delta \mathcal{L}_\nu^{(7)} &= x_{c} (\ell H\ell H)_{\MHMoreRep{1}{2}} \tilde{\xi}^2 /\Lambda^3  +\text{h.c.}\;. 
\end{align}
that contributes to the neutrino mass matrix as
\begin{align}
\delta M_\nu &=\frac{v^2 }{2\sqrt{3} \Lambda^3 } \tilde{c} \left( \begin{array}{ccc}
1 & 0 & 0\\
0&\omega& 0 \\
0 &0&\omega^2 \\
\end{array}\right)
\end{align}
with $\tilde{c}=x_c \tilde{w}^2$. A correction of this type leads to the so-called tri-maximal mixing pattern, which gives a good fit to the neutrino mixing data and predicts a testable correlation $a\approx-\frac12 r\cos\delta$\cite{King:2011zj}  between the deviation from TBM in the mixing angles $\sin\theta_{13}=\frac{r}{\sqrt{2}}$ and $\sin\theta_{23}=\frac{1}{\sqrt{2}}\left(1+a\right)$.  The purpose of this section is to demonstrate that the TMM VEV configuration can also be naturally obtained in the $Q_8\rtimes A_4$ model. At the renormalizable level the scalar potential for $\tilde{\xi}$ is given by 
\begin{align}
V_{ \tilde{\xi}}(\tilde{\xi})=\mu^2_4 \tilde{\xi}^* \tilde{\xi}+\lambda_{ \tilde{\xi} }  (\tilde{\xi}^* \tilde{\xi})^2
\end{align}
and the cross-coupling terms are
\begin{align}
V_{ \mathrm{cross}}= {\tilde{\xi}}^* \tilde{\xi}\left(\zeta_{14} (\phi_1 \phi_1)_{\MHMoreRep{1}{1}} +\zeta_{24} (\phi_2 \phi_2)_{\MHMoreRep{1}{1}} +\zeta_{34} (\chi \chi)_{\MHMoreRep{1}{1}}  \right).
\end{align}
Note that there are no non-trivial contractions between $\tilde{\xi}$ and the other flavons at the renormalizable level. Note further that this is a direct consequence of the model and that no additional symmetries have been required. It is now trivial to confirm that the number of independent minimization conditions matches the number of VEVs and that the TMM VEV pattern can be naturally realized.

\section{Conclusion}
We have presented a solution to the vacuum alignment problem from group theory. The essential point is to engineer particle content and symmetries of the model in such a way that there emerges an accidental symmetry at the renormalizable level under which the flavons of the charged lepton and neutrino sectors transform independently. We have presented the smallest groups that extend $A_4$, $S_4$ and $T^\prime$ and have discussed a model based on the smallest extension of $A_4$, $Q_8\rtimes A_4$. We have furthermore outlined how one could extend the model to account for the large value of $\theta_{13}$.  

\section*{Acknowledgments}
The author acknowledges support from the International Max-Planck Research School  Precision Tests of Fundamental Symmetries(IMPRS-PTFS).

\bibliography{holthausen}
\bibliographystyle{apsrev4-1}


%% file: Papers/martinjung.tex

\chapter[Determining Weak Phases from $B\to J/\psi P$ Decays (Jung)]{Determining Weak Phases from $B\to J/\psi P$ Decays}
\vspace{-2em}
\paragraph{M. Jung} 
\paragraph{Abstract}
Penguin pollution in the ``golden mode'' $B_d\to J/\psi K$ has gained importance due to the apparent smallness of new physics effects, together with the outstanding precision expected from present and future collider experiments. A very recent analysis is presented, which yields a stronger bound for the maximal influence of penguin contributions than previous analyses and shows the corresponding uncertainty to be reducible with coming data.\footnote{The main part of this text has been published as part of an article in the proceedings of ``The XIth International Conference on Heavy Quarks and Leptons'' \cite{Jung:2012pz}.}

\section{Introduction}
Roughly 40 years after its proposal \cite{Kobayashi:1973fv}, the Kobayashi-Maskawa mechanism continues to give a consistent interpretation of the available data on flavour observables and CP violation. This fact is reflected in successful fits to the Unitarity Triangle (UT) \cite{Charles:2004jd,Ciuchini:2000de},
where, despite the precision data which has become available during the last decade, still no clear sign of physics beyond the Standard Model (SM) is seen. However, in the extraction of the CKM angle $\beta$ ($\phi_1$) tensions have been present (see e.g. \cite{Deschamps:2008de,Feldmann:2008fb,Bona:2009cj,Lunghi:2009ke,Charles:2011va}).  
The main deviation used to be between the extractions using $B\to J/\psi K$ on the one hand and $B\to \tau\nu$ on the other. However, this effect got very recently significantly reduced by the new Belle result on $B\to\tau\nu$ \cite{Adachi:2012mm}, although  the resulting world average remains above the SM expectation. 
Other puzzles, like the difference between $|V_{ub}|$ extracted from inclusive and exclusive decays or the largish $\epsilon_K$ remain, but are less significant. The important lesson from these observations is that new physics (NP) effects in the related observables have to be small. This, together with the bright experimental prospects, renders precision predictions for the involved observables particularly important. This implies an increased interest in the so-called ``penguin pollution'' in $B\to J/\psi K$, which is one of the key observables in these analyses.

\section{Penguin Pollution in the Golden Modes}

The impressive precision obtained for $\beta$ became possible due to the fact that in the ``golden mode'', $B_d\to J/\psi K_S$, explicit calculation of the relevant matrix elements can be avoided once subleading doubly Cabbibo suppressed terms are assumed to vanish \cite{Bigi:1981qs}, in combination with a final state with a very clear experimental signature. However, given the discussion above on the size of NP effects and the precision the LHC experiments and planned next-generation $B$ factories are aiming at for this mode and related ones, a critical reconsideration of the used assumptions is mandatory. Estimates yield corrections to the famous relation $S_{J/\psi K_S}=\sin\phi_d$ of the order $\mathcal{O}(10^{-3})$, only \cite{Boos:2004xp,Li:2006vq,Gronau:2008cc}; 
it is, however, notoriously difficult to actually calculate the relevant matrix elements, and non-perturbative enhancements cannot be excluded.

To include these subleading contributions, the size of their matrix elements relative to the leading one has to be determined. An explicit calculation still does not seem feasible to an acceptable precision for the decays in question, which is why typically symmetry relations are used\footnote{For an approach using theory input to extract the $B_s$ mixing phase, see \cite{Lenz:2011zz}.}, i.e. $SU(3)$, relating up, down and strange quarks, or its subgroup $U$-spin, including only down and strange quark. These allow for accessing the unknown matrix element ratios via decays where their relative influence is larger (``control modes'') \cite{Fleischer:1999nz,Ciuchini:2005mg,Ciuchini:2011kd,Faller:2008zc,Faller:2008gt}. 
This method has the advantage of being a completely data-driven method, and the resulting value for the $B$ mixing phase provides improved access to NP in mixing once the SM value of this phase is determined independently. 

The main limitations of that approach were firstly the limited data for the control modes, as their rate is suppressed by $\lambda^2\sim5\%$ compared to the one of $B\to J/\psi K$, and secondly corrections to the symmetry limit. The first issue was already rendered less severe by recent data from CDF and LHCb \cite{Aaltonen:2011sy,Aaij:2012di,Aaij:2012jw} 
and will be resolved by LHC in combination with the planned Super Flavour Factories (SFF).  The second was addressed by a recent paper \cite{Jung:2012mp}. Here the idea is to include the symmetry-breaking corrections in a model-independent manner on a group-theoretical basis (for earlier applications of this method see e.g. \cite{Savage:1991wu,Gronau:1995hm,Grinstein:1996us,Jung:2009pb}). 
Extending furthermore the symmetry group from $U$-spin (used in \cite{Fleischer:1999nz,Ciuchini:2005mg,Ciuchini:2011kd,Faller:2008zc,Faller:2008gt}) 
to full $SU(3)$ then allows to relate a sufficiently large number of decay modes (the full set of $B\to J/\psi P$ modes, with $B\in\{B_u,B_d,B_s\}$ and $P\in \{\pi^+,\pi^0,K^+,K^0,\bar{K}^0\}$) to determine the parameters for the $SU(3)$ breaking as well as the penguin pollution from the fit, using mild assumptions which are mostly testable with data \cite{Jung:2012mp}.

Applying this method to presently available data for these decays \cite{Aaltonen:2011sy,Aaij:2012di,Aaij:2012jw,
Amhis:2012bh,Beringer:1900zz} 
shows clearly the importance of $SU(3)$-breaking effects. Even when allowing for huge values of the penguin parameters, the fit in the $SU(3)$ limit yields $\chi^2_{\rm min}/{\rm d.o.f.}=22.3(23.9)/5$, where the first number corresponds to using the former world average for the rate of $B^-\to J/\psi \pi^-$ (``dataset 1''), and the second to the new LHCb result (``dataset 2''), which yields a value about 3 standard deviations away from the former. This is why they are compared explicitly instead of averaging the results. 
Importantly, correlations to the measured branching ratios drive the shift $\Delta S=-S(B\to J/\psi K_S)+\sin\phi_d$ to relatively large values in this case, in the opposite direction of the tension observed in the UT fit.
It is furthermore interesting to note that 
the inclusion of neglected contributions 
does not 
improve the fit, confirming our choice to set them to zero. 
The same is true for factorizable $SU(3)$-breaking corrections, which were included in the fit for comparison purposes, only. 

In a next step, $SU(3)$-breaking contributions are included in the fit, while neglecting penguin pollution. This fit works rather well, yielding  $\chi^2_{\rm min}=9.4(6.0)$ for 7 effective degrees of freedom\footnote{\emph{Effective degrees of freedom} are defined here as number of observables minus the number of parameters which are effectively changing the fit.}.
The best fit point yields a ratio of the larger $SU(3)$-breaking matrix element with the leading one of $19(24)\%$, which is perfectly within the expectations for this quantity. Therefore the data can be explained with the expected amount of $SU(3)$ breaking and small penguin contributions.

Performing the full fit with both additional contributions, the fit improves slightly, to $\chi^2_{\rm min}=2.8(2.3)$ for 3 effective degrees of freedom, when we refrain from applying strong restrictions on the parameter values\footnote{We do not allow for ``exchanging roles'' though, i.e. we continue to assume the leading matrix element to be the one in the $SU(3)$ limit with no penguin contributions.}. In this fit, the $SU(3)$-breaking parameters allow to accommodate the pattern of branching ratios, while the penguin contributions are mainly determined by the CP and isospin asymmetries. The central values of the penguin parameters still tend to larger values than theoretically expected. This is not surprising, given the fact that the isospin asymmetry in $B\to J/\psi K$ has a central value about ten times larger than the naive expectation, however with large uncertainties. The corresponding branching ratios are predicted to be around one standard deviation higher (lower) for $\bar{B}^0\to J/\psi \bar{K}^0\,(B^-\to J/\psi K^-)$, making an  additional measurement of their ratio important, which correspondingly is predicted to take a significantly different central value than the one presently measured.
Restricting the fit parameters  to the expected ranges, i.e. at most an $SU(3)$ breaking of $r_{SU(3)}=40\%$, and a ratio of the penguin matrix element with the leading one of $r_{\rm pen}=50\%$, shows a preference for dataset 2, where the minimal $\chi^2$ remains basically unchanged, while for dataset 1 it approximately doubles. The new result for $BR(B^-\to J/\psi\pi^-)/BR(B^-\to J/\psi K^-)$ obtained by LHCb seems therefore favoured by this fit. While  it is too early to draw conclusions, this observation demonstrates once more the importance of precise branching ratio measurements in this context.

For both datasets, the shift $\Delta S$ now tends again to positive values, thereby lowering the corresponding tension in the UT fit. It is however still compatible with zero, in agreement with the above observation of a reasonable fit without penguin terms. The obtained ranges read
\begin{eqnarray}
\Delta S_{J/\psi K}^{\rm set\,1} &=& [0.001,0.005] ([-0.004,0.011])\,,\mbox{\quad and }\\
\Delta S_{J/\psi K}^{\rm set\,2} &=& [0.004,0.011] ([-0.003,0.012])\,,
\end{eqnarray}
for $68\%$ ($95\%$)~CL, respectively, where the preferred sign change compared to the $SU(3)$ limit is due to relaxed correlations between $S(B\to J/\psi \pi^0)$ and the branching ratios in the fit, because of the additional contributions. This underlines the necessity to treat $SU(3)$ breaking model-independently.
Note that $S(B_d\to J/\psi\pi^0)$ is predicted to lie below the present central value of the measurement, thereby supporting the Belle result \cite{Lee:2007wd} over the BaBar one \cite{Aubert:2008bs}, which indicates a very large value for this observable. 
\begin{figure}
\begin{center}
\includegraphics[width=7.2cm]{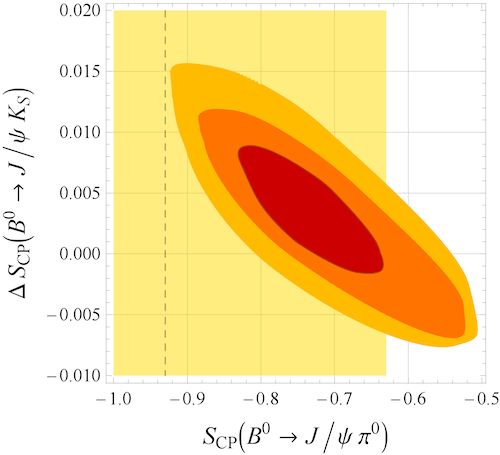}\hfill \includegraphics[width=7.2cm]{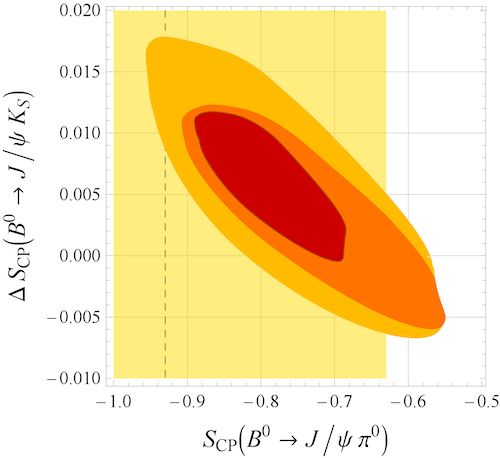}
\caption{\label{fig::resultsfullfita} Fit results for datasets 1 (left) and 2 (right), for $\Delta S$ versus $S_{\rm CP}(B^0\to J/\psi \pi^0)$, including all available data. The inner areas correspond to $68\%$ CL and $95\%$ CL with $r_{SU(3)}=40\%$ and $r_{\rm pen}=50\%$. The outer one is shown for illustration purposes, only, and corresponds to $95\%$ CL when allowing for up to $r_{SU(3)}=60\%$ and $r_{\rm pen}=75\%$. The light yellow area indicates the 2-$\sigma$ range of the $S(B^0\to J/\psi \pi^0)$ average, the dashed line its central value. Figure taken from  \cite{Jung:2012mp}.}
\end{center}
\end{figure}
These findings are illustrated in Fig.~\ref{fig::resultsfullfita}.
The same fit allows to predict the so far unmeasured CP asymmetries in $B_s\to J/\psi K$ decays: their absolute values lie for both datasets below approximately $30\%$ at $95\%$~CL. On the one hand this allows for a crosscheck for the description in the above framework, on the other hand it is clear that a measurement with a precision of $\sim10\%$ will already yield a significant additional constraint on the model parameters. Especially the dependence on the (already weak) theory assumptions will be further reduced with such a measurement \cite{Jung:2012mp}.

The mixing phase is extracted as $\phi_d^{\rm fit}=0.74\pm0.03$ (equal for both datasets), which is to be compared with $\phi_{d,{\rm naive}}^{\rm SM}=0.73\pm0.03$ when using the naive relation without penguin contributions. The inclusion of the correction therefore yields the same precision, but induces a shift of the central value. The same is true for future data, as shown in \cite{Jung:2012mp} by the consideration of several scenarios corresponding to additional data from the LHCb and SFF experiments. This implies the corresponding error to be reducible, and therefore ensures the golden mode to keep its special position among flavour observables.

In principle, the same approach can be used to constrain penguin pollution in the other ``golden mode'', $B_s\to J/\psi\phi$. Technical difficulties are the fact that the $\phi$ meson does not belong to a single representation, and the more complicated structure of the final state. The latter is also complicating the experimental analysis; so far only the $B\to J/\psi K^*$ decays have been measured, which are $b\to s$ transitions as well. If the $b\to d$ modes can be measured sufficiently precise to control the penguin pollution as well as the $SU(3)$ breaking is subject to further studies.

\section{Conclusions}
CP violation studies in heavy meson systems remain a very active field, and one of the main paths to discover NP. The general picture remains consistent with the KM mechanism as the only source of low-energy CP violation; in fact, the fits have improved very recently due to a new measurement for $B\to\tau\nu$. 

This -- in many ways unexpected -- situation requires a more precise knowledge of the corresponding SM expectations, as potential small NP contributions will compete with subleading SM ones. The ``golden modes'' $B_d\to J/\psi K$ and $B_s\to J/\psi\phi$ are examples where subleading contributions can affect the extraction of the mixing phase. For $B_d\to J/\psi K$, a new approach to control them has been advocated, allowing to take into account $SU(3)$ corrections model-independently, which were shown to affect the procedure severely. The main result is a new limit, $|\Delta S_{J/\psi K}|\lesssim 0.01$ ($95\%$~CL), which can additionally be improved by coming data.

In conclusion, the apparent smallness of NP effects in flavour observables poses a challenge to both theory and experiment. On the experimental side it is met by several high-luminosity collider experiments, both running and under construction, allowing for unprecedented precision. Also on the theory side the challenge is answered, by new strategies and adapting known ones to higher precision. Together, these developments make for an exciting way ahead.

\section*{Acknowledgements}
This work is supported by the Bundesministerium f\"ur Bildung und Forschung (BMBF).

\bibliography{martinjung}
\bibliographystyle{apsrev4-1}

%% file: Papers/AvihayKadosh.tex

%
%
%
%
%
%

\chapter[RS-A$_4$, $\theta_{13}$ and $\mu\to e,3e$ (Kadosh)]{RS-A$_4$, $\theta_{13}$ and $\mu\to e,3e$}
\vspace{-2em}
\paragraph{A. Kadosh}
\paragraph{Abstract}
In the first FLASY meeting I have introduced the RS-$A_4$ model,
aimed at a simultaneous explanation of quark and lepton masses and
mixings. The model was shown to produce realistic fermion masses
and mixing patterns and has been tested "successfully" against
various phenomenological constraints coming from electroweak
precision measurements (EWPM), rare decays and more. These
constraints allowed for a relatively low Kaluza-Klein (KK) mass
scale around 1.5TeV The recent measurement of
$\theta_{13}\simeq\theta_{C}/\sqrt{2}$ by RENO and Daya Bay
introduce a more stringent test to the model's ability to
naturally generate large enough deviations from tribimaximal (TBM)
mixing. We repeat the preliminary analysis and consider all higehr
order corrections to the PMNS matrix. Most importantly, we show
that the most significant constraint on RS-$A_4$ and similar
constructions come from the measurement of $BR(\mu\to e,3e)$ by
SINDRUM and the  future sensitivity of upcoming experiments

\section{Introduction}
Recently  we have proposed a model \cite{A4Warped}  based on a
bulk $A_4$ flavor symmetry \cite{a4} in warped geometry \cite{RS},
in an attempt to account for the  hierarchical charged fermion
masses, the hierarchical mixing pattern in the quark sector and
the large mixing angles and the mild hierarchy of masses in the
neutrino sector. In analogy with a previous RS realization of
A$_{4}$ for the lepton sector \cite{Csaki:2008qq}, the three
generations of left-handed quark doublets are unified into a
triplet of $A_4$; this assignment forbids tree level  FCNCs driven
by the exchange of KK gauge bosons. The scalar sector of the
RS-A${}_4$ model consists of two bulk flavon fields, in addition
to a bulk Higgs field. The bulk flavons transform as triplets of
$A_{4}$, and allow for a complete
 "cross-talk" \cite{Volkas} between the $A_{4}\to Z_{2}$
spontaneous symmetry breaking (SSB) pattern associated with the
heavy neutrino sector - with scalar mediator  peaked towards the
UV brane - and the $A_{4}\to Z_{3}$ SSB pattern associated with
the quark and charged lepton sectors - with scalar mediator peaked
towards the IR brane - and allows to obtain realistic masses and
almost realistic mixing angles in the quark sector. A bulk
custodial symmetry, broken differently at the two branes
\cite{Agashe:2003zs}, guarantees the suppression of large
contributions to electroweak precision observables
\cite{Carena:2007}, such as the Peskin-Takeuchi $S$, $T$
parameters. However, the mixing  between zero modes of the 5D
theory and their Kaluza-Klein (KK) excitations -- after 4D
reduction -- may still cause significant new physics (NP)
contributions to SM suppressed flavor changing neutral current
(FCNC) processes.

\noindent In general, when no additional flavor symmetries are
present and the 5D Yukawa matrices are anarchical, FCNC processes
are already generated at the tree level  by a KK gauge boson
exchange \cite{Agashe:2004cp}. Stringent constraints on the KK
scale come from the $K^{0}-\overline{K^{0}}$ oscillation parameter
$\epsilon_{K}$, the radiative decays $b\to s(d)\gamma$
\cite{Agashe:2004cp,Azatov}, the direct CP violation parameter
$\epsilon^\prime/\epsilon_K$ \cite{IsidoriPLB}, and especially the
neutron electric dipole moment (EDM) \cite{Agashe:2004cp}, also in
the presence of an RS-GIM suppression mechanism
\cite{rsgim1,Cacciapaglia:2007fw}. Conclusions may differ if a
flavor pattern of the Yukawa couplings is assumed to hold in the
5D theory due to bulk flavor symmetries. They typically imply an
increased alignment between the 4D fermion mass matrix and the
Yukawa and gauge couplings, thus suppressing the amount of flavor
violation induced by the interactions with KK states.

\noindent  The most relevant consequence of imposing an $A_4$
flavor symmetry is  the degeneracy of the left-handed fermion bulk
profiles $f_Q$, i.e. $diag(f_{Q_1,Q_2,Q_3})=f_Q\times \bf{1}$. In
addition, the distribution of phases, CKM and Majorana-like, in
the mixing matrices might induce zeros in the imaginary components
of the Wilson coefficients contributing to CP violating quantities
\cite{A4CPV}.

An additional important source of flavor violation arise from
anomalous off diagonal $Z$ couplings, which are a result of the KK
mixing of fermions and gauge bosons after electroweak symmetry
breaking (EWSB). In the following we study Tree level $Z$ exchange
contributions to lepton flavor violating (LFV) processes $\mu\to
e,3e$. In addition, we compare the RS-$A_4$ predictions for
$\theta_{13}$ to the new global fits, including RENO and Daya Bay.

\section{Higher order corrections to the PMNS matrix and
$\theta_{13}$}

The new measurements of $\theta_{13}$ by RENO and Daya Bay allows
one to rule out $\theta_{13}=0$ with a significance of more than
$10\sigma$. This situation ``poses a threat" to every model
predicting TBM at leading order and deviations from TBM should be
thoroughly tested. Within the RS-$A_4$ model, the dominant
cross-brane operator inducing deviations from TBM \cite{TBM}
mixing is $\overline{\ell}_L\chi H\nu_{R}$ \cite{A4Warped}. If the
bulk mass of $\chi (A_4\to Z_2)$ is vanishing, this operator is
suppressed only by $\epsilon^{\chi}\sim 0.05$, compared to the
leading order Dirac mass term.

We now recall   that in general the  effective Majorana Mass
matrix is a $3\times 3$ complex symmetric matrix and thus contains
$12$ parameters. These parameters are the $3$ masses, the $3$
mixing angles and $6$ phases, out of which $3$ can be absorbed in
the neutrino fields and the remaining are $2$ Majorana and $1$ KM
phase. Notice also that, in general, the various couplings are
complex. Thus, considering only the contributions of operators of
the form $\overline{\ell}_L\chi^m H\nu_{R}$, the
left-diagonalization matrix is now corrected to \cite{A4Warped}:
\begin{equation}
V_L^{\nu} = \left( \!\!\begin{array}{ccc} 1 & 0 & 0 \\ 0 & 1 & 0 \\
0 & 0 & e^{i\delta}
\end{array} \!\!\right)
\left( \!\!\begin{array}{ccc} 1/2(\sqrt{2}-\epsilon_{\chi}^2) & \ 0 & -(1/\sqrt{2}+\epsilon_{\chi}) \\
0 & \ 1 & \ 0
\\ 1/\sqrt{2}+\epsilon_{\chi} & \ 0 & \ 1/2(\sqrt{2}-\epsilon_{\chi}^2)
\end{array} \!\!\right)
\left(\!\! \begin{array}{ccc} e^{i\alpha_1} & 0 & 0 \\ 0 &
e^{i\alpha_2} & 0 \\ 0 & 0 & e^{i\alpha_3}
\end{array} \!\!\right),
\label{epschidef}
\end{equation}
where $\epsilon_{\chi} \sim{\cal O}(\chi_0/\Lambda_{5D}^{3/2})$
stands for contributions from $\epsilon_{13}^{\chi}$ and
$\epsilon_{11,22}^{\chi}$ in \cite{A4Warped} and we have omitted
terms of $\mathcal{O}(\epsilon_{\chi}^3)$ and higher. The phases
$\alpha_i$ can be absorbed in a rotation of the neutrino fields,
while the KM phase $\delta$, given by
\begin{equation}
\delta = {\rm Arg}(\tilde{M} + |\tilde{M}|\epsilon^{\chi *}_{11}))
- {\rm Arg}(\tilde{M} +
|\tilde{M}|\epsilon^{\chi}_{11}),\label{NuCP}
\end{equation}
will contribute to $CP$ violation in neutrino oscillations. The
MNSP matrix at $\cal{O}(\epsilon_{\chi})$ becomes:
\begin{eqnarray}
\noindent V_{MNSP}\!\!&=&\!\!U(\omega)^{\dagger} V_L^{\nu} \nonumber\\
\!\!&=& \!\!\frac{1}{\sqrt{6}} \left( \!\!\begin{array}{ccc}
1+e^{i\delta}+\epsilon_{\chi}(1-e^{i\delta})) &\!\! \sqrt{2} & e^{i\delta}-1-(e^{i\delta}+1)\epsilon_{\chi}^*) \\
1+\omega e^{i\delta}+(\omega e^{i\delta}-1)\epsilon_{\chi}) &
\!\!\sqrt{2}\omega^2 & \omega e^{i\delta}-1
-(\omega e^{i\delta}+1)\epsilon_{\chi}^*) \\
1+\omega^2 e^{i\delta}+(\omega^2 e^{i\delta}-1)\epsilon_{\chi} &
\!\!\sqrt{2}\omega & \omega^2 e^{i\alpha}-1-(\omega^2
e^{i\delta}+1)\epsilon_{\chi}^*)
\end{array} \!\!\right)\!.\nonumber\\\label{Nufix}
\end{eqnarray}
\noindent The middle column does not receive corrections. A non
zero $\theta_{13}$ is generated, and $\theta_{23}$ deviate from
its maximal value. Defining $\theta=\pi/4+\epsilon_{\chi}$, the
Jarlskog invariant turns out to be
\begin{equation}
{\rm Im}[V_{11}\, V_{12}^*\, V_{21}^*\, V_{22}] =
\frac{\sqrt{3}}{18} \left(\cos 2\theta - \sin 2\theta \sin\delta
\right)\,,
\end{equation}
where the $V_{ij}$ denote the entries of $V_{MNSP}$.

\begin{center}
\begin{figure}[h!]
\includegraphics[width=7.5cm]{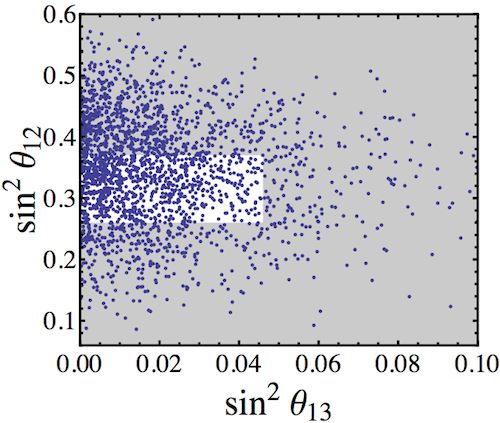}\quad\quad\!\!\!\!
\includegraphics[width=7.5cm]{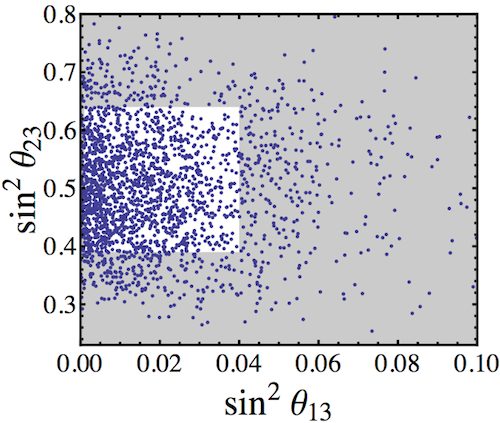}\caption{\it
Model predictions for $\theta_{13}$ vs. $\theta_{12}$ (left) and
$\theta_{23}$ (right) including all dominant higher order and
cross talk effects. The white rectangles represent the $3\sigma$
allowed regions from the global fit of
\cite{Fogli,Tortola}.}\label{Fig:AKNUAngles}
\end{figure}
\end{center}

 The global fits based on the recent indications of $\nu_\mu\rightarrow\nu_e$
appearance in the RENO, Daya Bay, T2K, MINOS and other
experiments, allow one to obtain a significance of $10\sigma$ for
$\theta_{13}>0$, with best fit points at around
$\theta_{13}\simeq0.15$, depending on the precise treatment of
reactor fluxes  \cite{Fogli,Tortola}. We wish the RS-$A_4$ higher
order corrections to the PMNS matrix to be such that the new fits
are still ``accessible" by a significant portion of the model
parameter space.

We are able to obtain analytic expressions for the corrected
diagonalization matrices of both charged leptons and neutrinos,
considering all dominant NLO effects. The resulting expressions
are incredibly long and depend on the $\tilde{x}_i^e,
\tilde{y}_i^e$, $y_\nu^{H\chi}$, $y_\nu^{\chi^2}$  parameters and
$C_\chi$ \cite{A4Warped}, which is also constrained by the quark
sector. Most importantly, these results do not depend on the LO
Yukawa couplings (Form diagonalizable LO rotation matrices). We
write below  approximate expressions for  $\theta_{12,13,23}$,
considering the dominant effects in the neutrino and charged
lepton sectors, parameterized by $(\delta,\epsilon_\chi\sim0.06)$
and $(\tilde{x}_i^\ell,\tilde{y}_i^\ell,\lambda_\ell\sim0.05)$,
respectively.

\begin{eqnarray}
\theta_{13}^{NLO}&\simeq&\frac{e^{i\delta}-1}{\sqrt{6}}-\frac{\epsilon_\chi}{\sqrt{3}}+\frac{1-\omega
e^{i\delta}
}{\sqrt{6}}(\tilde{x}_2^\ell+\tilde{y}_2^\ell+\omega\tilde{x}_3^\ell+\omega\tilde{y}_3^\ell)\lambda_\ell,\nonumber\\
\theta_{23}^{NLO}&\simeq&\frac{\omega
e^{i\delta}-1}{\sqrt{6}}-\frac{\epsilon_\chi}{\sqrt{3}}+\frac{e^{i\delta}-1}{\sqrt{6}}(\tilde{x}_2^{\ell
*} +\tilde{y}_2^{\ell
*})\lambda_\ell+\frac{1-\omega^2e^{i\delta}}{\sqrt{6}}(\tilde{x}_3^{\ell
*}+\omega^2\tilde{y}_3^{\ell *})\lambda_\ell,\nonumber\\
\theta_{12}^{NLO}&\simeq&\frac{1}{\sqrt{3}}-\frac{\omega^2}{\sqrt{3}}(\tilde{x}_2^\ell
+\tilde{y}_2^\ell)\lambda_\ell-\frac{\omega}{\sqrt{3}}(\tilde{x}_3^\ell+\tilde{y}_3^\ell)
\lambda_\ell.\label{NUexpressions} \end{eqnarray}

 We performed a scan over all NLO Yukawa couplings
in the range $[0.3,3]$ and with random complex phases. In
Fig.~\ref{Fig:AKNUAngles} we present the model predictions for
$\sin^2\theta_{13}$ vs. $\sin^2\theta_{12}$ (left) and
$\sin^2\theta_{23}$ (right) for a set of 3000 randomly generated
points, with the $3\sigma$ allowed ranges of \cite{Fogli,Tortola}
depicted as white rectangles.\\ We realize that the RS-A$_4$
predictions significantly overlap with the allowed ranges for the
neutrino mixing angles, which (re-)demonstrates the viability of
models predicting TBM at LO.

\section{Anomalous Z couplings and LFV}
As stated in the introduction the main source of charged lepton
flavor violation (cLFV) in RS-$A_4$ is anomalous Z couplings,
induced by KK mixing and generating Tree level Z exchange
contributions to $\mu\to e,3e$, $B_s\to\mu^+\mu^-$ and more. While
the effect of EWSB on the mixing of the Z boson with its KK
partners and those of the (custodial) $Z^\prime$, can be studied
directly from the equations of motion in the vicinity of an EWSB
IR boundary condition, the KK mixing of fermions has to be studied
directly from the mass matrix.

\begin{center}
\begin{figure}[h!]
\includegraphics[width=7.5cm]{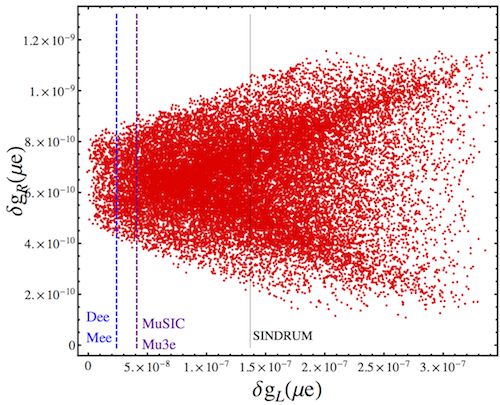}\quad\quad\!\!\!\!
\includegraphics[width=7.7cm]{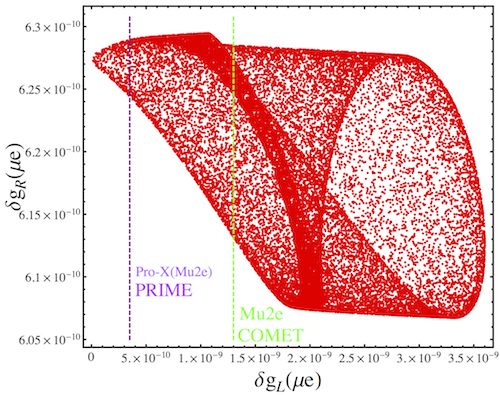}\caption{{\small \it
The RS-A$_4$ predictions for the anomalous  $Z\mu e$ couplings (LH
and RH) in the presence (left) or absence (right) of cross brane
interactions. Each point represents the contributions coming from
both gauge boson mixing and KK fermion mixing.  The dashed lines
represent the maximum sensitivities of past, present and future
LFV experiments taking place at PSI, FERMILAB and
J-PARC.}}\label{Fig:AKLFVscatter}
\end{figure}
\end{center}

To account for overlap effects and illustrate the flavor patterns
generated in RS-A$_4$ we write the LO  mass matrix for the first
generation in the down-type sector following
\cite{Azatov,IsidoriPLB}, including the zero modes and first level
KK modes and overlap effects \cite{A4CPV}
\begin{equation}\frac{\hat{\mathbf{M}}_\ell^{KK}}{(M_{KK})}=
\left(\begin{array}{c}
\bar{\ell}_L^{e(0)}\\\bar{e}_L^{(1^{--})} \\ \bar{\ell}^{e(1)}_L\\
\bar{\tilde{e}}_L^{(1^{+-})} \end{array} \right)^T \!\!\left(
\begin{array}{cccc}
\breve{y}_ef^{-1}_\ell f^{-1}_e r_{00} x & 0 &
\breve{y}_ef^{-1}_\ell r_{01} x &
\breve{y}_\nu f^{-1}_\ell r_{101} x \\
 0 &  \breve{y}_e^*r_{22} x & 1 & 0 \\
 \breve{y}_ef^{-1}_e r_{10} x & 1 &  \breve{y}_er_{11} x &  \breve{y}_\nu r_{111} x \\
 0 &  \breve{y}_\nu ^*r_{222} x & 0 & 1
\end{array}
\right)\left(\begin{array}{c}
e_R^{(0)}\\\ell_R^{e(1^{--})} \\ e^{(1)}_R\\
\tilde{e}_R^{(1^{-+})} \end{array} \right),\label{M4KK}
\end{equation}
where we factorized a common KK mass scale $M_{KK}$,
$\breve{y}_{e,\nu}\equiv 2y_{e,\nu}v_\Phi^{4D}e^{k\pi R}/k$ and
the perturbative expansion parameter is defined as $x\equiv
v/M_{KK}$ \cite{A4CPV}. In the above equation the various $r$'s
denote the ratio of the bulk and IR localized effective couplings
of the modes corresponding to the matrix element in question. For
simplicity, we define $r_{111}\equiv r_{11^{-+}}$, $r_{101}\equiv
r_{01^{-+}}$, $r_{22}\equiv r_{1^-1^-}$, $r_{222}\equiv
r_{1^{-}1^{+-}}$ and the notation for the rest of the overlaps is
straightforward. The corresponding Yukawa matrix,
$\hat{Y}^{e}_{KK}$  is obtained by simply eliminating $x$ and the
1's from the above matrix and it leads the  flavor structure of
the contributions of $(++)$, $(--)$, $(+-)$ and $(-+)$ KK modes.
The full three generation mass matrix will be $12\times 12$ and of
similar structure, which is modified mainly by the $A_4$ flavor
structure. Unfortunately, it can only be diagonalized numerically,
due to its dimension and the large number of input parameters. The
reason fermion KK mixing is so important for off diagonal $Z$
couplings is the presence of ``fake" custodial partners, which are
the $SU(2)_{R}$ partners of $(e_R,\mu_R,\tau_R)$ with $(-+)$ and
$(+-)$ boundary conditions. Recall that a 5D fermion corresponds
to two chiral fermions in 4D. As a result we will have LH states
which are charged under $Z$ as RH states and vice versa. Thus, the
$Z$ coupling matrix, which was proportional to the identity in the
zero mode approximation, contains instead both types of (diagonal)
entries $g_L^Z,g_R^Z$ and will thus acquire non vanishing off
diagonal elements, once rotated to the common mass basis of KK and
zero mode fermions. The effect of EWSB translates via the EOM to a
distortion of the $Z$ wave function near the IR brane, which
generate subdominant non universality in the interaction basis $Z$
coupling matrices. We performed a matching to the operators
contributing to $\mu\to e,3e$, scanned over the various parameters
(LO and NLO Yukawa couplings) and included both brane peaked and
cross-brane 5D interactions, to obtain $\delta g_{L,R}^{\mu eZ}$
for a random sample of 40000 points. Since $BR(\mu\to e,3e)\propto
(g_{L,R}^{Z\mu e})^2$ and $\delta g_L$ generally dominates, we can
plot the constraints coming from the various cLFV experiments as
vertical lines in the $\delta g_L^{\mu eZ}-\delta g_R^{\mu eZ}$
plane. The results are depicted in Fig.~\ref{Fig:AKLFVscatter}.
 When cross brane effects are neglected, corrections to
the TBM pattern can come only from higher order corrections to the
heavy Majorana mass matrix, for which achieving
$\theta_{13}\sim\theta_C/\sqrt{2}$ is slightly less natural. The
bound from SINDRUM \cite{sindrum} ``eliminates" around 50\% of the
cross-brane RS-A$_4$ setup and is easily and naturally satisfied.
We conclude that if no $\mu\to e,3e$ events will be actually
observed in Mu3e, MuSIC and Dee-Mee the ``cross talk" RS-$A_4$
model will be severely constrained and less appealing. The same
situation holds for the previous (brane localized) realizations of
RS-$A_4$, this time with the Mu2e, COMET and PRIME future
experiments.

\section{Conclusion}
 The RS-A$_4$ setup is one of the most elegant ways to simultaneously
 address the gauge and flavor hierarchy problems and generate
 realistic neutrino masses and mixing patterns. Furthermore, this
 construction was shown to relax the "little" CP problem
 associated with flavor anarchic RS setups. The most significant
 constraints come from the $Zbb$ anomalous coupling and $Z$
 mediated contributions to $\mu\to e,3e$. Hopefully, the situation
 of such models will be further clarified in the near future cLFV
 experiments mentioned above, although their results can serve
 only as an indirect test. The search for unique collider and
 experimental signatures is on the go, but the anticipated strength of the
 cLFV constraints still remains a very appealing feature of the
 RS-A$_4$ setup.

\section*{Acknowledgments}
I would like to thank the organizers of FLASY 2012 for the lovely
hospitality and atmosphere in Dortmund.

%% file: Papers/jernejkamenik.tex
\newcommand{\JFKslashed}{/ \hspace{-0.2cm}}

\chapter[Implications of $\Delta A_{CP}$ Measurement for New Physics (Kamenik)]{Implications of $\Delta A_{CP}$ Measurement for New Physics}
\label{chap:kamenik}
\vspace{-2em}
\paragraph{J. F. Kamenik}
\paragraph{Abstract}
I review the implications of recent measurements of CP violation in $D$ meson decays in the context of standard model extensions. 
Using effective theory methods, one can derive significant constraints on the possible non-standard contributions from measurements of $D^0-\bar D^0$ mixing and CP violation in kaon decays ($\epsilon'/\epsilon$). Due to an approximate universality of CP violation in  new physics scenarios which only break the $SU(3)_Q$ flavor symmetry of the standard model kinetic Lagrangian, such contributions are particularly constrained by $\epsilon'/\epsilon$. Explanations of the observed effect within several explicit well-motivated new physics frameworks are briefly discussed. Finally I comment on possible future experimental tests able to distinguish standard vs. non-standard explanations of the observed CP violation in the charm sector.

\section{Introduction}

CP violation in charm provides a unique probe of New Physics (NP). Not only is it sensitive to NP in the up sector, in the Standard Model (SM) charm processes are dominated by two generation physics with no hard GIM breaking, and thus CP conserving to first approximation. Until very recently, the common lore was that ``any signal for CP violation in charm would have to be due to NP". The argument was based on the fact the in the SM and in the heavy charm quark limit $m_c \gg \Lambda_{\rm QCD}$, CP violation in neutral $D$ meson mixing enters at $\mathcal O(|\lambda_b/\lambda_s|) \sim 10^{-3}$ ($\lambda_q \equiv V_{cq}V^*_{uq}$), while CP violating contributions to singly Cabibbo suppressed $D$ decays only appear at $\mathcal O(|\lambda_b/\lambda_s| \alpha_s(m_c)/\pi ) \sim 10^{-4}$~\cite{JFKGrossman:2006jg}.

\section{CP Violation in $D$ Decays: Experiment vs. SM Expectations}

CP violation in neutral $D$ meson decays to CP eigenstates $f$ is probed with  time-integrated CP asymmetries $(a_f)$. These can arise from interferences between decay amplitudes with non-zero CP odd ($\phi_f$) and even ($\delta_f$) phase differences  
\begin{equation}
a_f^{\rm dir} = -\frac{2 {r_f} \sin \delta_f \sin \phi_f}
  {1+2 r_f \cos \delta_f \cos \phi_f + r_f^2}\,,
\label{JFK_eq:adf}
\end{equation}
where $r_f$ is the absolute ratio of the two interfering amplitudes.  
Recently both the LHCb~\cite{JFKlhcb} and CDF~\cite{JFKCDF10784} collaborations reported evidence for a non-zero
value of the difference $\Delta a_{CP} \equiv a_{K^+ K^-} - a_{\pi^+ \pi^-}$. Combined with other measurements of these CP asymmetries~\cite{JFKHFAG}, the present world average is
\begin{equation}
\Delta a_{CP} = -(0.67\pm 0.16)\%\,.
\label{JFK_eq:acpExp}
\end{equation}

\begin{figure}
\begin{center}
  \includegraphics[width=0.6\textwidth]{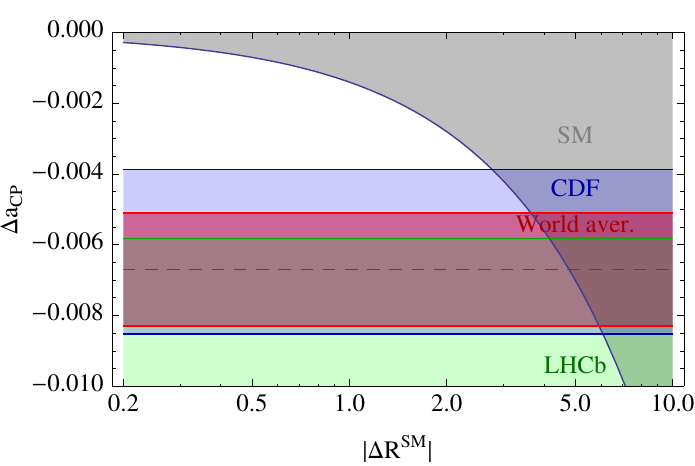}
\caption{ Comparison of the experimental $\Delta a_{CP}$ values with
the SM reach as a function of $|\Delta R^{\rm SM}|$. See text for details.}
\label{JFKfig:1}       
\end{center}
\end{figure}

This observation calls for a reexamination of theoretical expectations within the SM. 
Using CKM unitarity ($\sum_{q=d,s,b}\lambda_q = 0$), the relevant $D^0 \to K^+ K^-, \pi^+ \pi^-$ decay amplitudes ($A_{K,\pi}$) can be written compactly as
$A_{K,\pi} = \lambda_{s,d}(A_K^{s,d} - A_K^{d,s}) - \lambda_b  A_K^{d,s}$. In the isospin limit the two different isospin amplitudes in the first term provide the necessary condition for non-zero $\delta_{K,\pi}$, while $\phi^{\rm SM}_{K,\pi} = {\rm Arg}(\lambda_b/\lambda_{s,d}) \approx \pm 70^\circ$. On the other hand $r_{K,\pi}$ are controlled by  the CKM ratio $\xi= |\lambda_b/\lambda_s| \simeq |\lambda_b/\lambda_d| \approx 0.0007$. Parametrizing the remaining unknown hadronic amplitude ratios as $R_{K,\pi}^{\rm SM} \equiv -A_{K,\pi}^{d,s} / (A_{K,\pi}^{s,d} - A_{K,\pi}^{d,s})$, the SM contribution to $\Delta a_{CP}$ can be written as
\begin{equation}
\Delta a_{CP} \approx (0.13\%) {\rm Im} (\Delta R^{\rm SM})\,,
\end{equation}
where $\Delta R^{\rm SM} = R_K^{\rm SM} + R_\pi^{\rm SM}$\,. Comparison of this estimate with current experimental results is shown in Fig.~\ref{JFKfig:1}.
One observes that $|{\rm Im}(\Delta R^{\rm SM})| = \mathcal O (2-5) $ is needed to reproduce the experimental results in Eq.~\eqref{JFK_eq:acpExp}\,, in contrast to perturbative estimates in the heavy charm quark limit ($|R_{K,\pi}|\sim \alpha_s(m_c)/\pi \sim 0.1$) (see~\cite{JFKGrossman:2006jg} and the more recent analyses in Refs.~\cite{JFKotherSM1}). However, $\xi$ suppressed amplitudes in the numerator of $R_i$ cannot be constrained by rate measurements alone, and it has been pointed out a long time ago that ``$\Delta I=1/2$ rule" type enhancements are possible~\cite{JFKGolden:1989qx} (see also~\cite{JFKotherSM,JFKFeldmann:2012js}). Recently~\cite{JFKBrod:2011re}, an explicit estimate of potentially large $1/m_c$ suppressed contributions has been performed, yielding $\Delta a_{CP}^{\rm SM}$ $\lesssim 0.4~\%$\,. Although this is an order of magnitude above na\"ive expectations, the experimental value in Eq.~\eqref{JFK_eq:acpExp} cannot be reached.

\section{Implications of $\Delta a_{CP}$ for Physics Beyond SM}

In the following we will therefore assume the SM does not saturate the experimental value, leaving room for potential NP contributions. These can again be parametrized in terms of an effective Hamiltonian valid below the $W$ and top mass scales
\begin{equation}\label{JFK_eq:HNP}
\mathcal H^{\rm eff-NP}_{|\Delta c|=1} = \frac{G_F}{\sqrt 2} \sum_i C_i^{\rm NP(\prime)} \mathcal Q_i^{(\prime)}\,,
\end{equation}
where the relevant operators $\mathcal Q_i^{(\prime)}$ have been defined in~\cite{JFKDCPVD}.
Introducing also the NP hadronic amplitude ratios as $R^{{\rm NP},i}_{K,\pi} \equiv G_F \langle{K^+ K^-, \pi^+ \pi^-}|\mathcal Q^{(\prime)}_i|{D^0}\rangle / \sqrt 2 (A_{K,\pi}^{s,d} - A_{K,\pi}^{d,s})$ and writing $C_i^{\rm NP} = v_{\rm EW}^2 / \Lambda^2$, the relevant NP scale $\Lambda$ is given by
\begin{equation}
\frac{(10~{\rm TeV})^2}{\Lambda^2} = \frac{(0.61\pm 0.17)-0.12 {\rm Im}(\Delta R^{\rm SM})}{{\rm Im}(\Delta R^{{\rm NP},i})}\,.
\end{equation}

Comparing this estimate to the much higher effective scales probed by CP violating observables in $D$ mixing and also in the kaon sector, one first needs to verify, if such large contributions can still be allowed by other flavor constraints. Within the effective theory approach, this can be estimated via so-called ``weak mixing" of the effective operators.
In particular, time-ordered correlators of $\mathcal H^{\rm eff-NP}_{|\Delta c|=1}$ with the SM effective weak Hamiltonian can, at the one weak loop order, induce important contributions to CP violation in both $D$ meson mixing and kaon decays ($\epsilon'/\epsilon$). On the other hand, analogue correlators, quadratic in $\mathcal H^{\rm eff-NP}_{|\Delta c|=1}$ turn out to be either chirally suppressed and thus negligible, or yield quadratically divergent contributions, which are thus highly sensitive to particular UV completions of the effective theory~\cite{JFKDCPVD}. 

\subsection{Universality of CP Violation in $\Delta F=1$ processes}

The strongest bounds can be derived for a particular class of operators, which transform non-trivially only under the $SU(3)_Q$ subgroup of the global SM quark flavor symmetry $\mathcal G_F = SU(3)_Q \times SU(3)_U \times SU(3)_D$, respected by the SM gauge interactions. In particular one can prove that their CP violating contributions to $\Delta F=1$ processes have to be approximately universal between the up and down sectors~\cite{JFKUCPV}. Within the SM one can identify two unique sources of $SU(3)_Q$ breaking given by $\mathcal{A}_u \equiv (Y_u Y_u^\dagger)_{\JFKslashed{\mathrm{tr}}}$ and $\mathcal{A}_d \equiv (Y_d Y_d^\dagger)_{\JFKslashed{\mathrm{tr}}}$, where
$\JFKslashed{\mathrm{tr}}$ denotes the traceless part. Then in the two generation limit,  one can construct a single source of CP violation, given by $J\equiv i [\mathcal{A}_u,\mathcal{A}_d]$~\cite{JFKarXiv:1002.0778}. The crucial observation is that $J$ is invariant under $SO(2)$ rotations between the $\mathcal{A}_u$ and $\mathcal{A}_d$ eigenbases. Introducing now $SU(2)_Q$ breaking NP effective operator contributions  of the form $\mathcal  Q_L = \Big[(X_L)^{ij}\, \overline Q_i \gamma^\mu Q_j \Big] L_\mu$, where $L_\mu$ denotes a flavor singlet current, it follows that their CP violating contributions have to be proportional to $J$ and thus invariant under flavor rotations. The universality of CP violation induced by $\mathcal Q_L$ can be expressed explicitly as~\cite{JFKUCPV}
\begin{equation}\label{Uni2gen}
{\rm Im}(X^u_L)_{12} = {\rm Im}(X^d_L)_{12}\propto {\rm Tr} \left( X_L \cdot
J\right)\, .
\end{equation}

The above identity holds to a very good approximation even in the three-generation framework. In the SM, large values of $Y_{b,t}$ induce a $SU(3)/SU(2)$ flavor symmetry breaking pattern~\cite{JFKKagan:2009bn} which allows to decompose $X_L$ under the residual $SU(2)$ in a well defined way.  Finally, residual SM $SU(2)_Q$ breaking is necessarily suppressed by small mass ratios $m_{c,s}/m_{t,b}$, and small CKM mixing angles $\theta_{13}$ and $\theta_{23}$. 

The most relevant implication of Eq.~\eqref{Uni2gen} is that it predicts a direct correspondence between $SU(3)_Q$ breaking NP contributions to $\Delta a_{CP}$  and $\epsilon'/\epsilon$~\cite{JFKUCPV}. It follows immediately that stringent limits on possible NP contributions to the later, require $SU(3)_Q$ breaking contributions to the former to be below the per mile level (for $\Delta R^{{\rm NP},i}=\mathcal O (1)$).

~

The viability of the remaining 4-quark operators in $\mathcal H^{\rm eff-NP}_{|\Delta c|=1}$ as explanations of the $\Delta a_{CP}$ value in Eq.~\eqref{JFK_eq:acpExp},  depends crucially on their flavor and chiral structure. In particular, operators involving purely right-handed quarks are unconstrained in the effective theory analysis but may be subject to severe constraints from their UV sensitive contributions to $D$ mixing observables.  On the other hand, QED and QCD dipole operators are at present only weakly constrained by nuclear EDMs and thus present the best candidates to address the $\Delta a_{CP}$ puzzle~\cite{JFKDCPVD}.

\section{Explanations of $\Delta a_{CP}$ within NP Models}

Since the announcement of the LHCb result,  several prospective explanations of  $\Delta a_{CP}$ within various NP frameworks have appeared in the literature. In the following we briefly discuss $\Delta a_{CP}$ within some of the well-motivated beyond SM contexts. 

In the Minimal Supersymmetric SM (MSSM), the right size of the QCD dipole operator contributions  can be generated with non-zero left-right up-type squark mixing contributions $(\delta^u_{12})_{LR}$~\cite{JFKGrossman:2006jg,JFKGiudice:2012qq,JFKHiller:2012wf}.
 Parametrically such effects in $\Delta a_{CP}$ can be written as~\cite{JFKGiudice:2012qq}
\begin{equation}
|\Delta a_{CP}^{\rm SUSY}| \approx 0.6\% \times \left( \frac{|{\rm Im}(\delta_{12}^u)_{LR}|}{10^{-3}} \right) \left( \frac{\rm TeV}{\tilde m} \right)\,,
\end{equation}
where $\tilde m$ denotes a common squark and gluino mass scale. At the same time dangerous contributions to $D$ mixing observables are chirally suppressed.  It turns out however that even the apparently small $(\delta^u_{12})_{LR}$ value required implies a highly nontrivial flavor structure of the UV theory, in particular large trilinear ($A$) terms and sizable mixing among the first two generation squarks ($\theta_{12}$) are required~\cite{JFKGiudice:2012qq} 
\begin{eqnarray}
{\rm Im}(\delta^u_{12})_{LR} &\approx& \frac{{\rm Im}(A) \theta_{12} m_c}{\tilde m} \approx \left(\frac{{\rm Im}(A)}{3}\right) \left(\frac{\theta_{12}}{0.3}\right) \left(\frac{TeV}{\tilde m}\right) 0.5 \times 10^{-3}\,.\nonumber
\end{eqnarray}

Similarly, warped extra dimensional models~\cite{JFKRS1} that explain the quark spectrum through flavor anarchy~\cite{JFKRS1,JFKRS2} can naturally give rise to QCD dipole contributions 
affecting $\Delta a_{CP}$ as~\cite{JFKRS}
\begin{equation}
|\Delta a_{CP}^{\rm RS}| \approx 0.6\% \times \left( \frac{\mathcal O_\beta}{0.1} \right) \left( \frac{Y_5}{4} \right)^2 \left( \frac{3 \rm TeV}{m_{KK}} \right)^2\,,
\end{equation}
where $m_{KK}$ is the KK scale, $Y_5$ is the 5D Yukawa coupling in appropriate units of the AdS curvature and the function $\mathcal O_\beta$ parameterizes the Higgs profile overlap with
the fermion KK state wavefunctions.
Reproducing the experimental value of $\Delta a_{CP}$ requires near-maximal 5D Yukawa coupling, 
close to its perturbative bound~\cite{JFKAgashe:2008uz} of $4\pi/\sqrt{N_{KK}} \simeq 7$ for $N_{KK} = 3$ perturbative KK states. In term, this helps to suppress dangerous tree-level contributions to CP violation in $D-\bar D$ mixing~\cite{JFKDDRS}. This scenario can also be interpreted within the framework of partial compositeness in four dimensions, but generic composite models typically require smaller Yukawas to explain $\Delta a_{C P}$ and consequently predict sizable contributions to CP violation in $\Delta F = 2$ processes~\cite{JFKcomp}.  

On the other hand, in the SM extension with a fourth family of chiral fermions $\Delta a_{CP}$ can be affected by  $3\times 3$ CKM nonunitarity and $b'$ penguin operators
\begin{equation}
|\Delta a^{\rm 4th\, gen}_{CP}| \propto {\rm Im} \left( \frac{\lambda_{b'}}{\lambda_d - \lambda_s} \right)\,.
\end{equation}
However, due to the existing stringent constraints on the new CP violating phases entering $\lambda_{b'}$~\cite{JFK4gen}, only moderate effects comparable to the SM estimates are allowed~\cite{JFKFeldmann:2012js}.

\section{Prospects}

Continuous progress in Lattice QCD methods ({\it c.f.}~\cite{JFKHansen:2012tf}) gives hope that ultimately the role of SM long distance dynamics in $\Delta a_{CP}$ could be studied from first principles. In the meantime it is important to identify possible experimental tests able to distinguish standard vs. non-standard explanations of the observed value.

Explanations of $\Delta a_{CP}$ via NP contributions to the QCD dipole operators generically predict sizable effects in radiative charm decays~\cite{JFKradiative}. First, in most explicit NP models the short-distance contributions to QCD and EM dipoles are expected to be similar. Moreover, even assuming that only a non-vanishing QCD dipole  is generated at some high scale, the mixing of the two operators under the QCD renormalization group implies comparable size of the two contributions at the charm scale. Unfortunately, the resulting effects in the rates of radiative $D\to X \gamma$ decays are typically more than two orders of magnitude below the long-distance dominated SM effects~\cite{JFKRS}. This suppression can be partly lifted when considering CP asymmetries in exclusive $D^0\to P^+P^-\gamma$ transitions, where $M_{PP} = \sqrt{(p_{P^+}+p_{P^-})^2}$ is close to the $\rho,\omega,\phi$ masses~\cite{JFKradiative}. Related observables in rare semileptonic $D$ decays have also been studied recently~\cite{JFKFajfer:2012nr}.

An alternative strategy makes use of (sum rules of) CP asymmetries in various hadronic D decays (necessarily including neutral mesons). It is effective in isolating possible non-standard contributions to $\Delta a_{CP}$ if they are generated by effective operators with a $\Delta I = 3/2$ isospin structure~\cite{JFKGrossman:2012eb} (which unfortunately does not include the QCD dipoles).


Finally, correlations of non-standard contributions to $\Delta a_{CP}$ with other CP violating observables like electric dipole moments, rare top decays or down-quark phenomenology are potentially quite constraining but very NP model dependent~\cite{JFKGiudice:2012qq,JFKAltmannshofer:2012ur}.  

\section*{Acknowledgments}
The author would like to thank the organizers of {\it FLASY 2012} for the invitation to this fruitful workshop. This work is supported by the Slovenian Research Agency.

\bibliographystyle{apsrev4-1}


%% file: Papers/king.tex
%
%
%
%
%
%

\chapter[Flavour Symmetry Models after Daya Bay and RENO (King)]{Flavour Symmetry Models after Daya Bay and RENO}
\vspace{-2em}
\paragraph{S.F. King}
\paragraph{Abstract}
We discuss the impact of the recent measurements of the lepton mixing angle $\theta_{13}$
by the Daya Bay and RENO reactor experiments on neutrino mass models based on flavour
or family symmetry.
 

\section{Introduction}
\label{skIntroduction}

It is one of the goals of theories of particle physics beyond the
Standard Model to predict quark and lepton masses and mixings,
or at least to relate them. While the quark mixing angles are known to all be
rather small, by contrast two of the lepton mixing angles, the atmospheric
angle $\theta_{23}$ and the solar angle $\theta_{12}$, are identified as being
rather large.  Until recently the remaining reactor
angle $\theta_{13}$ was unmeasured.
Recently Daya Bay~\cite{An:2012eh} and RENO~\cite{Ahn:2012nd}, collaborations have measured $\sin^2(2\theta_{13})\approx 0.1$ corresponding to $\theta_{13}\approx 9^o$.

\begin{figure}[h!]
\begin{center}
 \includegraphics[width=15cm]{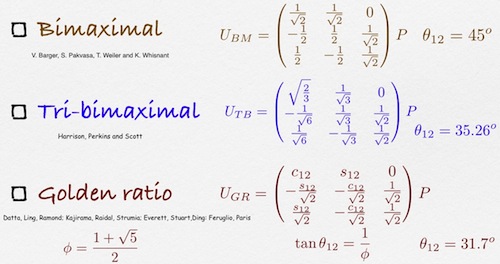}
 \caption{Simple lepton mixing patterns, all involving zero reactor angle and maximal
 atmospheric angle, and distinguished by solar angles as shown.
}\label{skfig4}
\end{center}
\end{figure}

From a theoretical or model building point of view, one significance of this
measurement is that it excludes the well known tri-bimaximal (TB) lepton mixing
pattern shown in Fig.\ref{skfig4}
in which the atmospheric angle is maximal, the reactor angle
vanishes, and the solar mixing angle is approximately $35.3^{\circ}$. 
When comparing global fits to TB mixing it is convenient to express the solar,
atmospheric and reactor angles in terms of deviation parameters ($s$, $a$ and
$r$) from TB mixing\cite{King:2007pr,Pakvasa:2007zj}: 
\begin{equation}
\label{skrsadef}
\sin \theta_{12}=\frac{1}{\sqrt{3}}(1+s),\ \ \ \ 
\sin\theta_{23}=\frac{1}{\sqrt{2}}(1+a), \ \ \ \ 
\sin \theta_{13}=\frac{r}{\sqrt{2}}.
\end{equation}
For example, the global fit in \cite{Fogli:2012ua} 
yields the $1\sigma$ ranges for the TB deviation parameters:
\begin{eqnarray}
\label{skrsafit}
\begin{array}{ccc}
-0.066\leq s\leq -0.013,&-0.146\leq a\leq -0.094,& 0.208\leq r\leq 0.231,
\end{array}
\end{eqnarray}
assuming a normal neutrino mass ordering.
As well as showing that TB is excluded by the reactor angle being non-zero, 
Eq.~\ref{skrsafit} shows a preference for the atmospheric angle to
be below its maximal value and also a slight preference for the solar angle to
be below its tri-maximal value. 
An interesting possibility consistent with the data is Tri-bimaximal-Cabibbo 
(TBC) mixing \cite{King:2012vj} with $a=s=0$ and $r$ set equal to the Wolfenstein parameter $\lambda$,
corresponding to $\theta_{13}=9.2^o$.

\begin{figure}[h!]
\begin{center}
 \includegraphics[width=15cm]{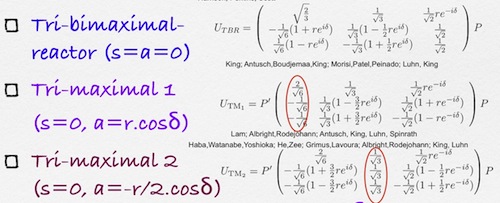}
 \caption{Simple variants of TB mixing, namely:
 tri-bimaximal-reactor (TBR) mixing; tri-maximal mixing with first column of TB form (TM1);
 tri-maximal mixing with second column of TB form (TM2).
The distinctive atmospheric sum rules are indicated in the notation of Eq.\ref{skrsadef}.
}\label{skfig6}
\end{center}
\end{figure}

As a result of the rapidly changing landscape of neutrino mixing parameters,
many models based on discrete family symmetry
which were proposed initially to account for TB mixing are
now either excluded, or have been subjected to
modification  \cite{Altarelli:2012ss}. This not only applies to TB mixing but also to 
other simple lepton mixing patterns as
shown in Fig.\ref{skfig4}, including bi-maximal (BM) and Golden ratio (GR).
All these simple mixing patterns can all be enforced by an underlying symmetry, as we will
shortly discuss. The fact that they 
are all excluded therefore calls into question the symmetry approach.
However it is worth noting at the outset that simple variants of TB mixing
are still viable such as those shown in Fig.\ref{skfig6}, and they may also arise from
family symmetry, as we shall discuss later.

\begin{figure}[h!]
\begin{center}
 \includegraphics[width=6cm]{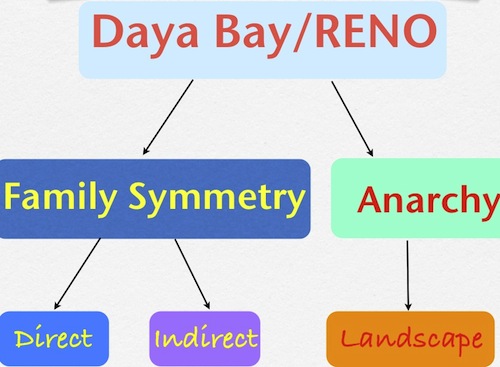}
  \includegraphics[width=9cm]{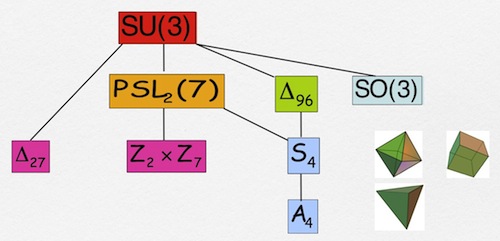}
 \caption{Left panel shows the simple choice facing theorists after Daya Bay and RENO.\newline
 The right panel shows some possible family symmetries.
 }\label{skfig1}
\end{center}
\end{figure}

Some authors regard the large reactor angle as signalling
an anarchical neutrino mass matrix \cite{deGouvea:2012ac}.
The basic choice facing theorists following Daya Bay and RENO is therefore: symmetry vs anarchy,
as shown in the left panel of Fig.~\ref{skfig1}. In this talk we shall continue to follow the symmetry approach, based on
family symmetries such as those shown in the right panel of Fig.~\ref{skfig1}, where the family symmetry may be implemented either directly or indirectly as also indicated in the left panel,
where this classification was introduced in \cite{King:2009ap}.

\begin{figure}[h!]
\begin{center}
 \includegraphics[width=10cm]{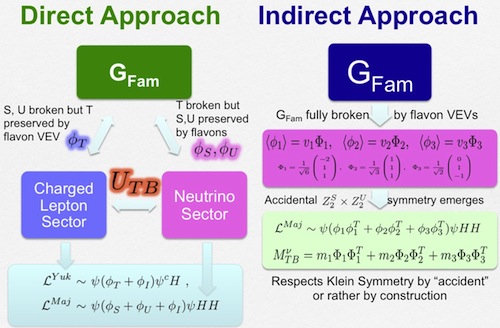}
 \caption{The direct vs indirect approach to family symmetry models of TB mixing before Daya Bay and RENO.
}\label{skfig11}
\end{center}
\end{figure}

It is worth recalling the situation before the measurement of the reactor angle.
For example, let us consider TB mixing. In this case, simple finite
family symmetries such as $A_4$ and $S_4$ were capable of embedding the Klein symmetry
of the TB neutrino mass matrix. For example $S_4$ contains the Klein generators
$S,U$, together with $T$ enforcing the diagonality of the charged lepton mass matrix in this basis.
In the direct approach to models of TB mixing, the family symmetry $G_F$ is broken by flavons such that the
$S,U$ preserving flavons $\phi_{S,U}$ only appear in the neutrino sector, while the $T$ preserving
flavon $\phi_T$ only appears in the charged lepton sector as shown in the left half of Fig.\ref{skfig11}.
Similar arguments may be applied to account for the other simple mixing patterns,
where BM mixing can also emerge from $S_4$,
while GR mixing may arise from $A_5$. All these possibilities BM, TB, GR involve zero reactor angle
and maximal atmospheric angle due to the 2-3 symmetry enforced by the $U$ generator of the Klein
symmetry.

%

%

Alternatively, in the indirect approach, the family symmetry $G_F$ is completely broken by three flavons whose VEVs
are aligned along the columns of the TB mixing matrix, but which appear quadratically in the neutrino sector, as shown in the right half of Fig.\ref{skfig11}.

\begin{figure}[h!]
\begin{center}
 \includegraphics[width=12cm]{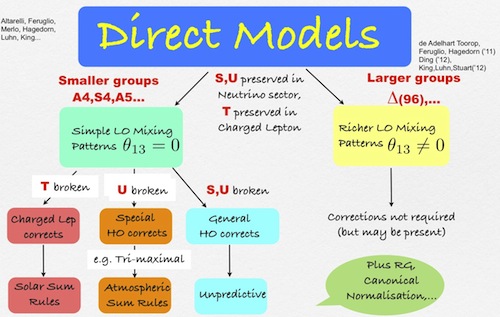}
 \caption{Possible strategies for direct models after Daya Bay and RENO.
}\label{skfig2}
\end{center}
\end{figure}

Following Daya Bay and RENO, the possible strategies for direct models are as shown in Fig.\ref{skfig2}.
For the smaller groups such as $A_4, S_4, A_5$, on the left-hand part of Fig.\ref{skfig2},
which all predict zero reactor angle at the leading order (LO),
the possible options are: break the T generator by invoking charged lepton corrections; break the $U$ generator
by some special higher order (HO) corrections, leaving the $S$ generator in tact, leading to special
mixing patterns such as tri-maximal mixing; or break both $S,U$ by general HO corrections, leading to a generally unpredictive scheme. We now consider each possibility in turn.

The case where only the $T$ generator is broken by a non-diagonal charged lepton mass matrix
leads to solar sum rules involving $\cos \delta$ as follows \cite{King:2005bj,Antusch:2005kw}:
\begin{eqnarray}
BM: &\theta_{12}& \approx 45^o+ \theta_{13}\cos \delta \nonumber \\
TB: &\theta_{12} &\approx 35.26^o+ \theta_{13}\cos \delta  \nonumber \\
GR: &\theta_{12} &\approx 31.7^o+ \theta_{13}\cos \delta .
\end{eqnarray}
Since $\theta_{13}\approx 9^o$ the requirement of a solar angle $\theta_{12}\approx 34^o$ leads to 
a distinctive prediction for $\cos \delta $ in each case. The basic assumption is that the charged lepton
correction is dominated by Cabibbo-like (1,2) mixing with a charged lepton mixing angle
equal to the Cabibbo angle, giving $\theta_{13}=\lambda/\sqrt{2}$ as in TBC mixing \cite{King:2012vj}.

\begin{figure}[h!]
\begin{center}
 \includegraphics[width=12cm]{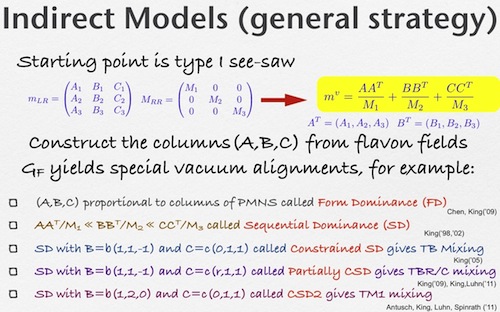}
 \caption{Possible strategies for indirect models before/after Daya Bay and RENO.
CSD yields TB mixing. CSD2 yields TM1 mixing as shown in Fig.\ref{skfig6}.
PCSD yields TBR mixing as shown in Fig.\ref{skfig6}.
If we set $r=\lambda$ then special case corresponds to TBC mixing.
}\label{skfig7}
\end{center}
\end{figure}

The case where only the $U$ generator is broken (with $S,T$ preserved) can lead to a simple pattern
of mixing, namely TM2 mixing, with examples of such models given in 
\cite{King:2011zj,Cooper:2012wf,Hagedorn:2012ut}.
Other TB variants in Fig.\ref{skfig6} can also arise from the indirect approach
as shown in Fig.\ref{skfig7}. For example, TBR and TM1 mixing can arise
from different kinds of sequential dominance (SD) with the alignments shown in  Fig.\ref{skfig7}.
CSD2 yields TM1 mixing as shown in Fig.\ref{skfig6} \cite{Antusch:2011ic} .
PCSD yields TBR mixing as shown in Fig.\ref{skfig6} \cite{King:2009qt,King:2011ab}.
If we set $r=\lambda$ then special case corresponds to TBC mixing  \cite{King:2012vj}.

Finally, if a larger family symmetry such as $\Delta(96) $ is assumed, 
as in the right-hand part of Fig.\ref{skfig2},
then it is possible
to have a different kind of Klein symmetry at the LO which already gives a reactor angle
of $\theta_{13}\approx 12^o$, together with $\theta_{12}\approx \theta_{23}\approx 36^o$,
closer to the desired value. However, in the framework of a GUT model, modest charged lepton
corrections of about $3^o$ can correct these angles to acceptable values of 
$\theta_{13}\approx 9.6^o$, together with $\theta_{12}\approx 33^o$ and $\theta_{23}\approx 37^o$
\cite{King:2012in}.

\section*{Acknowledgments}
SFK acknowledges partial support 
from the STFC Consolidated ST/J000396/1 and EU ITN grants UNILHC 237920 and INVISIBLES 289442 .

\bibliography{king}
\bibliographystyle{apsrev4-1}


%% file: Papers/krauss.tex

%
%
%
%
%
%

\chapter[Neutrino Mass Generation by Higher-Dimensional Effective Operators in GUTs (Krauss)]{Neutrino Mass Generation by Higher-Dimensional Effective Operators in GUTs}
\vspace{-2em}
\paragraph{M. B. Krauss} 
\paragraph{Abstract}
We will discuss neutrino mass generation by higher-dimensional effective operators in SUSY. As opposed to the standard type-I seesaw, these models are testable in experiments and have phenomenological implications at the LHC, such as processes with displaced vertices and lepton number violation. 
We will also discuss the possibility of embedding these effective operators into Grand Unified Theories and their agreement with Cosmological constraints.

\section{Introduction}
The standard type-I seesaw mechanism implies new physics close to the GUT scale, which is not testable in experiments. Therefore new models that introduce new particles with masses at the order of 1 TeV for the generation of neutrino mass have come into discussion recently. Due to the lower mass scale, these models have phenomenological implications at the LHC. Examples are radiative mass generation, where loop suppression factors enter, or models with a small lepton number violating contribution, such as the inverse seesaw mechanism or SUSY with R-parity violation. In a further group of models the generation of neutrino mass via a dimension 5 operator is forbidden so that the leading contribution is of higher dimension~\cite{Babu:1999me,Babu:2001ex,Chen:2006hn,Gogoladze:2008wz,Giudice:2008uua,Babu:2009aq,Gu:2009hu,Bonnet:2009ej,Picek:2009is,Liao:2010rx,Liao:2010ku,Liao:2010ny,Kanemura:2010bq}. In the following we want to discuss the latter option in the context of supersymmetry~\cite{Krauss:2011ur}.

\section{Neutrino mass and effective operators}

In general at low energies we can describe physics beyond the Standard Model (SM) by a tower of effective operators which are added to the SM Lagrangian:
\begin{align}
        \mathcal{L} = \mathcal{L}_{\rm SM} + \mathcal{L}^{d=5}_{\text{eff}} 
+ \mathcal{L}^{d=6}_{\text{eff}} + \cdots
\, , \quad \textrm{with} \quad \mathcal{L}^{d}_{\text{eff}} \propto \frac{1}{\Lambda_{\mathrm{NP}}^{d-4}} \, \mathcal{O}^{d}\,,
       \end{align}
where $d$ is the dimension of the operator, which is supressed by the new physics scale $\Lambda_\text{NP}$ to the power $d-4$.
The type-I seesaw, for example, becomes, after integrating out the heavy right handed neutrinos, the so called Weinberg operator~\cite{Weinberg:1979sa},
\begin{align}
\mathcal{L}^{d=5} \supset \frac{{Y_N}^2}{m_N} \langle H \rangle^2 \overline\nu^c \nu,
\end{align}
where $Y_N$ is the Yukawa coupling between the SM lepton doublet and the right handed neutrinos and $m_N$ the mass of the latter. Inserting the vacuum expectation value of the Higgs fields after electroweak symmetry breaking we obtain a small mass term for the left handed neutrinos, which is suppressed by $m_N$.
In models that include additional Higgs fields, such as Two Higgs Doublet Models and the (Next to) Minimal Supersymmetric Standard Model ((N)MSSM), higher contributions to the neutrino mass appear:
\begin{align}
m_\text{eff}^{d=6} &=\frac{1}{\Lambda^2} \langle H_u \rangle^2 \langle S \rangle, &m_\text{eff}^{d=7} &=\frac{1}{\Lambda^3} \langle H_u \rangle^2 \langle H_u \rangle \langle H_d \rangle, \ \dotsc
\end{align}
 If the $d=5$ operator is forbidden, the leading mass term will be of dimension $>5$ and therefore suppressed by higher powers of the new physics scale $\Lambda$. As a consequence $\Lambda$ can be at a significantly lower scale in order to obtain the small observed neutrino masses.
In a SUSY framework we can use, for example, a discrete $\mathbf{Z}_3$ symmetry to have the $d=7$ operator as leading contribution. (See also the discussion in Ref.~\cite{Bonnet:2009ej,Krauss:2011ur})

\section{A $d=7$ example}
The Weinberg operator can be decomposed into the type I, type II or type III seesaw model~\cite{Ma:1998dn}. Analogously there exist several possible fundamental theories, which lead to the same higher-dimensional operator after integrating out the heavy fields. One example of such a decomposition of a $d=7$ operator, which has been studied in Ref.~\cite{Krauss:2011ur}, is specified by the superpotential
     \begin{align}
      W = \ & W_\text{\tiny{NMSSM}}
      + Y_N \hat N \hat L \cdot \hat H_u
      - \kappa_1 \hat N' \hat \xi \cdot \hat H_d
      + \kappa_2 \hat N' \hat \xi' \cdot \hat H_u
      + m_N \hat N \hat N' 
      + m_\xi \hat \xi \cdot \hat \xi'\,,
    \end{align}
where the mediators $N$ and $N'$ are SM singlets and $\xi$ and $\xi'$ are doublets. 
From this superpotential we obtain the mass matrix for the neutral fermion fields. In the basis $f^0 = (\nu, N, N',\xi^0, {\xi'}^0)$ it reads
\begin{align}
 &M^0_f =
\left(\begin{array}{ccccc}
	0			& Y _N v_u	& 0			& 0			& 0			\\
	Y_N v_u		& 0			& m_N		& 0			& 0			\\
	0			& m_N		& 0			& \kappa_1 v_d	& \kappa_2 v_u	\\
	0			& 0			& \kappa_1 v_d	& 0			& m_\xi		\\
	0			& 0			& \kappa_2 v_u	& m_\xi		& 0			
\end{array}\right)\,.
\end{align}
By integrating out the heavy fields we obtain an effective mass matrix for the three SM neutrinos
\begin{align}
m_\nu = v_u^3 v_d Y_N^2 \frac{\kappa_1 \kappa_2}{m_\xi m_N^2},
\end{align}
where the couplings carry a flavor index. The flavor structure of the neutrino mass matrix can be obtained by choosing the coupling parameters accordingly. 

We can have masses of the mediator fields at the TeV scale for couplings $\mathcal O (10^{-3})$, which is in the range of the SM Yukawa couplings. Accordingly this model can be tested at the LHC. The SM singlet fields $N$ and $N'$ are only produced in small amounts due to the smallness of the Yukawa couplings. 
The SU(2) doublets $\xi$ and $\xi'$, however, can be produced in Drell-Yan processes, similarly to charginos and neutralinos, with a cross-section of up to $\sigma \sim 10^2$ fb. These particles will then decay into vector bosons and leptons. Since the mixing between the heavy and the light neutrinos is small, large decay length of up to several millimeters are expected. As an effect displaced vertices can be used to identify these processes. To establish the connection to neutrino physics, also lepton number violating signals have been studied. In a numerical calculation we obtained a cross-section for the lepton number conserving process $pp\rightarrow W\ell\ell$ of $\mathcal{O}(10^{2})$ fb whereas the corresponding lepton number violating process is suppressed due to pseudo-Dirac pairs ($<\mathcal{O}(10^{-9})$ fb). For $pp\rightarrow W \ell W \ell$, however, the lepton number violating process is larger than naively expected ($\mathcal{O}(10^{-2})$ fb).

\section{GUT completion}
Since the introduction of additional particles modifies the running of the gauge couplings, the presented model will spoil unification unless we add complete SU(5) multiplets to the (N)MSSM. The mediators are embedded as follows:
\begin{scriptsize}
\begin{align}
\begin{split}
 \overline{5}_M=
\left(
\begin{array}{c}
d_1^c \\
d_2^c \\
d_3^c\\
e^-\\
-\nu_e\\
\end{array}
\right)_L =
\left(
\begin{array}{c}
d^c_L \\L
\end{array}
\right)
\qquad 
\overline{5}_{\xi^\prime}=\left(
\begin{array}{c}
d_1^{\prime c} \\
d_2^{\prime c} \\
d_3^{\prime c}\\
\xi^{\prime -}\\
-\xi^{\prime 0}\\
\end{array}
\right)_L=
\left(
\begin{array}{c}
d_L^{\prime c} \\
\xi^\prime_L
\end{array}
\right)
 \qquad 
5_{\xi}=\left(
\begin{array}{c}
d_1^{''} \\
d_2^{''} \\
d_3^{''}\\
\xi^+\\
-\xi^0\\
\end{array}
\right)_R =
\left(
\begin{array}{c}
d_R^{''} \\
\xi_R
\end{array}
\right)  
\\ 
H_5=\left(
\begin{array}{c}
H_1 \\
H_2 \\
H_3\\
H_u^+\\
H_u^0\\
\end{array}
\right) =
\left(
\begin{array}{c}
H_{\rm col}  \\
H_u
\end{array}
\right)\qquad 
H_{\overline{5}}=\left(
\begin{array}{c}
H^\prime_1 \\
H^\prime_2 \\
H^\prime_3\\
H_d^-\\
H_d^0\\
\end{array}
\right) =
\left(
\begin{array}{c}
H^\prime_{\rm col}  \\
H_d
\end{array}
\right) \qquad N,N^\prime (S) {\rm~~fermionic~ singlets}\,.
\end{split}
\end{align}
\end{scriptsize}

The possible interaction terms which must be invariant under SU(5) can be realized either as an extension of the MSSM
\begin{eqnarray}
W&=& y_1\,N \, 5_\xi\,H_{\overline 5}+y_2\,N \, \overline{5}_{\xi^\prime}\,H_5+y_3\,N \,\overline{5}_M\,H_5
+\nonumber \\
&&y_1^\prime,N^\prime \, 5_\xi\,H_{\overline 5}+y_2^\prime\,N^\prime \, \overline{5}_{\xi^\prime}\,H_5+y_3^\prime\,N^\prime \,\overline{5}_M\,H_5 + \nonumber \\
&&m_{\xi^\prime}\, \overline{5}_M\, 5_\xi +
m_{\xi}\, \overline{5}_{\xi^\prime}\, 5_\xi + m_N N' N + m_{NN} N N + m_{N'N'} N' N' + \nonumber \\
&& y_d\,\overline{5}_M\,10\,H_{\overline 5} +  y_d^\prime\,
\overline{5}_{\xi^\prime}\,10\,H_{\overline 5} + y_u\,10\,10\,H_5\,
\end{eqnarray}
or the NMSSM
\begin{eqnarray}
W&=& y_1\,N \, 5_\xi\,H_{\overline 5}+y_2\,N \, \overline{5}_{\xi^\prime}\,H_5+y_3\,N \,\overline{5}_M\,H_5
+\nonumber \\
&&y_1^\prime,N^\prime \, 5_\xi\,H_{\overline 5}+y_2^\prime\,N^\prime \, \overline{5}_{\xi^\prime}\,H_5+y_3^\prime\,N^\prime \,\overline{5}_M\,H_5 + \nonumber \\
&&\lambda_{\xi^\prime}\, S\, \overline{5}_M\, 5_\xi +
\lambda_{\xi}\, S\, \overline{5}_{\xi^\prime}\, 5_\xi + \lambda_N S\, N' N + \lambda_{NN} S\, N N + \lambda_{N'N'} S\, N' N' +  \nonumber \\
&& y_d\,\overline{5}_M\,10\,H_{\overline 5} +  y_d^\prime\,
\overline{5}_{\xi^\prime}\,10\,H_{\overline 5} + y_u\,10\,10\,H_5\,.
\end{eqnarray}
In the NMSSM case the mediator mass scale is determined by $\langle S \rangle$, which is of order 1 TeV and results in the required mass scale for the light neutrinos. Furthermore the NMSSM avoids some problems of the $\mu$-term of the NMSSM (see also Ref.~\cite{Bonnet:2009ej,Krauss:2011ur}). We obtain, however, effective operators of the type
\begin{align*}
 &\frac{1}{\langle{S}\rangle} LLH_uH_u, \qquad
 &&\frac{1}{\langle{S}\rangle^3} (LLH_uH_d)(H_uH_d)\,.
\end{align*}
As a consequence $\langle S \rangle$ breaks any discrete symmetry under which it is charged. This prevents us from choosing a simple discrete symmetry group to avoid the $d=5$ contribution to neutrino mass.

As a further point, the promotion of the SM SU(2) doublets to 5-plets of SU(5) requires the introduction of additional d-quarks, which are expected to have masses of TeV and to be stable. Their stability is a consequence of the same symmetry that forbids the $d=5$ operator. Stable heavy d-quarks, however, cause conflicts with cosmological constraints. Their presence during Big Bang Nucleosynthesis would alter the observed abundances of the light elements in the universe (see, e.g., Ref.~\cite{Iocco:2008va} for a review). Further bounds come, e.g., from direct heavy element searches in water or from the stability of neutron stars.
When the heavy quarks are in thermal equilibrium in the early universe they have the possibility of pair annihilation via gluons into the lighter SM quarks or gluon pairs. This is, however, insufficient to lower the particle yield after freeze-out below observational constraints \cite{Nardi:1990ku}.
A decay of the heavy quarks via leptoquarks is possible and less suppressed then proton decay, due to the higher mass of the involved particles, but the lifetime of the heavy quarks is still not sufficiently small to avoid the given bounds.
For those reasons it will be studied if there are other ways to reduce the number density of additional d-quarks, or if these constraints are sufficient to rule out this particular model. Also a more systematic study how other decompositions of higher dimensional effective operators are affected by bounds will be done.

\section{Conclusion}
We have demonstrated that neutrino mass generation by higher-dimensional effective operators can lower the new physics scale to TeV. Phenomenological studies predict displaced vertices and lepton number violating signals at the LHC for certain decompositions of these effective operators. We have also discussed that in order to conserve gauge coupling unification the introduction of complete SU(5) multiplets is necessary. Additional heavy d-quarks that appear as a consequence, have cosmological constraints on their abundances.

\section*{Acknowledgments}
MBK would like to thank his collaborators Davide Meloni, Werner Porod and Walter Winter. He acknowledges support from Research Training Group 1147 ``Theoretical astrophysics and particle physics'' of Deutsche Forschungsgemeinschaft.

\bibliography{krauss.bib}
\bibliographystyle{apsrev4-1}

%% file: Papers/michalkreps.tex

%
%
%
%
%
%

\chapter[LHCb results now and tomorrow (Kreps)]{LHCb results now and tomorrow}
\vspace{-2em}
\paragraph{M. Kreps}


\paragraph{Abstract}
We discuss the status of the LHCb experiment in mid 2012 together with prospects for near and long term
future. 

\section{Introduction}

With start of the LHC and its dedicated quark flavour physics experiment LHCb we are entering a new
era in flavour physics. In this new era, we expect an unprecedent improvement of several key
measurements, which should reveal new physics or put stringent constraints on it. The length of any
proceedings is practically insufficient to cover any details of the existing measurements, so we will
only briefly summarize presented results and supplement this with information on running and
expectations for near and longer term future as such information is not always readily available in
a single place. When listing results, unless specified otherwise, the first uncertainty is always statistical and second
systematic.

\section{Current running}

The LHCb experiment started its data taking useful for flavour physics in 2010. During 2011 we
collected a dataset with an integrated luminosity of 1.0 fb$^{-1}$. This was achieved by running at
an instantaneous luminosity well above design, which brings an increased number of collisions within
a bunch crossing from the designed value of $\mu=0.4$ to $\mu=1.7$. While $\mu$ is significantly
above the design, physics results proved that the detector and reconstruction software can handle this
without major issues. Another significant achievement was the successful commissioning of luminosity
levelling, a process which allows LHCb to run at a constant instantaneous luminosity over an LHC
accelerator fill which should be contrasted with the general purpose experiments ATLAS and CMS. The
capability to run at the constant instantaneous luminosity not only allows us to maximize the amount of data
we collect, but also ensures quite extraordinary stability of running conditions, which practically
do not change. 
Since the restart of the data taking in April 2012, in about three months we collected 0.62 fb$^{-1}$.
Projection from this value into the end this year data taking suggests, that we will collect around
2.0 fb$^{-1}$ of data this year.

From a detector point of view, LHCb is running very well with an overall efficiency above 95\%. We not
only collect data with high efficiency, but we collect practically only good data, with only a tiny
fraction which has to be discarded from the data analysis.

\section{Preview of physics results}

Here we present a brief review of LHCb results prior to the ICHEP 2012 conference. As in time of the workshop
we could add less then half of the statistics of 2011, practically all results are obtained on the full
2011 dataset with exceptions to this explicitly marked. While there are many results available from
this dataset, we concentrate on those which are of highest interest for constraining new physics.

The first set of results to discuss are determinations of CKM angle $\gamma$. This  is the only angle in
the CKM matrix which can be determined from tree level processes and as such is expected to receive
a negligible contribution from possible new physics. Without it the \textit{CP}-violating phase in the
standard model is not well determined and one could always question whether new physics is just
hiding in the methods used to determine the parameters of the standard model. While important to define the standard
model, the angle $\gamma$ is not yet well determined experimentally. This is due to an interplay of small \textit{CP}
violation in decays which have relatively large rate and tiny rate in decays which have sizeable
\textit{CP} violation. Thus only in recent years experiments started to see significant signals
which can be used to extract the angle $\gamma$. We perform measurement in both ADS
and GLW decay modes and we obtain 5.8 $\sigma$ significance for direct \textit{CP} violation in
the $B^\pm\rightarrow DK^\pm$ decays \cite{Aaij:2012kz}. For the observables relevant for the $\gamma$ angle
extraction we obtain $R_{CP+}=1.007\pm0.038\pm0.012$, $A_{CP+}=0.145\pm0.032\pm0.010$,
$R_{ADS(K)}=0.0152\pm0.0020\pm0.0004$ and $A_{ADS(K)}=-0.52\pm0.15\pm0.02$. All those results are
limited by statistics with one of the largest systematic uncertainties stemming from the detector asymmetry,
knowledge of which will improve with increased statistics. 

An alternative way of extracting the CKM angle $\gamma$ is the use of $B_s\rightarrow D_s K$ decays. While
theoretically the method is very clean, the process involves $B_s$ mixing and as such could be possibly
affected by new physics. With 370 pb$^{-1}$ we observe about 400 signal events which are used to
measure the branching fraction of $(1.90\pm0.12\pm0.13^{+0.12}_{-0.14})\times 10^{-4}$ with the last
uncertainty coming from $f_s/f_d$ measurement~\cite{LHCb-PAPER-2011-018}. With full
statistics collected in 2011 we expect about 1200 signal events, which can be used for time
dependent analysis to extract the angle $\gamma$, results are expected in autumn 2012.

Moving on to measurements which are relevant for search for new physics we start with
the \textit{CP}-violating phase $\phi_S$, which is the phase between $B_s^0$ mixing diagrams and
$b\rightarrow c\bar{c}s$ decays. Traditionally this is measured using $B_s^0\rightarrow J/\psi\phi$
decays, which is a mixture of \textit{CP}-even and \textit{CP}-odd final state. To disentangle the two,
angular distributions have to be analyzed. With data collected in 2011, LHCb has about 21000 signal
events in this decay, which yields measurements of $\phi_S=-0.001\pm0.101\pm0.027$ and the decay
width difference $\Delta\Gamma_s=0.116\pm0.018\pm0.006$ ps$^{-1}$, both the most precise values
\cite{LHCb-CONF-2012-002}. With the unprecedent dataset collected by LHCb we are able to use also
$B_s^0\rightarrow J/\psi\pi^+\pi^-$ decays. Those provide a practically pure
\textit{CP}-odd final state to complement the $\phi_S$ measurement without need of the angular analysis. In case
of this decay, we constrain the mean decay width and decay width difference to the values obtained in the
analysis of $B_s^0\rightarrow J/\psi\phi$ decays and extract
$\phi_S=-0.019\,^{+0.173}_{-0.174}\,^{+0.004}_{-0.003}$ \cite{LHCb:2012ad}. Combination of the two measurements
yields $\phi_S=-0.002\pm0.083\pm0.027$. The dominant systematic uncertainty in these measurements comes
from assumptions of no \textit{CP} violation in mixing or decay and with an increased dataset we should
have enough sensitivity to remove those assumptions. 

While listing only a single solution, both
analyses above have a symmetrical solution with the opposite sign of $\Delta\Gamma_s$ and
\textit{CP}-violating phase $\pi-\phi_S$. To resolve the ambiguity, we turn to the $B_s^0\rightarrow
J/\psi K^+K^-$ decays, which are dominated by $B_s^0\rightarrow J/\psi\phi$, but have also a
contribution from $K^+K^-$ s-wave component. Following the  analysis in bins of the kaon pair invariant mass, we can see
variation of the \textit{CP}-conserving phase between $B_s^0\rightarrow J/\psi\phi$ and s-wave. From
this variation we can conclude that the physical solution is that with positive $\Delta\Gamma_s$ and thus
$\phi_S$ is close to its standard model expectation \cite{Aaij:2012eq}.

Significant attention has been received by $B\rightarrow hh'$ decays. These proceed through
gluonic $b\rightarrow s$ penguins and $b\rightarrow u$ trees. While they have sensitivity to a new physics, this
sensitivity is screened by  hadronic physics, so any interpretation is difficult or needs some
assumptions. One model independent test is the comparison of direct
\textit{CP} violation between $B^0\rightarrow K^+\pi^-$ and $B_s^0\rightarrow K^-\pi^+$ decays.
Based on the measured $A_{CP}$ in $B^0$ decay, one can in the standard model predict $A_{CP}$ for
$B^0_s$ \cite{Gronau:2000md}. At LHCb we have the worlds largest samples from which we measure
$A_{CP}(B^0)=-0.088\pm0.011\pm0.008$ \cite{Aaij:2012qe}. This result is consistent with
the average of other measurements \cite{Amhis:2012bh} and its
precision is close to that of the world average. For the $B_s^0$ decay we measure
$A_{CP}(B_s^0)=0.27\pm0.08\pm0.02$ with 3.3$\sigma$ significance for  \textit{CP} violation
\cite{Aaij:2012qe},
which is the first significant signal for any \textit{CP} violation in the $B_s^0$ system. The measured
value is also consistent with the standard model expectation \cite{Gronau:2000md}.

The next natural step is to exploit the large samples and move towards time dependent studies. The first result
to mention is a measurement of the effective lifetime in $B_s^0\rightarrow K^+K^-$ decays. Here, the $B_s^0$
decays in to the \textit{CP}-even final state and in the standard model lifetime measured in this decay
gives approximately the lifetime of light $B_s^0$ mass eigenstate. Even if new physics is present,
the effective
lifetime can be used to constrain  \textit{CP} violation and  $\Delta\Gamma_s$ in a global analysis. 
With dataset from 2011 collected by a dedicated decay time unbiased trigger
we measure $\tau_{\mathrm{eff}}=1.455\pm0.046\pm0.006$ ps \cite{Aaij:2012ns}.
A second time dependent analysis, which demonstrates the capabilities of LHCb, is the measurement of
 \textit{CP} violation in $B^0\rightarrow \pi^+\pi^-$ and $B_s^0\rightarrow K^+K^-$ decays. We
measure $A_{\pi\pi}^{dir}= 0.11 \pm 0.21 \pm 0.03$ with $A_{\pi\pi}^{mix}= -0.56 \pm 0.17 \pm 0.03$
for $B^0$ and $A_{KK}^{dir}= 0.02 \pm 0.18 \pm 0.04$ with $A_{KK}^{mix}=0.17\pm0.18\pm0.05$ for
$B_s^0$ \cite{LHCb-CONF-2012-007}. While not yet competitive with B-factories on $B^0$, with more data to come and hopefully
some improvements in the flavour tagging there are prospects on improving the knowledge from previous
experiments. For the $B_s^0$, while uncertainties are large, this is a first measurement and will remain
unique to the LHCb experiment.

\begin{figure}
  \centering
  \includegraphics[width=0.48\textwidth]{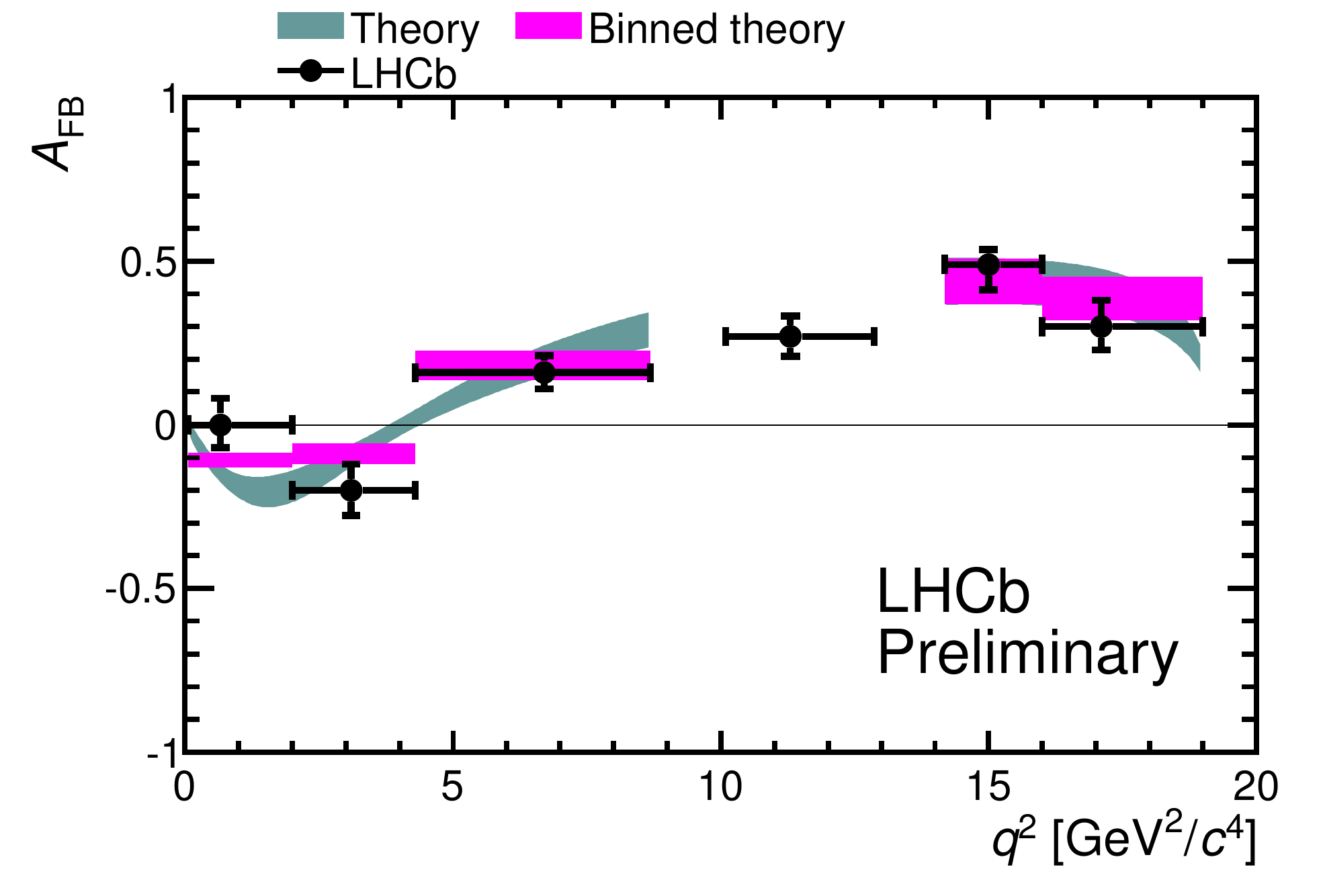}
  \includegraphics[width=0.48\textwidth]{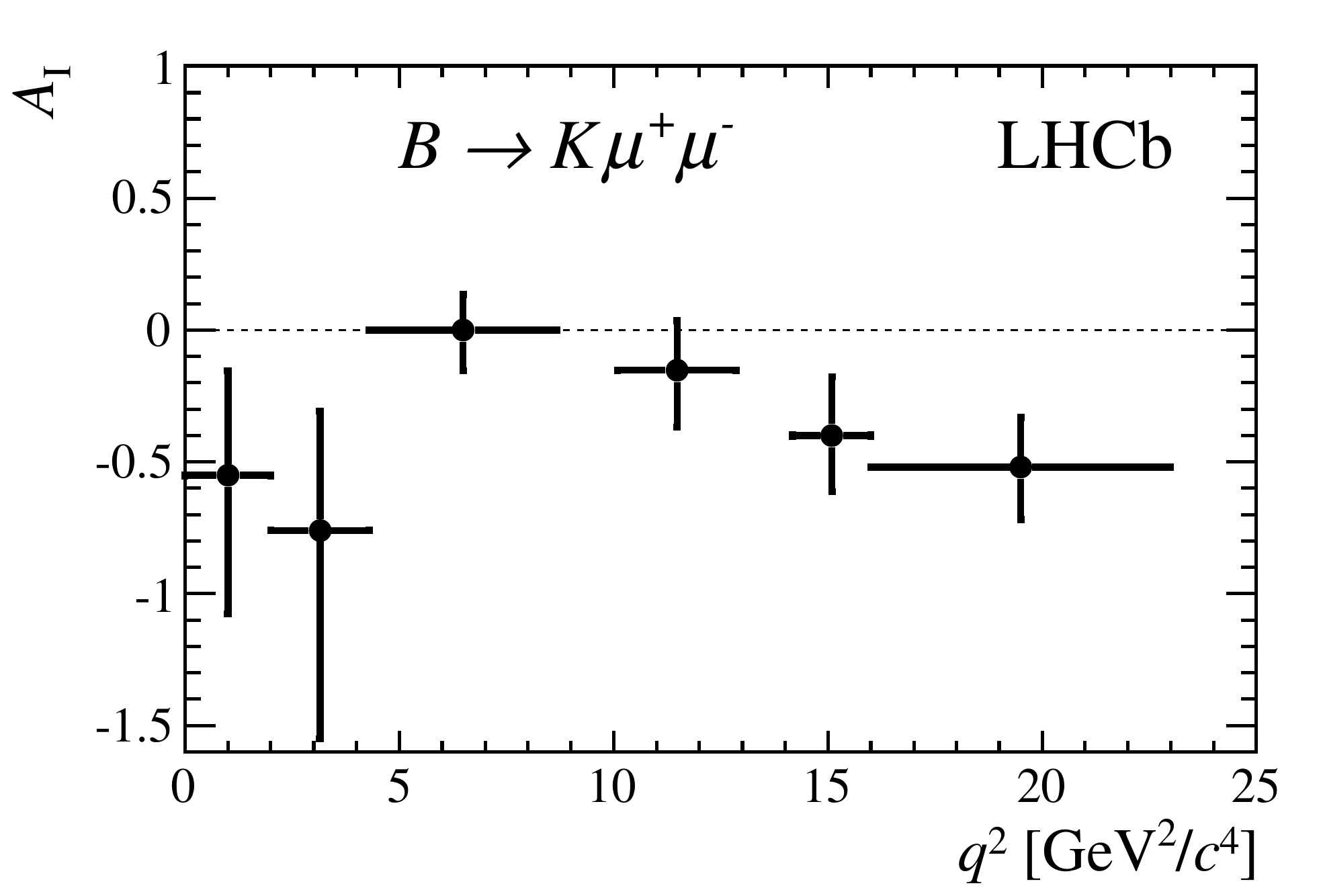}
  \caption{The forward-backward asymmetry $A_{FB}$ as a function of dimuon invariant mass squared
($q^2$) in $B^0\rightarrow K^*\mu^+\mu^-$ decays (left) and the isospin asymmetry in $B\rightarrow
K\mu^+\mu^-$ decays as a function of $q^2$.}
  \label{MK_fig1}
\end{figure}
The next class of decays to discuss are ones governed by the $b\rightarrow sll$ transition. Those
proceed through  electroweak penguin diagrams and provide a rich set of observables based on angular
distributions. The best studied decay in this class is $B\rightarrow K^*\mu^+\mu^-$. With about 900
$B^0\rightarrow K^{*0}\mu^+\mu^-$ events, LHCb has the worlds largest sample, which is used to make
th best
measurements of several quantities \cite{LHCb-CONF-2012-008}. One of the best known is the forward-backward asymmetry $A_{FB}$,
which we show in Fig.~\ref{MK_fig1}. The high statistics available directly reflects in the precision,
which is by far the best of all existing measurements. Additionally, for the first time, we also
measure the $q^2$ point at which $A_{FB}$ is crossing zero to be $q_0^2=4.9^{+1.1}_{-1.3}$ GeV$^2/c^4$.
In the determination of $q_0^2$, systematic uncertainties are completely negligible, so we quote only
statistical uncertainties. An other rather exciting result in this area is the measurement of isospin
asymmetry in $B\rightarrow K^{(*)}\mu^+\mu^-$ decays \cite{Aaij:2012cq}. The measurement is
essentially a comparison of the differential decay
widths between $B^0$ and $B^+$. For $B\rightarrow K^{*}\mu^+\mu^-$ the asymmetry across dimuon
invariant mass is found to be consistent with zero. For the $B\rightarrow K\mu^+\mu^-$ we show the asymmetry
in Fig.~\ref{MK_fig1}. We find an indication of a non-zero asymmetry at both low and high dimuon invariant
masses. The asymmetry itself appears to be caused mainly by a deficit of $B^0\rightarrow K_S^0\mu^+\mu^-$ decays
compared to the expectation. While exploitation of $b\rightarrow sll$ transitions has been pursued
by previous experiments, for
the first time LHCb could also detect a signal from $b\rightarrow dll$ transition. Specifically we
observe the $B^+\rightarrow\pi^+\mu^+\mu^-$ decay with 5.2$\sigma$ significance and measure its
branching fraction to be $(2.4\pm0.6\pm0.2)\times10^{-8}$ \cite{LHCb-CONF-2012-006}.

The above studies are supplemented by $b\rightarrow s\gamma$ transitions, which proceed purely
through electromagnetic penguins. While the LHCb calorimeter is not as performant as those at the B-factories it has
sufficient capabilities for significant improvement of the world knowledge on radiative decays. With
huge production rates of $B_s^0$ and baryons, we are unique in this sector. As an example we measure
$\mathcal{B}(B^0\rightarrow K^*\gamma)/\mathcal{B}(B_s^0\rightarrow\phi\gamma)$ to be
$1.12\pm0.08\,^{+0.06}_{-0.04}\,^{+0.09}_{-0.08}$ where first uncertainty is statistical, second
systematic and third  is due to the uncertainty on the production fraction ratio \cite{LHCb:2012ab}. This translates to
$\mathcal{B}(B_s^0\rightarrow\phi\gamma)=(3.9\pm0.5)\times 10^{-5}$. The large sample of
$B^0\rightarrow K^*\gamma$ decays is also used for the most precise measurement of  \textit{CP}
violation in this mode
with the obtained value $A_{CP}=0.008\pm0.017\pm0.009$ \cite{LHCb-CONF-2012-004}. 

The final result to mention is a search for the $B_s\rightarrow\mu\mu$ decay. This decay started to be
an effective hammer to new physics models already with limits several orders of magnitude away from
the standard model. This is mainly caused by the fact that in SUSY models the rate goes with a  high power of
$\tan\beta$ which for large $\tan\beta$ gives huge enhancements. A search
was performed on 2011 data with expected limit at 95\% C.L. of $3.4\times10^{-9}$ under pure background
hypothesis and $7.2\times10^{-9}$ assuming data have in addition to the background also standard model
signal \cite{Aaij:2012ac}. The observed limit is $4.5\times 10^{-9}$, which is the best in the world. Comparing
the observed limit to expectation we can conclude that there are likely to be some signal events in the
sample, but the total number of events is smaller than expected. With the limit being only about
$1.4$ times the  standard model and
with the good performance of the LHC this year, we should be closing this gap in the near future.
It should be noted that while at this stage we are limited by the available statistics, once we observe
the signal, a  precise determination of the fragmentation fraction ratio $f_s/f_d$ becomes important for
the measurement of the branching fraction.

\section{Short term expectation}
\vspace*{-0.3cm}

So what to expect in the near future? We are running very well and in mid-August 2012 we already
collected about 1.0 fb$^{-1}$ of data, the same as the 2011 total. While our goal was to collect 1.5
fb$^{-1}$ this year, with extension of the LHC run we are likely looking to a dataset above 2.0
fb$^{-1}$, which would triple the statistics from 2011. With such samples, practically all our results
can be meaningfully updated on the time scale of about one year. As in most cases we are not limited
by systematic uncertainties, and typically the dominant ones will improve with more data, there is no
real show stopper for improvements. The exception might be measurements of decay rates, which have
to be normalized and thus improvements in lattice calculations which enter the determination of
$f_s/f_d$ would make a difference. For the measurements of asymmetries, we are certainly far from the wall
of systematic uncertainties.

One thing of personal opinion I would like to stress, is the long shutdown of LHC in 2013 and 2014. If we
add year 2015 to collect data which would again allow a meaningful addition of the statistics to the
current round of analyses, there is some time when we will be looking to new ideas and new
measurements. If theory community has new ideas, I'm sure in a  year from now there will be lots of
eager people to translate them to new measurements not done before.

\vspace*{-0.3cm}
\section{Upgrade and long term expectation}
\vspace*{-0.3cm}

While we are taking data only for the third year, already now the time to double the dataset takes a significant
fraction of a year and soon it would take more than a year. Thus if we continue with existing LHCb detector
rather soon we have to wait too long for increase of statistics to make significant progress. One
obvious solution is to increase luminosity, but with the current detector and its readout, the increased
luminosity does not translate to an increase of yields, which is specially true for fully hadronic final
states. To overcome the limitation and to get to higher luminosities, LHCb collaboration proposed to
upgrade detector and run also during the next phase of LHC running \cite{CERN-LHCC-2011-001}. The idea was
well received and in response we worked out some more details in a  Framework TDR
\cite{Bediaga:1443882}. For theory community, one useful piece of information contained in it is an update on our
sensitivities on the scale of 2018 and then with 50 fb$^{-1}$ collected by the upgraded detector. Without listing
all details here, in most of the measurements we can do, our statistical precision will be in the
region of theory uncertainties. While most of the way to the upgrade is ahead of us, prospects for
the future are definitely bright.

\vspace*{-0.3cm}
\section{Conclusion}
\vspace*{-0.3cm}

In summary, the LHCb experiment is running extremely well and producing a large set of world best
measurements in important areas of quark flavour physics. With quickly increasing datasets our
sensitivity is constantly increasing. In addition we are on the road to upgrade the detector in
order to further increase the amount of data we can collect. This suggests a bright future for  quark
flavour physics and keeps us excited about the possibility to finally discover new physics beyond the standard
model.

\vspace*{-0.3cm}
\section*{Acknowledgments}
\vspace*{-0.3cm}

I would like to thanks to organizers of the Flasy 2012 workshop and my LHCb collaborators for
producing the interesting results I'm describing here.

%% file: Papers/kubo.tex
\chapter[Multi-Component Dark Matter System  with  non-standard annihilation processes of Dark Matter (Aoki, Duerr, \textit{Kubo}, Takano)]{Multi-Component Dark Matter System  with  non-standard annihilation processes of Dark Matter}
\vspace{-2em}
\paragraph{M. Aoki, M. Duerr, \textit{J. Kubo}, H. Takano}
\paragraph{Abstract}
We study multi-component DM systems
with conversions and  semi-annihilations
of dark matter (DM) particles 
in addition to the standard annihilation processes.
It is found  that the relic abundance of DM can be very sensitive to
these non-standard DM annihilation processes
even if the DM masses are not degenerate.
To consider a concrete model of a three-component DM system, we 
extend a radiative see saw model, so that 
the DM stabilizing symmetry is promoted to $Z_2 \times Z'_2$.
The semi-annihilation process in this model produces a 
monochromatic  left-handed neutrino.
We estimate the observation rates of 
the monochromatic neutrinos produced by  the semi-annihilation of the 
captured DM particles in the Sun.
Observations of high energy monochromatic neutrinos from the Sun may
indicate  a multi-component DM system.

\section{General Formalism}
DM particle can be made stable by an unbroken symmetry such as $Z_2$.
In this talk, which is based on the recent work \cite{Aoki:2012ub}, 
we consider multi-component DM systems.
In  multi-component DM systems, there are non-standard  DM annihilation 
processes that are different from the standard DM annihilation
processes.
The importance of the non-standard annihilation
processes such as the DM conversion
 \cite{D'Eramo:2010ep,Belanger:2011ww,Belanger:2012vp} and 
the semi-annihilation of DM 
 \cite{D'Eramo:2010ep,Belanger:2012vp} in  two-component
DM systems 
for the temperature evolution of the number density of DM has been recently reported.
Here we assume the existence of $K$ stable DM particles $\chi_i$ with mass $m_i$.
 To simplify the situation,  we restrict ourselves to   three types of processes
which enter the Boltzmann equation:
\begin{eqnarray}
& &\chi_i~\chi_i \leftrightarrow X_i~X'_i~,
\label{p0}\\
 & &  \chi_i~\chi_i  \leftrightarrow \chi_j~\chi_j~
 \mbox{(DM conversion)}~,
 \label{p1}\\
& & \chi_i~\chi_j \leftrightarrow \chi_k~X_{ijk}~
 \mbox{(DM semi-annihilation)},
\label{p2}
\end{eqnarray}
where 
$X$'s stand for  standard model (SM) particles in thermal equilibrium.

Using the dimensionless inverse temperature $x=\mu/T$
 ($1/\mu=(\sum_i m_i^{-1})$), 
 we obtain
 the Boltzmann equation for the number  per comoving volume $Y_i=n_i/s$
 ($n_i$ is the number density of $\chi_i$, and $s$  is the entropy density)  \cite{Aoki:2012ub}:
\begin{eqnarray}
& &\frac{d Y_i}{d x}=-0.264~ g_*^{1/2} \left[\frac{\mu M_{\rm PL}}{x^2} \right]
\Big\{~
<\!\sigma (ii;X_i X'_i) v\!>\left(  Y_i Y _i-\bar{Y}_{i}\bar{Y}_{i}\right)
\nonumber\\
& &\left. +
\sum_{i >j }<\!\sigma (ii;jj)v\!>\!\!\left(  Y_i Y _i-\frac{Y_j Y_j}
{\bar{Y}_{j}\bar{Y}_{j}} \bar{Y}_{i}\bar{Y}_{i}
\right)
\!-\!\sum_{j >i}<\!\sigma (jj;ii)v\!>\!\!
\left(  Y_j Y _j-\frac{Y_i Y_i}
{\bar{Y}_{i}\bar{Y}_{i}} \bar{Y}_{j}\bar{Y}_{j}
\right)~\right.~\nonumber\\
& &\!+\!\sum_{j,k}<\!\sigma (ij;k X_{ijk})v\!>\!\!
\left(  Y_i Y _j-\frac{Y_k}
{\bar{Y}_{k}} \bar{Y}_{i}\bar{Y}_{j}
\right)-\!\sum_{j,k}<\!\sigma (jk;i X_{jki})v\!>\!\!
\left(  Y_j Y _k-\frac{Y_i}
{\bar{Y}_{i}} \bar{Y}_{j}\bar{Y}_{k}
\right)\Big\},
\label{boltz}
\end{eqnarray}
where
 $g_*$ is the total number of effective
degrees of freedom, $ T$ and $M_{\rm PL}$ are the temperature and  the Planck mass,
respectively. $<\sigma (iiX_i X'_i) v>$ e.t.c. are
 thermally-averaged cross sections.
We have integrated this system of coupled non-linear
differential equations numerically in  fictive models with $K=2$ and $3$ and found
that
 the non-standard DM annihilation processes
 can influence
 the  relic abundance of DM significantly,
which has been  recently  also found for 
 two-component DM systems
in   \cite{D'Eramo:2010ep,Belanger:2011ww,Belanger:2012vp}. 

\section{A Model with three dark matter particles}
To consider  a concrete three-component DM system we   extend  \cite{Aoki:2012ub}
the  radiative seesaw model of \cite{Ma:2006km}
by adding  an extra Majorana fermion $\chi$ and an extra real scalar boson $\phi$.
The matter content is given in Table I.
\begin{table}
\caption{\footnotesize{The matter content of the model and
the corresponding quantum numbers. 
The quarks are suppressed.}}
\begin{center}
\begin{tabular}{|c|c|c|c|c|} 
\hline
field & $SU(2)_L$ & $U(1)_Y$ & $Z_2$ &$Z'_2$	\\ \hline\hline
$(\nu_{Li},l_i)$	& $2$ 	& $-1/2$ 	& $+$ & $+$	\\ \hline
$l_{i}^c$ 		& $1$ 	& $1$		& $+$ & $+$	\\ \hline
$N^c_{i}$			& $1$		& $0$ 	& $-$ & $+$	\\ \hline
$H=(H^+,H^0)$& $2$ 	& $1/2$ 	& $+$ & $+$	\\ \hline
$\eta=(\eta^+,\eta^0)$& $2$ 	& $1/2$ 	& $-$ & $+$	\\ \hline
$ \chi $       & $1$    & $0$        & $+$ & $-$	\\ \hline
$ \phi $     & $1$    & $0$        & $-$ & $-$	\\ \hline
\end{tabular}
\end{center}
\end{table}
The DM stabilizing symmetry is promoted to $Z_2\times Z'_2$,
 and we have assumed that 
$\eta_R^0$ (the CP even neutral component 
of the  inert Higgs $SU(2)_L$ doublet), $\chi$ and $\phi$ are DM particles.
The $\eta_R^0$ dark matter in the inert Higgs model without 
$\phi$ and $\chi$ has been studied in 
\cite{Barbieri:2006dq,LopezHonorez:2006gr,Dolle:2009fn},
while $N^c$ dark matter has been studied in \cite{Kubo:2006yx}.

\subsection{Relic abundance of dark matter}
To compute the relic abundance of DM we have to  take into account
 the following 
annihilation processes:
\begin{eqnarray}
& &\bullet \eta^0_R ~\eta^0_R \leftrightarrow \mbox{SMs}~,~
\bullet \phi ~\phi  \leftrightarrow \mbox{SMs}~
(\mbox{Standard annihilation})\\
& &
\bullet\eta^0_R~ \eta^0_R \leftrightarrow \phi ~\phi ~,~
\bullet \chi ~\chi\leftrightarrow  \phi~ \phi ~
(\mbox{Conversion})\\
& &
\bullet\eta^0_R ~\chi \leftrightarrow \phi ~\nu_L~,~
\bullet\chi ~\phi \leftrightarrow \eta^0_R ~\nu_L~,~
\bullet\phi ~\eta^0_R \leftrightarrow \chi ~\nu_L~
(\mbox{Semi-annihilation})~, \label{semiannihilation}
\end{eqnarray}
where $\eta_R^0~(\eta_I^0)$ is the real (imaginary) part of $\eta^0$, and
we  have neglected the coannihilations such as that of $\eta_R^0$ with
$\eta_I^0$ and $\eta^\pm$. 
The lower mass region
$60 ~\mbox{GeV} < m_{\eta_R^0} < 80~\mbox{GeV}$
is consistent  with all the experimental constraints  
in the absence of $\chi$ and $\phi$ \cite{Dolle:2009fn,Gustafsson:2012aj}, where
it is noted that there exists a marginal possibility to expend 
slightly this upper bound \cite{LopezHonorez:2010tb}.
But the elastic cross section
$\sigma(\eta_R^0) \simeq 7.9 \times 10^{-45}
(\lambda_L/0.05)^2 (60~\mbox{GeV}/m_{\eta_R^0})^2~\mbox{cm}^2$
with  $\lambda_L > 0.05$ in this mass range  
may exceed the upper bound of 
the XENON100 result \cite{Aprile:2011hi},
$ 2\times 10^{-45}~\mbox{cm}^2$ for the DM mass $55$ GeV
at $90$ \% C.L.
In the presence of $\chi$ and $\phi$  the situation changes.
The separation of two   allowed regions of $m_{\eta_R^0}$ 
\cite{Barbieri:2006dq,LopezHonorez:2006gr,Dolle:2009fn}
disappears, because  $\chi$ and $\phi$  also contribute to
the relic abundance.
To see how the allowed parameter space of the model without $\chi$ and $\phi$
changes, we have considered  a set of $(\delta_1=m_{\eta^\pm}-m_{\eta_R^0}~,~
\delta_2=m_{\eta_I^0}-m_{\eta_R^0})$, for which the
allowed parameter space without $\chi$ and $\phi$  is very small.
For  $(\delta_1=10~,~
\delta_2=10)$ GeV, for instance, there is no allowed 
range of $m_{\eta_R^0} < 500$ GeV \cite{Dolle:2009fn} without $\chi$ and $\phi$; 
the low mass  range of $m_{\eta_R^0}$, for  which
the relic abundance $\Omega_\eta h^2$ is consistent,
does not satisfy the LEP constraint. 
The LEP constraint implies
that the region satisfying $m_{\eta_R^0} >  80$ GeV
and $m_{\eta_I^0} > 100$ GeV
with $\delta_2 < 8$ GeV is excluded \cite{Dolle:2009fn}.
Therefore, for  $(\delta_1=10~,~
\delta_2=10)$ GeV we have to consider 
only $m_{\eta_R^0} > 80$ GeV.
Our calculations  \cite{Aoki:2012ub} have shown that  the region 
$m_{\eta_R^0} > 80$ GeV
indeed becomes an allowed area in the presence of  $\chi$ and $\phi$. 

 \subsection{Direct detection}
\begin{figure}
\hspace{4cm}  \includegraphics[width=10cm]{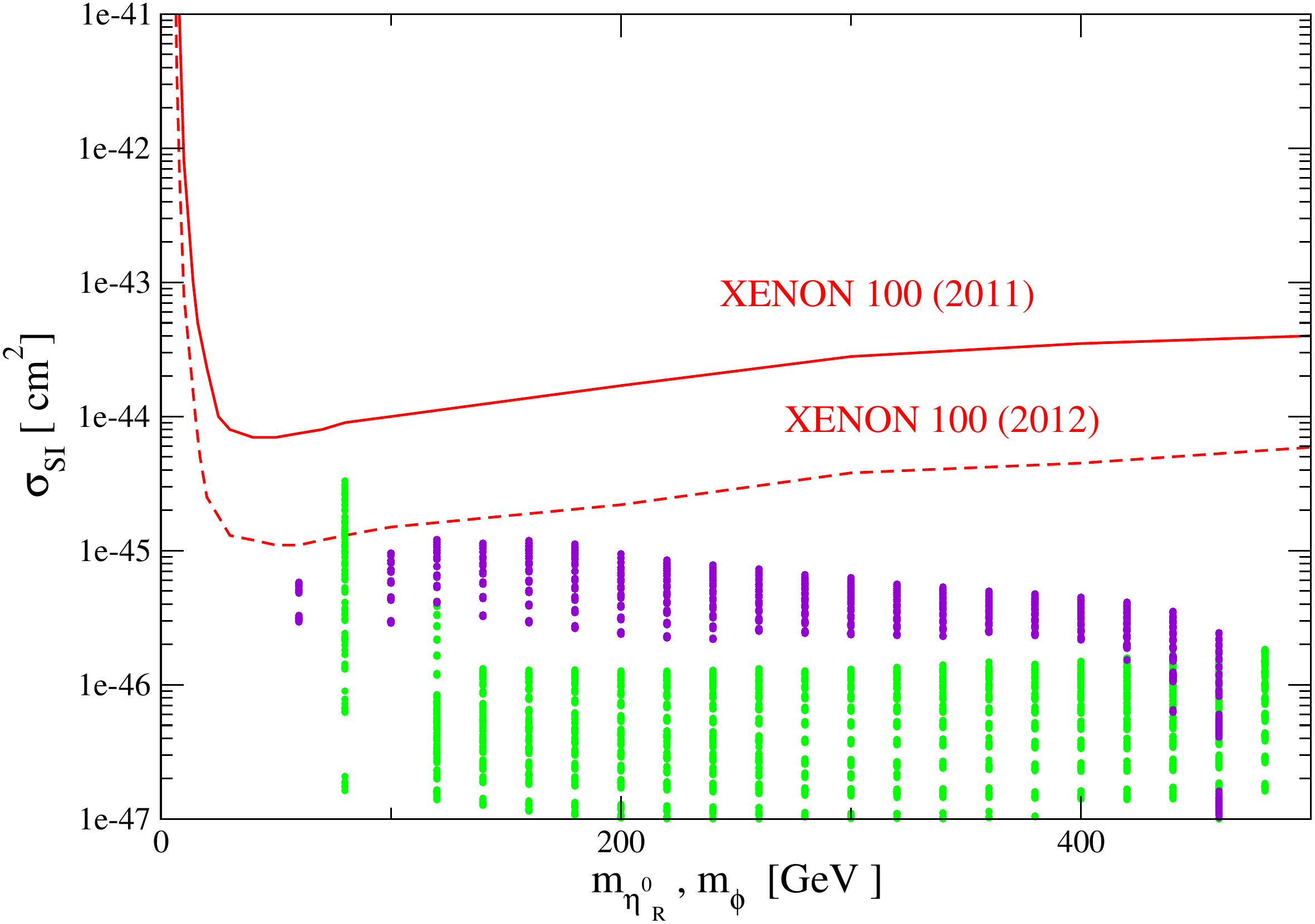}
\caption{\label{direct}\footnotesize
The spin-independent cross section
off the nucleon  is plotted as a function of the DM mass.
The green (violet) area stands for the $\eta_R^0$ ($\phi$ ) dark matter.}
\end{figure}

Fig.~\ref{direct} shows the spin-independent cross section off the nucleon
versus the DM mass of the present model;
the green area for the $\eta_R^0$ dark matter and the violet area for the $\phi$ dark matter.
(the spin-independent cross section for the $\chi$ dark matter
is suppressed, because it has no tree-level interaction with the nucleon.)
All the  constraints from  collider experiments,
and perturbativity,  in addition to those coming
from $\mu \to e\gamma$, $g-2$ of muon, the stability of the vacuum and 
the electroweak precision measurements are imposed,
where we have used:
$\delta_1=\delta_2=10$ GeV with
$m_\chi=m_{\eta_R^0}-10~\mbox{GeV}$,
$m_\phi=m_{\eta_R^0}-20~\mbox{GeV}$ and 
$M (\mbox{mass of right-handed neutrino})=1000$ GeV. 
We see from Fig.~\ref{direct} that the spin-independent cross sections are not only consistent
with the current bound of XENON100 \cite{Aprile:2011hi}, but also are within the accessible range of future experiments.
We  see here, too,  that the previously found 
separation \cite{Barbieri:2006dq,LopezHonorez:2006gr,Dolle:2009fn} of the allowed
parameter space in the low and high mass regions for $\eta_R^0$
disappears in the presence of $\chi$ and $\phi$  \cite{Aoki:2012ub}.

 \subsection{Indirect detection}
As we see from (\ref{semiannihilation}) the semi-annihilation 
produces  a left-handed neutrino.
Therefore,  it would be interesting to investigate
the neutrinos which are produced by the semi-annihilation of the captured DM  
in the Sun.
These neutrinos 
are monochromatic ($E_\nu\simeq m_{\eta^0_R}+m_\phi-m_\chi$  for instance)
and can be observed at neutrino telescopes 
\cite{Tanaka:2011uf,IceCube:2011aj}.
Diffuse neutrinos from the decay of $W$ and $b$ that  
 are produced by the standard annihilation of the captured DM in the Sun
 can also arrive at neutrino telescopes 
  \cite{Silk:1985ax,Griest:1986yu,Ritz:1987mh}
(see \cite{ Agrawal:2008xz} for the case of the inert Higgs model
without $\phi$ and $\chi$).

To compute  the DM numbers in the Sun, we have to solve their evolution equations  in the Sun.
 In contrast to the one-component DM case, they are coupled in the multi-component DM case
  \cite{Aoki:2012ub}:
\begin{eqnarray}
\dot{N}_\eta &=&
C_\eta-C_A(\eta \eta \leftrightarrow \mbox{SM}) N_\eta^2
-C_A(\eta \eta \leftrightarrow \phi \phi)  N_\eta^2
-C_A(\eta \chi\leftrightarrow \phi \nu_L) N_\eta  N_\chi \nonumber\\
& &
-C_A(\eta \phi\leftrightarrow \chi \nu_L)N_\eta N_\phi
+C_A(\phi \chi\leftrightarrow \eta\nu_L) N_\chi N_\phi~,\nonumber\\
\dot{N}_\chi&=&
C_\chi
-C_A(\chi \chi \leftrightarrow \phi \phi)  N_\chi^2
-C_A(\eta \chi\leftrightarrow \phi \nu_L) N_\eta  N_\chi \nonumber\\
& &
+C_A(\eta \phi\leftrightarrow \chi \nu_L)N_\eta N_\phi
-C_A(\phi \chi\leftrightarrow \eta\nu_L) N_\chi N_\phi~,
\label{evolution}\\
\dot{N}_\phi &=&
C_\phi-C_A(\phi \phi \leftrightarrow \mbox{SM}) N_\phi^2
+C_A(\eta \eta \leftrightarrow \phi \phi)  N_\eta^2
+C_A(\chi \chi \leftrightarrow \phi \phi)  N_\chi^2\nonumber\\
& &+C_A(\eta \chi\leftrightarrow \phi \nu_L) N_\eta  N_\chi 
-C_A(\eta \phi\leftrightarrow \chi \nu_L)N_\eta N_\phi
-C_A(\phi \chi\leftrightarrow \eta\nu_L) N_\chi N_\phi~,\nonumber
\end{eqnarray}
where the number of $ \eta(=\eta_R^0), \chi$ and $\phi$
in the Sun are denoted by $N_i$ with $i=\eta, \chi$ and $ \phi$.
The $C_i$'s are  the capture rates in the Sun and the $C_A$'s are given by 
\cite{Griest:1986yu}
\begin{eqnarray}
C_A(ij \leftrightarrow \bullet) &=& 
\frac{<\sigma(i j ;\bullet) v>}{V_{ij}}~,~
V_{ij}=5.7\times 10^{27} \left( \frac{100~\mbox{GeV}}{\mu_{ij}}
 \right)^{3/2} \mbox{cm}^3\, ,
\end{eqnarray}
with
$ \mu_{ij}=2 m_i m_j /(m_i+m_j)$ in the limit $v\to 0$.
As we see from (\ref{evolution})
there exist fixed points of  the evolution equations.
So, they  describe approaching
 equilibrium between the capture  and annihilation rates of DM, where
 we set the numbers $N_i$   equal to zero at the time of birth of the Sun.
The annihilation, conversion and semi-annihilation rates at $t=t_\odot$
(the age of the Sun $\simeq 4.5\times 10^9$ years) are then given by
\begin{eqnarray}
\Gamma(ij;\bullet ) &=& d_{ij}C_A(ij \leftrightarrow \bullet) 
N_i (t_\odot) N_j(t_\odot) ~,
\end{eqnarray}
where $d_{ii}=1/2$ and $d_{i j}=1$ for $i\neq j$.
We have solved (\ref{evolution}) numerically 
for a benchmark set of parameters and calculated
the annihilation rates:
 \begin{eqnarray}
\Gamma(\mbox{SM}) &=&
C_A(\eta \eta \leftrightarrow \mbox{SM}) N_\eta^2/2
+C_A(\phi \phi \leftrightarrow \mbox{SM}) N_\phi^2/2~,\\
\Gamma(\nu) &=&
C_A(\eta \phi\leftrightarrow\chi\nu) N_\eta N_\phi
+C_A(\eta \chi\leftrightarrow \phi\nu) N_\eta N_\chi
+C_A(\chi \phi\leftrightarrow \eta\nu) N_\chi N_\phi~,
\label{gamma-nu}\\
\Gamma(\nu\nu) &=&
C_A(\eta \eta \leftrightarrow \nu\nu) N_\eta^2/2~
\end{eqnarray}
 at $t=t_\odot$, where $\Gamma(\mbox{SM})$ can be related to the observation rates
of the diffuse neutrinos.
We have obtained   \cite{Aoki:2012ub}:
$0.28\times 10^{20}~\mbox{s}^{-1}$ for $\Gamma(\mbox{SM})$,
$ 1.1\times 10^{-3}\times 10^{20}~\mbox{s}^{-1}$ for $\Gamma(\nu)$, and
$1.3 \times 10^{-7}\times 10^{20}~\mbox{s}^{-1}$ for $\Gamma(\nu\nu)$.
$\Gamma(\mbox{SM})$  is consistent with the  recent upper limit 
$\sim 2.73 \times 10^{21}~\mbox{s}^{-1}$ 
for $m_{\rm DM}=250~\mbox{GeV}$
of  the AMANDA-II / IceCube neutrino
telescope  \cite{IceCube:2011aj}.
As for the monochromatic neutrinos we use
$  \Gamma_{\rm detect} = A P(E_\nu) \Gamma_{\rm inc}$ \cite{Halzen:2010yj}
to estimate the detection rate   $\Gamma_{\rm detect}$,
  where $A$ is the detector area facing 
  the incident beam, $P(E_\nu)$ 
  is the probability for detection as a function of the neutrino energy $P(E_\nu)$,
  and  $\Gamma_{\rm inc}=\Gamma/4 \pi R_\odot^2$ is 
the incoming neutrino flux
 ($ R_\odot$ is the distance to the Sun). We have obtained 
for the benchmark set of parameters that  $0.05$ events of 
monochromatic neutrinos with $\sim 200$ GeV per year 
may be detected at IceCube, where
we have used: $A=1\mbox{km}^2$, $L=1\mbox{km}$  
\cite{Halzen:2010yj,IceCube:2011aj}.
 $0.05$ events per year 
 may be too small to be  realistic.
 However, we would like to note that we have studied only one point in the whole parameter space.
We also note that if at least one of the fermionic DM in a multi-component
DM system has  odd parity of the discrete lepton number,
a monochromatic left-handed neutrino,
which is also odd,
 can be produced together with this fermionic  DM in a semi-annihilation of DM's.
 
\section*{Acknowledgments}
J.~K.\  would like to thank the organizers of FLASY2012 for their
efforts and hospitality.
The work of M.~D.\ is supported by the International Max Planck Research 
School for Precision Tests of Fundamental Symmetries. 
The work of M.~A.\ is supported in part by Grant-in-Aid for Scientific 
Research for Young Scientists (B) (No.22740137), 
and J.~K.\ is partially supported by Grant-in-Aid for Scientific
Research (C) from Japan Society for Promotion of Science (No.22540271).

%% file: Papers/ludl.tex
\chapter[The finite subgroups of $SU(3)$ (Ludl)]{The finite subgroups of $SU(3)$}
\vspace{-2em}
\paragraph{P. O. Ludl}
\paragraph{Abstract}
The finite subgroups of SU(3) are frequently used in particle physics. Though they
were classified already at the beginning of the 20th century, there have been many
new and interesting developments in the last few years. In this article we will list
the finite subgroups of SU(3) and summarize some of their properties.

\section{Introduction}

Particle physics offers a wide range of applications for the theory of finite groups, and
in particular the finite subgroups of SU(3) have been intensively studied in the past.
The wide range of applications of SU(3)-subgroups covers different fields such as
hadron physics and computational tools in lattice QCD.
The field of particle physics which has made the most intensive
use of the finite subgroups of SU(3) in the recent years is flavour physics, where finite
SU(3)-subgroups are frequently used as symmetries in the quark, lepton and scalar sector~\cite{Review_Altarelli,Smirnov_Discrete}.

The classification of the finite subgroups of SU(3) presented in this article is based on
the work of H.F.~Blichfeldt as published in the famous book~\cite{MillerBlichfeldtDickson}.
A short summary of the history of the contributions to the analysis of the finite subgroups of SU(3)
(from a physicist's perspective) can be found in the introduction of~\cite{Comments_SU3_Ludl}.

There is a lot of literature covering aspects of the finite subgroups of SU(3). Apart from the classic
textbook~\cite{MillerBlichfeldtDickson} we refer the reader to the review
articles~\cite{Ishimori, Review_Grimus_Ludl, Ludl_Diploma_Thesis} and references therein.

\section{The finite subgroups of SU(3)}\label{POL-section-SU3}

In 1916 H.F.~Blichfeldt classified the finite subgroups of SU(3) into
the following five classes~\cite{MillerBlichfeldtDickson}.
\begin{itemize}
 \item[(A)] Abelian groups.
 \item[(B)] Finite subgroups of SU(3) with faithful two-dimensional representations.
 \item[(C)] The groups $C(n,a,b)$ generated by the matrices
                \begin{equation}\label{POL_Cgenerators}
                E=\left(
                \begin{array}{ccc}
                0 & 1 & 0\\
                0 & 0 & 1\\
                1 & 0 & 0
                \end{array}
                \right),\quad
                F(n,a,b) = \mathrm{diag}(\eta^a,\,\eta^b,\,\eta^{-a-b}),
                \end{equation}
                where $\eta=\exp(2\pi i/n)$, $n\in\mathbb{N}\backslash\{0,1\}$ and $a,b\in\{0,...,n-1\}$.
 \item[(D)] The groups $D(n,a,b;d,r,s)$ generated by $E$, $F(n,a,b)$ and
                \begin{equation}\label{POL_Dgenerators}
                \widetilde{G}(d,r,s)=\left(
                \begin{array}{ccc}
                \delta^r & 0 & 0\\
                0 & 0 & \delta^s\\
                0 & -\delta^{-r-s} & 0
                \end{array}
                \right),
                \end{equation}
                where $\delta=\exp(2\pi i/d)$, $d\in\mathbb{N}\backslash\{0\}$ and $r,s\in\{0,...,d-1\}$.
 \item[(E)] Six exceptional finite subgroups of SU(3):
            \begin{itemize}
             \item $\Sigma(60)\cong A_5$, $\Sigma(168)\cong \mathrm{PSL(2,7)}$,
             \item $\Sigma(36\times 3)$, $\Sigma(72\times 3)$, $\Sigma(216\times 3)$
                   and $\Sigma(360\times 3)$,
            \end{itemize}
            as well as the direct products $\Sigma(60)\times\mathbb{Z}_3$ and $\Sigma(168)\times\mathbb{Z}_3$.
\end{itemize}

In the following we will go through these five types of groups and dwell a bit on the structures of their members.

\paragraph{(A) Abelian groups}

The possible structures of the Abelian finite subgroups of SU(3) are strongly
restricted by the following theorem~\cite{Comments_SU3_Ludl}.
\medskip
\\
\textbf{Theorem 1.} Every finite Abelian subgroup $\mathcal{G}$ of SU(3) is isomorphic to $\mathbb{Z}_m\times\mathbb{Z}_p$, where
        \begin{equation}
        m=\max_{a\in \mathcal{G}}\mathrm{ord}(a)
        \end{equation}
and $p$ is a divisor of $m$.
\medskip

Thus every finite Abelian subgroup of SU(3) is either a cyclic group or a direct product of two cyclic groups.
Examples for cyclic subgroups of SU(3) are the three-dimensional rotation groups about one axis. An example for a direct
product of two cyclic groups is Klein's four group
$\mathbb{Z}_2 \times \mathbb{Z}_2$.

\paragraph{(B) Groups with two-dimensional faithful representations}

Suppose we are given a finite group possessing a two-dimensional faithful representation (i.e. a finite
subgroup of U(2)), then via the homomorphism
    \begin{equation}
    A\mapsto
    \left(
    \begin{array}{cc}
    \mathrm{det} A^{\ast} &  0\\
    0 & A
    \end{array}
    \right)\in\mathrm{SU(3)}\quad\quad(A\in\mathrm{U(2)})
    \end{equation}
we can construct an isomorphic finite subgroup of SU(3)~\cite{Review_Grimus_Ludl}. In this way every finite subgroup of U(2)
can be interpreted as a finite subgroup of SU(3). Examples for SU(3)-subgroups possessing a two-dimensional
faithful representation are
\begin{itemize}
 \item the dihedral groups $D_n$ and
 \item the double covers $\widetilde{T}$, $\widetilde{O}$, $\widetilde{I}$, $\widetilde{D}_n$
       of the finite three-dimensional rotation groups.\footnote{The finite three-dimensional rotation groups (SO(3)-subgroups)
are~\cite{Hamermesh}: the rotation groups about one axis (cyclic groups), the dihedral groups $D_n$, the tetrahedral group $T\cong A_4$,
the octahedral group $O\cong S_4$ and the icosahedral group $I\cong A_5$.}
\end{itemize}

\paragraph{The groups of type (C)}

The groups $C(n,a,b)$ are generated by the permutation matrix $E$ and a diagonal
matrix $F(n,a,b)$--see equation~(\ref{POL_Cgenerators}). The subgroup $N(n,a,b)$ of all diagonal
matrices is generated by
\begin{equation}
F(n,a,b)\quad\mbox{and}\quad EF(n,a,b)E^{-1},
\end{equation}
from which follows that $N(n,a,b)$ is a normal subgroup of $C(n,a,b)$. Therefore the groups of type (C) have the
structure of a semi-direct product
\begin{equation}
C(n,a,b)\cong N(n,a,b)\rtimes\mathbb{Z}_3,
\end{equation}
where the $\mathbb{Z}_3$-subgroup is generated by $E$. Since $N(n,a,b)$ is an Abelian finite subgroup of SU(3), we
can use theorem 1 to arrive at
\begin{equation}\label{POL_Cstructure}
C(n,a,b)\cong (\mathbb{Z}_m\times\mathbb{Z}_p)\rtimes\mathbb{Z}_3.
\end{equation}
There are two important special cases emerging from~(\ref{POL_Cstructure}):
\begin{itemize}
 \item $p=1$ $\Rightarrow$ Groups of the type\footnote{$m$ must be a product of powers of primes of the
form $6k+1$, $k\in\mathbb{N}$~\cite{Ishimori}.} $T_m\cong \mathbb{Z}_m\rtimes\mathbb{Z}_3$.
 \item $p=m$ $\Rightarrow$ Groups of the type
$(\mathbb{Z}_m\times\mathbb{Z}_m)\rtimes\mathbb{Z}_3\cong \Delta(3m^2)$.
\end{itemize}

Examples for groups of type (C) are well-known groups such as $A_4\cong \Delta(12)$,
$\Delta(27)$, $T_7$ and $T_{13}$. However, there are also groups of type (C) which are neither of the form
$T_m$, nor of the form $\Delta(3m^2)$. The smallest example, which is not a direct product, is the group~\cite{Comments_SU3_Ludl}
\begin{equation}
C(9,1,1)\cong (\mathbb{Z}_9\times\mathbb{Z}_3)\rtimes\mathbb{Z}_3.
\end{equation}

\paragraph{The groups of type (D)}
The groups $D(n,a,b;d,r,s)$ are generated by the generators $E$ and $F(n,a,b)$ of (C) and the additional
generator $\widetilde{G}(d,r,s)$--see equation~(\ref{POL_Dgenerators}). 
It was shown in~\cite{Review_Grimus_Ludl} that by means of a unitary transformation one can obtain a different
set of generators consisting of three diagonal matrices and the two $S_3$-generators
\begin{equation}
                E=\left(
                \begin{array}{ccc}
                0 & 1 & 0\\
                0 & 0 & 1\\
                1 & 0 & 0
                \end{array}
                \right)\quad\mbox{and}\quad
                B=\left(
                \begin{array}{rrr}
                -1 & 0 & 0\\
                0 & 0 & -1\\
                0 & -1 & 0
                \end{array}
                \right).
\end{equation}
Thus, as in the case of (C), the subgroup $N(n,a,b;d,r,s)$ of diagonal matrices is
an invariant subgroup, and the structure of the groups of type (D) is found to be
\begin{equation}
D(n,a,b;d,r,s)\cong (\mathbb{Z}_m\times\mathbb{Z}_p)\rtimes S_3,
\end{equation}
where $N(n,a,b;d,r,s)\cong\mathbb{Z}_m\times\mathbb{Z}_p$ and $S_3$ is generated by $E$ and $B$.

For the special case of $p=m$ we obtain the groups $(\mathbb{Z}_m\times\mathbb{Z}_m)\rtimes S_3\cong\Delta(6m^2)$.
Thus the groups of type (D) comprise the well-known groups $S_4\cong \Delta(24)$, $\Delta(54)$ and $\Delta(96)$.
The smallest group of type (D), which is neither a direct product, nor of the form $\Delta(6m^2)$, is~\cite{Comments_SU3_Ludl}
\begin{equation}
D(9,1,1;2,1,1)\cong (\mathbb{Z}_9\times\mathbb{Z}_3)\rtimes S_3.
\end{equation}

\paragraph{(E) The exceptional finite subgroups of SU(3)} This set of groups collects all finite subgroups
of SU(3) which do not fall into one of the categories (A)--(D). Among these groups are the two simple groups
$\Sigma(60)\cong I\cong A_5$ and $\Sigma(168)\cong \mathrm{PSL}(2,7)$. For a detailed treatment of these two groups
we refer the reader to~\cite{Luhn_Nasri_Ramond}. Also the direct products $\Sigma(60)\times\mathbb{Z}_3$ and
$\Sigma(168)\times \mathbb{Z}_3$ are finite subgroups of SU(3).

The remaining four exceptional groups are $\Sigma(36\times 3)$, $\Sigma(72\times 3)$,
$\Sigma(216\times 3)$ and $\Sigma(360\times 3)$. Their generators can be found in~\cite{MillerBlichfeldtDickson}.
For a detailed study of the first three groups we refer the reader to~\cite{Grimus_Ludl_Principal}. The largest exceptional
finite SU(3)-subgroup $\Sigma(360\times 3)$ possesses only one non-trivial invariant subgroup, namely the
center $\{\mathbb{1},\,\omega\mathbb{1},\,\omega^2\mathbb{1}\}\cong\mathbb{Z}_3$ of SU(3) ($\omega=\exp(2\pi i/3)$).
The corresponding factor group $\Sigma(360)\equiv\Sigma(360\times 3)/\mathbb{Z}_3$ is isomorphic to the
permutation group $A_6$~\cite{Fairbairn_Fulton_Klink}. The character table of $\Sigma(360\times 3)$ can be found
in~\cite{Ludl_Diploma_Thesis}. $\Sigma(60)\cong I\cong A_5$ is a subgroup of $\Sigma(360\times 3).$

\section{Representations of the finite subgroups of SU(3)}

By definition a finite subgroup of SU(3) possesses at least one three-dimensional representation
of determinant one. This representation is not necessarily irreducible, however many SU(3)-subgroups possess
three-dimensional irreps.

\begin{itemize}
\item The groups of type (A) are Abelian, which implies that all their irreps are one-dimensional.

\item Much less is known about the representations of the groups of type (B), which are the finite subgroups of
U(2). By definition they possess at least one two-dimensional faithful representation. The dihedral groups $D_n$ and their
double covers possess only one- and two-dimensional irreps~\cite{Ramond}, while the double covers 
$\widetilde{T}$, $\widetilde{O}$ and $\widetilde{I}$
of the rotation groups $T$, $O$ and $I$ possess also three- and higher-dimensional irreps~\cite{Review_Grimus_Ludl}. 

\item It was shown in~\cite{Review_Grimus_Ludl} that the groups of type (C) possess only one- and three-dimensional
irreps.

\item Also the dimensions of the irreps of the groups of type (D) can be determined in general. A group of type (D)
can possess one-, two-, three- and six-dimensional irreps~\cite{Review_Grimus_Ludl}.

\item All of the exceptional finite subgroups (E) of SU(3) possess three-dimensional irreps. Since
$\Sigma(60)\cong I\cong A_5$ and $\Sigma(168)\cong\mathrm{PSL}(2,7)$ are simple, all their non-trivial
irreps are faithful. Detailed information on the irreps of $\Sigma(36\times 3)$, $\Sigma(72\times 3)$ and
$\Sigma(216\times 3)$ can be found in~\cite{Grimus_Ludl_Principal}. The largest exceptional SU(3)-subgroup
$\Sigma(360\times 3)$ possesses irreps of dimensions 1,3,5,6,8,9,10 and 15~\cite{Ludl_Diploma_Thesis}.
\end{itemize}

Let us finish this section with a theorem, which can be very helpful when one looks for
finite groups which possess irreps of a given dimension.
\medskip
\\
\textbf{Theorem 2.} The dimension of an irrep of a finite group is a divisor of the order of the group.
\medskip
\\
A collection of helpful theorems including references to their proofs
can be found in~\cite{Review_Grimus_Ludl}.

\section{Summary and outlook}

\begin{figure}[h]
\includegraphics[width=\textwidth]{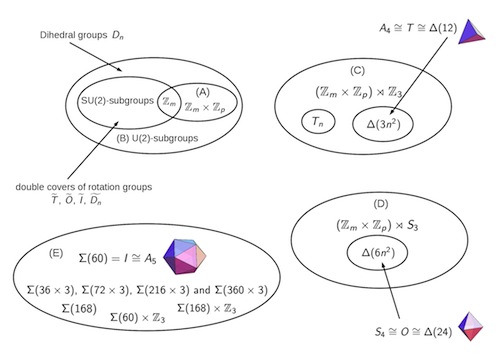}
\caption{The finite subgroups of SU(3) as presented in section~\ref{POL-section-SU3}.}\label{POL_SU3-subgroups}
\end{figure}

In this work we reviewed the structure of the finite subgroups of SU(3). A summary of these results can be
found in figure~\ref{POL_SU3-subgroups}, which shows the finite subgroups of SU(3) classified into the five types defined
in~\cite{MillerBlichfeldtDickson}.
Among the five types, we studied (A), (C) and (D) in more detail. The Abelian finite subgroups of SU(3) were found to have
the structure of a direct product $\mathbb{Z}_m\times\mathbb{Z}_p$. This insight allowed to determine the 
general structure of the hitherto not very well-known SU(3)-subgroups of type (C) and (D) which is
similar to the one of the well-known series $\Delta(3n^2)$ and $\Delta(6n^2)$, which are subseries of (C) and (D), respectively.

The finite subgroups of SU(3) comprise an interesting field of study, especially with respect to their application
as symmetries in particle physics.
An interesting question frequently arising in the context of flavour physics is the breaking of a group
to one of its subgroups. For works dealing with this question--especially in the context of
SU(3)-subgroups--we refer the reader to~\cite{Luhn_Pedestrian, Merle_Zwicky}.

Finally we would like to mention two very helpful tools for studying finite groups, namely the computer algebra system
GAP~\cite{GAP} and the SmallGroups library~\cite{SmallGroups}, a GAP-package which provides valuable
information on all finite groups up to order 2000.
Two examples for works where these tools have been successfully used are~\cite{Parattu-Wingerter,Ludl-U3-512}.

\section*{Acknowledgments}
The work of the author is supported by the Austrian Science Fund (FWF), Project No. P~24161-N16.

\bibliography{ludl}
\bibliographystyle{apsrev4-1}


%% file: Papers/Ma.tex

%
%
%
%
%
%

\chapter[$\textsf{A}_4$, $\theta_{13}$, and $\delta_{CP}$ (Ma)]{$\textsf{A}_4$, $\theta_{13}$, and $\delta_{CP}$}
\vspace{-2em}
\paragraph{E. Ma}
\paragraph{Abstract}
Since nonzero $\theta_{13}$ is now well established, tribimaximal neutrino mixing is no longer valid.  However, it only means that the previously very restrictive application of the non-Abelian discrete symmetry $A_4$ was a mistake.  A new simple application shows that for the current experimental central value of $\sin^2 2 \theta_{13} \simeq 0.1$, leptonic $CP$ violation is necessarily large, i.e.$|\tan \delta_{CP}| > 1.3$.

\section{Short History of A$_4$}
In 1978, soon after the putative discovery of the third family of leptons and 
quarks, it was conjectured by Cabibbo and Wolfenstein independently that 
\begin{equation}
U_{CW}^{l\nu} = \frac{1}{\sqrt{3}} \begin{pmatrix} 1 & 1 & 1 \\ 
1 & \omega & \omega^2 \\ 1 & \omega^2 & \omega \end{pmatrix},
\end{equation}
where $\omega = \exp (2 \pi i/3) = -1/2 + i \sqrt{3}/2$. This implies 
$\sin^2 \theta_{12} = \sin^2 \theta_{23} = 1/2$, $\sin^2 \theta_{13} = 1/3$, 
$\delta_{CP} = \pm \pi/2$, i.e. bibitrimaximal mixing.
In 2001, Ma and Rajasekaran showed that $U_{CW}$ occurs in $A_4$ which allows 
$m_{e,\mu,\tau}$ to be arbitrary, predicting also $\sin^2 2 \theta_{23} = 1$, 
$\theta_{13}=0$.  In 2002, Babu, Ma, and Valle showed how $\theta_{13} \neq 0$ 
can be radiatively generated in $A_4$ with $\delta_{CP} = \pm \pi/2$, i.e. 
maximal $CP$ violation.
In 2002, Harrison, Perkins, and Scott proposed the tribimaximal mixing 
matrix, i.e.
\begin{equation}
U_{HPS}^{l\nu} = \begin{pmatrix} \sqrt{2/3} & 1/\sqrt{3} & 0 \\ -1/\sqrt{6} & 
1/\sqrt{3} & -1/\sqrt{2} \\ -1/\sqrt{6} & 1/\sqrt{3} & 1/\sqrt{2} 
\end{pmatrix}.
\end{equation}
This means $\sin^2 2 \theta_{23} = 1$, $\tan^2 \theta_{12} = 1/2$, 
$\theta_{13} = 0$.
In 2004, I showed that this may be obtained in $A_4$, with
\begin{equation}
U^\dagger_{CW} {\cal M}_\nu U_{CW} = \begin{pmatrix} a+2b & 0 & 0 \\ 
0 & a-b & d \\ 0 & d & a-b \end{pmatrix}
\end{equation}
in the basis that ${\cal M}_l$ is diagonal.  At that time, SNO data gave 
$\tan^2 \theta_{12} = 0.40 \pm 0.05$, but it was changed in early 2005 
to $0.45 \pm 0.05$.  Tribimaximal mixing and $A_4$ then became part of 
the lexicon of the neutrino theorist.  After the 2005 SNO revision, two 
$A_4$ models quickly appeared. (1) Altarelli and Feruglio proposed 
\begin{equation}
U^\dagger_{CW} {\cal M}_\nu U_{CW} = \begin{pmatrix} a & 0 & 0 \\ 
0 & a & d \\ 0 & d & a \end{pmatrix},
\end{equation}
i.e. $b=0$, and (2) Babu and He proposed
\begin{equation}
U^\dagger_{CW} {\cal M}_\nu U_{CW} = \begin{pmatrix} a' - d^2/a' & 0 & 0 \\ 
0 & a' & d \\ 0 & d & a' \end{pmatrix},
\end{equation}
i.e. $d^2 = 3b(b-a)$.  The challenge, however, is to prove experimentally that 
$A_4$ exists.  If $A_4$ is realized by a renormalizable theory at the 
electroweak scale, then the extra Higgs doublets required will bear this 
information.  Specifically, $A_4$ breaks to the residual symmetry $Z_3$ in 
the charged-lepton sector, and all Higgs Yukawa interactions are determined 
in terms of lepton masses.  This notion of lepton flavor triality 
[Ma, Phys. Rev. D 82, 037301 (2010)] (exact if neutrino masses are zero) 
may be the key to such a proof, and these exotic Higgs doublets could be 
seen at the Large Hadron Collider (LHC) [Cao et al., Phys. Rev. Lett. 106, 
131801 (2011), Phys. Rev. D 83, 093012 (2011)].

\section{Nonzero $\theta_{13}$ in A$_4$}
There is now very strong experimental evidence for nonzero $\theta_{13}$. 
The Daya Bay collaboration reported $\sin^2 2 \theta_{13} = 0.089 \pm 0.010 \pm 
0.005$; the RENO collaboration $0.113 \pm 0.013 \pm 0.019$; and the Double 
CHOOZ collaboration $0.109 \pm 0.030 \pm 0.025$.  There is also some 
evidence for nonmaximal $\theta_{23}$, i.e. $\sin^2 2 \theta_{23} = 0.96 
\pm 0.04$ from the MINOS collaboration.  To understand this in the context 
of $A_4$, let
\begin{equation}
U^\dagger_{CW} {\cal M}_\nu U_{CW} = \begin{pmatrix} a & f & e \\ 
f & a & d \\ e & d & a \end{pmatrix},
\end{equation}
from 4 Higgs triplets $\sim \underline{1}, \underline{3}$ under $A_4$. 
The old idea was to enforce $e=f=0$ to obtain tribimaximal mixing. 
Technically this was very difficult (but not impossible) to do. 
Suppose $d,e,f$ are arbitrary (which is very easy to do), and let 
$b=(e+f)/\sqrt{2}$ and $c=(e-f)/\sqrt{2}$, then in the tribimaximal basis,
\begin{equation}
{\cal M}_\nu^{(1,2,3)} = \begin{pmatrix} a+d & b & 0 \\ 
b & a & c \\ 0 & c & a-d \end{pmatrix}.
\end{equation}
Note that the (1,3) and (3,1) entries are automatically zero.  If $a,b,c,d$ 
are all real, then
\begin{equation}
\sin^2 2 \theta_{23} \simeq 1 - 2 \sin^2 2 \theta_{13}.
\end{equation}
Since $\sin^2 2 \theta_{23} > 0.92$, it would predict $\sin^2 2 \theta_{13} 
< 0.04$ which is of course excluded by recent data.  This looks like bad 
news, but it is actually good news.

\section{Large  $\delta_{CP}$ in A$_4$}
In general, $a,b,c,d$ are not real, although $a$ may be chosen real by 
convention.  What the $A_4$ structure tells us is that there are 
relationships among the three masses, the three angles, and the three phases. 
To see how this works analytically, let us consider the simplifying case 
of $b=0$ (which may be maintained by an interchange symmetry), then 
${\cal M}_\nu^{(1,2,3)}$ can be diagonalized exactly by $U_\epsilon$ with an 
angle $\theta$ and a phase $\phi$.  Let $U' = U_{TB} U^T_\epsilon$, then
\begin{eqnarray}
&& U'_{e1} = \sqrt{\frac{2}{3}}, ~~~ U'_{e2} = \frac{\cos \theta}{\sqrt{3}}, ~~~ 
U'_{e3} = - \frac{\sin \theta}{\sqrt{3}} e^{-i\phi}, \\ 
&& U'_{\mu 3} = - \frac{\cos \theta}{\sqrt{2}} - \frac{\sin \theta}{\sqrt{3}} 
e^{-i\phi}, ~~~ U'_{\tau 3} = \frac{\cos \theta}{\sqrt{2}} - 
\frac{\sin \theta}{\sqrt{3}} e^{-i\phi}.
\end{eqnarray}
The angles $\theta_{12}, \theta_{23}, \theta_{13}$, and the phase $\delta_{CP}$ 
are extracted from $\tan^2 \theta_{12} = |U'_{e2}/U'_{e1}|^2$, 
$\tan^2 \theta_{23} = |U'_{\mu 3}/U'_{\tau 3}|^2$, and 
$\sin \theta_{13} e^{-i \delta_{CP}} = U'_{e3} e^{-i \alpha'_3/2}$, where 
$\alpha'_3$ depends on the specific values of the mass matrix. 
As a result,
\begin{eqnarray}
&& \tan^2 \theta_{12} = \frac{1-3\sin^2 \theta_{13}}{2}, \\ 
&& \tan^2 \theta_{23} = \frac{ \left( 1 - \frac{\sqrt{2} \sin \theta_{13} \cos 
\phi}{\sqrt{1-3\sin^2 \theta_{13}}} \right)^2 + \frac{2 \sin^2 \theta_{13} 
\sin^2 \phi}{1 - 3 \sin^2 \theta_{13}}}{ \left( 1 + \frac{\sqrt{2} 
\sin \theta_{13} \cos \phi}{\sqrt{1-3\sin^2 \theta_{13}}} \right)^2 
+ \frac{2 \sin^2 \theta_{13} \sin^2 \phi}{1 - 3 \sin^2 \theta_{13}}}.
\end{eqnarray}
Let $\sin \theta_{13} = 0.16$ (i.e. $\sin^2 2 \theta_{13} = 0.10$), then 
$\tan^2 \theta_{12} = 0.46$ which agrees well with data, but if $Im(c)=0$ 
as well, then $\phi=0$, and $\sin^2 2 \theta_{23} = 0.80$, which is ruled 
out.  Thus $\sin^2 2 \theta_{23} > 0.92$ implies $|\tan \phi| > 1.2$. 
In a full numerical analysis [Ishimori and Ma, arXiv:1205.0075], it is found 
that $b$ is indeed numerically very small and that $|\tan \delta_{CP}| > 1.3$ 
for $\sin^2 2 \theta_{23} > 0.92$.  The only solution in this case is for 
normal hierarchy of neutrino masses.

\section{Scotogenic Majorana Neutrino Mass}
In 2006, neutrino mass is linked to dark matter in a one-loop mechanism 
[Ma, Phys. Rev. D 73, 077301 (2006)] by having a second scalar doublet 
$(\eta^+,\eta^0)$ and three neutral fermion singlets $N_{1,2,3}$, all of 
which are odd under an exactly conserved $Z_2$ symmetry whereas all 
standard-model particles are even.  This may be called 'scotogenic' from 
the Greek 'scotos' meaning darkness,  The $\eta$ doublet was proposed later 
by itself [Barbieri, Hall, and Rychkov, Phys. Rev. D 74, 015007 (2006)] 
and became known as 'inert', although it has both gauge and scalar 
interactions.  Since the $(1/2) \lambda_5(\Phi^\dagger \eta)^2 + H.c.$ term 
is allowed, $\eta^0 = 
(\eta_R + i \eta_I)/\sqrt{2}$ is split so that $\eta_{R,I}$ have different 
masses.  The one-loop diagram for scotogenic Majorana neutrino mass is 
exactly calculable and is given by
\begin{equation}
({\cal M}_\nu)_{ij} = \sum_k \frac{h_{ik} h_{jk} M_k}{16 \pi^2} \left[ 
\frac{m_R^2}{m_R^2 - M_k^2} \ln \frac{m_R^2}{M_k^2} - 
\frac{m_I^2}{m_I^2 - M_k^2} \ln \frac{m_I^2}{M_k^2} \right].
\end{equation}
In the limit $m_R^2 - m_I^2 = 2 \lambda_5 v^2 << m_0^2 = (m_R^2 + m_I^2)/2 << 
M_k^2$, this reduces to the so-called radiative seesaw:
\begin{equation}
({\cal M}_\nu)_{ij} = \frac{\lambda_5 v^2}{8 \pi^2} \sum_k \frac{h_{ik}h_{jk}}
{M_k} \left[ \ln \frac{M_k^2}{m_0^2} - 1 \right].
\end{equation}

\section{Scotogenic Nonzero $\theta_{13}$ and Large $\delta_{CP}$ in A$_4$}
Let $(\nu_i,l_i) \sim \underline{3}$, $l^c_i \sim \underline{1}, 
\underline{1}', \underline{1}''$ as before.  Add $(\eta^+,\eta^0) \sim 
\underline{1}$, and $N_i \sim \underline{3}$, then $\nu_i$ is connected 
to $N_i$ by the identity matrix.  The structure of the $N_i N_j$ Majorana 
mass matrix is then communicated to $\nu_i$ through $U_{CW}$ to $l_j$. 
Assume the same form as Eq.~(6), i.e.
\begin{equation}
{\cal M}_N = \begin{pmatrix}A & F & E \\ F & A & D \\ E & D & A \end{pmatrix},
\end{equation}
with $F = -E$, which may be maintained by gauging $B-L$ with scalars 
$\sigma_0 \sim \underline{1}$ and $\sigma_i \sim \underline{3}$ 
under $A_4$, and then broken by soft terms respecting the interchange 
symmetry $\sigma_1 \to \sigma_1$, $\sigma_2 \to -\sigma_3$, 
$\sigma_3 \to -\sigma_2$.  In the tribimaximal basis,
\begin{equation}
{\cal M}_N^{(1,2,3)} =  \begin{pmatrix}A+D & 0 & 0 \\ 0 & A & C \\ 
0 & C & A-D\end{pmatrix},
\end{equation}
where $C = (E-F)/\sqrt{2} = \sqrt{2}E$.  Rescale $M_k$ so that 
\begin{equation}
m'_k = \frac{1}{M_k} \left( \ln \frac{M_k^2}{m_0^2} - 1 \right).
\end{equation}
Using the inputs $\Delta m^2_{21} = 7.59 \times 10^{-5}$ eV$^2$ and 
$\Delta m^2_{32} = 2.45 \times 10^{-3}$ eV$^2$, five representative solutions 
for $\sin^2 2 \theta_{23} = 0.96$ and $\sin^2 2 \theta_{13} = 0.10$ are 
obtained [Ma, Natale, and Rashed, arXiv:1206.1570].

\begin{table}
\begin{center}
\begin{tabular}{|c|c|c|c|c|}
\hline
solution & Im(D) & class & $|\tan \delta_{CP}|$ & $m_{ee}$ \\ 
\hline
I & 0 & IH & 2.05 & 0.020 \\
II & Re(D) & IH & 4.65 & 0.022 \\ 
III & 0 & NH & 3.59 & 0.002 \\ 
IV & 0 & QD & 2.20 & 0.046 \\ 
V & Re(D) & QD & 1.84 & 0.051 \\ 
\hline
\end{tabular}
\end{center}
\end{table}
In contrast to the simplest model presented earlier which admits only 
normal hierarchy (NH) of neutrino masses, inverted hierrachy (IH) is 
also possible in this case as well as quasidegenerate (QD) masses.  
The effective neutrino mass $m_{ee}$ (in eV) in neutrinoless double beta 
decay is also displayed.

\section{Conclusion}
With the new precise measurement of $\sin^2 2 \theta_{13}$, tribimaximal 
mixing is dead, but not $A_4$.  In fact, the original $A_4$ model had two 
important parts: (A) diagonalizing the charged-lepton mass matrix with 
$U_{CW}$ for arbitrary values of $m_{e,\mu,\tau}$, (B) allowing the neutrino 
mass matrix to be restricted.  The special case of tribimaximal mixing 
requires a condition which is very difficult to enforce theoretically. 
Relaxing (B) and keeping (A) do very well with present data. Predictions 
for necessarily large $|\tan \delta_{CP}|$ and their associated $m_{ee}$ 
are given in two $A_4$ models.

\section*{Acknowledgments}
This work is supported in part by the U.~S.~Department of Energy under 
Grant No.~DE-AC02-06CH11357.

%% file: Papers/Merle.tex
\chapter[On explicit and spontaneous symmetry breaking -- in regard to $SU(3)$ and its finite subgroups (\textit{Merle}, Zwicky)]{On explicit and spontaneous symmetry breaking -- in regard to $SU(3)$ and its finite subgroups}
\vspace{-2em}
\paragraph{\textit{A. Merle}, R. Zwicky}

\paragraph{Abstract}
We discuss the breaking of $SU(3)$ to is finite subgroups. We show that the explicit and spontaneous symmetry breaking are in one-to-one correspondence through the representation functions of $SU(3)$, called complex spherical harmonics. We review the formalism of the Molien function which serves to construct all types of invariants, called primary and secondary. Further aspects of Ref.~\cite{Merle:2011vy} are summarised, such as the necessary and sufficient conditions for breaking of $SU(3) \to H$ for a large number of subgroups.

\section{Introduction}

These proceedings aim at  giving a short, largely self-contained, summary of the somewhat extensive work presented~\cite{Merle:2011vy}, where the breaking of $SU(3) \to  ... \to  H $ in chains of (finite) subgroups is discussed. The finite subgroups of $SU(3)$ are, for example, of interest for family symmetries aiming to uncover structures in mass and mixing patterns of the Standard Model fermion sector. Throughout this write-up we shall refer to $SU(3)$ [or $SO(3)$] as the parent group and to $H$ as the target group. We shall mostly use a language familiar to the physics community, occasionally introducing and using mathematical terminology when it seems economic.

It is an indisputable fact that the description of  phenomena in terms of symmetries and symmetry-breaking is one of the most powerful tools in physics. Of which, 
the known types are \emph{explicit} (ESB), \emph{spontaneous} (SSB), and \emph{anomalous symmetry breaking}. Often physics or the physicists choose, initially, to assign a higher symmetry to the problem than is usually visible in the phenomena, necessitating the inclusion of SSB and ESB respectively. The first point we would like to make, c.f.\ Sec.~\ref{AM:relation}, is that there is a one-to-one relation between the \emph{vacuum expectation value} (VEV) in SSB and the breaking term in ESB even though at the level of physics the two are very different. Accepting the formal correspondence, we choose to discuss the problem in terms
of the language of ESB which corresponds to invariant polynomials of the subgroups, Sec.~\ref{AM:invariants}. The construction of the latter  can almost be cast in algorithmic form, yet the non-trivial problem is to find the necessary and sufficient conditions for breaking into a particular group, and not into one of its parent or subgroups. This is known as the \emph{little group problem}, and it is unsolved in the general case. In~\cite{Merle:2011vy}, we presented solutions for $SU(3)$ and its subgroups by resorting to explicit fundamental representations (rep's), c.f. Sec.~\ref{AM:little}.

In order to ease the presentation and  reading we shall adhere to the concrete example of breaking from the three-dimensional rotation group to the four permutation group, $SO(3) \to S_4$. The fundamental rep of the latter can be defined by all $3 \times 3$ matrices $O$ of unit determinant satisfying $O^T O = 1$, or by all linear operation acting on the vector $(x,y,z)$, which leave the polynomials $x^2 + y^2 + z^2$ and $(xyz)^2$ invariant. The group $S_4$ is algebraically defined, also known as the \emph{presentation}, by all words that can be formed out of the symbols $a$ and $b$ subject to the constraints: $\{ a^4 = 1, b^3 = 1, a b^2 a = b\}$. A three-dimensional irreducible representation (irrep) of the latter is given by~\cite{Ludl:2009ft}: 
\begin{equation}
  a= \left( \begin{array}{ccc}
	-1 & 0 & 0 \\ 0 & 0 & -1 \\ 0 & 1 & 0
	\end{array} \right), \quad
	b = \left( \begin{array}{ccc}
	0 & 0 & 1 \\ 1 & 0 & 0 \\ 0 & 1 & 0
	\end{array} \right)\;.
	\label{AM_eq:abex}
\end{equation}

\section{A snapshot of $SU(3)$ breaking to its finite subgroups}

\subsection{\label{AM:relation}One-to-one: explicit and spontaneous symmetry breaking}

SSB is described, formally, by collecting the elements of a given rep which leave a specific element of the representation space, the VEV $v$, invariant. Only a subset of the initial group, say $SO(3)$, called the \emph{little group} $H$, leaves the VEV invariant, and thus $SO(3) \to H$. It seems worthwhile to emphasise that in the fundamental rep of $SO(3)$, which selects one preferred direction, this necessarily breaks to $SO(2)$ as should be clear from three-dimensional visualisation. Thus, for breaking into discrete subgroups it is necessary resort to higher dimensional  reps. 

\begin{figure}[t]
  \centering
  \includegraphics[width=3.0in, trim=0.85mm 0.85mm 0.85mm 0.85mm, clip]{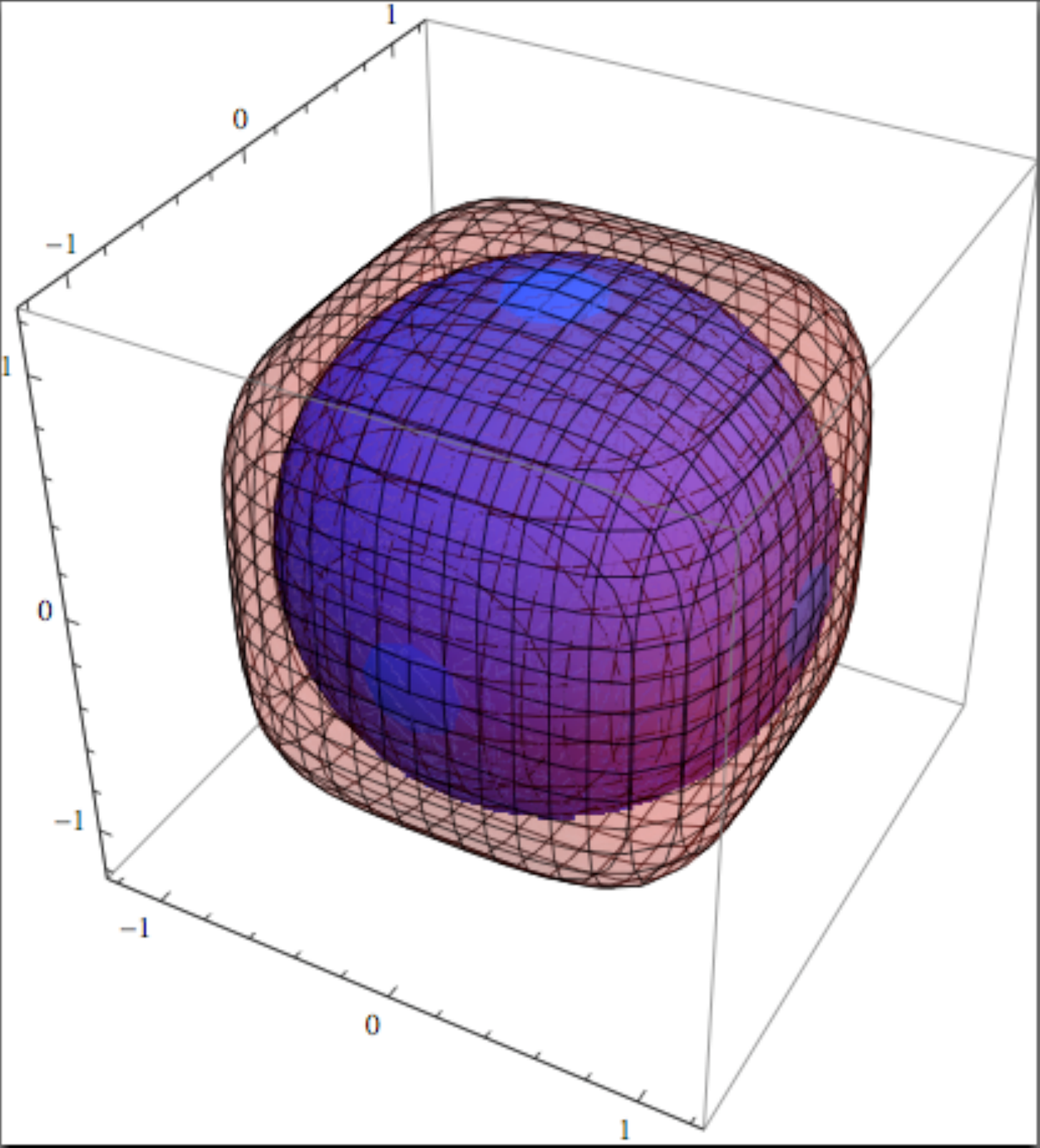}
  \includegraphics[width=3.0in, trim=0mm 0.85mm 0mm 0mm, clip]{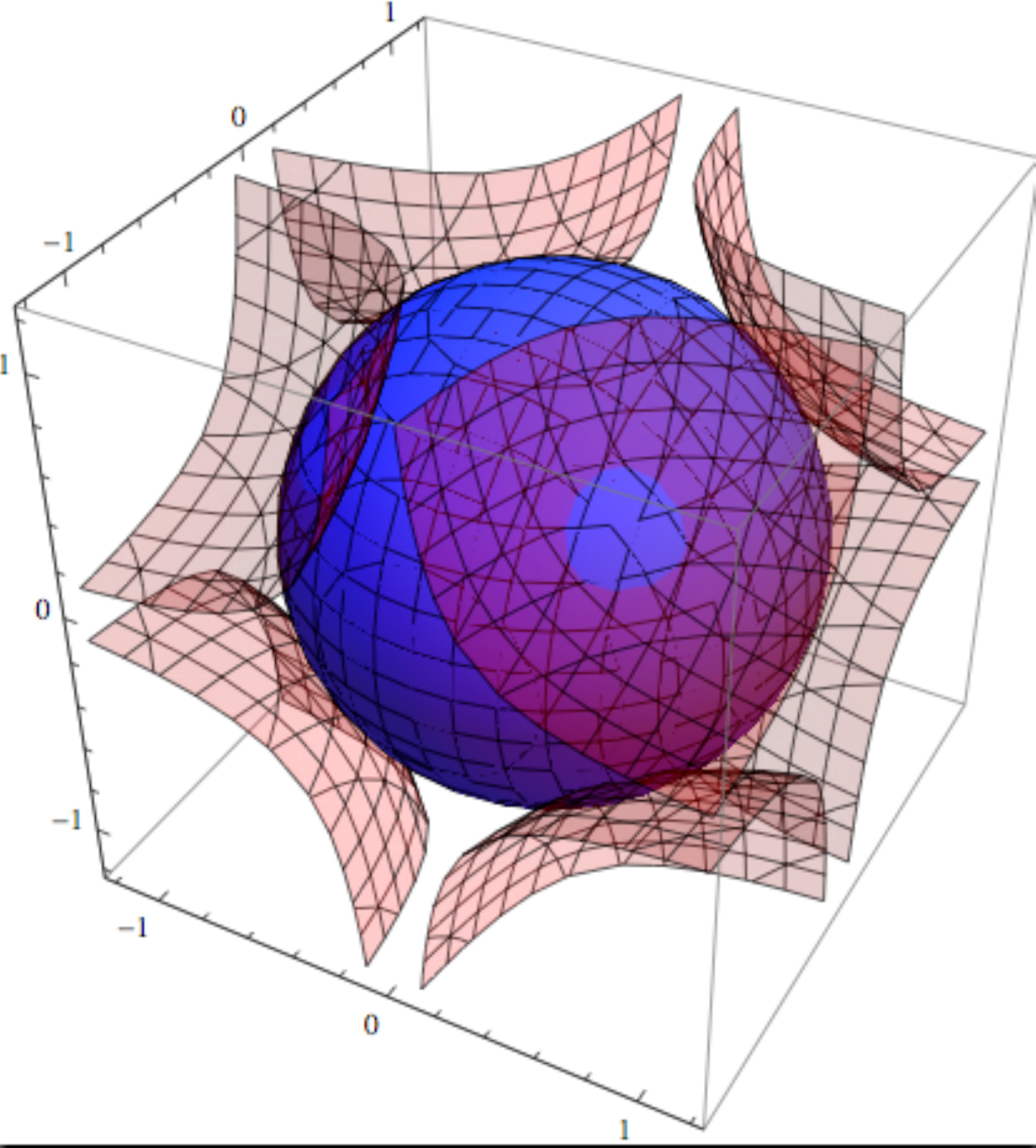} 
 \caption{\small The geometric illustration of the explicit breaking $SU(3) \to SO(3) \to  S_4$, by the invariant pairs $({\cal I}_2[S_4], {\cal I}_4[S_4])$ (left panel) and $({\cal I}_2[S_4], {\cal I}_6[S_4])$ (right panel), respectively. The invariant ${\cal I}_2[S_4]$~\eqref{AM_eq:IS4} breaks $SU(3) \to SO(3)$ and the further imposition of ${\cal I}_{4,6}[S_4]$, respectively, breaks $SO(3) \to S_4$. In both figures the two surfaces intersect at a finite number of objects, which can be scaled to points, and thus correspond to a finite group. In fact a cube (octahedron) and a hexagon emerge in the left and right panel respectively correspond to the geometric interpretation of $S_4$. Recall that the cube and the octahedron are dual under interchange of the number of edges and surfaces. (Figure taken from Ref.~\cite{Merle:2011vy}.)}
 \label{AM_fig:S4graphics}
\end{figure}

ESB is described by adding terms to a Lagrangian ${\cal L}$ which explicitly break the symmetry. In terms of the example $SO(3) \to S_4$:
\begin{equation}
 {\cal L}_{SO(3)}  \to  {\cal L}_{SO(3)} +  c {\cal I}_{S_4}  \;,  \qquad {\cal I}_4[S_4] =x^4 + y^4 + z^4 \;,
 \label{AM_eq:S4}
\end{equation}
where $c$ is a number and ${\cal I}_4[S_4]$ is an invariant of the $S_4$ irrep~\eqref{AM_eq:abex} but not of $SO(3)$. As we have shown in Ref.~\cite{Merle:2011vy}, ${\cal I}_4[S_4]$ is sufficient to break $SO(3) \to S_4$. This can be visually illustrated, cf.\ Fig.~\ref{AM_fig:S4graphics} (left). 

The key to the correspondence is to realise that the invariant polynomial ${\cal I}_4[S_4]$ can be expanded in representation functions of $SO(3)$, the celebrated \emph{spherical harmonics} $Y_{lm}$, which in turn furnish a rep space. Thus there is a one-to-one correspondence between the invariant polynomial ${\cal I}_4[S_4]$ and the $S_4$-VEV $v[S_4]$ of the nine-dimensional $l=4$ irrep of $SO(3)$:
\begin{eqnarray}
\label{AM_eq:Y4m}
& & \text{ESB:  }{\cal I}[S_4] = x^4 + y^4 + z^4  \; \sim \; 
 Y_{4-4} + \sqrt{\frac{14}{5}} Y_{4 0} +   Y_{44}  \; \quad  \longleftrightarrow  \nonumber  \\[0.1cm]
& & \text{SSB:  }  v[S_4] \sim (1,0,0,0, \sqrt{\frac{14}{5}},0,0,0,1) \;,
\end{eqnarray}
in a rep space ordered from $m = -4, ..,4$.\footnote{In terms of branching rules this reads:  ${\mathbf 9}_{SO(3)}|_{S_4} \to {\mathbf  1}_{S_4} + ..$, where
the dots stand for higher representations which we do not specify here.} This readily generalises to other Lie Groups, e.g.\ for $SU(3)$ the 5-parametric \emph{complex spherical harmonics} have to be considered~\cite{Merle:2011vy}.

\subsection{\label{AM:invariants}Invariant polynomials}

Once understood that invariants are essential to symmetry breaking, the question of how many invariants a group has got and how to construct them arises. Fortunately this is rather well-known territory in mathematics. According to Noether~\cite{noether}, there are exactly three algebraically independent invariants
for finite groups of three-dimensional irreps, and further information can be gained from the Molien function~\cite{Molien}.\footnote{An excellent and comprehensive treatment of this subject in mathematics can be found in Ref.~\cite{sturmfels}.} The latter is easily  computed  for a finite group $H$, by taking the average of the inverses of the characteristic polynomials of all matrices in a certain representation ${\cal R}(H)$ of $H$,
\begin{equation}
 M_{{\cal R}(H)}(P) \equiv \frac{1}{|{\cal R}(H)|} \sum_{h \in {\cal R}(H)} \frac{1}{\det(\mathbf{1}-P\, h)} = \sum_{m \geq 0} h_m  P^m \; .
 \label{AM_eq:Molien}
\end{equation}
According to the \emph{Molien theorem}~\cite{Molien} this function can  can be cast in a fraction of the following form
\begin{equation}
 M_{H(\text{\bf 3})}(P) = \frac{1 + \sum_{i \geq 1} a_{n_i} P^{n_i}}{(1-P^{m_1})(1-P^{m_2})(1-P^{m_3})} \; , \quad 
 a_{n_i} \in \text{positive integers,}
 \label{AM_eq:Molien_frac}
\end{equation}
for a three-dimensional irrep. The exponents $m_{1,2,3}$ in the denominator correspond to the degrees of the three so-called \emph{primary invariants} ${\cal I}_{m_{1,2,3}}$, while the exponents $n_i$ in the numerator signal the degrees of the \emph{secondary invariants} $\overline{{\cal I}}_{n_i}$.\footnote{The primary invariants are algebraically independent, whereas the secondary invariants obey relations of the form~\eqref{AM_eq:syzygy} called \emph{syzygies}.} The form of Eq.~\eqref{AM_eq:Molien_frac} has got to obey certain consistency conditions, such as $1 + \sum_i a_{n_i} = \frac{m_1\cdot m_2 \cdot m_3}{|H|}$~\cite{sturmfels}, which are of great help in practice to eliminate potential ambiguities, c.f.\ appendix~D.1 of~\cite{Merle:2011vy}. Note that the number of linearly independent invariants of degree, say $m$, corresponds to $h_m$ in the Taylor expansion of the Molien function, $M_{H(\text{\bf 3})}(P) = \sum_{m \geq 0} h_m P^m$. Knowing the degrees of the invariants,  the construction of the latter can be attempted by the \emph{Reynolds operator} (symmetrisation), 
\begin{equation}
 {\cal I}(x,y,z) =  \frac{1}{|{\cal R}(H)|} \sum_{h \in  {\cal R}(H) } f(h \circ x, h \circ y , h \circ z) \;,
 \label{AM_eq:genI}
\end{equation}
where $f(x,y,z)$ is some trial polynomial function, e.g., $f(x,y,z) = x^2 y z$ in case a four-dimensonial invariant is sought for. If by this procedure one is able to find a set of candidate primary and secondary invariants, there are only two things left to check: First the primary invariants must be algebraically independent, and second the squares of the secondary ivariants obey certain relations called \emph{syzygies}, of the following form:
\begin{equation}
\overline{{\cal I}}_{n_i}^{\;2} = f_0({\cal I}_{m_1},{\cal I}_{m_2},{\cal I}_{m_3}) + \sum_j f_1^{(j)}({\cal I}_{m_1},{\cal I}_{m_2},{\cal I}_{m_3}) \cdot \overline{{\cal I}}_{n_j}\; .
\label{AM_eq:syzygy}
\end{equation}
Once these two remaining points have been verified, a valid set of invariant polynomials for the group $H$ has been found.

For our example $S_4$, the Molien function  Eq.~\eqref{AM_eq:Molien} reads:
\begin{equation}
 M_{S_4}(P) =  \frac{1+P^9}{(1-P^2)(1-P^4)(1-P^6)}\;,
 \label{AM_eq:MolienS4}
\end{equation}
which suggests primary invariants of degrees $2$, $4$, and $6$, as well as one secondary invariant of degree $9$.\footnote{Strictly speaking, there is one more secondary invariant of degree 0, namely the trivial polynomial $1$.} Applying Eq.~\eqref{AM_eq:genI} immediately leads to primary and secondary invariants\footnote{Note that ${\cal I}_6[S_4]$ is an alternative to $ {\cal I}_4[S_4]$ in order to break from $SO(3) \to S_4$, cf.\ Fig.~\ref{AM_fig:S4graphics} (right). In SSB this correspond to $\mathbf{13}_{SO(3)}|_{S_4} \to {\mathbf  1}_{S_4} + ...$, where the 13-dimensional irrep orginates by $2l+1|_{l=6} = 13$ from the degree $l$ of the invariant polynomial.}:
\begin{eqnarray}
\label{AM_eq:IS4}
{\cal I}_2[S_4] &=& x^2 + y^2 + z^2\;, \quad {\cal I}_6[S_4] = (xyz)^2\;, \quad  {\cal I}_4[S_4] = x^4 + y^4 + z^4 \;, \nonumber \\[0.1cm]
\overline{{\cal I}}_9[S_4]  &=& xyz(x^2 - y^2)(y^2-z^2)(z^2-x^2) \; .
\end{eqnarray}
It is easy to show the algebraic independence of ${\cal I}_{2,4,6}$, and the syzygy 
\eqref{AM_eq:syzygy} reads explicitly:
\begin{equation}
 \overline{{\cal I}}_9^2  = {\cal I}_2^{\,4} {\cal I}_4 {\cal I}_6  -\frac{1}{4} {\cal I}_2^{\,6} {\cal I}_6 - \frac{5}{4} {\cal I}_2^{\,2} {\cal I}_4^{\,2} {\cal I}_6 + \frac{1}{2} {\cal I}_4^{\,3} {\cal I}_6 + 5 {\cal I}_2^{\,3} {\cal I}_6^{\,2} - 9 {\cal I}_2 {\cal I}_4 {\cal I}_6^{\,2} - 27 {\cal I}_6^{\,3} \;.
 \label{AM_eq:I9sq}
\end{equation}

\subsection{\label{AM:little}Little group problem: On  necessary \& sufficient conditions}

We shall first illustrate the little group problem and then discuss the solution for $SU(3)$ presented in~\cite{Merle:2011vy}. It is certain that a set of polynomials, say $ {\cal I}_{2,4,6}$~\eqref{AM_eq:IS4}, defines a group $H$. The subtle point is though that knowing a group or rep, say $H'$ or ${\cal R}(H')$, 
which leaves these polynomials invariant does necessarily mean that $H$ and $H'$ are  identical as they could be subgroups of each other: $H' \subset H$. For example $A_4 \subset S_4$, the group of even 4-permutations, leaves ${\cal I}_{2,4,6}$ invariant, too. Thus the problem can be reformulated as: finding the maximal group leaving a certain set of polynomials invariant. This problem has no general solution.\footnote{Let us note, though, that in the case where the parent group is finite all groups that lie within a sequence $P \subset .. H_i    \subset  H$ can be constructed algorithmically.}

The representation matrices of the finite $SU(3)$-subgroups have been known for a long time~\cite{Milleretal,FFK},\footnote{Modulo things as the topological structure such as (semi-) direct produtcs and the order of the groups. Progress in this direction is still ongoing, e.g.~\cite{Ludl:2011gn,Zwicky:2009vt}.} which can be cast in a small number of matrices with a few discrete parameters~\cite{Merle:2011vy,Ludl:2010bj}. Using these explicit reps it is possible to determine directly the maximal subgroup for a given set of polynomials. The explicit subgroup-tree, c.f.\ Fig.~4 and Tab.~6 in Ref.~\cite{Merle:2011vy}, is of great help in this endeavour. The robustness of the result with respect to the embedding is proven in Sec.~5 of~\cite{Merle:2011vy}.

\section{Epilogue}

After presenting a shortened version of the core of our work~\cite{Merle:2011vy} let us summarise what else has been done besides showing the equivalence of SSB and ESB, Sec.~\ref{AM:relation}, and the necessary and sufficient conditions, Sec.~\ref{AM:little}:

\begin{itemize}

\item Generalisations of the Molien function to tensor generating function, c.f.\ appendix~C of~\cite{Merle:2011vy}, from where the branching rules follow.

\item A Mathematica package, \emph{SUtree}, where the invariants, Molien and generating functions, syzygies, VEVs, branching rules, character tables, and Kronecker products can be obtained for the groups discussed above.
\begin{center}
{\tt http://theophys.kth.se/{$\sim$}amerle/SUtree/SUtree.html}
\end{center}

\item Computation of  all primary and secondary invariants for all proper finite subgroups of order smaller than $512$, 
and for the entire series of groups $\Delta(3n^2)$, $\Delta(6n^2)$ and all crystallgraphic groups, stored in the package \emph{SUtree} mentioned above.
\end{itemize}

As further possible extension we would like to mention that the breaking of $U(3)$ to its finite subgroups might be of interest as well, as the restriction to unit determinant is not necessary in general. This is however difficult as, to the best of our knowledge, no complete classification of finite $U(3)$-subgroups has been worked out. Let us emphasise that studying subgroups of $U(3) \simeq U(1) \times SU(3)$ is by far more complicated than studying the subgroups of $U(1)$ and $SU(3)$ individually as there can be, colloquially speaking, twists between them. A completely separate direction would be the study the potentials or their minima respectively that correspond to a certain VEV. One could for example write down generic potentials bounded by the degrees and then study the frequency of the VEVs under the variation of the parameters. 

Let us mention at last that finite or discrete groups can orginate from orbifolding (e.g.~\cite{Adulpravitchai:2009id,Adulpravitchai:2010na,Nilles:2012cy}) or string compactifications (e.g.~\cite{Nilles:2012cy,Kobayashi:2006wq}), as well as from a continuous group as discussed here. Examples of SSB from $SU(3)$ [$SO(3)$] to a specific subgroup have been discussed in~\cite{Etesi:1997jv,Koca:1997td,Koca:2003jy,Adulpravitchai:2009kd,Berger:2009tt,Luhn:2011ip} though the relation to ESB has not been made.

\section*{Acknowledgments}

We are indebted to Thomas Fischbacher for collaboration on this project in its early stages. AM acknowledges financial support by a Marie Curie Intra-European Fellowship within the 7th European Community Framework Programme FP7-PEOPLE-2011-IEF, contract PIEF-GA-2011-297557, and partial support from the European Union FP7 ITN-INVISIBLES (Marie Curie Actions, PITN-GA-2011- 289442). RZ gratefully acknowledges the support of an advanced STFC fellowship. AM would like to thank the participants of the FLASY12 Workshop for creating a nice atmosphere at the meeting and for many inspiring discussions. Along with several other participants of FLASY12, AM is grateful to Tom Weiler for locating one of the hidden gems among the restaurants in Dortmund.

\bibliography{Merle}
\bibliographystyle{apsrev4-1}


%% file: Papers/meroni.tex

\chapter[A SUSY $\boldsymbol{SU(5) \times T^{\prime}}$ Unified Model of Flavour with large $\boldsymbol{\theta_{13}}$ (Meroni)]{A SUSY $\boldsymbol{SU(5) \times T^{\prime}}$ Unified Model of Flavour with large $\boldsymbol{\theta_{13}}$}
\vspace{-2em}
\paragraph{A. Meroni}
\paragraph{Abstract}
We present a SUSY $SU(5) \times T^{\prime}$
unified flavour model with type I
see-saw mechanism of neutrino
mass generation with
$\theta_{13} \approx 0.14$
close to the recent results from the Daya
Bay and RENO experiments. The model predicts also
values of the solar and atmospheric neutrino
mixing angles, which are compatible
with the existing data.
The $T^{\prime}$ breaking leads to tri-bimaximal
mixing in the neutrino sector, which is
perturbed by sizeable corrections
from the charged lepton sector.
The model exhibits geometrical CP violation:
 all complex phases arise
from the complex Clebsch--Gordan coefficients (CGs) of $T^{\prime}$.
Both normal and inverted ordering are possible for the light
neutrino mass spectra. We give also predictions for the
$2\beta0\nu$-decay effective Majorana mass.

\section{Introduction}

The recent results of the short-baseline reactor experiments on
$\theta_{13}$, Daya Bay \cite{An:2012eh} and RENO \cite{Ahn:2012nd},
clearly indicate that the precise measurements era for neutrino
physics has started. A non zero value of $\theta_{13}$  has been
reported with an accuracy around 5$\sigma$ by both experiments. More
specifically,  Daya Bay and RENO measured $\sin^2 2\theta_{13} =
0.092 \pm 0.016 \pm 0.005$ and $\sin^22\theta_{13} = 0.113 \pm 0.013
\pm 0.019$, respectively. Motivated by the fact that at present we
know all three angles in the PMNS mixing matrix with a good
precision, we tried to construct a unified model of flavour, which
describes correctly the quark and charged lepton masses, the mixing
and CP violation in the quark sector, the mixing in the lepton
sector and predicts a value of the angle $\theta_{13}$ compatible
with the recent data (all the details in \cite{Meroni:2012ty}). The
model is supersymmetric and is based on two main ingredients: i) a
GUT embedding using $SU(5)$ as  gauge group; this may eventually
lead to a sizable $\theta_{13}$ \cite{Marzocca:2011dh} ii) a
discrete family symmetry $T^{\prime}$, double-valued group of the
tetrahedral symmetry T which is isomorphic to $A_4$; the latter has
three inequivalent spinorial unitary irreducible representations
which are relevant in the description of the quarks and lepton
mixing. Moreover the complex CGs of the $T^{\prime}$ group can be
source of CP violation, so-called ``geometrical'' CP violation.

We must notice that an interesting model  based on  $SU(5)\times
T^{\prime}$ as symmetry group was proposed in the  literature
\cite{Chen:2007afa} \cite{Chen:2009gf}, but it is now ruled out  by
the current data on $\theta_{13}$. In contrast, due to a
non-standard Higgs sector content
\cite{Antusch:2009gu,Marzocca:2011dh}, the model we are going to
describe predicts a value for this angle compatible with the recent
data.

The model presented in this talk includes three right-handed (RH) neutrino
fields $N_{lR}$, $l=e,\mu,\tau$,
which possess a Majorana mass term.
The light active neutrino masses
are generated by the type I see-saw
mechanism and are naturally small.
The corresponding Majorana mass term of
the left-handed flavour neutrino fields
$\nu_{lL}(x)$, $l=e,\mu,\tau$,
is diagonalized by the so-called tri-bimaximal unitary matrix:
\begin{equation}
U_{TBM} = \left(\begin{array}{ccc}
\sqrt{2/3} & \sqrt{1/3} & 0 \\
-\sqrt{1/6} & \sqrt{1/3} & -\sqrt{1/2} \\
-\sqrt{1/6} & \sqrt{1/3} & \sqrt{1/2}
\end{array}\right) \;.
\label{TBMM}
\end{equation}
Of course this mixing pattern has to be ``corrected'' in order to get a non zero value for $\theta_{13}$ in the standard PMNS mixing matrix, $U_{PMNS}$. Indeed from the simultaneous diagonalization of the neutrino and the charged lepton mass matrices
the PMNS mixing matrix reads (RL convention):
\begin{equation} U_{PMNS}=U_{eL}^{\dagger} U_{\nu} \label{mismatch}\end{equation}
Moreover,   a relation between the charged leptons and the down-type
quarks mass matrices is established through the SU(5) gauge symmetry
in such a way that the antireactor mixing angle $\theta_{13}$
results connected to the Cabibbo angle $\theta^c$: $
\sin^2\theta_{13}\cong C^2 (\sin^2\theta^c)/2$ where  $C\cong 0.9$
is a constant determined from the fit.

\section{Matter and Scalar Fields}

The model we proposed in \cite{Meroni:2012ty} is based on $SU(5)$ as gauge
group and $T^{\prime}$ as discrete family  symmetry plus an
extra shaping symmetry, $Z_{12} \times Z_8^3 \times Z_6^2 \times Z_4$, which is required  to
select the correct vacuum alignments and to forbid unwanted terms and
couplings in the superpotential.  We impose as well the $U(1)_R$ symmetry, the continuous
generalization of the usual $R$-parity.
The three generations of matter fields are defined in
the usual $\bar{\mathbf{5}}$ and $\mathbf{10}$,
representations of $SU(5)$, $\bar{F} = (d^c,L)_L$ and $T =
(q,u^c,e^c)_L$ and we introduce  three heavy right-handed Majorana neutrino fields
$N$ as singlets under $SU(5)$.
The Higgs sector is composed by
a number of  copies of Higgs fields in the $\mathbf 5$ and
$\bar{ \mathbf{5}}$ representation of $SU(5)$ which contain as
linear combinations the two Higgs doublets of the MSSM. To get
realistic mass ratios between down-type quarks and charged leptons
and to get a large reactor mixing angle
 we have introduced Higgs fields in the
adjoint representation of $SU(5)$, $\bar{ \mathbf{24}}$, which are as well responsible for
breaking the GUT group.
\begin{table}[h]
\centering
\begin{tabular}{c cccc ccccccccc}
\hline
& $T_3$ & $T_a$ & $\bar F$ & $N$ & $H_5^{(1)}$ & $H_5^{(2)}$ & $H_5^{(3)}$ & $\bar H_{5}^{(1)}$ & $\bar H_{5}^{(2)}$ & $\bar H_{5}^{(3)}$ & $\bar H_{5}^{\prime \prime}$  & $ H_{24}^{\prime \prime}$ & $ \tilde H_{24}^{\prime \prime}$  \\
\hline$SU(5)$ & $\mathbf{10}$ & $\mathbf{10}$ & $\mathbf{\bar 5}$
& $\mathbf{1}$ & $\mathbf{5}$ & $\mathbf{5}$ & $\mathbf{5}$ & $\mathbf{\bar 5}$
& $\mathbf{\bar 5}$ & $\mathbf{\bar 5}$ & $\mathbf{\bar 5}$ & $\mathbf{24}$ & $\mathbf{24}$  \\
$T^\prime$ & $\mathbf{1}$ & $\mathbf{2}$ & $\mathbf{3}$  &
$\mathbf{3}$ & $\mathbf{1}$ & $\mathbf{1}$ & $\mathbf{1}$
& $\mathbf{1}$ & $\mathbf{1}$ & $\mathbf{1}$
& $\mathbf{1^{\prime \prime}}$ & $\mathbf{1^{\prime \prime}}$ & $\mathbf{1^{\prime \prime}}$ \\
\hline
\end{tabular}
\caption{\label{tab:Matter+Higgs}
Matter and Higgs field content of the model including quantum numbers.}
\end{table}
In Tab. \ref{tab:Matter+Higgs}
we summarize the charge assignments of the matter and the Higgs fields under $SU(5)\times T^{\prime}$
(the charge assignments under the extra symmetries are given in full detail in \cite{Meroni:2012ty}).
Note that  the right-handed neutrinos $N$ are accommodated in $T^{\prime}$ triplets in such a way that the tri-bimaximal mixing pattern arises in the neutrino sector before considering corrections
from the charged lepton sector. Notice that
the complex CGs, involved whenever the spinorial representation is used,
is a source of CP violation in the quark and in the lepton
sector.

The scalar sector of fields related to the breaking of $T^{\prime}$ is composed by 13 flavons.
We introduce triplets with
two possible alignment in flavour space:
\begin{equation}
\label{eq:3FlavonAlignment}
\langle \phi \rangle = \begin{pmatrix} 0 \\0 \\ 1 \end{pmatrix} \phi_{0}\;, \quad
\langle \tilde\phi \rangle = \begin{pmatrix} 0 \\0 \\ 1 \end{pmatrix} \tilde\phi_{0}\;, \quad
\langle \xi \rangle = \begin{pmatrix} 1 \\ 1 \\ 1 \end{pmatrix} \xi_{0} \;.
\end{equation}
The  alignment $(0,0,1)$ is relevant for the quark and
the charged lepton sector while the $(1,1,1)$ alignment couples only
to the neutrino sector.
For the doublets we considered two possible orthogonal alignments:
\begin{equation}
\label{eq:2FlavonAlignment}
\begin{split}
\langle \psi^{\prime} \rangle &= \begin{pmatrix} 1 \\ 0 \end{pmatrix} \psi^{\prime}_{0}\sim 2^{\prime} \;, \quad
\langle \psi^{\prime \prime} \rangle = \begin{pmatrix} 0 \\ 1 \end{pmatrix} \psi^{\prime \prime}_{0} \sim 2^{\prime\prime}\;, \quad\\
\langle \tilde\psi^{\prime} \rangle &= \begin{pmatrix} 1 \\ 0 \end{pmatrix} \tilde\psi^{\prime}_{0}\sim 2^{\prime} \;, \quad
\langle \tilde\psi^{\prime \prime} \rangle = \begin{pmatrix} 0 \\ 1 \end{pmatrix} \tilde\psi^{\prime \prime}_{0}\sim 2^{\prime\prime}.
\end{split}
\end{equation}
Furthermore we have introduced six flavons
in one-dimensional representations of $T^{\prime}$
which receive all non-vanishing (and real) vevs
\begin{align}
\label{eq:1FlavonAlignment}
\langle \zeta^{\prime} \rangle &= \zeta^{\prime}_{0} \;, \quad
\langle \zeta^{\prime \prime} \rangle = \zeta^{\prime \prime}_{0} \;, \quad
\langle \tilde\zeta^{\prime}\rangle = \tilde\zeta^{\prime}_{0} \;, \quad
\langle \tilde\zeta^{\prime \prime} \rangle = \tilde \zeta^{\prime \prime}_{0} \;, \quad
\langle \rho \rangle = \rho_{0} \;, \quad
\langle \tilde \rho \rangle = \tilde \rho_{0} \;.
\end{align}
The primes indicates the type of singlets $1$, 1$^{\prime}$, $1^{\prime\prime}$.
We assume here that all flavon vevs are real i.e. we
considered as only source of CP violation the complex CGs
introduced geometrically by the group  $T^{\prime}$.
In the Appendix of \cite{Meroni:2012ty}
we show a superpotential that provides the desired flavon vev structure.
The latter is obtained adding the so called ``driving fields'', fields that are gauge singlets but transform non trivially  under $T^{\prime}$ and the extra shaping symmetry.

\section{Yukawa couplings}

When $T^{\prime}$ breaks and the flavons take their real vevs,
one can write down at GUT scale the Yukawa coupling
matrices (RL convention). In our model the elements of the Yukawa coupling matrices are
generated dynamically through a number of effective operators up to dimension seven which
structure is tightly related to the matter fields assignment under
the $T^{\prime}$ discrete symmetry. CP violation in the quark and
charged lepton sector is entirely due to the CGs of the
$T^{\prime}$ discrete group.  For the up-type quarks we find:
\begin{equation}
Y_u = \begin{pmatrix}
\bar{\omega} a_u & i b_u & 0 \\
i b_u &  c_u & \omega d_u \\
0 & \omega d_u  & e_u
\end{pmatrix} \;,
\end{equation}
while in the down-type sector and the charged lepton sector the Yukawas read:
\begin{equation}
Y_d = \begin{pmatrix}
  \omega \, a_d\, & i b_d^{\prime} & 0 \\
  \bar{\omega} \, b_d & c_d & 0 \\
 0 &  0 & d_d
\end{pmatrix} \;\;  \text{and} \quad Y_e = \begin{pmatrix}
 -\frac{3}{2} \, \omega \, a_d & \bar{\omega} \, b_d & 0 \\
 6 \, i \, b_d^{\prime} &  6\, c_d & 0 \\
 0 &  0 & -\frac{3}{2}\, d_d
\end{pmatrix} \;, \label{eq:Yd}
\end{equation}
where $\omega = (1+i)/\sqrt{2}$ and $\bar{\omega} =
(1-i)/\sqrt{2}$. The ten parameters appearing in the matrices are (real) functions
of the underlying parameters.
Notice that in this model
$b-\tau$ unification is not realized.
Indeed in order to get
fermion mass ratios compatible with experimental data we used
new relations that have been recently proposed in the literature
 \cite{Antusch:2009gu}, for instance
$y_\tau/y_b = -3/2$ and $y_\mu/y_s \approx 6$. Furthermore it was
shown in \cite{Marzocca:2011dh} (see also \cite{Antusch:2011qg})
that those new $SU(5)$ CGs  might also give a large
reactor neutrino mixing angle $\theta_{13}$. More importantly, due
to the $SU(5)$ symmetry of the model, $Y_d$ and $Y_e$ are related
(and therefore the corresponding down quark and charged lepton mass
matrices) and are expressed in terms of the same parameters. As a
consequence, since  $Y_e$ (and as well $Y_d$) is  a block
diagonal matrix in the  1-2 sector it is diagonalizable by a rotation of
angle $\theta^e_{12}$ (i.e. $U_{eL} \sim R_{12}(\theta^e_{12})$). In this way
$\theta^e_{12}$ is related to the Cabibbo angle $\theta^c \cong 0.226$.
Specifically using the results of a fit performed on the 10
parameters that appear in the Yukawas from GUT scale down to the
electroweak scale (more details in \cite{Meroni:2012ty})  we get:
$$|V_{us}| =\left| \frac{b_d}{c_d} \right|+ \mathcal{O}(a_d)$$
$$\theta_{12}^{e} =\left| \frac{6 i b_d'}{6c_d} \right|+ \mathcal{O}(a_d)=\left| \frac{b_d'}{ b_d}\right|\theta^c, \quad (b_d'=\,0.9\, b_d)$$
One can also get an expression for the angle
$\theta_{13}^{\text{PMNS}}$ using the equation (\ref{mismatch}):
$$\theta_{13}^{\text{PMNS}}  = \frac{1}{\sqrt{2}}  \theta_{12}^e = \frac{0.9}{\sqrt{2}} \theta^c.$$
This value is compatible with the recent Daya Bay and RENO results.

\section{Neutrino Sector}

The model includes three heavy right-handed Majorana
neutrino fields $N$ which are singlets under $SU(5)$
and form a triplet under $T^{\prime}$. Through the type I
seesaw mechanism we generate light
neutrino masses. The neutrino sector is described by
the following terms in the superpotential
\begin{equation}
\mathcal{W}_\nu = \lambda_1 N N \xi + N N (\lambda_2 \rho + \lambda_3 \tilde{\rho})
 + \frac{y_{\nu}}{\Lambda} (N \bar F )_1 (H_5^{(2)} \rho)_1
 + \frac{\tilde{y}_{\nu}}{\Lambda}  (N \bar F)_1 (H_5^{(2)} \tilde{\rho})_1
 \;, \label{eq:Wnu}
\end{equation}
%
where we have given the $T^{\prime}$ contractions
as indices at the brackets for non-renormalizable terms.
From the  superpotential we get the mass matrix
for the Majorana right-handed neutrinos and the Dirac neutrino
mass matrix
\begin{equation}
 M_R = \begin{pmatrix} 2 Z+X & -Z & -Z \\ -Z & 2 Z & -Z+X \\ -Z & - Z+X  & 2 Z \end{pmatrix} \;,\quad
 M_D = \begin{pmatrix} 1 & 0 & 0 \\ 0 & 0 & 1 \\ 0 & 1  & 0 \end{pmatrix}
\frac{\rho^{\prime}}{\Lambda}\; ,
\end{equation}
%
where $X$, $Z$ and $\rho^{\prime}$ are real
parameters.
The right-handed neutrino mass matrix $M_R$ is diagonalized
by the tri-bimaximal mixing (TBM) matrix
such that the heavy RH neutrino masses read:
$$
 U_{TBM}^T\, M_R\,U_{TBM} = D_{N} =
\mbox{diag}(3Z + X,X,3Z - X)
$$
It is more convenient to change parametrization and to use
$\alpha\equiv|3Z/X|
> 0$ and $\phi\equiv{\rm arg}(Z)-{\rm arg}(X)$ so the diagonal
Majorana mass matrix becomes :
$$ \begin{pmatrix} 3Z + X & 0 & 0\\
0& X & 0 \\ 0 & 0 & 3Z - X \end{pmatrix} \longrightarrow
|X| \begin{pmatrix} |1+\alpha e^{i \phi}|e^{i \phi_1} & 0 & 0\\
0& e^{i \phi_2} & 0 \\ 0 & 0 & |1- \alpha e^{i \phi}|e^{i \phi_3}
\end{pmatrix}
$$
where $\phi_1=\phi_2=\phi_3=0, \pi$.
The light neutrino Majorana mass term  is obtained via type I
see-saw mechanism:
$$\label{mnuLO}
M_\nu  \;=\;- M_D^T\, M_R^{-1}\,M_D =
U^*_{\nu}\mbox{diag}\left(m_1,m_2,m_3\right)U^\dagger_{\nu}\,,
$$
where the unitary matrix $U_\nu$ that diagonalize the Majorana light
mass matrix is proportional to $U_{TBM}$, precisely:
$$ U_{\nu} = i \, U_{TBM} \,\mbox{diag}\left(e^{i \phi_1/2},e^{i
\phi_2/2}, e^{i \phi_3/2}\right).$$ The masses
of the light neutrinos result:
$$
 m_i = \left(\frac{\rho^{\prime}}{\Lambda}\right )^2\,
\frac{1}{M_i}\,,\,\,\,i=1,2,3\,\quad m_i>0
\label{numasses}
$$
The value of the phase $\phi$ defines the type of the neutrino mass
spectrum in the model since one can show that:
\begin{equation}
\Delta m^2_{31} \equiv \Delta m^2_{A} =
\frac{1}{|X|^2}\,\left(\frac{\rho^{\prime}}{\Lambda}\right )^4\,
\frac{4\alpha\,\cos\phi} {\left |1 + \alpha\,e^{i \phi}\right |^2
\left |1 - \alpha\,e^{i \phi}\right |^2}\,. \label{Dm231}
\end{equation}
%
Thus, for $\cos\phi = +1$, we get $\Delta m^2_{31} > 0$, i.e., a
neutrino mass spectrum with normal ordering (NO), while for
$\cos\phi = -1$ one has  $\Delta m^2_{31} < 0$, i.e., neutrino mass
spectrum with inverted ordering (IO). In order to find the numerical
values of the light masses one can use  as input parameters the
experimental values of $\Delta m^2_{21}$ and $r = \frac{\Delta
m^2_{\odot}}{|\Delta m^2_{A}|}= 0.032 \pm 0.006$.
For a given type of neutrino mass spectrum, i.e., for a fixed $\phi
= 0~{\rm or}~\pi$, one can find a value of the parameter $\alpha$.
It is easy in this way to get the value of the lightest neutrino
mass, which together with the data on  $\Delta m^2_{21}$ and $\Delta
m^2_{31(32)}$ allows to obtain the values of the other two light
neutrino masses. Knowing the latter one can find also the two ratios
of the heavy Majorana neutrino masses. In the case of NO neutrino
mass spectrum ($\phi = 0$), there are two possible values of
$\alpha$ and so there are two possible spectra (solution A and B):
\begin{equation}
m_1 \cong 4.44\times 10^{-3}~{\rm eV}\,, m_2 \cong 9.77\times
10^{-3}~{\rm eV}\,, m_3 \cong 4.89\times 10^{-2}~{\rm eV}\,,~{\rm
solution~A~(NO)}\,. \label{eq:m123A}
\end{equation}
\begin{equation}
m_1 \cong 5.89\times 10^{-3}~{\rm eV}\,, m_2 \cong 1.05\times
10^{-2}~{\rm eV}\,, m_3 \cong 4.90\times 10^{-2}~{\rm eV}\,,~{\rm
solution~B~(NO)}\,. \label{eq:m123B}
\end{equation}
 The ratios of the heavy Majorana neutrino masses read for the solution (A) $M_1/M_3
\cong 11.0$ and $M_2/M_3 \cong 5.0$.  For solution B we find:
$M_1/M_3 \cong 8.33$ and $M_2/M_3 \cong 4.67$. In both cases we have
$M_3 < M_2 < M_1$. For the IO spectrum ($\phi = \pi$), we find only
one possible value for $\alpha$ and in this case the light neutrino
masses read:
\begin{equation}
m_1 \cong 5.17\times 10^{-2}~{\rm eV}\,, m_2 \cong 5.24\times
10^{-2}~{\rm eV}\,, m_3 \cong 1.74\times 10^{-2}~{\rm eV}\,,~~{\rm
(IO)}\,, \label{eq:m123IO}
\end{equation}
%
i.e., the light neutrino mass spectrum is not hierarchical
exhibiting only partial hierarchy. For the heavy Majorana neutrino
mass ratios we obtain: $M_1/M_2 \cong 1.014$ and $M_3/M_2 \cong
3.01$. Thus, in this case $N_1$ and $N_2$ are quasi-degenerate in
mass: $M_1 \cong M_2 < M_3$.

In this model is possible to predict also the values of  observables such as the
fundamental parameter of $2\beta0\nu$-decay, the Majorana effective
mass $\langle m \rangle$. At this purpose one has to find the values of the angles
and phases of the PMNS mixing matrix and this can be done with standard formulae
(see \cite{Marzocca:2011dh} for instance) recasting the PMNS mixing matrix
in the standard parametrization.  We list in table \ref{Tab:NuData} the numerical values of the angles and phases found in our analysis. We found that the Dirac phase is $\delta \cong 84.3^{\circ}$
and the values of the Majorana phases
in the standard parametrization
are not CP conserving.
As one can see the value of $\delta$ predicted by the model
is close to $\pi/2$: this implies
that the magnitude of the CP violation effects
in neutrino oscillations, is predicted
to be  relatively large. The rephasing invariant associated
with the Dirac phase reads
$J_{\text{CP}} = {\rm Im}(U^*_{e1}U_{\mu 1}U_{e3}U^*_{\mu3})= 0.0324$.
\begin{table}[h!]
\centering
\begin{tabular}{ccc}
\hline
Quantity & Experiment (2$\sigma$ ranges) & Model  \\
\hline
$\sin^2 \theta_{12}$ & 0.275 -- 0.342 & 0.340    \\
$\sin^2 \theta_{23}$ & 0.36 -- 0.60   & 0.490     \\
$\sin^2 \theta_{13}$ & 0.015 -- 0.032 & 0.020    \\
$\delta$     & - & 84.3$^\circ$    \\
 $\beta_1$ &  - & 337.1$^\circ$  + $\phi_3$\; \\
$\beta_2$ &  - & 11.5$^\circ$ + $\phi_3$ - $\phi_2$\;\\
\hline
\end{tabular}\caption{Numerical results for the neutrino sector. The experimental results
are taken from \protect\cite{Fogli:2011qn} apart from
the value for $\theta_{13}$ which is the DayaBay result
\cite{An:2012eh}. \label{Tab:NuData}}
\end{table}
Finally we are able to compute the value of the Majorana effective mass $\langle m \rangle$ for NO:
$$
\langle m \rangle =
4.90 \, (7.95)\times 10^{-3}~\text{eV}\;,~~{\rm solution~A~(B)}\,,
\label{meffA}
$$
and for IO:
$$
\langle m \rangle =
2.17\times 10^{-2}~\text{eV}.
$$
\section*{Acknowledgments}
The author would like to thank the Organizers for the possibility to
present this work and for creating a pleasant working atmosphere
during the Workshop.

\bibliography{meroni}
\bibliographystyle{apsrev4-1}

%% file: Papers/mondragon.tex

%
%
%
%
%
%

\chapter[The $S_3$ flavour symmetry: quarks, leptons and Higgs sectors. (González Canales, A. Mondragón, \textit{M. Mondragón}, Saldaña Salazar)]{The $S_3$ flavour symmetry: quarks, leptons and Higgs sectors.}
\vspace{-2em}
\paragraph{F. González Canales, A. Mondragón, \textit{M. Mondragón}, U. Saldaña Salazar}
\paragraph{Abstract}
We present a brief overview of the minimal $S_3$ extension of the
Standard Model, in which the concept of flavour is extended to the
Higgs sector by introducing in the theory three Higgs fields which are
$SU(2)$ doublets. In both the quark and lepton sectors the mass
matrices are reparametrized in terms of their eigenvalues, thus
allowing to express the mixing angles in terms of mass ratios. In the
leptonic sector, the $S_3 \times Z_2$ symmetry implies a non-vanishing
$\theta_{13}$, which is different from zero and in very good agreement
with the latest experimental data.

\section{The $S_3$ symmetry}
The success of the Standard Model (SM) in describing the fundamental
particles and their interactions has been confirmed with the recent
observation of a state compatible with a (SM-like) Higgs boson at the
LHC \cite{:2012gk,:2012gu}. Despite this success, the SM leaves many
open questions and has too many free parameters whose value can only
be determined by the experiment.  Among these open questions is the
whole subject of flavour physics, namely the origin of the masses and
mixings of quarks and leptons.

It is possible to ask if the data on quark and lepton masses suggests
already a flavour symmetry at the Fermi scale. The fact that the third
generation is heavier than the first two already proposes a
path to follow: if instead of looking at the masses alone we look at
the mass ratios obtained by dividing the masses of each type of quarks
and leptons by the heaviest of each sector, then a clear pattern
emerges. The first two generations belong to a doublet representation
and the third generation to a singlet one.  The simplest flavour
symmetry with doublet and singlet irreducible representations is the
permutational flavour symmetry $S_3$
\cite{Pakvasa:1977in,Mondragon:1999jt,Mondragon:1998gy}.

In the Standard Model analogous fermions in different generations have
identical couplings to all gauge bosons of the strong, weak and
electromagnetic interactions. Prior to electroweak symmetry breaking,
the Lagrangian is chiral and invariant 
under the action of the group of permutations acting on the flavour
indices of the matter fields.  Since the Standard Model has only one
Higgs $SU(2)_{L}$ doublet, which can only be an $S_{3}$ singlet, it is
necessary to break the $S_3$ symmetry in order to give different
masses to all quarks and leptons.  Hence, in order to impose $S_{3}$
as a fundamental symmetry, unbroken at the Fermi scale, we are led to
extend the Higgs sector of the theory
\cite{Kubo:2003iw,Kubo:2005sr,Felix:2006pn,Mondragon:2007af,Mondragon:2007nk,
  Mondragon:2007jx,Mondragon:2008gm}.  The quark, lepton and Higgs
fields are $ Q^T=(u_L,d_L)$, $u_R$, $d_R$, $L^T=(\nu_L,e_L)$, $e_R$,
$\nu_R$, and $H_S$, $H_1$ and $H_2$, in an obvious notation. All these
fields have three species, and we assume that each one forms a
reducible representation ${\bf 1}_S\oplus{\bf 2}$ of the $S_3$ group.
The doublets carry capital indices $I$ and $J$, which run from $1$ to
$2$ and the singlets are denoted by
$Q_3,~u_{3R},~d_{3R},~L_3,~e_{3R},~\nu_{3R}$ and $~H_S$. Note that the
subscript $3$ denotes the singlet representation and not the third
generation. The most general renormalizable Yukawa interactions for
the lepton sector of this model are given by $ {\cal L}_Y = {\cal
  L}_{Y_E}+{\cal L}_{Y_\nu}$, where
   \begin{equation}\label{lage}
    \begin{array}{lll}
     {\cal L}_{Y_E} &=& -Y^e_1\overline{ L}_I H_S e_{IR} - Y^e_3 \overline{ L}_3 H_S e_{3R} 
     - Y^{e}_{2}[~ \overline{ L}_{I}\kappa_{IJ}H_1  e_{JR} + \overline{ L}_{I} \eta_{IJ} H_2  e_{JR}~]\\
     &  & -Y^e_{4}\overline{ L}_3 H_I e_{IR} - Y^e_{5} \overline{ L}_I H_I e_{3R} +~\mbox{h.c.},
    \end{array}
   \end{equation}
   \begin{equation}\label{lagnu}
    \begin{array}{lcl}
     {\cal L}_{Y_\nu} &=& -Y^{\nu}_1\overline{ L}_I (i \sigma_2)H_S^* \nu_{IR} 
     -Y^\nu_3 \overline{ L}_3(i \sigma_2) H_S^* \nu_{3R} -Y^\nu_{4}\overline{ L}_3(i \sigma_2) H_I^* \nu_{IR} \\
     &  &   -Y^{\nu}_{2}[~\overline{ L}_{I}\kappa_{IJ}(i \sigma_2)H_1^*  \nu_{JR}
     + \overline{ L}_{I} \eta_{IJ}(i \sigma_2) H_2^*  \nu_{JR}~]
     -Y^\nu_{5} \overline{ L}_I (i \sigma_2)H_I^* \nu_{3R}+~\mbox{h.c.},
    \end{array}
   \end{equation}
and
   \begin{equation}\label{mon_kappa}
    \kappa = 
    \left( \begin{array}{cc}
     0 & 1 \\ 
     1 & 0 \\
    \end{array} \right) 
    ~~\mbox{and}~~
    \eta = 
     \left( \begin{array}{cc}
     1 & 0 \\
     0 & -1 \\
    \end{array} \right).
   \end{equation} 
   We also add to the Lagrangian the Majorana mass terms for
   the right-handed neutrinos, ${\cal L}_{_M} = - \nu_{ R }^T C {\bf
     M}_{\nu_{R}} \nu_{ R }\,$,
   where $C$ is the charge conjugation matrix and ${\bf M}_{\nu_{R}} =
   \textrm{diag} \left\{ M_1 , M_2, M_3 \right\} $ is the mass matrix
   for the right-handed neutrinos.  The Lagrangian for the quark
   sector of this model has a similar form \cite{Kubo:2005sr}.

Due to the presence of the three Higgs
fields, the Higgs potential 
is more complicated than that of the Standard Model
\cite{Kubo:2005sr,EmmanuelCosta:2007zz,Beltran:2009zz,Teshima:2012cg}.
The $S_3$ Higgs potential was first analyzed by Pakvasa and
Sugawara~\cite{Pakvasa:1977in} who found that in addition to the
$S_{3}$ symmetry, it has an accidental permutational symmetry
$S_{2}^{\prime}$: $H_{1}\leftrightarrow H_{2}$, which is not a
subgroup of the flavour group $S_{3}$.
With these assumptions,the Yukawa interactions, eqs.~(\ref{lage})-(\ref{lagnu}) yield mass 
   matrices, for all fermions in the theory, of the general
   form~\cite{Kubo:2003iw}
   \begin{equation}\label{matGene}
    {\bf M} = 
    \left( \begin{array}{ccc}
     \mu_{1} +\mu_{2} & \mu_{2} & \mu_{5} \\ 
     \mu_{2} & \mu_{1}-\mu_{2} & \mu_{5}   \\ 
     \mu_{4} & \mu_{4}&  \mu_{3}
    \end{array}\right). 
   \end{equation}
   
In principle, all entries in the mass matrices can be complex since
there is no restriction coming from the flavour symmetry
$S_{3}$. 
\section{Quarks}

Quark models using the $S_3$ permutational symmetry as the flavour
symmetry have been already explored, see for instance 
\cite{Pakvasa:1977in,Derman:1978rx,Wyler:1978fj,Yahalom:1983kf,Mondragon:1999jt,Mondragon:1998gy,Kubo:2003iw,Teshima:2012cg}.
Some of these studies have been done through numerical approaches, and
all of them show very good agreement with the experimental
data.  
Recently we have been able to reparametrize the quark mass
matrices in terms of their eigenvalues, which allowed us to  express the mixing
angles in terms of quark mass ratios, an independent phase parameter,
and a free parameter, when we do not consider the accidental $S_2'$
symmetry  \cite{Pakvasa:1977in}, for more details see 
\cite{PASCOS:2012aa}.

An interesting result from this analysis of the $S_3$ quark model is
the fact that, after electroweak symmetry breaking, the generic mass
matrix  that comes from the $S_3$-invariant Yukawa interactions is
equivalent, when demanding hermiticity, to the four-zero Fritzsch-like
texture and, when hermiticity is not demanded, to the Nearest
Neighbour Interaction mass matrix form (NNI), both of which have been
found not only to have a viable phenomenology, but also to allow a
unified treatment of both quarks and leptons
\cite{Canales:2011ug,Barranco:2010we}.  The mass matrices are,
   \begin{eqnarray}
		{\bf{M}}_{Hermitian}^{f} = 
				\begin{pmatrix}
	 0 & |\mu_2^{f}|{\text {sin}}\theta {\cos}\theta(3-{\tan}^2\theta) & 0 \\
	|\mu_2^{f}|{\text {sin}}\theta {\cos}\theta(3-{\tan}^2\theta) & -2|\mu_2^{f}|{\cos}^2\theta(1-3{\tan}^2\theta) & \mu_8^{f}{{\sec}\theta} \\
		0 & \mu_8^{f*}{{\sec}\theta} & |\mu_3^{f}|-\Delta_f
              \end{pmatrix}, 
\end{eqnarray}
\begin{eqnarray}
              {\bf{M}}_{NNI}^{f} =
				\begin{pmatrix}
	 0 & \dfrac{2}{\sqrt{3}}\mu_2^{f} & 0 \\
	{\bf +}\dfrac{2}{\sqrt{3}}\mu_2^{f} & 0 & \dfrac{2}{\sqrt{3}}\mu_7^{f} \\
		0 & \dfrac{2}{\sqrt{3}}\mu_8^{f} & \mu_3^{f}-\Delta_f
				\end{pmatrix},
  \end{eqnarray}
  where $\Delta_f << 1$, $\tan\theta = w_1/w_2$, $\mu_2^{f} \equiv
  {Y}_{3}^{f} w_2$, $\mu_3^{f} \equiv 2 {Y}_{1}^{f} v_S$, $\mu_7^{f}
  \equiv \sqrt{2} {Y}_{5}^{f} w_2$, $\mu_8^{f} \equiv
  \sqrt{2}{Y}_{6}^{f} w_1$, where $Y_i^f$ are the complex Yukawa
  couplings, and $v_S$, $w_1$ and $w_2$ are the vacuum expectation
  values of the three Higgs $SU(2)_L$ doublets, $H_S$ and $H_D = (H_1,
  H_2)^T$, singlet and doublet of $S_3$, respectively \cite{PASCOS:2012aa}.
     
\section{Lepton masses and mixings}
A further reduction of the number of parameters in the leptonic sector
may be achieved by means of an Abelian $Z_{2}$ symmetry
\cite{Kubo:2003iw}, which forbids the following Yukawa couplings $
Y^e_{1} = Y^e_{3}= Y^{\nu}_{1}= Y^{\nu}_{5}=0$. Therefore, the
corresponding entries in the mass matrices vanish. 
The resulting expression for ${ \bf M}_e$, reparametrized in terms of
its eigenvalues and written to order
$\left(m_{\mu}m_{e}/m_{\tau}^{2}\right)^{2}$ and
$x^{4}=\left(m_{e}/m_{\mu}\right)^4$, is
\begin{equation}\label{emass}
    { \bf M}_{e}\approx m_{\tau} \left( 
    \begin{array}{ccc}
    \frac{1}{\sqrt{2}}\frac{\tilde{m}_{\mu}}{\sqrt{1+x^2}} & \frac{1}{\sqrt{2}}\frac{\tilde{m}_{\mu}}{\sqrt{1+x^2}}&   
    \frac{1}{\sqrt{2}} \sqrt{\frac{1+x^2-\tilde{m}_{\mu}^2}{1+x^2}} \\ \\
    \frac{1}{\sqrt{2}}\frac{\tilde{m}_{\mu}}{\sqrt{1+x^2}} &-\frac{1}{\sqrt{2}}\frac{\tilde{m}_{\mu}}{\sqrt{1+x^2}}& 
    \frac{1}{\sqrt{2}} \sqrt{\frac{1+x^2-\tilde{m}_{\mu}^2}{1+x^2}} \\ \\
    \frac{\tilde{m}_{e}(1+x^2)}{\sqrt{1+x^2-\tilde{m}_{\mu}^2}}e^{i\delta_{e}} & 
    \frac{\tilde{m}_{e}(1+x^2)}{\sqrt{1+x^2-\tilde{m}_{\mu}^2}}e^{i\delta_{e}} & 0
    \end{array} \right). 
\end{equation}
This approximation is numerically exact up to order $10^{-9}$ in units
of the $\tau$ mass. Notice that this matrix has no free parameters
other than the Dirac phase $\delta_e$ \cite{Kubo:2003iw,Canales:2012dr}.

The mass matrix of the left-handed Majorana neutrinos, ${\bf
  M}_{\nu_L}$, is generated by the type~I seesaw mechanism.  The mass
matrix ${\bf M}_{\nu_L}$ takes the
form~\cite{1742-6596-378-1-012014,Canales:2012dr}:
\begin{equation}\label{NeuMajorana}
  {\bf M_{\nu_L}} =
  \left( \begin{array}{ccc}
   \frac{ 2 \left( \mu^{\nu}_{2} \right)^{2} }{ \overline{M} } & 
   \frac{ 2 \lambda \left( \mu^{\nu}_{2} \right)^{2} }{ \overline{M} } & 
   \frac{ 2 \mu^{\nu}_{2} \mu^{\nu}_{4} }{ \overline{M} } \\ 
   \frac{ 2 \lambda \left( \mu^{\nu}_{2} \right)^{2} }{ \overline{M} } &  
   \frac{ 2 \left( \mu^{\nu}_{2} \right)^{2} }{ \overline{M} } &
   \frac{ 2 \lambda \mu^{\nu}_{2} \mu^{\nu}_{4} }{ \overline{M} } \\  
   \frac{ 2 \mu^{\nu}_{2} \mu^{\nu}_{4} }{ \overline{M} } & 
   \frac{ 2 \lambda \mu^{\nu}_{2} \mu^{\nu}_{4}  }{ \overline{M} } &
   \frac{ 2 \left( \mu^{\nu}_{4} \right)^{2} }{ \overline{M} } + 
   \frac{ \left( \mu^{\nu}_{3} \right)^{2} }{ M_{3} }
  \end{array} \right), \, \textrm{with} \quad 
  \begin{array}{l}
   \lambda = \frac{1}{2} \left( \frac{ M_2 - M_1 }{ M_1 + M_2 } \right),   \\\\
   \overline{M} = 2\frac{ M_1 M_2 }{ M_2 + M_1  }.
  \end{array}
\end{equation} 
where $M_{i}$~$\left(~i=1,2,3~\right)$ are the masses of right-handed
neutrinos, whereas the $\mu_{2}$, $\mu_{3}$ and $\mu_{4}$ are
complex parameters that come from the mass matrix of the Dirac neutrinos, which
is obtained from the $S_{3} \otimes Z_{2}$ flavour
symmetry~\cite{1742-6596-378-1-012014,Canales:2012dr}.
    
The identity of the leptons is encoded in the mass matrices ${\bf
  M}_{e}$ and ${\bf M_{\nu_L}}$, for charged leptons and left-handed
neutrinos, respectively. Nevertheless, these matrices are basis
dependent, since given any pair ${\bf M}_{e}$, ${\bf M_{\nu_L}}$ one
can obtain other pairs of matrices through a unitary rotation, without
affecting the physics. On the other hand, the phase factors may be
factored out of ${\bf M}_{\nu_L}$ if $\phi_{2} = \arg \left\{
  \mu^{\nu}_{2} \right\}$ and $\phi_{3} = \arg \left\{ \mu^{\nu}_{3}
\right\}$ satisfy the relation $\phi_{2} =
\phi_{3}$~\cite{1742-6596-378-1-012014}. Hence, the mass matrix of the
left-handed Majorana neutrinos can be written as:
\begin{equation}
  {\bf M}_{\nu_L} = 
  {\bf Q} {\bf \cal U}_{_\frac{\pi}{4}} \left( \mu_{_0} { \bf I}_{ _{3 \times 3} } 
  + { \bf \widehat{M} } \right)  {\bf \cal U}_{_\frac{\pi}{4}}^{\dagger} { \bf Q}, 
\end{equation}   
where 
 $\mu_{_0} = 2 | \mu^{\nu}_{2} |^{2} | \overline{M} |^{-1} \left( 1 - \left| \lambda \right|  
 \right)$, ${ \bf Q} = e^{i \phi_{2} } \textrm{diag} \left\{ 1 ,1, e^{i \delta_{\nu} } \right\}$ 
 with $\delta_{\nu} = \arg \left\{ \mu^{\nu}_{4} \right\} - \arg \left\{ \mu^{\nu}_{2} \right\}$,
 \begin{equation}
  {\bf \cal U}_{_\frac{\pi}{4}} =
  \left( \begin{array}{ccc}
   \frac{1}{\sqrt{2}} & 0 & \frac{1}{\sqrt{2}} \\
   -\frac{1}{\sqrt{2}} & 0 & \frac{1}{\sqrt{2}} \\
   0 & 1 & 0
  \end{array}\right), 
  \quad \textrm{and} \quad  
  { \bf \widehat{M} } = 
  \left( \begin{array}{ccc}
   0 & A & 0 \\
   A & B & C \\
   0 & C & 2d
  \end{array} \right) ~,
\end{equation}
with $A = \sqrt{2} |\mu^{\nu}_{2}| |\mu^{\nu}_{4}| \left(1 -
  \left|\lambda\right| \right) | \overline{M}|^{-1}$, $B = 2
|\mu^{\nu}_{4}|^{2} |\overline{M}|^{-1} + |\mu^{\nu}_{3}|^{2}
M_{3}^{-1} - 2 |\mu^{\nu}_{2}|^{2} |\overline{M}|^{-1} \left( 1 - |
  \lambda | \right)$, $ C = \sqrt{2} |\mu^{\nu}_{2}| |\mu^{\nu}_{4}|
|\overline{M}|^{-1} \left( 1 + \left|\lambda\right| \right)$ and $d =
2 |\lambda| | \mu^{\nu}_{2} |^{2} |\overline{M}|^{-1}$.  The
diagonalization of ${ \bf M}_{\nu_L}$ is reduced to the
diagonalization of matrix ${ \bf \widehat{M} }$, which is a matrix
with two texture zeroes of class~I~\cite{Canales:2011ug}. For this we
use all the information  we already have about the diagonalization of
a matrix with two texture
zeroes~\cite{Canales:2011ug,Barranco:2010we,Mondragon:1998gy,Mondragon:1999jt}.
Thus, the matrix ${\bf M}_{\nu_L}$ is diagonalized by a unitary matrix
$  {\bf U}_{\nu} = {\bf Q}^{\nu} {\bf \cal U}_{_\frac{\pi}{4}}{\bf O}^{^{N[I]}}_{\nu} $
where the orthogonal matrix ${\bf O}^{^{N[I]}}_{\nu}$ reparametrized
in terms of the neutrino masses is given
by~\citep{1742-6596-378-1-012014}: { \footnotesize 
\begin{equation}
 \left( \begin{array}{ccc} \label{ec:10} \sqrt{ \frac{ [-1]\left( m_{\nu_{3}} -
            \mu_{_{0}}\right) \left( m_{\nu_{2}} - \mu_{_{0}} \right)
          f_{1} }{ {\cal D}_{1}^{^{N[I]}} } } & \sqrt{ \frac{ \left(
            m_{\nu_{3[1]}} - \mu_{_{0}} \right) \left( \mu_{_{0}} -
            m_{\nu_{1[3]}} \right) f_{2}^{^{N[I]}} }{ {\cal
            D}_{2}^{^{N[I]}} } } & - \sqrt{ \frac{ [-1] \left(
            \mu_{_{0}} - m_{\nu_{1}} \right)
          \left( m_{\nu_{2}}  - \mu_{_{0}}\right) f_{3}^{^{N[I]}} }{ {\cal D}_{3}^{^{N[I]}} } } \\
      \sqrt{ \frac{ [-1 ] 2d \left( \mu_{_{0}} - m_{\nu_{1}} \right)
          f_{1} }{ {\cal D}_{1}^{^{N[I]}} } } & \sqrt{ \frac{ 2d
          \left( m_{\nu_{2}} - \mu_{_{0}} \right) f_{2}^{^{N[I]}} }{
          {\cal D}_{2}^{^{N[I]}} } } & \sqrt{ \frac{ [-1 ] 2d \left(
            m_{\nu_{3}} - \mu_{_{0}} \right) f_{3}^{^{N[I]}} }{
          {\cal D}_{3}^{^{N[I]}}  } } \\
      - \sqrt{ \frac{ [-1] \left( \mu_{_{0}} - m_{\nu_{1}} \right)
          f_{2}^{^{N[I]}} f_{3}^{^{N[I]}} }{ {\cal D}_{1}^{^{N[I]}} }
      } & \sqrt{ \frac{ \left( m_{\nu_{2}} - \mu_{_{0}} \right) f_{1}
          f_{3}^{^{N[I]}} }{ {\cal D}_{2}^{^{N[I]}} } } & - \sqrt{
        \frac{ \left( m_{\nu_{3}} - \mu_{_{0}} \right) f_{1}
          f_{2}^{^{N[I]}} }{ {\cal D}_{3}^{^{N[I]}} } }
 \end{array} \right), 
\end{equation} }

\noindent where {\small $f_{1} = \left( 2d + \mu_{ _{0} } - m_{\nu_{1}}  \right)$}, 
{\small $f_{2}^{^{N[I]}} = [-1] \left( 2d + \mu_{ _{0} } - m_{\nu_{2}}  \right)$},
{\small $f_{3}^{^{N[I]}} = [-1] \left( m_{\nu_{3}} - \mu_{ _{0} } - 2d \right)$},
{\small $\;{\cal D}_{1}^{^{N[I]}} = 2d \left( m_{\nu_{2}} - m_{\nu_{1}} \right) 
  \left( m_{\nu_{3[1]}} - m_{\nu_{1[3]}} \right)$},~~ 
{\small ${\cal D}_{2}^{^{N[I]}} = 2d \left( m_{\nu_{2}} - m_{\nu_{1}} \right) 
  \left( m_{\nu_{3[2]}} - m_{\nu_{2[3]}} \right)$} and 
{\small ${\cal D}_{3}^{^{N[I]}} = 2d \left( m_{\nu_{3[1]}} - m_{\nu_{1[3]}} \right)
  \left( m_{\nu_{3[2]}} - m_{\nu_{2[3]}} \right)$}. The values allowed for the parameters  
$\mu_{_{0}}$ and $ 2d + \mu_{_{0}} $ are in the following ranges:
{\small $ m_{\nu_{2[1]}} > \mu_{_{0}} >  m_{\nu_{1[3]}}$} and 
{\small $m_{\nu_{3[2]}} >  2d + \mu_{_{0}}  > m_{\nu_{2[1]}}$}.
The superscripts $N$ and $I$ denote the normal and inverted
 hierarchies, respectively.
    
\begin{flushleft}
 { \bf The neutrino mixing matrix}
\end{flushleft}
The neutrino mixing matrix ${ \bf V}_{_{PMNS}}$ is the product ${ \bf
  U}_{eL}^{\dagger} {\bf U}_{\nu} {\bf K}$, where ${\bf K}$ is the
diagonal matrix of the Majorana phase factors, defined by ${\bf
  K}=diag(1,e^{i\alpha},e^{i\beta})$~\cite{Beringer:1900zz}. The
theoretical expression for the lepton mixing matrix, ${\bf
  V}^{^{th}}_{_{PMNS}}$ is:
\begin{equation}
   \left(  \begin{array}{ccc}
    \frac{ \tilde{ m }_{e}  }{ \tilde{m}_{\mu} } O_{11}^{^{N[I]}} - O_{21}^{^{N[I]}} 
    e^{ i \delta_{l} }  & 
    \left( \frac{ \tilde{ m }_{e}  }{ \tilde{m}_{\mu} } O_{12}^{^{N[I]}} - O_{22}^{^{N[I]}} 
    e^{ i \delta_{l} } \right) e^{ i \alpha } &  
    \left(  \frac{ \tilde{ m }_{e}  }{ \tilde{m}_{\mu} } O_{13}^{^{N[I]}} - O_{23}^{^{N[I]}}
    e^{ i \delta_{l} } \right)  e^{ i \beta } \\
    - O_{11}^{^{N[I]}} -\frac{ \tilde{ m }_{e}  }{ \tilde{m}_{\mu} } O_{21}^{^{N[I]}} 
    e^{ i \delta_{l} } & \left(- O_{12}^{^{N[I]}} - \frac{ \tilde{ m }_{e}  }{ \tilde{m}_{\mu} }
    O_{22}^{^{N[I]}} e^{ i \delta_{l} } \right) e^{ i \alpha } &  \left( - O_{13}^{^{N[I]}} 
    -\frac{ \tilde{ m }_{e}  }{ \tilde{m}_{\mu} } O_{23}^{^{N[I]}} e^{ i \delta_{l} } \right) 
    e^{ i \beta } \\
     O_{31}^{^{N[I]}} & O_{32}^{^{N[I]}} e^{ i \alpha } & O_{33}^{^{N[I]}}  e^{ i \beta } 
    \end{array} \right) \,
\end{equation}
where the elements $O^{^{N[I]}}$ are given in eq.~(\ref{ec:10}) and 
   $\delta_{l} = \delta_{\nu} - \delta_{e}$.
  
\begin{flushleft}
   { \bf The reactor mixing angle }
\end{flushleft}
In the case of an inverted neutrino mass hierarchy $( m_{ \nu _{ 2 } }
> m_{ \nu _{ 1 } } > m_{ \nu _{ 3} } )$, the theoretical expression
for the neutrino reactor mixing angle as function of the ratios of the
lepton masses and, in a preliminary analysis, the numerical values are:
\begin{equation}
   \begin{array}{ll}
   \sin^{2}{ \theta_{13}^{ l } } \approx  
     \frac{ \left( \mu_{_{0}} + 2d - m_{\nu_{3}} \right) \left( \mu_{_{0}} -  m_{\nu_{3}} \right) 
     }{ 
     \left( m_{\nu_{1}} - m_{\nu_{3}}  \right) \left( m_{\nu_{2}} - m_{\nu_{3}}  \right) }, & 
    \sin^{2}{ \theta_{13}^{ l } } \approx 0.029  \longrightarrow 
    \theta_{13}^{ l } \approx 9.8^{\circ}, 
   \end{array}
\end{equation}
with the following values for the neutrino masses $m_{\nu_{2}} =
0.056$~eV, $m_{\nu_{1}}=0.053$~eV and $m_{\nu_{3}}=0.048$~eV, and the
parameter values $\delta_{l}=\pi/2$, $\mu_{0}=0.049$~eV and $d = 8
\times 10^{-5}$~eV, we get $\theta_{13}$ in very good agreement with experimental
data~\cite{An:2012eh,Ahn:2012nd}.

\section{The Higgs potential}
As already mentioned, the Higgs potential of this model is more
complicated than the one of the SM, because of the extended Higgs
sector. The correct computation of the potential is important for a
meaningful and significative comparison of theory and
experiment. Thus, any conclusions on the Higgs phenomenology depend
strongly on the form of the Higgs potential. Work on the phenomenology
implied by the Higgs potential has been done before
\cite{Pakvasa:1977in,Wyler:1978fj,Derman:1978rx,Yahalom:1983kf,Koide:1999mx,Kubo:2004ps,Chen:2004rr,Kimura:2005sx,Koide:2005ep,
  Beltran:2009zz,Meloni:2010aw,Bhattacharyya:2010hp,Bhattacharyya:2012ze}.
Nevertheless, it has not been made clear how the $S_3$ symmetry should
be used among the different and independent terms of the Higgs
potential expression, such that it has the highest degree of flavour
symmetry.  In this sense, we have explored the possibility that the
different combinations of $SU(2)_L$ indices contractions belonging to
the same $S_3$-invariant term are assigned to the same self coupling
parameter. A preliminary analysis following this proposal can be found
in ref.~\cite{1742-6596-378-1-012027}.
		
\section{Conclusions}
The permutational group $S_3$ is well motivated by the data on quarks
and lepton masses as an underlying flavour symmtery. To be able to
impose $S_{3}$ as a fundamental, unbroken symmetry, the Higgs sector
of the theory has to be extended with two extra Higgs $SU(2)$
doublets.  Both in the quark and lepton sectors it is possible to
reparametrize the mass matrices in terms of their eigenvalues, thus
allowing to express the mixing angles in terms of mass ratios.  The
resulting form of the quark and lepton mixing matrices allows for a
unified treatment of quarks and leptons under the flavour symmetry.
In the quark sector it is possible to express the mixing angles in
terms of the quark mass ratios, an independent phase parameter, and a
free parameter.  In the leptonic sector, the flavour symmetry,
together with the seesaw mechanism, imply a non-vanishing reactor
mixing angle $\theta_{13}$, and from a preliminary numerical analysis,
gives results for all mixing angles in very good agreement with the
most recent experimental data.  Finally, from symmetry arguments
regarding the gauge and flavour symmetries, it is possible to find the
most general form of the Higgs potential with the highest degree of
flavour symmetry.

\section*{Acknowledgements}
This work was supported by a PAPIIT grant IN113712.
\newpage
\bibliography{mondragon}
\bibliographystyle{apsrev4-1}

%% file: Papers/morisi.tex
\chapter[Predictive Discrete Dark Matter Model (Morisi)]{Predictive Discrete Dark Matter Model}
\vspace{-2em}
\paragraph{S. Morisi}
\paragraph{Abstract}

  We propose a type-II seesaw
  model where left-handed matter transforms nontrivially under the
  flavor group $\Delta(54)$, providing correlations between neutrino
  oscillation parameters, consistent with the recent Daya-Bay and RENO
  reactor angle measurements, as well as lower bounds for neutrinoless
  double beta decay.  Dark Matter stability is achieved through a 
  partial breaking of  a flavor symmetry and its phenomenology 
  is provided by a Higgs-portal.

\vskip10.mm

The discovery of neutrino oscillations and the growing
evidence for the existence of dark matter
provide strong indications for the need of physics beyond Standard
Model (SM). However the detailed nature of the new physics remains
elusive.
On the one hand, the typology of mechanism responsible for neutrino
mass generation and its flavor structure, as well as the nature of the
associated messenger particle are unknown.
Consequently the nature of neutrinos, their mass and mixing parameters
are all unpredicted.

Likewise the nature of Dark Matter (DM) constitutes one of the most
challenging questions in cosmology since decades, though recently some
direct and indirect DM detection experiments are showing tantalizing
hints favoring a light WIMP-like DM candidate 
opening hopes for an imminent detection.

Linking neutrino mass generation to dark matter, two seemingly
unrelated problems into a single framework is not only theoretically
more appealing, but also may bring us new insights on both issues.

Among the requirements a viable DM candidate must pass, stability has
traditionally been ensured through the \textit{ad hoc} imposition of a
stabilizing symmetry; usually a parity. Clearly a top-down approach
where stability is naturally achieved is theoretically more appealing.
This is what motivated attempts such gauged as
$U(1)_{B-L}$~\cite{Hambye:2010zb}, gauged discrete
symmetries~\cite{Batell:2010bp} and the recently proposed discrete
dark matter mechanism
(DDM)~\cite{Hirsch:2010ru,Meloni:2011cc,Boucenna:2011tj,Meloni:2010sk},
where stability arises as a remnant of a suitable flavor
symmetry~\footnote{For other flavor models with DM candidates see
  \cite{Meloni:2011cc,Kajiyama:2011gu,Kajiyama:2011fx,Daikoku:2011mq,Kajiyama:2010sb,Haba:2010ag}}.
The DDM scenario provides a potential link between DM and the neutrino 
sector through the stability issue.  
The main idea behind DDM is outlined below.

Consider the group of the even permutations of four objects $A_4$.  It
has one triplet and three singlet irreducible representations, denoted
${\bf 3}$ and ${\bf 1,1',1''}$ respectively.  $A_4$ can be broken
spontaneously to one of its $Z_2$ subgroups. Two of the components of
any $A_4$ triplet are odd under such a parity, while the $A_4$ singlet
representation is even. This residual $Z_2$ parity can be used to
stabilize the DM which, in this case, must belong to an $A_4$ triplet
representation, taken as an $SU(2)_L$ scalar Higgs doublet, $\eta\sim
{\bf 3}$
\cite{Hirsch:2010ru,Meloni:2011cc,Boucenna:2011tj,Meloni:2010sk}.
Assuming that the lepton doublets $L_i$ are singlets of $A_4$ while
right-handed neutrinos transform as $A_4$ triplets $N\sim{\bf 3}$, the
contraction rules imply that the DM couples only to Higgses and heavy
right-handed neutrinos $\overline{L}_i \, N\, \tilde{\eta}$.
In this case $\eta$ and $N$ have even as well as odd-components while
$L_i$ are even so that $\overline{L}_i \, N\, \tilde{\eta}$
interaction preserves the $Z_2$ parity.  Invariance under $Z_2$
implies that $N$ components odd under $Z_2$ are not mixed with the
$Z_2$-even light neutrinos $\nu_i$. This forbids the decay of the
lightest $Z_2$-odd component of $\eta$ to light neutrinos through the
heavy right handed neutrinos, ensuring DM stability.
However, simplest schemes of this type lead to $\theta_{13}=0$ as a
first-order prediction~\cite{Hirsch:2010ru}, at variance with recent
reactor results.

In contrast, assigning the three left-handed leptons to a
flavor-triplet implies that the ``would-be'' DM candidate decays very
fast into light leptons, through the contraction of the triplet
representations, see general discussion in
ref.~\cite{Kajiyama:2011gu}.
This problem has been considered by Eby and Framptom \cite{Eby:2011qa}
using a $T'$ flavor symmetry. While the suggested model has the merit
of incorporating quarks nontrivially, it requires an ``external''
$Z_2$ asymmetry in order to stabilize dark matter.
In fact this observation lead ref.~\cite{Lavoura:2011ry} to claim that
a successful realization of the DDM scenario requires the lepton
doublets to be in three inequivalent singlet representations of the
flavour group.

Here we provide an explicit example of a model based on a $\Delta(54)$
flavour symmetry in which left-handed leptons are assigned to
nontrivial representations of the flavour group, with a viable stable
dark matter particle and a nontrivial inclusion of quarks. 
We demosntrate the phenomenological viability of the scenario by
performing a fit of the neutrino oscillation parameters, taking into
account the recent reactor angle
measurements.

We search for a group $G$ that contains at least two irreducible
representations of dimension larger than one, namely $r_a$ and $r_b$
with dim$(r_{a,b})>1$. We also require that all the components of
$r_a$ transform trivially under an abelian subgroup of $G\supset Z_N$
(with $N=2,3$) while at least one component of $r_b$ is charged with
respect to $Z_N$. The stability of the lightest component of $r_b$ is
guaranteed by $Z_N$ giving a potential~\footnote{Of course,
  other requirements are necessary in order to have a viable DM
  candidate, such as neutrality, correct relic abundance, and
  consistency with constraints from DM search experiments.} DM
candidate.

The simplest group we have found with this feature is $\Delta(54)$,
isomorphic to $(Z_3\times Z_3) \rtimes S_3$.  In addition to the
irreducible triplet representations, $\Delta(54)$ contains four
different doublets ${\bf 2_{1,2,3,4}}$ and two irreducible singlet
representations, ${\bf 1_{\pm}}$.  The product rules for the doublets
are ${\bf 2_k}\times {\bf 2_k}= {\bf 1_+}+{\bf 1_-}+{\bf 2_k}$ and
${\bf 2_1}\times {\bf 2_2}={\bf 2_3}+{\bf 2_4}$. Of the four doublets
${\bf 2_1}$ is invariant under the $P\equiv (Z_3\times Z_3)$ subgroup
of $\Delta(54)$, while the others transform nontrivially, for example
${\bf 2_3} \sim (\chi_1,\chi_2)$, which transforms as
$\chi_1\,(\omega^2,\omega)$ and $\chi_2\,(\omega,\omega^2)$
respectively, where $\omega^3=1$
\cite{Ishimori:2010au}. We can see that by taking $r_a
= {\bf 2_1}$ and $r_b = {\bf 2_3}$ that $\Delta(54)$ is perfect for
our purpose.


Let us now turn to the explicit model, described in table~\ref{tab1},
where $L_D \equiv (L_\mu ,L_\tau) $ and $l_D \equiv (\mu_R, \tau_R)$.
There are 5 $SU(2)_L$ doublets of Higgs scalars: the $H$ is a singlet
of $\Delta(54)$, while $\eta=(\eta_1,\eta_2) \sim {\bf 2_3}$ and
$\chi=(\chi_1,\chi_2) \sim {\bf 2_1}$ are doublets.  In order to
preserve a remnant $P$ symmetry, the doublet $\eta$ is not allowed to
take vacuum expectation value (vev).
Such a prescription is not necessary for $H$, $\chi_1$ and $\chi_2$
since these are all invariant under $P$.  We also need to introduce an
$SU_L(2)$ Higgs triplet scalar field $\Delta \sim{\bf 2_1}$ whose vev
will induce neutrino masses through the type-II seesaw
mechanism.
Regarding dark matter, note that the lightest $P$-charged particle in
$\eta_{1,2}$ can play the role of ``inert'' DM~\cite{barbieri:2006dq},
as it has no direct couplings to matter. The conceptual link
between dark matter and neutrino phenomenology arises from the fact
that the DM stabilizing symmetry is a remnant of the underlying flavor
symmetry which accounts for the observed pattern of oscillations.  See
phenomenological implications below.
\begin{table}[h]
\begin{center}
\begin{tabular}{ccccc|cccc}
\hline
\hline
& $\overline{L}_e$ & $\overline{L}_D$ & $e_R$ & $l_{D}$   
& $H$  & $\chi $ & $\eta $ &$\Delta$\\
\hline 
$SU(2)$ & $2$ & $2$ & $1$ & $1$  &$2$ & $2$ & $2$ & $3$\\
$\Delta (54)$ & $\bf 1_+$ & $\bf 2_1$ & $\bf 1_+$ & $\bf 2_1$    & $\bf 1_+$& $\bf 2_1$ & $\bf 2_3$ & $\bf 2_1$ \\
\hline
\hline
\end{tabular}
\caption{Lepton and higgs boson assignments of the model. }\label{tab1}
\end{center}
\end{table}

The lepton part of the Yukawa Lagrangian is given by  
\begin{eqnarray}
\mathcal{L}_{\ell}&=&y_1 \overline{L}_e e_R H+
y_2 \overline{L}_e l_D \chi+ 
y_3 \overline{L}_D e_R\,\chi + \\
&&+ y_4\overline{L}_Dl_D H + y_5 \overline{L}_Dl_D \chi \nonumber\\
&&\nonumber\\
\mathcal{L}_{\nu}&=&
y_b\overline{L}_D \overline{L}_D\Delta +
y_a\overline{L}_DL_e \Delta 
\end{eqnarray}
After electroweak symmetry breaking the first term
$\mathcal{L}_{\ell}$ gives the following charged lepton mass matrix
\begin{equation}\label{Ml}
M_\ell =
\begin{pmatrix}
a & br & b \\
cr & d & e \\
c & e & dr
\end{pmatrix}
\end{equation}
where $a=y_1 \SMvev{H}$, $b= y_2\SMvev{\chi_1}$, $c=y_3 \SMvev{\chi_1}$,
$d=y_5 \SMvev{\chi_1}$, $e=y_4 \SMvev{H}$, and
$$r=\SMvev{\chi_2}/\SMvev{\chi_1}.$$

On the other hand the $\mathcal{L}_{\nu}$ is the term responsible for
generating the neutrino mass matrix.  Choosing the solution
$\SMvev{\Delta}\sim (1,1)$ and $\SMvev{\chi_1} \ne \SMvev{\chi_2}$,
consistent with the minimization of the scalar potential one finds
that
\begin{equation}\label{mnu}
M_\nu \propto 
\begin{pmatrix}
0 & \delta  & \delta   \\
\delta & \alpha & 0 \\
\delta  & 0  & \alpha
\end{pmatrix},
\end{equation}
where $\delta=y_a\SMvev{\Delta}$, $\alpha= y_b \SMvev{\Delta}$.

Our model corresponds to a flavour-restricted realization of the {\it
  inert dark matter} scenario proposed in~\cite{barbieri:2006dq}. As
such, it has nontrivial consequences for neutrino phenomenology, which
we now study in detail.
As seen in eq.~(\ref{mnu}) the neutrino mass matrix depends only on
two parameters, $\delta$ and $\alpha$ (taken to be real), which can be
expressed as a function of the measured squared mass differences, as
follows
\begin{equation}
m_{1,3}^\nu=\frac{\alpha \mp \sqrt{8 \delta^2+\alpha^2}}{2}, \: m_2^\nu=\alpha.
\label{massesin}
\end{equation}

\begin{figure}[h!]
\begin{center}
 \includegraphics[width=9cm]{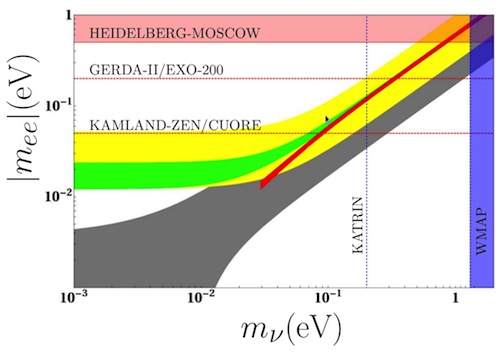}
 \caption{Effective neutrinoless double beta decay parameter $m_{ee}$
   versus the lightest neutrino mass. 
 }\label{figbb}
\end{center}
\end{figure}

For simplicity, we fix the intrinsic neutrino
CP--signs~\cite{Schechter:1981hw} as $\eta=diag(-,+,+)$, where $\eta$
is defined as $U^\star=U \eta$, $U$ being the lepton mixing matrix.
It is easy to check that, in this case, only a normal hierarchy
spectrum is allowed.
In contrast, a different permutation of the eigenvalues corresponding
to our $\eta$ matrix, namely $(1,2,3) \to (1,3,2)$ in
Eq.\,\ref{massesin}, gives only inverse hierarchy spectrum.
Moreover, notice that the masses in eq.~(\ref{massesin}) obey a
neutrino mass sum rule of the form $m_1^\nu+m_2^\nu=m_3^\nu$ which has
implications for the neutrinoless double beta decay
process~\cite{Dorame:2011eb}, as illustrated in Fig.~(\ref{figbb}).

\begin{figure}[h!]
\begin{center}
 \includegraphics[width=7.5cm]{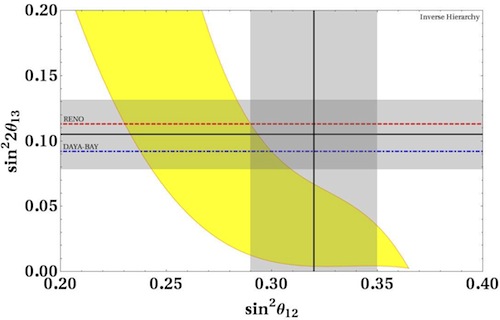}
 \includegraphics[width=7.5cm]{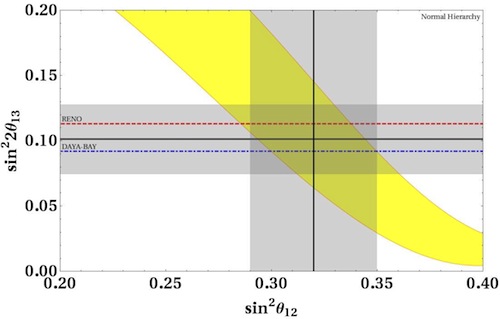}
 \caption{
({\it Left})-- The shaded (yellow) curved band gives the predicted
   correlation between solar and reactor angles when $\theta_{23}$ is
   varied within 2$\sigma$ for the normal hierarchy spectrum.
({\it Right})-- Same  for the inverse hierarchy case.}\label{figci}
\end{center}
\end{figure}

We now turn to the second prediction. Although in our scheme
neutrino mixing parameters in the lepton mixing matrix are not
strictly predicted, there are correlations between the
reactor and the atmospheric angle, as illustrated in
Figs.~\ref{figci}~\footnote{There is also a second band allowed in
  this case which is, however, experimentally ruled out by the
  measurents of $\theta_{12}$ and $\theta_{13}$.} 
for the cases of normal and inverse mass hierarchies, respectively.

While the solar angle is clearly unconstrained and can take all the
values within in the experimental limits, correlations
exist with the reactor mixing angle, indicated by the curved yellow
bands in Fig.~\ref{figci}. These correspond to
$2\sigma$ regions of $\theta_{23}$ as determined in
Ref.~\cite{Schwetz:2011zk}.  The horizontal lines give the best global
fit value and the recent best fit values obtained in Daya--Bay and
RENO reactors.

Now we turn to quarks. In
Ref.~\cite{Hirsch:2010ru,Meloni:2011cc,Boucenna:2011tj,Meloni:2010sk}
quarks were singlets of the flavor symmetry to guarantee the stability
of the DM. Consequently the generation of quark mixing was difficult
\cite{deAdelhartToorop:2011ad}.  This problem has been recently
considered in \cite{Eby:2011qa} using $T'$ flavor symmetry.

A nice feature of our current model is that with $\Delta(54)$ we can
assign quarks to the singlet and doublet representations as shown in
table\,\ref{tab2}. This opens new possibilities to fit the CKM mixing
parameters.
Indeed, as shown in table~\ref{tab2} quarks transforming nontrivially
under the flavor symmetry can be consistently added in our picture.
\begin{table}[h!]
\begin{center}
\begin{tabular}{cccccccc}
\hline
\hline
$ $ & $Q_{1,2}$ & $ Q_{3}$ & $(u_{R},c_R)$ &$t_R$ &$d_R$ & $s_R$ &$b_R$\\
\hline
$SU(2)$ & $2$ & $2$ & $1$ & $1$  &$1$ & $1$ & $1$ \\
\hline
$\Delta(54)$ & ${\bf 2_1}$& ${\bf 1_+}$& ${\bf 2_1}$& ${\bf 1_+}$& ${\bf 1_-}$& ${\bf 1_+}$& ${\bf 1_+}$\\
\hline
\hline
\end{tabular}\caption{Quark gauge and flavour representation assignments.}
\label{tab2}
\end{center}
\end{table}

The resulting up- and down-type quark mass matrices in our model are
given by
\begin{equation}
M_d=
\begin{pmatrix}
ra_d & rb_d & rd_d \\
-a_d & b_d & d_d \\
0 & c_d & e_d
\end{pmatrix},\quad 
M_u=
\begin{pmatrix}
ra_u & b_u & d_u \\
b_u & a_u & rd_u \\
c_u & rc_u & e_u
\end{pmatrix}.
\end{equation}

Note that the Higgs fields $H$ and $\chi$ are common to the lepton and
the quark sectors and in particular the parameter $r$.  
Assuming for simplicity real couplings we have 11 free parameters
characterizing this sector, 10 Yukawa couplings plus the ratio of the
the isodoublet vevs, $r$, introduced earlier in the neutrino
sector.
We have verified that we can make a fit of all quark masses and
mixings provided $r$ lies in the range of about $ 0.1<r<0.2$. We do
not extend further the discussion on the quark interactions which can
be easily obtained from table\,\ref{tab2} (a full analysis of the
quark phenomenology is beyond the scope of this paper and will be
taken up elsewhere).


Notice that our scalar Dark matter candidate $\eta_1$ has quartic
couplings with the Higgs scalars of the model such as
$\eta^\dagger\eta \,H^\dagger H$ and $\eta^\dagger\eta \,\chi^\dagger
\chi$. These weak strength couplings provide a Higgs portal production
mechanism, and ensure an adequate cosmological relic abundance.
Direct and indirect detection prospects are similar to a generic WIMP
dark matter, as provided by multi-Higgs extensions of the SM.

In short we have described how spontaneous breaking of a $\Delta(54)$
flavor symmetry can stabilize the dark matter by means of a residual
unbroken symmetry. In our scheme left-handed leptons as well as quarks
transform nontrivially under the flavor group, with neutrino masses
arising from a type-II seesaw mechanism.  We have found 
lower bounds for neutrinoless double beta decay, even in the case of
normal hierarchy, as seen in Fig.~\ref{figbb}. In addition, we have
correlations between solar and reactor angles consistent with the
recent Daya-Bay and RENO reactor measurements, see Fig.~\ref{figci}.

Unfortunately, however, the DM particle is not directly involved as
messenger in the neutrino mass generation mechanism. This issue will
be considered elsewhere.

\section*{Acknowledgments}

This work was supported by the Spanish MEC
under grants FPA2011-22975 and MULTIDARK CSD2009-00064
(Consolider-Ingenio 2010 Programme), by Prometeo/2009/091 (Generalitat
Valenciana), by the EU ITN UNILHC PITN-GA-2009-237920.

\bibliography{morisi}
\bibliographystyle{apsrev4-1}


%% file: Papers/soumitra.tex

%
%
%
%
%
%

\chapter[Extraction of the CP phase and the life time difference from penguin free tree level $B_s$ decays (Nandi)]{Extraction of the CP phase and the life time difference from penguin free tree level $B_s$ decays}
\vspace{-2em}
\paragraph{S. Nandi}
\paragraph{Abstract}
In this talk I present alternative methods for the extraction of the CP phase $2\beta_s$ and lifetime difference $\Delta\Gamma_s$ using penguin-free tree level
two body $B_s \to D^0_{CP} \phi$ and three body $B_s ({\bar B_s}) \to D^0_{CP} K K$ decays.

\section{Introduction}
Apart from the direct searches at colliders, low energy observables in flavor physics play an essential role for an indirect search of NP;
 in this respect FCNC processes are important. The data from the decays of K, D and B mesons have so far been consistent with
the Cabbibo-Kobayashi-Maskawa (CKM) paradigm of Standard Model (SM), however the flavor changing neutral current
(FCNC) processes involving $b\to s$ transitions are expected to be sensitive to many sources of new physics (NP) since FCNC decays
are rare (i. e. loop-suppressed) in the SM \cite{Cheng:2005ug,Buchalla:2005us,Lunghi:2009sm}.

In light of this, it is particularly important to study $b\to s$
transitions and look for new-physics (NP) effects. Now, if NP is
present in $\Delta B = 1$ $b\to s$ decays, it would be highly unnatural
for it not to also affect the $\Delta B=2$ transition, in particular
$\SNbs$-$\SNbsbar$ mixing. At the same time, we do hope wealth of data on $B_s$
system from LHCb.

In order to see where NP can enter, we briefly review the mixing. 
Effective Hamiltonian for $B_q-\bar{B_q}$ mixing
\SNbea
{H_{eff}=\left (\begin{array}{cc}
M_{11q} -\textstyle{\frac{i}{2}} \Gamma_{11q} & M_{12q}-\textstyle{\frac{i}{2}}\Gamma_{12q}  \\
M^\ast_{12q}-\textstyle{\frac{i}{2}}\Gamma^\ast_{12q} & M_{11q}-\textstyle{\frac{i}{2}}\Gamma_{11q} \end{array} \right)},
\nonumber
\SNeea
where $M=M^\dagger$ and $\Gamma=\Gamma^\dagger$ correspond
respectively to the dispersive and absorptive parts of the mass
matrix. The off-diagonal elements, $M^s_{12}=M_{21}^{s*}$ and
$\Gamma^s_{12}=\Gamma_{21}^{s*}$, are generated by $\SNbs$-$\SNbsbar$
mixing.

\begin{figure}
\centering
\includegraphics[width=0.69\linewidth]{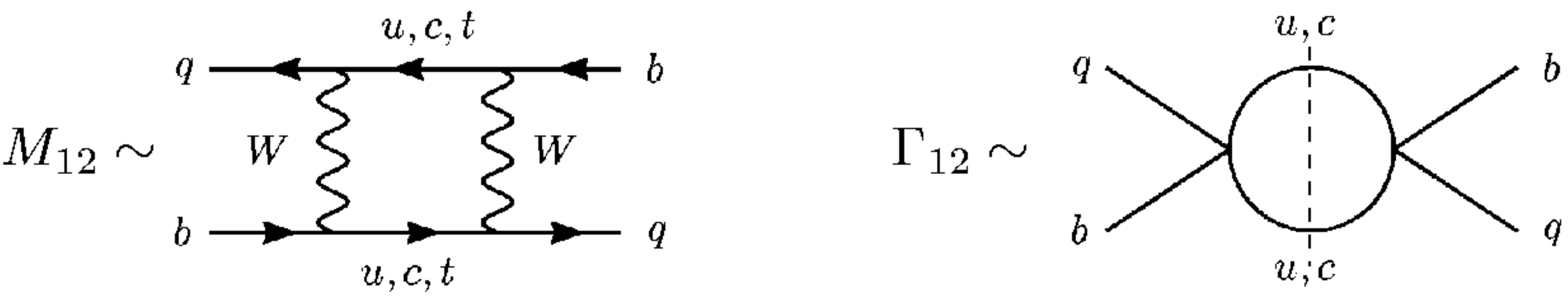}
\caption{Diagrams contribute to $M_{12}$ and $\Gamma_{12}$.}
\label{fig1}
\end{figure}

We define
\SNbea
\Gamma_s \equiv \frac{\Gamma_H
  + \Gamma_L}{2}, \quad \Delta M_s \equiv   M_H - M_L, \quad \Delta \Gamma_s \equiv   \Gamma_L -
\Gamma_H ~,
\SNeea
where $L$ and $H$ indicate the light and heavy states, respectively.
${M_{L,H}}$ and $\Gamma_{L,H}$ are the masses and decay widths of the light and heavy
mass eigenstates respectively. Mass difference and width difference can be calculated from the dispersive and absorptive part of the box diagram
shown in~\ref{fig1}

Expanding the mass eigenstates, we find, to a very good approximation \cite{Dunietz:2000cr},
\SNbeq
\Delta M_s = 2 |M_{12}^s| ~, \hskip 0.2cm
 \Delta \Gamma_s =  2 |\Gamma_{12}^s| \cos \phi_s  ~, \hskip 0.2cm
 \frac{q}{p} =  e^{-2i\beta_s}\left[1-\frac{a}{2}\right] ~,
\label{eq:mix_para}
\SNeeq
where $\phi_s \equiv \arg(-M_{12}^s/\Gamma_{12}^s)$ is the CP phase in
$\Delta B=2$ transitions. In Eq.~(\ref{eq:mix_para}) the small
expansion parameter $a$, the semileptonic asymmetry, is given by
\SNbeq
a = a^s_{sl} = \frac{|\Gamma^s_{12}|}{|M^s_{12}|}\,\sin\phi_s ~.
\label{asl}
\SNeeq
This is expected to be $\ll 1$, and hence can be neglected in the
definition of $q/p$. The weak phase $2\beta_s$ appears in the indirect (mixing-induced) CP
asymmetries. 
%
%

The SM predictions of all these observables are given by \cite{Lenz:2011ti}
\SNbea
\Delta M_s &=& (17.3 \pm 2.6)\,ps^{-1}, \hskip 1cm \Delta\Gamma_s = (0.087 \pm 0.021)\,ps^{-1} \nonumber \\
2\beta_s &\approx& 2^{\circ}, \hskip 1cm \phi_s = 0.22^{\circ},  \hskip 1cm a^s_{sl} = (1.9 \pm 0.3)\times 10^{-5}
\label{obssm}
\SNeea
The present world averages are given by \cite{CDF:2011af,Abazov:2011ry,LHCb:2011aa,Amhis:2012bh,Xie:2009fs}
\SNbea
\Delta M_s &=& (17.69 \pm 0.08) ps^{-1}, \hskip 1cm  \Delta\Gamma_s = ( 0.103 \pm 0.014 ) ps^{-1} \nonumber \\
2\beta_s &=& 0.14^{+0.11}_{-0.16}, \hskip 1cm a^s_{sl} = - 0.0105 \pm 0.0064
\label{obsexp}
\SNeea
Therefore, the present data still allow 20\% to 30\% CP-violating NP effects.
There is no separate measurement on $\phi_s$ and it is not wise to consider $\phi_s = 2\beta_s$, as
we can see from Eq.~(\ref{obssm}), even in the SM they are not equal. However, it is possible
to constrain $\phi_s$ along with $\Gamma^s_{12}$ from the measurement of $\Delta\Gamma_s$ and
semileptonic asymmetry $a_{sl}^s$. Combining Eqs.~(\ref{eq:mix_para}) and (\ref{asl}) we obtain
\SNbea
\tan\phi_s &=& \frac{a_{sl}^s\,\Delta M_s}{\Delta\Gamma_s} = -1.80 \pm 1.12 ~, \nonumber \\
|\Gamma_{12}^s| &=& \frac{\sqrt{{\Delta\Gamma_s}^2 + {a_{sl}^s}^2\,{\Delta M_s}^2}}{2} = 0.106 \pm 0.051 ~.
\label{phis}
\SNeea
The constrained values of the phase $\phi_s$ and $|\Gamma^s_{12}|$ are consistent with the SM within the error bar, however,
significant deviations can not be ruled out. Once we include $a^s_{sl} = (-1.81 \pm 1.06)\%$, the data
provided by D\O\ from dimuon asymmetry measurement \cite{:2012he} , the situation will be further worsen with respect
to SM predictions. Therefore, more precise measurements of $\Delta\Gamma_s$ and $a^s_{sl}$ are essential.

So far $2\beta_s$ seems to be SM like, however, there are facts to remember. Extraction of
$2\beta_s$ from $B_s \to J/\psi M$ decays is theoretically clean, provided the subleading
terms are assumed to vanish. In the next few years, with the LHCb we are entering the
era of high precision physics. For example, the CP asymmetry $S_{\psi\phi}$ in $\SNbs \to J/\psi \phi$
decay will be measured with 3\% accuracy. Hence, subleading SM contributions will become important.
On the other hand due to our poor understanding of low energy QCD it is extremely hard to
estimate/calculate reliably the ratio of leading to the subleading contributions \cite{Jung:2012mp}.
The problem lies with the evaluation of the hadronic matrix element. At the same time, the
possibility of NP in $b\to c{\bar c}s$ decays can not be ruled out \cite{Chiang:2009ev}.
Therefore,  it is worthwhile to look for a process in which NP in the decay can
essentially be neglected, and permits the determination of $2\beta_s$ and $\Delta\Gamma_s$
without any ambiguity. In this regard, tree level $B_s$ decays via $ b\to c {\bar u} s$ and
$ b\to u {\bar c} s$ transitions may play an interesting role. In the following sections, we discuss the 
extraction of $2\beta_s$ and $\Delta\Gamma_s$ from two and three body $B_s$ decays.

\section{Two body decays: $\SNbs (\SNbsbar) \to D^0 \phi, {\bar D^0} \phi $}

We consider first the two body decays via $ b\to c {\bar u} s$ and $ b\to u {\bar c} s$ transitions,
and try to see what can we learn from such decays. Consider a final state $f$ to which both
$\SNbs$ and $\SNbsbar$ can decay, and the decay amplitudes are dominated by a single weak phase.
\SNbeq
\frac{\Gamma(\SNbs(t) \to f) - \Gamma(\SNbsbar(t) \to f)}
{\Gamma(\SNbs(t) \to  f) + \Gamma(\SNbsbar(t) \to f)} =
\frac{ C \cos \Delta m_s t -  S \sin \Delta m_s t}
{ \cosh(\Delta \Gamma_s t /2) - {{\cal A}}_{\Delta\Gamma} \sinh(\Delta \Gamma_s t /2)} ~.
\label{indirectCPA2}
\SNeeq
Therefore, the following interesting observables can be extracted
\SNbeq
C \equiv \frac{1 - |\lambda_f|^2}{1 + |\lambda|^2}, \hskip 0.5cm
S \equiv \frac {2 \, {\rm Im} \lambda_f} {1 + |\lambda_f|^2}, \hskip 0.5cm
{\cal A}_{\Delta\Gamma} \equiv \frac {2 \, {\rm Re} \lambda_f} {1 + |\lambda_f|^2},
\label{obsp}
\SNeeq
where
$\lambda_f \equiv \frac{q}{p} \frac{{\bar A}_f}{A_f}  = |\lambda_f| e^{- i (\phi^{mix}_s + \theta - \delta)}$,
$\phi^{mix}_s$ is the mixing phase and $\theta-\delta = - Arg \left[ \frac{{\bar A}_f}{A_f} \right]$.
The weak and strong phase difference between the decay amplitudes
${\bar A}_f = \SNbsbar \to f$ and $ A_f = \SNbs \to f$ are given by $\theta$ and $\delta$ respectively.
Similarly, for the final state $\bar f$ we get
\SNbeq
{\bar S} \equiv \frac {2 \, {\rm Im} {\bar\lambda_f}} {1 + |{\bar\lambda_f}|^2}, \hskip 0.5cm
{\bar{\cal A}}_{\Delta\Gamma} \equiv \frac {2 \, {\rm Re} {\bar\lambda_f}} {1 + |{\bar\lambda_f}|^2},
\label{obspa}
\SNeeq
with
${\bar\lambda_f} \equiv \frac{p}{q} \frac{A_{\bar f}}{{\bar A}_{\bar f}} = \frac{1}{|\lambda_f|} 
e^{- i ( \phi^{mix}_s + \theta + \delta)}$. The various combination of the these observables are useful to
extract the CP phase.

In the SM, the amplitude of the $\SNbs \to D^0\phi$ and $\SNbsbar \to D^0\phi$ decays are
of the same order, hence, leads to interference effects between $\SNbs$-$\SNbsbar$ mixing and
decay process. By measuring the time dependence of the decays, one can obtain $S$,
$\bar S$, $A_{\Delta\Gamma}$ and ${\bar A}_{\Delta\Gamma}$ as given in
Eqs.~(\ref{obsp}) and (\ref{obspa}), for detail see Ref.~\cite{Nandi:2011uw}. Using these observables we
extract $\sin(2\beta_s + \gamma + \delta_{\phi})$, $\sin(2\beta_s + \gamma - \delta_{\phi})$,
$\cos(2\beta_s + \gamma + \delta_{\phi})$, $ \cos(2\beta_s + \gamma - \delta_{\phi})$, which
allows us to obtain $2\beta_s + \gamma$ with a twofold ambiguity; similar information as in
$\SNbs (\SNbsbar) \to D_s^{\pm} K^{\mp}$ decays \cite{Fleischer:2003yb,Nandi:2008rg}.

The advantage of these decays is that there is a third decay which is
related: $\SNbs (\SNbsbar) \to D^0_{CP} \phi$, where $D^0_{CP}$ is a CP
eigenstate (either CP-odd or CP-even). In our analysis we consider
$D^0_{CP}$ as the CP-even superposition $(D^0 + {\bar D^0})/\sqrt{2}$.
In this case, time dependent decay distributions allow to extract two more functions
$\cos(\gamma + \delta_{\phi})$ and $\cos(\gamma - \delta_{\phi})$. Therefore,
various algebraic combination of all these functions allow one to determine
$\sin2\beta_s$, $\cos2\beta_s$, $\sin(2\beta_s+ 2\gamma)$, $ \cos(2\beta_s+ 2\gamma)$.
Hence, unambiguous determinations of $2\beta_s$ and $2\gamma$ is possible.

\section{Three body $\SNbs(\SNbsbar) \to D^0_{CP} K {\bar K}$ decays: Dalitz analysis}
In the previous section we discussed two-body ${\bar b} \to {\bar c} u
{\bar s}$/${\bar b} \to {\bar u} c {\bar s}$ decays; in this section
we examine the corresponding three-body decays. In recent years, various tests of the SM,
as well as the extraction of weak phases, have been examined in the context of $B \to K \pi \pi$, $B \to K {\bar
  K} K$, $B \to \pi {\bar K} K$ and $B \to \pi \pi \pi$ decays \cite{Ciuchini:2006st,Gronau:2006qn}, which uses
Dalitz-plot analyses. The extra piece of information available in $B_s$ decays, due to the 
sizeable lifetime difference $\Delta\Gamma_s$, can provide important insights into the 
CP violation studies of three body Dalitz analysis.
The $\SNbs ({\SNbsbar}) \to D^0_{CP} K {\bar K}$ 
decays receive a tree contribution. The CKM matrix elements of these decays are the same as in the 
corresponding two-body decay modes, and will therefore exhibit very similar time-dependent CP asymmetries.

In the following, we perform a time-dependent Dalitz-plot analysis of the
$\SNbs(\SNbsbar) \to D^0_{CP} K {\bar K}$ decays, which can decay either via
intermediate resonances ($\phi$, $f_0$ etc.) or non-resonant contributions.
This permits the measurement of each of the contributing amplitudes, as
well as their relative phases. In the isobar model, the individual terms are
interpreted as complex production amplitudes for two-body resonances, and one also includes a term
describing the non-resonant component. The amplitude is then given by
\SNbeq
{\cal A}(s^+,s^-) = \sum_j a_j F_j(s^+,s^-), \hskip 1cm  {\bar {\cal A}}(s^+,s^-) = \sum_j {\bar a_j} {\bar F_j}(s^-,s^+)
\SNeeq
where the sum is over all decay modes (resonant and non-resonant).
Here, the $a_j$ are the complex coefficients describing the magnitudes
and phases of different decay channels, while the $F_j(s_{12},s_{13})$
contain the strong dynamics. It takes different (known) forms for the various contributions.

The time-dependent decay rates for decay to the same final state $f$, are given by \cite{Nandi:2011uw}
\SNbea
\Gamma(\SNbs(t) \to f) &\! \sim \!& \frac12 e^{-\Gamma_s t}
\Big[A_{ch}(s^+,s^-) \cosh(\Delta \Gamma_s t /2) - A_{sh}(s^+,s^-) \sinh(\Delta \Gamma_s t /2)
  \SNnn \\
&& \hskip0.4truein
 +~A_{c}(s^+,s^-) \cos(\Delta m_s t) - A_{s}(s^+,s^-) \sin(\Delta m_s t)\Big] ~, \SNnn\\
\Gamma(\SNbsbar(t) \to f) &\! \sim \!& \frac12 e^{-\Gamma_s t}
\Big[A_{ch}(s^-,s^+) \cosh(\Delta \Gamma_s t /2) - A_{sh}(s^-,s^+) \sinh(\Delta \Gamma_s t /2)
  \SNnn \\
&& \hskip0.4truein
 -~A_{c}(s^-,s^+) \cos(\Delta m_s t) + A_{s}(s^-,s^+) \sin(\Delta m_s t)\Big] ~.
\label{timedalitz}
\SNeea
Here
\SNbea
A_{ch}(s^+,s^-) &=& |{\cal A}(s^+,s^-)|^2 + |{\bar {\cal A}}(s^+,s^-)|^2 ~, \SNnn \\
A_{c}(s^+,s^-) &=& |{\cal A}(s^+,s^-)|^2 - |{\bar {\cal A}}(s^+,s^-)|^2 ~, \SNnn \\
A_{sh}(s^+,s^-) &=& 2 {\rm Re} \left( e^{-2 i \beta_s} {\bar {\cal A}}(s^+,s^-) {\cal A}^{\ast}(s^+,s^-) \right) ~,  \SNnn \\
A_{s}(s^+,s^-) &=&  2 {\rm Im} \big( e^{-2 i \beta_s} {\bar {\cal A}}(s^+,s^-) {\cal A}^{\ast}(s^+,s^-) \big) ~.
\label{cdalitz}
\SNeea
Maximum likelihood fit over the entire Dalitz plot, allows to extract the magnitudes
and relative phases of the $a_j$ or ${\bar a}_j$.

As mentioned before the $\SNbs(\SNbsbar) \to D^0_{CP} K {\bar K}$ decays can proceed
via various two body resonances, here for simplicity we consider only the
interefernce of two such resonances. Maximum likelihood fit to the Dalitz-plot
PDFs allows to extract $ \tan\gamma$ without ambiguity from
$A^{DKK}_{c}$ and $A^{DKK}_{ch}$, 
\SNbea
A^{DKK}_{c} &=&  \sum_{i=\phi,f_0} \left[ \left(|A_i|^2 - |{\bar A_{i}}|^2\right) + 2 {\rm Re}  \big( A_{\phi} A^{\ast}_{f_0}
- {\bar A_{\phi}} {\bar A^{\ast}_{f_0}} \big) \right] ~, \SNnn \\
A^{DKK}_{ch} &=& \sum_{i=\phi,f_0} \left[ \left(|A_i|^2 + |{\bar A_{i}}|^2\right) + 2 {\rm Re}  \big( A_{\phi} A^{\ast}_{f_0}
+ {\bar A_{\phi}} {\bar A^{\ast}_{f_0}} \big) \right] ~,
\label{adkkc}
\SNeea
for detail see~\cite{Nandi:2011uw}. Hadronic uncertainties cancel, theoretically clean determination of the CKM angle
$\gamma$ is possible.

From the interference of two resonances in $A^{DKK}_s$, 
\SNbeq
A^{DKK}_s = {\rm Im}\left[ e^{-2i\beta_s} {\cal A}^{\ast} {\bar{\cal A}}\right] = {\rm Im} \left[
e^{-2i\beta_s} \big( A^{\ast}_{\phi} \bar A_{\phi} +
A^{\ast}_{\phi} \bar A_{f_0} + A^{\ast}_{f_0} \bar A_{\phi} + A^{\ast}_{f_0} \bar A_{f_0} \big) \right] ~,
\SNeeq
we extract $\sin2\beta_s$, $\sin(2\beta_s + \gamma \pm \delta_i)$, $\cos(2\beta_s + \gamma \pm \delta_i)$,
$\sin(2\beta_s + 2\gamma \pm \delta_{ij})$, $ \cos(2\beta_s + 2\gamma \pm \delta_{ij})$, where $i = \phi$ or $f_0$. 
In the above functions $\delta_i$ is the strong phase difference between the amplitudes of the $B_s$ and $\bar B_s$
decay to the final state $i$. From these trigonometric functions, it is straightforward to find expressions for 
$\tan2\beta_s$ and $\tan\gamma$.
We can extract $\sin2\beta_s$ along with constraining $\tan2\beta_s$, hence, an unambiguous
determination of $2\beta_s$ is possible. The tagged analysis alone allows the extraction of
$2\beta_s$ without ambiguity \cite{Nandi:2011uw}.

The time dependent untagged differential decay distribution is given by
\SNbeq
\Gamma_{untagged}(D^0_{CP} K^+ K^-,t) = e^{-\Gamma_s t}\left[ A^{DKK}_{ch} \cosh(\Delta \Gamma_s t /2)
 + A^{DKK}_{sh}  \sinh(\Delta \Gamma_s t /2) \right].
\label{untagged}
\SNeeq
For a single resonance, say $\phi$,
\SNbea
A^{DKK}_{ch} &=& A^2_{\phi} + {\bar A}^2_{\phi}, \SNnn \\
A^{DKK}_{sh} &=& {\rm Re}\left[e^{-2i\beta_s} |C^{\phi}_2|^2 |F_{\phi}|^2 \big\{1 + {r_{\phi}}^2 e^{-2i\gamma} + r_{\phi}
(e^{-i(\gamma + \delta_{\phi})} + e^{-i(\gamma - \delta_{\phi})})\big\}\right] ~.
\label{Ashdef}
\SNeea
$A^{DKK}_{ch}$ is fully known from the CP-averaged branching fraction of the intermediate resonance
$\phi$. Fit to the tagged decay rate distribution determines: $2\beta_s$, $(2\beta_s + \gamma \pm \delta_{\phi})$
and $\cos(2\beta_s + 2\gamma)$ without ambiguity, hence, $A^{DKK}_{sh}$ can be fully obtained
Therefore, $\Delta\Gamma_s$ is the only unknown in the untagged decay rate distribution given in Eq.~(\ref{untagged}),
it can be determined from the fit, for detail see \cite{Nandi:2011uw}.

\section{Conclusion}
We are entering a new era of high precision studies, the CP phase $2\beta_s$
and $\Delta\Gamma_s$ will be measured with better accuracy. Extraction of same 
observable from various processes are always encouraging, in particular from those modes which
are theoretically clean, as was done in $B_d$ decays. In this regard, the tree level processes 
via $b\to c {\bar u} s$ and $b\to u {\bar c} s$ transitions may play an interesting role.
Combining tagged and untagged measurements of $ \SNbs (\SNbsbar) \to (D^0,{\bar D^0}, D^0_{CP}) \phi$ decays,
we can extract $2\beta_s$ without any ambiguity. Time dependent Dalitz analysis of the 
$ \SNbs (\SNbsbar) \to D^0_{CP} K K$ allows us to extract $2\beta_s$ (from tagged) and
$\Delta\Gamma_s$ (from untagged) without any ambiguity. In addition, this processes 
allow a theoretically clean determination of the CKM angle $\gamma$.

\section*{Acknowledgments}
I would like to thank David London for fruitful collaboration.

\bibliographystyle{apsrev4-1}
\bibliography{soumitra}


%% file: Papers/omura.tex

%
%
%
%
%
%

\chapter[Top FB asymmetry and charge asymmetry in chiral $U(1)$ flavor models (Ko, \textit{Omura}, Yu)]{Top FB asymmetry and charge asymmetry in chiral $U(1)$ flavor models}
\vspace{-2em}
\paragraph{P. Ko, \textit{Y. Omura}, C. Yu}
\paragraph{Abstract}
We study the flavor-dependent chiral U(1)$^\prime$ model where only
the right-handed up-type quarks are charged under U(1)$^\prime$ and
additional Higgs doublets with nonzero U(1)$^\prime$ charges are 
introduced to give proper Yukawa couplings.
We find that some parameter regions could achieve not only the top forward-backward asymmetry
at the Tevatron, but also the charge asymmetry at the LHC without exceeding the upper limit of the same-sign top-quark pair production at the LHC.

\section{Introduction}
The top forward-backward asymmetry ($A_\textrm{FB}^t$) is one of the most
interesting observables because there exists discrepancy
between theoretical predictions in the standard model (SM) 
and experimental results at the Tevatron. The most recent measurement
for $A_\textrm{FB}^t$ at CDF is 
$A_\textrm{FB}^t=0.162\pm 0.047$ in the letpon+jets channel with a full
set of data~\cite{cdfnew}, which is consistent with
the previous measurements at CDF and D0 within uncertainties~\cite{oldafb}. 
The SM predictions are between $0.06$ and $0.09$~\cite{smafbac,smafb},
so that the deviation is around $2 \sigma$.

If the discrepancy in $A_\textrm{FB}^t$ is generated by new physics,
the new physics model would be tested at the LHC. 
One of the good measurements is the charge asymmetry $A_C^y$, which is defined by
the difference of numbers of events with the positive and negative 
$\Delta |y|=|y_t|-|y_{\bar{t}}|$ divided by their sum.
The current values for $A_C^y$ are
$A_C^y=-0.018\pm 0.028\pm 0.023$ at ATLAS~\cite{atlasacy} and 
$A_C^y=0.004\pm 0.010\pm 0.012$ at CMS~\cite{cmsacy}, respectively, which are
consistent with the SM prediction $\sim 0.01$~\cite{smafbac}. 
Another interesting observable at the LHC is the cross section for
the same-sign top-quark pair 
production, $\sigma^{tt}$, which is not allowed in the SM. 
The current upper bound on $\sigma^{tt}$ is about 17 pb at CMS~\cite{cmssame}
and 2 pb or 4 pb at ATLAS depending on the model~\cite{atlassame}.
\footnote{Very recently it is updated by CMS~\cite{Chatrchyan:2012sa}.}
Some models which were proposed 
to account for $A_\textrm{FB}^t$ at the Tevatron,
predict large $A_C^y$ and/or $\sigma^{tt}$ so that they are already disfavored
by present experiments at the LHC.

In this work, we examine the $\textit{so-called}$ 
chiral U(1)$^\prime$ model with flavored Higgs doublets
and flavor-dependent U(1)$^\prime$ charge assignment~\cite{u1models}.
This model is an extension of a $Z^\prime$ model with off-flavor-diagonal
interactions~\cite{zprime}, 
which is excluded by $A_C^y$ and $\sigma^{tt}$ at the LHC.
In the Refs.~\cite{u1models}, the authors 
proposed a model with chiral U(1)$^\prime$ symmetry for the construction
of a realistic $Z^\prime$ model with flavor-off-diagonal couplings,
where only the right-handed up-type quarks are charged under U(1)$^\prime$.
Then, in order to construct a realistic Yukawa interactions,
additional Higgs doublets with U(1)$^\prime$ charges should be introduced.
New chiral fermions should also be introduced in order to cancel
the gauge anomaly. For more details of the chiral U(1)$^\prime$ models,
we refer readers to Refs.~\cite{u1models}. We point out that the simple 
$Z^\prime$ model may be disfavored by the experiments at the LHC, but
if one considers more complete model, then the extended model could be
revived.

In this proceeding, we consider two scenarios of the chiral U(1)$^\prime$
model. One is a light $Z^\prime$ boson case 
with a neutral scalar Higgs boson $H$
and a pseudoscalar Higgs boson $a$. The other is a light Higgs boson $h$ case
with a heavier $Z^\prime$ boson, a heavier Higgs boson $H$, and 
a pseudoscalar Higgs boson $a$, motivated by the recent observation
of a SM-like Higgs boson at the LHC~\cite{higgs}. 
The other particles newly
introduced in the model are assumed to be heavy or have small couplings
so that they are decoupled from top physics.

\begin{figure*}[t]
\centering
\includegraphics[width=135mm]{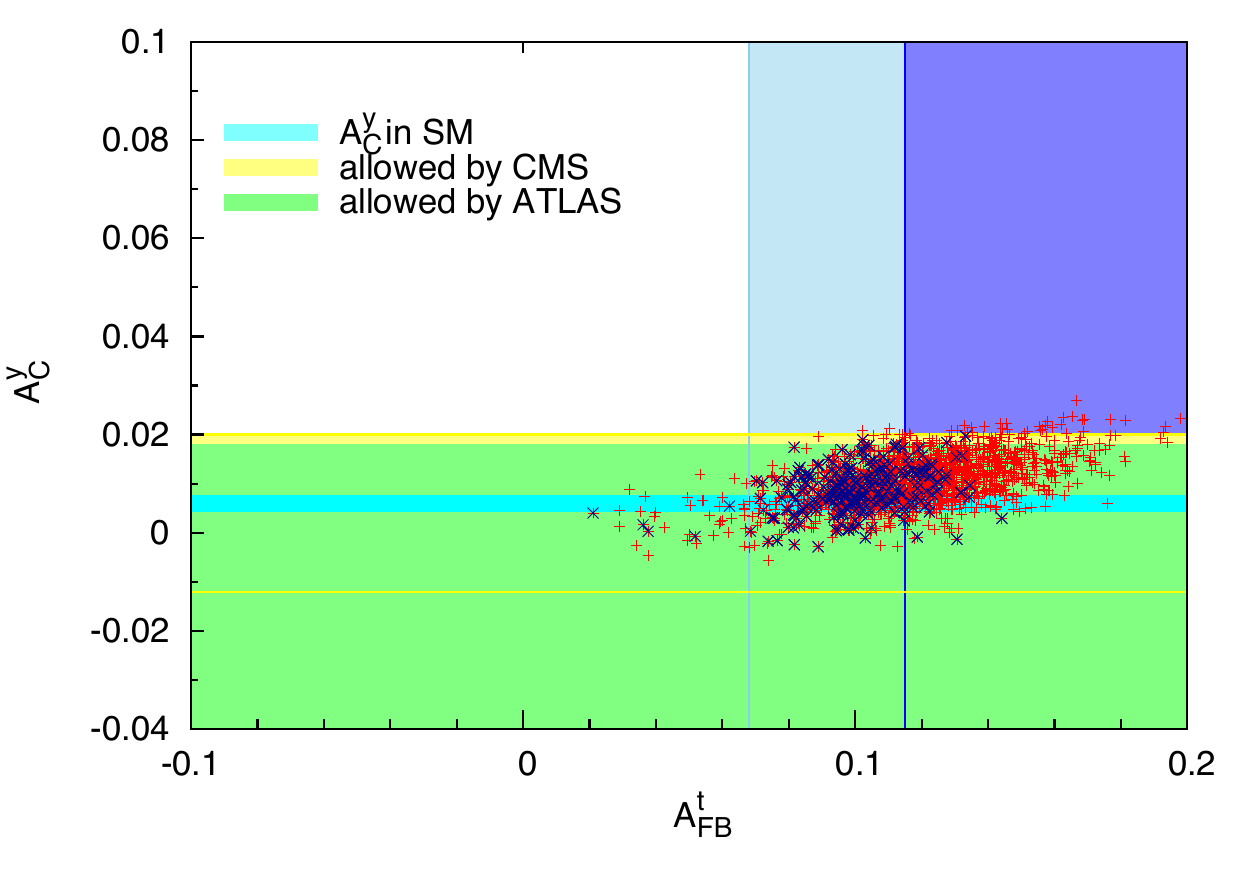}
\caption{$A_{FB}^t$ at the Tevatron and $A_C^y$ at the LHC
for $m_Z^\prime=145$ GeV.} 
\label{omura_fig1}
\end{figure*}

\section{Light Z$^\prime$ case}
In this section, we consider a light $Z^\prime$ boson case with a mass
$m_{Z^\prime}=145$ GeV. In order to suppress the non-SM decay of the top quark,
we require that the Higgs bosons $H$ and $a$ are heavier than the top quark.
However, this requirement might be inconsistent with the recent observation
of an SM-like Higgs boson at the LHC~\cite{higgs} 
and with non-observation of a Higgs-like signal in a large region 
between 130 GeV and 600 GeV~\cite{nonhiggs}. 
In order to accommodate these results, we assume that lightest Higgs $h$
has a zero $u$-$t$-$h$ coupling, and its branching ratio is SM-like.

In this model, the $Z^\prime$ boson can contribute to the top-quark pair
production through its $s$-channel and $t$-channel exchanges
in the $u\bar{u}\to t\bar{t}$ process. While the Higgs bosons contribute
to the top-quark pair production
only in the $t$ channel because the diagonal elements of their Yukawa couplings to light quarks
are negligible.
We scan the following parameter regions:
$180~\textrm{GeV} \le m_{H,a} \le 1~\textrm{TeV}$,
$0.005 \le \alpha_x \le 0.012$,
$0.5 \le Y_{tu}^{H,a} \le 1.5$, and
$(g_R^u)_{tu}^2=(g_R^u)_{uu} (g_R^u)_{tt}$, 
where $\alpha_x \equiv (g_R^u)_{tu}^2 g'^2/(4 \pi)$ is defined and $Y_{tu}^{H,a}$ are flavor-off-diagonal 
Yukawa couplings. One can also consider the case where the $s$-channel exchange
of the $Z^\prime$ boson is negligible by setting the coupling $(g_R^u)_{uu}=0$,
but the numerical results are not so different.

In Fig.~\ref{omura_fig1}, we show the scattered plot for $A_\textrm{FB}^t$ at the
Tevatron and $A_C^y$ at the LHC.
The green and yellow regions are consistent with $A_C^y$ at ATLAS and CMS
in the $1\sigma$ level, respectively. The blue and skyblue regions are
consistent with $A_{FB}^t$ in the lepton+jets channel at CDF in the
$1\sigma$ and $2\sigma$ levels, respectively.
The red points are in agreement with the cross section 
for the top-quark pair production
at the Tevatron in the $1\sigma$ level and the blue points are consistent with
both the cross section for the top-quark pair production at the Tevatron
in the $1\sigma$ level and the upper bound on the same-sign top-quark pair
production at ATLAS. We find that a lot of parameter points can explain all
the experiments: the total cross section, the forward-backward asymmetry,
the same-sign top-pair production, and the top charge asymmetry, which are
considered in this work. We emphasize that the simple $Z^\prime$ model
is excluded by the same-sign top-quark pair production, but
in the chiral U(1)$^\prime$ model, this strong bound could be evaded
due to the destructive interference between the $Z^\prime$ boson and Higgs
bosons.

\begin{figure*}[t]
\centering
\includegraphics[width=135mm]{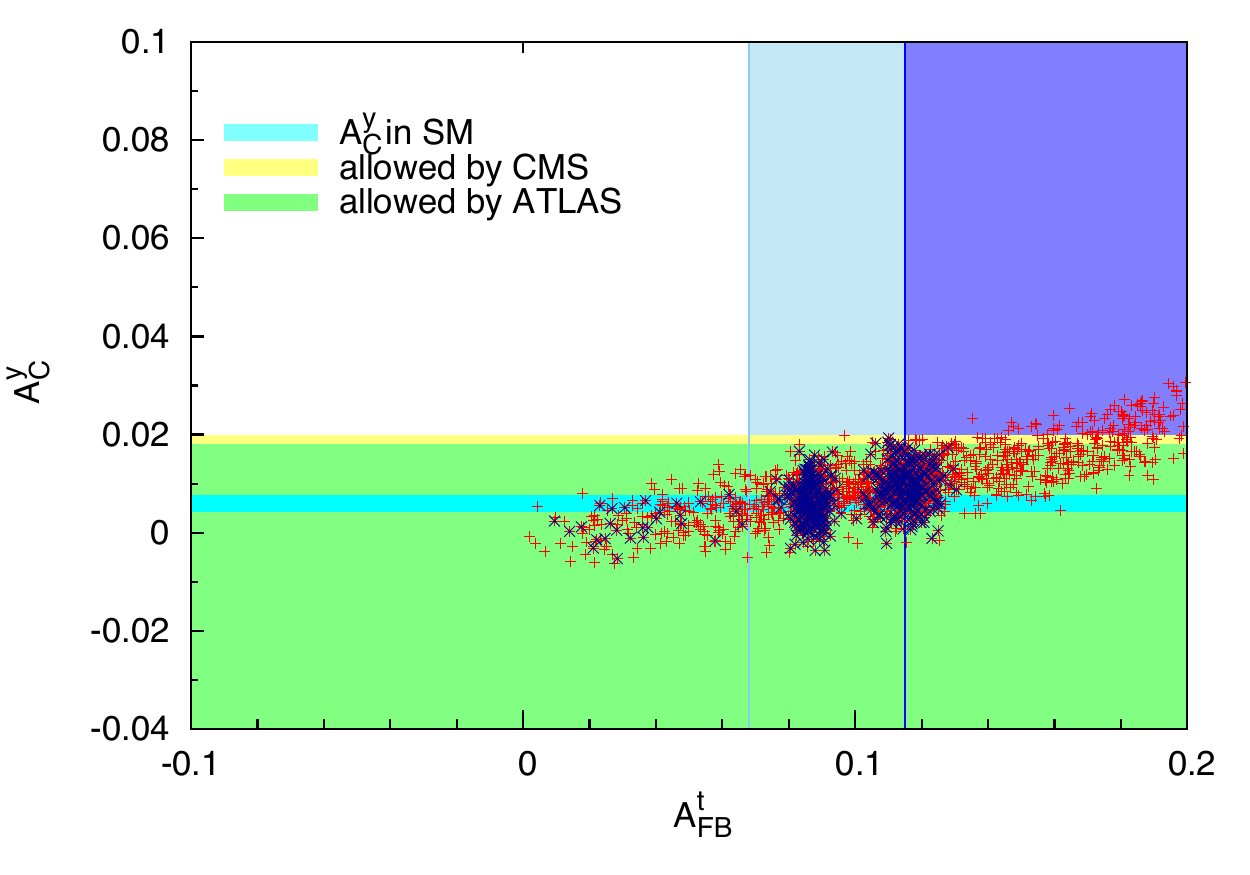}
\caption{$A_{FB}^t$ at the Tevatron and $A_C^y$ at the LHC
for $m_h=125$ GeV.} 
\label{fig2}
\end{figure*}

\section{Light Higgs boson case with heavier Z$^{\prime}$}
In this section, we discuss the scenario that
 a light Higgs boson $h$ with $m_h=125$ GeV, motivated by the recent observation of an SM-Higgs like scalar boson 
at the LHC~\cite{higgs}, also has a nonzero $Y^h_{tu}$. 
In this case, the $Z^\prime$ boson and Higgs bosons
$h$, $H$, and $a$ contribute to the top-quark pair production.
In order to suppress the exotic decay of the top quark into $h$ and $u$, 
we set the Yukawa
coupling of $h$ to be $Y_{tu}^h \le 0.5$ and masses of $Z^\prime$, $H$,
and $a$ are larger than the top-quark mass or approximately equal to the
top-quark mass. We scan the following parameter regions:
$160~\textrm{GeV} \le m_{Z^\prime} \le 300~\textrm{GeV}$,
$180~\textrm{GeV} \le m_{H,a} \le 1~\textrm{TeV}$,
$0 \le \alpha_x \le 0.025$,
$0 \le Y_{tu}^{H,a} \le 1.5$, $0 \le Y_{tu}^{h} \le 0.5$ and 
$(g_R^u)_{tu}^2=(g_R^u)_{uu} (g_R^u)_{tt}$. 
The mass region of the $Z^\prime$ boson
is taken to avoid the constraint from the $t\bar{t}$ invariant mass
distribution at the LHC. If $(g_R^u)_{uu}\simeq0$ 
and the $s$-channel contribution of the $Z^\prime$ could be ignored,
the mass region of the $Z^\prime$ boson could be enlarged.

In Fig.~\ref{fig2}, we show the scattered plot for $A_\textrm{FB}^t$ 
at the Tevatron and $A_C^y$ at the LHC for $m_h=125$ GeV.
All the legends on the figure are the same as those in Fig.~\ref{omura_fig1}.
We find that there exist parameter regions which agree with all the experimental
constraints considered in this work. We emphasize that in some parameter spaces
$\sigma^{tt}$ is less than 1 pb.

\section{Summary}
The top forward-backward asymmetry at the Tevatron
is the only quantity which has 
deviation from the SM prediction in the top quark sector up to now.
A lot of new physics models have been introduced to 
account for this deviation
or it has been analyzed in a model-independent way~\cite{modelindep}.
Some models have already been disfavored by experiments at the LHC and
some new observables are introduced to test the models~\cite{modelindep}.
In this work, we investigated the chiral U(1)$^\prime$ model with flavored
Higgs doublets and flavor-dependent couplings. Among possible scenarios,
we focused on two scenarios: a light $Z^\prime$ boson case and 
a light Higgs boson case. We found that both scenarios can be accommodated with
the constraints from the same-sign top-quark pair production and
the charge asymmetry at the LHC as well as the top forward-backward asymmetry
at the Tevatron.

The chiral U(1)$^\prime$ model has a lot of new particles except for
the $Z^\prime$ boson and neutral Higgs bosons. The search 
for exotic particles
may constrain our model severely. For example, our model is strongly
constrained by search for the charged Higgs boson in the
$b\to s\gamma$, $B\to \tau \nu$, and $B\to D^{(\ast)}\tau \nu$ decays.
In order to escape from such constraints, we must assume a quite heavy charged
Higgs boson or it is necessary to study our model more carefully
by including all the interactions which have been neglected in this work.
Search for the dijet resonances would also give strong constraints 
on the $Z^\prime$ boson. If the $s$-channel contribution is not negligible,
the coupling $(g_R^u)_{uu}$ is constrained by the search for the dijet 
resonances. 
New chiral fermions must also be included
for the anomaly cancellation. Then, search for the exotic fermions like
the 4th generation fermions would also constrain our model. Furthermore,
there could be cold dark matter candidates in our model
so that the dark matter experiments could be discussed. 
The most severe constraints arise from the search
for the Higgs boson. In this work, 
we discussed the cases where $m_h=125$ GeV, but
we did not take into account its branching ratios.
If the branching ratios of the SM-like Higgs boson observed at the LHC
settle down at the present values, our model will severely be constrained.
We emphasize that this study is not complete yet, because there are extra
fields which are subdominant in the top-quark production.
To a complete study, we need to consider especially the Higgs phenomenology more carefully.

\section*{Acknowledgments}
This work is supported in part by Basic Science Research Program
through NRF 2011-0022996 and in part by NRF Research Grant
2012R1A2A1A01006053.


\bibliographystyle{apsrev4-1}


%% file: Papers/peinado.tex

%
%
%
%
%
%

\chapter[Reactor angle and flavor symmetries (Peinado)]{Reactor angle and flavor symmetries}
\vspace{-2em}
\paragraph{E. Peinado}
\paragraph{Abstract}
Recently experiments on neutrino mixings indicate large reactor neutrino mixing angle $\theta_{13}$. Here we discuss the possibility to achieve large $\theta_{13}$ within the T2K region with maximal atmospheric mixing angle, $\sin^2\theta_{23}= 1/2$, and
trimaximal solar mixing angle, $\sin^2\theta_{12} = 1/3$, through the deviation from the exact tri-bimaximal mixing.
\section{Introduction}
Recently with the new neutrino data {\it (see M. Tortola in this proceedings)}~\cite{Fogli:2012ua,Tortola:2012te,GonzalezGarcia:2012sz}, the neutrino mixing angles are very different to the ones we have before and Tri-bimaximal mixing (TBM) is far from the best fit values. This is challenge for the flavor symmetries point of view. If we want to explain the neutrino mixing pattern from flavor symmetries scenarios, the general perspective change. Basically we have three paths to this:
\begin{itemize}
\item Tri-bimaximal mixing, Bi-maximal mixing, golden-ratio or other patterns predicting zero reactor mixing angle\footnote{ and also probably maximal atmospheric} can be modified by means of non-diagonal charged lepton mass matrix or by modification of the neutrino mass matrix itself {\it(see the contribution by E. Ma in this proceeding)}.
\item Start with another option like TBM-reactor {\it(see the contribution by S. F. King in this proceeding)}\cite{King:2009qt,Morisi:2011pm}
\item Start with something less trivial and make a fit of the neutrino mixing angles and see what kind of correlation the flavor symmetry gives, see for instance~\cite{Meloni:2010aw,Meloni:2012sy}.
\end{itemize}
\begin{figure}[h]
\centering
\includegraphics[width=0.7\linewidth]{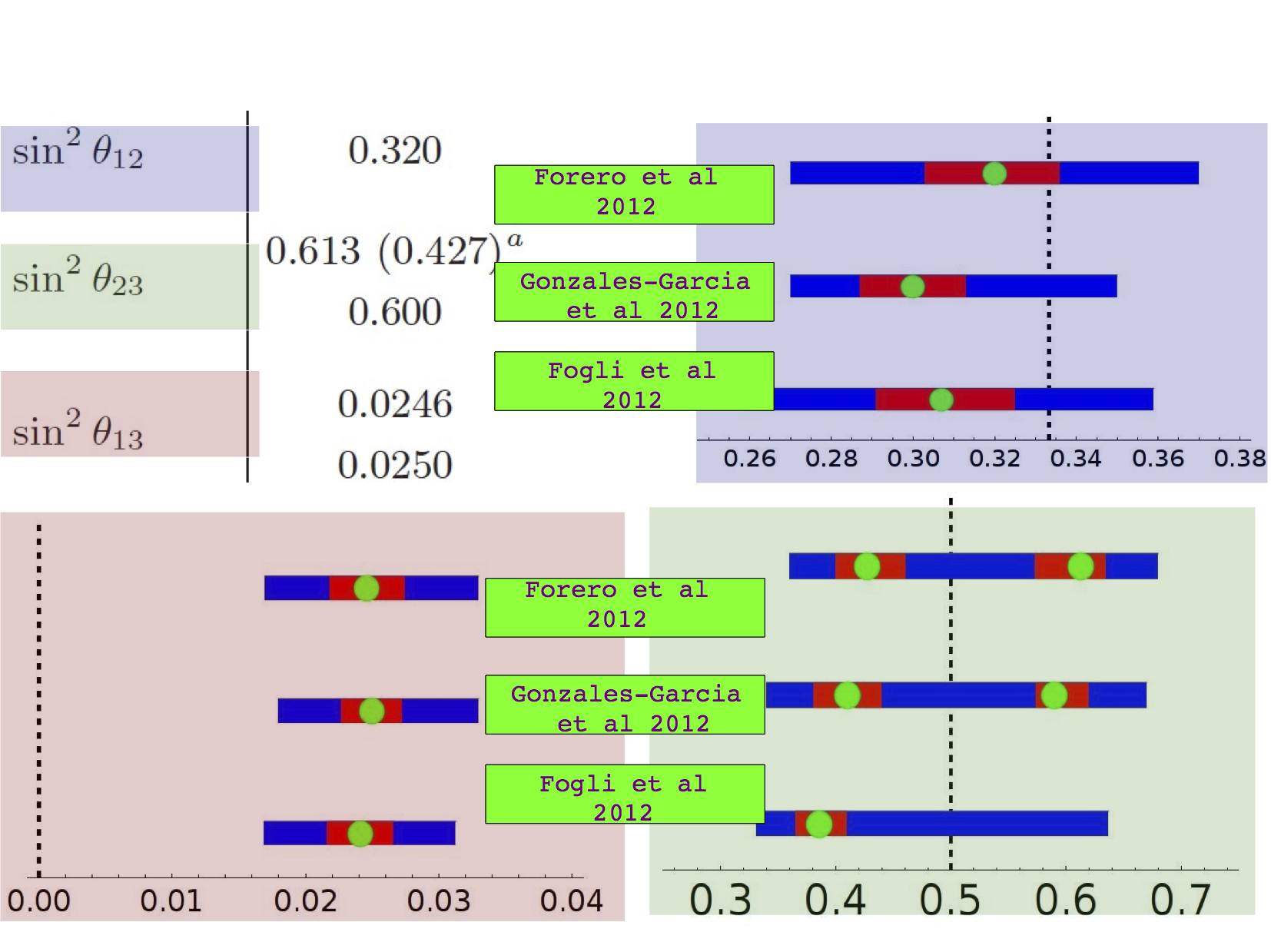}
\caption{Pictorial representation of the three mixing angles, in blue (right up) we present the solar, the reactor in light-red (bottom left) and in green the atmospheric (bottom right). The dark blue bands are for the 3$\sigma$ for the different fitting groups, the red band is for the 1$\sigma$ and the green dot is for the best fit value. The vertical lines are for the corresponding TBM values.}
\label{EPmixingpicture}
\end{figure}
\begin{center}{\bf A model for Tri-bimaximal reactor}\end{center}
Lets see the structure of the neutrino mass matrix (in the diagonal basis of the
charged leptons) that gives maximal atmospheric angle $\theta_{23}= \pi/4$, trimaximal solar angle $\sin\overline{\theta}_{12}= 1/\sqrt{3}$ and an arbitrary reactor angle $\theta_{13}=\lambda$. In the standard PDG \cite{Beringer:1900zz} parametrization, the lepton mixing matrix with the above values of mixing angles is given by \cite{King:2009qt,Morisi:2011pm} 
\begin{equation}\label{EPLRTB}
U_{\text{TBR}}=R_{23} (\frac{\pi}{4})\,R_{13}(\lambda) \,R_{12}(\overline{\theta}_{12})=
\left(
\begin{array}{ccc}
\sqrt{\frac{2}{3}}&\frac{1}{\sqrt{3}}& \lambda\\
-\frac{1}{\sqrt{6}}+\frac{\lambda}{\sqrt{3}} &\frac{1}{\sqrt{3}}+\frac{\lambda}{\sqrt{6}} &
-\frac{1}{\sqrt{2}}\\
-\frac{1}{\sqrt{6}}-\frac{\lambda}{\sqrt{3}} &\frac{1}{\sqrt{3}}-\frac{\lambda}{\sqrt{6}} &
\frac{1}{\sqrt{2}}\\
\end{array}
\right)+\mathcal{O}(\lambda^2).
\end{equation}
We do not consider the CP violation in the lepton sector assume that the above
parameters are real for simplicity. The neutrino mass matrix diagonalized by\,(\ref{EPLRTB}) is given by
\begin{equation}\label{EPmLRTB}
m_\nu^{\text{TBR}}=U_{\text{TBR}}\cdot m_\nu^{\text{diag}} \cdot U_{\text{TBR}}^T= m_\nu^{TB}+\delta
m_\nu
\end{equation}
where $m_\nu^{\text{diag}}$ is a diagonal matrix with the neutrino mass eigenvalues, $m_{\nu_1}$,
$m_{\nu_2}$ and $m_{\nu_3}$. This leads to the following structure of the neutrino mass matrix
\begin{equation}\label{EPmntbm}
m_\nu^{\text{TB}}=
\left(
\begin{array}{ccc}
2y-x & x&x\\
x & y+z&y-z\\
x & y-z&y+z
\end{array}
\right),
\end{equation}
where $x=(m_2-m_1)/3$, $y=(m_1+2m_2)/6$ and $z=m_3/2$ and
\begin{equation}\label{EPdeviation}
\delta m_\nu=\lambda
\left(
\begin{array}{ccc}
0 & \alpha_1& -\alpha_1\\
\alpha_1 & \beta_1&0\\
-\alpha_1 & 0& -\beta_1
\end{array}
\right)+
\lambda^2 \left(
\begin{array}{ccc}
\gamma & \alpha_2 & \alpha_2\\
\alpha_2 & \beta_2&-\beta_2\\
\alpha_2 & -\beta_2& \beta_2
\end{array}
\right)+
\sum_{n\ge 3} \lambda^n\left(
\begin{array}{ccc}
0 & \alpha_n& (-1)^n\alpha_n\\
\alpha_n & 0&0\\
(-1)^n\alpha_n & 0& 0
\end{array}
\right),
\end{equation}
with $\alpha_1=-(x-2y+2z)/\sqrt{2}$, $\beta_1=\sqrt{2}x$, $\alpha_2=-x/2$, $\beta_2=-(x-2y+z)/2$
and $\gamma=x-2y+2z$. Note that $\beta_2$ can be reabsorbed into the TB term $m_\nu^{\text{TB}}$.
The above form of neutrino mass matrix predicts maximal atmospheric mixing angle and trimaximal solar mixing angle if all the terms with all powers of $\lambda$ are taken into account.
If one truncates the series in eq.\,(\ref{EPdeviation}) at $n < 3$, the neutrino mass matrix then
implies
\begin{itemize}
\item (A) negligible deviations from maximality in the atmospheric mixing angle;
\item (B) small deviation from trimaximality in the solar mixing angle;
\item (C) prediction of $0\nu\beta\beta \propto \lambda^2$.
\end{itemize}
The prediction (C) is evident from eq.\,(\ref{EPdeviation}). We observe that the main structure of the deviation $\delta
m_\nu$ of order $\lambda$ in eq.\,(\ref{EPdeviation}) is $\mu$-$\tau$ antisymmetric,
see\,\cite{Grimus:2006jz}\footnote{Note that the main structure of the deviation $\delta
m_\nu$ of order $\lambda$ in eq.\,(\ref{EPdeviation}) is similar to the one found in \cite{Araki:2011wn} where (contrary with respect to us) the solar angle is not fixed to be the trimaximal one.}. Therefore a possible flavor symmetry with neutrino mass matrix texture
(\ref{EPmLRTB}) must contain the group $S_2$ of the $\mu$-$\tau$ permutation and must be
compatible with tri-bimaximal in the unperturbed limit. One possible flavor symmetry with such features is $S_4$ which contains $S_2$ as a subgroup and leads to tri-bimaximal mixing.
%
We assume $S_4$ flavor symmetry and extra Abelian $Z_N$ symmetry in order to
separate the charged leptons from the neutrino sector as usual in models for TB mixing, 
In order to simplify the model as much as possible and to render
more clear the main features of the model, we do not enter into the details of the particular $Z_N$ symmetry required in this model. Our purpose is to show that the neutrino mass matrix (\ref{EPmLRTB}) with the structure given by (\ref{EPmntbm}) and (\ref{EPdeviation}) can be obtained from symmetry principle. We assume that light neutrino masses arise from both type-I and type-II seesaw and introduce only one right-handed neutrino. The matter content of our model is given in
table\,\ref{EPtabpeinado}.

\begin{table}[h!]\begin{center}
\begin{tabular}{|c|c|c|c|c||c|c||c|c|}
\hline
 & $L$ & $l_R$&$\nu_R$  & $h$ & $\Delta_{}$ &  $\phi_{}$  & $\varphi_l$ & $\xi_l$ \\
\hline
$SU_L(2)$ & 2 & 1 & 1 & 2 & 3  & 1 &1&1\\
\hline
$S_4$ & $3_1$ & $3_1$ & $1_1$ & $1_1$ & $3_1$  & $3_1$ & $2$&$1_1$\\
\hline
\end{tabular}
\caption{Matter content of the model giving TB mixing at the leading order}\label{EPtabpeinado}\end{center}
\end{table}
In the scalar sector, we have one $SU_L(2)$ triplet $\Delta$ and one singlet $\phi$ in the
neutrino sector transforming both as $3_1$ of $S_4$. We have two electroweak singlets $\varphi_l$
and $\xi_l$ in the charged lepton sector, transforming as doublet and singlet of $S_4$
respectively.  The Yukawa interaction of the model is
\begin{eqnarray}
-\mathcal{L}_l   &=& \frac{1}{\Lambda} y_1 (\overline{L}l_R)_{1_1}h\xi_l + \frac{1}{\Lambda} y_2
(\overline{L}l_R)_2h\varphi_l+h.c.\\
-\mathcal{L}_\nu &=& y_a LL\Delta + \frac{y_b}{\Lambda} (\overline{L}
\phi)_{1_1}\tilde{h}\nu_R+\frac{1}{2}M\nu^c\nu^c+h.c.
\end{eqnarray}
where $\Lambda$ is an effective scale. We assume the following $S_4$ alignment in the vacuum
expectation values (vevs) of the scalar fields.
\begin{eqnarray}
\langle\Delta^0\rangle=v_\Delta (1,1,1)^T,\quad \langle \phi\rangle=v_\phi (0,1,-1)^T,\quad \langle\varphi\rangle =
(v_1,v_2)^T,
\end{eqnarray}
where $v_1\ne v_2$. Using the product rules shown in appendix A, one can easily see that the
charged lepton mass matrix is diagonal and the lepton masses can be fitted in terms of three free
parameters $y_1$, $v_1$ and $v_2$, see \cite{Morisi:2011ge} for details.

The type-II seesaw gives a contribution to the neutrino mass matrix with zero diagonal entries
and equal off diagonal entries since it arises from the product of three $S_4$ triplets. Since
we introduced only one right-handed neutrino, Dirac neutrino mass matrix is a column
$m_D\sim (0,1,-1)^T$ and the light-neutrino mass matrix from seesaw relation is given by
\begin{equation}\label{EPmntbm2}
m_\nu^{\text{type-I}}= \frac{1}{M}m_D\,m_D^T\sim
\left(
\begin{array}{ccc}
0 & 0&0\\
0& 1&-1\\
0&-1&1
\end{array}
\right)
\end{equation}
Considering both the type-I and type-II contributions, we have the light neutrino mass
matrix which can be diagonalized by TBM matrix,
\begin{equation}\label{EPmn0}
m_\nu^{\text{TB}} =
\left(
\begin{array}{ccc}
0&a&a\\
a&b&a-b\\
a&a-b&b
\end{array}
\right),
\end{equation}
The mass eigenvalues of the above matrix are $m_1=-a$, $m_2=2a$ and $m_3=-a+2b$.
Here $a= y_a v_\Delta $ and $b=y_b^2 v_h^2 v_\phi^2/(\Lambda^2 M)$ where $v_h=\langle h^0\rangle$. This neutrino mass matrix is compatible with the normal hierarchy only and predicts zero neutrinoless double beta decay $m_{ee}=0$.

In order to reproduce deviations like eq.\,(\ref{EPdeviation}) in the neutrino mass matrix, we
introduce in the scalar sector one Higgs triplet $\Delta_d$ that transforms as a doublet
under $S_4$ and an electroweak singlet $\phi_d$ that transforms as a triplet $3_1$ under $S_4$.
With inclusion of these fields, the Yukawa interaction Lagrangian $\mathcal{L}_\nu$ contains also
the terms
\begin{equation}\label{EPlag2}
-\mathcal{L}_\nu\supset y_\beta LL\Delta_d+ \frac{y_\alpha}{\Lambda}
(\overline{L}\phi_d)_{1_1} \tilde{h}\nu_R+h.c.
\end{equation}
We assume that $\Delta_d$ and $\phi_d$ take vevs along the following directions
\begin{equation}\label{EPall2}
\langle\Delta_d^0\rangle=v_d (1,0)^T,\quad  \langle\phi_d\rangle=u_d(1,0,0)^T.
\end{equation}
Here we also assume that $y_{\alpha,\beta}\ll y_{a,b}$. This can be realized assuming that
$\Delta_d$ and $\phi_d$ are charged under some extra Abelian symmetry like $Z_N$ or $U_{FN}(1)$.

After electroweak symmetry breaking and integrating out the right-handed neutrino, eq. (\ref{EPlag2}) gives the following contribution to the neutrino mass matrix
\begin{equation}\label{EPeff}
\frac{y_by_\alpha v_h^2}{\Lambda^2 M}(\nu \phi)_{1_1}(\nu \phi_{\text{d}})_{1_1}+
\frac{y_\alpha^2 v_h^2}{\Lambda^2 M}(\nu \phi_{\text{d}})_{1_1}(\nu \phi_{\text{d}})_{1_1}.
\end{equation}
The second term in eq.\,(\ref{EPeff}) is smaller with respect to the first since we have assumed $y_\alpha \ll y_b$. In particular assuming $y_b\sim 1$ and $y_\alpha\sim\lambda$ the first term is proportional to $\lambda$ and the second term is proportional to $\lambda^2$.
The extra contributions to the neutrino mass matrix from the type-I see-saw are as follows
\begin{equation}
\delta m_\nu^{\text{type-I}}\sim
c_1\lambda \left(
\begin{array}{ccc}
0 & 1& -1\\
1 & 0&0\\
-1 & 0& 0
\end{array}
\right)+
c_2\lambda^2  \left(
\begin{array}{ccc}
1 & 0& 0\\
0 & 0&0\\
0 & 0& 0
\end{array}
\right).
\end{equation}
where, $c_1$ and $c_2$ are coefficients of order ${\mathcal O}(1)$. From the extra type-II seesaw
term in eq. (\ref{EPlag2}) and using the vev alignments as in (\ref{EPall2}), the additional
contribution to the perturbed neutrino mass matrix will be proportional to
$\nu_1\nu_1-\nu_2\nu_2$, therefore the contribution to the neutrino mass matrix coming from
Type-II see-saw is
\begin{equation}
\delta m_\nu^{\text{type-II}}
\sim \left(
\begin{array}{ccc}
0 & 0& 0\\
0 & 1&0\\
0 & 0& -1
\end{array}
\right).
\end{equation}
Putting all these results together, the structure of the deviation in neutrino mass matrix
can be written is
\begin{equation}\label{EPdmn}
\delta m_\nu=
\left(
\begin{array}{ccc}
\gamma' & \alpha'& -\alpha'\\
\alpha' & \beta'&0\\
-\alpha' & 0& -\beta'
\end{array}
\right),
\end{equation}
where $\alpha'=y_by_\alpha v_h^2 v_\phi u_d/(\Lambda^2 M)$, $\beta'=y_\beta v_d$,
$\gamma'=y_\alpha^2 v_h^2 v_\phi u_d/(\Lambda^2 M)$.
The deviation obtained in our model  equal to the neutrino mass deviation in eq.\,(\ref{EPdeviation})
truncated at $\lambda^2$ with $\alpha_2=0$. Such a difference does not modify significantly the
prediction of maximal atmospheric angle and trimaximal solar angle. In the next section, we study
the phenomenological implication of our neutrino mass texture.

\section{Conclusion}
We found the structure for the deviation in the neutrino mass matrix from the
well known TB pattern in such a way that the lepton mixing matrix has large
atmospheric mixing angle and trimaximal solar mixing angle with an arbitrary large reactor
angle. The deviation must be approximately $\mu$-$\tau$ antisymmetric. This fact suggests us that
the flavor symmetry could be some permutation symmetry containing $S_2$ ($\mu$-$\tau$ exchange)
subgroup. $S_3$ is too small since it does not give the TB mixing. The smallest permutation group
with this property is $S_4$. We provide a candidate model based on $S_4$ where in the unperturbed
limit the neutrino mass matrix is TB. Then assuming extra scalar fields we show the possibility to
generate deviations from the TB that give a large $\theta_{13}$ in agreement with T2K result,
maximal atmospheric mixing angle and trimaximal solar mixing angle in good agreement with neutrino
data.

\section*{Acknowledgments}
I thank my collaborators Stefano Morisi and Ketan M. Patel for the work presented here. This work was supported by the Spanish MEC under grants FPA2011-22975 and MULTIDARK CSD2009-00064 (Consolider-Ingenio 2010 Programme), by Prometeo/2009/091 (Generalitat Valenciana), by the EU ITN UNILHC PITN-GA-2009-237920. and by CONACyT (Mexico).
\bibliography{peinado}
\bibliographystyle{apsrev4-1}

%% file: Papers/FilippoSala.tex

%
%
%
%
%
%

\chapter[Flavour physics from an approximate $U(2)^3$ symmetry (Sala)]{Flavour physics from an approximate $U(2)^3$ symmetry}
\vspace{-2em}
\paragraph{F. Sala}
\paragraph{Abstract}
The approximate $U(2)^3$ symmetry exhibited by the quark sector of the Standard Model, broken in specific directions dictated by minimality (Minimal $U(2)^3$), can explain the current success of the CKM picture of flavour and CP violation while allowing for large deviations form it at foreseen experiments. On top of this, one can consider all the possible breaking terms appearing in the quark Yukawas (Generic $U(2)^3$), and derive the most relevant bounds on these new parameters. In this extended framework, if needed, one could account for the recently observed CP asymmetry in $D \to \pi \pi, K K$ decays, while being consistent with all the other constraints.

\section{Introduction}
The Standard Model (SM) description of flavour and CP violation (CPV) in the quark sector, encoded in the Cabibbo Kobayashi Maskawa (CKM) matrix, is substantially in excellent agreement with any experimental data, leaving in several cases little room for new physics (NP) contributions. In other words if NP effects in flavour and CP violation are parameterized via the effective Lagrangian
\begin{equation}
 \Delta \mathcal{L} = \sum_i \frac{1}{\Lambda^2_i} \mathcal{O}_i\,+ \text{h.c.}\,,
\end{equation}
where $\mathcal{O}_i$ are generic dimension 6 gauge invariant operators obtained by integrating out the new degrees of freedom appearing above the scale $\Lambda_i$, one finds that lower limits on the scales $\Lambda_i$ are in many cases of the order of $10^3 \div 10^4$ TeV \cite{Isidori:2010kg}. If one believes some new physics has to appear at a scale $\Lambda_{\text{NP}}$ in the TeV range, whatever the reason for this belief is (naturalness of the Fermi scale, Dark Matter, $\dots$), then the flavour and CP structure of the NP theory have to be highly non trivial. A possibility is the requirement for this new theory to respect some flavour symmetry, so that the effective Lagrangian
\begin{equation}
 \Delta \mathcal{L} = \sum_i \frac{c_i}{\Lambda_{\text{NP}}^2} \xi_i \,\mathcal{O}_i\,+ \text{h.c.}\,,
\label{FS_LNPsimm}
\end{equation}
where $\xi_i$ are small parameters controlled by the symmetry, is in agreement with all current data for coefficients $c_i$ of $O(1)$.

The most popular attempt in this direction is the so called Minimal Flavour Violation paradigm \cite{Chivukula:1987py,Hall:1990ac,D'Ambrosio:2002ex}: the Yukawa couplings are promoted to \textit{spurions} transforming as $Y_u \sim (3,\bar{3},1)$ and $Y_d \sim (3,1,\bar{3})$ under $U(3)^3 = U(3)_q \times U(3)_u \times U(3)_d$, so that the SM is formally invariant under this symmetry, then also NP effects are assumed to be formally invariant via the only use of the spurions $Y_{u,d}$. In this way one obtains parameters $\xi_i$ equal to some power of the CKM matrix elements (depending on the specific operator $\mathcal{O}_i$), in such a way that a scale $\Lambda_{\text{NP}}\sim$ a few TeV is in agreement with all experimental data.
Why then the need to go beyond this paradigm? First of all, the $U(3)^3$ symmetry is already badly broken in the SM, so that it appears more natural to take as a starting point a symmetry which is instead preserved to a good level of approximation. One can think of other reasons for not being satisfied with $U(3)^3$, for example the lack of an explanation for the non observation of electric dipole moments (EDMs), if not setting all the flavour blind NP phases to zero. An attempt to address these issues already pursued in the literature is the reduction of $U(3)^3$ to a $U(2)$ acting on the first two generations of quarks, irrespective of their chiralities \cite{Pomarol:1995xc,Barbieri:1995uv}. While providing a rationale for explaining both the quark's hierarchies and the smallness of EDMs, this framework does not yield to enough suppression of right-handed currents contribution to the $\epsilon_K$ parameter \cite{Barbieri:2011ci,Barbieri:1997tu}. The considerations developed so far motivate us to study the 
flavour symmetry $U(2)^3 = U(2)_q \times U(2)_u \times U(2)_d$, exhibited by the SM if one neglects the masses of the quarks of the first two generations, as well as their mixing with the third generation ones. The subjects presented here are mainly based on \cite{Barbieri:2012uh,Barbieri:2012bh}, where we built on previous work in the specific context of Supersymmetry \cite{Barbieri:2011ci,Barbieri:2011fc}\footnote{For a recent analysis of some aspects of the phenomenology of Minimal $U(2)^3$ see also \cite{Buras:2012sd} and the talk by Jennifer Girrbach during this workshop (chapter \ref{chap:girrbach}).}.

\section{Construction of the framework}
The logic we follow is assuming that some small parameters in the Yukawa matrices have definite transformation properties under $U(2)^3$, and control at the same time its breaking in any extension of the SM. In order to reproduce the light quark masses, we introduce two spurions $\Delta Y_u \sim (2,\bar{2},1)$ and $\Delta Y_d \sim (2,1,\bar{2})$. We introduce another spurion $\boldsymbol{V} \sim (2,1,1)$ to let the first two generations communicate with the third one. This is the minimal choice that reproduces the correct pattern of masses and mixing angles of the quark sector of the SM, we call it Minimal $U(2)^3$. One can complete the list of spurions to all the possible breaking terms entering the quark masses, by adding $\boldsymbol{V_u} \sim (1,2,1)$ and $\boldsymbol{V_d} \sim (1,1,2)$. We call this Generic $U(2)^3$. An exhaustive list of the $U(2)^3$ breaking (but formally invariant) terms appearing in the quark Yukawa matrices then is
\begin{equation}
\lambda_t (\boldsymbol{\bar q}_{\boldsymbol{L}} \boldsymbol{V})t_R, \quad \lambda_t \boldsymbol{\bar q}_{\boldsymbol{L}} \Delta Y_u \boldsymbol{u}_{\boldsymbol{R}},
\quad \lambda_t \bar{q}_{3L} (\boldsymbol{V_u}^\dagger \boldsymbol{u}_{\boldsymbol{R}}),
\label{FS_Yuk_u}
\end{equation}
\begin{equation}
\lambda_b(\boldsymbol{\bar q}_{\boldsymbol{L}} \boldsymbol{V})b_R, \quad \lambda_b \boldsymbol{\bar q}_{\boldsymbol{L}} \Delta Y_d \boldsymbol{d}_{\boldsymbol{R}},
\quad \lambda_b \bar{q}_{3L} (\boldsymbol{V_d}^\dagger \boldsymbol{d}_{\boldsymbol{R}}),
\label{FS_Yuk_d}
\end{equation}
where $\boldsymbol{q}_{\boldsymbol{L}}, \boldsymbol{u}_{\boldsymbol{R}}, \boldsymbol{d}_{\boldsymbol{R}}$  stand for doublets under $U(2)_Q, U(2)_u, U(2)_d$ respectively\footnote{In \eqref{FS_Yuk_d} we have extracted out the bottom Yukawa coupling $\lambda_b$ as a common factor, which in principle requires an explanation due to its smallness. One could consider for this pourpose a symmetry, either continuous or discrete, acting in the same way on all the right-handed down-type quarks, broken by the small parameter $\lambda_b$.}.

We perform appropriate $U(2)^3$ transformations to put the spurions in the forms
\begin{equation}\label{FS_genericV}
\boldsymbol{V} = \begin{pmatrix}0\\ \epsilon_L\end{pmatrix},\qquad \boldsymbol{V_u} = \begin{pmatrix}0\\ \epsilon^u_R\end{pmatrix},\qquad \boldsymbol{V_d} = \begin{pmatrix}0\\ \epsilon_R^d\end{pmatrix},
\end{equation}
\begin{align}\label{FS_genericY}
\Delta Y_u &= L_{12}^u\,\Delta Y_u^{\rm diag}\,\Phi_R^u R_{12}^u, & \Delta Y_d &= \Phi_L L_{12}^d\,\Delta Y_d^{\rm diag}\,\Phi_R^d R_{12}^d,\\
\Phi_L &= {\rm diag}\big(e^{i\phi},1\big), & \Phi_R^{u, d} &= {\rm diag}\big(e^{i\phi_1^{u,d}}, e^{i\phi_2^{u,d}}\big),
\end{align}
where $\epsilon_L$ and $\epsilon_R^{u,d}$ are real parameters, $L_{12}^{u,d}$ and $R_{12}^{u,d}$ are two dimensional rotations of angles $\theta_L^{u,d}$ and $\theta_R^{u,d}$ respectively, and for convenience we define $s_{L,R}^{u,d} = \sin \theta_{L,R}^{u,d}$. Note that in Minimal $U(2)^3$ one has $\epsilon_R^{u,d} = s_R^{u,d} = \phi_{1,2}^{u,d} = 0$. 
The diagonalization to the mass basis, which is done perturbatively by taking into account the smallness of the parameters, results in a unique form for the CKM matrix $V$ which depends only on the 4 Minimal $U(2)^3$ parameters
 \begin{equation}\label{FS_CKM}
V = \begin{pmatrix}
c^u_L c^d_L & \lambda & s^u_L s\,e^{-i\delta}\\
-\lambda & c^u_L c^d_L & c^u_L s\\
-s^d_L s\,e^{i(\delta - \phi)} & - c^d_L s & 1
\end{pmatrix},
\end{equation}
where $s\sim O(\epsilon_L)$, $c^{u,d}_L=\cos{\theta_L^{u,d}}$ and
$s^u_L c^d_L - s^d_L c^u_L e^{i\phi} = \lambda e^{i\delta}$. Then a direct fit to tree level observables, assumed not to be influenced by NP, fixes the values of $s$, $s_L^{u,d}$ and $\phi$, while the extra parameters of Generic $U(2)^3$ remain uncostrained.

\section{Effective Field Theory analysis}
In the mass basis we give as an example, to a sufficient level of approximation, the form of some relevant flavour-violating operators, which again we build using the spurions in order to be consistent with the $U(2)^3$ symmetry (in parentheses the processes they contribute to):
\begin{align}
&c_{LL}^K \left( V_{ts}V_{td}^*\right)^2 \left(\bar{d}_L \gamma_\mu s_L\right)^2 & \big[K -\bar{K}\; \text{mixing}\big], \label{FS_epsL}\\
&c_{LL}^B e^{i \phi_B} \left( V_{tb}V_{ti}^*\right)^2 \left(\bar{d}_L^i \gamma_\mu b_L\right)^2 & \big[B_{d,s} - \bar{B}_{d,s} \;\text{mixing}\big],  \label{FS_BB}\\
&c_{7 \gamma} e^{i \phi_{7\gamma}} m_b V_{tb} V_{ti}^* \left( \bar{d}_L^i \sigma_{\mu\nu} b_R\right) e F_{\mu\nu}&  \big[b\to s(d) \gamma\big],  \label{FS_Bsgamma}\\
&c_D e^{i \phi_D} m_t \frac{\epsilon_R^u}{\epsilon_L} V_{ub} V_{cb}^* \left( \bar{u}_L \sigma_{\mu\nu} T^a c_R\right) g_s G_{\mu\nu}^a & \big[D\to \pi\pi,KK\big],\label{FS_Dop}\\
&c_{LR}^K e^{i \phi_R^K} \frac{s_R^d}{s_L^d} \left(\frac{\epsilon_R^d}{\epsilon_L}\right)^2 \left( V_{ts}V_{td}^*\right)^2 \left(\bar{d}_L \gamma_\mu s_L\right)\left(\bar{d}_R \gamma_\mu s_R\right) & \; \big[ K -\bar{K}\; \text{mixing}\big] \label{FS_epsR},
\end{align}
where $c_{LL}^K$, $c_{LL}^B$, $c_{7 \gamma}$, $c_D$ and $c_{LR}^K$ are real coefficient, in principle of order one, and in each operator we understood a factor $1/\Lambda^2$. Some remarks are in order: (i) exactly as in MFV, flavour-violating operators are suppressed by products of the CKM matrix elements; (ii) conversely, the above operators are more constrained in $U(3)^3$, where $c_{LL}^B = c_{LL}^K$ and $\phi_B = 0$. This extra freedom of $U(2)^3$ can be used to solve the CKM unitarity fit tensions; (iii) in Minimal $U(2)^3$, as well as in MFV, the operators \eqref{FS_Dop} and \eqref{FS_epsR} are absent.

\begin{figure}[tb]
\includegraphics[width=1\textwidth]{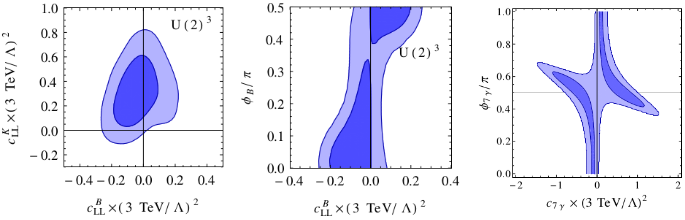}
\caption{68 and 95\% C.L. allowed regions for some $\Delta F = 2$ (left and centre) and $\Delta F =1$ (right) coefficients in $U(2)^3$.}
\label{FS_MU2}
\end{figure}

In Minimal $U(2)^3$ we performed a global fit to the experimental data, the resulting allowed regions for the coefficients of the operators defined in \eqref{FS_epsL}, \eqref{FS_BB} and \eqref{FS_Bsgamma} are shown in fig. \ref{FS_MU2}. In this derivation we fixed the energy scale $\Lambda$ to the value of 3 TeV $\simeq 4 \pi v$, which can be either the typical scale of a new strong interaction or the one associated with new weakly interacting particles of mass $\simeq v = 246$ GeV circulating in loops.
The constraints on the other operators are similar to those we showed, so that the picture that emerges is consistent with an effective Lagrangian like \eqref{FS_LNPsimm}, with the coefficients $|c_i|$ ranging between 0.2 and 1.

\begin{figure}[bt]
\centering
\includegraphics[width=0.8\textwidth]{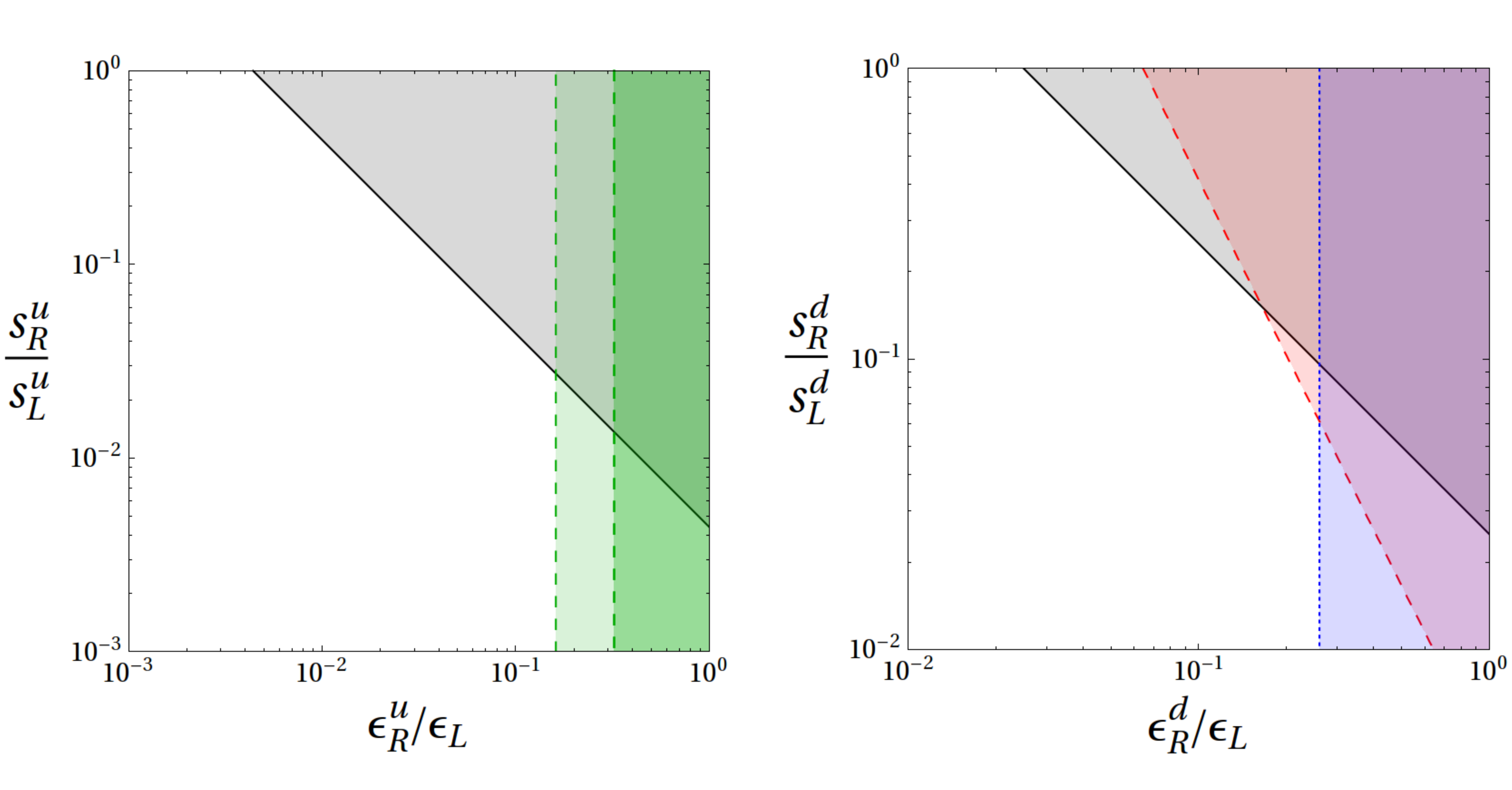}
\caption{Shaded: 90\% C.L. disfavoured regions for the free parameters of Generic $U(2)^3$, normalized to those of Minimal $U(2)^3$
. The bounds come from the neutron EDM (both plots, black solid line), $\epsilon_K$ and $\epsilon_K'$ (right-hand plot, red dashed and blue dotted lines respectively). 
The green dashed lines in the left-hand plot correspond to a NP contribution to $\Delta A_{CP}$ of 50\% and 100 \% respectively of the measured central value, so that in the lighter green region NP could account for the large experimental value.}
\label{FS_Genbounds}
\end{figure}

An analogous fit can be performed for the Generic $U(2)^3$ case, in fig.\ref{FS_Genbounds} we show the allowed regions for the real parameters $\epsilon_R^{u,d}$ and $s_R^{u,d}$ coming from  the most stringent (to the best of our knowledge) observables, assuming all the real coefficients $c_i^\alpha$ to be unity and all the phases to maximize the corresponding bounds, to be conservative. If needed, the recently measured \cite{Aaij:2011in,CDF-Note-10784} CP asymmetry in D decays, $\Delta A_{CP} = A_{CP}(K^+K^-)-A_{CP}(\pi^+\pi^-)$, could be accounted for by new physics compatible with $U(2)^3$ without violating all the other constaints. Independently of this, both Generic and Minimal $U(2)^3$ are not expected to give rise to any sizeable effect neither in CPV in $D-\bar{D}$ mixing nor in flavour changing neutral current (FCNC) top decays at near future experiments. We stress that the bounds shown in fig. \ref{FS_Genbounds} carry order one uncertainties due to the lack of theoretical control over the SM long 
distance 
contributions.

\section{Summary and conclusions}
A suitably broken $U(2)^3$ flavour symmetry acting on the first two generations of quarks\footnote{For a possible extension of $U(2)^3$ to the charged lepton sector, both from an EFT point of view and in composite Higgs models, see \cite{Barbieri:2012uh}. For a discussion in Supersymmetry with the inclusion of both charged leptons and neutrinos see \cite{Blankenburg:2012nx} and the talk by Gianluca Blankenburg during this workshop (chapter \ref{chap:blankenburg}).} could be consistent with the SM explanation of current experimental data, while allowing from sizeable deviations from it at near future experiments. A qualitative summary of FCNC and CPV effects in $U(3)^3$ and in Minimal and Generic $U(2)^3$ is given in Table \ref{FS_summa}.
\begin{table}[h!]
\renewcommand{\arraystretch}{1.5}
 \begin{center}
\begin{tabular}{lcccccc}
&\multicolumn{2}{c}{Chirality conserving} & \multicolumn{2}{c}{Chirality breaking}\\
\hline
& $\Delta B = 1,2$ & $\Delta S = 1,2$ & $\Delta B = 1$ & $\Delta C = 1$ 
\\\hline
$U(3)^3$ moderate $t_\beta$ & \multicolumn{2}{l}{\hspace{.61cm}
\fbox{$\mathbb{R}\qquad \qquad\; \; \; \, \mathbb{R}$}} & $ \mathbb{C}$ & 0 
\\
Minimal $U(2)^3$, $U(3)^3$ large $t_\beta$  & $ \mathbb{C}$ & $ \mathbb{R}$ & $ \mathbb{C}$ & 0 
\\
Generic $U(2)^3$ & $ \mathbb{C}$ & $ \mathbb{C}$ & $ \mathbb{C}$ & $ \mathbb{C}$ 
\\\hline
 \end{tabular}
 \end{center}
\caption{Expected new physics effects in $U(3)^3$ and both Minimal and Generic $U(2)^3$, for $\Delta F=1,2$ FCNC operators in the $B$, $K$, $D$ systems. $\mathbb{R}$ denotes possible effects, but aligned in phase with the SM, $\mathbb{C}$ denotes possible effects with a new phase, and 0 means no or negligible effects. In $U(3)^3$ with moderate $\tan\beta$ an additional feature is that the effects in $b\to q$ ($q = d, s$) and $s\to d$ transitions are perfectly correlated.}
\label{FS_summa}
\end{table}
\\Quantitatively, in $U(3)^3$ with moderate $\tan\beta$ the effects are smaller than in the other cases, because of the stronger constraints due to the extra correlation of some coefficients.

Suppose now some NP effect is observed at foreseen experiment: how could one tell if it is compatible with an approximate $U(2)^3$ symmetry of Nature? The most promising way would be to look for correlations in $d$ and $s$ final states of $B$ decays, which would have to be SM-like. This could be actually reproduced by $U(3)^3$ in the presence of two Higgs doublets and at large values of $\tan \beta$ \cite{Feldmann:2008ja,Kagan:2009bn}, but it would in turn imply other peculiar effects which are not necessarily present in $U(2)^3$. In the absence of an extra Higgs doublet (or in the case of small $\tan \beta$) MFV could be distinguished by $U(2)^3$ by means of new CP violating effects in $B$ decays, and/or non SM-like correlations between semileptonic $B$ and $K$ decays.

We conclude by mentioning the possible embedding of $U(2)^3$ in concrete extensions of the SM, like Supersymmetry or composite Higgs models. We stress two important differences with respect to the MFV case: (i) thanks to the freedom to separate the NP energy scale associated with the third generation from the one associated with the first two, in both cases this embedding leaves space to satisfy collider constraints without spoiling significantly the naturalness of the theory; (ii) in Supersymmetry there is now the possibility to suppress the EDMs via the heaviness of the first two generation squarks, while keeping lighter stops and sbottoms.
For a thorough discussion of the relevant features in Supersymmetry and in composite Higgs models we refer to \cite{Barbieri:2011ci,Barbieri:2011fc} (see \cite{Sala:2012ib} for a summary) and to \cite{Barbieri:2012uh} respectively
.

\section*{Acknowledgments}
I thank R. Barbieri, D. Buttazzo and D.M. Straub for the pleasent collaboration, and D.M.Straub for comments on the manuscript. I also thank the organizers of the FLASY12 workshop for the opportunity to give a talk.
\bibliographystyle{apsrev4-1}
\bibliography{FilippoSala}


%% file: Papers/schacht.tex

%
%
%
%
%
%

\chapter[Squark Flavor Implications from $\bar{B}\rightarrow \bar{K}^{(*)}l^+l^-$ (Schacht)]{Squark Flavor Implications from $\bar{B}\rightarrow \bar{K}^{(*)}l^+l^-$}
\vspace{-2em}
\paragraph{S. Schacht}
\paragraph{Abstract}
We present new results on supersymmetric flavor from the recently improved constraints on 
$\bar{B}\rightarrow \bar{K}^{(*)}l^+l^-$.
In part of the parameter space the bound on the scharm-stop left-right mixing is as strong as 
$\left(\delta_{23}^u\right)_{LR} \lesssim 10\%$. We inspect the reach of Supersymmetry (SUSY) models
with flavor violation and present implications for models based on Radiative Flavor Violation. 

\section{Introduction}

\paragraph{SM and SUSY Flavor Puzzle} 

In the Standard Model (SM) the question for the origin of the hierarchy of the Yukawa couplings is 
not answered: The only natural Yukawa coupling is the one of the top quark $\lambda_{\text{Top}} \sim 1$ 
which is of the order of the gauge couplings. The other Yukawa couplings are small yet hierarchical, 
which forms the SM flavor puzzle.
When we switch on Supersymmetry (SUSY) we do not only have a puzzle but a serious problem: SUSY itself says 
nothing about flavor violation in SUSY breaking so generically SUSY flavor violation can be $\sim\mathcal{O}(1)$. 
On the other hand, flavor changing neutral current (FCNC) data partly drastically constrains SUSY flavor violation, 
so also here, in the SUSY breaking, a non-generic structure is necessary. 

The many new sources of flavor violation in SUSY can be parametrized by $6\times 6$ squark mass matrices 
that are in general not diagonal. Commonly, one normalizes the off-diagonal elements of these matrices 
to the average of the diagonal elements in form of mass insertion (MI) parameters $ \delta_{ij} =  \Delta_{ij} /  M^2_{av}$.
In writing so, we use the super-CKM basis which is determined by rotating the squarks and quarks in parallel while 
diagonalizing the quark Yukawa couplings. The bounds on the MI parameters by FCNC data are partly as strong as $\lesssim 10^{-4}$ \cite{Buchalla:2008jp}.

Here, we present a recent study of the implications of improved constraints on $\bar{B}\rightarrow\bar{K}^{(*)}l^+ l^-$ on SUSY flavor \cite{Behring:2012mv}.
In this channel we are especially sensitive to the scharm-stop left-right mixing $\left(\delta_{23}^u\right)_{LR}$. This MI parameter does not have strong bounds at present. Without the new semileptonic data it could be $\sim\mathcal{O}(1)$.

\paragraph{Low Energy Effective Field Theory}

An inevitable tool for benefiting from data on rare decays is the effective field theory framework.
In order to describe the semileptonic decay $b~\rightarrow~sl^+l^-$ we use the 
$\Delta B = 1$-Hamiltonian $\mathcal{H}_{\mathrm{eff}} \propto \sum_i C_i(\mu) O_i(\mu)$. 
For $b~\rightarrow~sl^+l^-$ the most important operators are the electromagnetic dipole operator
\begin{align}
O_7 &= \frac{e}{16\pi^2} m_b \left(\bar{s}_{L} \sigma_{\mu\nu}  b_{R} \right) F^{\mu\nu}
\end{align} 
and the 4-fermion semileptonic operators 
\begin{align}
O_9 &= \frac{e^2}{16\pi^2} \left(\bar{s}_{L} \gamma_\mu  b_{L}\right) \left(\bar{l} \gamma^\mu l\right) 
\, ,&
O_{10} &= \frac{e^2}{16\pi^2} \left(\bar{s}_{L} \gamma_\mu  b_{L} \right) \left(\bar{l} \gamma^\mu \gamma_5 l \right).  
\end{align}
In the Minimal Supersymmetric Standard Model (MSSM) we get additional contributions to the Wilson coefficients of the latter operators
$C_i = C_i^{\mathrm{SM}} + C_i^{\mathrm{NP}}, i=7,9,10$, 
with the new physics (NP) contributions $C_i^{\mathrm{NP}}$. For the Wilson coefficients we employ the results from \cite{Bobeth:1999mk, Cho:1996we}. We use here the SM operator basis as we only take into account small values of $\tan\beta\lesssim 15$. In this
regime we can neglect the contributions of the additional scalar operators which contribute at higher $\tan\beta$ \cite{Carena:2000uj}.
A study of flavor-diagonal SUSY at larger values of $\tan\beta$ can be found in \cite{Mahmoudi:2012un}. 
Preceding SUSY studies of $b\rightarrow sl^+l^-$ can also be found in \cite{Hewett:1996ct,Lunghi:1999uk,Ali:2002jg}. 
The calculation is done by extending \texttt{EOS}, a tool for the calculation of flavor observables \cite{EOS::2011}.

\section{Comparison of SUSY Predictions with Data}

In order to confront SUSY predictions with data it is useful to compare directly the possible spread of SUSY models 
in the planes of the Wilson coefficients with the associated model independent bounds on the latter. 
As is well-known, the radiative decay $\bar{B}\rightarrow X_s \gamma$ gives quite strong constraints on $C_7$.
Recent experimental data \cite{Aaltonen:2011ja,LHCb-CONF-2011-038} on the semileptonic process $\bar{B}\rightarrow \bar{K}^{(*)}l^+l^-$ gives in addition to that new model independent constraints on 
$C_9$, $C_{10}$ \cite{Bobeth:2010wg,Bobeth:2011gi,Bobeth:2011nj,Altmannshofer:2011gn}.
Therefore, it is very interesting to study the reach of flavor violating SUSY models in the plane of the Wilson coefficients 
$C_9$ and $C_{10}$. We do so before actual applying the bound from $\bar{B}\rightarrow \bar{K}^{(*)}l^+l^-$ in order to see 
how much influence the observables of this particular channel have.

We perform a scan of the SUSY parameters at the electroweak (EW) scale, not specifying a dedicated mechanism of 
SUSY breaking and thus keeping a more model independent perspective. Consequently, we allow for a light stop quark 
$m_{\tilde{t}_1}\geq 100$ GeV \cite{Abazov:2008rc} which is without further model assumptions not excluded by data at present. 
Other existing recent stronger bounds on the light stop mass \cite{Aaltonen:2010uf,ATLAS-CONF-2012-036,CMS-PAS-SUS-11-020} 
are model dependent. Vice versa, light stop masses are also an important part of many SUSY models 
\cite{Papucci:2011wy,Csaki:2012fh,Craig:2012yd,Craig:2012di}.

\begin{figure}[t]
\begin{center}
\includegraphics[width=0.32\textwidth]{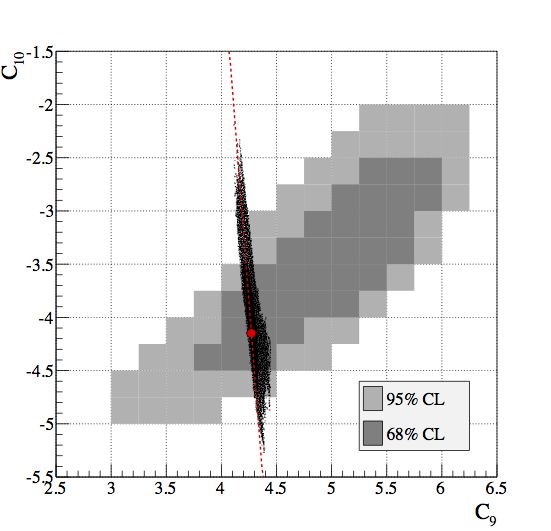}
\includegraphics[width=0.32\textwidth]{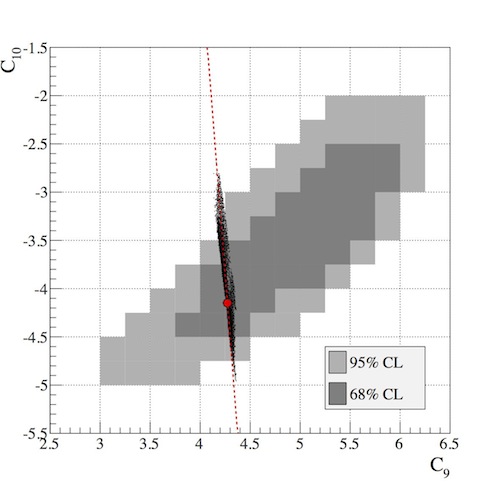}
\includegraphics[width=0.32\textwidth]{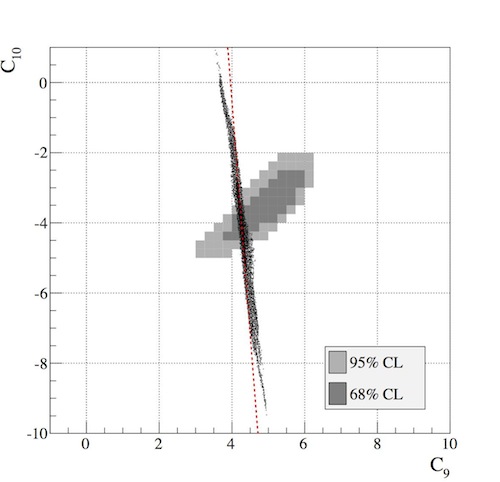}
\end{center}
\caption{Spread of the 4-fermion semileptonic Wilson coefficients at $\mu_b=4.2$ GeV for SUSY models including flavor violation in $\left(\delta_{23}^u\right)_{LR}\neq 0$. Gray: Model independent bounds from \cite{Bobeth:2011nj}. 
Red line: Z penguin dominance. 
Red dot: SM value. Left figure, taken from \cite{Behring:2012mv}: $m_{h_0}\geq 114.4$ GeV. Right: Taking into account points with large $\left(\delta_{23}^u\right)_{LR}$ not allowed by \texttt{FeynHiggs}. Middle: Applying the cuts $100 \, \mathrm{GeV} \leq m_{\tilde{t}_1} \leq 250$ GeV and $120 \, \mathrm{GeV} \leq m_{h_0} \leq 130$ GeV. \label{SS_fig_scan}}
\end{figure}

The result of the scan is shown in Figure \ref{SS_fig_scan}. In addition to the bound on the 
squark masses we also include here the other bounds from direct searches. 
Especially, we account for the bound on the Higgs mass by calculating $m_{h_0}$ including the effects 
from flavor violation using \texttt{FeynHiggs} 
\cite{Heinemeyer:1998yj, Heinemeyer:1998np,Degrassi:2002fi,Frank:2006yh,AranaCatania:2011ak}. 
Very recently, the finding of a new scalar boson has been reported which can be interpreted as the 
Higgs boson \cite{:2012gk,:2012gu}. We show therefore how the spread of SUSY models in the plane of the 
Wilson coefficients depends on these exciting news. 

We start on the left hand side of Figure \ref{SS_fig_scan} where we show the result of the SUSY scan using a Higgs mass bound of 
$m_{h^0}\geq 114.4$ GeV \cite{Nakamura:2010zzi}. Note that solutions for $C_7>0$ are not shown here as 
they are disfavored by the zero of $A_{\mathrm{FB}}(\bar{B}\rightarrow K^*\mu^+\mu^-)$ \cite{LHCb-CONF-2012-008}. 
Furthermore, there is a strong correlation between $C_9$ and $C_{10}$ which is due to the Z penguin dominance 
which leads to $C_{10}^{\mathrm{SUSY}} / C_9^{\mathrm{SUSY}} \simeq 1 / (4s_w^2-1)$.
As a consequence of this strong correlation the bounds in the MSSM are stronger than the model independent ones. 

The NP effect can be measured by the ratio of Wilson coefficients ${R_i} \equiv \left| C_i^{\mathrm{NP}} / C_i^{\mathrm{SM}} \right|$.
By applying the semileptonic bounds the maximal range of $R_{10}$ on the left hand side of Figure \ref{SS_fig_scan} is reduced from  
$R_{10}(\mu_b) \lesssim 47\%$ to $R_{10}(\mu_b) \lesssim 16\%$ (at $68\%$ C.L.),
i.e. the semileptonic bounds cut deeply into the parameter space of the MSSM. 

Now, on the right hand side of Figure \ref{SS_fig_scan} we also include points with large $\left(\delta_{23}^u\right)_{LR}$
that were not accepted by \texttt{FeynHiggs}. These points still fulfill all the other constraints including 
the $\bar{B}\rightarrow X_s \gamma$ constraint. It is clearly visible that the inclusion of the Higgs mass constraint 
has a huge impact.

With a Higgs mass of $m_{h_0}\sim 126$ GeV, global CMSSM fits give quite large results for the light stop mass
\cite{Bechtle:2012zk,Buchmueller:2012hv}. It is therefore interesting if there are parameter points in the $C_9$-$C_{10}$ plane 
which are not only near the measured Higgs mass but in addition to this also include a light 3rd generation squark. 
In order to inspect this in the middle of Figure \ref{SS_fig_scan} we apply the cuts 
$100 \, \mathrm{GeV} \leq m_{\tilde{t}_1} \leq 250$ GeV and $120 \, \mathrm{GeV} \leq m_{h_0} \leq 130$ GeV.
One observes two things: Firstly, the characteristics of the SUSY scan persist also in light of the new results on the mass
of a scalar boson. Secondly, the parameter space subset where a light stop is still allowed has a significant spread in the $C_9$-$C_{10}$ plane. 
We note however, that larger values for the trilinear coupling $A_t$ are also needed for that. 

\begin{figure}[t]
\begin{center}
\includegraphics[width=0.45\textwidth]{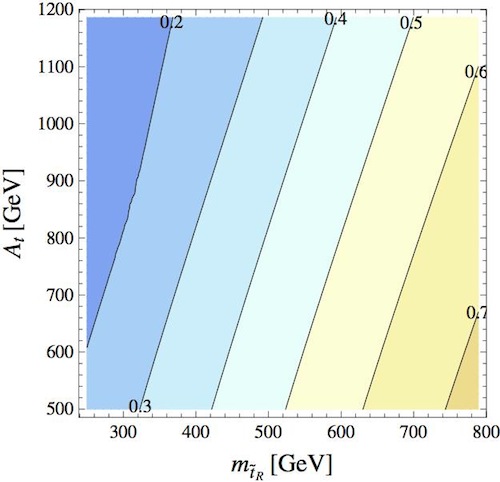}
\qquad
\includegraphics[width=0.45\textwidth]{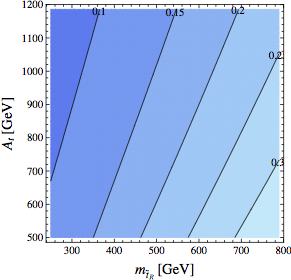}
\end{center}
\caption{Bound on $\vert \left(\delta_{23}^u\right)_{LR}\vert$ for the SUSY example point given in Table \ref{SS_table_SUSY_point} before (left) and after (right) applying the semileptonic bounds at 68\% C.L. Figures taken from \cite{Behring:2012mv}. \label{SS_fig_bound}}
\end{figure}

Finally, one can translate the semileptonic bounds shown in Figure \ref{SS_fig_scan} into an improved bound on 
$\vert\left(\delta_{23}^u\right)_{LR}\vert$. This can be  
seen in Figure \ref{SS_fig_bound} in the $A_t$ vs. $m_{\tilde{t}_R}$ plane for the SUSY example point given in Table 
\ref{SS_table_SUSY_point}. In part of the parameter space the bound has significantly improved, to 
$\vert \left(\delta_{23}^u\right)_{LR}\vert \lesssim 10\%$. Regarding the dependence of the bounds on the other parameters, for $\vert\mu\vert\gg M_2$ they get stronger and for larger $\tan\beta$ they get somewhat weaker. Note that in Figure \ref{SS_fig_bound} the Higgs bounds are not taken into account as we only want to show the effect of the flavor bounds here.

\begin{table}[h]
\centering
\begin{tabular}{|c|c|c|c|c|c|c|c|c|} \hline
$m_{H^\pm}$ &
$\tan\beta$ &
$M_2$ &
$\mu$ &
$m_{\tilde{t}_R}$ &
$m_{\tilde{q}}$ &
$A_t$ &
$m_{\tilde{\nu}}$ &
$m_{\tilde{g}}$
\\\hline
300 &
4 &
150 &
$-300$ &
300 &
1000 &
1000 &
100 &
700
\\\hline
\end{tabular}
\caption{Example SUSY point at $\mu_0= 120$ GeV, all masses in GeV. 
\label{SS_table_SUSY_point} }
\end{table}

\section{Implications for SUSY flavor models}

In SUSY flavor models, the expectations for $\left(\delta_{23}^u\right)_{LR}$ are commonly rather small:
In MFV models the trilinear couplings can be written as an expansion in the Yukawa coupling matrices
$A_u=A\left(a 1 + b Y_d Y_d^\dagger\right)Y_u$ with $a$ and $b$ $\sim\mathcal{O}(1)$. 
From this expansion it follows $\left(\delta_{23}^u\right)_{LR} \sim {\lambda_b^2 V_{cb}} V_{tb}^* (m_t/m_{\tilde{q}})$, 
i.e. there is not only a suppression by $\lambda_b^2$ but also by $V_{cb}$ \cite{D'Ambrosio:2002ex,Hiller:2008wp}. 
In models with horizontal flavor symmetries \cite{Nir:1993mx} one gets $\left(\delta_{23}^u\right)_{LR} \sim {V_{cb}} (m_t/m_{\tilde{q}})$. Also here the predictions are an order of magnitude below the limits.

A model with rather large predictions for $\left(\delta_{23}^u\right)_{LR}$ is Radiative Flavor Violation (RFV) \cite{Weinberg:1972ws, Crivellin:2008mq, Crivellin:2011sj}. In this framework one traces back the SM flavor puzzle to the SUSY flavor puzzle. 
One supposes that the bare CKM matrix is as simple as possible, being just the unit matrix.
The small off-diagonal elements of the CKM matrix then stem from quantum corrections through non-diagonal 
trilinear SUSY breaking couplings. These must have just the right values in order to generate the CKM matrix 
elements through loop diagrams. 

\begin{figure}[tb]
\begin{center}
\includegraphics[width=0.32\textwidth]{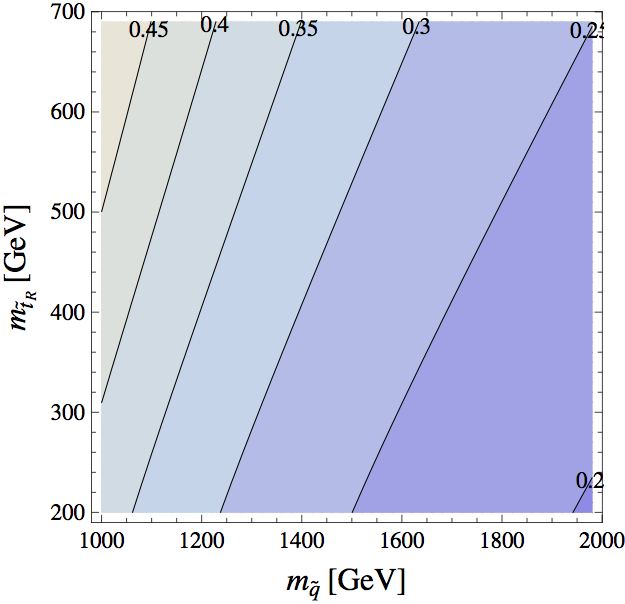}
\includegraphics[width=0.32\textwidth]{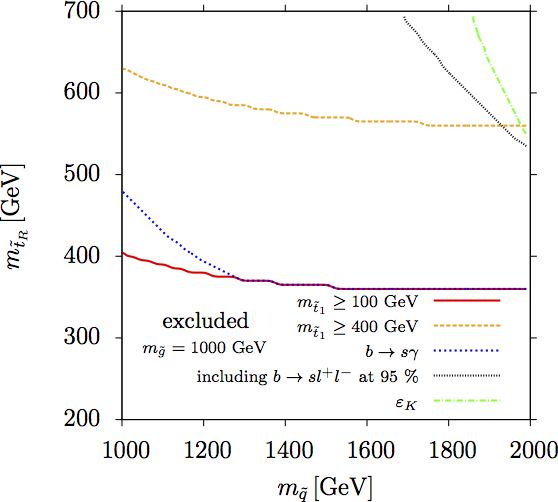}
\includegraphics[width=0.32\textwidth]{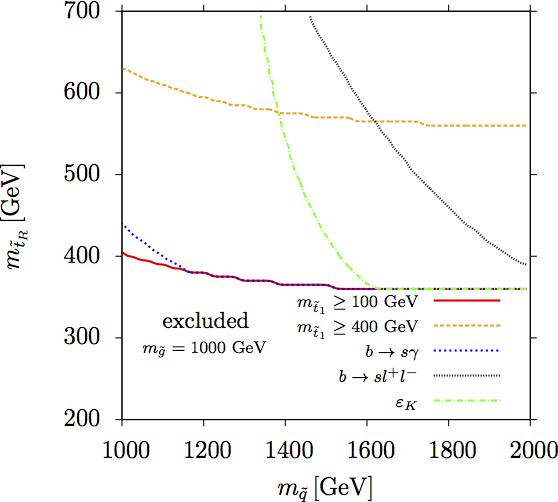}
\end{center}
\caption{Left: Requisite $\left(\delta_{23}^{u}\right)_{LR}$ for generating $V_{cb}$ for $m_{\tilde{q}} = 1000$ GeV, Figure taken from \cite{Behring:2012mv}. Middle and Right: Interplay of constraints on the RFV parameter space. Red: Stop mass limit. Orange: Hypothetical stop mass limit. Blue: Bound from $b\rightarrow s\gamma$. Green: Bound from $\varepsilon_K$ through $\left(\delta_{23}^u\right)^*_{LR}\left(\delta_{13}^u\right)_{LR}$. Black: Bound from $b\rightarrow sl^+ l^-$. Right: $M_2=800$ GeV, Figure taken from \cite{Behring:2012mv}. Middle: $M_2=500$ GeV. \label{SS_fig_interplay_constraints}}
\end{figure}

The requisite $\left(\delta_{23}^{u}\right)_{LR}$ for the generation of $V_{cb}$ through quantum corrections in the up-sector is shown on the left hand side in Figure \ref{SS_fig_interplay_constraints}. One can recognize that relatively large 
values of $\left(\delta_{23}^u\right)_{LR}$ are needed. 
In the middle and on the right side of Figure \ref{SS_fig_interplay_constraints} we study how this translates into a constraint on 
the RFV parameter space. 
The strongest bounds come both from Kaon mixing and from $b\rightarrow sl^+l^-$ data.
The former is due to chargino contributions from double mass insertions $\left(\delta_{23}^u\right)^*_{LR}\left(\delta_{13}^u\right)_{LR}$ \cite{Colangelo:1998pm,Buras:1999da,Ciuchini:1998ix}. 
As can be seen in the middle of Figure \ref{SS_fig_interplay_constraints} for small values of $M_2$ the bound from 
Kaon mixing dominates. As $M_2$ gets larger the semileptonic bounds get in part of the parameter space 
even stronger than the bounds from Kaon mixing. 
For large $m_{\tilde{t}_R}$ this effect is again washed out by the Glashow-Iliopoulus-Maiani (GIM) mechanism.
Altogether, from Figure \ref{SS_fig_interplay_constraints} we conclude that the spectrum of RFV with up-sector CKM
generation must be $\gtrsim 1~\mathrm{TeV}$.

\section{Conclusion}
New results on $\bar{B} \rightarrow \bar{K}^{(*)} l^+ l^-$ give improved constraints on squark flavor violation 
through large chargino contributions to the Wilson coefficients $C_9$, $C_{10}$. 
Depending on the parameter space the bound on scharm-top left-right mixing is as low as 
$\left(\delta_{23}^u\right)_{LR}\lesssim 10\%$. This gives bounds on RFV models that partly are even sharper then 
from $\varepsilon_K$. For lighter stops the bounds get stronger. This is reflected in the large spread of SUSY models in the 
$C_9$-$C_{10}$ plane. In the future, even more precise measurements are to come for the LHCb 
roadmap channel $B\rightarrow K^{*0} l^+l^-$ so that we expect more statistics and additional observables \cite{Bediaga:2012py}.

\section*{Acknowledgments}
StS is happy to thank his collaborators Arnd Behring, Christian Gross and Gudrun Hiller. The work presented here is supported in part by the \emph{German-Israeli Foundation for Scientific Research and Development (GIF)}.

\bibliography{schacht}
\bibliographystyle{apsrev4-1}


%% file: Papers/DanielSchmidt.tex

%
%
%
%
%
%

\chapter[Direct Detection of Leptophilic Dark Matter in a Model with Radiative Neutrino Masses (Schmidt)]{Direct Detection of Leptophilic Dark Matter in a Model with Radiative Neutrino Masses}
\vspace{-2em}
\paragraph{D. Schmidt}
\paragraph{Abstract}
Direct detection of fermionic Dark Matter (DM) enabled at the 
1-loop level is discussed for the Ma model of radiative neutrino
mass generation. Effectively, there are charge-charge, dipole-charge
and dipole-dipole interactions. The parameter space consistent with
constraints from neutrino masses and mixing, charged lepton-flavor violation,
perturbativity, and the thermal production of the correct DM abundance
is investigated and the expected event rate in DM direct detection experiments
is calculated. Current data from XENON100 start to constrain certain regions
of the allowed parameter space, whereas future data from XENON1T has the potential
to significantly probe the model. This talk is based on~\cite{Schmidt:2012yg}
where detailed calculations and relevant references can be found if they are not included here.

\section{Introduction}
Astroparticle physics meets two main challenges, namely
-Neutrinos have mass and the Universe has Dark Matter (DM)-,
which only physics beyond the Standard Model (SM) can cope with.
The two challenges are established by
neutrino flavor oscillation experiments 
and by cosmological observations ranging from
small to large scales.\\
Neutrino masses and DM must not be related to each other at all,
however, it is tempting to link a neutrino mass generation mechanism with a DM particle.
Among the many ideas on such a common framework,
those generating neutrino masses radiatively are appealing because the
loop suppression factors of $16\pi^2$ make the relevant physical scale at which
neutrino masses and DM appear accessible at the TeV range, e.g.,~\cite{Lindner:2011it} and references therein.\\
We elaborate on the model proposed by Ma~\cite{Ma:2006km}, in
which neutrino masses are generated through 1-loop interactions and
the particles which propagate in the loop can be DM candidates, being
leptophilic by construction. The DM phenomenology of the model and 
extended versions thereof has been studied in the literature.
In this talk, the lightest right handed neutrino is considered to be the the DM candidate which is assumed
to be almost degenerated with the second lightest right handed neutrino.
Under this situation, inelastic
scattering induced by a lepton-loop coupled to the photon gives the
dominant contribution to the event rate in direct detection
experiments. We calculate the event rate in the model and compare it
with XENON100~\cite{Aprile:2011hi}, KIMS~\cite{Lee.:2007qn} and DAMA~\cite{Bernabei:2010mq} data.

\section{The Model}
The invariant Lagrangian is 
\begin{equation}
\mathcal{L}_N=\overline{N_i}i\slashed{\partial}P_RN_i
+\left(D_\mu\eta\right)^\dag\left(D^\mu\eta\right)
-\frac{M_i}{2}\overline{N_i\:\!^c}P_RN_i+h_{\alpha
 i}\overline{\ell_\alpha}\eta^\dag P_RN_i+\text{h.c.}-\mathcal{V}(\phi,\eta),
\label{eq:lg}
\end{equation}
with the scalar potential
\begin{multline}
\mathcal{V}(\phi,\eta)=
m_\phi^2\phi^\dag\phi+m_{\eta}^2\eta^\dag\eta
+\frac{\lambda_1}{2}\left(\phi^\dag\phi\right)^2
+\frac{\lambda_2}{2}\left(\eta^\dag\eta\right)^2\nonumber\\
+\lambda_3\left(\phi^\dag\phi\right)\left(\eta^\dag\eta\right)
+\lambda_4\left(\phi^\dag\eta\right)\left(\eta^\dag\phi\right)
+\frac{\lambda_5}{2}\left(\phi^\dag\eta\right)^2+\text{h.c.},
\end{multline}
where $\phi$ is the SM Higgs doublet and the new Yukawa couplings are
$h_i=|h_i|e^{i\varphi_i}$ including the phases $\varphi_i$.
The vacuum expectation value (VEV) of $\eta$ is assumed to be zero, so
that Dirac neutrino masses are not generated through the
Yukawa couplings in Eq.~(\ref{eq:lg}).
However, Majorana neutrino masses are generated
radiatively involving the DM candidate, such that
neutrino physics and the existence of DM are correlated to each other.
Due to $\theta_{13}\neq0$~\cite{An:2012eh}, the flavor structure 
for the Yukawa couplings $h_{\alpha i}$ (rows are labeled by $\alpha=e,\mu,\tau$ and 
columns by $i=1,2,3$) can be chosen as
\begin{equation}
h_{\alpha i}=\left(
\begin{array}{ccc}
\epsilon_1 & \epsilon_2 & h_3'\\
h_1 & h_2 & h_3\\
h_1 & h_2 & -h_3
\end{array}
\right)+\mathcal{O}(\epsilon^2),
\label{eq:ykw-ex}
\end{equation} 
where $\epsilon_1$ and $\epsilon_2$ are small perturbations allowing 
$\sin\theta_{13}=\epsilon_3\neq0$ with
\begin{align}
\epsilon_1h_1+\epsilon_2h_2
&=
\sqrt{2}\left(h_1^2+h_2^2\right)
\frac{\left(h_1^2+h_2^2\right)\Lambda_1-\sec^2\theta_{12}h_3^2\Lambda_3}
{\left(h_1^2+h_2^2\right)\Lambda_1-h_3^2\Lambda_3}\epsilon_3\equiv P\epsilon_3.
\end{align}
The branching ratio for $\mu\to e\gamma$ is sensitive to $\theta_{13}\neq0$:
\begin{equation}\label{eq:m2eg}
\text{Br}\left(\mu\to e\gamma\right)=
\frac{3\alpha_{\text{em}}}{64\pi G_F^2M_\eta^4}
\left|
P\epsilon_3F_2\left(\frac{M_1^2}{M_\eta^2}\right)
+\sqrt{2}\tan\theta_{12}|h_3|^2F_2\left(\frac{M_3^2}{M_\eta^2}\right)
\right|^2.
\end{equation}
\begin{figure}
    \centering
\includegraphics[width=0.44\linewidth]{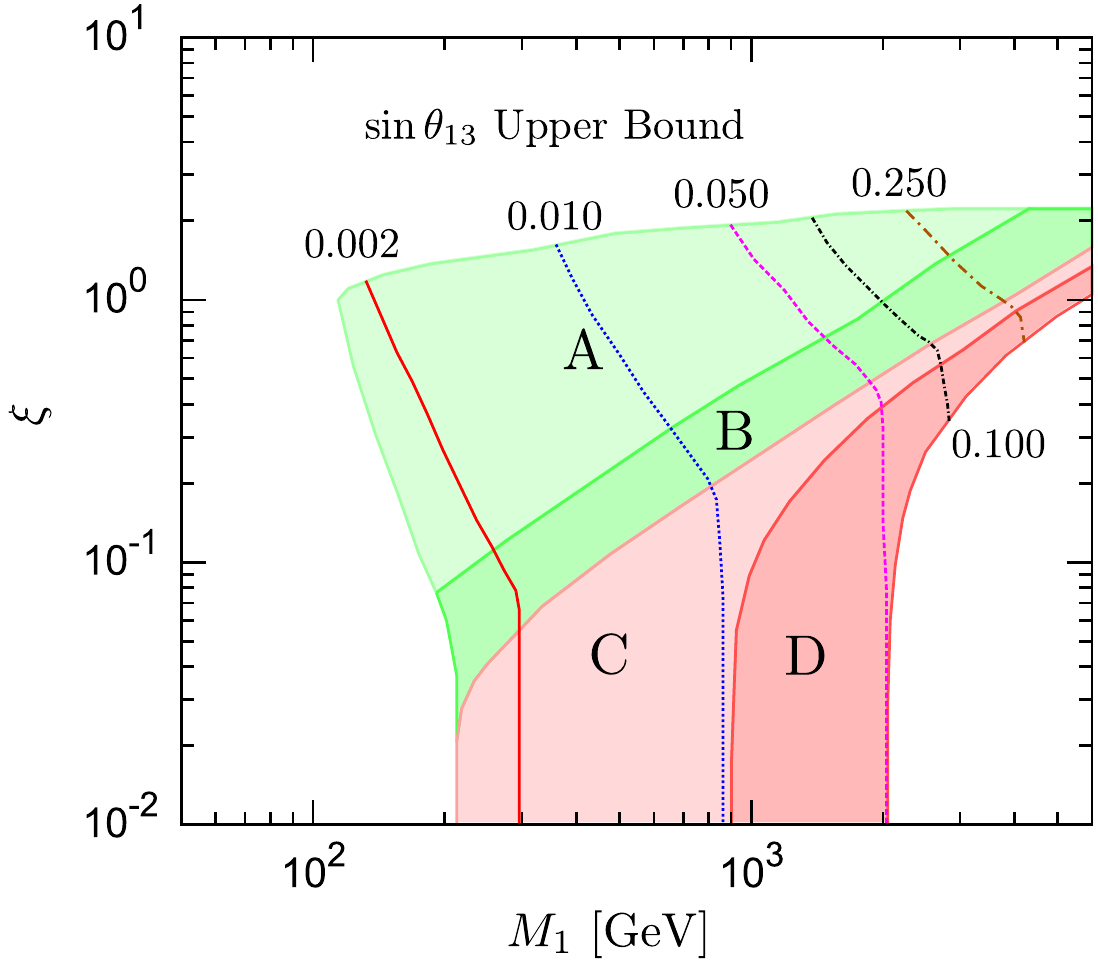} 
\caption{Region in the space of DM mass $M_1$ and
  $\xi = |h_1h_2|\sin(\varphi_1 - \varphi_2)$.}
\label{fig:relic}
\end{figure}
Given $\text{Br}(\mu\to e\gamma)<2.4\times
10^{-12}$~\cite{Adam:2011ch}, we take as benchmark points
$M_3=6000$\,GeV and $|h_3|=0.3$.
Using the correlations between neutrino oscillation data and
lepton-flavor structure the independent parameters are $M_\eta,\, M_1 ,\, \delta \equiv M_2 - M_1$ and $\xi \equiv {\textrm{Im}}(h_2^*h_1)$.
The region of the parameter space consistent with neutrino data,
lepton-flavor violation, perturbativity and DM relic density is 
shown in Figure~\ref{fig:relic}, where A, B, C, D, correspond to
different assumptions on $M_\eta$, with A: $2.0 < M_\eta/M_1 <
{9.8}$, B: $1.2 < M_\eta/M_1 < 2.0$, C: $1.05 < M_\eta/M_1 < 1.20$,
D: $1.0 < M_\eta/M_1 < 1.05$. The curves show the
upper bound on $\sin\theta_{13}$ from $\mu\to e\gamma$
with respect to Eq.~(\ref{eq:m2eg}).

\section{Direct Detection}
Inelastic scattering of DM off nuclei is realized by photon exchange in
1-loop processes involving charged leptons and the charged component 
of the inert doublet.
The 3-point vertex effective interactions of $N_1$, $N_2$ and $\gamma$
which give a dominant contribution are written as
\begin{equation}
\mathcal{L}_{\text{eff}}=
ia_{12}\overline{N_2}\gamma^{\mu}N_1\partial^{\nu}F_{\mu\nu}
+i\left(\frac{\mu_{12}}{2}\right)\overline{N_2}\sigma^{\mu\nu}N_1F_{\mu\nu}
+ic_{12}\overline{N_2}\gamma^{\mu}N_1A_{\mu},
\label{eq:eff}
\end{equation}
where the factor $i$ is a conventional factor to obtain real couplings
$a_{12}$, $c_{12}$ and $\mu_{12}$, and $F_{\mu\nu}$ is the electromagnetic
field strength.  The coefficient $\mu_{12}$ is known as the transition
magnetic moment between $N_1$ and $N_2$.
To our knowledge, the relevant loop processes for inelastic scattering
of fermionic DM in the Ma model~\cite{Ma:2006km} have not been previously calculated.
The effective interactions yield three types of
differential scattering cross sections with a nucleus which has atomic
number $Z$, mass number $A$, mass $m_A$, spin $J_A$ and magnetic moment $\mu_A$. They are called charge-charge (CC), dipole-charge (DC), and dipole-dipole (DD) couplings:
\begin{figure}
    \centering
\includegraphics[width=0.44\linewidth]{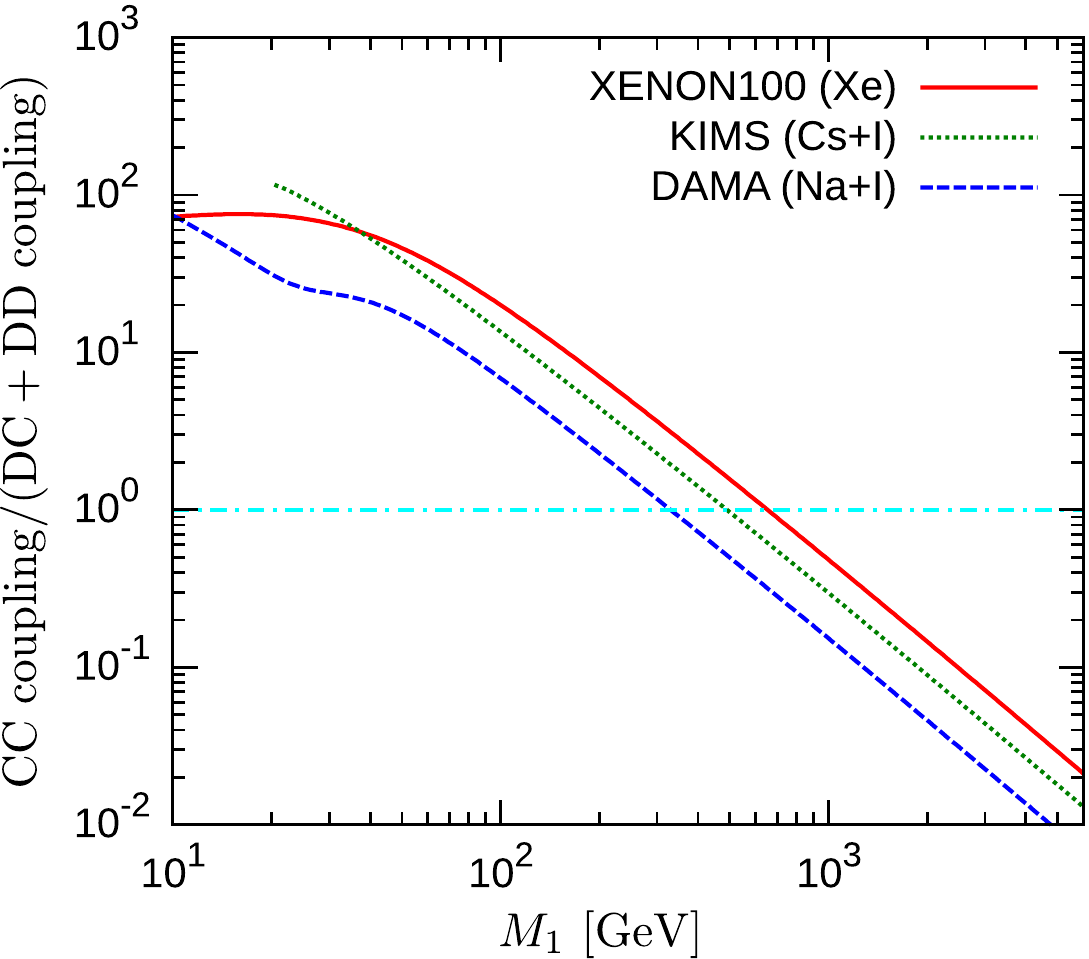}
\includegraphics[width=0.44\linewidth]{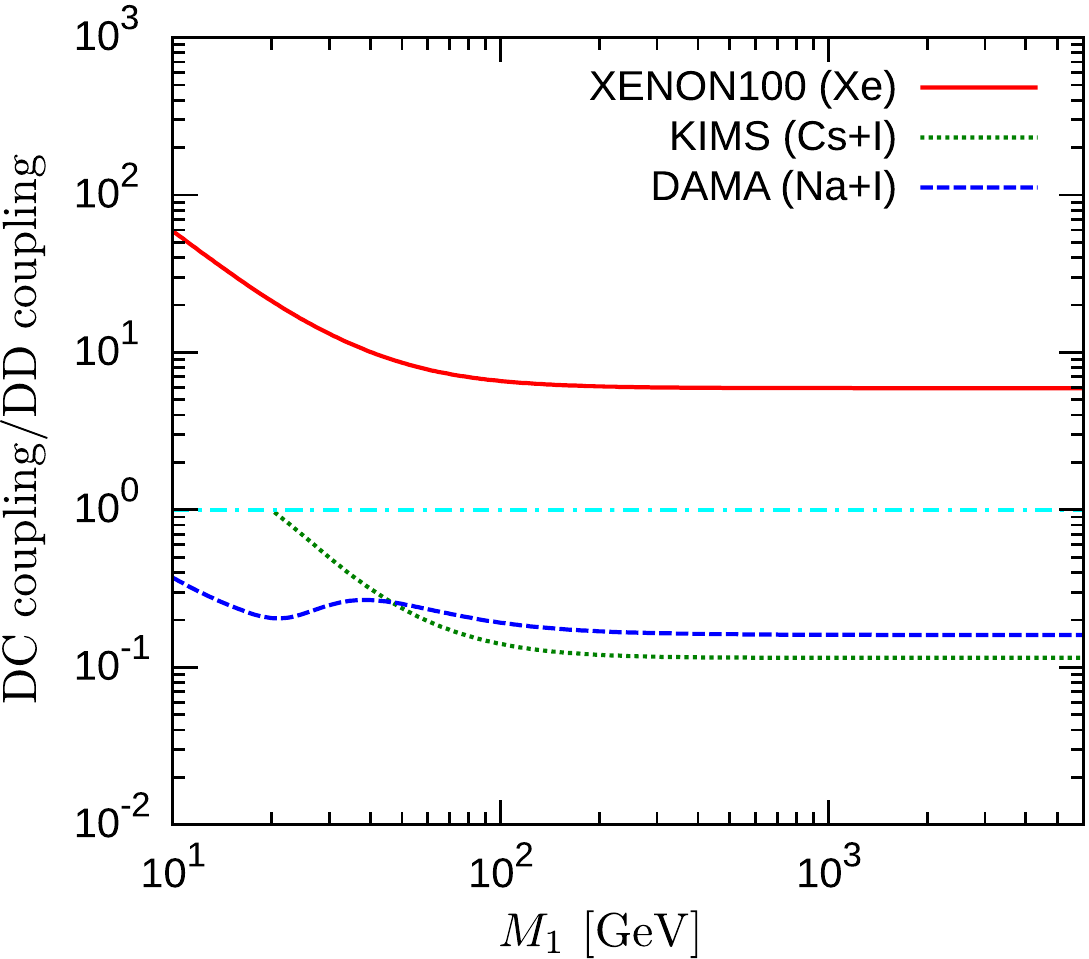}   
  \caption{Effective interactions. $M_\eta / M_1 = 1.5$ and $\delta =
    0$ is assumed. \textit{Left}: Contribution from CC relative to the sum of DC+DD. \textit{Right}: Ratio of the DC and DD
    contributions.}
\label{fig:ratio}
\end{figure}
\begin{align}
\frac{d\sigma_{\text{CC}}}{dE_R}&=\frac{Z^2b_{12}^2m_A}{2{\pi}v^2}F^2(E_R),
\label{eq:zz}\\
\frac{d\sigma_{\text{DC}}}{dE_R}&=\frac{Z^2\alpha_{\text{em}}\mu_{12}^2}{E_R}
\left[1-\frac{E_R}{v^2}\left(\frac{1}{2m_A}+\frac{1}{M_1}\right)
-\frac{\delta}{v^2}\frac{1}{\mu_{\text{DM}}}
-\frac{\delta^2}{v^2}\frac{1}{2m_AE_R}\right]F^2(E_R),
\label{eq:dz}\\
\frac{d\sigma_{\text{DD}}}{dE_R}&=\frac{\mu_A^2\mu_{12}^2m_A}
{{\pi}v^2}\left(\frac{J_A+1}{3J_A}\right)
F_D^2(E_R),
\label{eq:dd}
\end{align}
with the coefficient $b_{12}=(a_{12}+c_{12}/q^2)e$, the nuclear form factor $F(E_R)$ and the nuclear magnetic form factor $F_D(E_R)$, both of which depending on the recoil energy $E_R$.
Typically CC interactions are more important for small masses $M_1$,
which follows from the different dependence on the DM mass of $b_{12}$
and $\mu_{12}$. As can been seen in Figure~\ref{fig:ratio}, for XENON100~\cite{Aprile:2011hi} the DC coupling is more important, whereas for KIMS~\cite{Lee.:2007qn} and
DAMA~\cite{Bernabei:2010mq} DD dominates, because of the large magnetic moments of iodine and
sodium.
\begin{figure}
\centering
\includegraphics[width=0.49\linewidth]{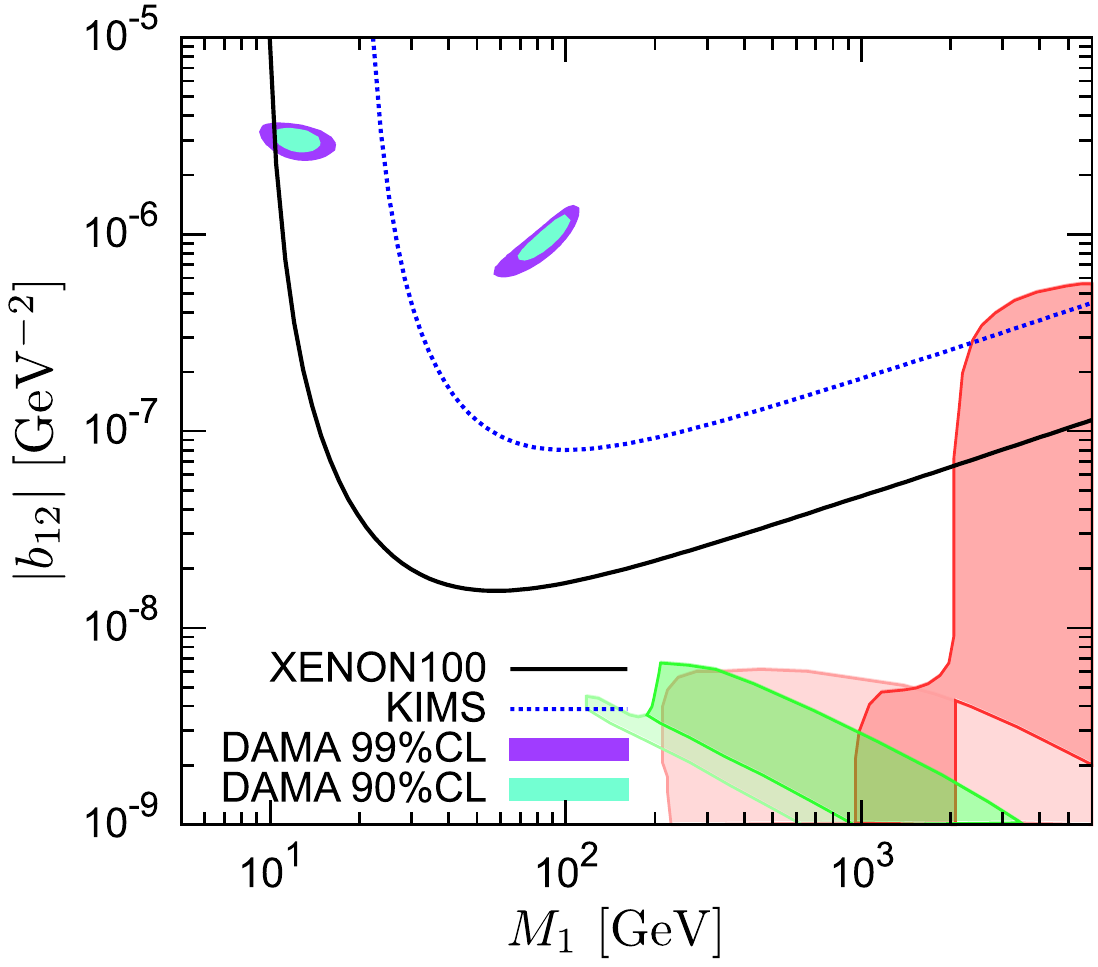}~ 
\includegraphics[width=0.49\linewidth]{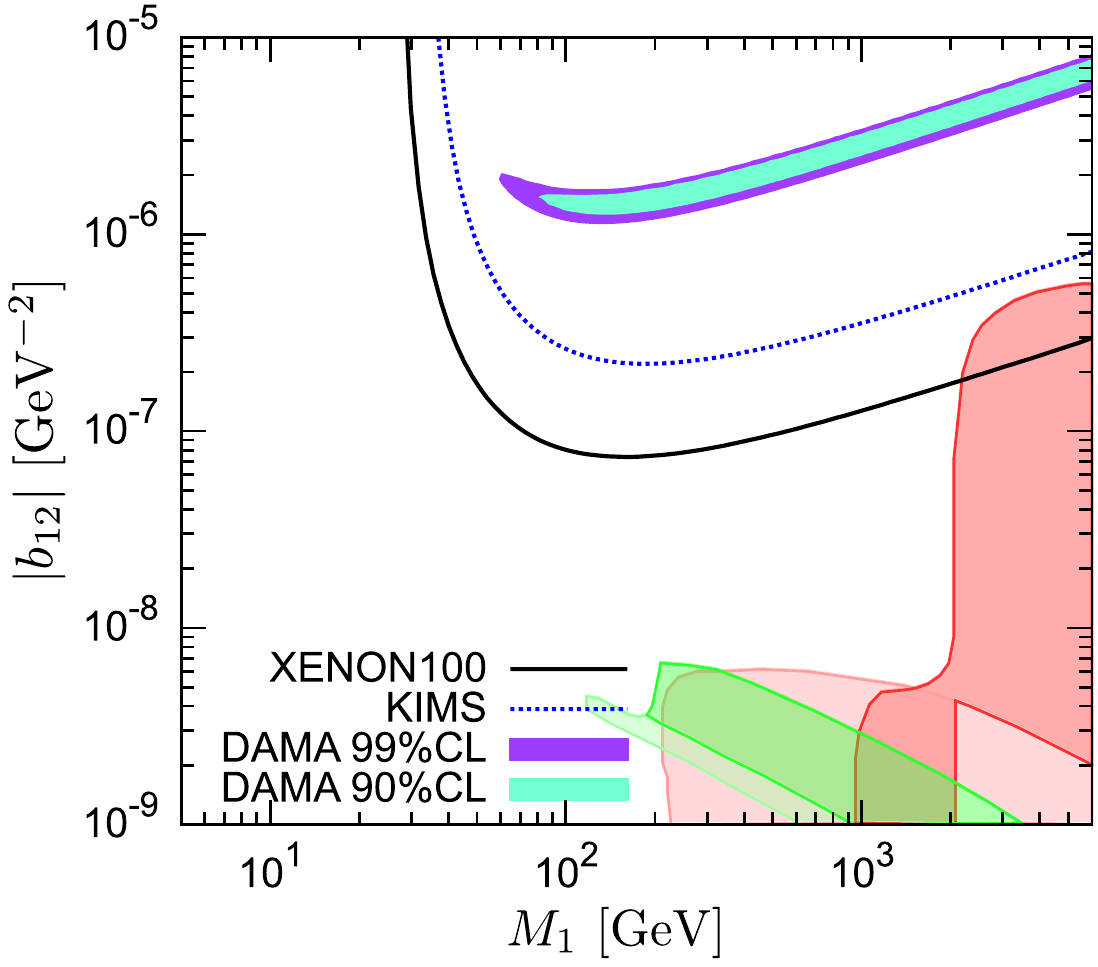}\\ 
\includegraphics[width=0.49\linewidth]{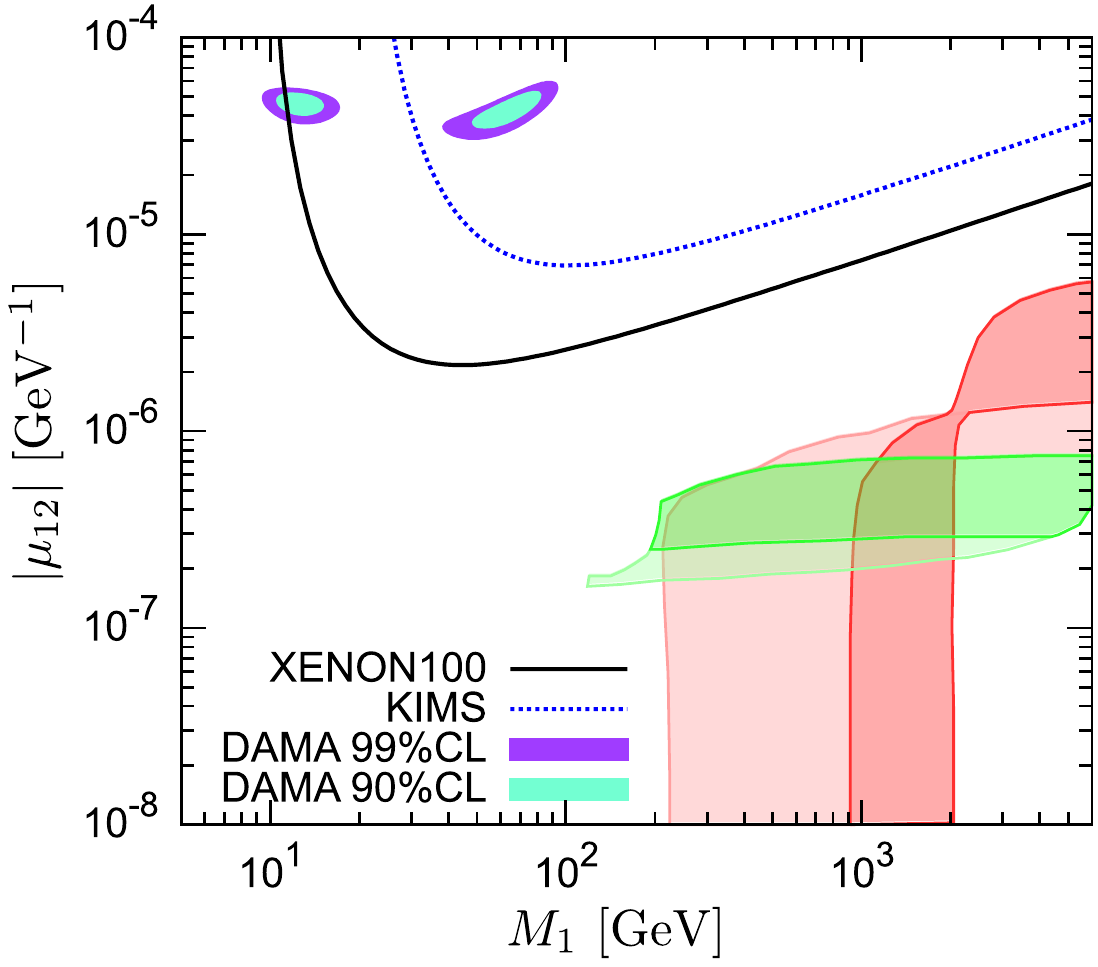}~
\includegraphics[width=0.49\linewidth]{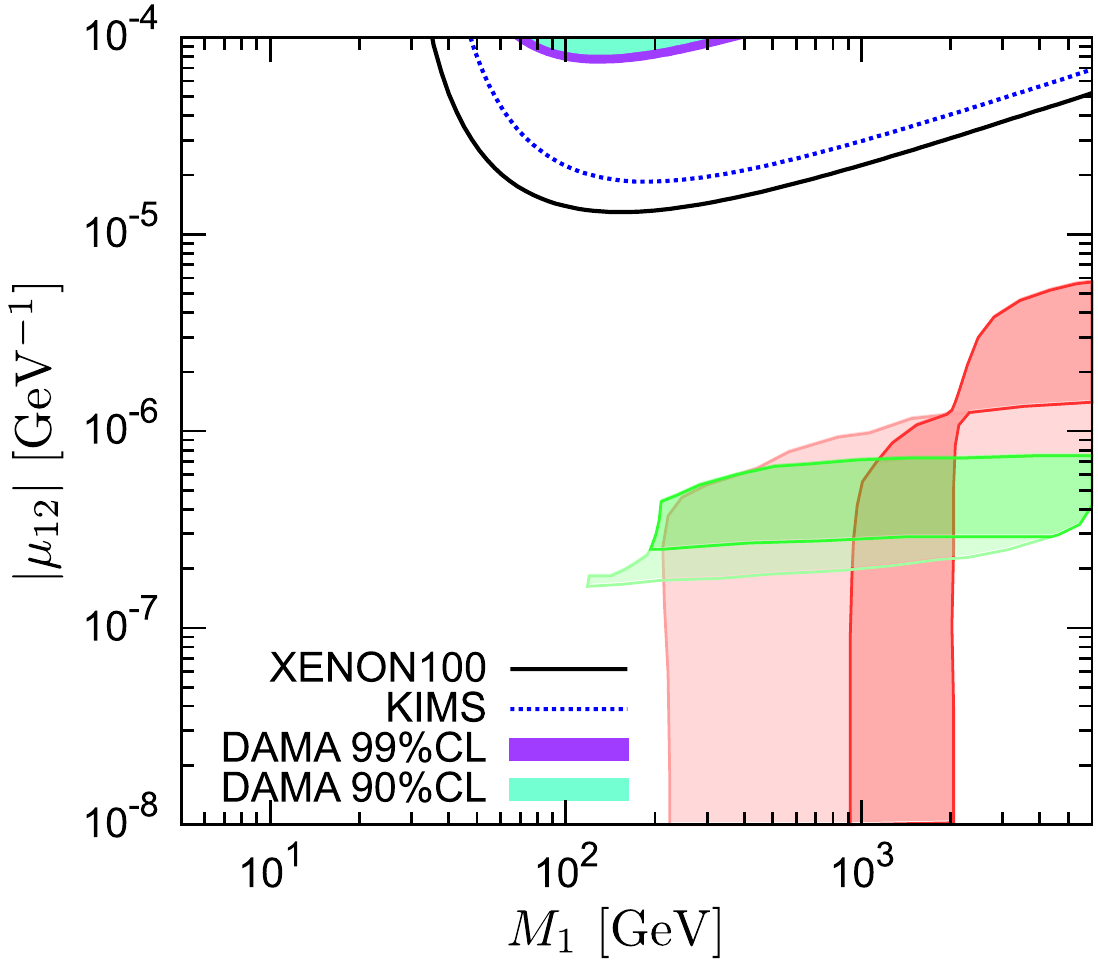}\\
\includegraphics[width=0.49\linewidth]{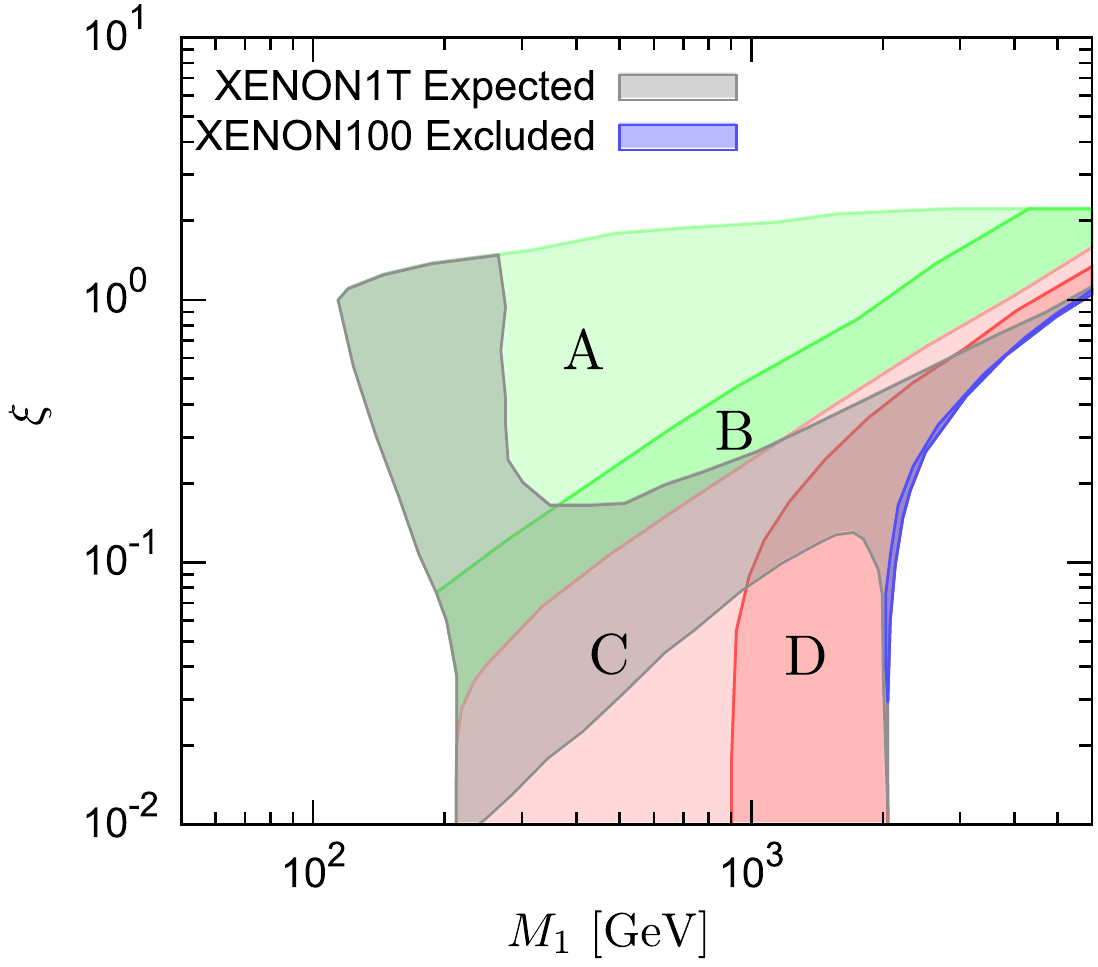}~
\includegraphics[width=0.49\linewidth]{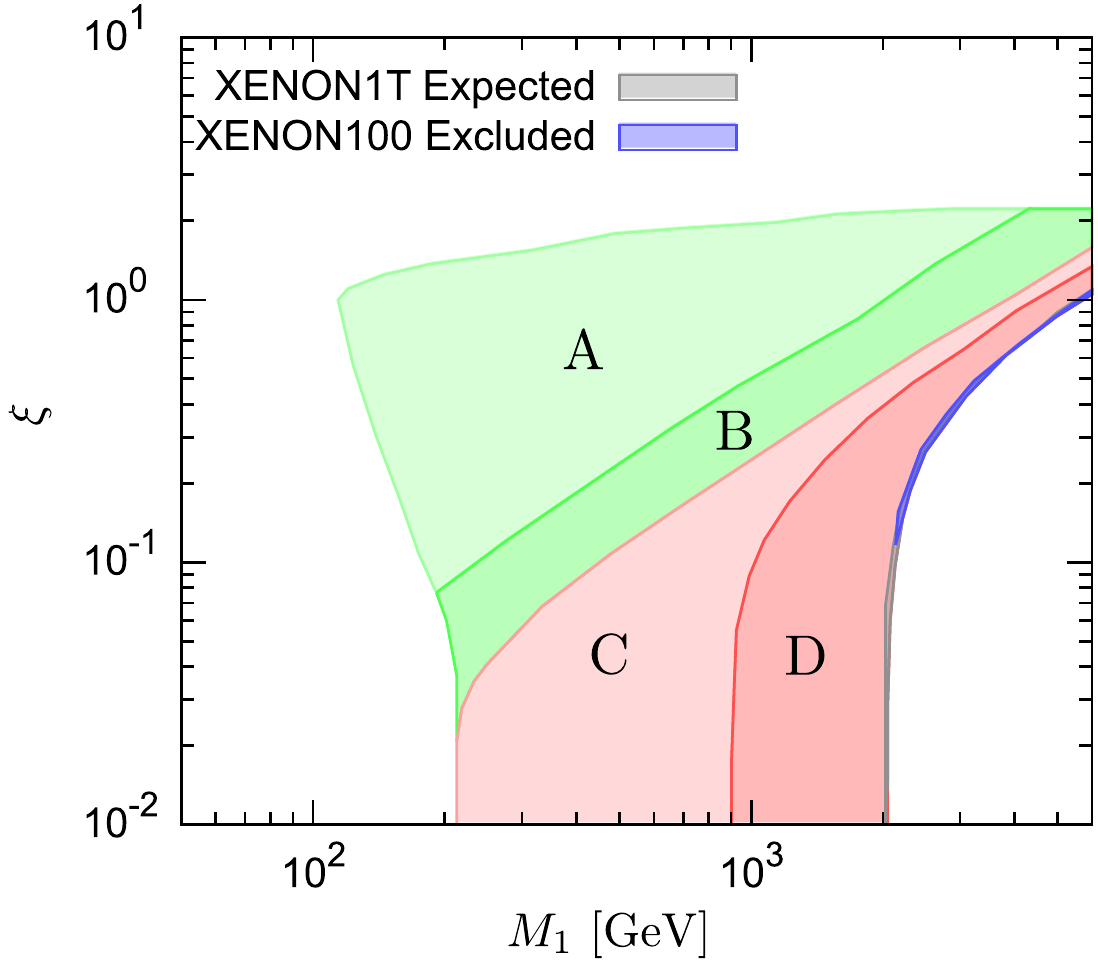} 
\caption{Results for CC (\textit{up}), DC+DD (\textit{middle}) interactions and Yukawa couplings (\textit{down}). \textit{Left}: $\delta=0$\,keV. \textit{Right}: $\delta=80$\,keV.  Regions A, B, C and D as described above.}
\label{fig:xenon-kims-dama}
\end{figure}text
From the effective interactions we calculate the total event rate $R$.
The constraints on the model are obtained by demanding that the predicted rate is less
than the observed rate in each experiment, respectively.
Our results are shown in Figure~\ref{fig:xenon-kims-dama}.

\section{Conclusion}
We have studied the model proposed by Ma~\cite{Ma:2006km},
which adds three right handed neutrinos $N_i$ $(i=1,2,3)$ and
one inert Higgs doublet $\eta$ to the SM.
Given that new particles are $\mathbb{Z}_2$-odd we have assumed 
that the lightest right handed neutrino $N_1$ is the lightest of
the $\mathbb{Z}_2$-odd particles, and hence it serve as
the DM candidate.
Our main conclusion is that direct detection of leptophilic DM is
possible by photon exchange which we have calculated for the first time to our knowledge.
To obtain a sizable scattering rate $N_1$ and $N_2$ have to be highly
degenerate.  Although the scattering cross section in this model
is too small to account for the DAMA annual modulation signal, we find
that current data from
the XENON100 experiment start to exclude certain regions of the
parameter space. The predicted event rate for XENON100 is dominated by
the charge-charge interaction. Future data, for example from XENON1T,
will significantly dig into the allowed parameter space and provide a
stringent test for the model provided $\delta$ is small enough.

\section*{Acknowledgments}
I thank Thomas Schwetz and Takashi Toma for collaboration.
I am grateful to Takashi Toma, who proposed this work, for instructive discussions. 
I acknowledge support by the International Max Planck Research School for
Precision Tests of Fundamental Symmetries.

\bibliography{DanielSchmidt}
\bibliographystyle{apsrev4-1}


%% file: Papers/Hserodio.tex

%
%
%
%
%
%

\chapter[Spontaneous leptonic CP violation and $\theta_{13}$ (Ser\^{o}dio)]{Spontaneous leptonic CP violation and $\theta_{13}$}
\vspace{-2em}
\paragraph{H. Ser\^{o}dio}
\paragraph{Abstract}
In this work one reviews a simple scenario~\cite{Branco:2012vs} where the above three aspects (leptogenesis, leptonic mixing and spontaneous CP violation) are related. To this aim, we shall add to the Standard Model (SM) a minimal particle content: two Higgs triplets $\Delta_a\, (a=1,2)$ with unit hypercharge and a complex scalar singlet $S$ with zero hypercharge.

\section{The model}
The SM is extended with two Higgs triplets $\Delta_a\, (a=1,2)$ of unit hypercharge and a complex scalar singlet $S$ with zero hypercharge. In the $SU(2)$ representation:
\begin{align}
\Delta_a=\begin{pmatrix}
         \Delta^0_a & -\Delta^+_a/\sqrt{2} \\
         -\Delta^+_a/\sqrt{2} & \Delta^{++}_a \\
       \end{pmatrix}.
\end{align}

CP invariance is imposed at the Lagrangian level and a $Z_4$ symmetry is introduced under which the scalar and lepton fields transform as indicated in Table.~\ref{HS_reps} 

The most general scalar potential invariant under the above symmetries can be written as
\begin{align}\label{HS_pot}
    V^{CP\times Z_4}=V_S+V_\phi+V_\Delta+V_{S\phi}+V_{S\Delta}+V_{\phi \Delta}+V_{S\phi \Delta},
\end{align}
where each terms are presented in~\cite{Branco:2012vs}. Since CP invariance has been imposed at the Lagrangian level, all the parameters are assumed to be real. This symmetry can be spontaneously broken by the complex VEV of the scalar singlet $S$. To show that this is indeed the case, let us analyze the scalar potential for $S$. The tree-level potential then reads
\begin{align}\label{HS_potV0}
 V_0=m_S^2 v_S^2 + \lambda_S v_S^4 + 2 \left(\mu_S^2+\lambda_S^{\prime\prime} v_S^2\right) v_S^2\,\cos{(2\alpha)}+2 \lambda_S^\prime v_S^4\,\cos{(4\alpha)}\,,
\end{align}
with $\langle S\rangle=v_S e^{i\alpha}$. Besides the trivial solution $v_S =0$, which leads to $V_0=0$, there are other three possible solutions to the minimization problem with $v_S\neq 0$: (i) $\alpha=0, \pm\pi$\,, (ii) $\alpha\pm\frac{\pi}{2}$ and (iii) $\cos(2 \alpha) = -\dfrac{\mu_S^2+\lambda_S^{\prime\prime}v_S^2}{4\lambda_S^\prime v_S^2}$.

Only the last solution is of interest to us since it leads not only to the spontaneous breaking of the CP symmetry but also to a non-trivial CP-violating phase in the one-loop diagrams relevant for leptogenesis, for a review on this subject see~\cite{Branco:2011zb}. 

In order to generate a realistic lepton mixing pattern we shall also impose an $A_4$ discrete symmetry at high energies. We recall that, in a particular basis, the Clebsch-Gordan decompositions of the $A_4$ group are can be made with real coefficients. The spontaneous breaking of the $A_4$ symmetry is then guaranteed by adding to the theory two extra heavy scalar fields, $\Phi$ and $\Psi$, with a suitable VEV alignment. The complete symmetry assignments of the fields under $A_4 \times Z_4$ and $SU(2)_L \times U(1)_Y$ are given in Table~\ref{HS_reps}.

Below the cut-off scale $\Lambda$, the flavour dynamics is encoded in the relevant effective Yukawa Lagrangian $\mathcal{L}$, which contains the lowest-order terms\footnote{In principle, one could also include the renormalizable 4-dimension term $\Delta_2 L^T L$. This term is however easily removed by imposing an additional shaping $Z_4$ symmetry.} in an expansion in powers of $1/\Lambda$,
\begin{align}
\begin{split}
\mathcal{L}=&\frac{y^\ell_e}{\Lambda} \left(\overline{L} \Phi\right)_{\mathbf{1}}\phi e_R+\frac{y^\ell_\mu}{\Lambda}  \left(\overline{L} \Phi\right)_{\mathbf{1^{\prime\prime}}}\phi\mu_R+\frac{y^\ell_\tau}{\Lambda}  \left(\overline{L} \Phi\right)_{\mathbf{1^{\prime}}}\phi\tau_R\\
&+\frac{y_{2}}{\Lambda}\, \Delta_2\left(L^TL\Psi\right)_{\mathbf{1}}+\frac{1}{\Lambda} \,\Delta_1\left(L^TL\right)_{\mathbf{1}}\left(y_{1} S+y^{\prime}_{1} S^\ast\right)+\text{H.c.}\,.
\end{split}
\label{HS_LYukA4}
\end{align}
As soon as the heavy scalar fields develop VEVs along the required directions, namely,
\begin{align}
\langle\Phi\rangle=(r,0,0)\;,\;\langle\Psi\rangle=(s,s,s)\,,
\label{HS_vevalign}
\end{align}
and the scalar singlet $S$ acquires a complex VEV, $\langle S \rangle = v_S\,e^{i\alpha}$, the Yukawa matrices become
\begin{align}\label{HS_Ymatrices}
\mathbf{Y}^e=
\begin{pmatrix}
y_e&0&0\\
0&y_\mu&0\\
0&0&y_\tau
\end{pmatrix}\,,\quad \mathbf{Y}^{\Delta_1}=y_{\Delta_1}
\begin{pmatrix}
1&0&0\\
0&0&1\\
0&1&0
\end{pmatrix}\,,\quad  \mathbf{Y}^{\Delta_2}=\frac{y_{\Delta_2}}{3}
\begin{pmatrix}
2&-1&-1\\
-1&2&-1\\
-1&-1&2
\end{pmatrix},
\end{align}
and
\begin{align}
y_{e, \mu, \tau} = \frac{r}{\Lambda} y^\ell_{e, \mu, \tau}\,, \quad  y_{\Delta_1}=\frac{v_S}{\Lambda^\prime}\left(y_{1} e^{i\alpha}+y^{\prime}_{1} e^{-i\alpha}\right), \quad y_{\Delta_2}=\frac{y_{2}}{\Lambda}\,s.
\end{align}
Notice that the Yukawa matrices $\mathbf{Y}^{\Delta_1}$ and $\mathbf{Y}^{\Delta_2}$ exhibit the so-called $\mu-\tau$ and magic symmetries, respectively.
\begin{center}
\begin{table}[t]
\caption{\label{HS_reps} Representations of the fields under the $A_4 \times Z_4$ and $SM=SU(2)_L \times U(1)_Y$ symmetries.}
\begin{tabular}{ccccccccc}
Field &$L$&$e_R,\mu_R,\tau_R$&$\Delta_1$&$\Delta_2$&$\phi$&$S$&$\Phi$&$\Psi$\\
\hline\hline
$A_4$&$\mathbf{3}$&$\mathbf{1}$, $\mathbf{1^{\prime}}$,
$\mathbf{1^{\prime\prime}}$&$\mathbf{1}$&$\mathbf{1}$&$\mathbf{1}$&$\mathbf{1}$&
$\mathbf{3}$&$\mathbf{3}$\\
$Z_4$&$i$&$-i$&$1$&$-1$&$i$&$-1$&$i$&$1$\\
$SM$&$(2,-1/2)$&$(1,-1)$&$(3,1)$&$(3,1)$&$(2,1/2)$&$(1,0)$&$(1,0)$&$(1,0)$\\
\hline
\end{tabular}
\end{table}
\end{center}

\section{Low-energy phenomenology}
In the present framework, neutrinos acquire masses through the well-known type II seesaw mechanism due to the tree-level exchange of the heavy scalar triplets $\Delta_a$. The unitary mixing matrix $\mathbf{U}$ is given by
\begin{align}
\mathbf{U}= e^{-i\sigma_1/2}\,
\begin{pmatrix}
\frac{2}{\sqrt{6}}&\frac{1}{\sqrt{3}}&0\\
-\frac{1}{\sqrt{6}}&\frac{1}{\sqrt{3}}&-\frac{1}{\sqrt{2}}\\
-\frac{1}{\sqrt{6}}&\frac{1}{\sqrt{3}}&\frac{1}{\sqrt{2}}
\end{pmatrix}\text{diag}\left(1,e^{i\gamma_1},e^{i\gamma_2}\right)\,,
\end{align}
where
\begin{align}
\begin{split}
\gamma_1 &= (\sigma_1-\beta)/2, \quad \gamma_2=(\sigma_1-\sigma_2)/2, \quad \sigma_{1,2}=\text{arg}\left(z_2 \pm z_1 e^{i\beta}\right).
\end{split}
\end{align}
Hereafter we consider the relevant CP-violating phase as being $\beta$. Since at this point there is no Dirac-type CP violation ($\mathbf{U}_{13}=0$), the Majorana phases $\gamma_{1,2}$ are the only source of CP violation in the lepton sector.

\begin{figure}[t]
\begin{center}
\includegraphics[width=10cm]{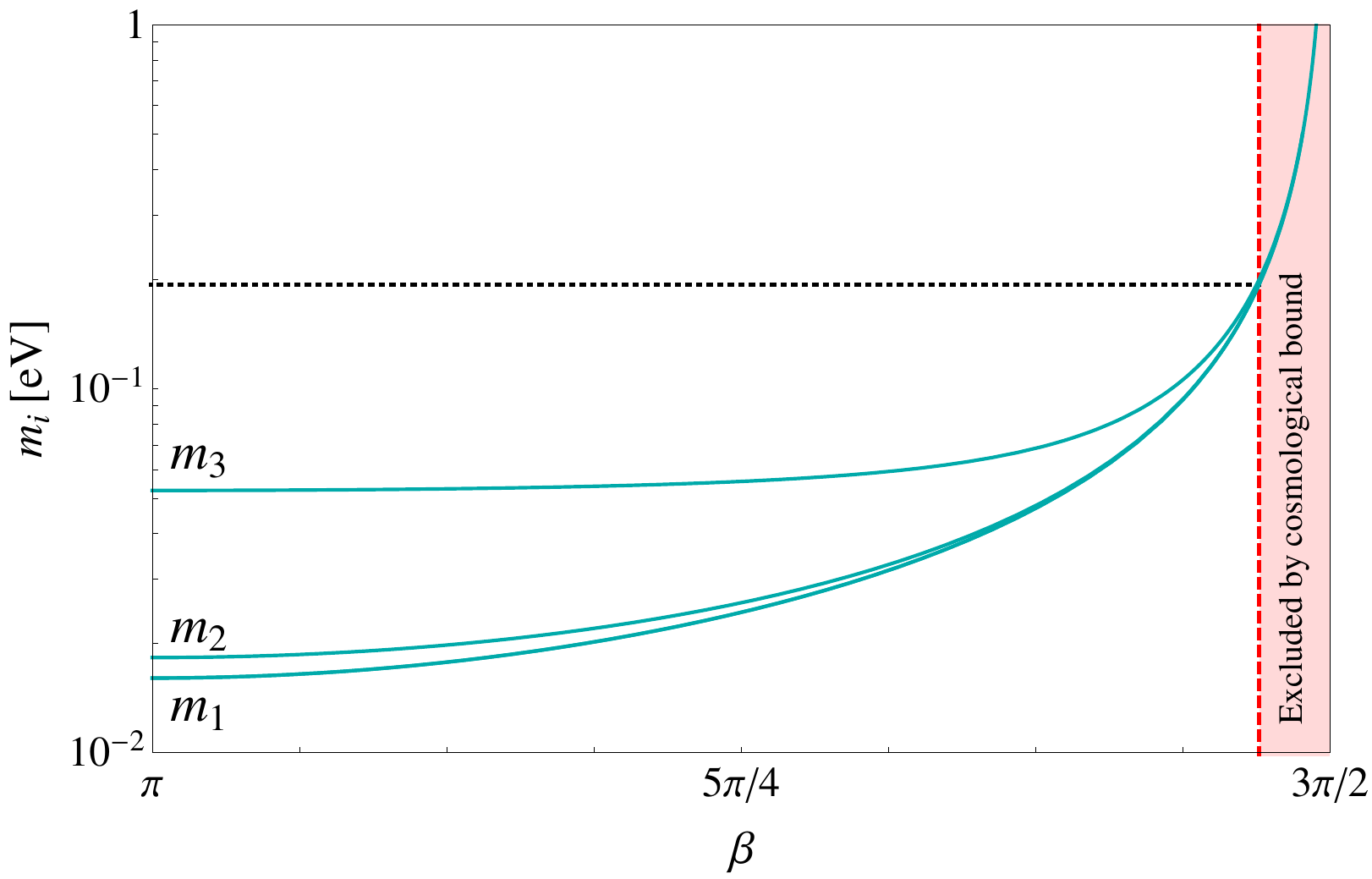}
\caption{\label{HS_fig1} Neutrino masses $m_i$ as a function of the high-energy phase $\beta$ in the exact TBM case.}
\end{center}
\end{figure}

At $1\sigma$ confidence level, the neutrino mass squared differences are~\cite{Schwetz:2011zk}
\begin{align}\label{HS_expdmass}
\Delta m^2_{21} =\left(7.59^{+0.20}_{-0.18}\right)\times 10^{-5}\,\text{eV}^2,\quad \Delta m^2_{31}=\left(2.50^{+0.09}_{-0.16}\right)\left[-2.40^{+0.09}_{-0.08}\right]
\times 10^{-3}\,\text{eV}^2,
\end{align}
for the normal [inverted] neutrino mass hierarchy. The present model cannot accommodate an inverted hierarchy for the neutrino mass spectrum. The dependence of neutrino masses on the high-energy phase $\beta$ is presented in Fig.~\ref{HS_fig1} for the exact TBM case. The light red shaded area is currently disfavoured by the recent WMAP seven-year cosmological observational data~\cite{Komatsu:2010fb}.

The T2K~\cite{Abe:2011sj} and MINOS~\cite{Adamson:2011qu} neutrino oscillation data imply for the $\theta_{13}$ mixing angle
\begin{align}
 \sin^2\theta_{13}=0.013^{+0.007}_{-0.005}\left(^{+0.015}_{-0.009}\right)\left[^{+0.022}_{-0.012}\right]\,,
\end{align}
at $1\sigma(2\sigma)[3\sigma]$. Recently, through the observation of electron-antineutrino disappearance, the Daya Bay Reactor Neutrino Experiment has also measured the non-zero value~\cite{DayaBay}:
\begin{equation}
\sin^2(2\theta_{13})=0.092\pm0.016({\rm stat})\pm 0.005({\rm syst})\,,
\label{HS_DBresults}
\end{equation}
with a significance of $5.2\sigma$. In the light of these results, models that lead to tribimaximal mixing appear to be disfavored. Here we shall consider small perturbations around the TBM vacuum-alignment conditions~(\ref{HS_vevalign}). We consider two distinct cases (with $|\varepsilon_{1,2}|\ll 1$.):
\vspace{0.5cm}

{\bf CASE A} - Small perturbations around the flavon VEV $\left<\Phi\right>=(r,0,0)$ of the form $\left<\Phi\right>=r(1,\varepsilon_1,\varepsilon_2)$;\\

Due to the new form of $\left<\Phi\right>$, the charged lepton Yukawa matrix is
\begin{equation}
\mathbf{Y}^\ell=
\begin{pmatrix}
y_e&y_\tau\varepsilon_1&y_\mu\varepsilon_2\\
y_\tau \varepsilon_2&y_\mu&y_e\varepsilon_1\\
y_\mu\varepsilon_1&y_e\varepsilon_2&y_\tau
\end{pmatrix}\,,
\end{equation}
which implies $\mathbf{U}_\ell\neq 1\!\!1$, where $\mathbf{U}_\ell$ is the unitary matrix which rotates the left-handed charged-lepton fields to the their physical basis. The new lepton mixing matrix $\mathbf{U}=\mathbf{U}_\ell^\dagger \mathbf{U}_{\rm TBM}$ yields  the perturbed mixing angles
\begin{equation}
\sin^2\theta_{12}\simeq\frac{1}{3}\left[1-2(\varepsilon_1+\varepsilon_2)\right]\,,\quad \sin^2\theta_{23}\simeq\frac{1}{2}(1+2\varepsilon_1)\,,\quad
\sin^2\theta_{13}\simeq \frac{\left(\varepsilon_1-\varepsilon_2\right)^2}{2}\,,
\label{HS_aps}
\end{equation}
at lowest order in $\varepsilon_{1,2}$. Obviously, the rotation of the charged lepton fields does not affect the neutrino spectrum nor generate a Dirac-type CP-violating phase. Since the flavon fields are real, the Majorana phases $\gamma_{1,2}$ also remain unaltered.
\vspace{0.5cm}

{\bf CASE B} - Small perturbations around the flavon VEV $\left<\Psi\right>=s(1,1,1)$ of the form $\left<\Psi\right>=s(1,1+\varepsilon_1,1+\varepsilon_2)$;

The Yukawa couplings $\mathbf{Y}^{\Delta_{2}}$ contributing to the neutrino mass matrix are now given by
\begin{equation}
\mathbf{Y}^{\Delta_{2}}=\frac{y_{\Delta_2}}{3}
\begin{pmatrix}
2&&-1-\varepsilon_2&&-1-\varepsilon_1\\
-1-\varepsilon_2&&2+2\varepsilon_1&&-1\\
-1-\varepsilon_1&&-1&&2+2\varepsilon_2
\end{pmatrix}.
\end{equation}
Consequently, at first order in $\varepsilon_{1,2}$, the neutrino mass spectrum get small corrections. Still, as in the unperturbed case, it can be shown that an inverted neutrino hierarchy is not allowed. In the present case, the approximate analytic expressions for the mixing angles are
\begin{equation}
\sin^2\theta_{12}\simeq\frac{1}{3}+\frac{2}{9}(\varepsilon_1+\varepsilon_2),\quad  \sin^2\theta_{13}\simeq\frac{(\varepsilon_1-\varepsilon_2)^2}{72\cos^2\beta},\quad
\sin^2\theta_{23}\simeq\frac{1}{2}+\frac{1}{6}(\varepsilon_1-\varepsilon_2)\,,
\label{HS_apsB}
\end{equation}
while for the Dirac-type CP-violating invariant $J_{\rm CP}$ we have
\begin{equation}
J_{\rm CP}={\rm Im}\left[\,\mathbf{U}_{11} \mathbf{U}_{22}
\mathbf{U}_{12}^\ast \mathbf{U}_{21}^\ast\,\right] \nonumber \\
\simeq \frac{\varepsilon_2-\varepsilon_1}{36}\tan\beta\,.
\label{HS_JCP}
\end{equation}

\begin{figure}[t]
\begin{center}
\begin{tabular}{cc}
\includegraphics[width=7.6cm]{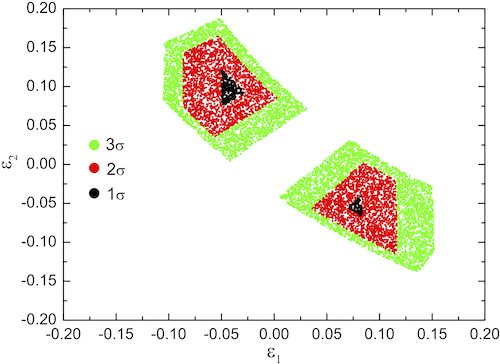}&
\includegraphics[width=7.6cm]{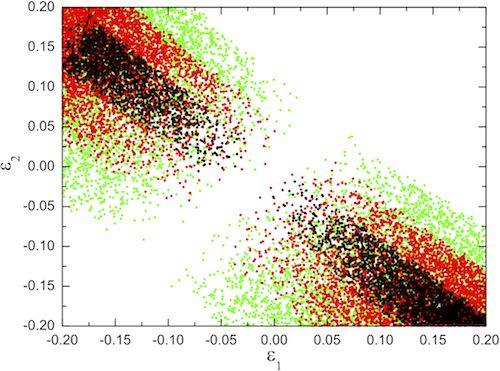}
\end{tabular}
\end{center}
\vspace{-0.5cm}

\caption{Allowed regions in the $(\varepsilon_1,\varepsilon_2)$ plane corresponding to the VEV perturbations of the flavon field $\left<\Phi\right>=r(1,\varepsilon_1,\varepsilon_2)$ in case A (left panel) and $\left<\Psi\right>=s(1,1+\varepsilon_1,1+\varepsilon_2)$ in case B (right panel). The scatter points were obtained considering the $1\sigma$ (black), $2\sigma$ (red) and
$3\sigma$ (green) neutrino oscillation data.\label{HS_fig2}}
\end{figure}
\vspace{0.5cm}

We now comment on the possibility of reproducing the recent Daya Bay $\theta_{13}$ value (\ref{HS_DBresults}) in our framework. In the absence of a 3-neutrino global analysis of the oscillation data including the Daya Bay results, we take the $1\sigma$ values for $\theta_{12}$, $\theta_{23}$ and $\Delta m^2_{21,31}$ obtained in \cite{Schwetz:2011zk}. One can see that the new Daya Bay value for $\theta_{13}$ is not compatible with the remaining mixing angles for case A. Instead, for case B we get a perfect agreement with all data~\cite{Branco:2012vs}. 

\section{Higgs triplet decays and leptogenesis}

The mechanism of leptogenesis can be naturally realized in the present model due to the presence of the scalar triplets $\Delta_1$ and $\Delta_2$. In the presence of CP-violating interactions, the decay of $\Delta_a$ into two leptons generates a nonvanishing leptonic asymmetry for each triplet component ($\Delta^0_a, \Delta^+_a, \Delta^{++}_a$). Assuming $M_a \ll M_b$, the flavoured CP asymmetry given for each triplet component can be rewritten as
\begin{align}\label{HS_CPasymm1}
\epsilon_{a}^{\alpha\beta}=c_{\alpha\beta}\,\mathbf{P}_{\alpha\beta}^a\,\epsilon_a^0\,,\quad \epsilon_a^0=\frac{1}{3\pi} \frac{z_a\,z_b\,|u_a|^2\,M_a^2\,\sin\beta}{z_a^2\,t_a v^4 + 4\,|u_a|^4 M_a^2}\,,
\end{align}
where $c_{\alpha\beta}$ is $2-\delta_{\alpha\beta}$ for $\Delta_a^0\,,\,\Delta_a^{++}$ and $1$ for $\Delta_a^+$, and
with $t_1=3$ and $t_2=2$. The matrix $\mathbf{P}^a$ is given by
\begin{align}
\mathbf{P}^a=\frac{(-1)^a}{2}\begin{pmatrix}
         -2(1+\varepsilon_1+\varepsilon_2) & \varepsilon_1-\varepsilon_2 & \varepsilon_2-\varepsilon_1\\
         \varepsilon_1-\varepsilon_2 &  4\left(\varepsilon_1+\varepsilon_2\, y_\mu^2/y_\tau^2\right) & 1+\varepsilon_1+\varepsilon_2\\
         \varepsilon_2-\varepsilon_1 & 1+\varepsilon_1+\varepsilon_2 & -4\left(\varepsilon_1+\varepsilon_2\, y_\mu^2/y_\tau^2\right)
       \end{pmatrix},
\end{align}
for case A, while
\begin{align}
\mathbf{P}^a=(-1)^a\left[\frac{1}{2} +\delta_{a2}\frac{v^4z_2^2(\varepsilon_1+\varepsilon_2)}{18M_2^2u_2^4 + 9v^4z_2^2}\right]\left(\begin{array}{rrr}
         -2 &\quad 0 &\quad 0\\
         0 &  0 & 1\\
         0 & 1 & 0
       \end{array}\right),
\end{align}
in case B. Obviously, in the TBM limit ($\varepsilon_{1,2}=0$), there is a unique matrix $\mathbf{P}^a$. In this case, the flavour structure of $\mathbf{P}^a$ dictates that the only allowed decay channels of $\Delta_a$ are into the $ee$ and $\mu\tau$ flavours. Once the VEV perturbations are introduced, new decay channels are opened in case A with the corresponding CP asymmetries suppressed by $\mathcal{O}(\varepsilon)$ factors.

Maximizing $\epsilon_a^0$ with respect to the VEV of the decaying scalar triplet $u_a$, one obtains
\begin{align}\label{HS_epsmax}
    \epsilon^0_\text{1,max} \simeq \frac{M_1 \sqrt{\Delta m^2_{31}}}{12\sqrt{6} \pi v^2} \sin\beta, \quad \epsilon^0_\text{2,max} \simeq \frac{M_2 \sqrt{\Delta m^2_{31}}}{48 \pi v^2} \tan\beta\,.
\end{align}
One can see from the above equations that sufficiently large values of the CP asymmetries can be obtained in the flavoured regime, i.e. $M_a<10^{12}$. Therefore, unlike the type-I seesaw framework~\cite{Branco:2009by,Bertuzzo:2009im,AristizabalSierra:2009ex,Felipe:2009rr}, imposing to the Lagrangian a discrete symmetry do not necessarily leads to a vanishing leptonic CP asymmetry in the type II seesaw case~\cite{deMedeirosVarzielas:2011tp}.

\section{Conclusion}
A simple scenario where spontaneous CP violation, leptonic mixing and  thermal leptogenesis are related was presented. We added a minimal particle content to the SM, namely, two Higgs triplets $\Delta_{1,2}$ and a complex scalar singlet $S$. In this framework, a single phase connects low- and high-energy CP violation.
\bibliographystyle{apsrev4-1}


%% file: Papers/Simoes.tex

\chapter[The quark NNI textures rising from $SU(5)\times Z_4$ symmetry (Sim\~oes)]{The quark NNI textures rising from $SU(5)\times Z_4$ symmetry}
\vspace{-2em}
\paragraph{C. Sim\~oes}

\paragraph{Abstract}
In this work we explored the consequences of the $SU(5)\times Z_4$ symmetry, where the 
quark mass matrices are in the Nearest-Neighbor-Interaction, on the leptonic sector. 
The model is based on the minimal $SU(5)$ Grand Unification~(GUT) model with three right-handed 
neutrinos and two Higgs quintets. Due to the $SU(5)$ 
symmetry, the charged lepton mass matrix gets the same NNI form as the quark sector. However, in the 
context of the type-I seesaw mechanism, the effective neutrino mass 
matrix can have six different textures, of which only two are compatible 
with the leptonic experimental data.

\section{Introduction}

One of the open questions in particle physics is the explanation of the observed pattern of fermion masses 
and their mixings. One way to study this puzzle is for example by playing with texture zeroes; one example is the 
Nearest-Neighbour-Interaction~(NNI)~\cite{Branco:1988iq}. It has zeroes on the (1,1), 
(1,3), (2,2) and (3,1) elements and together with the hermiticity condition leads to the well known Fritzsch form~\cite{Fritzsch:1977vd,Li:1979zj,Fritzsch:1979zq}. 

It was shown in Ref.~\cite{Branco:2010tx} that it is possible to obtain the quark mass 
matrices in the NNI form, in the context of the two-Higgs doublet model, through the implementation of a 
$Z_4$ flavour symmetry.

The goal of this work is to extend the idea developed in Ref.~\cite{Branco:2010tx} 
to $SU(5)$ and study the consequences of such an implementation 
on the leptonic sector. This work is organised as follows: in section~\ref{CS_sec:model} 
we present the model and discuss how to obtain the quark mass matrices in the NNI form; 
in section~\ref{CS_sec:matrices} we explore the leptonic sector concerning the viability 
of the mass matrices, then we conclude.

\section{The model}
\label{CS_sec:model}

The model is based on the minimal $SU(5)$~\cite{Georgi:1974sy} with three generations of 10 and 
$5^\ast$ fermionic representations as $10_i=(Q,u^c,e^c)_i$ and $5^{\ast}_i=(L,d^c)_i$. In order to generate neutrino masses we 
have introduced three right-handed neutrinos, $\nu^c_{1,2,3}$, singlets of $SU(5)$ that acquire 
mass via type-I seesaw mechanism~\cite{Minkowski:1977sc,Yanagida:1979as,GellMann:1980vs,Mohapatra:1979ia}.

The Higgs sector is composed by one 24 dimensional representation, $\Sigma$, and two 
quintets $H_1$ and $H_2$. The adjoint Higgs representation $\Sigma$ is introduced to 
break the $SU(5)$ down to the standard model~(SM) gauge group~($SU(3)_c\times SU(2)_L\times U(1)_Y$) 
through the vacuum expectation value~(VEV), $\langle\Sigma\rangle=\sigma\text{diag}(2,2,2,-3,-3)$
where $\sigma= \frac{a}{2\lambda}\frac{1 \,+\, \sqrt{1 \,+\,4\,\xi\,(60\eta\,+\,7)}} { 60\eta\, +\,7}$ 
(see Ref.~\cite{EmmanuelCosta:2011jq}). 
The quintets break the SM gauge group down to $SU(3)_c\times U(1)_{em}$ when the neutral 
component of each doublet acquires a VEV v$_1$, v$_2$ such that 
$\text{v}^2\,\equiv\, \left|\text{v}_1\right|^2 + \left|\text{v}_2\right|^2=(246.2\,\text{GeV})^2$ and 
generate the fermion masses via the Yukawa interactions. Note that at low energy scale one 
falls into a two Higgs doublet model.

The most general Yukawa Lagrangian is given by,
\begin{equation}
\label{CS_eq:lagr}
\begin{aligned} 
-\mathcal{L}_\text{Y} =&\frac{1}{4}\left(\Gamma^1_u\right)_{ij}10_i10_jH_1
+\frac{1}{4}\left(\Gamma^2_u\right)_{ij}10_i10_jH_2\\&+
\sqrt2\left(\Gamma^1_d\right)_{ij}10_i5^{\ast}_jH_1^{\ast} +
\sqrt2\left(\Gamma^2_d\right)_{ij}10_i5^{\ast}_jH_2^{\ast}\\&+
\left(\Gamma^1_D\right)_{ij}\,5^{\ast}_i\nu^c_jH_1+
\left(\Gamma^2_D\right)_{ij}\,5^{\ast}_i\nu^c_jH_2
+\frac{1}{2}\left(M_R\right)_{ij}\nu^c_i\nu^c_j+\text{H.c.}\,
\end{aligned}
\end{equation} 
where $\Gamma^{1,2}_u$ and $\Gamma^{1,2}_d$ are the up- and down-quark Yukawa matrices, $\Gamma^{1,2}_D$ and $M_R$ 
are the Dirac Yukawa and Majorana matrices for neutrinos. The up- and down-quark mass matrices are then given by 
$M_u = \text{v}_1\,\Gamma^1_u+\text{v}_2\,\Gamma^2_u$ and $M_d = \text{v}_1^{\ast}\,\Gamma^1_d+\text{v}_2^{\ast}\,\Gamma^2_d$.

The NNI form of the quark mass matrices is achieved through the introduction of a $Z_n$ discrete 
flavour symmetry. Under this $Z_n$ symmetry all 
fields except the adjoint Higgs field are charged. The two quintets $H_1$, $H_2$ carry 
charges $\phi_1$, $\phi_2$, the fermionic fields $10_i$, $5^\ast_i$ and $\nu^c_i$ carry 
charges $q_i$, $d_i$ and $n_i$.

In order to have mass matrices with the NNI form one should ensure that the zero entries 
in the mass matrices correspond to a non zero $Z_n$ charge of the trilinear 
terms and vice-versa. Following the method on Ref.~\cite{Branco:2010tx} and 
choosing that the (3,3) entry of $M_u$ does not vanish
we obtain $\phi_2=-2q_3$, which leads to the following $Z_n$ charges,
\begin{equation}
\label{CS_eq:charrep}
\mathcal{Q}(10_i)=(3q_3+\phi_1,-q_3-\phi_1,q_3),\qquad 
Q(5^{\ast}_i)=(q_3+2\phi_1,-3q_3,-q_3+\phi_1)\,.
\end{equation}
The charge matrix of the up- and down-quark bilinears is then given by,
{\scriptsize
\begin{equation}
\label{CS_eq:charbilin}
Q(10_i\,10_j)=
\begin{pmatrix} 
6q_3+2\phi_1 & 2q_3 & 4q_3+\phi_1 \\ 
2q_3 & -2\phi_1-2q_3 & -\phi_1\\ 
4q_3+\phi_1 & -\phi_1 & 2q_3 
\end{pmatrix}\,,\quad
Q(10_i\,5^{\ast}_j)=
\begin{pmatrix} 
4q_3+3\phi_1 & \phi_1 & 2q_3+2\phi_1 \\ 
\phi_1 & -\phi_1-4q_3 & -2q_3\\ 
2q_3+2\phi_1 & -2q_3 & \phi_1
\end{pmatrix}\,,
\end{equation}}
from which we conclude that $\phi_1$ must be different from $\phi_2$ and that the 
minimal realization of $Z_n$ that makes the NNI structure possible is $Z_4$ as in 
Ref.~\cite{Branco:2010tx}.  
  
The up- and down-quark mass matrices in terms of the Yukawa matrices are given as
{\scriptsize
\begin{equation}
M_u=\text{v}_1\begin{pmatrix} 
0 & 0 & 0\\ 
0 & 0 & b_{u}\\ 
0 & b_{u}& 0 
\end{pmatrix}+\text{v}_2
\begin{pmatrix} 
0 & a_{u} & 0\\ 
a_{u} & 0 &0\\ 
0 & 0 & c_{u} 
\end{pmatrix}\,,\quad
M_d=\text{v}^{\ast}_1
\begin{pmatrix} 
0 & a_{d} & 0\\ 
a^{\prime}_{d} & 0 &0\\ 
0 & 0 & c_{d} 
\end{pmatrix}+ \text{v}^{\ast}_2 
\begin{pmatrix} 
0 & 0 & 0\\ 
0 & 0 & b_{d}\\ 
0 & {b^{\prime}}_{d}& 0 
\end{pmatrix}\,.
\end{equation}
}

This being a GUT model some comments on proton decay and unification are in order. 

\paragraph{$\mathbf{M_e=M_d^T}$ Relation}
As a consequence of the $SU(5)$ symmetry the charged-lepton mass matrix is equal to the 
down-type quark mass matrix transposed which is not compatible with the down-type 
quark and charged-lepton masses hierarchies observed at low energy scale. One possibility 
to correct this relation is to introduce non-renormalisable higher dimension 
operators~\cite{Bajc:2002bv,EmmanuelCosta:2003pu} due to physics at $\Lambda'$ scale above the GUT scale.
For instance, dimension 5 operators contribute as
\begin{equation}
\label{CS_eq:massdiff}
\sum_{n=1,2}\frac{\sqrt{2}}{\Lambda'}
      \left(\Delta_n\right)_{ij}\, H^{\ast}_{n\,a}\, 10_i^{ab}\,
      \Sigma_b^c\, 5^{\ast}_{jc}\,,
\end{equation}
leading to the mass difference,
$M_d-M_{e}^{\top}=5\frac{\sigma}{\Lambda^{\prime}}(\text{v}_1^{\ast}\,\Delta_1+\text{v}_2^{\ast}\,\Delta_2)$
without destroying the NNI structure once $\Sigma$ is trivial under $Z_4$. 
Another alternative to correct the relation $M_e=M_d^T$ is to substitute the 
second Higgs quintet by a 45 dimensional Higgs representation~\cite{Perez:2007rm}. 
In this case the mass difference will be given by $M_d-M_{e}^{\top}=8\,\Gamma^2_d\,\text{v}^{\ast}_{45}$, 
where v$_{45}$ is the VEV of the 45.
In any of those situations the up-quark mass matrix is no longer symmetric, which is the 
reason why we have considered arbitrary NNI mass matrices in section ~\ref{CS_sec:matrices}.

\paragraph{Proton Decay}

The proton decay can occur through the exchange of X and Y heavy gauge bosons or the 
exchange of the colour Higgs triplets, $T_1$ and $T_2$ contained in the quintets.

For the proton decay via the exchange of heavy gauge bosons 
the decay width can be estimated~\cite{Langacker:1980js} as $\Gamma\approx \alpha_U^2 \frac{m_p^5}{M_V^4}$. 
Using the partial proton lifetime~\cite{Nakamura:2010zzi} $\tau(p\rightarrow \pi^0 e^+)>8.2\times10^{33}$ years
 the mass of the heavy gauge bosons is estimated as $M_V>(4.0-5.1)\times 10^{15}$ GeV for a 
unified gauge coupling in the range $\alpha_U^{-1}\approx 25 - 40$.

Concerning the proton decay via the exchange of the colour Higgs triplets, the dimension 
6 operators contributions at tree-level are given by
\begin{equation}
\label{CS_eq:contr1}
\sum_{n=1,2}
\frac{\left(\Gamma^n_u\right)_{ij}\left(\Gamma^n_d\right)_{kl}}{M^2_{T_n}}
\left[\frac{1}{2} (Q_iQ_j)(Q_kL_l)+(u^c_ie^c_j)(u^c_kd^c_l)\right]\,,
\end{equation}
that in fact vanish due to the Yukawa matrices form.

\paragraph{Unification}
We have found unification of the gauge couplings at two-loop level without considering the 
threshold effects and performing the splitting between the masses of the $\Sigma_3$ and $\Sigma_8$. 
In our computation, we have set the fields X, Y, $T_1$, $T_2$ 
at GUT scale, $\Lambda$, and $H_1$, $H_2$ around electroweak scale. We found a GUT 
scale around $\Lambda\approx(1.3-2.4)\times 10^{14}$ GeV and the masses of the $\Sigma_3$ 
and $\Sigma_8$ components of $\Sigma$ in the range $M_Z \leq  M_{\Sigma_3}\leq 1.8\times 10^4\,\text{GeV}$ and
$5.4\times 10^{11}\,\text{GeV}\leq  M_{\Sigma_8}\leq 1.3\times 10^{14}\,\text{GeV}$. 
Unfortunately, the unification scale found is smaller than what we expect from the computation of 
the proton decay through the exchange of the heavy X and Y gauge bosons and 
the mass splitting between $M_{\Sigma_3}$ and $M_{\Sigma_8}$ is unnaturally large. This discrepancy 
can be avoid by the introduction of a 24 fermionic representation \cite{Bajc:2006ia}. In such case the 
neutrino masses will get contributions also from type-III seesaw mechanism 
in addition to the usual type-I.

\section{Mass matrices}
\label{CS_sec:matrices}

In Ref.~\cite{Branco:2010tx} we have shown that the quark mass matrices 
in the NNI form accommodate all observed up- and down-quark masses and 
the CKM mixing matrix. 
As a consequence of $SU(5)$ symmetry and since the NNI form has zeroes in 
symmetric positions, the charged lepton mass matrix, $M_e$, has also NNI 
form. 

Both quark and charged-lepton mass matrices can be written as,
\begin{equation}
M_{x}=\begin{pmatrix}
0&A_x(1-\epsilon_a^x)&0\\
A_x(1+\epsilon_a^x)&0&B_x(1-\epsilon_b^x)\\
0&B_x(1+\epsilon_b^x)&C_x
\end{pmatrix}\,,
\end{equation}
where $x=u,\,d,\,e$ and $\epsilon$ measures the deviation from the Hermiticity; 
a global measurement of the asymmetry in the quark, $\varepsilon_q$, and leptonic, $\varepsilon_{\ell}$, 
sectors is given by
\begin{equation}
\varepsilon_q\equiv\frac{1}{2}\sqrt{\epsilon^{u\,2}_a+\epsilon^{u\,2}_b+\epsilon^{d\,2}_a+\epsilon^{d\,2}_b}\,\quad \text{and}\quad 
\varepsilon_{\ell}\equiv\sqrt{\frac{\epsilon^{e\,2}_a+\epsilon^{e\,2}_b}2}\,.
\end{equation}
For $\varepsilon_q=\varepsilon_e=0$ one recovers the Fritzsch
form~\cite{Fritzsch:1977vd,Li:1979zj,Fritzsch:1979zq}. 

The fact that the $Z_4$ neutrino charges are free parameters obliges 
us to scan all charge combinations and select the viable textures by 
confronting them with neutrino experimental data. 
\begin{table}
\scriptsize
\begin{center}
\begin{tabular}{ccc}
\hline
parameters & NH  & IH\\
\hline\\[-1ex]
$\Delta m^2_{21}\,\left(\times 10^{-5}\,\text{eV}^2\right)$ & \multicolumn{2}{c}{$7.62\pm 0.19$}\\[2ex]
$\left|\Delta m^2_{31}\right|\,\left(\times10^{-3}\,\text{eV}^2\right)$ & $2.53^{+0.08}_{-0.10}$
& $2.40^{+0.10}_{-0.07}$
\\[2ex]
$\sin^2\theta_{12}$ & \multicolumn{2}{c}{$0.320^{+0.015}_{-0.017}$}
\\[2ex]
$\sin^2\theta_{23}$ &
$0.49^{+0.08}_{-0.05}$ &
$0.53^{+0.05}_{-0.07}$
\\[2ex]
$\sin^2\theta_{13}$ &
$0.026^{+0.003}_{-0.004}$ &
$0.027^{+0.003}_{-0.004}$\\
\hline
\end{tabular}\quad
\begin{tabular}{l}
$m_e(M_Z)=0.486661305\pm{0.000000056}\text{ MeV}$\,, \\[2ex]
$m_{\mu}(M_Z)=102.728989\pm{0.000013}\text{ MeV}$\,,\\[2ex]
$m_{\tau}(M_Z)=1746.28\pm{0.16}\text{ MeV}$\,,
\end{tabular}
\caption{\label{CS_tab:data} The three-flavour oscillation parameters with 1 $\sigma$ 
errors, from Ref.~\cite{Schwetz:2011zk}, for normal hierarchy (NH) and inverted hierarchy 
(IH) (on the left) and the charged-lepton mass 
at $M_Z$ scale (on the right)~\cite{EmmanuelCosta:2011jq}.} 
\end{center}
\end{table}
The effective neutrino mass matrix is given by the type-I seesaw 
formula~\cite{Minkowski:1977sc}
$m_{\nu}=-m_D\,M^{-1}_R\,m^{\top}_D$ to an excellent approximation ($m_D\ll M_R$).
Performing the scan of all $Z_4$ charges for $\phi_1$, $q_3$ and neutrinos one is 
able to determine the shape of the effective neutrino mass matrix. After its 
analysis one concludes that among the six different possibilities (see Ref.~\cite{EmmanuelCosta:2011jq}) 
only two textures are viable: II and II$_{(12)}$ where $\text{II}=P_{12}^{\top}\text{II}_{(12)}\,P_{12}$.
\begin{equation}
\label{CS_eq:textures}
\text{II}=\begin{pmatrix}
  0 & \ast & 0 \\
  \ast & \ast & \ast\\
  0 & \ast & \ast
\end{pmatrix}\,,\qquad \qquad \text{II}_{(12)}=\begin{pmatrix}
  \ast &  \ast & \ast \\
  \ast & 0 & 0\\
  \ast & 0 & \ast
 \end{pmatrix}\,.
\end{equation}

In order to confront the predictions from $M_e$ and $m_\nu$ with the neutrino oscillation 
data at $M_Z$ energy scale one needs to diagonalize both $M_e$ and $m_\nu$ and compute 
the Pontecorvo-Maki-Nakagawa-Sakata~(PMNS) matrix~\cite{Pontecorvo:1957cp,Pontecorvo:1957qd,Maki:1962mu}. The leptonic mixing matrix 
is given by $U_{PMNS}=U^{\top}_{\ell}\,P_{12}\,U_{\nu}$ where $U_{\ell}$ and $U_\nu$ are the 
diagonalizing matrices of charged-leptons and neutrinos respectively. 

In our numerics we have varied all charged-lepton masses and neutrino mass differences 
within their allowed range (see Table~\ref{CS_tab:data}), scanned the mass of the lightest neutrino for 
different magnitudes below 2 eV and computed the other two masses through $\Delta m^2_{ij}\equiv m _i^2-m _j^2$ 
the mass squared difference, using the actual neutrino oscillation 
data~\cite{Schwetz:2011zk}; the free parameters of $M_e$ and $m_\nu$ were also properly 
taken into account (see Ref.~\cite{EmmanuelCosta:2011jq}).

We have considered as additional constraints the effective Majorana mass~\cite{Pascoli:2001by,Pascoli:2002xq,Pascoli:2003ke} $m_{ee}\equiv\sum_{i=1}^3 m_i\,U^{\ast2}_{1i}$; the constraint from Tritium 
$\beta$ decay \cite{Nakamura:2010zzi} $m_{\nu_e}^2\equiv\sum_{i=1}^3\,m_i^2|U_{1i}|^2<\left(2.3\,\text{eV}\right)^2$ at 95\% C.L. 
and constraints on the sum of light neutrino masses from cosmological and astrophysical 
data \cite{Spergel:2006hy} $\mathcal{T}\equiv\sum_{i=1}^3 m_i < 0.68\,\text{eV}$ at 95\% C.L..

We got that texture II is compatible just with normal hierarchy~(NH) while texture 
II$_{(12)}$ is compatible just with inverted hierarchy~(IH). 
\begin{figure}[h]
\centering
\includegraphics[scale=0.35]{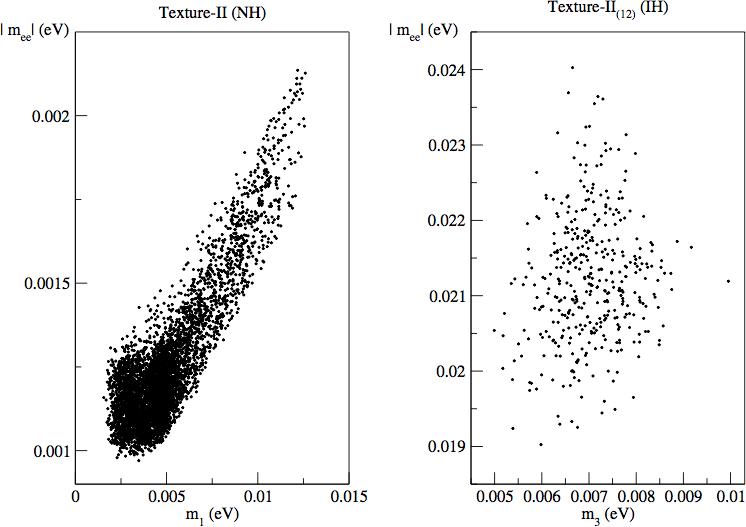}
\caption{\label{CS_fig1} Plot of the effective majorana mass, $|m_{ee}|$, as a function 
of the lightest neutrino mass $m_1$ for the Textures-II (NH) (left) and  $m_3$ for Texture-II$_{(12)}$ (IH) (right).}
\end{figure}
For texture II and normal hierarchy we found that the lightest neutrino mass varies in the range 
$m_1=\left[0.0015, 0.013\right]$ eV while the global deviation is $\varepsilon_{\ell}>0.005$; the effective 
Majorana mass found was $0.00097\,\text{eV} <|m_{ee}| <0.0021\,\text{eV}$.

Concerning texture II$_{(12)}$ where inverted hierarchy applies we found the lightest neutrino mass to be in the 
range $m_3=\left[0.005, 0.010\right]$ eV, the global deviation is $\varepsilon_{\ell}>0.003$ and the $|m_{ee}|$ 
parameter is given by $0.015\,\text{eV}<|m_{ee}|<0.021\,\text{eV}$.

\section{Conclusion}

In this work we showed that it is possible to implement a $Z_4$ flavour symmetry, in 
the context of $SU(5)$ with minimal fermionic content plus three right-handed neutrinos 
and two Higgs quintets, that leads to quark mass matrices in the NNI form. We have studied 
the implications of this $SU(5)\times Z_4$ symmetry on the leptonic sector and found that, 
among the six possible textures for the effective neutrino mass matrix, only two are 
phenomenologically viable and it is possible to distinguish them by the light neutrino 
mass spectrum hierarchy.

\section*{Acknowledgments}
I would like to thank the organizers of FLASY12 - Workshop on Flavor Symmetries  for the opportunity to participate and 
present this work. I would like to thank David Emmanuel-Costa for the encouragement to do 
the presentation and the proceeding. This work was partially supported by Funda\c{c}\~ao para a Ci\^encia e
a Tecnologia (FCT, Portugal) through the contract SFRH/BD/61623/2009 and the projects CERN/FP/123580/2011, 
PTDC/FIS/098188/2008 and CFTP-FCT Unit 777 which are partially funded through POCTI (FEDER). 

\bibliography{Simoes}
\bibliographystyle{apsrev4-1}

%% file: Papers/spinrath.tex

%
%
%
%
%
%

\chapter[Two Approaches for Flavour Models with Large $\theta_{13}$ (Spinrath)]{Two Approaches for Flavour Models with Large $\theta_{13}$}
\vspace{-2em}
\paragraph{M. Spinrath}
\paragraph{Abstract}
The recent experimental confirmation of a large reactor mixing angle
makes a careful revision of flavour models necessary. Many models are
actually ruled out by this observation and the popular bimaximal and
tri-bimaximal mixing patterns are challenged. We will present here two possible
scenarios with a large $\theta_{13}$. First, corrections from
the charged lepton sector might modify the (tri-)bimaximal mixing pattern
to generate an effectively large $\theta_{13}$ which is well motivated
in GUT frameworks. Second, (tri-)bimaximal mixing itself might need to
be modified, e.g.\ by a modified vacuum structure of the family symmetry
breaking flavon fields. We present examples for both possibilities and
show implications for the CP violation in the lepton sector.

\section{Introduction}

In the last year we have seen a tremendous progress in the experimental
determination of the neutrino mixing parameters due to the precise determination
of the last missing neutrino mixing angle $\theta_{13}$ by the Daya Bay \cite{An:2012eh}
and RENO \cite{Ahn:2012nd} collaborations. A recent global fit \cite{Tortola:2012te} gives
$\sin^2 \theta_{13} = 0.026^{+0.003}_{-0.004}$ which deviates from zero by more
than $6\sigma$. First evidence for a non-vanishing reactor angle was already
announced last year by the T2K collaboration \cite{Abe:2011sj}. This result
was in tension with the very popular bimaximal \cite{Vissani:1997pa, Barger:1998ta,
Baltz:1998ey, Georgi:1998bf, Stancu:1999ct} and tri-bimaximal mixing schemes
\cite{Harrison:2002er, Harrison:2002kp, Xing:2002sw, He:2003rm, Wolfenstein:1978uw}
which both predict a vanishing $\theta_{13}$. Subsequently many attempts were
made to explain this large value and we will present here two of them. In the first
approach \cite{Marzocca:2011dh} we start with bimaximal or tri-bimaximal mixing in the neutrino sector and
correct the wrong prediction for $\theta_{13}$ with sizeable corrections from the
charged lepton sector in the context of an $SU(5)$ Grand Unified Theory (GUT) and
in the second approach \cite{Antusch:2011ic} we give a model for an alternative mixing scheme called
trimaximal mixing \cite{Haba:2006dz, He:2006qd, Grimus:2008tt, Ishimori:2010fs,
Shimizu:2011xg, He:2011gb, Lam:2006wm, Albright:2008rp, Albright:2010ap}.

\section{Possibility I: Charged Lepton Corrections}

\begin{table}
\centering
\begin{tabular}{lc} \hline
$\{ \alpha, \beta, \beta', \gamma \}$ & $\sin \theta_{13}$  \\ \hline
$\{ -, -1/2, 6, 6 \}$ &  $ 0.164 \pm 0.013 $  \\
$\{-3/2, -3, -3, -3\} $ & $ 0.164 \pm 0.007 $ \\
$\{-18, 9/2, 9/2, 9/2\} $ & $ 0.149 \pm 0.003 $\\
\hline
\end{tabular}
\caption{The three combinations of Clebsch--Gordan
coefficients taken from \cite{Marzocca:2011dh} which
survive after the recent global fit results from
\cite{Tortola:2012te}. \label{MSTab:CG_combinations}}
\end{table}

Charged lepton corrections to mixing schemes
like tri-bimaximal mixing are a natural feature in GUT flavour
models. Additionally, in this class of models the Yukawa couplings
for different fermion species are usually related to each other.
We focus here on $SU(5)$ GUTs where the charged lepton masses and down-type
quark masses are related. We assume bimaximal or tri-bimaximal
mixing in the neutrino sector and the following
structure of the Yukawa matrices of the charged leptons and
down-type quarks
\begin{equation}
\hat{\lambda}^D_{[12]} = 
\begin{pmatrix}
a & b' \\ b & c
\end{pmatrix}
\qquad
\hat{\lambda}^E_{[12]} = 
\begin{pmatrix}
\alpha a & \beta b \\ \beta' b' & \gamma c
\end{pmatrix} \;,
\end{equation}
where $a$, $b$, $b'$, $c$ are free parameters and $\alpha$,
$\beta$, $\beta'$, $\gamma$ are group theoretical Clebsch--Gordan
coefficients. A list of them including non-standard ones from non-renormalisable
operators is given in \cite{Antusch:2009gu}.
We assume a hierarchical Yukawa structure and hence
mixing with the third generation can be neglected and
we discuss only the 1-2 block. The reactor
mixing angle (for $\theta_{13}^\nu = 0$ and $\theta_{23}^\nu = \pi/4$) is then approximately
\begin{equation}
\sin \theta_{13} \approx \sin \theta_{12}^e \sin \theta_{23}^\nu \approx \frac{1}{\sqrt{2}} \frac{\beta'}{\gamma} \frac{b'}{c} \;.
\end{equation}
Given a set of Clebsch--Gordan coefficients we can determine the
parameters from the fermion masses and can predict the reactor
neutrino mixing angle. In \cite{Marzocca:2011dh} we have done
this for some well motivated cases and from the cases discussed
there only three remain according to the recent fit results
from \cite{Tortola:2012te} which are collected in
Tab.\ \ref{MSTab:CG_combinations}.

\begin{figure}
 \centering
 \includegraphics[scale=0.68]{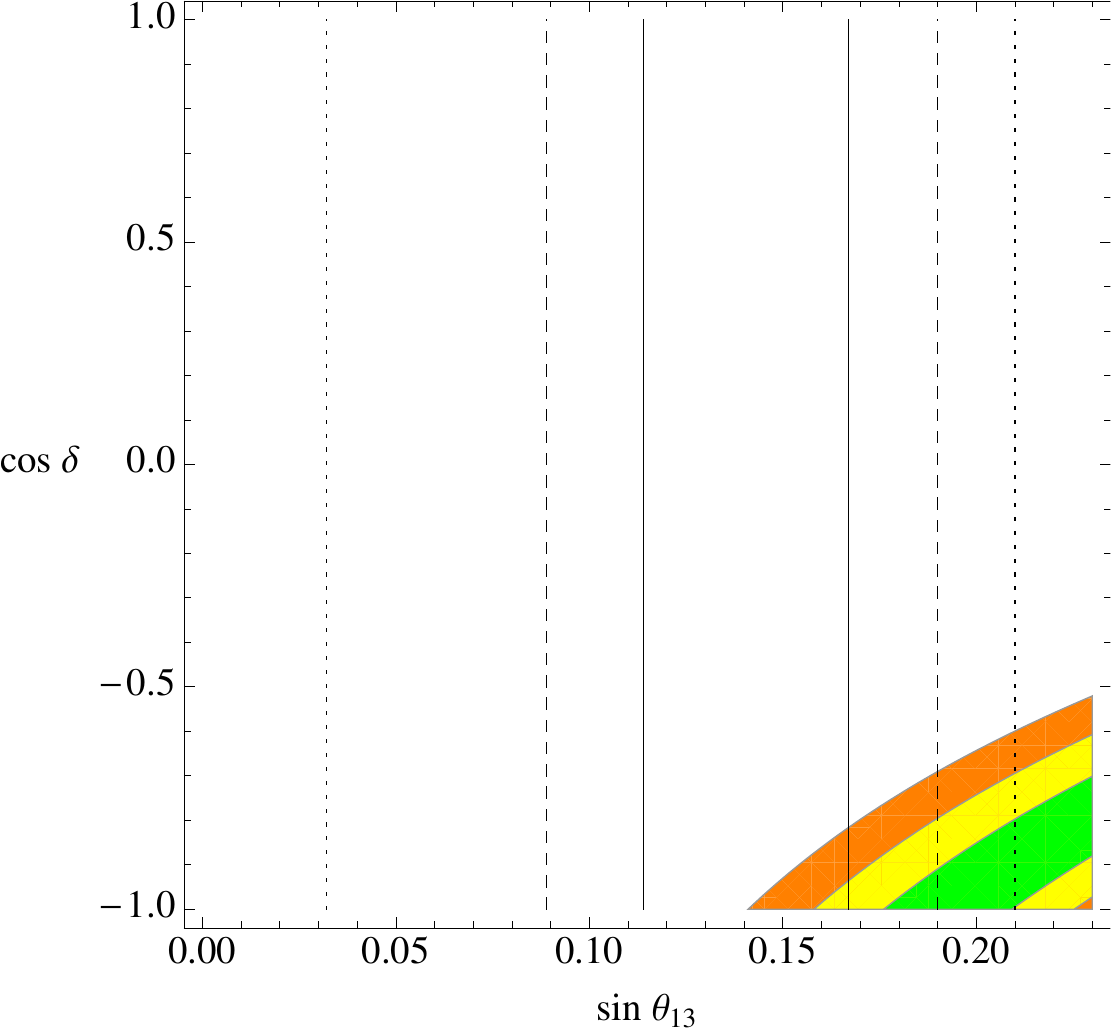} \hspace{1ex}
 \includegraphics[scale=0.68]{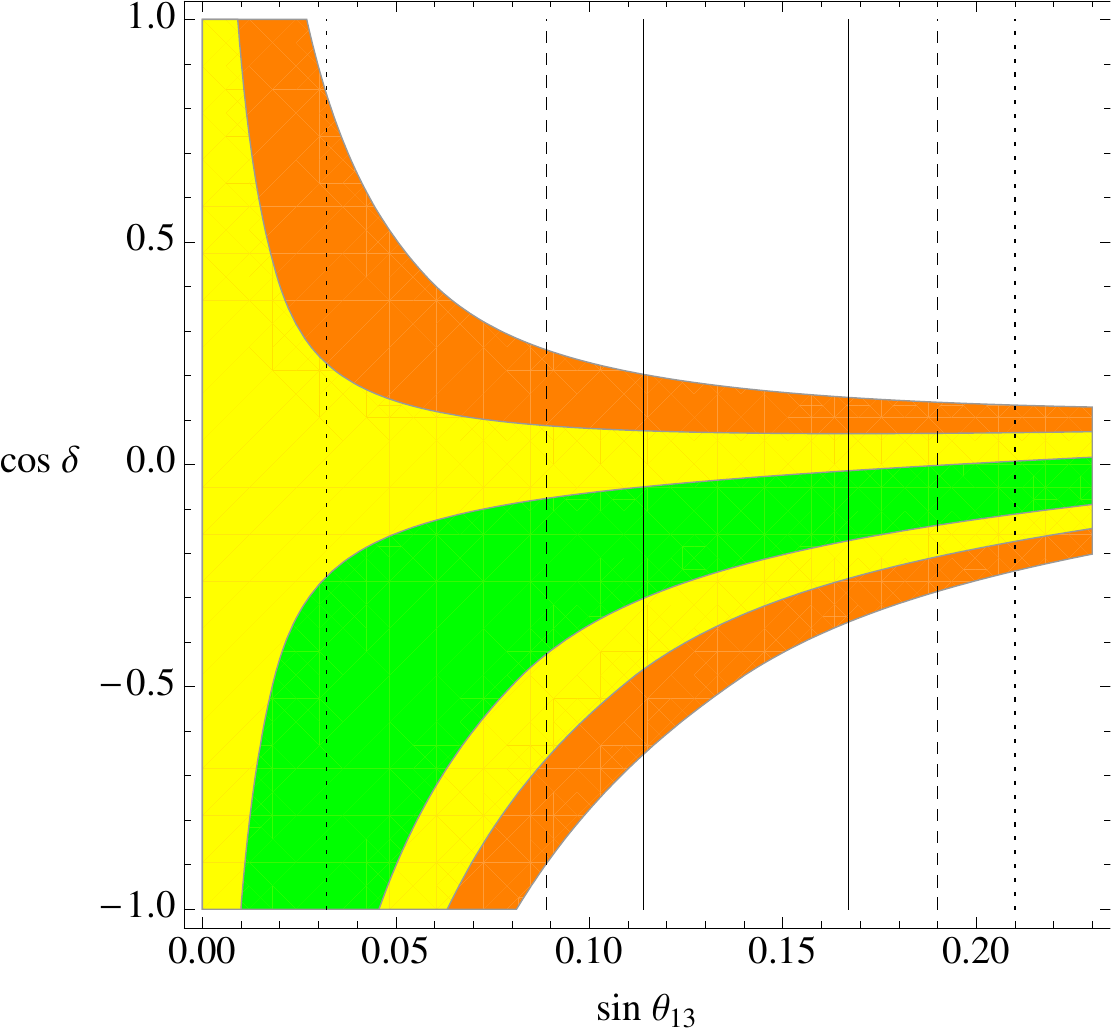} \\
 \includegraphics[scale=0.68]{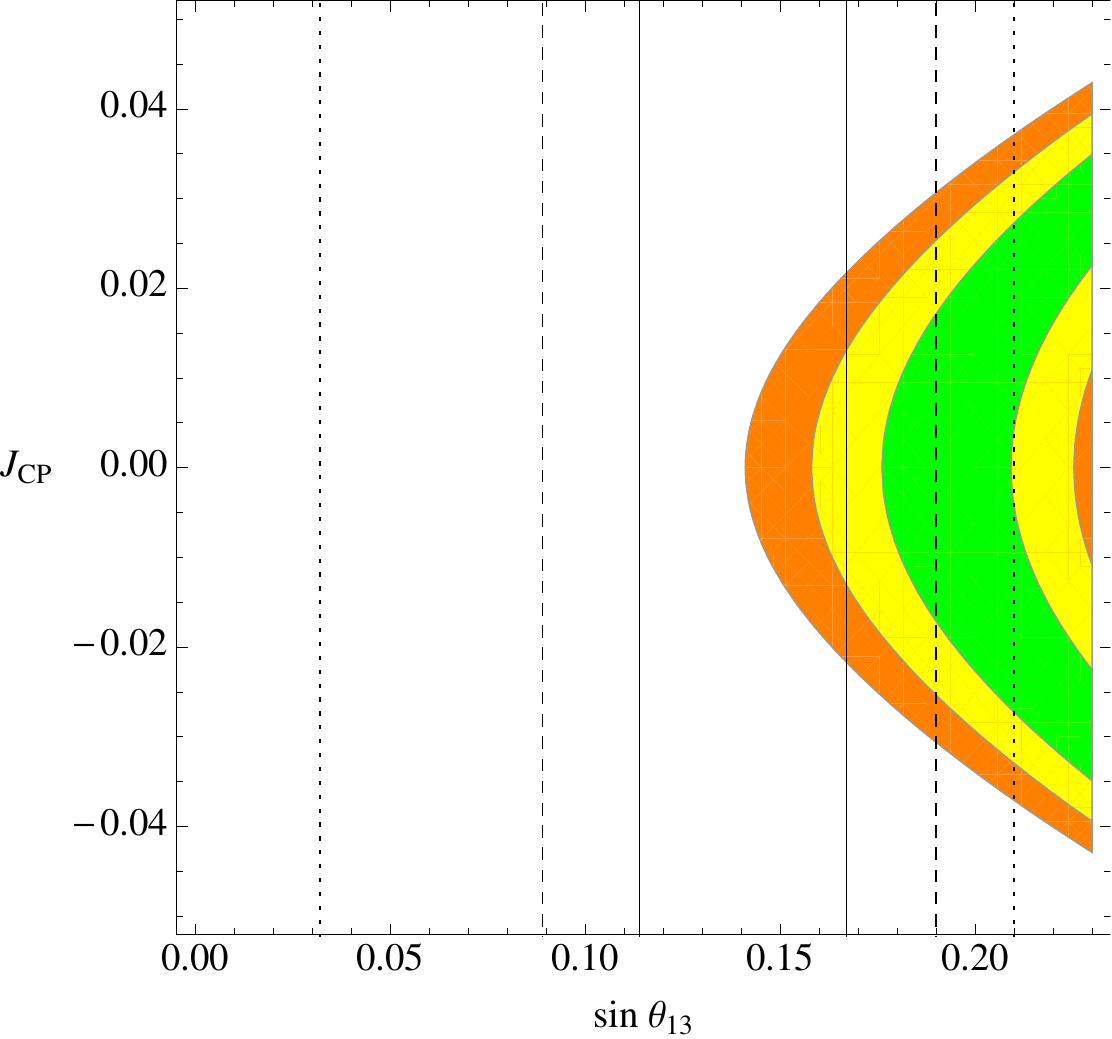} \hspace{1ex}
 \includegraphics[scale=0.68]{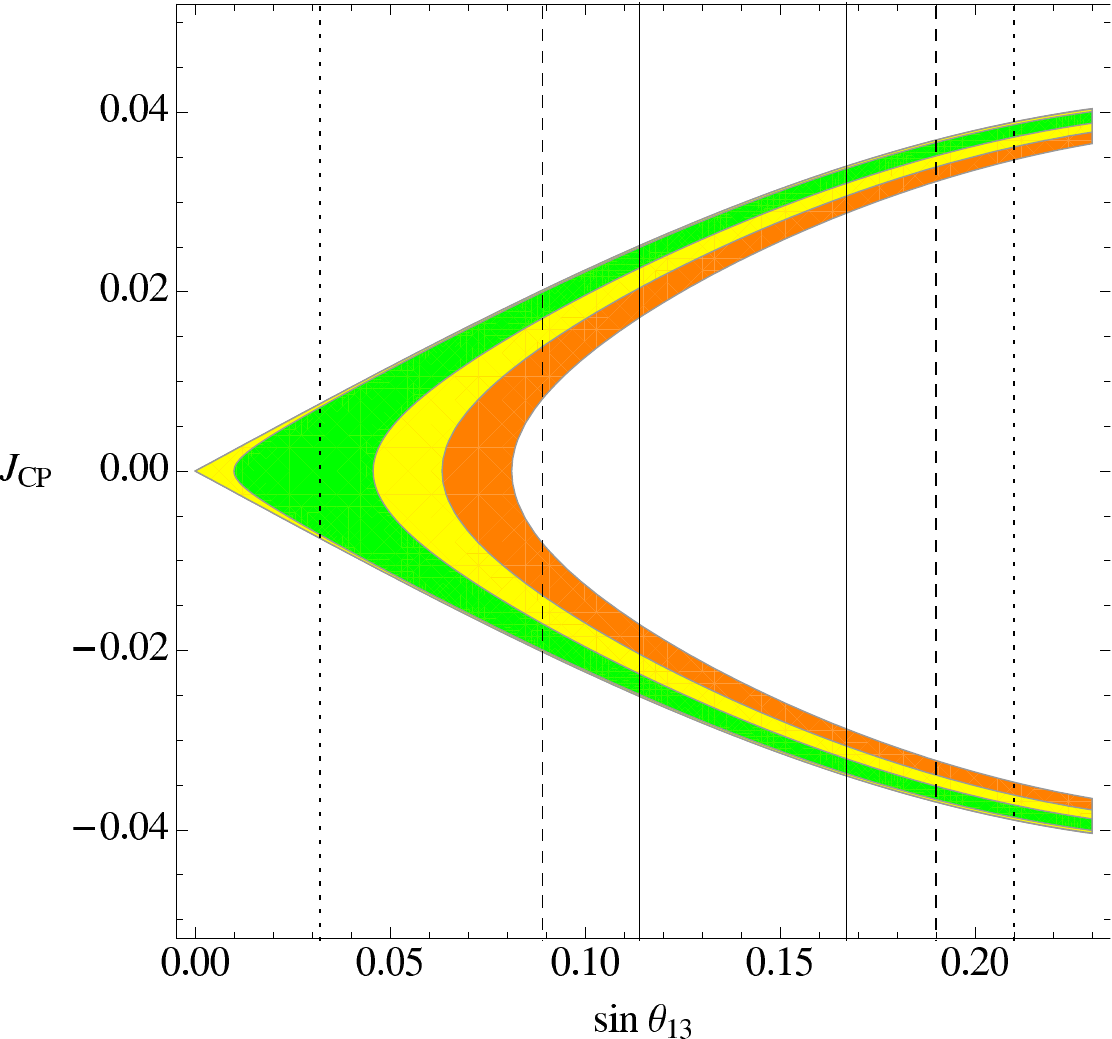}
 \caption{The cosine of the Dirac CP phase $\delta$ (top)
   and the rephasing invariant $J_{\text{CP}}$ (bottom) as a function
   of $\sin \theta_{13}$ in the cases of bimaximal (left) and
   tri-bimaximal (right) mixing from the neutrino sector. The green,
   yellow, orange regions correspond to the 1, 2, 3$\sigma$
   allowed ranges of $\sin^2 \theta_{12}$. The vertical straight,
   dashed, dotted lines show the  1, 2, 3$\sigma$ allowed ranges
   of $\sin \theta_{13}$. 
   The values of $\sin^2 \theta_{12}$ and  $\sin \theta_{13}$ 
   are from \cite{Fogli:2011qn} and the plots from
   \cite{Marzocca:2011dh} (see text for details).
   \label{MSFig:CP}}
\end{figure}

In this setup it is furthermore possible to constrain the
amount of CP violation in the lepton sector because the physical
mixing angles $\theta_{12}$, $\theta_{13}$ and the leptonic Dirac
CP phase $\delta$ are related to each other via
\begin{align}
\text{Bimaximal:} \quad\sin^2 \theta_{12} &\approx \frac{1}{2} + \sin \theta_{13} \cos \delta \;,\\
\text{Tri-Bimaximal:} \quad \sin^2 \theta_{12} &\approx \frac{1}{3} + \frac{2 \sqrt{2}}{3} \sin \theta_{13} \cos \delta \;,
\end{align}
for references and details see \cite{Marzocca:2011dh}.
The mixing angles are experimentally determined such that
we get a constrain on the phase $\delta$ and the Jarlskog
invariant $J_{\text{CP}}$, see Fig.\ \ref{MSFig:CP}.
In the bimaximal case we need $\delta$ to be close to $180^\circ$
($\cos \delta \approx -1$) to get the right result while
in the tri-bimaximal case $\delta$ should be close to $90^\circ$.
In other words for the
tri-bimaximal case we expect a large amount of CP violation
(a large $J_{\text{CP}}$) while for the bimaximal case it
should be rather small.

\section{Possibility II: Trimaximal Mixing}

The second possibility is based on the framework of sequential
dominance \cite{King:1998jw, King:1999cm, King:1999mb,
King:2002nf, Antusch:2010tf}. The right-handed neutrino mass
matrix $M_R$ is set to be diagonal and the columns of the neutrino
Yukawa matrix are labelled $A$, $B$ and $C$,
\begin{equation}
Y_\nu = (A,B,C) \quad \text{and} \quad M_R = \text{diag}(M_A,M_B,M_C) \;.
\end{equation}
The effective low energy neutrino mass matrix is then given as
\begin{equation}
M_\nu = \frac{v^2 A A^T}{M_A} + \frac{v^2 B B^T}{M_B} +\frac{v^2 C C^T}{M_C} \;,
\end{equation}
and sequential dominance assumes a hierarchy $A^2/M_A \gg B^2/M_B \gg C^2/M_C$.
In a minimal
setup with only two right-handed neutrinos (which
corresponds to neglecting terms of the order of $C^2/M_C$)
 tri-bimaximal mixing
is governed, e.g.\ by $A \propto (0,1,-1)$ and $B \propto (1,1,1)$
which is dubbed constrained
sequential dominance \cite{King:2005bj} (CSD).

If we replace the $B$ column of tri-bimaximal mixing
with $B \propto (1,0,2)$ we get a
different mixing pattern, a special kind of trimaximal mixing
\cite{Lam:2006wm, Albright:2008rp, Albright:2010ap}
which was dubbed CSD2 \cite{Antusch:2011ic}
to distinguish it from the original CSD.

\begin{figure}
\centering
\includegraphics[scale=0.5]{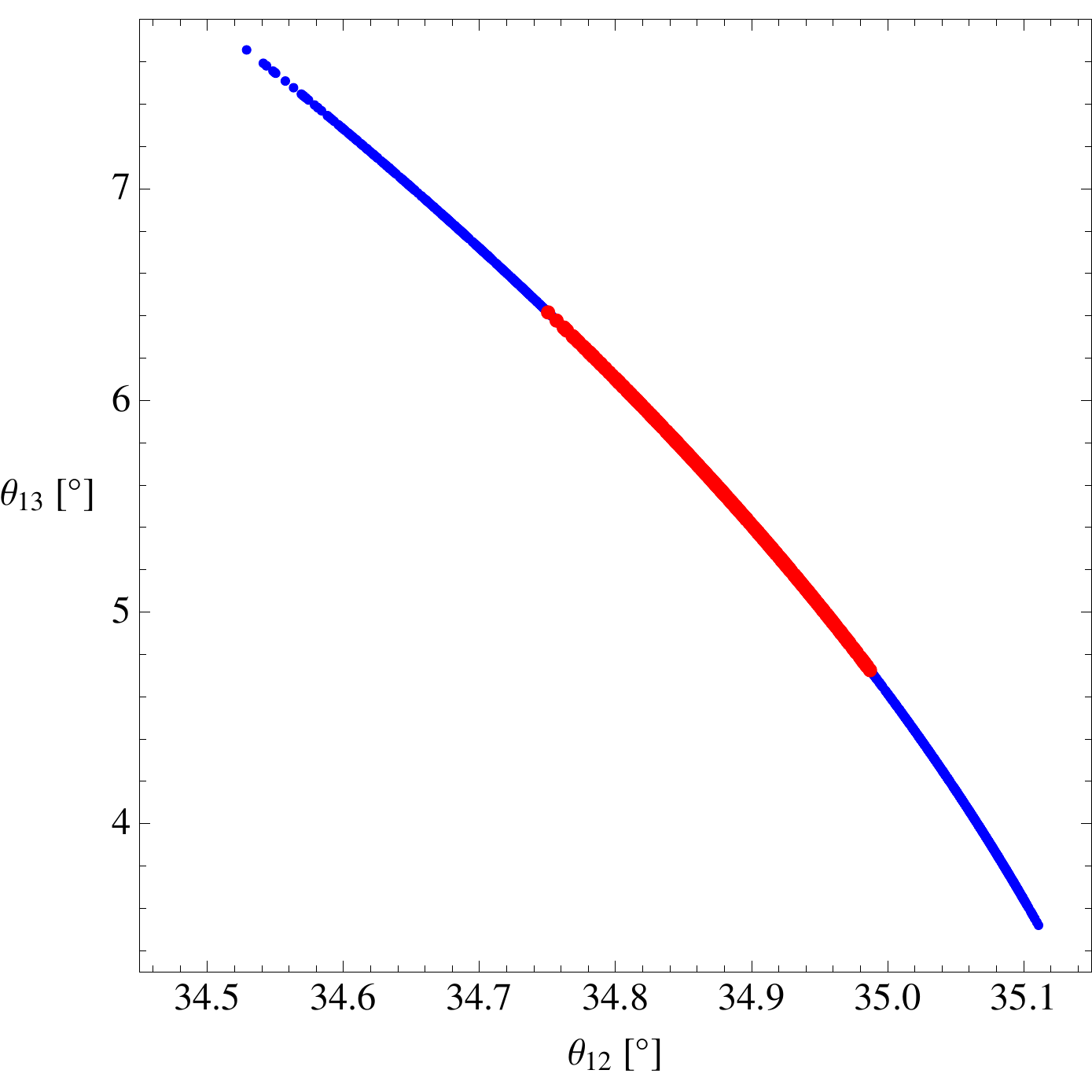} \hspace{3ex}
\includegraphics[scale=0.5]{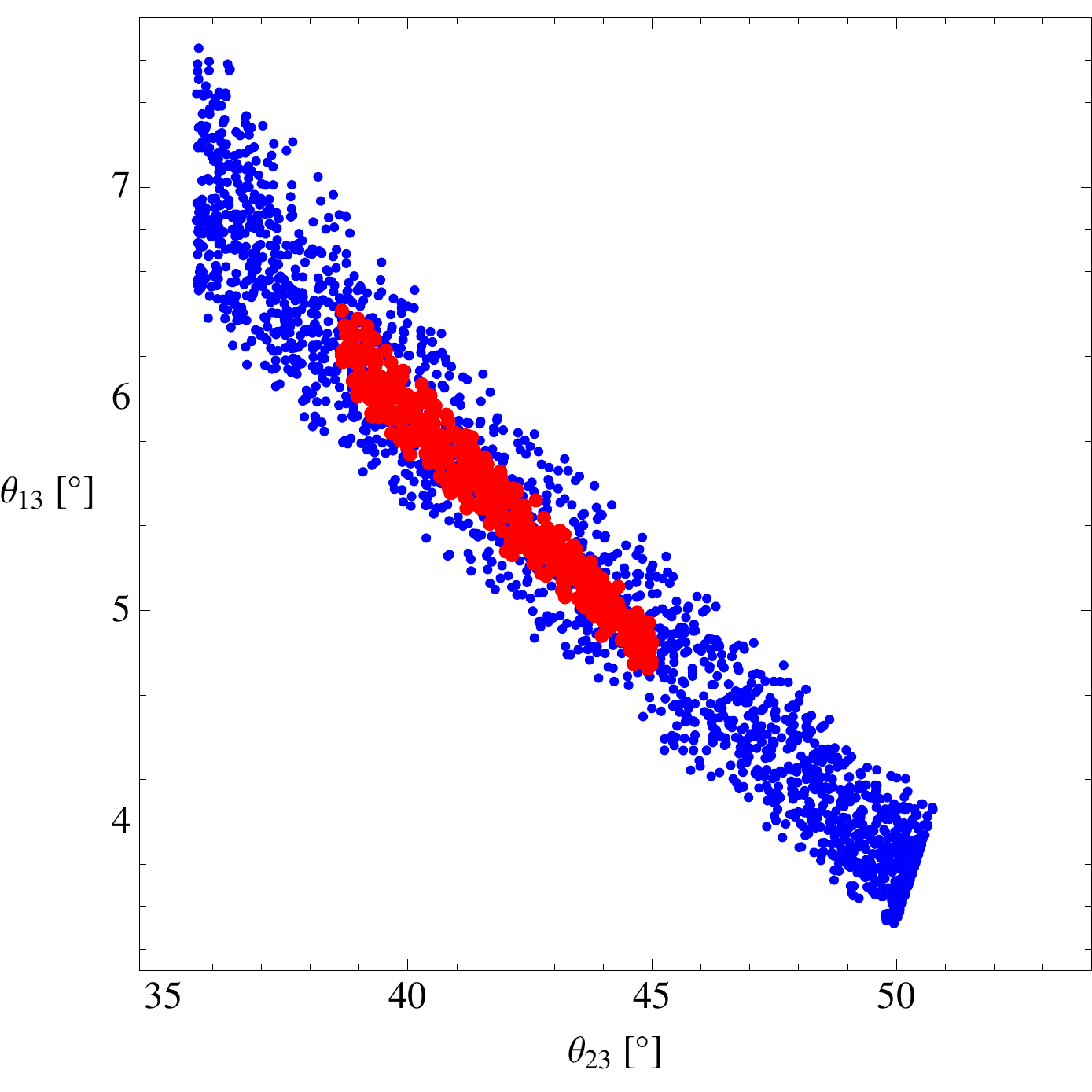} \\
\includegraphics[scale=0.5]{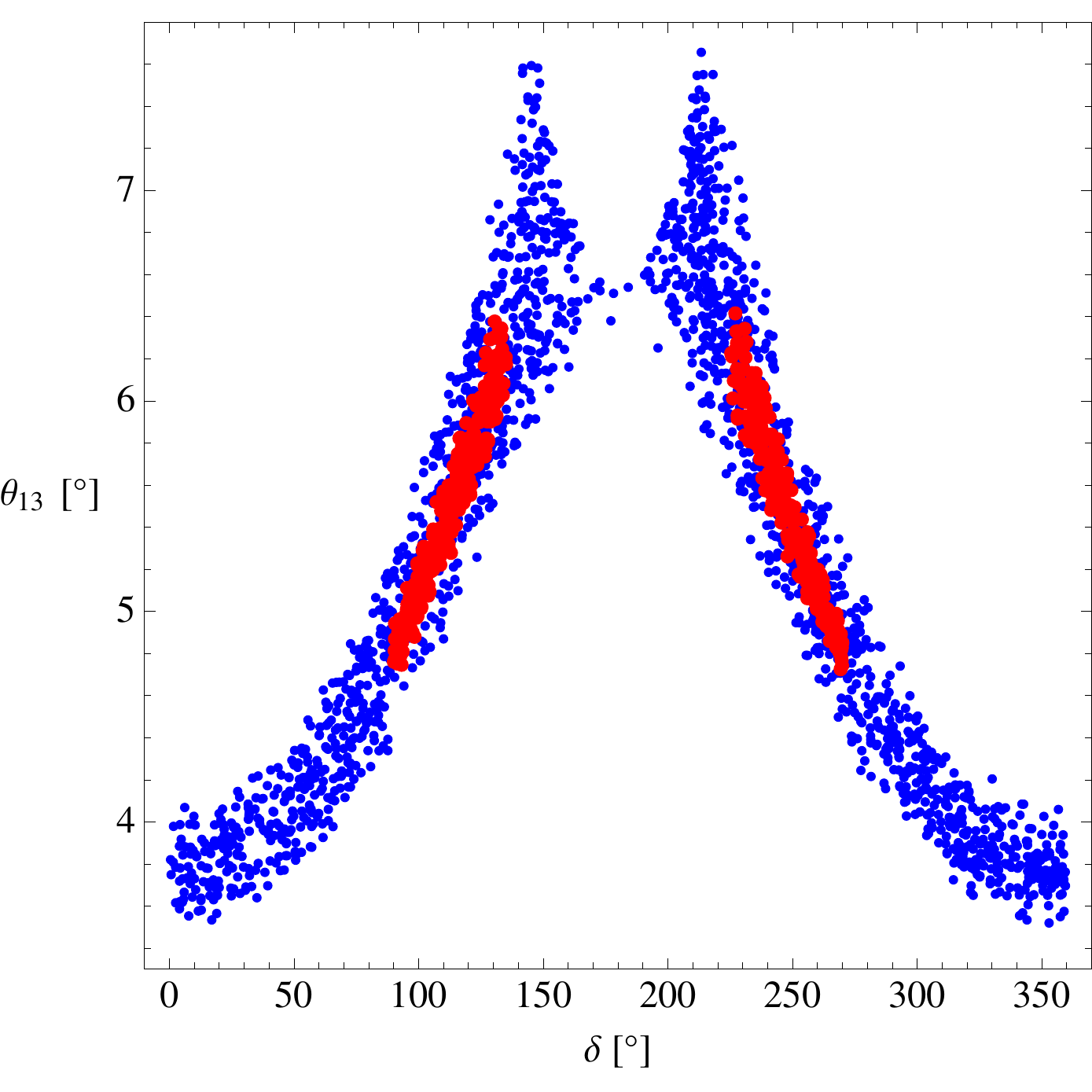} \hspace{3ex}
\includegraphics[scale=0.5]{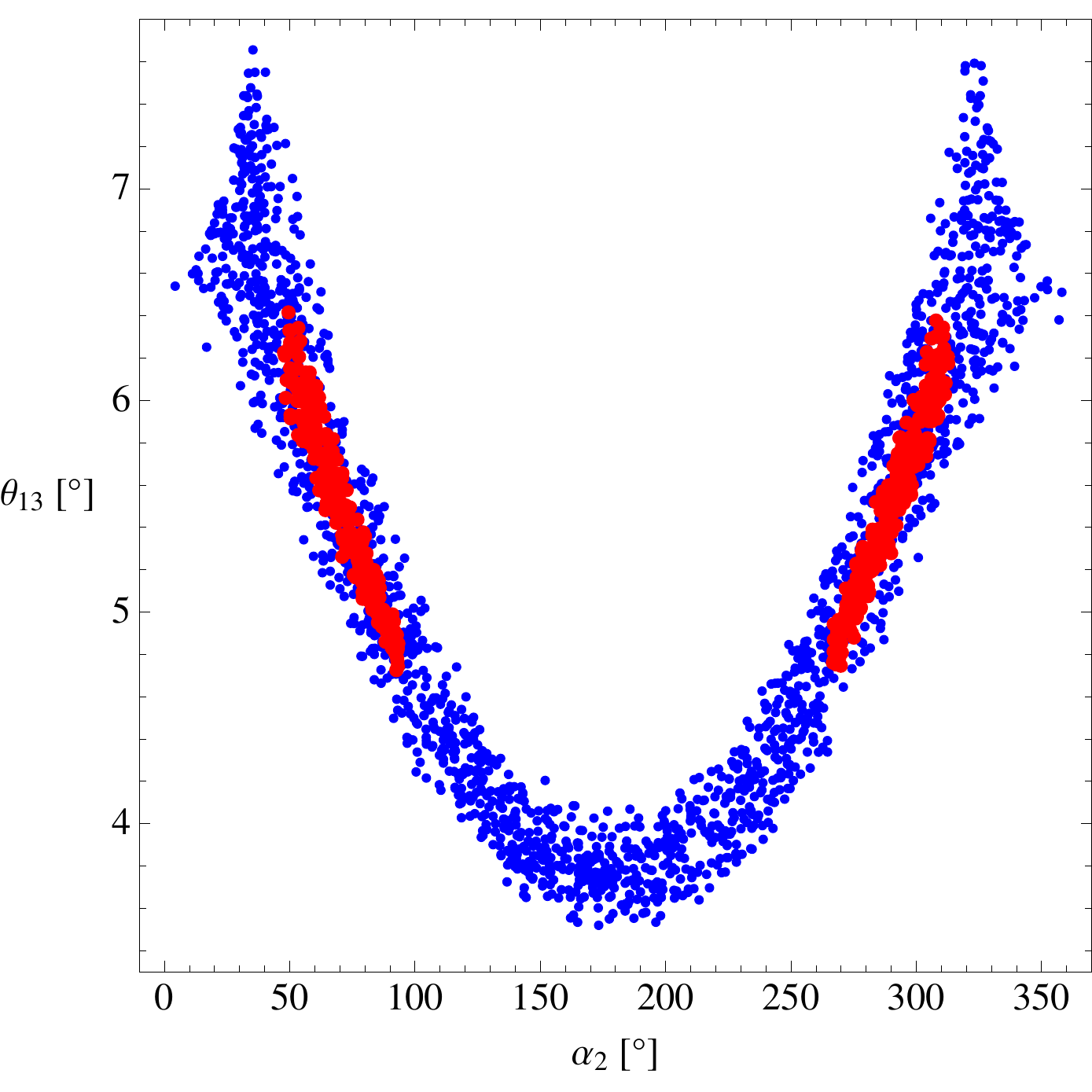}
\caption{The correlations between $\theta_{13}$ and the other physical mixing parameters in  CSD2 for the  $(1,0,2)^T$ alignment. Regions compatible with the 1$\sigma$ (3$\sigma$) ranges of the atmospheric   and solar neutrino mass squared differences and mixing angles, taken from  \cite{Fogli:2011qn}, are depicted by the red (blue) points. The plots are taken from \cite{Antusch:2011ic}.
\label{MSFig:Correlations}}
\end{figure}

To get this column in the Yukawa matrix we assume an
underlying model of flavour with triplet representations
such that the columns are given by the vacuum expectation
values of family symmetry breaking flavon fields.
With standard vacuum alignment tools one can
get two sets of flavons with the alignments
\begin{align}
\langle \phi^e_{1}\rangle &\propto \begin{pmatrix} 1 \\ 0 \\  0 \end{pmatrix} , \quad 
\langle \phi^e_{2}\rangle \propto \begin{pmatrix} 0 \\ 1 \\ 0 \end{pmatrix}  , \quad  
\langle \phi^e_{3}\rangle \propto \begin{pmatrix} 0 \\ 0 \\ 1 \end{pmatrix}  ,\\
\langle \phi^{\nu}_{1}\rangle &\propto \begin{pmatrix} 0 \\ 1 \\ -1 \end{pmatrix} , \quad 
\langle \phi^{\nu}_{2}\rangle \propto \begin{pmatrix} 1 \\ 1 \\ 1 \end{pmatrix}  , \quad 
\langle \phi^{\nu}_{3}\rangle \propto \begin{pmatrix} -2 \\ 1 \\ 1 \end{pmatrix} .
\end{align}
The new $(1,0,2)$ alignment can then be derived
straightforwardly by the superpotential
\begin{equation}
\mathcal{W} = O_1 (\phi_2^e \cdot \phi_{102}) + O_2 (\phi_3^\nu \cdot \phi_{102}) \;,
\end{equation}
where $O_1$ and $O_2$ are driving fields, singlets under the family
symmetry. Solving the $F$-term equations of these two fields ($F_{O_i} = 0$) enforces
$\phi_{102}$ to be orthogonal to $\phi_2^e$ and $\phi_3^\nu$
and hence $\phi_{102} \propto (1,0,2)$.
For a full $A_4$ model including a choice of additional
discrete shaping symmetries see the original CSD2 paper
\cite{Antusch:2011ic}.

The effective low energy neutrino mass matrix in this setup
\begin{equation}
M_\nu = m_a \begin{pmatrix}
 \eta & 0 & 2 \eta \\
 0 & 1 & -1 \\
 2 \eta & -1 & 1 + 4 \eta
\end{pmatrix} \;, \;\; \eta = \epsilon \text{ e}^{\text{i} \, \alpha} \;,
\end{equation}
 can be described in terms of only
three parameters, the neutrino mass scale
$m_a$, the modulus $\epsilon$ and the relative phase
$\alpha$. For a full list of relations between these
three parameters and the observables in the neutrino
sector see \cite{Antusch:2011ic}. We only want to
highlight here two of them.
First it is interesting to note that in this setup the reactor mixing angle
is related to the ratio of the neutrino masses
\begin{equation}
\theta_{13} = \frac{\sqrt{2}}{3} \frac{m_2^\nu}{m_3^\nu} \;,
\end{equation}
which is nevertheless too small (of the order of $6^\circ$
instead of $9^\circ$). This can be lifted in the context of
a GUT embedding where additional corrections from the
charged lepton sector are present.
Second the leptonic Dirac CP violation is directly
related to the phase difference $\alpha$
\begin{equation}
\delta = \pi + \alpha - \epsilon\, \frac{5}{2}\, \sin\alpha\ .\\
\end{equation}
As we can see from Fig.\ \ref{MSFig:Correlations} certain
values for $\alpha$ are preferred which
could be accomodated in the context of CP violation from
discrete symmetries \cite{Antusch:2011sx}.
Note also that we can have a significant deviation
from maximal atmospheric mixing in the right direction
and we find a preferred range for the physical Majorana phase,
see Fig.\ \ref{MSFig:Correlations}.

\section{Summary and Conclusions}

The confirmation of a non-vanishing, sizeable reactor
mixing angle by the Daya Bay \cite{An:2012eh} and RENO
\cite{Ahn:2012nd} experiments have ruled out many flavour
models and disfavoured the popular bimaximal and
tri-bimaximal mixing patterns in its exact forms.
This raises the question in which direction to go in the future.

We have presented here two possibilities. The first one
keeps a bimaximal or tri-bimaximal mixing pattern in the
neutrino sector alone, which is well motivated by symmetries,
and then adds corrections from the charged lepton sector,
which are well motivated in GUTs, to get the right value
for $\theta_{13}$ \cite{Marzocca:2011dh}. The second possibility, CSD2,
is based on a flavour model leading to a different mixing pattern,
dubbed trimaximal, which gives a sizeable $\theta_{13}$ and a significant
deviation from maximal atmospheric mixing \cite{Antusch:2011ic}.

Although neutrino physics has made a huge progress in the
last year there are still many open questions left. What is the mass
hierarchy and mass scale of the neutrinos? Are they Majorana particles? How
large is the amount of CP violation in the lepton sector?
How close is $\theta_{23}$ to maximal?
The prospects for answering at least some of this questions in the
near future are good and every answer will again rule out a lot
of flavour models. This will focus the attention on the surviving
models which can then be studied in greater details which
will hopefully shed some light on the origin of flavour and
CP violation.

\bibliographystyle{apsrev4-1}
%


%% file: Papers/tanimoto.tex

%
%
%
%
%
%

\chapter[Relating neutrino mixing angles to neutrino masses (Tanimoto)]{Relating neutrino mixing angles to neutrino masses}
\vspace{-2em}
\paragraph{M. Tanimoto}
\paragraph{Abstract}
The observation of $\theta_{13}$ suggests us that
we should not persist in the paradigm of the tri-bimaximal mixing.
We present the $A_4$ model with   $1'$  and $1''$ flavons, which 
predicts $\sin\theta_{13}\simeq 0.15$.
There is another aspect of the large flavor mixing of neutrions.
We  show  that our  minimal texture   describes  all known
empirical values. The magnitude of $\theta_{13}$ is predicted to be  in the
middle of the range of the experimental data.

\section{Introduction}

The flavor symmetry is expected to explain  the mass spectrum  and
the mixing matrix of both  quarks and leptons. 
Especially, the non-Abelian discrete symmetry~
\cite{Ishimori:2010au,Altarelli:2010gt}
has been studied  intensively in the lepton sectors.
Actually, the three flavor analyses of the neutrino mixing
\cite{Schwetz:2008er,Fogli:2008jx}
have suggested the tri-bimaximal mixing pattern of leptons 
\cite{Harrison:2002er}.
This simple mixing is at first understood 
based on the non-Abelian finite group 
$A_4$~\cite{Ma:2001dn}$^-$\cite{Ma:2005mw}.
The tri-bimaximal mixing gives the vanishing $\theta_{13}$.
However, Daya Bay experiment reported  non-vanishing $\theta_{13}$
\cite{An:2012eh}, and then the RENO experiment
 also presented the same magnitude \cite{Ahn:2012nd}.
 Now, we do not need to persist in 
the paradigm of the tri-bimaximal mixing.
The tri-bimaximal structure is  broken.

It should be emphasized that the $A_4$ flavor symmetry does not necessarily
give the tri-bimaximal mixing at the leading order even 
if the relevant alignments 
of the vacuum expectation values (VEVs) are realized.
Certainly, 
the $A_4$ symmetry can give the mass matrix with
  $(1,3)$ or  $(1,2)$ off diagonal matrices  
  at the leading order if $1'$ 
and $1''$ flavons exist~\cite{Brahmachari:2008fn}:
\begin{equation}
\begin{pmatrix}
0 & 0 & 1 \\
0 & 1 & 0 \\
1 & 0 & 0 
\end{pmatrix} \quad {\rm for} \,\, 1', 
\quad\quad
\begin{pmatrix}
0 & 1 & 0 \\
1 & 0 & 0 \\
0 & 0 & 1 
\end{pmatrix} \quad {\rm for} \,\, 1''. 
\label{dterms}
\end{equation}
The tri-bimaximal mixing is broken at the leading order in such a case
\cite{Haba:2006dz,Ishimori:2010fs,Grimus:2008tt}.

As a concrete realization of such a pattern,
 we discuss an $A_4$ flavor model \cite{Shimizu:2011xg},  
which is a modified version of the Altarelli and Feruglio model
~\cite{Altarelli:2005yp,Altarelli:2005yx}.
We find that  $\theta_{12}$ and  $\theta_{23}$ are not so different compared 
with the tri-bimaximal mixing, but  $\sin \theta_{13}$ is expected to be 
around $0.15$ if  the neutrino mass spectrum  is    normal hierarchical.  
Our model is completely consistent with the data by 
Daya Bay  and   RENO experiments.


 On the other hand, there is another aspect of large flavor mixing.
 The  neutrino mass ratios may suggest the large flavor mixing angles.
 the large mixing angles could be  realized in the  specific textures of 
the neutrino mass matrix.
 We show a simple texture 
\cite{FTY93,FTY03,Fukugita:2012jr}.

\section{Neutrino mass matrix breaking the  tri-bimaximal mixing}
Let us discuss the breaking the  tri-bimaximal mixing.
The neutrino mass matrix which leads to  the tri-bimaximal mixing of
flavor is given by 
\begin{equation}
M_{\rm TBM}=\frac{m_1+m_3}{2}
\begin{pmatrix}
1 & 0 & 0 \\
0 & 1 & 0 \\
0 & 0 & 1 
\end{pmatrix}+ \frac{m_2-m_1}{3}
\begin{pmatrix}
1 & 1 & 1 \\
1 & 1 & 1 \\
1 & 1 & 1
\end{pmatrix}+\frac{m_1-m_3}{2}
\begin{pmatrix}
1 & 0 & 0 \\
0 & 0 & 1 \\
0 & 1 & 0 
\end{pmatrix}.
\label{tribimass}
\end{equation}
Here $m_1$, $m_2$, and $m_3$ are  neutrino masses, while
the charged lepton mass matrix is diagonal. Certainly,
 the $A_4$ symmetry can realize the mass matrix 
in Eq.~(\ref{tribimass}).
Additional   matrices in Eq.(\ref{dterms}) are added 
at the leading
 order  in the flavor model with the non-Abelian discrete symmetry.
For example, such extra terms appear in the $A_4$ flavor model 
if $1'$ and $1''$ flavons couple to the $A_4$ triplet neutrinos such as 
$ 3\times 3 \times 1'$ and $3\times 3 \times 1''$ as discussed later.

It is noticed that the additional  two terms in Eq.~(\ref{dterms}) 
are not independent  from each other.
Thus we can consider the neutrino mass matrix, which 
breaks the tri-bimaximal mixing,
\begin{equation}
M_{\nu}=a
\begin{pmatrix}
1 & 0 & 0 \\
0 & 1 & 0 \\
0 & 0 & 1 
\end{pmatrix}+b
\begin{pmatrix}
1 & 1 & 1 \\
1 & 1 & 1 \\
1 & 1 & 1
\end{pmatrix}+c
\begin{pmatrix}
1 & 0 & 0 \\
0 & 0 & 1 \\
0 & 1 & 0 
\end{pmatrix} +d
\begin{pmatrix}
0 & 0 & 1 \\
0 & 1 & 0 \\
1 & 0 & 0 
\end{pmatrix},
\label{generalmass}
\end{equation}
without loss of generality.
Here the parameters $a$, $b$, $c$ and $d$ are arbitrary in general.
The neutrino masses $m_1$, $m_2$ and $m_3$ are given in terms of these four
parameters.

By factoring out the tri-bimaximal mixing matrix $V_\text{tri-bi}$ 
\begin{equation}
V_\text{tri-bi}=
\begin{pmatrix}
\frac{2}{\sqrt{6}} & \frac{1}{\sqrt{3}} & 0 \\
-\frac{1}{\sqrt{6}} & \frac{1}{\sqrt{3}} & -\frac{1}{\sqrt{2}} \\
-\frac{1}{\sqrt{6}} & \frac{1}{\sqrt{3}} & \frac{1}{\sqrt{2}} \\
\end{pmatrix},
\end{equation}
the left-handed neutrino mass matrix~(\ref{generalmass}) is written as 
\begin{equation}
M_\nu =V_\text{tri-bi}
\begin{pmatrix}
a+c-\frac{d}{2} & 0 & \frac{\sqrt{3}}{2}d \\
0 & a+3b+c+d & 0 \\
\frac{\sqrt{3}}{2}d & 0 & a-c+\frac{d}{2}
\end{pmatrix}V_\text{tri-bi}^T\ .
\end{equation}
At first, suppose the parameters $a, b, c, d$ to be real
in order to see the effect of the non-vanishing $d$ clearly.
Then,  we have the mass eigenvalues of the left-handed neutrinos as
\begin{equation}
a+\sqrt{c^2+d^2-cd}, \, \qquad   
a+3b+c+d, \qquad 
a-\sqrt{c^2+d^2-cd} .
\label{tanimoto_masses}
\end{equation}


As the charged lepton mass matrix is diagonal, the  mixing matrix 
$U_\text{MNS}$ is
\begin{equation}
U_\text{MNS}=V_\text{tri-bi}
\begin{pmatrix}
\cos \theta & 0 & \sin \theta \\
0 & 1 & 0 \\
-\sin \theta & 0 & \cos \theta 
\end{pmatrix},  \quad\qquad
\tan 2\theta =\frac{\sqrt{3}d}{-2c+d}\ .
\end{equation}
The relevant mixing matrix elements of $U_\text{MNS}$ are given as 
\begin{eqnarray}
\left |U_{e2}\right |=\frac{1}{\sqrt{3}}\ ,\quad \left |U_{e3}\right |=\frac{2}{\sqrt{6}}\left |\sin \theta \right |, \quad
\left |U_{\mu 3}\right |=\left |-\frac{1}{\sqrt{6}}\sin \theta -\frac{1}{\sqrt{2}}\cos \theta \right |\ ,
\label{mixingrelation}
\end{eqnarray}
which is the trimaximal lepton mixing.


Let us discuss
a concrete example of the flavor model with $A_4$, which is modified version
of the model proposed by Altarelli and 
Feruglio~\cite{Altarelli:2005yp,Altarelli:2005yx}.
We introduce an $A_4$ singlet  $\xi'$, which is a $1'$ flavon,
in addition to $\phi_l$, $\phi_\nu$, and $\xi$ as shown
in Table~1.
\begin{table}[h]
\begin{center}
\begin{tabular}{|c|cccc||c||cccc|}
\hline
& $(l_e,l_\mu ,l_\tau )$ & $e^c$ & $\mu ^c$ & $\tau ^c$ & $h_{u,d}$ & $\phi _l $ & $\phi _\nu $ & $\xi $ & $\xi '$ 
\\ \hline 
$SU(2)$ & $2$ & $1$ & $1$ & $1$ & $2$ & $1$ & $1$ & $1$ & $1$ \\
$A_4$ & $\bf 3$ & $\bf 1$ & $\bf 1''$ & $\bf 1'$ & $\bf 1$ & $\bf 3$ & $\bf 3$ & $\bf 1$ & $\bf 1'$ \\
$Z_3$ & $\omega $ & $\omega ^2$ & $\omega ^2$ & $\omega ^2$ & $1$ & $1$ & $\omega $ & $\omega $ & $\omega $ \\
\hline
\end{tabular}
\caption{Assignments of $SU(2)$, $A_4$, and $Z_3$ representations, where $\omega = e^{\frac{2\pi i}{3}}$.}
\end{center}
\end{table}

The relevant Yukawa interaction which respects   
the flavor symmetry is described by
\begin{align}
\mathcal{L}_\ell = y^ee^cl\phi _lh_d/\Lambda +y^\mu \mu ^cl\phi _lh_d/\Lambda +y^\tau 
\tau ^cl\phi _lh_d/\Lambda
\ +(y_{\phi _\nu }^\nu \phi _\nu +y_{\xi }^\nu \xi +y_{\xi '}^\nu \xi ')
llh_uh_u/\Lambda ^2\ .
\end{align}
The VEVs $\langle h_{u,d}\rangle =v_{u,d}$, 
$\langle \xi \rangle =\alpha _\xi \Lambda $, 
and $\langle \xi '\rangle =\alpha _{\xi '}\Lambda $ and vacuum alignment  
\begin{equation}
\langle \phi _l\rangle =\alpha _l\Lambda (1,0,0)\ ,\quad \langle \phi _\nu \rangle =
\alpha _\nu \Lambda (1,1,1)\,
\end{equation}
lead to the diagonal charged lepton mass matrix  
and effective neutrino mass matrix,
\begin{equation}
M_l=\alpha _lv_d
\begin{pmatrix}
y^e & 0 & 0 \\
0 & y^\mu & 0 \\
0 & 0 & y^\tau 
\end{pmatrix}  , \ \ \
M_\nu=a
\begin{pmatrix}
1 & 0 & 0 \\
0 & 1 & 0 \\
0 & 0 & 1 
\end{pmatrix}+b
\begin{pmatrix}
1 & 1 & 1 \\
1 & 1 & 1 \\
1 & 1 & 1
\end{pmatrix}+c
\begin{pmatrix}
1 & 0 & 0 \\
0 & 0 & 1 \\
0 & 1 & 0 
\end{pmatrix}+d
\begin{pmatrix}
0 & 0 & 1 \\
0 & 1 & 0 \\
1 & 0 & 0 
\end{pmatrix},
\label{a4matrix}
\end{equation}
where 
\begin{equation}
a=\frac{y_{\phi _\nu }^\nu \alpha _\nu  v_u^2}{\Lambda },
\qquad b=-\frac{y_{\phi _\nu }^\nu \alpha _\nu  v_u^2}{3\Lambda },
\qquad c=\frac{y_\xi ^\nu \alpha _\xi  v_u^2}{\Lambda },
\qquad d=\frac{y_{\xi '}^\nu \alpha _{\xi '} v_u^2}{\Lambda }.
\label{a4parameters}
\end{equation}
The non-vanishing $d$ is generated through the coupling $l l \xi'h_u h_u$.
Since there is the relation  $a=-3b$,
 we can predict  $\theta_{13}$.
In the case where the parameters $a,c,d$ are real, they are fixed 
by the three neutrino masses $m_1$, $m_2$ and  $m_3$.
We  predict  $\theta_{13}$ by taking input data  at the 90\% confidence level 
as \cite{Schwetz:2008er,Fogli:2008jx}
\begin{eqnarray}
&&\Delta m_{\rm atm}^2=(2.24-2.65)\times 10^{-3}~{\rm eV}^2, \quad 
\Delta m_{\rm sol}^2=(7.29-7.92)\times 10^{-5}~{\rm eV}^2,  \nonumber \\
&&\sin ^2\theta _{23}=0.40-0.62, \quad
\sin ^2\theta _{12}=0.29-0.34. 
\end{eqnarray}

In Figure 1, we show the predicted $\sin\theta_{13}$ versus $\sum m_i$,
where the normal hierarchy of the neutrino masses is taken.
In the case of  $m_3\gg m_2, m_1$, that is   
$\sum m_i\simeq 0.05$~eV, $\sin\theta_{13}$ is expected to be
around $0.15$, which is completely consistent with the experimental data.

The magnitude of $\sin^2 \theta_{23}$ is correlated with
 the magnitude of   $\sin \theta_{13}$ as seen  in Eq.(\ref{mixingrelation}).
By putting the data of  $\sin \theta_{13}$,
 we have predicted value  $\sin^2 \theta_{23}=0.40\sim 0.42$.

\begin{figure}[ttb]
\begin{minipage}[]{0.4\linewidth} 
\includegraphics[width=1\textwidth ]{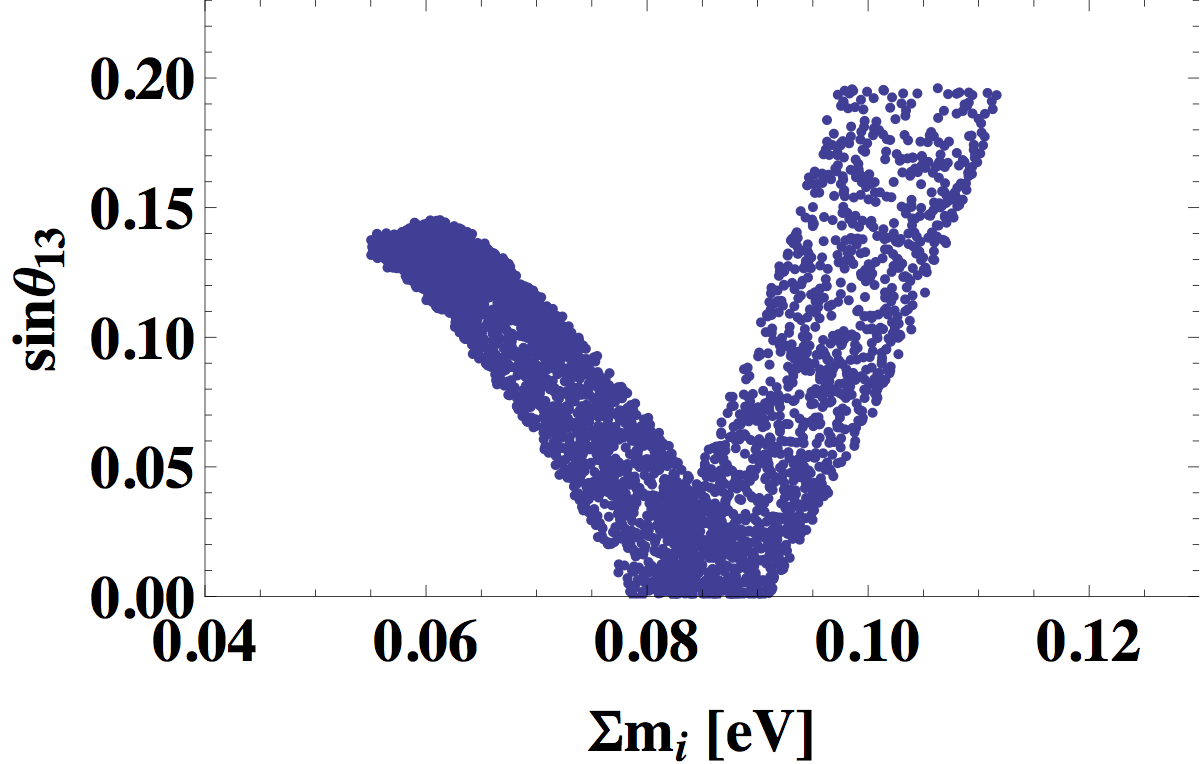}
\caption{$\sin\theta _{13}$ versus  $\sum m_i$ 
for the normal  mass  hierarchy.}
\end{minipage}
\hspace{1.5cm}
\begin{minipage}[]{0.4\linewidth} 
\includegraphics[width=6.5 cm]{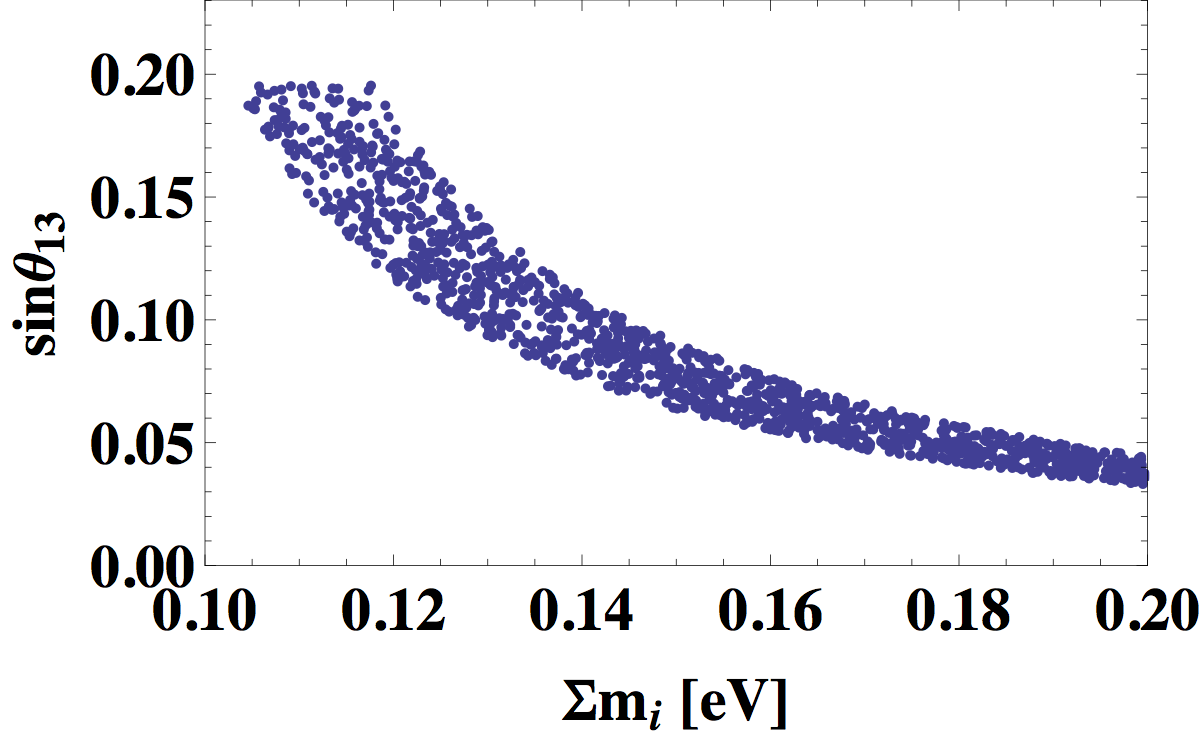}
\caption{$\sin\theta _{13}$ versus  $\sum m_i$ 
for the inverted  mass  hierarchy.}
\end{minipage}
\end{figure}
For the case of the inverted hierarchy of  neutrino masses,
we also show the  predicted value of  $\sin\theta_{13}$ versus  $\sum m_i$
in Figure 2.  In the hierarchical limit of
  $m_2\simeq  m_1 \gg  m_3$, we predict   $\sin\theta_{13}\simeq 0.2$,
which is somewhat larger than the experimental data.

\section{Large $\theta_{13}$ and the neutrino mass ratio}

 We discuss another  aspect of large flavor mixing.
Here, we introduce  our minimal texture hypothesis
~\cite{FTY93,FTY03,Fukugita:2012jr} and the resulting consequences.
Our hypothesis consists of 
the mass matrices of the charged leptons, the Dirac neutrinos
of the form \cite{Fritzsch} and  the right-handed Majorana 
mass matrix 
\begin{eqnarray}
m_E = 
\begin{pmatrix}
0 & A_\ell & 0 \cr A_\ell & 0 & B_\ell \cr
                     0 & B_\ell & C_\ell
  \end{pmatrix}
  \ ,\qquad
m_{\nu D} = 
\begin{pmatrix}
 0 & A_\nu & 0 \cr A_\nu & 0 & B_\nu \cr
                     0 & B_\nu & C_\nu  
  \end{pmatrix}   ,  \qquad  M_R = M_0 {\bf I},
\end{eqnarray}
where each entry is complex. 
We obtain the three light neutrino mass matrix as
\begin{equation}
m_\nu=m_{\nu D}^TM_R^{-1}m_{\nu D}.
\end{equation}
The lepton mixing matrix is given by
\begin{eqnarray} 
 U = U_\ell^\dagger \ Q \ U_\nu ,
\qquad\quad  Q= 
\begin{pmatrix}
1 & 0 & 0 \cr 0 & e^{i \sigma} & 0\cr  0 & 0 & e^{i \tau}
 \end{pmatrix} ,
\end{eqnarray}
where the expressions of  $U_\ell$ and  $\ U_\nu$ 
 are given in \cite{FTY03}, and  $Q$ 
 is a reflection of phases contained in the 
charged lepton mass matrix and the Dirac mass matrix of neutrinos.

Since the charged lepton masses are known, the
number of parameters  contained in our model is six: 
$m_{1D},\ m_{2D}, \ m_{3D}$, $\sigma$, $\tau$ and $M_0$.
They are to be determined by empirical
neutrino masses and mixing angles.

The  relevant lepton mixing matrix elements are written approximately,
\begin{eqnarray}
U_{e2} &\simeq& -\left({m_1 \over m_2}\right)^{1/4}+
\left({m_e\over m_\mu}\right)^{1/2}e^{i\sigma} , \cr
U_{\mu 3} &\simeq&  \left({m_2\over m_3}\right)^{1/4}e^{i\sigma}
-\left({m_\mu\over m_\tau}\right)^{1/2} e^{i\tau} , \cr 
U_{e3} &\simeq&  \left({m_e\over m_\mu}\right)^{1/2}U_{\mu 3}+ 
\left({m_2\over m_3}\right)^{1/2} \left({m_1\over m_3}\right)^{1/4}, 
\label{appromixing}
\end{eqnarray}
where charged lepton mass is denoted as $m_e$, $m_\mu$ and $m_\tau$, and
$(m_e/m_\tau)^{1/2}$ is  neglected.
Rough characteristics of mixing angles can be seen from these expressions.

For instance, these equations allow us to see the relation between 
neutrino masses and mixing angles, roughly as 
\begin{equation}
|U_{e2}|\approx \left (\frac{m_1}{m_2}\right )^{1/4},
\quad |U_{\mu 3}|\approx \left (\frac{m_2}{m_3}\right )^{1/4},
\quad |U_{e3}|\approx \left (\frac{m_2}{m_3}\right )^{1/2}\left (\frac{m_1}{m_3}\right )^{1/4}.
\end{equation}
The relation among the mixing angles are 
\begin{equation}
|U_{e3}|\approx |U_{\mu 3}|^2|U_{e2}U_{\mu 3}|=|U_{\mu 3}|^3|U_{e2}|.
\end{equation}
With $|U_{\mu 3}|\sim 1/\sqrt{2}$ and $|U_{e2}|\sim 1/\sqrt{3}$  
we see that 
$|U_{e3}|\sim 1/(2\sqrt{6})\simeq 0.2$.
We emphasize that only the normal neutrino mass hierarchy is allowed 
in our model, which allows us to predict uniquely the effective mass
 that appear in double beta decay.


We now present the numerical results using the accurate expression of the
lepton mixing angles given in \cite{FTY03}.
Figures 3  shows  $|U_{e3}|=\sin \theta _{13}$ versus $\sin^22\theta _{23}$.
The range of  $\sin^22\theta _{12}$ and $\sin^22\theta _{23}$ 
are cut at the boundary of the region 
experimentally allowed at the $90\%$ confidence level.
Our predicted value of  $\sin\theta _{13}$ falls in the
middle of the range of the experimental data. 

We also  see that the maximum mixing $\theta_{23}=\pi/4$ 
($\sin^2 2\theta _{23}=1$) is excluded.

\begin{figure}[h!]
\begin{minipage}[]{0.45\linewidth}
\includegraphics[width=7.5cm]{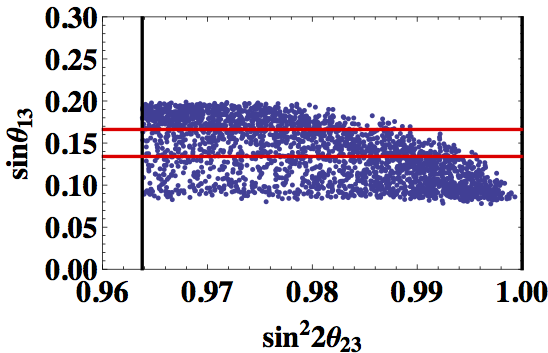}
\caption{Predicted $\sin \theta _{13}$ versus $\sin^22\theta _{23}$.}
\end{minipage}
\hspace{5mm}
\begin{minipage}[]{0.45\linewidth}
\vspace{-4mm}
\includegraphics[width=7.5cm]{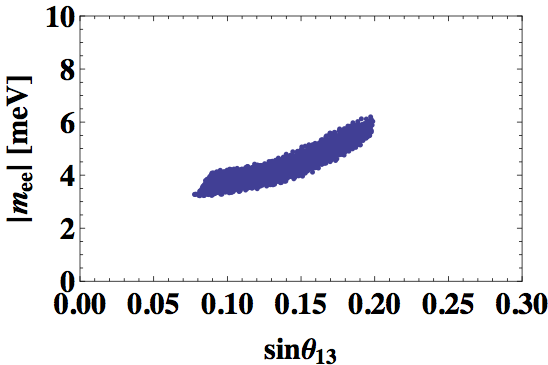}
\caption{Predicted $m_{ee}$ versus  $\sin\theta _{13}$.}
\end{minipage}
\end{figure}

We also predict the effective electron neutrino mass $|m_{ee}|$ that appears
in neutrinoless double beta decay.
Figures 4  shows the predicted $|m_{ee}|$ versus $\sin\theta_{13}$, where
the allowed effective mass for double beta decay is 
$|m_{ee}|=3.7-5.6 \ {\rm meV}$.

\section{Conclusion}

The $A_4$ model by 
  Altarelli and Feruglio can be simply modified by
introducing   $1'$  and $1''$ flavons, and then, it predicts
 $\sin\theta_{13}\simeq 0.15$ 
for the normal hierarchical neutrino masses.

On the other hand, 
there is another aspect of the large flavor mixing of neutrions.
Our minimal texture describes all known
empirical values, without adding any further matrix elements or
extending the assumptions. The magnitude of $\theta_{13}$ falls in the
middle of the range of the experimental data.  The matrix only allows
the normal hierarchy of the neutrino mass, excluding either inverse
hierarchy or degenerate mass cases.  This predicts the effective mass
of double beta decay to lie within the range $m_{ee}=3.7-5.6$ meV. 




\bibliographystyle{apsrev4-1}

%% file: Papers/tortola.tex
\chapter[2012 status of neutrino oscillation parameters: $\theta_{13}$ and beyond. (T{\'o}rtola)]{2012 status of neutrino oscillation parameters: $\theta_{13}$ and beyond.}
\vspace{-2em}
\paragraph{M. T{\'o}rtola}
\paragraph{Abstract}
We present an updated global fit of neutrino oscillations including
the most recent reactor antineutrino disappearance data from Double
Chooz, Daya Bay and RENO, together with the latest MINOS and T2K
long-baseline appearance and disappearance results.
The highlights of the updated analysis are, on the one hand, the large
value of $\theta_{13}$ implied by the new reactor data (with
$\theta_{13} = 0$ excluded at more than 10$\sigma$) and, on the other
hand, the non-maximal value of $\theta_{23}$ preferred by the new
long-baseline results.

\section{Introduction}

Last year there were some indications for a non-zero $\theta_{13}$
mixing angle coming from the observation of electron neutrino
appearance on a muon neutrino beam at the accelerator oscillation
experiments T2K~\cite{Abe:2011sj} and
MINOS~\cite{Adamson:2011qu}. Together with the hints from the solar
and atmospheric neutrino data samples, the global analysis of neutrino
oscillation data reported indications of non-zero $\theta_{13}$ at the
level of 3-4$\sigma$ (see Refs \cite{Schwetz:2011qt,Schwetz:2011zk}
for more details).
Along this year, these hints have been largely confirmed thanks to the
first measurements of $\theta_{13}$ reported by the reactor
experiments Double Chooz~\cite{Abe:2011fz}, Daya Bay~\cite{An:2012eh}
and RENO~\cite{Ahn:2012nd}.
These new generation of reactor experiments look for the disappearance
of reactor antineutrinos over baselines of the order of 1~km with very
large statistics and, most importantly, with several detectors located
at different distances from the reactor core, to reduce the systematic
errors relative to the neutrino flux normalization.

Besides the new reactor neutrino data, the global fit presented in
this work considers also the most recent long-baseline neutrino data
from the MINOS~\cite{new-MINOS} and T2K~\cite{Abe:2012gx,new-T2K-app}
experiments presented at the Neutrino 2012 Conference.
The new long-baseline data imply some improvements with respect
to the previous MINOS and T2K results in
Refs. ~\cite{Adamson:2011ig,Adamson:2011fa,Adamson:2012rm,Abe:2011sj,Adamson:2011qu}.
First, the new results on $\nu_\mu \to \nu_e$ appearance
searches allow a better determination of the $\theta_{13}$ mixing
angle, although its current determination is fully dominated by the
Daya Bay reactor data.
And second, for the first time they show a preference for a
non-maximal $\theta_{23}$ in the $\nu_\mu$ and $\bar\nu_\mu$
disappearance channels.

\section{Global analysis of neutrino oscillation data}

\subsection{Neutrino data samples}

Here we will give a brief description of the neutrino data samples
considered in our global fit. For more details see
Ref.~\cite{Tortola:2012te}.
For the solar neutrino sector we include the most recent solar
neutrino data from the radiochemical experiments
Homestake~\cite{Cleveland:1998nv}, Gallex/GNO~\cite{Kaether:2010ag}
and SAGE~\cite{Abdurashitov:2009tn}, as well as the latest data from
Borexino~\cite{Bellini:2011rx}, and the three phases of
Super-Kamiokande~\cite{hosaka:2005um,Cravens:2008aa,Abe:2010hy} and the
Sudbury Neutrino Experiment SNO~\cite{aharmim:2008kc,Aharmim:2009gd}.
For the KamLAND reactor experiment we consider the most recent results
corresponding to a total livetime of 2135 days~\cite{Gando:2010aa}.
In the atmospheric sector we use the atmospheric neutrino analysis
done by the Super-Kamiokande Collaboration~\cite{Wendell:2010md}, and
we also include the most recent results from the
MINOS~\cite{new-MINOS} and T2K~\cite{Abe:2012gx,new-T2K-app}
long-baseline experiments released last June at the Neutrino 2012
Conference, either for the appearance and disappearance channels and
for the neutrino and antineutrino runs in the case of MINOS.
Finally, we include in our global analysis the latest results released
by the new generation of reactor experiments. We consider the total
event rate measured by Double Chooz~\cite{new-DChooz,Abe:2012tg} with
an exposure of 227.93 live days and the far to near event ratio
observed at the Daya Bay~\cite{new-DayaBay} and RENO~\cite{Ahn:2012nd}
experiments.  The more recent Daya Bay results presented at the
Neutrino 2012 conference with 2.5 times more statistics than their
previous data release allow a very strong rejection
for $\theta_{13}$ = 0 that now is excluded at almost 8$\sigma$ by Daya
Bay alone.

\subsection{Results}

Here we summarize the main results for the neutrino oscillation parameters
obtained in our global neutrino analysis. A more detailed description
is given in Ref.~\cite{Tortola:2012te}.

\begin{figure}
  \centering
 \includegraphics[width=0.48\textwidth]{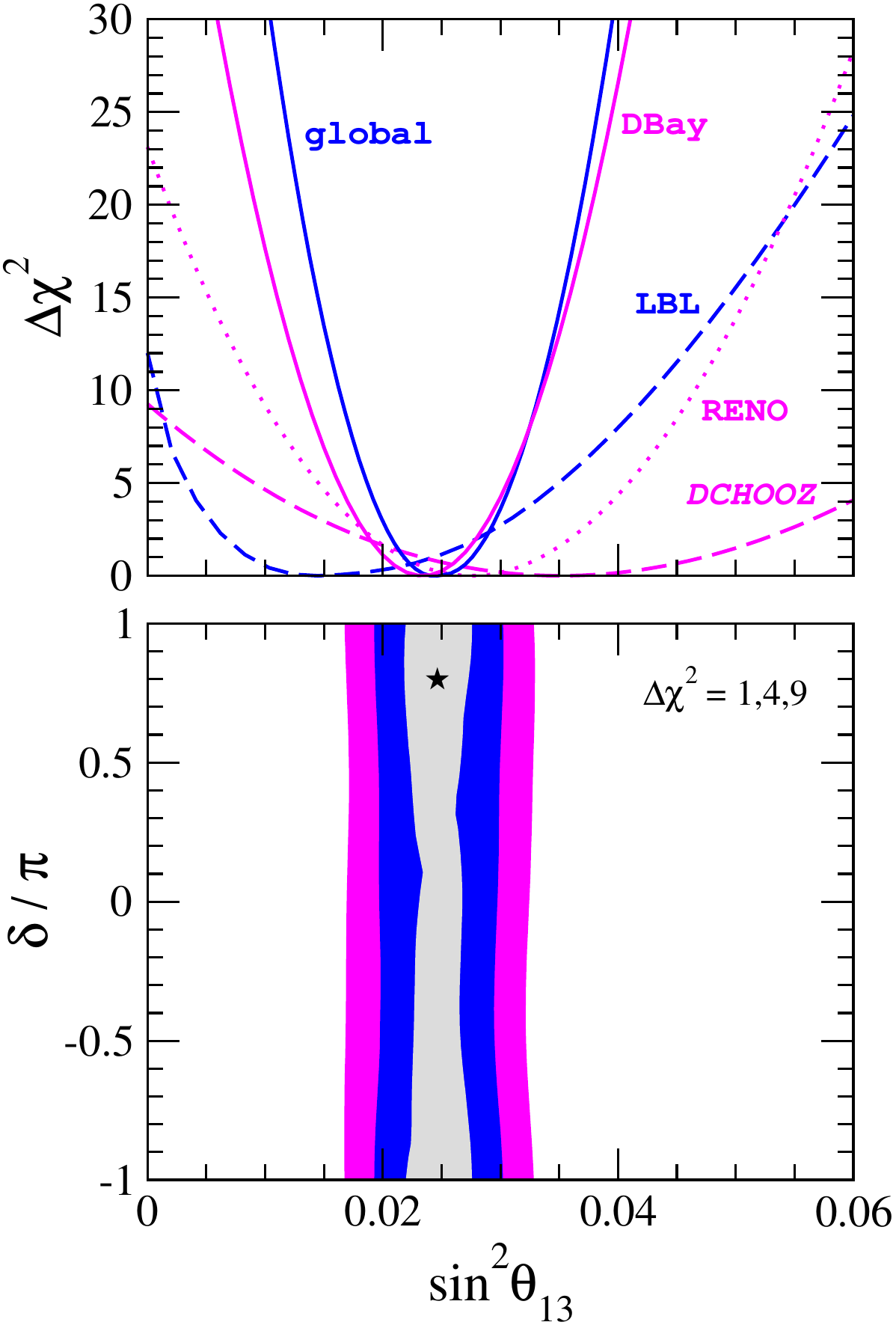}
  \includegraphics[width=0.48\textwidth]{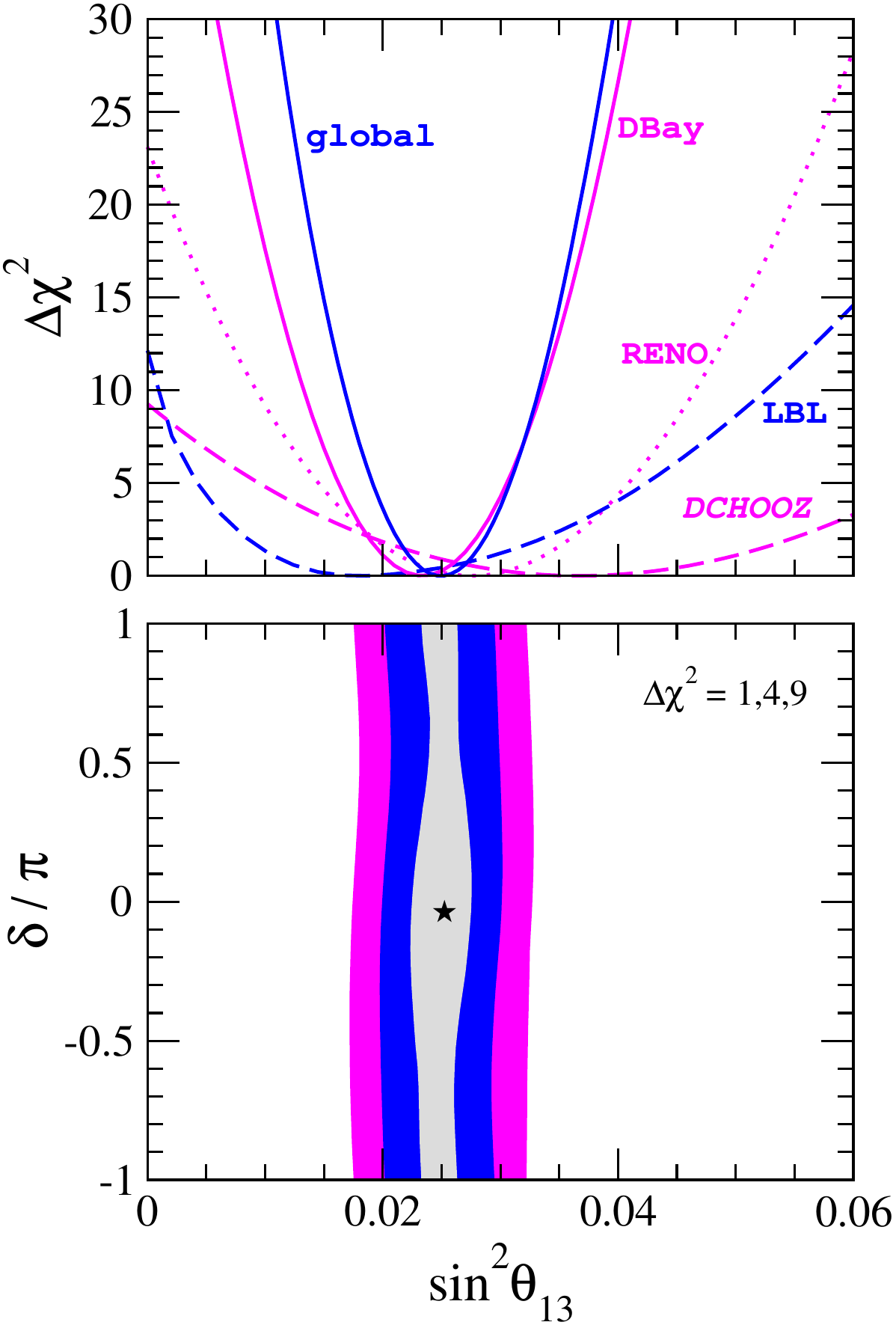}
  \caption{Upper panels: $\Delta\chi^2$ versus $\sin^2\theta_{13}$
    from the analysis of reactor (magenta/light lines), long-baseline
    (dashed blue/dark line) and global neutrino data (solid blue/dark
    line).  Lower panels: contours of $\Delta\chi^2=1,4,9$ in the
    $\sin^2\theta_{13}-\delta$ plane from the global fit to the
    data. Left (right) panels are for normal (inverted) neutrino mass
    ordering.}
\label{MT_fig:t13-delta}
\end{figure}

Fig.~\ref{MT_fig:t13-delta} summarizes the results obtained for
$\sin^2\theta_{13}$ and $\delta$.  The upper panels show the
$\Delta\chi^2$ profile as a function of $\sin^2\theta_{13}$ for normal
(left panel) and inverted (right panel) neutrino mass orderings for
the individual reactor data samples, the combination of all
long-baseline data and the global analysis, as indicated.  One sees
that the current global constraint on $\theta_{13}$ is largely
dominated by the recent Daya Bay results.
For both neutrino mass hierarchies we find that $\theta_{13} = 0$ is now
excluded at $10.2\sigma$.
The lower panels of Fig.~\ref{MT_fig:t13-delta} show the contours
of $\Delta\chi^2=1,4,9$ in the $\sin^2\theta_{13}-\delta$ plane from
the global fit to the neutrino oscillation data.
As shown in the figure, the sensitivity of the current neutrino data
to the CP phase $\delta$ is still very poor. Marginalizing over the CP
phase and all the remaining oscillation parameters we get the
following results for the best fit and one-sigma $\sin^2\theta_{13}$
errors:
\begin{equation}
\begin{array}{c@{\qquad}l}
\sin^2\theta_{13} = 0.0246^{+0.0029}_{-0.0028}\,, & \text{(normal hierarchy),} \\
\sin^2\theta_{13} = 0.0250^{+0.0026}_{-0.0027}\,, & \text{(inverted hierarchy).} 
\end{array}
\end{equation}

\begin{figure}
  \centering
 \includegraphics[width=0.85\textwidth]{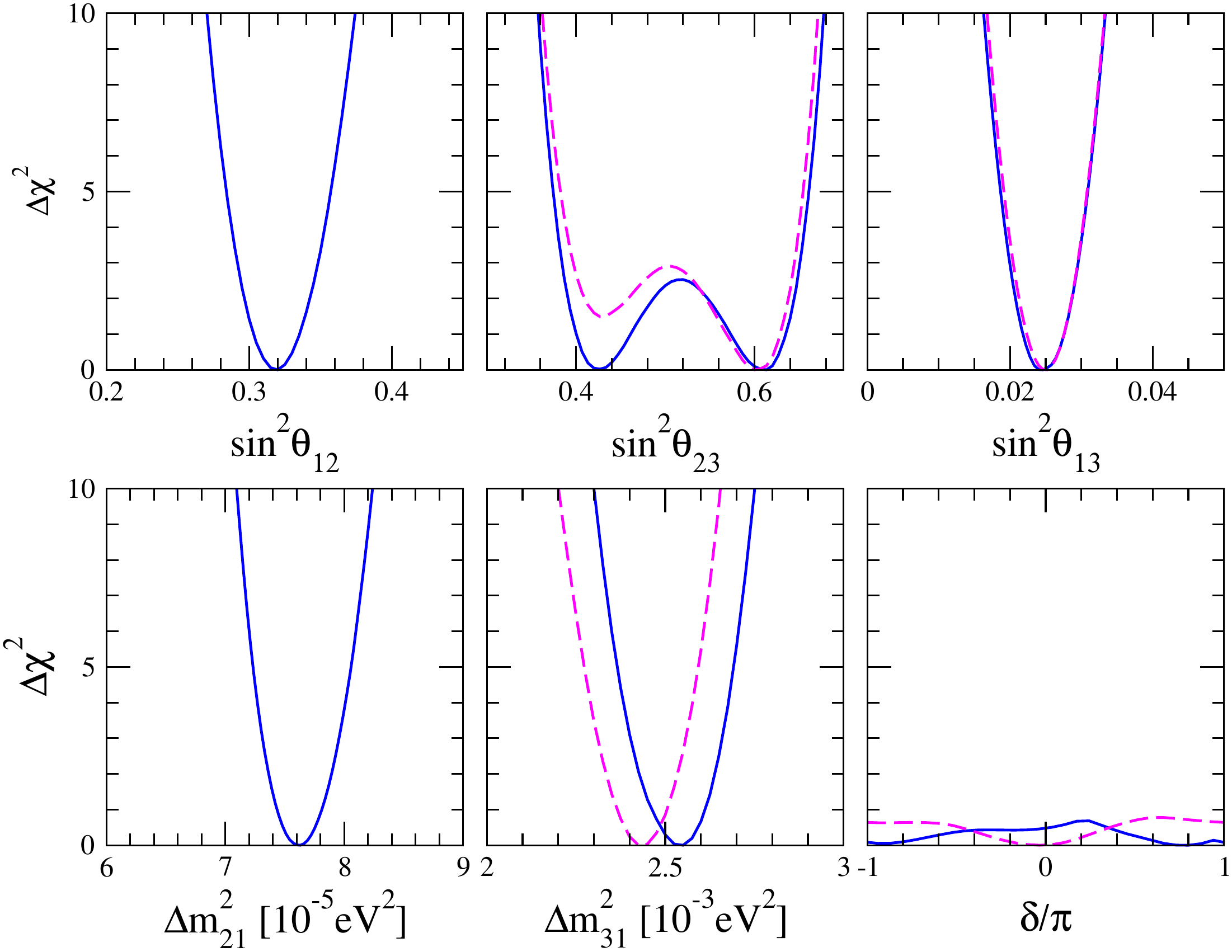}
 \caption{$\Delta\chi^2$ profiles for all the neutrino oscillation
   parameters. Solid lines correspond to the case of normal mass
   hierarchy and dashed lines to inverted mass hierarchy.}
\label{MT_fig:summary}
\end{figure}

\begin{table}[t]\centering
  \catcode`?=\active \def?{\hphantom{0}}
  \begin{tabular}{|c|c|c|c|}
    \hline
    parameter & best fit & $1\sigma$ range&  3$\sigma$ range
    \\
    \hline
    $\Delta m^2_{21}\: [10^{-5}\text{eV}^2]$
    & 7.62 & 7.43--7.81  & 7.12--8.20 \\[3mm] 
    $|\Delta m^2_{31}|\: [10^{-3}\text{eV}^2]$
    &
    \begin{tabular}{c}
      2.55\\
      2.43
    \end{tabular}
    &
    \begin{tabular}{c}
      2.46--2.61\\
      2.37--2.50\\
    \end{tabular}
    &
    \begin{tabular}{c}
      2.31--2.74\\
      2.21--2.64
    \end{tabular}
    \\[6mm] 
    $\sin^2\theta_{12}$
    & 0.320 & 0.303--0.336 &  0.27--0.37\\[3mm]  
    $\sin^2\theta_{23}$
    &
    \begin{tabular}{c}
      0.613 (0.427)\footnote{This is a local minimum in the first octant of
        $\theta_{23}$ with $\Delta\chi^2 = 0.02$ with respect to the
        global minimum}\\
      0.600
    \end{tabular}
    &
    \begin{tabular}{c}
      0.400--0.461 \& 0.573--0.635\\
      0.569--0.626\\
    \end{tabular}
    &
    \begin{tabular}{c}
      0.36--0.68\\ 
      0.37--0.67 
    \end{tabular}
    \\[5mm] 
    $\sin^2\theta_{13}$
    &
    \begin{tabular}{c}
      0.0246\\
      0.0250
    \end{tabular}
    &
    \begin{tabular}{c}
      0.0218--0.0275 \\
      0.0223--0.0276
    \end{tabular}
    &
    0.017--0.033 \\
    $\delta$
   &
   \begin{tabular}{c}
     ?0.80$\pi$\\
     -0.03$\pi$
   \end{tabular}
   &
   0--2$\pi$
   &
   0--2$\pi$ \\
       \hline
     \end{tabular}
     \caption{ \label{MT_tab:summary} Neutrino oscillation parameters
       summary. For $\Delta m^2_{31}$, $\sin^2\theta_{23}$, $\sin^2\theta_{13}$, 
       and $\delta$ the upper (lower) row corresponds to normal (inverted)
       neutrino mass hierarchy. \newline{\footnotesize $^1$ Local minimum in the first octant of
       $\theta_{23}$ with $\Delta\chi^2 = 0.02$ with respect to the global minimum.}}
\end{table}

The global fit results for all the other neutrino oscillation
parameters are summarized in Fig.~\ref{MT_fig:summary} and
Table~\ref{MT_tab:summary}.
The inclusion of the new reactor and long-baseline data does not have
a strong impact on the solar neutrino parameter determination, since
they are already quite well determined by solar and KamLAND reactor
data.
Concerning atmospheric neutrino parameters, the new MINOS
disappearance data in Ref.~\cite{new-MINOS} have shifted the best fit
value of $\Delta m^2_{31}$ to slightly larger values. The accuracy in
the determination of the mass splitting has also been improved thanks
to the new long-baseline data.
For the atmospheric angle we found that maximal mixing
$\theta_{23} = \pi/4$ is disfavoured at $\sim$ 90\% C.L.  Our global
fit shows a weak preference for the mixing angle in the second octant,
somewhat steeper for the inverse hierarchy case.
The preference for non-maximal values of $\theta_{23}$ appears as a
consequence of the new MINOS results, while the choice of a particular
octant comes from the interplay of long-baseline, reactor and
atmospheric neutrino data.
A more detailed discussion about the impact of the new long-baseline
data and the atmospheric neutrino analysis in the determination of the
atmospheric mixing angle as well as a comparison with other recent
global analysis can be found in Ref.~\cite{Tortola:2012te}.

\section{Summary and outlook}

In this work we have summarized the current status of the
three-neutrino oscillation parameters, including the most recent
reactor antineutrino data reported by Double Chooz, Daya Bay and RENO
as well as the latest MINOS and T2K appearance and disappearance
results, as presented at the Neutrino 2012 conference.
From the global fit to neutrino data we found a best fit value of
$\sin^2\theta_{13} = 0.0246 (0.0250)$ for normal (inverted) neutrino
mass hierarchy, with $\sin^2\theta_{13} = 0$ excluded at 10.2$\sigma$.
Concerning the atmospheric neutrino sector, we find a best fit value
of the atmospheric mixing angle $\sin^2\theta_{23}$ in the
second-octant for the two neutrino mass orderings.
This preference, however, is still marginal and first octant values of
$\theta_{23}$ are allowed at the 1$\sigma$ level.
The new official Super-Kamiokande analysis in Ref.~\cite{new-SK-atm}
with three flavour effects points to a weak preference for non-maximal
$\theta_{23}$ mixing, together with a correlation between the neutrino
mass ordering and the preferred octant for $\theta_{23}$, in
qualitative agreement with our results.
Conversely, the analyses of atmospheric neutrino data in
Refs.~\cite{Fogli:2012ua,nufitcoll} obtain a preference for mixing
angle in the first octant for both mass hierarchies. The origin of
this discrepancy between the different analyses is not yet clear.
The impact of the new reactor and long-baseline accelerator
measurements upon the solar neutrino oscillation parameters is
completely marginal, the results are summarized in
Table~\ref{MT_tab:summary}.
No significant sensitivity to the CP-violating phase $\delta$ or the
neutrino mass ordering has been found by the combination of all
current neutrino data.

After the accurate measurements of the neutrino oscillation parameters
presented in last section, it is time now for precision measurements.
In particular, a precise determination of $\theta_{13}$ will be
crucial to perform the proposed CP violation searches in neutrino
oscillations~\cite{Bandyopadhyay:2007kx} and will be very helpful also
to determine the neutrino mass hierarchy.
In this respect, further improvement will be obtained after the
completion of the Daya Bay detector site along this year and, in 3
years of operation the Daya Bay uncertainties on $\sin^22\theta_{13}$
will be reduced from 20\% to 4-5\% \cite{talk:Cao}.
The installation of the near detector in Double Chooz expected by the
end of 2013 will also help understanding the spectral distortions in
the reactor neutrino spectrum.
Muon-neutrino disappearance results from the T2K and NO$\nu$A
long-baseline accelerator experiments will improve the determination
of the atmospheric mass splitting $|\Delta m^2_{31}|$ at the level of
a few percent~\cite{Abe:2011ks, Ayres:2004js,Huber:2004ug}. The
atmospheric mixing angle $\theta_{23}$ is also likely to be measured
with improved precision.
Regarding the deviations of $\theta_{23}$ from maximal mixing, a
comparative study of the different atmospheric neutrino analysis
together with the combination of future accelerator and reactor
neutrino data may help to clarify the ambiguity between the two
octants~\cite{Huber:2009cw}.
Several ideas have been proposed to address the issue of the neutrino
mass ordering.
One of them~\cite{Blennow:2012gj} exploits the sensitivity to matter
effects at the accelerator experiment NO$\nu$A and the atmospheric
neutrino observations at the future India-based Neutrino Observatory
(INO)~\cite{INO}. A combined analysis of the two
experiments would allow a 2$\sigma$ rejection of the wrong mass
hierarchy by 2020.
Likewise, the possibility of identifying the neutrino mass hierarchy with a
reactor neutrino experiment at an intermediate baseline ($\sim$ 60 km)
has ben discussed in Refs.~\cite{Petcov:2001sy,Zhan:2008id}.

\section*{Acknowledgments}

M.T.\ acknowledges financial support from CSIC under the JAE-Doc
programme, co-funded by the European Social Fund. This work was also
supported by the Spanish MINECO under grants FPA2011-22975 and
MULTIDARK CSD2009-00064 (Consolider-Ingenio 2010 Programme), by
Prometeo/2009/091 (Generalitat Valenciana).

\newpage

\bibliography{tortola}
\bibliographystyle{apsrev4-1}


%% file: Papers/turczyk.tex

%
%
%
%
%
%

\chapter[Constraining CP violation in neutral meson mixing with theory input (Freytsis, Ligeti, \textit{Turczyk})]{Constraining CP violation in neutral meson mixing with theory input}
\vspace{-2em}
\paragraph{M. Freytsis, Z. Ligeti, \textit{S. Turczyk}}
\paragraph{Abstract}
There has been a lot of recent interest in experimental hints of $CP$ violation in $B_{d,s}^0$ mixing. The D\O\ measurement of the semileptonic $CP$ asymmetry would - with higher significance - be a clear signal of beyond the standard model physics. We present a relation \cite{Freytsis:2012ja} for the mixing parameters, which allows clearer interpretation of the data in models in which new physics enters in $M_{12}$ and/or $\Gamma_{12}$. This result implies that the central value of the D\O\ measurement in $B_{d,s}^0$ decay is not only in conflict with the standard model, but in a stronger tension with data on $\Delta\Gamma_s$ than previously appreciated. After we derive the relation between the theoretical prediction of $|\Gamma_{12}|$  and measurements of $\Delta M$, $\Delta \Gamma$ and $A_\text{SL}$, we will explain how the result can help to better constrain $\Delta\Gamma$ or $A_\text{SL}$, whichever is less precisely measured.

\section{Introduction}

The D\O\ measurement of the $CP$ asymmetry in decays of a $b\bar b$ pair to two same-sign muons~\cite{Abazov:2011yk} hinted towards $CP$ violation in $B$\,--\,$\bar B$ mixing, which would be a clear sign of new physics~\cite{Laplace:2002ik, Ligeti:2010ia}
\begin{equation}\label{expD0}
    A_\text{SL}^b = -[7.87 \pm 1.72\, \mbox{(stat)} \pm 0.93\, \mbox{(syst)}] \times 10^{-3} \,.
\end{equation}
The time evolution of the flavor eigenstates is determined by
\begin{equation}
    i\, \frac{\text{d}}{\text{d} t} 
  \begin{pmatrix} |B^0(t)\rangle \\ |\bar B^0(t)\rangle \end{pmatrix} 
    = \left(M - \frac{i}{2}\,\Gamma\right)
  \begin{pmatrix}|B^0(t)\rangle\\ |\bar B^0(t)\rangle \end{pmatrix} \,, \label{timedep}
\end{equation}
where $M$ and $\Gamma$ are $2\times2$ Hermitian matrices. $CPT$ invariance implies $M_{11} = M_{22}$ as well as $\Gamma_{11} = \Gamma_{22}$. The physical eigenstates, in the notation customary to $B$ physics, are given by
\begin{equation}\label{physeigen}
    |B_{H,L}\rangle = p\, |B^0\rangle \mp q\, |\bar B^0\rangle \,,
\end{equation}
where we chose $|p|^2+|q|^2=1$. $CP$ violation in mixing occurs if the mass and
$CP$ eigenstates do not coincide, $ \delta \equiv \langle B_H | B_L \rangle   = (|p|^2-|q|^2)/(|p|^2+|q|^2) \neq 0$. The solution for the mixing parameters satisfies the relation $q^2 / p^2 = (2M_{12}^* - i\Gamma_{12}^*)/(2M_{12} - i\Gamma_{12})$. In the small $\delta$ limit, $A_\text{SL} \approx 2 \delta $ is a very good approximation on the $B_{d/s}$-systems. In the $|\Gamma_{12}/ M_{12}| \ll 1$ limit, which applies model independently for the $B_{d,s}^0$ systems,
\begin{align}\label{approx}
    \Delta m &\approx 2\, |M_{12}|\, ,\quad  \Delta\Gamma \approx 2\, |\Gamma_{12}|\, \cos [\text{arg}(-\Gamma_{12}/M_{12})] \,,\,\quad  A_\text{SL} \approx \textrm{Im}\,(\Gamma_{12}/M_{12})\,.
\end{align}
In this limit $q/p$ is a pure phase to a good approximation, determined by $M_{12}$, which has good a sensitivity to NP. 

The width difference, defined as $\Delta\Gamma_s \equiv \Gamma_L-\Gamma_H$, has been obtained with a reasonable uncertainty in the $B_s$ system, and has not been observed for the $B_d$ meson yet
\begin{align}\label{deltaGs}
    \Delta\Gamma_s &= (0.068 \pm 0.027)\, \text{ps}^{-1}\,, \qquad
  \mbox{CDF~\cite{CDF:2011af}}\,, \quad \Delta\Gamma_s &= (0.163^{+0.065}_{-0.064})\, \text{ps}^{-1}\,, \qquad  \mbox{D\O~\cite{Abazov:2011ry}}\nonumber\\
    \Delta\Gamma_s &= (0.116 \pm 0.019)\, \text{ps}^{-1}\,, \qquad
  \mbox{LHCb~\cite{LHCb-CONF-2012-002}} \,.
\end{align}
For our numerical analysis, we use the most precise single measurement from LHCb in the lack of a world average. For $\Delta m_s$ we take the average of the CDF~\cite{Abulencia:2006ze} and LHCb~\cite{LHCb-CONF-2011-005,LHCb-CONF-2011-050} measurements and $\Delta m_d$ from HFAG~\cite{Asner:2010qj}
\begin{align}
    \Delta m_s \equiv m_H-m_L = (17.719 \pm 0.043)\, \text{ps}^{-1} \,,\quad  \Delta m_d = (0.507 \pm 0.004)\,\text{ps}^{-1} \label{deltamd}\,.
\end{align}

The measurement in Eq.~\eqref{expD0} is a linear combination of the two individual asymmetries, as both $B^0_d$ and $B^0_s$ are produced at the Tevatron
\begin{equation}\label{weight}
    A_\text{SL}^b = (0.594\pm0.022)\, A_\text{SL}^d  + (0.406\pm 0.022)\, A_\text{SL}^s \,.
\end{equation}
The individual asymmetries have been measured at the $e^+e^-$ $B$                                                                          
factories~\cite{Asner:2010qj} and at D\O~\cite{Abazov:2009wg} and are compatible with the Standard Model (SM) prediction\cite{Lenz:2011ti} $A_\text{SL}^d = - (0.5 \pm 5.6) \times 10^{-3}$ and  $A_\text{SL}^s = - (1.7 \pm 9.2) \times 10^{-3}$.

One naturally should ask, if there are any non-trivial constraints on the mixing parameters, besides the requirement of having positive mass and width eigenvalues for the physical states.

\section{Theoretical Constraints on the Mixing Parameters}

The unitarity bound~\cite{Bell:1990tq,Lee:1965hi} is a requirement on the mixing parameters, which constrains the eigenvalues of $\Gamma$ to be positive independent of the physical eigenvalues, or equivalently
\begin{equation}\label{unitbound}
\delta^2 < \frac{\Gamma_H \Gamma_L} {(m_H-m_L)^2 + (\Gamma_H + \Gamma_L)^2/4}
  = \frac{1-y^2}{1+x^2}\,.
\end{equation}
Here we define, using $\Gamma = (\Gamma_H+\Gamma_L)/2$, the quantities $ x = (m_H - m_L)/\Gamma$ and $y = (\Gamma_L - \Gamma_H)/(2\Gamma)$. $x$ is positive by definition, while $y \in (-1,\, +1)$. To derive this bound on a mathematical basis, we define the complex quantities
\begin{equation}\label{uboundvec}
a_i = \sqrt{2\pi \rho_i}\, \langle f_i | \mathcal{H} | B \rangle \,, \qquad
\bar{a}_i = \sqrt{2\pi \rho_i}\, \langle f_i | \mathcal{H} | \bar B \rangle \,,
\end{equation}
with $\rho_i$ denoting the phase space density for final state $f_i$. If we
treat $a_i$ and $\bar{a}_i$ as vectors in a complex $N$-dimensional vector
space, then taking the standard inner product on complex vector spaces, and
using the optical theorem~\cite{Bell:1990tq}, amounts to the relations
\begin{equation}\label{ubound12}
  a_i^*\, a_i = \Gamma_{11}\,, \qquad
  \bar{a}_i^*\, \bar{a}_i = \Gamma_{22} \,, \qquad
  \bar{a}_i^*\, a_i = \Gamma_{12} \,,
\end{equation}
where $CPT$ fixes $\Gamma_{11} = \Gamma_{22} = \Gamma$. Applying the
Cauchy-Schwarz inequality to the vectors $a_i$ and $\bar{a}_i$
implies~\cite{Bell:1990tq}
\begin{equation}\label{gammapos}
  |\Gamma_{12}| \leq \Gamma_{11}\,.
\end{equation}
This is equivalent to the statement that the eigenvalues of the
$\Gamma$ matrix must be positive in addition to the physical width $\Gamma_{H,L}>0$.

To see that this is also equivalent to the unitarity bound of
Eq.~\eqref{unitbound}, we use Eq.~\eqref{physeigen} to define $a_{Hi}$ and $a_{Li}$ analogously to the physical states, and proceed with similar steps as above. The unitarity bound in Eq.~\eqref{unitbound} then arises from using these expressions for $\Gamma_{11}$ and $\Gamma_{12}$ in Eq.~\eqref{gammapos}.

\subsection{Deriving a Relation using Theoretical Input}
In the kaon system, for which this bound was originally derived, the assumption in  Eq.~\eqref{gammapos} was a necessity due to the dominance of long-distance physics in the result. For $B_{d,s}$ mesons, the large mass scale $m_b \gg \Lambda_\text{QCD}$ allows $\Gamma_{11}$ and $\Gamma_{12}$ to be calculated in an operator product expansion, and at leading order $|\Gamma_{12} / \Gamma_{11}| = {\cal O}[(\Lambda_\text{QCD}/m_b)^3\, (16\pi^2)]$. We extend the preceding derivation with assuming additional theoretical knowledge in Eq.~\eqref{gammapos}, and define
\begin{equation}\label{epsdef}
y_{12} = |\Gamma_{12}| \,\big/\, \Gamma \,.
\end{equation} 
Thus we obtain as the solution an exact relation instead of the inequality
\begin{equation}\label{unitrel}
\delta^2 = \frac{y_{12}^2-y^2}{y_{12}^2+x^2} =
  \frac{|\Gamma_{12}|^2-(\Delta\Gamma)^2/4}{|\Gamma_{12}|^2+(\Delta m)^2}\,. 
\end{equation}
This equation also follows from the solution of the eigenvalue problem, and was
previously derived in Ref.~\cite{Branco:1999fs} with the resulting bound on
$\delta$ noted. It also appears in related forms in Refs.~\cite{Ciuchini:2007cw,Kagan:2009gb} and follows from Eqs.~(9) and (12) in~\cite{Grossman:2009mn}.  

For fixed $x$ and $y$, $\delta^2$ is monotonic in $y_{12}$, so an upper bound on $y_{12}$ gives an upper bound on $|\delta|$ and with the requirement $y_{12} \leq 1$ the usual unitarity bound in Eq.~\eqref{unitbound} is recovered.

A better understanding of the physical situation can be gained, by obtaining  Eq.~\eqref{unitrel} from a scaling argument: As
$\delta$ only depends on mixing parameters, it is independent of the value of
$\Gamma$. One can then scale $\Gamma$ by $y_{12}$, which cannot affect $\delta$
but changes $x\to x/y_{12}$ and $y \to y/y_{12}$. The exact relation Eq.~\eqref{unitrel} follows then from this argument and Eq.~\eqref{unitbound}. The derivation above makes the physical origin of this relation clear and also holds in the CPT violating case for $|\delta|^2$, as $\delta$ can become complex in this case.

Even if a precise calculation of $\Gamma_{12}$ is not possible or one assigns a
very conservative uncertainty to it, an upper bound on $y_{12}$ implies an upper
bound on $|\delta|$, which is stronger than Eq.~\eqref{unitbound}. For
small values of $y_{12}$, as in the $B_d$ system, this bound can be much
stronger.

\subsection{Application to Recent Data}

We will now compare the absolute value of the semi-leptonic asymmetry with other mixing parameters using the result implied by the relation above. First we will apply the relation to the two same-sign muon result from D\O\, and then to the individual asymmetries.
\begin{figure}[htp]
 \centering\includegraphics[scale=1.25]{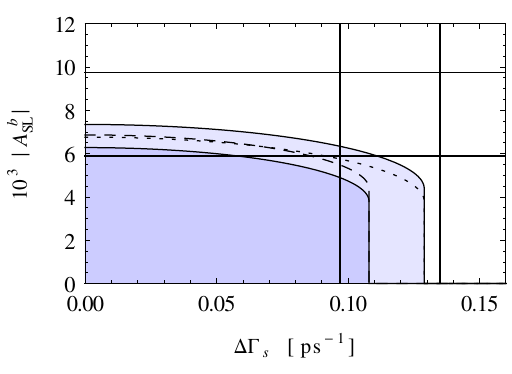}
    \caption{Upper bounds on $A_\text{SL}^b$ as a function of $\Delta\Gamma_s$,
setting $\Delta\Gamma_d = 0$, description is in text.}\label{1Dplot}
\end{figure}
We can only compare the absolute value of $A_\text{SL}^b$ measured by D\O\ with the
result implied by the relation above, since Eq.~\eqref{unitrel} only bounds $|\delta|$. Thus the bound on $A_\text{SL}^b$ is not sensitive to possible cancellations between $A_\text{SL}^d $ and $A_\text{SL}^s$.  Denoting this upper bound by $\delta^{d,s}_{\text{max}}$ and using the weight factors from Eq.~\eqref{weight},
$ |A_\text{SL}^b| \le (1.188\pm 0.044)\, \delta^d_{\text{max}} + (0.812\pm0.044)\, \delta^s_{\text{max}}$.  

As $\Delta m_{d,s}$ are precisely known, we plot the bound as a function of the width differences $\Delta\Gamma_{d,s}$. Because $\Delta\Gamma_{d}$ has not been measured yet, we set this to zero as the most conservative choice. If LHCb measures the difference $A_{\text{SL}}^s - A_{\text{SL}}^d$~\cite{Calvi:2011qq}, then the above bound with modified coefficients apply for that measurement as well.

In Fig.~\ref{1Dplot}, the darker shaded region shows the upper bound on $|A_\text{SL}^b|$ using the $1\sigma$ ranges for $|\Gamma_{12}^{d,s}|$ in the
SM~\cite{Lenz:2011ti}, and the lighter shaded region includes both
$2\sigma$ regions, with
\begin{equation}
 2|\Gamma_{12}^s| = (0.087 \pm 0.021)\, \text{ps}^{-1} \quad \text{and} \quad 2|\Gamma_{12}^d| = (2.74 \pm 0.51)\times 10^{-3}\, \text{ps}^{-1}\,.
\end{equation}
The dashed [dotted] curve shows the impact of using the $2\sigma$ region for
$\Gamma_{12}^d$ [$\Gamma_{12}^s$].  The vertical boundaries of the shaded regions arise because
$|\Delta\Gamma_s| > 2\, |\Gamma_{12}^s|$ is unphysical. A tension between
the $A_\text{SL}^b$ measurement and the bound is visible, independent of
the discrepancy between the D\O\ result and the global fit to the
latest available experimental data~\cite{Lenz:2012az}.

\begin{figure}[htp]
\centering  \includegraphics[scale=0.75]{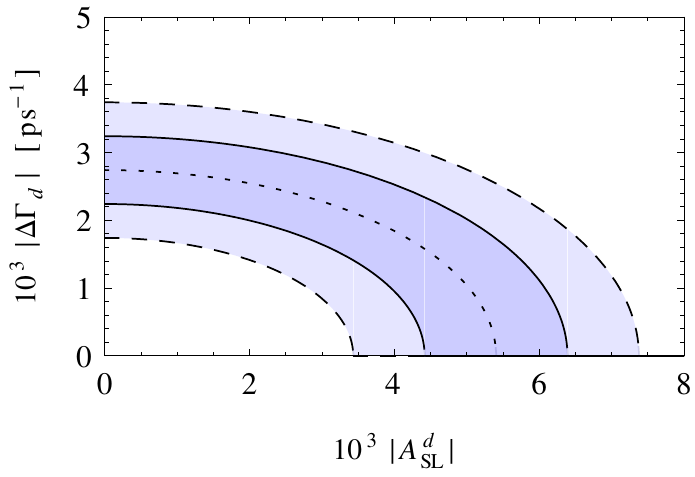}\hspace{1cm} \includegraphics[scale=0.75]{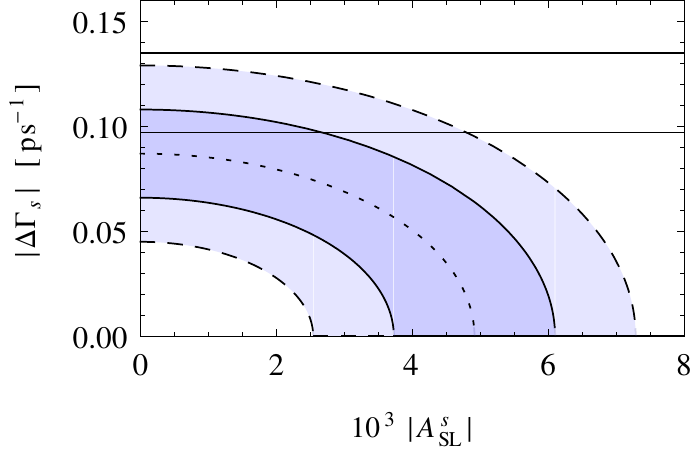}
    \caption{Left plot: the region allowed by Eq.~(\ref{unitrel}) in the $A_\text{SL}^d - \Delta\Gamma_d$ plane.  The SM calculation of $|\Gamma_{12}^{d,s}|$ at $1\sigma$ [$2\sigma$] gives the darker [lighter] shaded region. Right plot: same for $A_\text{SL}^s - \Delta\Gamma_s$; the straight lines show the $1\sigma$ range of the LHCb result for $\Delta\Gamma_s$.}\label{2Dplot}
\end{figure}

Of course, independent of the D\O\ measurement of $A_\text{SL}^b$, we can also
compare the bound implied by our relation to the individual best bounds on the
semi-leptonic asymmetries. To this end, in Fig.~\ref{2Dplot}
we plot $A_\text{SL}^d$ vs. $\Delta\Gamma_d$ (and similarly for $B_s$) allowed
by Eq.~\eqref{unitrel} and the $1\sigma$ and $2\sigma$ ranges of the SM
calculation of $|\Gamma_{12}|$~\cite{Lenz:2011ti}. Here, there have been no
discrepancies claimed between the theory predictions and measurements, but our
relation allows us to place a bound tighter than the current experimental
constraints which is more robust than the purely theoretical SM calculation as
outlined above. A non-zero observation of  $\Delta\Gamma_d$ will strengthen the upper bound on $A_\text{SL}^d$ obtained with the relation \eqref{unitrel}.

\subsection{No-Go theorem}

In a future publication \cite{FreytsisLigetiTurczyk} we will show, that besides the trivial conditions on the physical mixing parameters, algebraical relations and the unitarity bound no other consistency relations from physical considerations can appear. The no-go theorem does not apply for the relation presented in this talk, because we additionally assume knowledge about a parameter of the underlying Hamiltonian. We now sketch a physical understanding for the proof of this theorem.

From Eqs.~\eqref{ubound12} and \eqref{gammapos} it is obvious that the unitarity bound is exactly satured, if the two vectors of $B_0$ and $\bar B_0$, equivalently $B_L$ and $B_H$, are aligned: $\langle f | \mathrm{T} | B_H \rangle \propto  \langle f | \mathrm{T} | B_L \rangle$.

Without loss of generality we can start with an arbitrary, generic decaying two-state system in the Wigner-Weisskopf approximation, i.e. no strong interactions obscure the situation. We therefore choose an orthogonal, non CP violating system, which obeys the above required alignment of states as a starting point, which has $\delta \equiv 0$.

By adding arbitrary new UV physics, which does not necessarily need to be compatible with data, we can change $M_{12}$ independently of $\Gamma_{12}$, introducing a non-vanishing $\delta$. We can then vary $M_{12}$, however have to keep the mass and width of the physical states positive.

We can always saturate the unitarity bound, leaving no room for stronger constraints than the unitarity bound in any parameter space. In other words by relaxing the constraint of having no CP violation $\text{Arg } M_{12}=\text{Arg } \Gamma_{12} $, this gets replaced by a new constraint, the unitarity bound. Thus the total number of relations is conserved and without assuming knowledge no further bound or relation can be obtained. 

\section{Discussion}
We derived not an absolute bound in the fashion of the unitarity bound but a
relation between calculable and measured quantities.  It is thus worth
clarifying the relationship of our result to the stated $3.9\,\sigma$
disagreement of $A_\text{SL}^b$ with the SM reported in~\cite{Abazov:2011yk}. 

The SM prediction of $A_\text{SL}$ uses the calculation of $\Gamma_{12}$, and
$|\Gamma_{12}|$ also enters our bound; the discrepancies are thus correlated. 
Although the calculation of $|\Gamma_{12}|$ and $\text{Im}(\Gamma_{12})$ both
rely on the same operator product expansion and perturbation theory, the
existence of large cancellations in ${\rm Im}(\Gamma_{12})$ may lead one to
think that the uncertainties could be larger in its SM calculation than what is
tractable in the behavior of its next-to-leading order
calculation \cite{Ciuchini:2003ww, Beneke:2003az}. The sensitivity of
$\Gamma_{12}$ to new physics is generally weaker than that of $M_{12}$
(see \cite{Bai:2010kf, Bobeth:2011st} for other options).  Thus, it is
interesting to determine $\delta$ from Eq.~\eqref{unitrel}, besides its direct
calculation.

The relation  \eqref{unitrel} is a monotonic function in $y_{12}$ and thus an upper bound on this theory prediction implies an upper bound on $|\delta|$. Therefore this relation is much stronger for small values of $y_{12}$, as is e.g. present in the $B_d$ system. A non-zero observation of the width difference does improve the upper bound as well.

Now we are in the position to present numerical upper bounds for the individual asymmetries. We use $\Delta\Gamma_s$ from LHCb in Eq.~\eqref{deltaGs} and neglect
$\Delta\Gamma_d$ and find the $2 \sigma$ bounds
\begin{equation}\label{newbound}
  |A_{\text{SL}}^d| < 7.4 \times 10^{-3}, \qquad 
  |A_{\text{SL}}^s| < 4.2 \times 10^{-3}\,. 
\end{equation}
While this bound on $A_{\text{SL}}^s$ may seem to disagree with
Fig.~\ref{2Dplot}, note that in the plot the uncertainties of $\Gamma^s_{12}$
and $\Delta\Gamma_s$ are not combined. Propagating the uncertainties,
$|\Gamma_{12}^s|^2-(\Delta\Gamma_s)^2/4 < 0$ and thus $\delta^2$  is negative at the $1\sigma$ level, which is an unphysical result. Hence we compute the $2\sigma$ bounds in Eq.~\eqref{newbound}.  

The bound on $A_\text{SL}^s$ is better than the bounds of the measurements in section 1 by more than a factor of 3, while that for $A_{\text{SL}}^d$
is comparable. However, in the case of $B_d$ this is driven primarily by the
uncertainty in the lifetime difference. If a non-zero value of $\Delta\Gamma_d$
were observed, a better bound could be derived. It is worth emphasizing that
this implication goes in both directions, given that an observation of
$A_{\text{SL}}^d \neq 0$ may happen before that of $\Delta\Gamma_d \neq 0$.  Due
to Eq.~\eqref{unitrel}, as soon as one of the two is measured to be nonzero, the
other is constrained to be significantly smaller at worst and given a definite
prediction at best.

\section*{Acknowledgments}
S.T. thanks the organizers for their effort to host this conference and for providing financial support. ST~is supported by a DFG Forschungsstipendium under contract no.~TU350/1-1. 
\bibliography{turczyk}
\bibliographystyle{apsrev4-1}


%% file: Papers/vicente.tex

%
%
%
%
%
%

\chapter[Enhancing lepton flavor violation with the $Z$-penguin (Vicente)]{Enhancing lepton flavor violation with the $Z$-penguin}
\vspace{-2em}
\paragraph{A. Vicente}
\paragraph{Abstract}
We show that $Z$-penguin diagrams can give an enhancement for $l_i
\to 3 l_j$ decays and $\mu-e$ conversion in nuclei in many extensions
of the Minimal Supersymmetric Standard Model (MSSM). We demonstrate
this effect in two models, namely, the supersymmetric inverse seesaw
and R-parity violating supersymmetry. The non-decoupling behavior of
the $Z$-penguins is also briefly discussed.

\section{Introduction}
Flavor violation in the neutral lepton sector has been firmly
established by neutrino oscillation experiments. In contrast, in what
concerns to the charged lepton sector, no evidence has been found so
far, and only upper limits are known. This is the case of observables
such as $\mu \to e \gamma$, $\mu \to 3 e$ and $\mu-e$ conversion in
nuclei \cite{Abada:2011rg,Hoecker:2012nu,Deppisch:2012vj}.

In the case of supersymmetry (SUSY), one expects potentially large LFV
effects. Regarding $l_i \to 3 l_j$ decays, in the Minimal
Supersymmetric Standard Model (MSSM) it has been shown
\cite{Hisano:1995cp,Arganda:2005ji} that the photonic penguin diagram
gives the dominant contribution in large regions of parameter
space. In fact, the so-called \emph{dipole dominance} has been part of
the common lore regarding LFV in SUSY theories. In this case the
3-body decays $l_i \to 3 l_j$ have rates roughly a factor $\alpha$
maller than those of the 2-body channel $l_i \to l_j \gamma$. Thus,
usually it is concluded that the decays $l_i \to l_j \gamma$ are more
constraining than the decays $l_i \to 3 l_j$.

However, it has recently been noticed~\cite{Hirsch:2012ax} that this
expectation does not hold in extended models where new couplings are
present or where the particle content is larger than that of the
MSSM. In such scenarios, the $Z$-penguin contributions provide the
dominant contributions to LFV processes such as the 3-body decays $l_i
\to 3 l_j$ and $\mu - e$ conversion in nuclei. In the following, we
will review the physics behind the $Z$-penguin enhancement and present
two examples where this fact has been demonstrated numerically: (i)
trilinear R-parity violating SUSY~\cite{Dreiner:2012mx}, and (ii) the
supersymmetric inverse seesaw~\cite{Abada:2012cq}.

\section{Enhancing LFV with the $Z$-penguin}
\label{secvicente:ZLFV}

In order to understand the r\^ole of the $Z$-boson contributions in
the MSSM we begin with some simple mass scaling considerations:
consider the chargino-sneutrino 1-loop diagrams leading to $l_i \to 3
l_j$. The photon and $Z$-boson contributions can be written as
\begin{equation} \label{Fchar}
A_a^{(c)L,R} = \frac{1}{16 \pi^2 m_{\tilde{\nu}}^2} {\cal O}_{A_a}^{L,R}s(x^2) \quad \text{and}  \quad F_{X} = \frac{1}{16 \pi^2 g^2 \sin^2 \theta_W m^2_{Z}}{\cal O}_{F_X}^{L,R}t(x^2)\, ,
\end{equation}
respectively. Here $X=\left\{LL,LR,RL,RR\right\}$, ${\cal
  O}_{y}^{L,R}$ denote combinations of rotation matrices and coupling
constants and $s(x^2)$ and $t(x^2)$ represent loop functions which
depend on $x^2 = m_{\tilde{\chi}^-}^2/m_{\tilde{\nu}}^2$ (for exact
definitions see \cite{Arganda:2005ji}). The only mass scale involved
in the $A$ form factors is $m_{SUSY}$ (the photon being massless),
whereas the mass scale in the $F_X$ form factors is set by the
$Z$-boson mass, $m_Z$.  This implies the scalings $A \sim
m_{SUSY}^{-2}$ and $F \sim m_{Z}^{-2}$.  Assuming that all loop
functions, mixing matrices and coupling constants are of the same
order, one can estimate $F/A \sim 500$ for the arbitrary value
$m_{SUSY} = 300$ GeV. Therefore, from these considerations one
concludes that, in principle, $Z$-boson contributions should dominate.

However, a subtle cancellation between the different diagrams
contributing to the leading $Z$-contribution~\cite{Hirsch:2012ax}
imply that, in the MSSM, the photon penguin is found to be
dominant\cite{Hisano:1995cp,Arganda:2005ji}. The dominant contribution
to $l_i \to 3 l_j$ comes from diagrams where the leptons in the
external legs are left-handed (other cases are suppressed by the
Yukawa couplings of the charged leptons). This is given by the form
factor $F_{LL} \propto F_L$, where $F_L$ is the $Z-l_i-l_j$ 1-loop
effective vertex, with $i \ne j$. $F_L$ receives contributions from
many different 1-loop diagrams. However, let us focus on the
chargino-sneutrino contribution, typically the dominant one. Expanding
in the chargino mixing angle, $\theta_{\tilde{\chi}^\pm}$, one can
write
\begin{equation}
F_L = F^{(0)}_L + \frac{1}{2} \theta_{\tilde{\chi}^\pm}^2 F^{(2)}_L + \dots \quad .
\end{equation}
Notice that there is no term in the expansion involving
$\tilde{H}^\pm$ at the leading order, nor at the 1st order, since
there is no $\tilde{H}^\pm-\tilde{\nu}_L-l_L$ coupling. For this
reason, only the wino contributes at the zeroth order in
$\theta_{\tilde{\chi}^\pm}$. $F^{(0)}_L$ can be written as
$\left(F^{(0)}_L\right)_{ij} \equiv F^{(0)}_L = - \frac{1}{16 \pi^2}
\left( M^{ij}_{\text{wave}} + M^{ij}_{\text{p1}} + M^{ij}_{\text{p2}}
\right)$, where the three addends come from different types of
diagrams: wave function diagrams ($M_{\text{wave}}$) and penguins with
the $Z$-boson attached to the chargino line ($M_{\text{p1}}$) or to
the sneutrino line ($M_{\text{p2}}$).  The sum exactly vanishes, as
can be verified by grouping the different terms as
\begin{equation} \label{MSSM-cancellation}
F^{(0)}_L = \frac{g^2}{32 \pi^2} \left( g \, c_W Z_V^{xi*} Z_V^{xj} X_1^x 
+ g' s_W Z_V^{xi*} Z_V^{xj} X_2^x \right)
\end{equation}
with $X_1^x$ and $X_2^x$ different combinations of loop functions, in
principle dependent on the ratio
$x=m_{\tilde{\chi}^-}/m_{\tilde{\nu}}$, see~\cite{Hirsch:2012ax} for
exact definitions. Moreover, $Z_V$ is a $3 \times 3$ unitary matrix
that diagonalizes the mass matrix of the sneutrinos. We also use the
notation $c_W = \cos \theta_W$ and $s_W = \sin \theta_W$. Using the
exact expressions for the loop functions~\cite{Arganda:2005ji}, one
finds that the masses cancel out in $X_1^x$ and $X_2^x$ and these
combinations become independent of
$x=m_{\tilde{\chi}^-}/m_{\tilde{\nu}}$: $X_1^x = X_1 = - \frac{3}{4}$
and $X_2^x = X_2 = - \frac{1}{4}$, $\forall x$. Therefore, one is left
with $F^{(0)}_L \propto \sum_x Z_V^{xi*} Z_V^{xj} = \left( Z_V^\dagger
Z_V \right)^{ij}$, which vanishes for $i \ne j$ due to unitarity of
the $Z_V$ matrix. In conclusion, the first non-vanishing term in the
expansion appears at 2nd order in the chargino mixing angle, which
naturally suppresses the $Z$-mediated contributions. This is the
reason why the photon contributions turn out to be dominant in the
MSSM.

However, there are many cases where the cancellation of the zeroth
order term in the expansion no longer holds~\cite{Hirsch:2012ax}. For
instance, the introduction of new interactions for the leptons
modifies the previous conclusion and enhances $F_L$ by a huge
factor. Furthermore, although the previous discussion has been focused
on $l_i \to 3 l_j$, the same enhancement in the $Z-l_i-l_j$ effective
vertex also affects other observables which are mediated by $Z$-boson
exchange.  This is the case for $\mu-e$ conversion in nuclei
\cite{Arganda:2007jw} and $\tau \to P l_i$, where $P$ is a
pseudoscalar meson \cite{Arganda:2008jj}.

\section{Beyond the MSSM}
\label{subsecvicente:beyond}

We turn now to particular examples of models where the cancellation in
the $Z$-boson zeroth order contribution does not hold.

\subsection{Trilinear R-parity violation}
\label{subsecvicente:RPV}

The impact of the $Z$ penguins in the MSSM extended with trilinear
R-parity violation (RPV) was considered in~\cite{Dreiner:2012mx}.  The
superpotential of the model includes the lepton number violating terms
\cite{Hall:1983id}
\begin{equation}
\label{eq:SuperpotentialRPV}
{\mathcal W}_{\text{RPV}} = \frac{1}{2} \lambda_{ijk} \widehat{L}_i
\widehat{L}_j \widehat{E}^c_k + \frac{1}{2} \lambda^{'}_{ijk} \widehat{L}_i
\widehat{Q}_j \widehat{D}^c_k \, .
\end{equation}
Bounds for these trilinear couplings have been set so far not only by
using lepton flavor violating decays, but also $\mu - e$ conversion in
nuclei or cosmological observations.  This lead to limits on
individual couplings or specific products of couplings
\cite{Barbier:2004ez}. However, the $Z$-penguins were not considered
in most of the studies. In contrast, the computations in
Ref.~\cite{Dreiner:2012mx} include all contributions: photonic and
$Z$-penguins as well as Higgs penguins and box diagrams.

\begin{table} 
\centering
\begin{tabular}{|c||c|c|c|c|} 
\hline 
Coupling & $l_i \to l_j  \gamma$ & $l_i \to 3 l_j$ & $ \tau \to l_i P / \mu-e $ & $ Z^0 \to l_i l_j $  \\  
\hline 
 $|\lambda_{123} \lambda_{133}|$ & $1.8\times 10^1$ & $1.2\times 10^{-2}$  & $2.8\times 10^{-2}$  & $1.4\times 10^1$  \\ 
 $|\lambda_{123} \lambda_{233}|$ & $1.3\times 10^1$ & $1.4\times 10^{-2}$  & $2.4\times 10^{-2}$  & $4.\times 10^1$  \\ 
 $|\lambda_{132} \lambda_{232}|$ & $2.4\times 10^{-1}$ & $3.5\times 10^{-5}$  & $8.4\times 10^{-6}$  & $1.7\times 10^1$  \\ 
 $|\lambda_{133} \lambda_{233}|$ & $1.7\times 10^{-3}$ & $3.5\times 10^{-5}$  & $8.4\times 10^{-6}$  & $1.7\times 10^1$  \\ 
 $|\lambda_{231} \lambda_{232}|$ & $9.5\times 10^{-4}$ & $2.5\times 10^{-5}$  & $5.7\times 10^{-6}$  & $2.3\times 10^1$  \\ 
 $|\lambda'_{122} \lambda'_{222}|$ & $4.5\times 10^{-4}$ & $3.9\times 10^{-5}$  & $9.3\times 10^{-6}$  & $7.5\times 10^{-1}$  \\ 
 $|\lambda'_{123} \lambda'_{223}|$ & $4.6\times 10^{-4}$ & $3.9\times 10^{-5}$  & $9.3\times 10^{-6}$  & $7.5\times 10^{-1}$  \\ 
 $|\lambda'_{132} \lambda'_{232}|$ & $4.9\times 10^{-4}$ & $4.1\times 10^{-5}$  & $9.8\times 10^{-6}$  & $1.4$  \\ 
 $|\lambda'_{133} \lambda'_{233}|$ & $4.9\times 10^{-4}$ & $4.1\times 10^{-5}$  & $9.8\times 10^{-6}$  & $1.4$  \\ 
 $|\lambda'_{133} \lambda'_{333}|$ & $1.3\times 10^{-1}$ & $1.6\times 10^{-2}$  & $2.8\times 10^{-2}$  & $3.3$  \\ 
 $|\lambda'_{233} \lambda'_{333}|$ & $1.5\times 10^{-1}$ & $1.4\times 10^{-2}$  & $3.3\times 10^{-2}$  & $3.6$  \\ 
 \hline 
\end{tabular} 
\caption{Limits for the benchmark point 10.4.1 of
  Ref.~\cite{AbdusSalam:2011fc} ($m_0 = 750$ GeV, $M_{1/2} = 350$ GeV,
  $\tan \beta = 10$, $A_0 = 0$, $\mu>0$) on different combinations of
  $LLE$ and $LQD$ operators derived from experimental limits on LFV
  observables.}
\label{tab:loopResults_bp2} 
\end{table} 

As explained in section \ref{secvicente:ZLFV}, the relative importance
of the $Z$-penguin contributions increases for large supersymmetric
masses. Therefore, one does not expect a great improvement in the
bounds by including $Z$-mediated observables when SUSY is
light~\cite{Dreiner:2012mx}. In contrast, when supersymmetric
particles are heavy one expects LFV to be clearly dominated by
$Z$-boson exchange.  This is due to the fact that photonic penguins
scale as $m_{SUSY}^{-4}$ \cite{Hirsch:2012ax}, whereas $Z$-penguins
are much less sensitive to the SUSY scale. In order to show this
explicitly, we have considered the Constrained Minimal Supersymmetric
Standard Model (CMSSM) benchmark point 10.4.1 of
Ref.~\cite{AbdusSalam:2011fc}, defined by the set of parameters $m_0 =
750$ GeV, $M_{1/2} = 350$ GeV, $\tan \beta = 10$, $A_0 = 0$ and
$\mu>0$. This parameter point is perfectly valid regarding bounds from
direct SUSY searches at the LHC. After the calculation of the
resulting MSSM spectra, we switched on the different combinations of
the RPV couplings which can open flavor violating transitions and
calculated the different observables.

Our results
confirm the theoretical expectation, see table
\ref{tab:loopResults_bp2}. In this case the enhancement given by the
$Z$-penguins leads to a scenario where the most stringent bounds are
obtained from $Z$-mediated observables. In particular, $\mu-e$
conversion in nuclei turns out to be very constraining, with bounds
derived from this observable orders of magnitude better than those
obtained from $\mu \to e \gamma$. Similarly, $l_i \to 3 l_j$ gets also
large enhancements, and is only a bit less constraining than $\mu-e$
conversion in nuclei. The main reason for this is the very good
experimental limit for $\mu-e$ conversion in gold.

In conclusion, we have shown that the $Z$-penguins dominate in most
parts of parameter space, and especially for heavy SUSY spectra, the
amplitudes for $l_i \to 3 l_j$, $\tau \to l_i P$ and $\mu-e$
conversion in nuclei. Therefore, the limits on combinations of
$\lambda$ and $\lambda'$ couplings given by these observables can be
improved by several orders of magnitude with respect to the bounds
already present in the literature.

\subsection{Inverse Seesaw}
\label{subsecvicente:inverse}

In the supersymmetric inverse seesaw \cite{Mohapatra:1986bd} three
pairs of singlet superfields, $\widehat{\nu}^c_i$ and $\widehat{X}_i$
($i=1,2,3$) with lepton numbers assigned to be $-1$ and $+1$,
respectively, are added to the superfield content. The superpotential
can be written as
\begin{equation}
{\mathcal W}_{\text{ISS}}= {\mathcal W}_{\text{MSSM}} + Y^{ij}_\nu
\widehat{\nu}^c_i \widehat{L}_j \widehat{H}_u
+M_{R_{ij}}\widehat{\nu}^c_i\widehat{X}_j+\frac{1}{2}\mu_{X_{ij}}\widehat{X}_i\widehat{X}_j
\, ,
\label{eq:SuperPotISS}
\end{equation}
where $i,j = 1,2,3$ are generation indices.

The active neutrinos mix with the singlets through the Dirac mass term
$m_D= \frac{1}{\sqrt 2} Y_\nu v_u$. Assuming $m_D,\mu_X \ll M_R$, the
diagonalization of the full $9\times9$ mass matrix leads to an
effective Majorana mass matrix for the light neutrinos, $m_\nu = m_D^T
{M_R^{T}}^{-1} \mu_X M_R^{-1} m_D = \frac{v_u^2}{2} Y^T_\nu M^{-1}
Y_\nu$, where, in analogy to a type-I seesaw, we define $M^{-1} =
{M_R^{T}}^{-1} \mu_X M_R^{-1}$ as an effective right-handed neutrino
mass matrix\footnote{In our numerical analysis we always consider
  scenarios where $M =
  \text{diag}(\hat{M},\hat{M},\hat{M})$. Therefore, for the sake of
  brevity, we use the simple notation $M$ both for the matrix and its
  eigenvalues.}. The smallness of the light neutrino masses is
controlled by the size of $\mu_X$.  Hence the lepton number conserving
mass parameters $m_D$ and $M_R$ can easily accommodate large Yukawa
couplings and a right-handed neutrino mass scale around the TeV. These
features lead to a very rich LFV
phenomenology~\cite{Deppisch:2004fa,Deppisch:2005zm,Hirsch:2009ra,Abada:2011hm,Hirsch:2012kv}.

\begin{figure}
\centering
\includegraphics[width=0.45\linewidth]{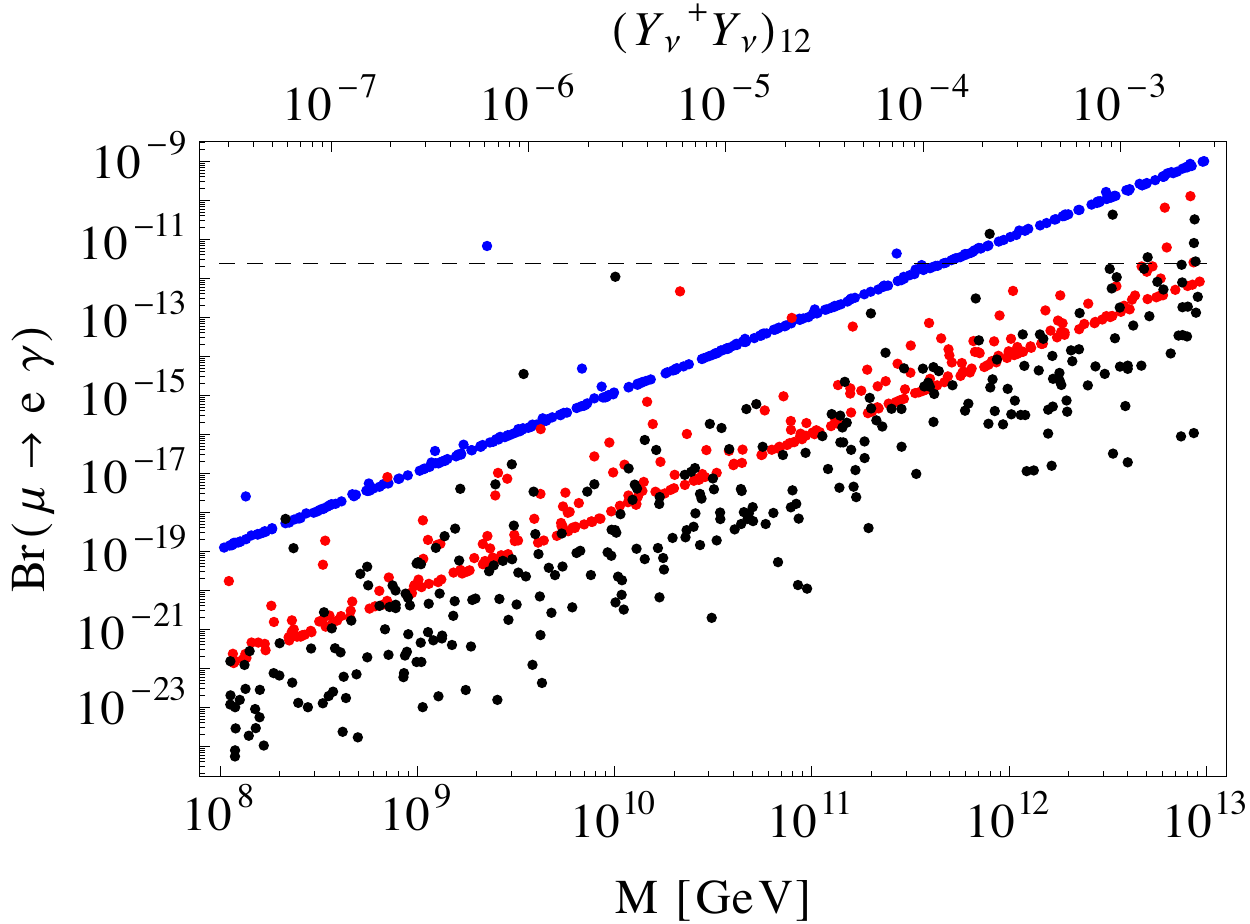}
\includegraphics[width=0.45\linewidth]{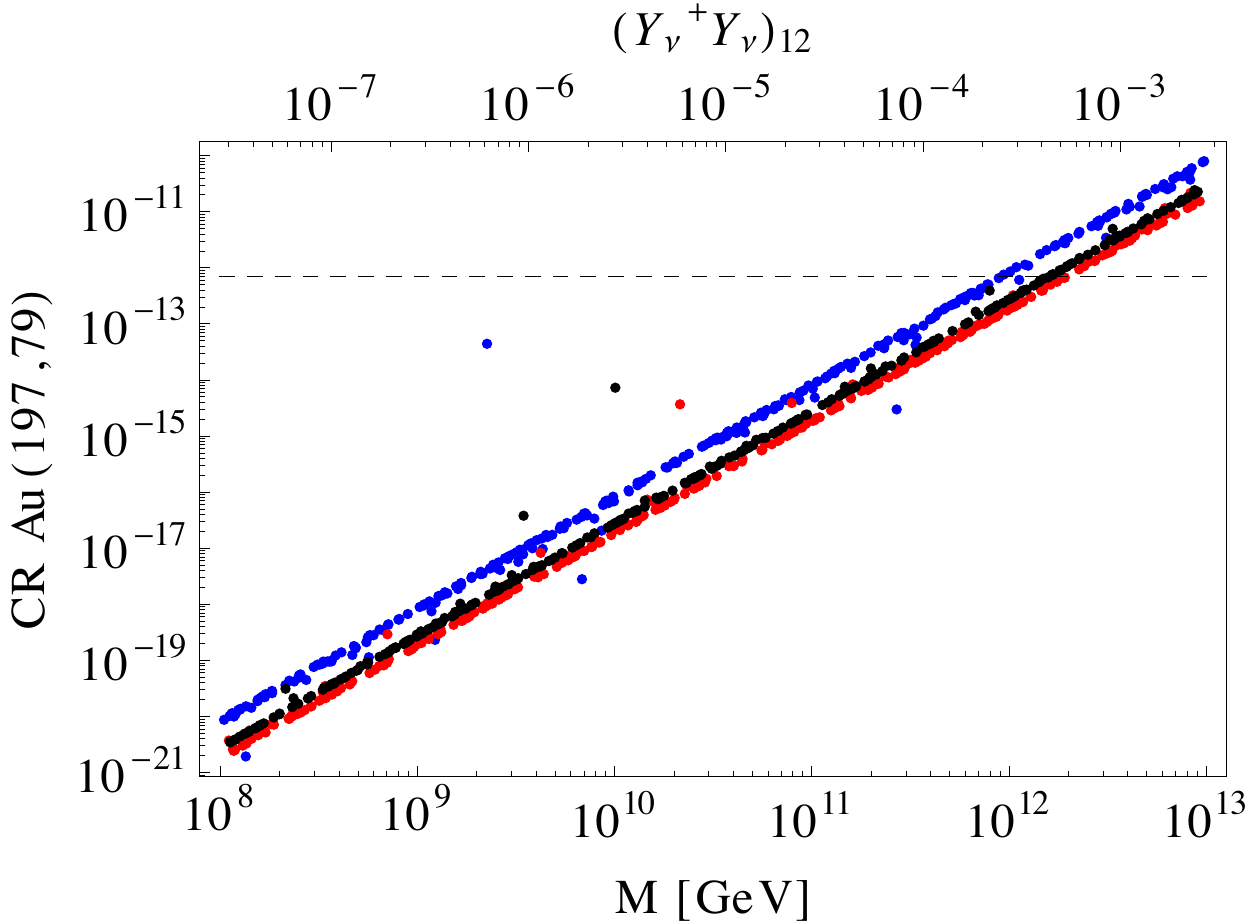}
\caption{$\text{Br}(\mu \to e \gamma)$ (left) and $\mu-e$ conversion
  rate in $^{197}_{79}\text{Au}$ (right), as a function of $M$ and
  $(Y_\nu^\dagger Y_\nu)_{12}$. The horizontal dashed lines represent
  the current experimental bounds.}
\label{vicenteinversefig}
\end{figure}

Although in principle the SUSY inverse seesaw allows for $Y_\nu\sim
{\mathcal{O}}(1)$, LFV sets very important constraints on the
(off-diagonal) Yukawa couplings~\cite{Abada:2012cq}. This is
demonstrated in figure \ref{vicenteinversefig}, where we show our
numerical results for two LFV observables, $\text{Br}(\mu \to e
\gamma)$ (left) and $\mu-e$ conversion rate in $^{197}_{79}\text{Au}$
(right), as a function of $M$ and $(Y_\nu^\dagger Y_\nu)_{12}$. A
degenerate right-handed neutrino spectrum has been assumed at
$m_{SUSY}$, with $M_R = \hat{M}_R \, \mathbb{1}$ and $\mu_X =
\hat{\mu}_X \, \mathbb{1}$, where $\mathbb{1}$ is the $3 \times 3$
identity matrix. Three different values of $\hat{M}_R$ are shown in
figure \ref{vicenteinversefig}: $\hat{M}_R = 100$ GeV (blue),
$\hat{M}_R = 1$ TeV (red) and $\hat{M}_R = 10$ TeV (black). Moreover,
CMSSM-like boundary conditions have been assumed at the grand
unification (GUT) scale. We set $A_0 = -300$ GeV, $\tan \beta = 10$,
$\text{sign}(\mu) = +$ and we randomly vary $m_0$ and $M_{1/2}$ in the
range [$0,3$ TeV].

Let us first concentrate on $\text{Br}(\mu \to e \gamma)$. For low
$\hat{M}_R$, this observable has very little dependence on $m_0$ and
$M_{1/2}$, whereas for large $\hat{M}_R$, one can find very large
variations due to the different values of the SUSY masses. The
non-SUSY contributions become relevant only for $\hat{M}_R < 1$ TeV
and, in fact, for $\hat{M}_R = 100$ GeV they totally dominate, so that
all dependence on $m_0$, $M_{1/2}$ and on the rest of CMSSM parameters
disappears \cite{Deppisch:2004fa}. In contrast, for large $\hat{M}_R$,
SUSY contributions dominate, and the usual $m_{SUSY}^{-4}$ dependence
is found.

In what concerns $\mu-e$ conversion in gold, the behavior is quite
different. Note that there is a sharp correlation with $M$, hardly
distorted by the changes in $m_0$ and $M_{1/2}$. For the red and black
dots (associated to $\hat{M}_R \ge 1$ TeV), this is a consequence of
$Z$-boson dominance, that shows very little dependence on $m_{SUSY}$.
It is remarkable that, apart from the case $\hat{M}_R = 100$ GeV, the
limits on $(Y_\nu^\dagger Y_\nu)_{12}$ (or $M$) obtained from $\mu-e$
conversion in nuclei are more stringent than those obtained from $\mu
\to e \gamma$.

Finally, we emphasize one crucial aspect of the $Z$-penguins. As
observed in the numerical results, $Z$-mediated processes
exhibit a non-decoupling behavior and large supersymmetric masses do
not suppress the charged lepton flavor violating signatures induced by
$Z$-boson exchange. One can confirm this result by analytical
computation of the 1-loop $Z-l_i-l_j$ effective vertex, $F_L$, with $i
\neq j$. In case of the supersymmetric inverse seesaw one
finds~\cite{Abada:2012cq}
\begin{equation}\label{FL0final}
F_L \simeq F^{(0)}_L = \frac{g}{8 c_W} \left( Y_\nu^\dagger Y_\nu \right)_{ij} \left( c_W^2 - \frac{1}{2} \right).
\end{equation}
This result does not contain any dependence on supersymmetric
parameters, thus providing an analytical cross-check of the
non-decoupling behavior found in the numerical results.

\section{Conclusion}
It has been shown that the $Z$-penguin can give the dominant
contribution to lepton flavor violating processes in many models. As
numerical examples, this has been explicitly demonstrated in
supersymmetry with trilinear R-parity violation and the supersymmetric
inverse seesaw. Finally, the non-decoupling behavior of the
$Z$-penguins has been briefly discussed.

\section*{Acknowledgments}
I am very grateful to Asmaa Abada, Debottam Das, Herbi Dreiner, Martin
Hirsch, Kilian Nickel, Florian Staub and C\'edric Weiland for fruitful
collaboration on the works this talk is based on. I also acknowledge
support by the ANR project CPV-LFV-LHC {NT09-508531}.

\bibliography{vicente}
\bibliographystyle{apsrev4-1}


%% file: Papers/JavierVirto.tex

%
%
%
%
%
%

\chapter[New Physics constraints from optimized observables in $B\to K^* \mu^+\mu^-$ at large recoil (Descotes-Genon, Matias, \textit{Virto})]{New Physics constraints from optimized observables in $B\to K^* \mu^+\mu^-$ at large recoil}
\vspace{-2em}
\paragraph{S. Descotes-Genon, J. Matias, \textit{J. Virto}}
\paragraph{Abstract}

$B\to K^* \mu^+\mu^-$ angular observables have become a key ingredient in global model-independent analyses of $b\to s$ transitions. However, as experimental precision improves, the use of theoretically clean quantities becomes a crucial issue. Global analyses that use \emph{clean} observables integrated in small bins are already a reality, opening up a new chapter in our quest for New Physics.


\section{Status of $\ b\to s\ \ell^+\ell^-$ decays}

During the last few years, intensive theoretical work and impressive experimental results --and even more impressive prospects-- have pushed our understanding on $b\to s\ \ell^+\ell^-$ decays far beyond expectations. Among the large set of inclusive and exclusive $b\to s\ \ell^+\ell^-$ modes, a considerable attention has been put into $B\to K^{(*)}\mu^+\mu^-$, where experimental analyses are based on ${\cal O}(10^2)$ events in the case of the B-factories and CDF \cite{0904.0770,1204.3933,1108.0695} and ${\cal O}(10^3)$ events at LHCb \cite{LHCbinned}. Also, recent bounds on the $B_s\to \mu^+\mu^-$ branching ratio \cite{Bsmumu} are very close to the SM prediction, taking into account the correction from the $B_s$ width difference to branching ratio measurements at LHCb \cite{1111.4882,1204.1735+1204.1737}. These modes are very sensitive to New Physics contributing to right-handed currents (since transverse asymmetries in $B\to K^{(*)}\mu^+\mu^-$ measure indirectly the polarization of the virtual photon) and to scalar and pseudo-scalar operators (specially in the case of $B_s\to \mu^+\mu^-$).
 
 The theoretical description of $b\to s\ \ell^+\ell^-$ decays within and beyond the SM is given by the $\Delta B=-\Delta S=1$ effective Hamiltonian ${\cal H}_{\rm eff}=\sum_i C_i {\cal O}_i$ \cite{Chetyrkin:1996vx,Bobeth:1999mk}. In the SM the relevant operators are the electromagnetic dipole and semileptonic operators ${\cal O}_7$, ${\cal O}_9$ and ${\cal O}_{10}$. The 4-quark current-current ${\cal O}_{1,2}^{u,c}$ and QCD-penguin ${\cal O}_{3,4,5,6}$ operators and the chromo-magnetic dipole operator ${\cal O}_8$ are involved at higher orders in perturbation theory. Beyond the SM, non-standard operators become important, such as the chirality-flipped operators ${\cal O}_{7,9,10}'$, scalar ${\cal O}_{S^{(\prime)}}$, pseudo-scalar ${\cal O}_{P^{(\prime)}}$ and tensor ${\cal O}_{T,T5}$ operators.

 The theoretical description of the $B\to K^{*}\mu^+\mu^-$ decay becomes uncontrollable when the invariant dilepton mass $q^2$ approaches the threshold of $q\bar q$ resonance production. This happens predominantly in the vicinity of the $\psi$ and $\psi'$ $c\bar c$ states, around $q^2\sim 8-15$ GeV$^2$. The theoretical methods used to describe the regions below (low-$q^2$) and above (large-$q^2$) the vetoed range are different. The effect of a finite width of the $K^*$ including two scalar resonances has been addressed in Ref.~\cite{Becirevic:2012dp}, pointing to a non-negligible impact in some observables at low-$q^2$. However, an experimental fit to certain \emph{folded} angular distributions can decouple these effects \cite{mat}.
 
 At large recoil, the transversity amplitudes are computed in QCD factorization in the large energy limit of the $K^*$ \cite{0106067,0412400}. In this limit, symmetry relations between the seven heavy-to-light form factors allow one to express the amplitudes in terms of two soft form factors $\xi_{\|,\bot}$, distribution amplitudes and calculable hard kernels up to ${\cal O}(\alpha_s,\Lambda_{QCD}/m_b)$ \cite{Charles:1998dr} (see also \cite{Beneke:2000wa}). Soft gluon contributions at the tail of $c\bar c$ resonances in the low-$q^2$ region have been computed in Ref.~\cite{1006.4945}. Corrections of order ${\cal O}(\Lambda_{QCD}/m_b)$ are unknown, and include symmetry-breaking contributions to form factor relations and non-factorizable contributions from distribution amplitudes.

 Currently, the main uncertainties in the prediction of angular observables in $B\to K^{*}\mu^+\mu^-$ are due to unknown ${\cal O}(\Lambda_{QCD}/m_b)$ corrections and hadronic uncertainties in form factor computations from light-cone sum rules \cite{1006.4945,ffs}. The efforts to reduce these uncertainties in phenomenological applications have  led to the identification of \emph{clean} or \emph{optimized} observables, defined as ratios where most of the dependence on form factors cancels. A complete list of such observables is given by $A_T^{(2,3,4,5)}$ \cite{AT},  $A_T^{\rm (re, im)}$ \cite{1106.3283}, $P_{1,2,3}$, $P_{4,5,6}^{(\prime)}$, $M_{1,2}$ and $S_{1,2}$ \cite{1202.4266,1207.2753} at low-$q^2$ and $H_T^{(2,3,4,5)}$ \cite{1006.5013} at high-$q^2$. Experimental analyses have focused on the measurements of the branching ratio $BR$, the forward-backward asymmetry $A_{\rm FB}$, the longitudinal polarization fraction $F_L$, $A_{im}$ and $S_3$ (see \cite{0811.1214}), always integrated in a series of $q^2$ bins. CDF has measured directly the optimized observable $A_T^{(2)}\equiv P_1$ \cite{1108.0695}, while the observables $P_{1,2,3}$ can be obtained indirectly from the LHCb results (see Ref.~\cite{1207.2753} and Table \ref{JVPis}).

A wealth of model-independent combined analyses of $b\to s\gamma$ and $b\to s\ell^+\ell^-$ decays have appeared recently in the literature \cite{1207.2753,1104.3342+1202.2172,1105.0376+1111.2558,1111.1257+1206.0273,1205.1838,1206.1502}. The differences include the statistical treatment, the set of observables included in the analysis, and the NP scenarios considered. All in all, the data is compatible with the SM, as well as with the flipped-sign point $C_{7,9,10}=-C_{7,9,10}^{\rm \scriptscriptstyle SM}$. For a more thorough status review of $b\to s\ \ell^+\ell^-$ see Refs.~\cite{1208.3057,1208.3355}.

\section{Clean observables in $B\to K^* \mu^+\mu^-$ at large recoil}

\begin{table}
\centering
\begin{tabular}{lcrcr}
\hline\hline
Observable && Experiment && SM prediction \\
\hline\hline
$\langle P_1 \rangle_{[2,4.3]}$ && $-0.19 \pm 0.58$ && $-0.051 \pm 0.050$ \\
\hline
$\langle P_1 \rangle_{[4.3,8.68]}$ && $0.42 \pm 0.31$ && $-0.117\pm 0.059$ \\
\hline
$\langle P_1\rangle_{[1,6]}$ && $0.29 \pm 0.47$ && $-0.055 \pm 0.051$\\
\hline\hline
$\langle P_2\rangle_{[2,4.3]}$ && $0.51 \pm 0.27$ && $0.232 \pm 0.069$\\
\hline
$\langle P_2\rangle_{[4.3,8.68]}$ && $-0.25 \pm 0.08$ && $-0.405\pm 0.064$ \\
\hline
$\langle P_2\rangle_{[1,6]}$ && $0.35 \pm 0.14$ && $0.084 \pm 0.066$\\
\hline\hline
$\langle P_3\rangle_{[2,4.3]}$ && $0.08 \pm 0.35$ && $-0.004 \pm 0.024$\\
\hline
$\langle P_3\rangle_{[4.3,8.68]}$ && $-0.05 \pm 0.16$ && $-0.001\pm 0.027$ \\
\hline
$\langle P_3\rangle_{[1,6]}$ && $-0.21 \pm 0.21$ && $-0.003\pm 0.024$\\
\hline\hline
\end{tabular}
\caption{Experimental values for the clean observables $P_1$, $P_2$ and $P_3$ within different $q^2$-bins, extracted from the measurements of $S_3$, $A_{\rm im}$, $A_{\rm FB}$ and $F_L$, and their SM predictions.}
\label{JVPis}
\end{table}

Based on the symmetries of the angular distribution discussed in Ref.~\cite{1005.0571}, the minimum number of observables needed to describe the full $B\to K^{*}\mu^+\mu^-$ angular distribution can be inferred, which varies depending on whether mass and/or scalar effects are considered. A \emph{basis} of angular observables can then be identified, with the property of containing a minimum number of observables from which \emph{any} other observable can be obtained.
The basis is not unique, but there is a subset of bases with a quality feature: they contain a maximum number of \emph{clean} observables. One such bases has been constructed and studied in detail in Refs.~\cite{1202.4266,1207.2753}:
\begin{equation}
O=\Big\{\frac{d\Gamma}{dq^2},A_{\rm FB},P_1,P_2,P_3,P'_4,P'_5,P'_6,M_1,M_2,S_1,S_2\Big\}\ .
\end{equation}
All but $d\Gamma/dq^2$ and $A_{\rm FB}$ are clean observables, whereas\footnote{The scalar observables $S_{1,2}$ here should not be confused with the observables in Ref.~\cite{0811.1214}.} $S_i$ vanish in the absence of contributions from scalar operators, and $M_i$ go to zero in the limit of zero lepton masses. While the observables $M_2$ and $S_{1,2}$ are very much constrained by the $B_s\to \mu^+\mu^-$ branching ratio, the rest shows a good sensitivity to New Physics, especially $P_{1,2}$, $P'_{4,5}$ \cite{1202.4266,1207.2753}. From the latest LHCb measurements of $B\to K^{*}\mu^+\mu^-$ \cite{LHCbinned}, experimental results can be derived for $P_{1,2,3}$ integrated in the bins $[2,4.3]$, $[4.3,8.68]$ and $[1,6]$ GeV$^2$. The experimental numbers together with the theoretical predictions are displayed in Table \ref{JVPis}.

\begin{figure}
\centering
\includegraphics[height=5cm,width=7cm]{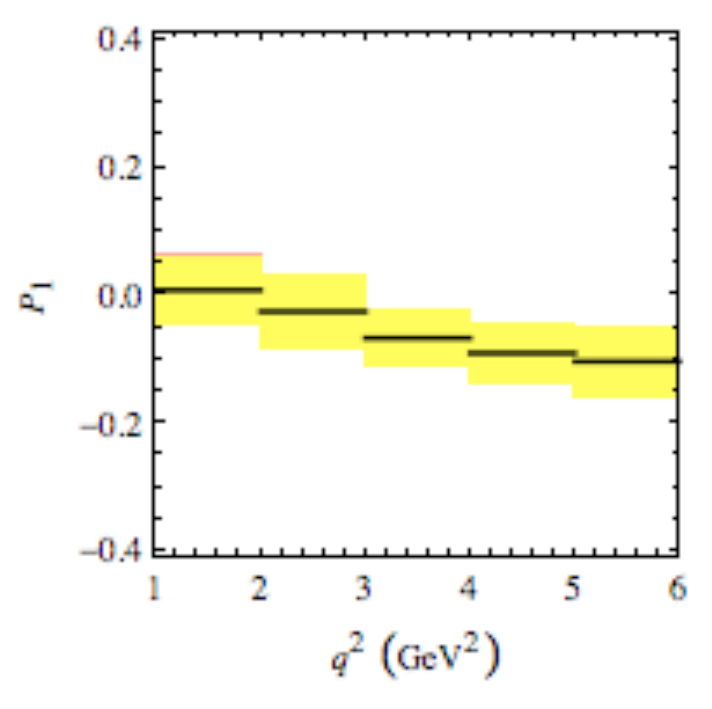}\hspace{0.5cm}
\includegraphics[height=5cm,width=7cm]{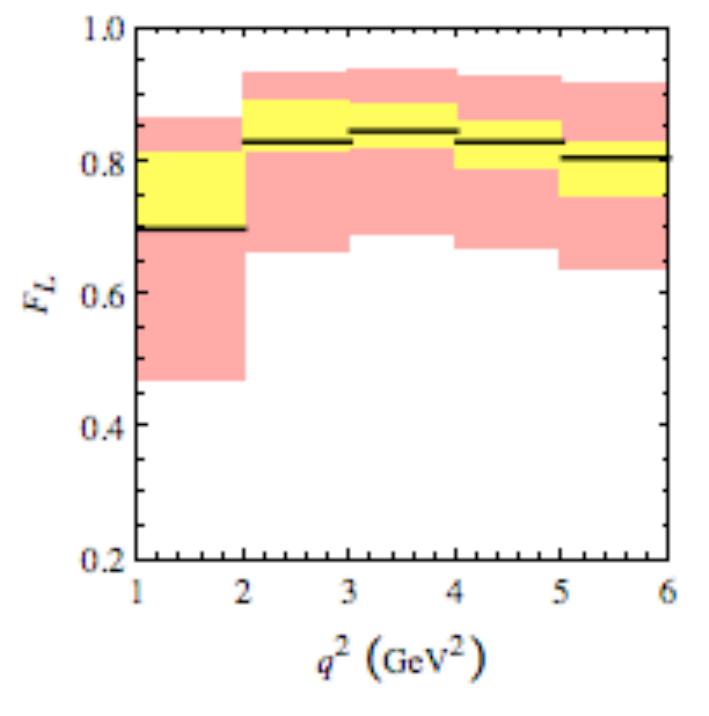}\\
\includegraphics[height=5cm,width=6.5cm]{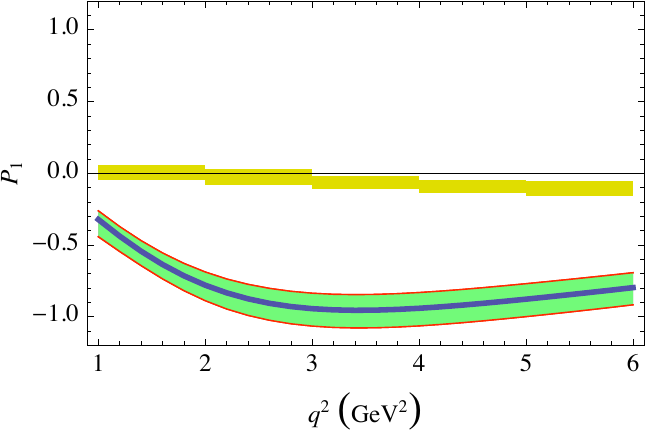}\hspace{1cm}
\includegraphics[height=5cm,width=6.5cm]{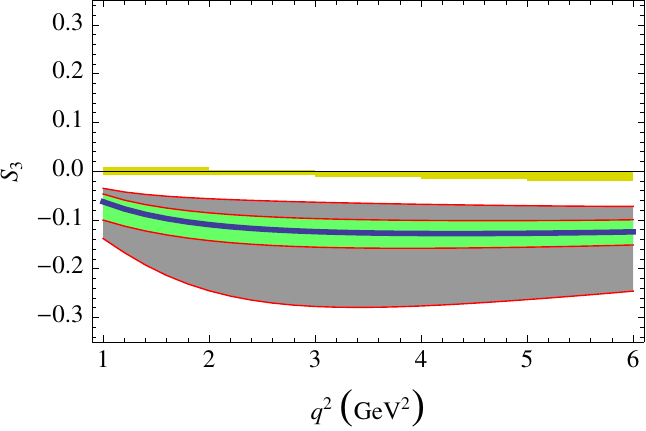}
\caption{Comparison between the observables $P_1$, $F_L$ and $S_3$ concerning their dependence on hadronic uncertainties. See the text for details.
}
\label{JVP1S3fig}
\end{figure}

The suppressed dependence on hadronic uncertainties of the clean observables $P_i$ compared to other observables can be checked directly. In the upper plots in Fig.~\ref{JVP1S3fig}, we show the SM predictions for $P_1$ and $F_L$, including all uncertainties, with the form factors taken from Ref.~\cite{ffs} (yellow) and from Ref.~\cite{1006.4945} (red) --this last reference is more conservative in the error treatment--. The conclusion is that, while $P_1$ is basically insensitive to this choice, the theoretical error in $F_L$ can vary by more than a factor of 2, with uncertainties up to a 30\%. In the lower plots in Fig.~\ref{JVP1S3fig}, a similar comparison is performed between $P_1$ and the corresponding observable $S_3$ of Ref.~\cite{0811.1214}. In this case the yellow boxes are the SM predictions, the blue curve is a NP benchmark point consistent with all other data (benchmark point `$b2$' in \cite{1207.2753}), with the green band corresponding to the total uncertainty taken the form factors in \cite{ffs} and the gray band for the form factors in \cite{1006.4945}. While the observable $S_3$ is protected from form factor uncertainties near the SM point, we can see that this is no longer true around other allowed regions in the parameter space. While this benchmark point is clearly discernible from the SM measuring $P_1$ with a 20\% error, a measurement of $S_3$ will hardly bring a definite conclusion. These examples demonstrate the importance of focusing on clean observables.

\begin{figure}
\centering
\includegraphics[height=7.2cm]{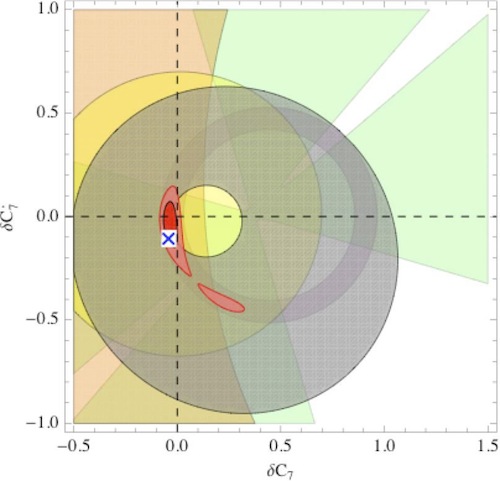}\hspace{0.1cm}
\includegraphics[height=7.2cm]{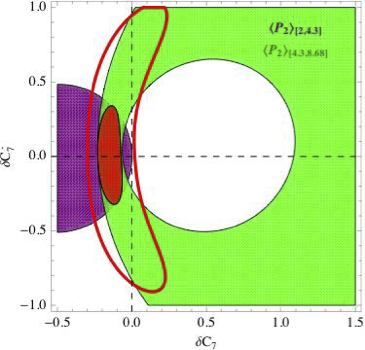}
\caption{Left: 68.3\% and 95.5\% C.L. constraints on $C_7,C_7'$ from $BR(B\to X_s\gamma)$, $S_{K^*\gamma}$, $A_I(B\to K^*\gamma)$, $BR(B\to X_s\mu^+\mu^-)$, $\langle A_{\rm FB} \rangle_{[1,6]}$ and $\langle F_L \rangle_{[1,6]}$. Right: 68.3\% and 95.5\% C.L. constraints on $C_7,C_7'$ from $\langle P_2 \rangle_{[2,4.3]}$ and $\langle P_2 \rangle_{[4.3,8.68]}$. The notation is $C_7=C_7^{\rm SM}+\delta C_7$, and similarly for $C_7'$.
}
\label{JVcons}
\end{figure}

\section{Model-Independent constraints}

A combined model-independent analysis including constraints from $BR(B\to X_s\gamma)$, $S_{K^*\gamma}$, $A_I(B\to K^*\gamma)$, $BR(B\to X_s\mu^+\mu^-)$, together with binned observables in $B\to K^* \mu^+\mu^-$ at low-$q^2$ has been presented in Ref.~\cite{1207.2753}. $B\to K^* \mu^+\mu^-$ observables include the forward-backward asymmetry, $F_L$, and $P_{1,2,3}$.

In Fig.~\ref{JVcons} (left) we show the 68.3\% and 95.5\% C.L. combined constraints on $C_7,C_7'$ from $BR(B\to X_s\gamma)$, $S_{K^*\gamma}$, $A_I(B\to K^*\gamma)$, $BR(B\to X_s\mu^+\mu^-)$, $\langle A_{\rm FB} \rangle_{[1,6]}$ and $\langle F_L \rangle_{[1,6]}$. In the right plot of the same figure, the constraints from $\langle P_2 \rangle_{[2,4.3]}$ and $\langle P_2 \rangle_{[4.3,8.68]}$ are shown. While the experimental numbers for $\langle P_2\rangle_{\rm bin}$ must be still improved considerably (the values used do not include correlations), the constraints from $\langle P_2\rangle_{\rm bin}$ are already interesting in comparison with the combined constraints from the other observables. Both bins point towards negative $\delta C_7$. This result is not affected by form factor uncertainties.

\begin{figure}
\centering
\includegraphics[height=8cm]{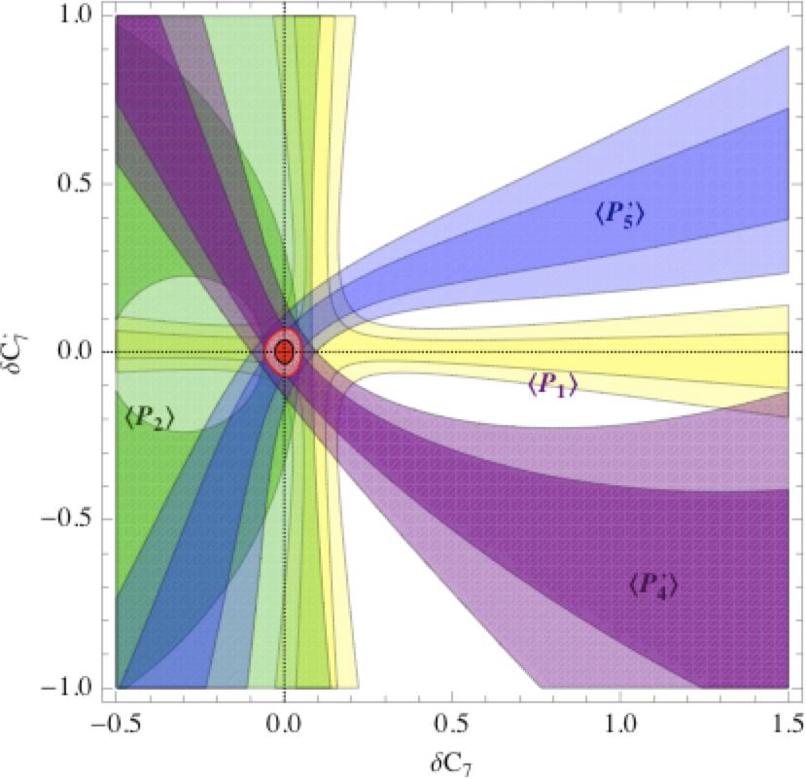}
\caption{Future scenario for constraints from $P_i$ observables.}
\label{JVfut}
\end{figure}

To finish, we comment on the prospects for constraints in the $C_7-C_7'$ plane from $\langle P_i\rangle_{\rm bin}$ observables. We consider the situation in which $\langle P_1\rangle_{[2,4.3]}$, $\langle P_2\rangle_{[2,4.3]}$, $\langle P'_4\rangle_{[2,4.3]}$ and $\langle P'_5\rangle_{[2,4.3]}$ are measured, with central values equal to their SM predictions and experimental uncertainties of $\sigma_{exp}=0.10$ (note that this experimental precision is feasible soon). In Fig.~\ref{JVfut} we show the 68.3\% and 95.5\% C.L. combined constraints on $C_7,C_7'$ from these observables. Comparing this plot with Fig.~\ref{JVcons} we can see that the observables $\langle P_i\rangle$ will play a very important role in the future.

\section*{Acknowledgments}

It is a pleasure to thank the organizers of the FLASY12 conference for the arrangement of a very stimulating workshop in the beautiful city of Dortmund.

\bibliographystyle{apsrev4-1}

%% file: Papers/wang.tex




%
%
%
%
%
%

\chapter[Semileptonic  $B \to K^{(*)} \ell^{+} \ell^{-}$ decays at large hadronic recoil (Wang)]{Semileptonic  $B \to K^{(*)} \ell^{+} \ell^{-}$ decays at large hadronic recoil}
\vspace{2em}
\paragraph{Y.-M. Wang}

\paragraph{Abstract}

I report the QCD calculation of hadronic amplitudes responsible for the  FCNC decays
$B\to K^{(*)} \ell^+\ell^-$ at large hadronic recoil, emphasizing on the non-form-factor
type contributions. The factorization properties of various non-local matrix elements
are presented. Finally, I discuss the access to the hadronic decay amplitude
at  time-like $q^2$ with the hadronic dispersion relation  and report the first calculation of 
isospin asymmetry for  $B\to K^{} \ell^+\ell^-$ decay.

\section{Introduction}

Exclusive semi-leptonic  $B \to K^{(\ast)}  l^{+} l^{-}$ decays are of a great importance for the
precision test of the Standard Model  as well as for the search
for new physics \cite{Ali:1991is,Ali:1999mm, Beneke:2001at, Bediaga:2012py}.  The leading contributions to the transition amplitudes of these FCNC  processes
can be reduced to the heavy-to-light $B \to K^{(\ast)}$ form factors. On the one hand, the lattice QCD
simulation and QCD sum rules allow the calculations of these form factors in the  large and small $q^2$
regions respectively, with growing accuracy.  On the other hand, different factorization theorems
have been constructed for the heavy-to-light form factors at large recoil, in the frameworks of
both collinear factorization \cite{Beneke:2002ph,Beneke:2003pa} and $k_T$ factorization \cite{Li:2010nn,Li:2012nk}.
One major issue for the factorization approaches is how to implement the  power corrections which  come from many
different sources, e.g., the power suppressed configurations of hadronic states and the effective operators of higher powers.

The hadronic form factors are, however, not sufficient to describe the strong interaction dynamics involved in
the decay amplitude of the FCNC processes $B \to K^{(\ast)}  l^{+} l^{-}$
\begin{eqnarray}
 A(B \to K^{(*)} \ell^+\ell^-)=
-\langle K^{(*)} \ell^+\ell^-\mid H_{eff}\mid B\rangle \,.
 \label{eq:amplA}
\end{eqnarray}
The non-local effects from the electromagnetic correction to the four-quark operators
do not necessarily  factorized into the convolution of Wilson coefficients and hadronic form factors.
The factorizations of hard spectator scattering amplitudes have been achieved by separating the nonperturbative dynamics
parameterized by the hadronic distribution amplitudes from the short-distance dynamics calculable in the perturbative  theory.
Moreover, the hadronic matrix elements  describing the soft gluon radiation contributions
cannot be computed in the QCD factorization  and some non-perturbative approach in QCD,
e.g., light-cone sum rules (LCSR), is in demand. Conceptually, the non-local effects in  $B \to K^{(\ast)}  l^{+} l^{-}$ decays
are  only accessible in QCD at space-like $q^2$ due to the breakdown of operator product expansion (OPE) for the quark loops
and the emergence of  end-point singularity in weak annihilation contribution.
Our strategy of computing the physical amplitudes of $B \to K^{(\ast)}  l^{+} l^{-}$
is to construct the relevant hadronic dispersion relations, inspired from the fact
that the  hadronic matrix elements are analytical functions of $q^2$.

The layout of this talk is as follows: I explain the calculations of
hadronic amplitudes for $B \to K^{(\ast)}  l^{+} l^{-}$  at space-like $q^2$ in section 2,
focusing on the factorization properties of hard contributions and the light-cone OPE for
soft gluon radiation from the charm loop. In Section 3,  I discuss the  hadronic
amplitudes at time-like $q^2$  and, in particular, the modeling of  continuum integrals
 for the background contributions. The isospin asymmetry for  $B\to K^{} \ell^+\ell^-$
decay is also presented here. Section 4 is reserved for the concluding discussion  and outlook.

\section{Hadronic $B \to K^{(\ast)} \ell^{+} \ell^{-}$ amplitudes  in space-like $q^2$ region }

\begin{figure}
\centering
\includegraphics[width=0.30 \linewidth]{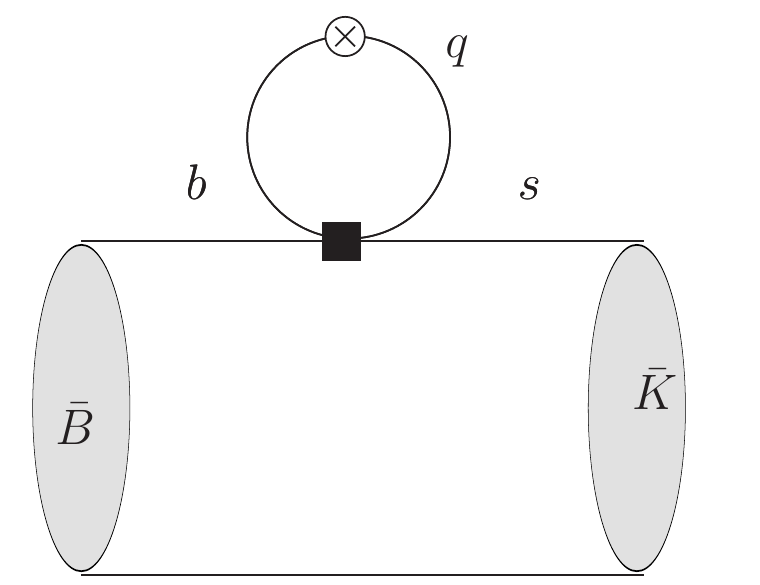}
\includegraphics[width=0.30 \linewidth]{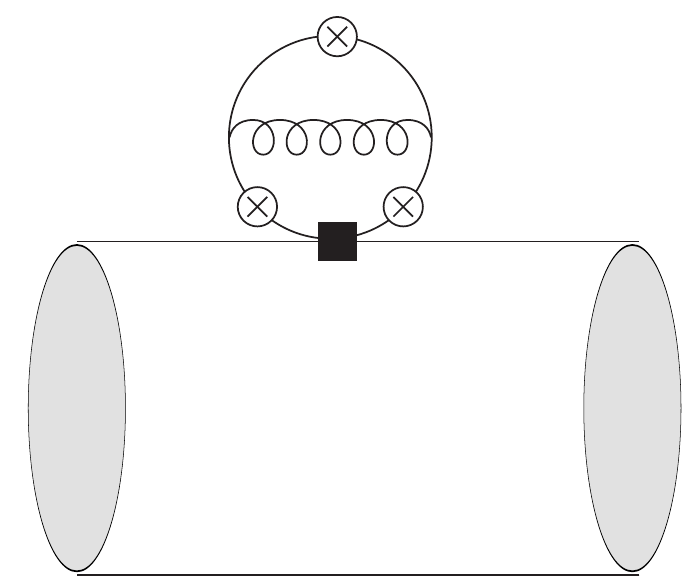}
\includegraphics[width=0.30 \linewidth]{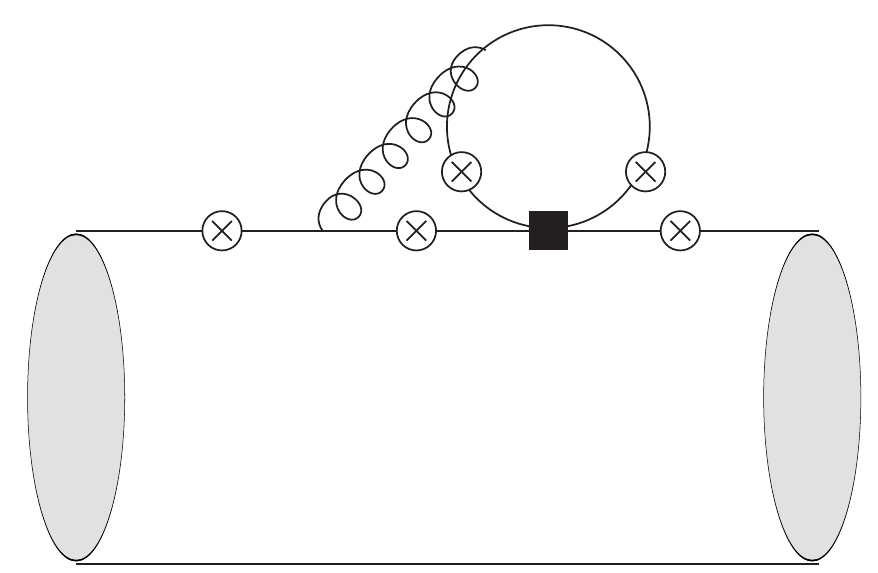}\\
\includegraphics[width=0.30 \linewidth]{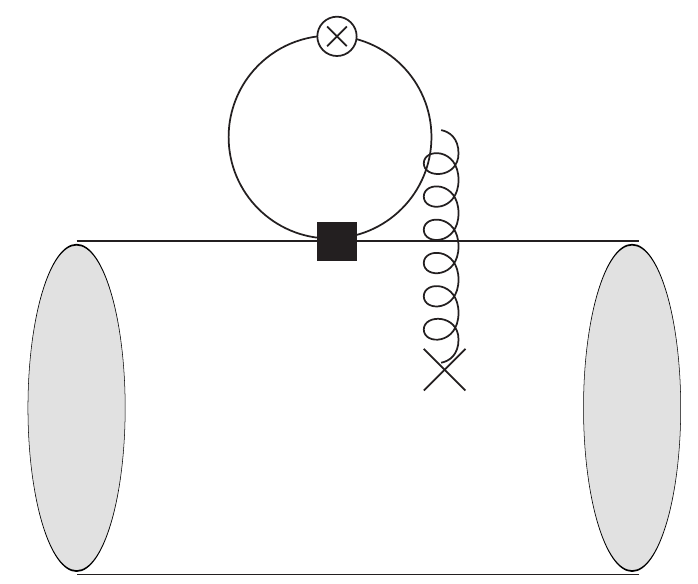}
\includegraphics[width=0.30 \linewidth]{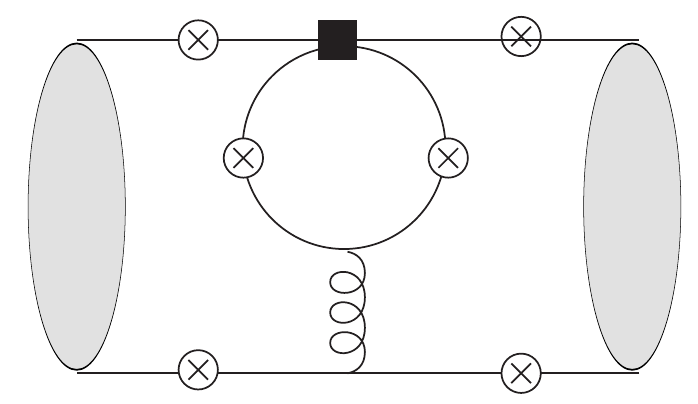}
\caption{Factorizable and nonfactorizable quark-loop contributions
 to $B\to K^{\ast)} \ell^+\ell^-$ amplitudes.}
\label{wang:fig1}
\end{figure}

\begin{figure}
\centering
\includegraphics[width=0.30 \linewidth]{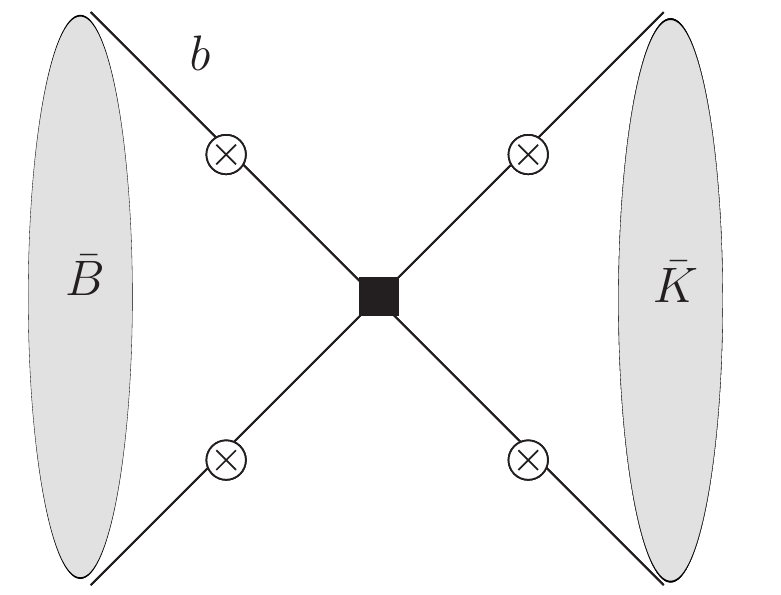}
\includegraphics[width=0.30 \linewidth]{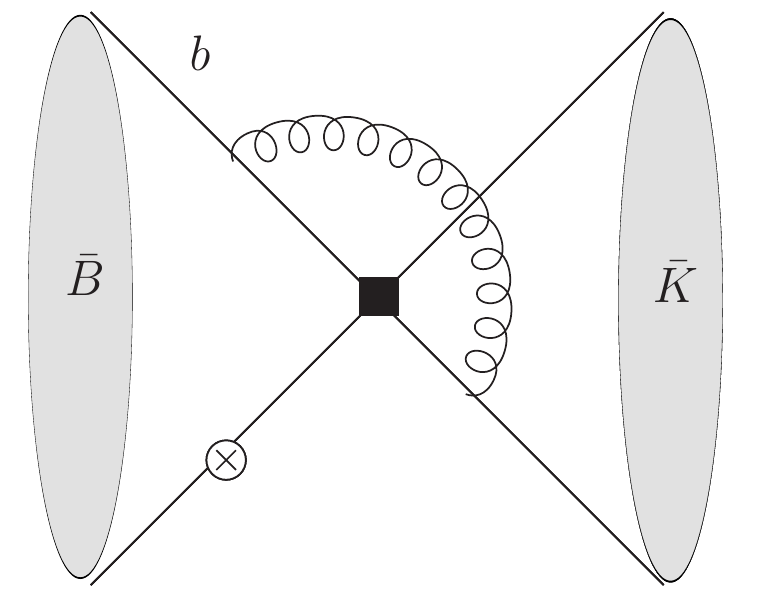}
\caption{Weak annihilation contribution to $B\to K^{(\ast)} \ell^+\ell^-$ amplitude.}
\label{wang:fig2}
\end{figure}

\begin{figure}
\centering
\includegraphics[width=0.30 \linewidth]{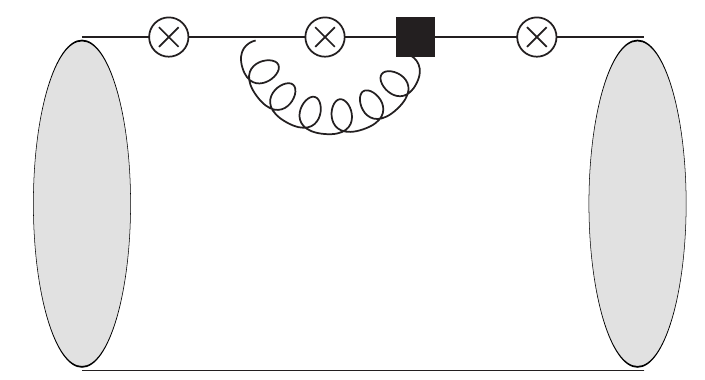}
\includegraphics[width=0.30 \linewidth]{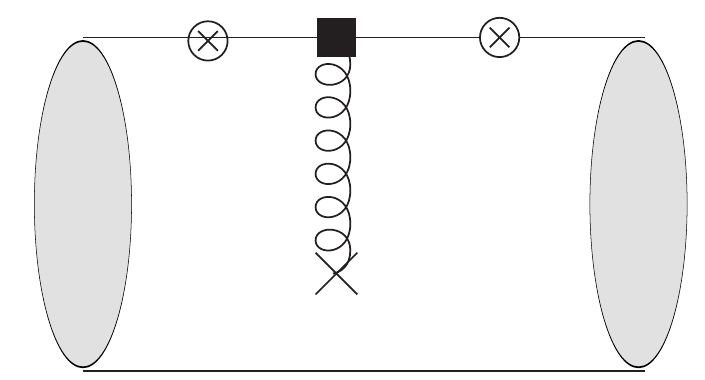}
\includegraphics[width=0.30 \linewidth]{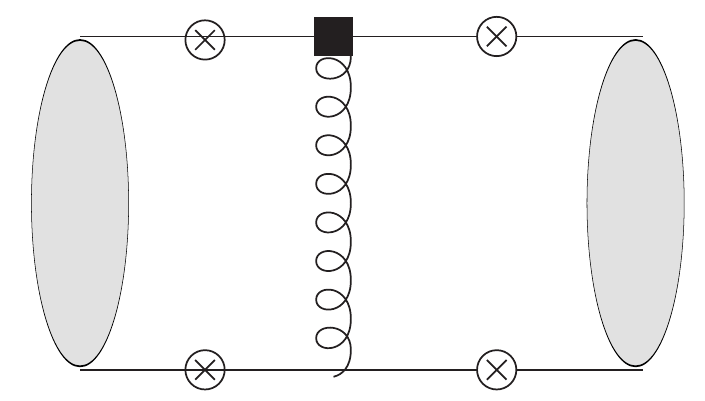}
\caption{Color-magnetic penguin operator $O_8$  contribution to $B\to K^{\ast)} \ell^+\ell^-$ amplitude.}
\label{wang:fig3}
\end{figure}

In addition to the leading contributions from the semileptonic
and electromagnetic penguin operators, the non-local effects
due to electromagnetic corrections are represented by the figures \ref{wang:fig1}-\ref{wang:fig3},
grouped in terms of the topologies of the diagrams.  Below, I will
summarize the calculations of these diagrams in space-like $q^2$ region
one by one, with  particular  attention  to the  limitations of the current
theoretical  tools.

The first class of diagrams, generated by the four-quark operators, are of emission topology
as displayed in figure \ref{wang:fig1}. The amplitudes of the first three diagrams  can be still factorized
as the convolution of $B \to K^{(\ast)}$ form factors and short distance coefficient functions,
with the assumption of parton-hadron duality. As discussed in \cite{Khodjamirian:2010vf},
the factorizable charm-loop amplitude develops an imaginary part for $q^2 \geq 4 m_c^2$,
which indicates that the charm-quark pair gradually evolve into the charmonium resonances
and the applicability  of OPE  does not hold anymore. The situation becomes even worse for
the amplitudes of light-quark loops, where the appearance of light vector meson resonances (for instance,
 $\rho$, $\omega$ and $\phi$) implies the breakdown of OPE starting from very low $q^2$.
Phenomenologically,  the contributions from the light-quark loops are suppressed,  either  by the  Wilson
coefficients of penguin operators or by the CKM matrix elements of $|V_{ub} V_{us}^{\ast}|$.
This can be also understood from the fact that the background effect generated by the light vector mesons
is negligible compared to that from the charmonium resonances. Another remark concerning the factorizable
charm loop is that the resulting amplitude is suppressed by the color factor and both power correction and
perturbative correction could be  sizeable. In fact,  the large branching ratios of color suppressed
tree channels $B \to J/\psi K^{(\ast)}$ remain an unresolved  puzzle in heavy flavor physics for more than two decades.
I also mention by passing that the strong phase originated from the form-factor type  correction to the quark loop
is tiny and this could justify the OPE calculation of two-loop $b \to s \ell^+\ell^-$  transition form factors.

The soft gluon radiation from the charm-quark loop has been computed in \cite{Khodjamirian:2010vf}
with  OPE controlled dispersion relation. I will briefly review the calculation of such effect here.
One can firstly  isolate the time-ordered product of two charm quark currents and then derive the revelent
effective non-local operator $\widetilde{{\cal O}}_\mu(q)$ in terms of light-cone OPE.
It needs to be pointed out  that the local OPE fails for the  calculation of soft gluon
radiation contribution, as the expansion parameter
$q \cdot k / (4m_c^2-q^2)$ ($k$ being the four-momentum of soft gluon) is not small numerically.
Compared to the leading-order factorizable charm-loop,   the amplitude of soft non-factorizable
charm-loop is suppressed by one power of $\Lambda^2_{QCD}/(4 m_c^2-q^2)$, however, it is enhanced by
a color factor $2 C_1(\mu)/(C_1(\mu)/3+C_2(\mu))$. Following a similar argument, each extra soft gluon radiation
from the charm-loop will bring about an additional suppression factor $\Lambda^2_{QCD}/(4 m_c^2-q^2)$
with respect to the leading one-gluon term. To estimate the hadronic matrix elements of the above-mentioned
effective  non-local operator $\widetilde{{\cal O}}_\mu(q)$,  we construct the vacuum-to-$B$-meson correlation function
with the time-ordered product of a local current interpolating the $K^{(\ast)}$ state
and the effective transition operator. Matching the QCD and hadronic representations of one and the same
correlator and performing the Borel transformation, one can derive a sum rule for the hadronic matrix element
$\langle  K^{(*)}(p)|\widetilde{{\cal O}}_\mu(q)| B(p+q)$, where the non-perturbative dynamics is parameterized
by the three-particle distribution amplitudes of $B$-meson in HQET.  For the phenomenological convenience, one
can absorb  the  charm-loop effect into the effective Wilson coefficient $C_{9}^{eff}$ in a $q^2$- 
and process- dependent way.   Numerically,  the nonfactorizable charm-loop
amplitude in $B \to K \ell^{+} \ell^{-}$ at $q^2 \ll 4 m_c^2$ amounts to  several percent
of the factorizable one, with a  different sign. However, the soft gluon radiation effect is more pronounced
in $B \to K^{*} \ell^{+} \ell^{-}$ at small $q^2$,
for a transverse polarized $K^{\ast}$ meson.

The spectator scattering diagrams generated by the four-quark operators can be decomposed into
two different subgroups, depending on whether the virtual photon is radiated from the external
quark lines or from the internal quark loop. For the photon radiation from the external legs,
it is obvious that the leading power  contribution is given by a single diagram with the photon
emission from the spectator quark of $B$-meson. The amplitude of this diagram develops an end-point
divergence in the soft $q^2$ limit, as firstly observed in \cite{Beneke:2001at}, and such effect
also contributes to the isospin symmetry breaking of $B\to K^{(\ast)} \ell^+\ell^-$ decays.
For the photon emission from the quark loop, the leading contribution to the quark loop amplitude
can be extracted by setting the gluon momentum flowing into the loop as the momentum of the
spectator quark in the $K^{(\ast)}$ meson.  The momentum of spectator quark in  the $B$-meson
only show up in the propagator of hard-collinear gluon. Hence, the soft dynamics inside $B$-meson
decouples from the dynamics of quark loop at the scales $q^2 \geq m_b \Lambda_{QCD}$ and $m_b^2$.
This observation is essential to justify the factorization theorem shown in Eq. (15) of \cite{Beneke:2001at}.
It is also true that one can also further factorize the dynamics at the scale $m_b^2$  from that
of the scale $q^2$. The factorization formula for the contribution of these four-point diagrams
can be also confirmed by an analysis based upon the $B$-meson LCSR in soft-collinear effective theory.
Numerical analysis shows that  these two diagrams generate sizeable strong phase
even in the space-like $q^2$ region. In the hadronic level, this can be understood from the initial state
re-scattering effect of  $B \to D^{(*)} D_s^{(*)}$ and $D^{(*)} D_s^{(*)} \to \gamma^{\ast}  K^{(\ast)}$.
In other words, the analytical structure of hadronic dispersion relation for the non-local
contribution to the decay amplitude  $A(B \to K^{(*)} \ell^+\ell^-)$ in the variable $q^2$ is
considerably complicated at $O(\alpha_s)$ order, since the
residuals of charmonium resonance contributions as well as the hadronic
spectral density could be expressed as  dispersion integrals in
the variable $p_B^2$ and the double dispersion relations of
non-local matrix elements would be probably needed.

The second class of diagrams are of annihilation topology as presented in figure \ref{wang:fig2}.
The leading-order weak annihilation amplitude was computed in the framework of
QCD factorization. As discussed in  \cite{Beneke:2001at}, such effect does not vanish
in the leading power in the heavy quark limit for the  final state being a kaon  or
a longitudinal polarized $K^{\ast}$  meson. Another comment concerning the weak
annihilation effect is that the leading contribution is from the diagram with photon
radiation from the spectator quark of $B$-meson, however, only the sum of four diagrams
respects the Ward identity as a consequence of charge conservation in the weak interaction.
Weak annihilation contribution is numerically insignificant due to the suppression of
either the Wilson coefficient or the CKM matrix element. From the phenomenological aspect,
it is safe to drop out the perturbative correction to the weak annihilation displayed
in the right planar of figure \ref{wang:fig2}.  The calculation of these diagrams is, however, important
to the understanding of renormalization  property of $B$-meson distribution amplitudes 
in HQET \cite{Lange:2003ff,Braun:2003wx,Bell:2008er,Li:2012md}.
Moreover, it is also interesting to investigate the weak annihilation contribution within the
$B$-meson LCSR and examine the factorization formulae in the heavy quark limit.

The third class of diagrams are generated by the colormagnetic operator $O_8$
as shown in figure \ref{wang:fig3}.  Similar to the weak annihilation diagrams,  the spectator
scattering amplitude contributed from the  operator $O_8$ also produces
end point singularity for $q^2 \sim \Lambda_{QCD}^2$. The soft gluon radiation
from the vertex of $O_8$ is  calculated in the LCSR approach with $B$-meson distribution
amplitudes.  However, such effect is suppressed by both one power of $\Lambda_{QCD}^2/m_b^2$
and the small Wilson coefficient $O_8$ and hence it is irrelevant numerically.
The factorizable contribution from the operator $O_8$ was firstly computed in
\cite{Asatrian:2001de} in expanded form and in \cite{Beneke:2001at} in compact form.

\section{Hadronic $B \to K^{(\ast)} \ell^{+} \ell^{-}$ amplitudes  in time-like $q^2$  region }

Following the discussion in previous sections,  a conservative way to compute the
 $B \to K^{(*)} \ell^{+} \ell^{-}$ amplitudes is to employ local/light-cone OPE
and QCD factorization theorem (also with implicit use of parton-hadron duality )
in the space-like $q^2$ region.  The next question is,  how can one access the  hadronic
amplitudes at  time-like $q^2$? The strategy discussed in this talk,
following  \cite{Khodjamirian:2010vf}, is to construct the relevant hadronic dispersion
relations for the non-local contributions,  using the fact that
the hadronic matrix element is an analytical function of  $q^2$.
For the convenience of exploring the isospin asymmetry in   $B \to K^{(\ast)} \ell^{+} \ell^{-}$
decays, one can isolate the contributions of  $u$ and $d$ flavors involved in the
electromagnetic current of the non-local matrix element, from that of
$b$, $c$ and $s$ flavors, and write down separate
dispersion relation for each term.  For the amplitude contributed from $u$ and $d$ quarks,
the corresponding dispersion relation reads \cite{KMW:2012}
\begin{eqnarray}
{\cal H} ^{(BK)}_{ud}(q^2)={\cal H} ^{(BK)}_{ud}(q_0^2)+
(q^2-q_0^2) \Big[ \sum_{h=\rho,\omega} \frac{\kappa_h f_h  |A_{B h
K}| e^{i \varphi_h} }{(m_h^2-q_0^2)(m_h^2-q^2-
im_h\Gamma^{tot}_h)}
\nonumber\\
+\int_{s_0^h}^{\infty} ds
\frac{\rho(s)}{(s-q_0^2)(s-q^2-i\epsilon)}\Big]\,,
\end{eqnarray}
where one substraction for the amplitude
has been performed at $q_0^2=-1 {\rm GeV^2}$,
$\kappa_{\rho}=1/\sqrt{2}$,
$\kappa_{\omega}=1/(3 \sqrt{2})$.  One can write down the hadronic dispersion relation
for the contribution from $b$, $c$ and $s$ quarks in a similar manner.
The decay constant $f_h$ and the hadronic $B$ decay amplitude
$|A_{B h K}|$ can be extracted from the experimental data.
However, the continuum integral accumulating the  contribution from excited states and
continuum cannot be constrained from the experimental side, due to the absence of the
measurements of three body decays $B \to K \pi \pi$,  $B \to K K \bar{K}$  and $B \to K D \bar{D}$.
Therefore, one has to parameterize the continuum integral in a model-dependent way.
It is extremely difficult to avoid the model dependence here, if it is possible conceptually.
The unknown parameters involved in the model of continuum integral can be determined by
matching the hadronic dispersion relation  to the  calculated hadronic amplitude at space-like
$q^2$ from QCD.  Numerical analysis indicates that  different parameterizations of the
continuum integral does not bring about  distinct discrepancy for
the physical observables of $B \to K \ell^{+} \ell^{-}$.

The predicted isospin asymmetry of
$B \to K \ell^{+} \ell^{-}$ below charmonium  threshold has been collected in Table \ref{wang:table 1},
where the measurements from BaBar, Belle and LHCb collaborations are also presented for a comparison.
The central values of experimental data on the isospin asymmetry reveal large derivation from zero,
however, all of these measurements  suffer from significant uncertainties.
Albeit with the tiny isospin asymmetry of $B \to K \ell^{+} \ell^{-}$ from theoretical side,
no conclusive statement about the emergence of new physics in exclusive FCNC transition
can be made at present, without an enormous improvement of the accuracy of the experimental
measurements.

\begin{table}[h]
\caption{ Isospin asymmetry $a_I(B\to K\ell^+\ell^-)$
integrated over $1.0<q^2<6.0$ GeV$^2$. Taken from \cite{KMW:2012}.}
\begin{center}
  \begin{tabular}{|c|c|c|c|c|}
\hline
 &&& \\
 Belle \cite{Wei:2009zv}   &  BaBar \cite{:2012vwa}   &  LHCb
\cite{Aaij:2012cq}  &  this work  \\[1mm]
\hline
&&& \\[-1mm]
$-0.41^{+0.25}_{-0.20} \pm
0.07$ & $ -0.41 \pm 0.25 \pm 0.01$  & $-0.35^{+0.23}_{-0.27}$    &
$-0.01^{+0.02}_{-0.00}$        \\
&&& \\  \hline
\end{tabular}
\label{wang:table 1}
\end{center}
\end{table}

\section{Summary and outlook}

I  summarize the current status of  QCD computations of
the $B \to K^{(*)} \ell^{+} \ell^{-}$ amplitudes at large hadronic recoil 
\footnote{Extensive studies of $B \to K^{(*)} \ell^{+} \ell^{-}$ decay at low hadronic recoil 
have been available in the literature \cite{lowrecoil},  based upon the local OPE in the HQET limit, with different 
treatment of the charm-quark field. }.
Detailed discussion on the non-local effects due to electro-magnetic correction
to the four-quark operators and color-magnetic operator is  presented in the
framework of OPE and factorization theorem. The strategy to access the hadronic
amplitude at time-like $q^2$ using the hadronic dispersion relation is also reviewed.
Phenomenological aspects of the $B \to K^{(*)} \ell^{+} \ell^{-}$ decays are not the
focus of this talk, only the isospin asymmetry of $B \to K \ell^{+} \ell^{-}$ decay is briefly discussed.
One can apply the similar procedure to the calculation of  FCNC transition of $\Lambda_b$
baryon in the Standard Model \cite{Lambda_b:SM} and beyond \cite{Lambda_b:NP}.
Lastly, I emphasize again that understanding the power correction and perturbative correction
is essential to search for the new physics in heavy flavor physics.

\section*{Acknowledgments}

I am  grateful to Alexander Khodjamirian, Thomas Mannel and Alexey Pivovarov for a fruitful
collaboration on this topic.




%% file: Papers/tomweiler.tex

%
%

%
%
%
%
%

\chapter[New Classically-Stable, Closed Timelike Curves (CTCs) (Ho, \textit{Weiler})]{New Classically-Stable, Closed Timelike Curves (CTCs) }
\vspace{-2em}

\paragraph{C.M. Ho, \textit{T. Weiler}}
\paragraph{Abstract}
A new class of closed timelike curves (CTCs) using a compactified extra dimension are constructed.
Non-physical requirements that plague previously conjectured CTCs do not apply here.
Our CTCs are physical, and classically stable, in that:
(i) no matter distributions of infinite extent are required;
(ii) there is no need of negative energy densities -- in fact, no need of matter distributions at all
(all energy conditions -- null, weak, strong and dominant -- are satisfied); and
\,(iii) the energy of a time-traveling particle is conserved.
An example of a particle which may time-travel is the ``fourth-flavored" neutrino,
the ``sterile" neutrino.

\section{Introduction}
\label{sec:introduction}

It is well known that closed timelike curves (CTCs) are allowed solutions of general
relativity, and so time travel is theoretically possible. For decades, many proposals for CTCs have been discussed
in the literature. These include van Stockum's rotating cylinder~\cite{vanStockum} (extended later by Tipler~\cite{Tipler}),
G\"{o}del's rotating universe \cite{Godel},
Wheeler's spacetime foam \cite{Wheeler},
the region between the two horizons of the Kerr and Kerr-Newman rotating black holes~\cite{Kerr},
Morris, Thorne and Yurtsever's traversable wormholes \cite{MTY},
Gott's pair of spinning cosmic strings \cite{Gott},
Alcubierre's warp drive \cite{warp}, Ori's vacuum torus \cite{Ori} and a few more recent proposals \cite{Gron}.
All of these CTCs are constructed in our 4D (``brane") universe.
An excellent overview is available in~\cite{Visser}.

The success of large \cite{ADD} and warped \cite{RS} extra dimensions has led many people to think of gravitons or gauge singlets
taking ``shortcuts" through the extra dimensions (``bulk")~\cite{shortcut1,shortcut2,shortcut3,shortcut4,shortcut5}.
For instance, a graviton or gauge-singlet ``sterile" neutrino may take a ``shortcut" from one point on the brane through the bulk
and back to the brane at a different point,
with a shorter transit time than that for a photon traveling along a brane geodesic between the same two points .
But note that, although a ``shortcut" allows for superluminal communication,
it nevertheless obeys time-ordering and so does not constitute a CTC --
a shorter time-of-flight is not the same as time evolving backwards.
However, using the idea of asymmetrically warped extra dimensions \cite{csaki},
it has been shown how paths can be constructed to form CTCs~\cite{Tom}.
These constructed paths are not solutions of geodesic equations,
and so would not be the paths traversed by physical particles.
Also, the paths constructed in~\cite{Tom} require some negative-energy
matter distribution in the bulk for their stabilization.

The purpose of this article is to highlight the very recent proposal of Ho and Weiler~\cite{HoWeiler}
for CTCs which are solutions to the geodesic equations of a certain class of 5D metric with a compactified extra dimension.
These new CTCs  have no  classical pathologies.

\section{The 5D Metric}

As inspired by the idea of large extra dimensions \cite{ADD} and guided by
analogy with
G\"{o}del's rotating universe~\cite{Godel} and the CTCs therein,
we are led to consider a metric off-diagonal in compactified extra dimension ($u$, with size $L$) and time $t$.
The periodic boundary condition requires the point $u+L$ to be identified with $u$.
With simplicity in mind, we consider the following time-independent (``stationary'') metric:
\TWbea
\label{metric}
d\tau^2= \eta_{ij}dx^{i}dx^{j}+dt^2+ 2 \,g(u)\,dt\,du -h(u)\,du^2\,,
\TWeea
where $i,\,j=1,\,2,\,3$, and $ \eta_{ij}$ is the spatial part of the Minkowski metric.
The 4D metric induced from this 5D metric is completely Minkowskian.
%
The determinant of our metric is
$
\rm{Det}[g_{\mu\nu}] = g^2+h$.\,
The spacelike nature of the $u$ coordinate requires $\rm{Det} > 0 $ for the entire 5D metric,
which in turn requires that $g^2 +h >0 $ for all $u$.
It is desirable to maintain a Minkowski metric as the brane is approached;
thus, we set $\rm{Det}(u=0) =g_0^2 +h_0=+1$, where $g_0\equiv g(0)$, etc.
We will also assume, for definiteness, that $h_0 \ge 0$, which implies that $|g_0| \le 1$.

The metric tensor must reflect the $S^1$ topology of the compactified extra dimension.
Thus, $g(u)$ and $h(u)$ must be periodic functions of $u$ with period $L$. We expand $g(u)$
in terms of Fourier modes:
\TWbea
\label{genmetric}
g(u) = g_0+A  -\sum_{n=1}^\infty \left\{
   a_n\,\cos\left(\,\frac{2\pi\,n\,u}{L}\,\right)
   +b_n\,\sin\left(\,\frac{2\pi\,n\,u}{L}\,\right)
   \right\}\,,
\TWeea
where $g_0=g(0)$ and $A\equiv \sum_{n=1}^\infty a_n$ are constants.
A similar expression can be written down for $h(u)$, but it will not be needed.
For use later in this report, we note here that the value of $g(u)$ averaged over the compact dimension
is ${\overline g}=g_0 +A$.


The next task is to obtain the geodesic equations of motion and solve for their solutions.
Since the metric~(\ref{metric}) is completely Minkowskian
on the brane, the geodesic equations of motion along the brane are just $\ddot{\vec r}=0$,
where the over-dot denotes differentiation with respect to the proper time, $\tau$.
Solutions to these geodesic equations are simply
$\dot{\vec r}=\dot{\vec r}_0$,\;or ${\vec r}={\vec r}_0\,\tau$.
%

The geodesic equations for $t$ and $u$ are more interesting.
Due to the time-independence of the metric, there exists a timelike Killing vector;
the corresponding conserved quantity is
\TWbea
\label{timeconst}
\dot{t}+g(u)\,\dot{u}=\gamma_0+g_0\,\dot{u}_0\,,
\TWeea
where we have evaluated the right-handed side at its initial ($\tau=0$) value.
Given this conserved quantity, it is almost evident that time will run backwards ($\dot{t}<0$), provided that
the condition $g(u)\,\dot{u}>\gamma_0+\dot{u}_0\,g_0$ is consistent with the geodesic equation for $u$.

The geodesic equation for $u$ is
\TWbea
\label{2nd_geodesic}
2\,(g\,\ddot{t} - h\,\ddot{u}) - h'\,\dot{u}^2=0\,;
\TWeea
we use the superscript ``prime" to denote differentiation with respect to $u$.
We can eliminate $\ddot{t}$ and $\ddot{u}$ from Eq.~(\ref{2nd_geodesic}).
First, we take the dot-derivative of Eq.~(\ref{timeconst}).
Then we rewrite Eqs.~(\ref{timeconst}) and (\ref{2nd_geodesic}) as
\TWbea
\label{geodesic_eqns}
\ddot{t}(\tau) &=&  \frac{1}{2} \,\frac{-2g' h+gh'}{g^2+h}\,\dot{u}^2 \,,\\
\label{geodesic_u}
\ddot{u}(\tau) &=& -\frac{1}{2} \,\frac{2gg'+ h'}{g^2+h} \,\dot{u}^2=-\frac{1}{2} \,\ln' (g^2+h)\,\dot{u}^2 \,.
\TWeea
Inspection of these two geodesic equations suggests
that it prove fruitful to fix the determinant to be
\TWbea
\label{parameterize}
\rm{Det}(u)=g^2(u)+h(u) = 1 \,,\quad\ \forall\ u\,.
\TWeea
For simplicity, we do so.
Once the metric function $g(u)$ is given by the Fourier series of Eq.~\eqref{genmetric},
then the second metric function $h(u)=1-g^2(u)$ is automatically determined.
Substituting Eq.~\eqref{parameterize} into Eq.~\eqref{geodesic_u} immediately leads to
\TWbea
\label{proper_velocity}
\dot{u}(\tau) = \dot{u}_0\,,
\quad{\rm \ and \ }
u(\tau) = \dot{u}_0\, \tau\,,~~~~~~ (\,\textrm{mod}~ L\,)\,.
\TWeea
Integrating Eq.~\TWrf{timeconst} yields
$t(\tau)= (\gamma_0+g_0\,\dot{u}_0)\,\tau -\int^{u(\tau)}\, du\;g(u)$,
which  we rewrite, using Eq.~\eqref{proper_velocity}, in a form more useful for later discussions:
\TWbea
\label{u}
t(u)=\left( g_0 +\frac{1}{\beta_0}\right)\,u - \int_0^u du\,g(u)\,.
\TWeea
Here we have introduced the symbol
$\beta_0=\frac{{\dot u}_0}{\gamma_0}=\left(\frac{du}{dt}\right)_0$\,
for the initial velocity of the particle along $u$-direction,
as measured by a stationary observer on the brane.
Analogous to those historical CTCs arising from metrics describing rotation, we will
call a particle ``co-rotating'' if  $\beta_0>0$, and ``counter-rotating'' if $\beta_0<0$.

\section{Closed Timelike Curves}
\label{sec:CTCpossibility}

Closed timelike curves, by definition, are geodesics that return a particle to the same space coordinates from which it left,
but with a negative time so that its arrival equates to or precedes its departure.
Due to the periodic boundary condition from
the compactified extra dimension, a particle created on the brane but propagating into the bulk
will necessarily come back to the brane.
So the ``closed'' condition for a CTC is satisfied automatically by a compactified metric.
We note that the motion along the brane is trivially ${\dot{\vec r}}=$ constant.
When this is added to the geodesic solution for $u(\tau)$, it leads to a helical particle motion which periodically
intersects the brane.

To ascertain whether the travel time can be negative (the ``timelike'' condition for a CTC),
we must solve the geodesic equation for time, Eq.~\eqref{u}.
With the general $g(u)$ given by Eq.~\eqref{genmetric},
we perform the integration in Eq.~\eqref{u} to obtain
\TWbea
\label{tu}
t(u) = \left(\frac{1}{\beta_0}-A\right)\:u
   + \left(\frac{L}{2\pi}\right)\sum_{n=1}^\infty \left(\frac{1}{n}\right)
   \left\{
     a_n\,\sin\left(\frac{2\pi\,n\,u}{L}\right) + b_n\,\left[1-\cos\left(\frac{2\pi\,n\,u}{L}\right)\right]
   \right\}\,.
\TWeea
Due to the periodic boundary condition, the particle returns to the brane at $u=\pm \,N\,L,\;N=1,2,\dots$,
after traversing $N$ times around the extra dimension
(with $\pm$~signs for co-rotating and counter-rotating particles, respectively).
At the $N^{th}$ return, the time measured by a stationary clock on the brane,
given by Eq.~\TWrf{tu}, is
\TWbea
\label{tNL}
t_N\equiv t(u=\pm \,N\,L)=\pm \,\left( \,\frac{1}{\beta_0}-A\,\right)\,N\,L\,.
\TWeea
Interestingly, $t_N$ depends on the Fourier modes only through $A=\sum_{n=1} a_n$,
and is completely independent of the $b_n$.
Thus, the potential for a CTC arises only from the cosine modes\footnote
{In fact, we can show that a single mode from the set $\{a_n\}$ is sufficient to admit a CTC.
}.

To have a viable CTC, we require \,$t_N <0$,
which is satisfied
for a co-rotating ($\beta_0>0$) particle only if
\TWbea
\label{Acondition}
A > \frac{1}{\beta_0}\,.
\TWeea
For a counter-rotating ($\beta_0<0$) particle, $t_N <0$ is satisfied only if
\TWbea
\label{Acondition2}
A,\,\beta_0 < 0 \quad{\rm and}\quad  |A|\,>\,\left|\frac{1}{\beta_0}\right|\,.
\TWeea
Thus, a viable CTC requires $\textrm{sign}(A)$ to be the same as $\textrm{sign}(\beta_0)$
in either case of co-rotating or counter-rotating particles.
Once Nature chooses the constant $A$ with a definite sign,
these CTC conditions for co-rotating and counter-rotating particles are not compatible.
For definiteness, we will assume that $A>\frac{1}{\beta_0}$ is satisfied for some $\beta_0$,
so that only the co-rotating particles can traverse the CTC backwards in time.
We note that the conditions \eqref{Acondition} and \eqref{Acondition2} can be satisfied even if $|\beta_0| < 1$.
This means that Nature does not need superluminal speeds to realize CTCs.

The parameter conditions for a CTC are the following:
from Eq.~\eqref{Acondition} we have $A\ge 1$;
from the form of our metric as the brane ($u=0$) is approached, we have
$|g_0|\le 1$.  It turns out~\cite{HoWeiler} that periodicity of the particle's quantum mechanical wave function
around the compact dimension requires a generalization of the latter condition to
$|{\overline g}|=|g_0+A| \le 1$.
Thus the parameter regions allowing a CTC are two, given in Fig. 1.
the important feature is that these CTC-admitting regions are nonzero!
%
\begin{figure*}[ht]
\centerline{\includegraphics[height=3cm,width=9.5cm]{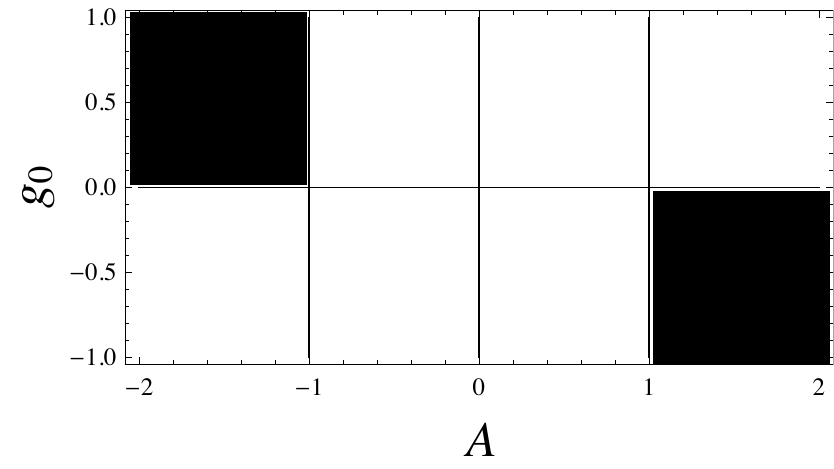}}
\caption
{The two regions ion the $g_0$-$A$ plane for which CTCs are possible.
}
\label{fig:CTCplane}
\end{figure*}

\section{Comparison of 5D CTC and 4D Spinning String}
\label{sec:compared2spinning}

Our class of 5D metrics admitting CTCs resembles in some ways the well-studied metric
for a 4D spinning cosmic string \cite{DJtH-Annals,DJ-Feinberg}:
\TWbea
\label{spinstr}
d\tau^2_{\rm\stackrel{spinning}{string}}= (dt+ 4\,G\,J \,d\theta)^2-dr^2-(1-4\,G\,m)^2\,r^2\,d\theta^2-dz^2\,,
\TWeea
where $G$ is Newton's constant, $J$ is the angular momentum,
and $m$ is the mass per unit length of the cosmic string.
In three spacetime dimensions, the Weyl tensor vanishes, and so any region without a gravitational source must be flat.
Consequently, in the region outside the spinning string, the local Minkowski coordinates may be extended to
cover the entire region: $\tilde{t} = t+ 4\,G\,J \,\theta $ and\, $\varphi =(1-4\,G\,m)\,\theta $ such that
the metric becomes Minkowskian, with the conformal factor being unity.
Similar to $\theta$, \,$\varphi$ is periodic, subject to the identification
$\varphi \sim \varphi + 2\,\pi- 8\,\pi \,G\,m$, and
the wedge $\Delta\varphi=8\,\pi \,G\,m$ is removed from the plane, leaving behind a cone.
While these coordinate transformations apparently lead to simplicity, in fact
$\tilde{t}$ is a pathological coordinate, a linear combination of a non-compact variable $t$ and a compact variable $\theta$.
For a fixed $\theta$ (or $\varphi$), \,$\tilde{t}$ is a smooth and continuous variable. But
for a fixed $t$, one needs the identification $\tilde{t} \sim \tilde{t}+ 8\,\pi\,G\,J$ to avoid a ``jump'' in the new variable.
A a result, the singularity at $g_{\theta\theta}=0$, which occurs at $r=4GJ/(1-4Gm)$,
is in effect encoded in the pathological coordinate $\tilde{t}$~\cite{DJtH-Annals}.

In the $(t,\ u)$-plane, our metric has the form
$d\tau^2 = (\,dt+g(u)\,du\,)^2 - du^2$.
where we have used the simplifying condition in Eq. \eqref{parameterize}. This appears similar to the 4D spinning-string metric.
Analogously, we can define a new exact differential $d\bar{t}\equiv dt+g(u)\,du$ to put our metric into the diagonal ``Minkowskian'' form:
$d\tau^2 = \eta_{ij}\,dx^{i}dx^{j}+ d\bar{t}^2 - du^2$.
This nontrivial coordinate transformation defines a new time variable $\bar{t}= t+ \int_0^{u(t)}\,du\;g(u)$\,
which is measured in the frame that ``co-rotates" with the circle $S^1$.
Since the equivalent metric is locally Minkowskian everywhere, the entire 5D spacetime is flat.
This is consistent with the theorem which states that
any two-dimensional (pseudo) Riemannian metric, whether in a source-free region or not, is conformal to a
Minkowski metric.
However, similar to the case of the spinning string, the topology of our 5D spacetime is non-trivial.
The new time variable $\bar{t}$
is ill-defined globally,
a pathological combination of a non-compact $t$ coordinate and a compact $u$ coordinate.

We remark that the time measured by an observer (or experiment) on our brane should just be
given by $t$. The reason is that the constraint equation that reduces the 5D metric to the induced 4D metric
is simply $u(x^\mu)=0$, and taking the differential gives $du=0$. When the latter result is substituted into the 5D metric in Eq.~\TWrf{metric},
the standard 4D Minkowski metric with timed $t$ is induced.


Indeed, the metric for the 4D spinning string leads to CTCs \cite{DJtH-Annals}.
However, this metric has been criticized in that the definition of spin becomes singular at the string's center.
There is no analogous problem in our compactified 5D metric (Eq.~\eqref{metric}),
because the ``center'' of the periodic $u$-space is not part of the spacetime.
An improved CTC, making use of a pair of infinitely-long cosmic strings with a relative velocity, was proposed by~\cite{Gott}.
In his scheme, spin angular-momentum is replaced by orbital angular-momentum of a two-string system.
He showed that there exists a `figure-eight'' CTC geodesic encircling the strings and crossing between them.
However, the non-trivial topology in Gott's spacetime results in non-linear energy-momentum addition rules.
While each of the spinning cosmic strings carries a timelike energy-momentum vector, the two-string center-of-mass energy-momentum vector
turns out to be spacelike, leading to
violations of null, strong and dominant energy conditions \cite{Hooft,Shore,Carroll,Tye}.
Another vulnerability of Gott's CTC is the increasing energy of the particle traversing the CTC~\cite{Tye}.
Since the particle can traverse the CTC infinitely many times, it can be infinitely blue-shifted,
all while keeping the time elapsed negative~\cite{Hawking,Carroll}.
This implies that the total energy of the pair of the cosmic strings would have been infinitely dissipated even before the particle
enters the CTC for the first time.
This simply means that the CTC cannot form in the first place.

We may contrast the string CTCs, Gott or no Gott, with the results of our metric.
it is easily verified that all the components of the 5D curvature tensor $R_{ABCD}$ and Ricci tensor $R_{AB}$
derived from the metric of Eq.~(\ref{metric}) are identically zero.
Thus, by the Einstein field equation, the energy-momentum tensor $T_{AB}$ is also vanishing,
and so our 5D spacetime automatically satisfies all of the
standard null, weak, string and dominant energy conditions.
%
%
%
In addition, particles traversing the compactified 5D CTCs conserve energy.
The contravariant momentum is defined as
$p^A \equiv m\,(\dot{t},\,\dot{\vec{r}},\,\dot{u})$,
for a particle with mass $m$.
Correspondingly, the covariant five-momentum is given by
\TWbea
\label{PdnA}
p_A = G_{AB}\,p^B = m\,\left(\,\dot{t}+ g \,\dot{u},\;-\dot{\vec{r}},\;g\,\dot{t}-h\,\dot{u}\,\right)\,.
\TWeea
From Eq.~\TWrf{timeconst}, it is clear that the quantity $p_0= m \,(\dot{t}+ g \,\dot{u})$ is covariantly conserved
along the geodesic on and off the brane, a result traceable to the time-independence of the metric $G_{AB}$.
We can therefore identify this conserved quantity as the energy $E$ of the time-traveling particle.

\section{Discussions and Conclusions}
\label{sec:conclusion}

We have constructed a new class of CTCs that are physical and classically stable.
Since it is the compactified extra dimension that enables the CTCs,
only the Kaluza-Klein (KK) particle modes can traverse through these CTCs and go backwards in time.
In the framework of large extra dimensions \cite{ADD} where Standard Model particles are confined to
our familiar 4D brane, we may still anticipate that the KK modes of gauge singlets
(gravitons, sterile neutrinos, higgs singlets, etc.) propagate through these CTCs,
provided that our specific metric in Eq.~\eqref{metric} is realized by Nature.

Finally we mention that our derivation has been purely classical.
Whether or not our results survive in a quantum mechanical picture is a story yet to be written.

\section*{Acknowledgments}
This work
was supported in part by
Department of Energy Grant DE-FG05-85ER40226.

\bibliographystyle{apsrev4-1}

{}


%% file: Papers/yamamoto.tex

%
%
%
%
%
%

\chapter[Squark flavor mixing and CP violation of neutral B mesons at LHCb (Yamamoto)]{Squark flavor mixing and CP violation of neutral B mesons at LHCb}
\vspace{-2em}
\paragraph{K. Yamamoto}
\paragraph{Abstract} 

 We study the contribution of the squark flavor mixing 
from  the $LR(RL)$ component of the squark mass matrices
to the direct CP violation of the $b\to s\gamma$ decay and
the CP asymmetry of $B_d \to K^* \gamma$ decay and  the non-leptonic decays of 
$B$ mesons.
The magnitude of the  $LR(RL)$ component is constrained
by the branching ratio and the direct CP violation  of $b\to s\gamma$.
We predict the time dependent CP asymmetries of the $B$ decays.

\section{Introduction}

Recently LHCb has reported new data of the CP asymmetries of $B_s$ mesons.
They measured the time dependent  CP asymmetry $S_f$ of 
 $B_s \to J / \psi \phi$ and $B_s \to J / \psi f_0(980)$ decays \cite{:2012dg}. 
The CP violation in the $K$ and $B_d$ meson decays has been successfully explained 
 within the framework of the standard model (SM),
 so called Kobayashi-Maskawa (KM) model \cite{Kobayashi:1973fv}.
However, there are a possibility of new sources of the CP violation
 if the SM is extended to the supersymmetric (SUSY) models.
Therefore, we expect the SUSY contribution to the CP violation in the $B$ meson decays.

The typical contribution of SUSY is the gluino-squark mediated flavor changing process \cite{King:2010np}-\cite{Ishimori:2011nv}. 
We predict the time dependent CP asymmetries of $B_d^0\to \phi K_S$ and $B_d^0\to \eta 'K^0$  decays 
 which are deviated from the SM predictions in the framework of the SUSY.
In this regard we consider constraints from the branching ratio and the direct CP violation of $b\to s\gamma$.

In that framework of the SUSY, 
 the asymmetries of 
 $B_d^0\to \phi K_S$ and $B_d^0\to \eta 'K^0$ are deviated from the
SM predictions \cite{Hayakawa:2012ua,Shimizu:2012ru}.
Then, these contributions  of the new physics 
are  correlated with the direct CP violation  of the  $b\to s\gamma$ decay.
In this work, we present the numerical analyses in the  case that 
$LR$ and $RL$ components of squark mass matrices dominate the penguin decays.

\section{CP violation in $B$ meson decays}

Let us discuss the effect of the new physics 
 in the non-leptonic decays of $B$ mesons. 
The contribution of new physics to the dispersive part $M_{12}^q(q=d,s)$ 
is parameterized as 
\begin{equation}
M_{12}^q=M_{12}^{q,\text{SM}}+M_{12}^{q,\text{SUSY}}=
M_{12}^{q,\text{SM}}(1+h_qe^{2i\sigma _q})~, \quad (q=d,s)
\label{M12}
\end{equation}
where $M_{12}^{q,\text{SUSY}}$ is the SUSY contribution, 
and $M_{12}^{q,\text{SM}}$ is  the SM contribution \cite{sanda}.

The time dependent  CP asymmetry $S_f$ decaying into the final state $f$
 is defined as \cite{Aushev:2010bq}
\begin{equation}
\mathcal{S}_{f}=\frac{2\text{Im}\lambda _{f}}{|\lambda_{f}|^2+1}
, \qquad
\lambda_{f}=\frac{q}{p} \bar \rho\ , \qquad 
\frac{q}{p}=\sqrt{\frac{M_{12}^{q*}-\frac{i}{2}\Gamma_{12}^{q*}}
{M_{12}^{q}-\frac{i}{2}\Gamma_{12}^{q}}}, \qquad
\bar \rho \equiv
\frac{\bar A(\bar B_q^0\to f)}{A(B_q^0\to f)}.
\label{sf}
\end{equation}
In the decay of  $B_d^0\to J/\psi  K_S$, the new physics parameters
$h_d$ and $\sigma_d$ appear in
\begin{equation}
\lambda_{J/\psi  K_S}=
-e^{-i\phi _d},\quad \phi _d
=2\beta_d+\text{arg}(1+h_de^{2i\sigma _d}),
\label{new}
\end{equation}
by putting  $|\bar \rho |=1$ and $q/p\simeq \sqrt{M_{12}^{q*}/M_{12}^q}$,
where the phase $\beta_d $ is given  in the SM.

The CKMfitter provided the allowed region of $h_d$ and $\sigma_d$,
where the central values are 
$h_d \simeq 0.3,\sigma _d\simeq 1.8 \ {\rm rad}
$\cite{CKMfitter,Ligeti}.

In the decay of  $B_s^0\to J/\psi\phi$, we have 
\begin{equation}
\lambda _{J/\psi \phi }=
e^{-i\phi _s},\qquad \phi _s
=-2\beta_s+\text{arg}(1+h_se^{2i\sigma _s}),
\end{equation}
where $\beta _s$ is given  in the SM.
Recently the  LHCb has presented the observed 
 CP-violating phase $\phi _s$ in $\bar B_s^0\to J/\psi \pi^+\pi^- $ 
decay~\cite{:2012dg}.
This result leads to
$
\phi _s=-0.019^{+0.173+0.04}_{-0.174-0.03} ~\text{rad},
$
which is consistent with the SM prediction
$
\phi _s^{J/\psi \phi ,SM}=-2\beta_s=-0.0363\pm 0.0017~\text{rad}
$ \cite{CKMfitter}.

Taking  account of these data, the CKMfitter has presented
the allowed values of $h_s$ and $\sigma _s$ ~\cite{CKMfitter,Ligeti}.
We take the central values
$h_s\simeq 0.1, \sigma _s \simeq 0.9 - 2.2 \ \text{rad}$
as a typical parameter set.

Since the $B_d^0\to J/\psi K_S$ process occurs at the tree level in  SM, 
 the CP-violating  asymmetry  originates   from  $M_{12}^d$.
Although the $B_d^0\to \phi K_S$ and $B_d^0\to\eta 'K^0$ decays 
are penguin dominant ones,
their asymmetries also come from  $M_{12}^d$.
Then,  asymmetries of
 $B_d^0\to J/\psi K_S$,  $B_d^0\to \phi K_S$ and 
$B_d^0\to \eta 'K^0$ are expected to be same magnitude in SM.

On the other hand, 
 if the squark flavor mixing  contributes to the decay 
 at the one-loop level, its magnitude could be  comparable 
to the SM penguin one
 in  $B_d^0\to \phi K_S$ and $B_d^0\to \eta 'K^0$, 
but it is tiny in $B_d^0\to J/\psi K_S$. 
Endo, Mishima and Yamaguchi proposed the possibility to
 find the SUSY contribution  in these asymmetries 
\cite{Endo:2004dc}.

The new physics contribute to the $b\to s\gamma$ process.
The observed $b\to s\gamma$ branching ratio (BR) is 
$(3.60\pm  0.23)\times 10^{-4}$ \cite{PDG}, on the other hand 
the SM prediction is given as $(3.15\pm  0.23)\times 10^{-4}$
at ${\cal{O}}(\alpha_s^2)$
\cite{Buras:1998raa,Misiak:2006zs}.
Therefore, the contribution of the new physics should be suppressed
compared with the experimental data.
The new physics is also constrained
 by the direct CP violation  
\begin{equation}
A_{\text{CP}}^{b\to s\gamma}
\equiv
\frac{\Gamma (\bar{B} \to X_{s}\gamma )
-\Gamma (B \to X_{\bar{s}}\gamma )}
{\Gamma (\bar{B} \to X_{s}\gamma )
+\Gamma (B \to X_{\bar{s}}\gamma )} .
\end{equation}
Since the SM prediction   $A_{\text{CP}}^{b\to s\gamma}\simeq 0.005$
is tiny \cite{Kagan:1998bh}, the new physics may appear in this CP asymmetry.
The present data $A_{\text{CP}}^{b\to s\gamma}=-0.008\pm 0.029$ \cite{PDG}
has large error bar, so the constraint of the new physics is not so severe.
However improved data will provide the crucial test for the new physics.
We also discuss the time dependent CP asymmetry of  $B_d \to K^* \gamma$.

\section{Squark flavor mixing in $B$ meson decays}

Let us consider the flavor structure of squarks
in order to estimate the CP-violating asymmetries of $B$ meson decays.
We take the most popular anzatz,  
a degenerate SUSY breaking mass spectrum for down-type squarks.
Then, in the super-CKM basis, we can parametrize 
 the soft scalar masses squared 
$M^2_{\tilde d_{LL}}$, $M^2_{\tilde d_{RR}}$, 
$M^2_{\tilde d_{LR}}$, and $M^2_{\tilde d_{RL}}$ for the down-type squarks.
For example,
\begin{align}
M^2_{\tilde d_{LR}}&=(M_{\tilde d_{RL}}^2)^\dagger =m_{\tilde q}^2
\begin{pmatrix}
(\delta _d^{LR})_{11} & (\delta _d^{LR})_{12} & (\delta _d^{LR})_{13} \\
(\delta _d^{LR})_{21} & (\delta _d^{LR})_{22} & (\delta _d^{LR})_{23} \\
(\delta _d^{LR})_{31} & (\delta _d^{LR})_{32} & (\delta _d^{LR})_{33}
\end{pmatrix},
\end{align}
where  $m_{\tilde q}$ is the average squark mass, and
$(\delta _d^{LR})_{ij}$ and $(\delta _d^{RL})_{ij}$ are  called as
the  mass insertion (MI) parameters.
The MI parameters are supposed to be much smaller than $1$.

The SUSY contribution by the gluino-squark box diagram to the dispersive part of the effective Hamiltonian for 
the $B_q$-$\bar B_q$ mixing is written as \cite{Hayakawa:2012ua,Gabbiani:1996hi,Altmannshofer:2009ne} 
\begin{align}
M_{12}^{q,SUSY}&=A_1^q\Big [A_2\left \{ (\delta _d^{LL})_{ij}^2+
(\delta _d^{RR})_{ij}^2\right \} +
A_3^q(\delta _d^{LL})_{ij}(\delta _d^{RR})_{ij} \nonumber \\
&+A_4^q\left \{ (\delta _d^{LR})_{ij}^2+(\delta _d^{RL})_{ij}^2\right \} +A_5^q(\delta _d^{LR})_{ij}(\delta _d^{RL})_{ij}\Big ],
\label{bbbarmixing}
\end{align}
where $A_i^q$ is a function of $x = m_{\tilde{g}}^2 / m_{\tilde{q}}^2$.

The squark flavor mixing can be tested in
 the CP-violating  asymmetries of $B$ meson. 
Let us present our framework.
The effective Hamiltonian for $\Delta B=1$ 
process is defined as 
\begin{equation}
H_{eff}=\frac{4G_F}{\sqrt{2}}\left [\sum _{q'=u,c}V_{q'b}V_{q's}^*\sum _{i=1,2}C_iO_i^{(q')}-V_{tb}V_{ts}^*
\sum _{i=3-6,7\gamma ,8G}\left (C_iO_i+\widetilde C_i\widetilde O_i\right )\right ],
\end{equation}
where $O_i$'s are the local operators \cite{Hayakawa:2012ua}.
The Wilson coefficient $C_i$ includes both SM contribution and gluino  one,
such as  $C_i=C_i^{\rm SM}+C_i^{\tilde g}$, where
$C_i^{\text{SM}}$ and $C_{7 \gamma}^{\tilde g}$ and $C_{8 G}^{\tilde g}$ are given in Ref.~\cite{Buchalla:1995vs,Endo:2004fx}.

The CP-violating  asymmetries $\mathcal{S}_f$ in Eq.~(\ref{sf}) are 
 calculated by using $\lambda_f$, which is given 
 for  $B_d^0\to \phi K_S$ and $B_d^0\to \eta 'K^0$ as follows:
\begin{align}
\lambda_{\phi K_S,\  \eta 'K^0}&=-e^{-i\phi _d}\frac{\displaystyle \sum _{i=3-6,7\gamma ,8G}
\left (C_i^\text{SM}\langle O_i \rangle+C_i^{\tilde g}\langle O_i \rangle+
\widetilde C_i^{\tilde g}\langle \widetilde O_i \rangle \right )}
{\displaystyle \sum _{i=3-6,7\gamma ,8G}
\left (C_i^{\text{SM}*}\langle O_i \rangle+C_i^{{\tilde g}*}
\langle O_i \rangle+\widetilde C_i^{{\tilde g}*}\langle\widetilde O_i \rangle \right )}~.
\label{asymBd}
\end{align}
It is noticed that  
$\langle\phi K_S|O_i|B_d^0\rangle=\langle\phi K_S|
\widetilde O_i|B_d^0\rangle $
and $\langle\eta' K^0|O_i|B_d^0\rangle=-\langle\eta' K^0|
\widetilde O_i|B_d^0\rangle$
because of  the parity of the final state.
We estimate each  hadronic matrix elements
by using the factorization relations in Ref.~\cite{Harnik:2002vs}. 

The $b\to s\gamma $ decay is a typical process to investigate the new physics.
We can discuss the direct CP violation  $A_{\text{CP}}^{b\to s\gamma}$
in the $b\to s\gamma $ decay, which  is given as \cite{Kagan:1998bh}:

\begin{equation}
\begin{split}
A_{\text{CP}}^{b \to s \gamma} 
&=
\frac{\alpha_s (m_b)}
{|C_{7\gamma}|^2}
\Big [
\frac{40}{81}
\text{Im} \small[ C_2 C_{7\gamma}^* \small]
-
\frac{8 z}{9} \small[v(z)+b(z, \delta)\small ]
\text{Im}\Big[\left (1+\frac{V_{us}^* V_{ub}}{V_{ts}^* V_{tb}}\right  )
C_2 C_{7\gamma}^*\Big] \nonumber \\
&
-\frac{4}{9}
\text{Im} \small[ C_{8G} C_{7\gamma}^* \small]
+
\frac{8z}{27}
b(z,\delta) \text{Im} \Big[ \left(1+\frac{V_{us}^* V_{ub}}{V_{ts}^* V_{tb}} \right ) C_2 C_{8G}^* \Big]
\Big ],
\end{split}
\end{equation}
where $v(z)$ and $b(z,\delta)$ are explicity given in  \cite{Kagan:1998bh}.

We also discuss the time dependent CP asymmetry $S_{K^* \gamma}$ of $B_d \to K^* \gamma$ decay, 
 which is given as \cite{Endo:2004fx} 
\begin{equation}
S_{K^* \gamma}
=
\frac{2 {\rm Im}(e^{2 i \phi_1}\tilde{C}_{7 \gamma}(m_b)/C_{7 \gamma}(m_b))}
{|\tilde{C}_{7 \gamma}(m_b)/C_{7 \gamma}(m_b)|^2 +1} .
\end{equation}

Let us set up the framework of our calculations.
Suppose that $\mu\tan\beta$ is at most ${\cal} O(1)$TeV.
Then, magnitudes of $(\delta _d^{LL})_{23}$ and $(\delta _d^{RR})_{23}$
are constrained by $M_{12}^s$ as seen in Eq.(\ref{bbbarmixing}).
Taking account of $h_s=0.1$ ,
  we obtain  $|(\delta _d^{LL})_{23}|\simeq |(\delta _d^{RR})_{23}|\simeq 0.02$
 in our  previous work \cite{Hayakawa:2012ua}.
Then,   these contributions to  $C_{7\gamma}^{\tilde g}$ and 
$C_{8G}^{\tilde g}$ are minor.
On the other hand,  $(\delta _d^{LR})_{23}$ and $(\delta _d^{RL})_{23}$
are severely  constrained by  $C_{7\gamma}^{\text{eff}}$ and 
$C_{8G}^{\text{eff}}$ independent of  $\mu\tan\beta$.
We show the constraint for   
$(\delta _d^{LR})_{23}$ and $(\delta _d^{RL})_{23}$ 
in our following calculations.
In our convenience,  we suppose 
$|(\delta _d^{LR})_{23}|=|(\delta _d^{RL})_{23}|$.
Then, we can   parametrize the MI parameters  as follows:
\begin{equation}
(\delta _d^{LR})_{23}=|(\delta _d^{LR})_{23}|e^{2i\theta _{23}^{LR}},
\qquad (\delta _d^{RL})_{23}=|(\delta _d^{LR})_{23}|e^{2i\theta _{23}^{RL}}.
\label{MILR}
\end{equation}

\section{Numerical results}

We show the numerical analyses of the CP violation
 in the $B$ mesons. 
In our following numerical calculations,
 we fix the squark mass and the gluino mass as
$m_{\tilde q}=1000~\text{GeV}$ and $m_{\tilde g}=1500~\text{GeV}$,
which are consistent with recent lower bound of these masses at LHC
\cite{Aad:2011ib}.

At first, we discuss  the $b\to s\gamma$ decay.
The observed $b\to s\gamma$ branching ratio is 
$(3.60\pm  0.23)\times 10^{-4}$ \cite{PDG}, on the other hand 
the SM prediction is given as $(3.15\pm  0.23)\times 10^{-4}$
at ${\cal{O}}(\alpha_s^2)$
\cite{Buras:1998raa,Misiak:2006zs}.
The branching ratio gives the constraint for the magnitude 
 of $(\delta _d^{LR})_{23}$. 
The direct CP violation of the $b\to s\gamma$ is also
useful to constraint  $(\delta _d^{LR})_{23}$. 

We show the  $|(\delta _d^{LR})_{23}|$ dependence of the branching ratio
taking accont of the constraint of $A_{\text{CP}}^{b \to s \gamma}$ in Figure 1,  
where the upper and lower bounds of the experimental data with $90\%$ C.L.
 are denoted red lines.
As the magnitude of    $(\delta _d^{LR})_{23}$ increases,
 the predicted  region of the branching ratio splits into the larger
 region and smaller one.
The excluded region around ${\rm BR}=3\times 10^{-4}$ is due to 
the constraint of $A_{\text{CP}}^{b \to s \gamma}$.
Then,  
 the predicted branching ratio becomes inconsistent with
 the experimental data at $|(\delta _d^{LR})_{23}|\geq 5.5\times 10^{-3}$.

In Figure 2,
we plot the allowed region 
of the  $\theta _{23}^{LR}-|(\delta _d^{LR})_{23}|$ plane
by putting  the experimental data at $90\%$ C.L. of
the branching ratio and the direct CP violation
$A_{\text{CP}}^{b \to s \gamma}$.
 The $|(\delta _d^{LR})_{23}|$ is cut at $5.5\times 10^{-3}$,
where   $\theta _{23}^{LR}$ is tuned around $\pi/2$.
Around $\pi/4$ and  $3\pi/4$,  $A_{\text{CP}}^{b \to s \gamma}$
give the severe constraint.
This CP-violating phase
  also contributes on the CP-violating asymmetry
 of the non-leptonic decays of  $B_d^0$ and $B_s^0$ mesons.

In addition to the direct CP violation of $b \to s \gamma$,
 we predicted the time dependent CP asymmetry $S_{K^* \gamma}$ of $B_d \to K^* \gamma$ decay in Figure 3.
The experimental upper and lower bounds with $90\%$ C.L. are denoted by the red lines
 and the case of $1 \sigma$ is denoted by the pink lines.
We find that the constraint from $S_{K^* \gamma}$ is not severe at present. 

Let us discuss ${\mathcal S}_{f}$, which is the measure
 of the CP-violating asymmetry, 
for  $B_d^0\to {J/\psi  K_S}, \ {\phi K_S}$ and ${\eta' K^0}$.
As discussed in Section 2, 
these   ${\mathcal S}_{f}$'s are predicted
 to be same ones in the SM.
On the other hand, 
 if the squark  flavor mixing contributes to the decay process 
 at the one-loop level, these asymmetries are different
 from among as seen in  Eq.(\ref{asymBd}).
We present 
the predicted region of
 the $\mathcal{S}_{\eta 'K^0}$-$\mathcal{S}_{\phi K_S}$ plane in Figure 4,
the black line denotes the SM prediction 
$\mathcal{S}_{J/\psi K_S}=\mathcal{S}_{\phi K_S}=\mathcal{S}_{\eta 'K}$,
where the observed value
 $\mathcal{S}_{J/\psi K_S}=0.671\pm 0.023$ is put.
 The experimental data is denoted by red lines at $90\%$ C.L. and
 we  fix  $|(\delta _d^{LR})_{23}|=10^{-4}$(orange)
and $10^{-3}$(blue) for typical values.
The reduction of the experimental error of
 $A_{\text{CP}}^{b \to s \gamma}$ will give us severe predictions for 
$\mathcal{S}_{\phi K_S}$ and $\mathcal{S}_{\eta 'K^0}$.

\begin{center}
\begin{figure}[t]
\begin{minipage}{.45\textwidth}
\begin{center}
\includegraphics[width=1\textwidth]{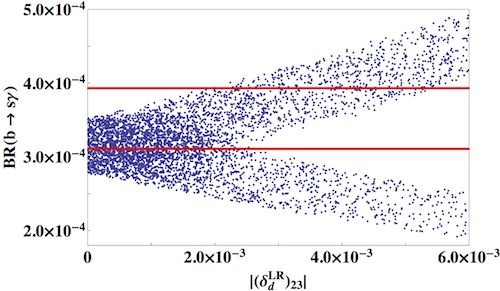}
\caption{The predicted branching ratio of  $b\to s\gamma$ 
versus $|(\delta _d^{LR})_{23}|$.}
\end{center}
\end{minipage}
~~~~~
\begin{minipage}{.45\textwidth}
\begin{center}
\includegraphics[width=1\textwidth]{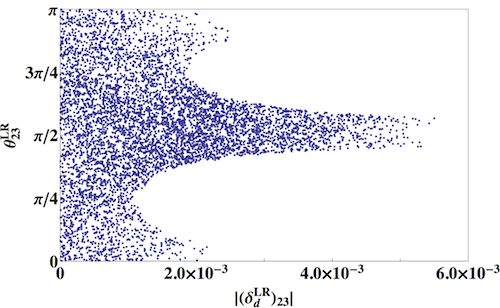}
\caption{The allowed region of $\theta _{23}^{LR}-|(\delta _d^{LR})_{23}|$ plane.}
\end{center}
\end{minipage}
\end{figure}
\end{center}
\begin{center}
\begin{figure}[h]
\begin{minipage}{.45\textwidth}
\begin{center}
\includegraphics[width=1\textwidth]{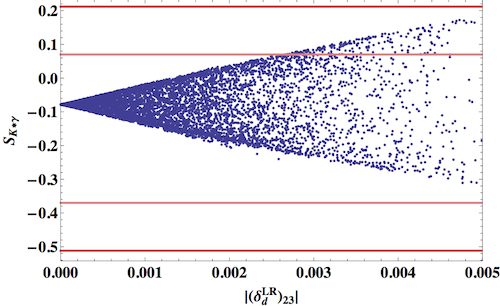}
\caption{The allowed region of $S_{K^* \gamma}$ - $|(\delta _d^{LR})_{23}|$plane.}
\end{center}
\end{minipage}
~~~~~
\begin{minipage}{.45\textwidth}
\begin{center}
\includegraphics[width=1\textwidth]{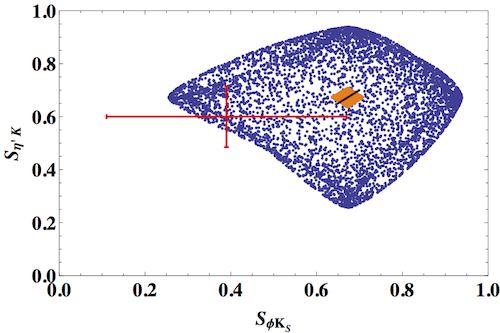}
\caption{The predicted region of $\mathcal{S}_{\eta 'K^0}$-$\mathcal{S}_{\phi K_S}$ plane.}
\end{center}
\end{minipage}
\end{figure}
\end{center}

\section{Conclusion}

We have discussed the contribution of the squark flavor mixing 
from  $(\delta _d^{LR})_{23}$ and $(\delta _d^{RL})_{23}$
on the direct CP violation of the $b\to s\gamma$ decay and
the CP-violating asymmetry in the non-leptonic decays of 
$B_d^0$ meson.
The magnitude of the  $|(\delta _d^{LR})_{23}|$ 
is constrained by the branching ratio of  $b\to s\gamma$ 
 with  the constraint of $A_{\text{CP}}^{b \to s \gamma}$.
 The predicted branching ratio becomes inconsistent with
 the experimental data at $|(\delta _d^{LR})_{23}|\geq 5.5\times 10^{-3}$.
We have obtained  the allowed region 
on the $\theta _{23}^{LR}$-$|(\delta _d^{LR})_{23}|$ plane.
 
Based on this result, we have predicted $\mathcal{S}_{f}$ of the $B_d^0$ and $B_s^0$ decays.
These CP-violating asymmetries could deviate from
the SM predictions.

 In the near future,
 the precise data of 
the direct CP violation and  CP-violating asymmetries
 in the non-leptonic decays of 
$B_d^0$ and $B_s^0$ mesons  give us the crucial  test
 for  our  framework of the squark flavor mixing.
 



\bibliographystyle{apsrev4-1}

%% file: Papers/yang.tex

%
%
%
%
%
%

\chapter[RG effects on the CEDM RG effects on the CEDM via CP violating four-Fermi operators (Yang)]{RG effects on the CEDM RG effects on the CEDM via CP violating four-Fermi operators}
\vspace{-2em}
\paragraph{M. J. S. Yang}
\paragraph{Abstract}
  In this study, the renormalization-group equations for the
  (flavor-conserving) CP-violating interaction are derived up to the
  dimension six, including all the four-quark
  operators, at one-loop level. We apply them to the models with the
  neutral scalar boson  that have CP-violating Yukawa
 interactions with quarks, and discuss the neutron electric dipole moment in
  this models.

\section{Introduction}

The electric dipole moment (EDM) for neutrons is sensitive to CP
violation in physics beyond the standard model (SM). This is because, while the CP phase in the
Cabibbo-Kobayashi-Maskawa (CKM) matrix is $O(1)$, the CKM contribution
to the neutron EDM is too much suppressed \cite{Shabalin:1978rs} to be
observed in near future. (The recent evaluation of the CKM contribution
to the neutron EDM is given in Refs.~\cite{Mannel:2012qk}.)  

The (flavor-conserving) CP-violating effective operators at parton
level up to the dimension six are the QCD theta term, the EDMs and the
chromoelectric dipole moments (CEDMs) of quarks, the Weinberg's
three-gluon operator~\cite{Weinberg:1989dx} and the four-quark
operators. In the evaluation of the neutron EDM, the CP-violating
four-quark operators tend to be ignored since the four-light quark
operators suffer from chiral suppression in many models. However, the
four-quark operators including heavier ones, such as bottom/top
quarks, may give sizable contributions to the neutron EDM.  The EDMs,
CEDMs, and the three-gluon operator are radiatively generated from
the four-quark operators by integrating out heavy quarks.

In the multi-Higgs models, the Barr-Zee diagrams are known to give the
sizable contribution to the neutron EDM \cite{Barr:1990vd}. In the
Barr-Zee diagrams, the heavy-quark loops are connected to light-quark
external lines by the neutral scalar boson exchange so that the CEDMs 
for light quarks are generated at two-loop level at $O(\alpha_s)$. However, 
it is unclear which renormalization scale should be chosen for
$\alpha_s$. 

In this study, in order to answer those questions, we derive the
renormalization-group equations (RGEs) for the Wilson coefficients for
the CP-violating effective operators up to the dimension six at
one-loop level, including operator mixing \cite{Hisano:2012cc}.  The RGEs for the EDMs and
CEDMs for quarks and the three-gluon operator have been derived in
Ref.~\cite{Shifman:1976de, Dai:1989yh, Boyd:1990bx}. The next-leading order
corrections to them are also partially included \cite{Degrassi:2005zd}. We
include the four-quark operators in the calculation at the leading
order. Using the derived RGEs, we evaluate the EDMs, and CEDMs for 
light quarks and the three-gluon operators induced by the
neutral scalar boson exchange including the QCD correction. 

This proceeding is organized as follows. In next section, we review the
neutron EDM evaluation from the parton-level effective Lagrangian at
the hadron scale. In Section 3, we derive RGEs for the Wilson
coefficients for the CP-violating effective operators up to the
dimension six at one-loop level. In Section 4, we show the effect of
the running $\alpha_s$ on the evaluation of the Wilson coefficients,
assuming the neutral scalar boson  exchange induces the CP-violating
effective operators. Section 5 is devoted to conclusion.

\section{Neutron EDMs}

First, we review about evaluations of the neutron EDM from the low-energy
effective Lagrangian at parton level. The CP-violating interaction at
parton level around the hadron scale ($\mu_H^{}=1$~GeV) is given by,
\begin{align}
 {\cal L}_{\rm CPV}
 =&~ \theta \frac{\alpha_s}{8\pi}G^A_{\mu\nu}\widetilde{G}^{A\mu\nu} 
  -\frac{i}{2}\sum_{q=u,d,s}d_q\, \overline{q} ~F_{\mu\nu}\sigma^{\mu\nu}~ \gamma_5q \nonumber \\
-&~\frac{i}{2}\sum_{q=u,d,s}\tilde{d}_q\, \overline{q}g_s ~G^A_{\mu\nu}\sigma^{\mu\nu}T^A~ \gamma_5q
+\frac{1}{3}w f_{ABC}G^{A}_{\mu\nu}\tilde{G}^{B\nu\lambda}
G^{C\mu}_\lambda.
\label{Lagrangian}
\end{align}
Here, $F_{\mu\nu}$ and $G^A_{\mu\nu} (A=1$--$8)$ are the electromagnetic and
gluon field strength tensors, $g_s$ is the strong coupling constant
($\alpha_s=g_s^2/4\pi)$ , and $\tilde{G}^A_{\mu\nu}\equiv
\frac{1}{2}\epsilon_{\mu\nu\rho\sigma}G^{A\rho\sigma}$ with
$\sigma^{\mu\nu}=\frac{i}2[\gamma^\mu,\gamma^\nu]$ and $\epsilon^{0123}=+1$. 
The matrix $T^A$ denotes the generators in the SU(3)$_C$ algebra, 
and $f^{ABC}$ is the structure constant. The first, second, third and forth terms in
Eq.~(\ref{Lagrangian}) are called the QCD $\theta$ term, the EDM and the CEDM 
for quarks, and the three-gluon operator, respectively. In Eq.~(\ref{Lagrangian}), we
ignore the CP-violating four-quark operators, since their coefficients
are often proportional to the light quark masses in typical models, as
mentioned in Introduction.

The neutron EDM is evaluated from various methods. The evaluation in term of the QCD sum rules is more
systematic than the others, at least for the contributions from the
QCD theta term, and the quark EDMs and CEDMs to the neutron EDM
\cite{Pospelov:2005pr, Pospelov:2000bw, Pospelov:1999mv, Pospelov:1999ha}.
The recent evaluation of the neutron EDM with the QCD sum rules is given by \cite{Hisano:2012sc} 
\begin{align}
 d_n \simeq 2.9\times 10^{-17}\bar{\theta}~[e~{\rm cm}]+0.32d_d -0.08d_u +
e(+0.12\tilde{d}_d-0.12\tilde{d}_u-0.006\tilde{d}_s)~.
\label{sumrule}
\end{align}
In the evaluation, the recent QCD lattice result is used for the
low-energy constant $\lambda_n$, which is defined by $ \langle 0 |
\eta_n(x)|N(\vec{p},s)\rangle = \lambda_n u_n(\vec{p}, s)$ with
$\eta_n(x)$ the neutron-interpolating field.
If a value of $\lambda_n$ evaluated with the QCD sum rules is used,
the neutron EDM is enhanced by about five times compared with Eq.~(\ref{sumrule}). 

The contribution from the three-gluon operator might be comparable to
the quark EDMs and CEDMs. The quark EDMs and CEDMs are proportional
to the quark masses, while the three-gluon operator does not need to
suffer from chirality suppression. However, the size of the contribution
from the three-gluon operator depends on the methods of
the evaluation. In Ref.~\cite{Demir:2002gg} the authors compare the
several evaluations and propose
\begin{align}
d_n(w)\sim& ({\rm 10-30})\, {\rm MeV} \times  e w \,.
\label{weinbergopn}
\end{align}

\section{Operator Bases and Anomalous Dimension Matrix} 

We would like to introduce heavy quarks in the low-energy effective theory 
and evaluate their contributions to the neutron EDM. 
In this section, we show the one-loop RGEs for the
Wilson coefficients for the CP-violating effective operators up to the
dimension six, including heavy quarks.

First, we define the operator bases for the RGE analysis.  The
flavor-conserving effective operators for the CP violation
in QCD are given up to the dimension six as
\begin{align}
{\cal L}_\text{CPV} =& 
\sum_{i=1,2,4,5} \sum_q C_i^q(\mu) {\mathcal O}_i^q(\mu)
+C_3(\mu) {\mathcal O}_3(\mu) \nonumber \\
&+\sum_{i=1,2} \sum_{q'\ne q} \widetilde{C}_i^{q'q}(\mu) \widetilde{\mathcal O}_i^{q'q}(\mu)
+\frac12\sum_{i=3,4} \sum_{q'\ne q} \widetilde{C}_i^{q'q}(\mu) \widetilde{\mathcal O}_i^{q'q}(\mu)\,,
\label{effop}
\end{align}
where the sum of $q$ runs not only light quarks but also heavy ones,  and
we ignore the QCD theta term since it is irrelevant to our discussion here. 
The effective operators are defined as 
\begin{align}
{\mathcal O}_1^q
=& -\frac{i}2 m_q \bar{q} \,eQ_q (F\cdot 
\sigma) \gamma_5q\, , 
~~~ 
{\mathcal O}_2^q
= -\frac{i}2 m_q \overline{q} \,g_s (G\cdot\sigma) \gamma_5q \, , \nonumber\\
{\mathcal O}_3
=& -\frac16 g_s f^{ABC} \epsilon^{\mu\nu\rho\sigma} 
G^A_{\mu\lambda} {G^B}_{\nu}^{~\lambda} G^C_{\rho\sigma}\,,
\end{align}
and 
\begin{align}
\begin{array}{ll}
{\mathcal O}_4^q
=\, \overline{q_\alpha} q_\alpha \overline{q_\beta} \,i\gamma_5 q_\beta\,, 
&
{\mathcal O}_5^q
=\, \overline{q_\alpha} \sigma^{\mu\nu} q_\alpha \overline{q_\beta} \,i\sigma_{\mu\nu}\gamma_5 q_\beta\,, \nonumber\\
\widetilde{\mathcal O}_1^{q'q}
=\, \overline{q'_\alpha} q'_\alpha \overline{q_\beta} \,i\gamma_5 q_\beta\,, 
&
\widetilde{\mathcal O}_2^{q'q}
=\, \overline{q'_\alpha} q'_\beta \overline{q_\beta} \,i\gamma_5 q_\alpha\,, \nonumber\\
\widetilde{\mathcal O}_3^{q'q}
=\, \overline{q'_\alpha} \sigma^{\mu\nu} q'_\alpha \overline{q_\beta} \,i\sigma_{\mu\nu}\gamma_5 q_\beta\,, 
&
\widetilde{\mathcal O}_4^{q'q}
=\, \overline{q'_\alpha} \sigma^{\mu\nu} q'_\beta \overline{q_\beta} \,i\sigma_{\mu\nu}\gamma_5 q_\alpha \,.
\label{4foperator}
\end{array}
\end{align}
Here, $m_q$ are masses for quark $q$. 
In Eq.~(\ref{4foperator}) we explicitly show the color indices, $\alpha$ and $\beta$. 
A factor of $1/2$ appears in front of the fourth term of Eq.~(\ref{effop}), 
since the term is symmetric under the exchange of $q'$ and $q$. 
The Wilson coefficients in Eq.~(\ref{effop}) are related to the parameters in Eq.~(\ref{Lagrangian}) as 
\begin{align}
d_q =&\, { m_{q}} \,e Q_q\, C^q_1(\mu_H^{})\, , ~~~
\tilde{d}_q =\, {m_{q}}\, C^q_2(\mu_H^{})\, , ~~~
w= -\frac12 g_s\, C_3(\mu_H^{})\, .
\end{align}

The RGEs for the Wilson coefficients of these operators 
and the anomalous dimension matrix are given as follows,
\begin{align}
\mu \frac{\partial}{\partial\mu}{\bf C}= {\bf C}{\bf \Gamma},
~~~~~~~
{\bf \Gamma} = \begin{bmatrix}
\frac{\alpha_s}{4\pi} \gamma_s & {\bf 0}                        & {\bf 0} \\
\frac1{(4\pi)^2} \gamma_{sf}   & \frac{\alpha_s}{4\pi} \gamma_f & {\bf 0} \\
\frac1{(4\pi)^2} \gamma'_{sf}  & {\bf 0}                        & \frac{\alpha_s}{4\pi} \gamma'_f
\end{bmatrix}. \label{Eq:Gamma}
\end{align}

Here, the Wilson coefficients are written in a column vector as 
\begin{align}
{\bf C}=(C_1^q,C_2^q,C_3,C_4^q,C_5^q,
\widetilde{C}_1^{q'q},\widetilde{C}_2^{q'q},
\widetilde{C}_1^{qq'},\widetilde{C}_2^{qq'},
\widetilde{C}_3^{q'q},\widetilde{C}_4^{q'q}).
\end{align}
and the explicit forms of the components in the matrix are in the Ref. \cite{Hisano:2012cc}.

\section{Neutral Scalar Boson Exchange}

In multi-Higgs models, a color-singlet neutral scalar boson $\phi$ may have
the CP-violating Yukawa coupling with quarks.  If the Yukawa
interaction violates the CP invariance, the CP-violating four-quark
operators are induced at tree level, after integrating the neutral scalar boson out, as
\begin{align}
C_4^q = \sqrt2G_F 
\frac{m^2_q}{m_\phi^2}f_S^qf_P^q\,, ~~~
\widetilde{C}_1^{q'q} = \sqrt2G_F 
\frac{m_qm_{q'}}{m_\phi^2}f_S^qf_P^{q'}\,, ~~~
\widetilde{C}_1^{qq'} = \sqrt2G_F 
\frac{m_qm_{q'}}{m_\phi^2} f_S^{q'}f_P^q\,,
\end{align}
where we assume that $\phi$  is heavier than heavy quarks ($m_\phi\gg m_q,m_{q'}$).
Here, $f_S^q$ and $f_P^q$ are the CP-even and odd Yukawa coupling
constants, respectively,  defined as
\begin{align}
{\mathcal L}_\phi 
&= 2^{1/4}G_F^{1/2} m_q
\overline{q_\alpha} (f_S^q + i\,f_P^q \gamma_5) q_\alpha \phi,
\label{yukawa}
\end{align}
where $\phi$ is a (CP even) real scalar field, and 
$G_F$ is the Fermi constant. We parametrize the
Yukawa coupling constants as they are proportional to the quark masses. 
For the SM Higgs boson, the Yukawa 
coupling constants are of $f_S^q=1$ and $f_P^q=0$.

It is known that, in these models, the EDMs and CEDMs for light
quarks are generated by the Barr-Zee diagrams at two-loop level, and
the three-gluon operator is also induced by the heavy-quark loops at
two-loop level.  Here, we compare values of the EDM and CEDM operators 
for down quark and the three-gluon operators including and not
including the renormalization group evolution of the strong coupling constant. We assume that
the Yukawa coupling constants for down and bottom quarks with
$\phi$ are nonzero in Eq.~(\ref{yukawa}) and then
\begin{align}
\widetilde{C}_1^{bd}(m_\phi)\ne0, \quad C_4^b(m_\phi)\ne0\,, ~~~
C_1^b(m_\phi)=C_2^b(m_\phi)= - \frac{3}{16\pi^2}C_4^b(m_\phi) \,.
\end{align}
The last assumption comes from the matching condition between
 the explicit one-loop calculation and the result of one-loop RGEs. 
 This initial condition is interpreted as the short-distance contribution 
 in which the loop momentum is around $m_\phi$. 

In Fig.~1 the CEDM for down quark, $\tilde{d}_d$, (a) 
and the coefficient of the three-gluon operators, $w$, (b) 
at the hadron scale ($\mu=\mu_H^{}=1$~GeV) 
are shown as functions of $m_\phi^{}$ with
$f_S^d=f_P^d=1$ and $f_S^b=f_P^b=1$.  
Here, we ignore the contributions from the top quark, and other short-distance effects.
If the scalar mass $m_\phi^{}$ is larger than the top quark mass ($m_\phi^{}> m_{t}$), 
the RGEs are solved using $\beta_{0}$ with $n_{f} = 6$, or if not, with $n_{f} = 5$.
When bottom quark is integrated out, the Wilson coefficient of Weinberg operator emerges.
Then the RGEs are solved using $\beta_{0}$ with $n_{f} = 4$ to the scale $\mu = m_{c}$, and 
with $n_{f} = 3$ to the scale $\mu = 1$ GeV. 
We use $m_d(\mu_H)=9$~MeV,
$m_c(m_c)=1.27$~GeV, $m_b(m_b)=4.25$~GeV, $m_t(m_t)=172.9$~GeV, and $\alpha_s(m_Z^{})=0.12$.
For the coefficient $w$, we multiply 10~MeV in the figure, which is a
factor in Eq.~(\ref{weinbergopn}), so that one may estimate the
contribution to the neutron EDM. It is from
Eqs.~(\ref{sumrule},\ref{weinbergopn}) found that the three-gluon
operator might be comparable to the CEDM when $f_{S/P}^d\sim f_{S/P}^b$.

In Fig.~2 the ratios of the CEDM for down quark (a) and the three-gluon
operator (b) at $\mu = m_b$ between including the running effect of
$\alpha_s$ and not including it (using the constant coupling
$\alpha_s=\alpha_s(m_Z^{})$), are shown as functions of $m_\phi^{}$. It is found
that the  running coupling $\alpha_s(\mu)$ changes the CEDM
by about 20\% while the three-gluon operator is changed by at most 10
\%. This results come from inclusions of the four-quark operators to
the RGEs for the Wilson coefficients.

\begin{figure}
\begin{center}
\begin{tabular}{cc}
   \includegraphics[width=9cm,bb = 0 0 400 300]{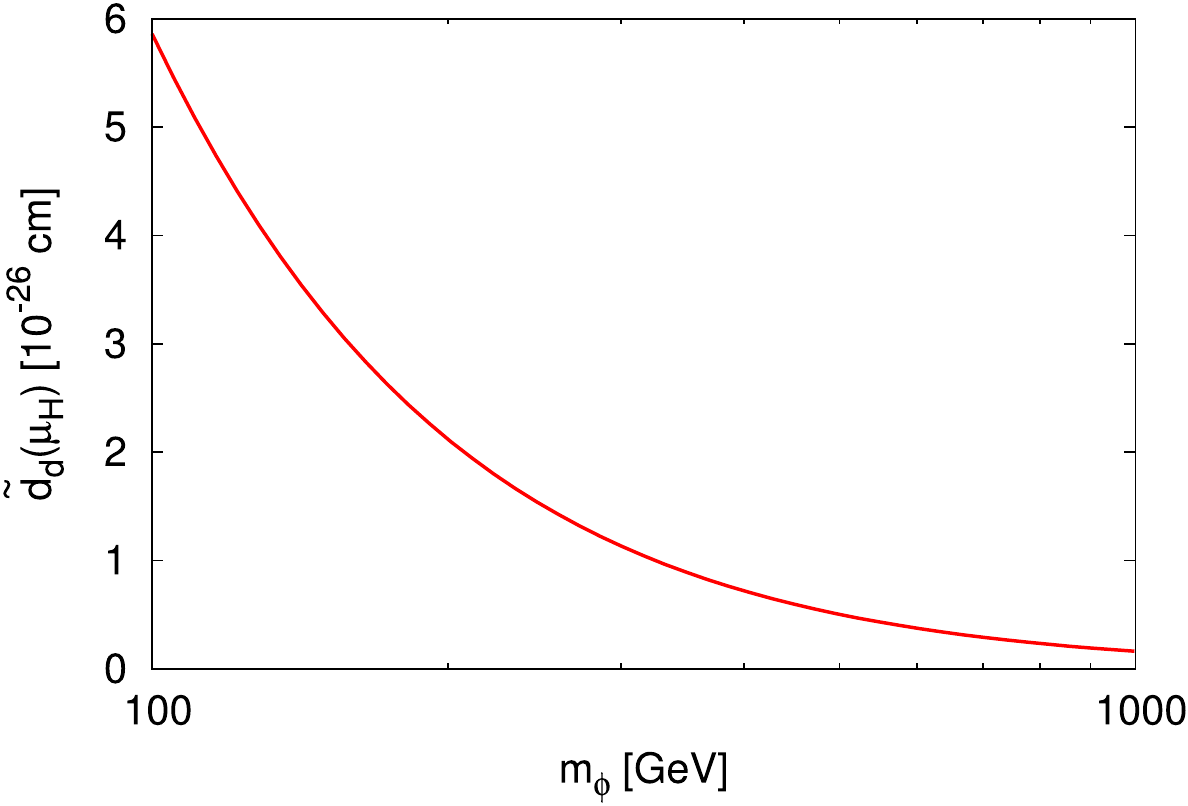} &
   \includegraphics[width=9cm,bb = 0 0 400 300]{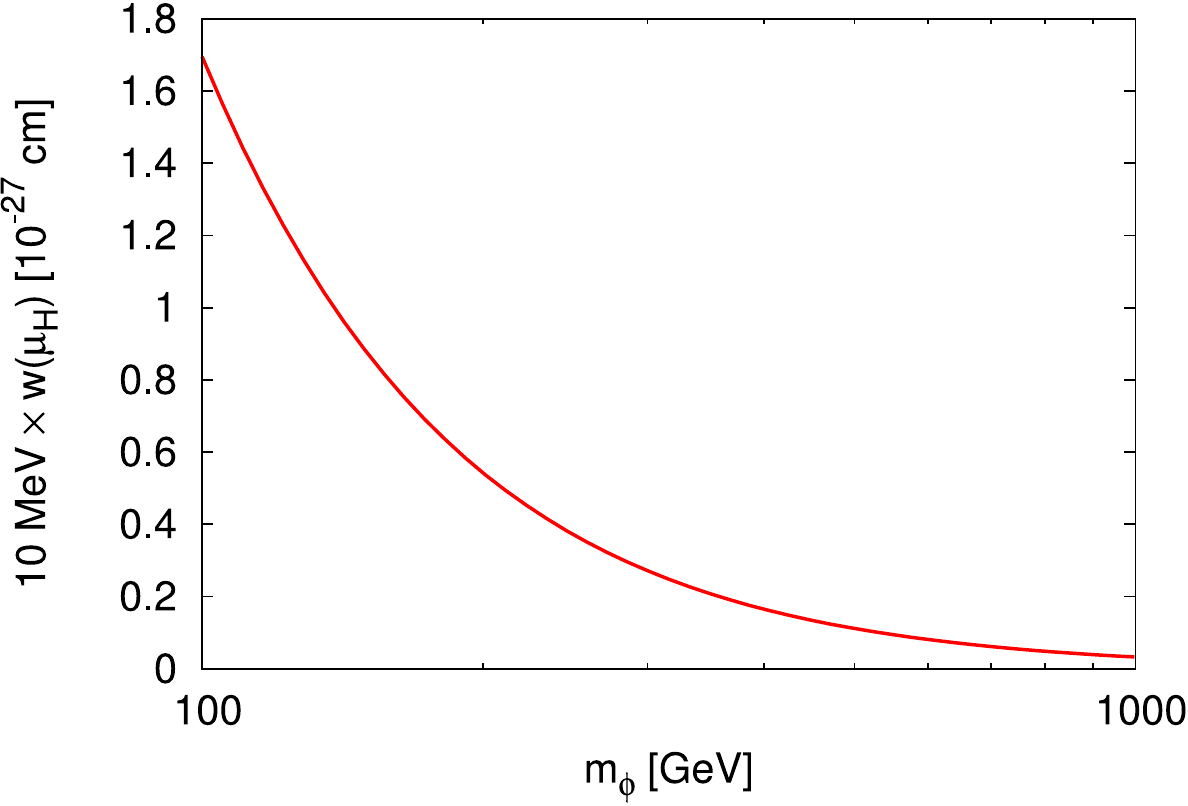} \\
   (a) & (b) 
\end{tabular} 
\caption{(a)  CEDM for down quark, $\tilde{d}_d$, 
  and  (b) coefficient of three-gluon operator, $w$, 
at hadron scale as functions of $m_\phi^{}$.}
\end{center}
\end{figure}

\begin{figure}
\begin{center}
\begin{tabular}{cc}
   \includegraphics[width=9cm,bb = 0 0 400 300]{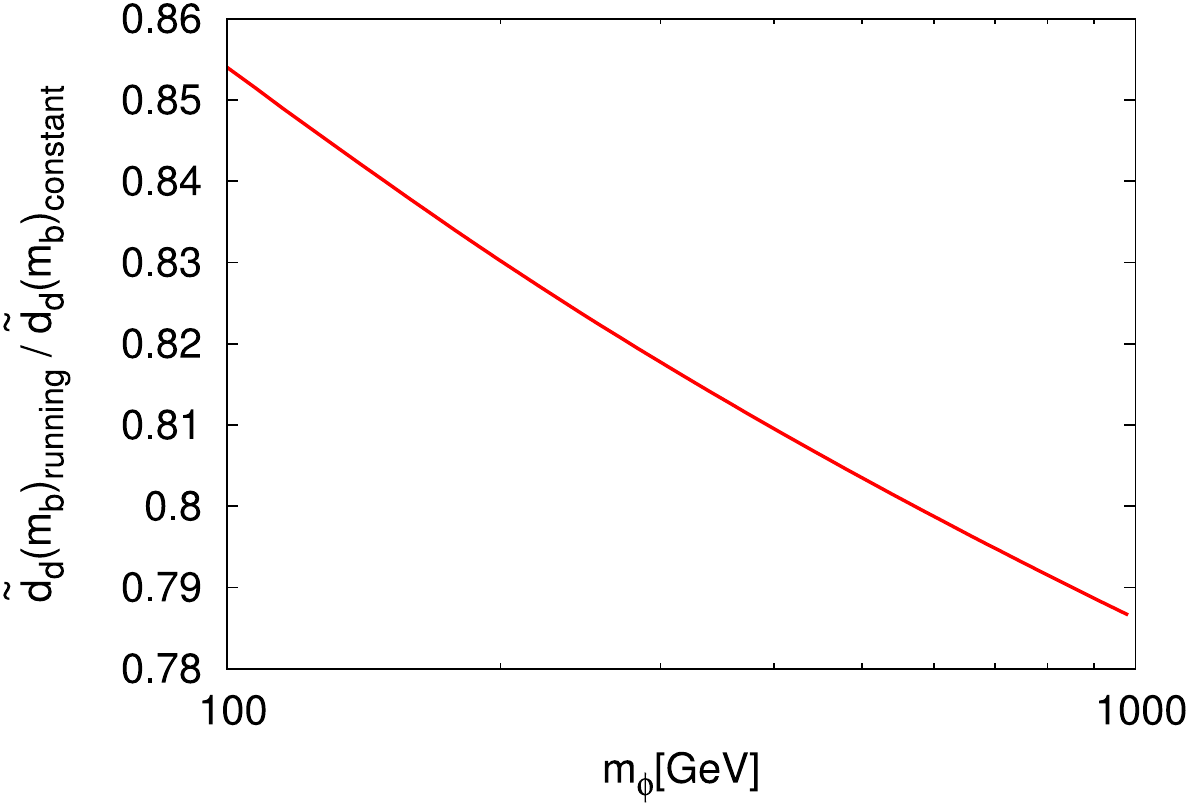} &
   \includegraphics[width=9cm,bb = 0 0 400 300]{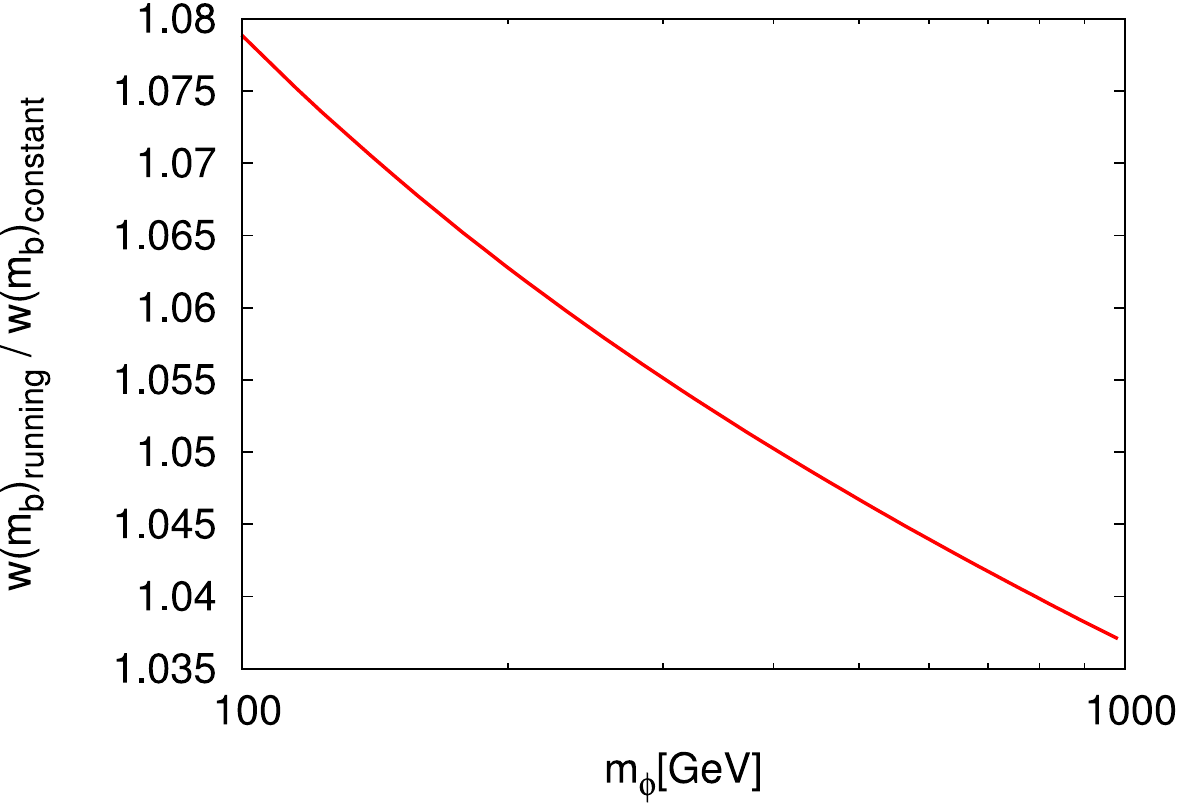} \\
   (a) & (b) 
\end{tabular} 
\caption{(a) : Ratio of the CEDM for down quark, $\tilde{d}_d$, at $\mu=m_b$
  between including and not including running of the strong coupling
  constant, as a function of $m_\phi^{}$. (b) The same ratio for coefficient of  
  three-gluon operator, $w$.}
\end{center}
\end{figure}

\section{Conclusion}

In this study, we have derived the renormalization-group equations for
the CP-violating interaction including the quark EDMs and  CEDMs and the
Weinberg's three-gluon operator as well as all the flavor-conserving
four fermion operators.  

Assuming the CP-violating Yukawa interactions for the neutral scalar
bosons, it is known that the CEDMs for light quarks are generated
from the diagrams with heavy-quark loops, called as the Barr-Zee
diagrams. We show that when the neutral scalar boson is much heavier than
heavy quarks, the Barr-Zee diagrams are systematically evaluated
with the RGEs of the CP-violating interaction.  We also show that 
the running effect of the strong coupling constant gives corrections to
the contribution with more than 20 \% compared with assuming the
constant coupling. 
The uncertainties in the calculation of the neutron EDM have been estimated in the literature~\cite{Leinweber:1995fn}. It gives about 50 \% error for the QCD sum rule, while 40 \% error for the low-energy constant evaluated from the lattice QCD calculation.
Therefore, hadronic uncertainties would overcome the QCD corrections from the renormalization group evolution 
at this moment. 
We hope that the lattice QCD simulation will improve and reduce uncertainties significantly~\cite{Aoki:2008gv,Shintani:2008nt,Shintani:2006xr,Berruto:2005hg,Shintani:2005xg}. 
%

\section*{Acknowledgments}

This work is supported in part by JSPS Research Fellowships for Young Scientists.

\bibliography{yang}
\bibliographystyle{apsrev4-1}